%% file: phd_thesis.tex
\documentclass[10pt]{book}
\usepackage[T1]{fontenc}
\usepackage[sc]{mathpazo}
\usepackage[english]{babel}
\usepackage{graphics}
\usepackage[pdftex]{graphicx}
\usepackage{amsmath}
\usepackage{amssymb}
\usepackage{wrapfig}
\usepackage{epsfig}
\usepackage{epic}
\usepackage[small,font=normalsize,bf,up]{caption}

\usepackage{fancyhdr}
\usepackage{latexsym}
\usepackage{lscape}
\usepackage{rotating}
\usepackage[a4paper, twoside,
  left=4cm, right=3cm, bottom=3cm, top=3cm, marginparwidth=2cm]{geometry}
\usepackage[usenames, pdftex, dvipsnames]{color}
\usepackage{framed}
\usepackage{units}
\usepackage{multirow}
\usepackage{tabularx}
\usepackage[Lenny]{fncychap}
\ChNameVar{\fontsize{18}{20}\usefont{T1}{ugm}{m}{n}\selectfont\sc} 
\ChNumVar{\fontsize{70}{72}\usefont{T1}{ugm}{m}{n}\selectfont}
\ChNameVar{\fontsize{18}{20}\selectfont\sc} 
\usepackage{lettrine}
\usepackage{makeidx}
\usepackage{tocbibind}

\usepackage{hyphenat}
\usepackage{array,booktabs}
\usepackage[pdftex=true, colorlinks=true, hyperindex, citecolor=blue]{hyperref}
\usepackage{multicol}
\usepackage{verse}

\fancypagestyle{plain}{
  \fancyhead{}
  
  \fancyfoot{}
  
  \fancyfoot[OR,EL]{\bfseries \thepage}
}

\fancypagestyle{intro}{
  \fancyfoot{}
  
  \fancyfoot[OR,EL]{\bfseries \thepage}
  \fancyhead{}
  
  \fancyhead[OR]{\bfseries \leftmark}
  \fancyhead[EL]{\bfseries \rightmark}
}

\fancypagestyle{myfancy}{
\fancyfoot{}

\fancyfoot[OR,EL]{\bfseries \thepage}
\fancyhead{}

\fancyhead[OR]{\bfseries \nouppercase\leftmark}
\fancyhead[EL]{\bfseries \nouppercase\rightmark}
}

\newenvironment{shadefundtheory}[0]
{ 
\definecolor{shadecolor}{rgb}{0.35,1.0,1.0} 
\begin{shaded}
}
{\end{shaded}}

\newenvironment{shadefundnumber}[0]
{ 
\definecolor{shadecolor}{rgb}{1.0,0.8,0.0} 
\begin{shaded}
}
{\end{shaded}}

\newenvironment{shademinornumber}[0]
{ 
\definecolor{shadecolor}{rgb}{0.85,1.0,1.0} 
\begin{shaded}
}
{\end{shaded}}

\DeclareMathOperator{\Ree}{Re}
\DeclareMathOperator{\Imm}{Im}
\DeclareMathOperator{\tr}{Tr}
\DeclareMathOperator{\de}{d}

\DeclareMathOperator{\sinc}{sinc}
\DeclareMathOperator{\diag}{diag}

\newcommand{\imag}{\imath}
\newcommand{\beq}{\begin{equation}}
\newcommand{\eeq}{\end{equation}}
\newcommand{\beqnn}{\begin{equation*}}
\newcommand{\eeqnn}{\end{equation*}}
\newcommand{\besplt}{\begin{split}}
\newcommand{\esplt}{\end{split}}
\newcommand{\betab}{\begin{table}}
\newcommand{\eetab}{\end{table}}

\newcommand{\accPSDunit}{\unitfrac{m}{s^{2}\sqrt{Hz}}}
\newcommand{\secminsqunit}{\unitfrac{1}{s^{2}}}
\newcommand{\metresunit}{\unit{m}}
\newcommand{\secminoneunit}{\unit{s^{-1}}}

\newcommand{\vc}[1]{{\boldsymbol{#1}}}

\title{LTP and LISA\\
or: ``look how far we have to go just to please Herr Einstein!''}
\author{Michele Armano}

\makeindex

\makeatletter
\renewcommand{\@tocrmarg}{2cm}
\renewcommand{\@pnumwidth}{1.5cm}
\makeatother

\begin{document}
\frontmatter
\pagenumbering{Roman}

\pagestyle{empty}
\include{chapters/frontpage}
\cleardoublepage

\include{chapters/citation}
\cleardoublepage
\pagestyle{intro}

\pagestyle{myfancy}
\renewcommand{\chaptermark}[1]{\markboth{#1}{}}
\renewcommand{\sectionmark}[1]{\markright{\thesection\ #1}}

\include{chapters/introduction}
\include{chapters/fundconstants}
\include{chapters/acronyms}
\tableofcontents
\listoffigures
\listoftables
\newcounter{endofintro}
\setcounter{endofintro}{\value{page}}
\ifthenelse{\isodd{\value{page}}}{%
\addtocounter{endofintro}{2}}{%
\addtocounter{endofintro}{3}
}
\mainmatter

\include{chapters/refsys}

\include{chapters/ltp2}
\include{chapters/noise}

\include{chapters/experiment}

\include{chapters/g-phys-with-ltp}
\appendix
\include{chapters/gwtheory}

\include{chapters/geodesics}


\vfill
\backmatter
\pagenumbering{Roman}
\setcounter{page}{\value{endofintro}}

\include{chapters/conclusions}
\include{chapters/colophon}
\include{chapters/homage}

%

\bibliography{phd_thesis_bibliography,lisapfdesign}
\bibliographystyle{utphys}




\end{document}

%% file: chapters/frontpage.tex
\begin{center}
{\sc University of the Studies of Insubria\\
Seat of Como}

\includegraphics[width=5cm]{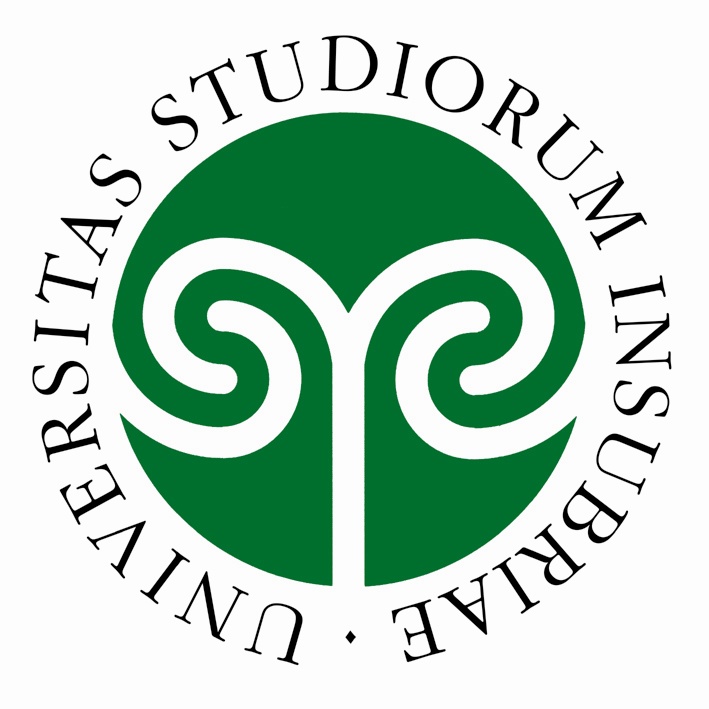}

{\large The LTP Experiment on LISA Pathfinder:

Operational Definition of TT Gauge in Space.}
\vspace{.5 cm}

{\sc D i s s e r t a t i o n}

to Partial Fulfillment of the Work\\
for the Academic Degree of Philosophy Doctor in Physics

{\sc S u b m i t t e d}

\end{center}

\begin{flushright}
by the candidate:

{\it\bf Michele Armano}

Matricula: R00101\\
Cycle: XVIII
\end{flushright}

\begin{flushleft}
Internal Tutor:\\
{\it Prof. Francesco Haardt}\\
Professor at University of Studies of Insubria

External Tutor:\\
{\it Prof. Stefano Vitale}\\
Professor at University of the Studies of Trento
\end{flushleft}

\begin{center}
In Collaboration with the University of Trento

\includegraphics[width=3cm]{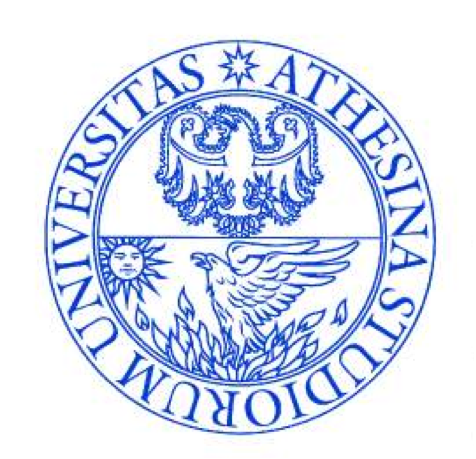}

\vfill
Academic Year 2004-2005

{\sc Como},
Thursday, September 26th , 2006 A.D.
\end{center}

%% file: chapters/citation.tex
\vspace*{2cm}\hspace*{5cm}\begin{minipage}[b]{0.6\linewidth}

\begin{verse}
``Ella sen va notando lenta lenta;\\
rota e discende, ma non me n'accorgo\\
se non che al viso e di sotto mi venta.''
\footnote{\cite{Alighieri:1982kx} Inferno, Canto XVII, vv 115-117, \cite{Ricci:2005fk}.}
\end{verse}

\vspace*{1cm}
``Addo etiam, quod satis absurdum videretur, continenti sive locanti motum adscribi, et non potius contento et localto,
quod est Terra. Cum denique manifestum sit, errantia sydera propinquiora fieri Terrae ac remotiora,
erit tum etiam, qui circa medium, quod volunt esse centrum Terrae, a medio quoque et ad ipsum
unius corporis motus. Oportet igitur motum, qui circa medium est, generalis accipere, ac satis esse,
dum unusquisque motus sui ipsius medio incumbat.''
\footnote{\cite{Copernico:1543uq} Cap. VIII}

\vspace*{1cm}
``Die allgemeinen Naturgesetze sind durch Geichungen auszudr\"uchen, die f\"ur alle Koordinatensysteme
gelten, d.h. die beliebingen Substitutionen gegen\"uber kovariant (allgemeinen kovariant) sind.''
\footnote{\cite{Einstein:1916kx}, A.3., \cite{einsteinprinc}}

\vspace*{1cm}
``Conditions that are observed in the universe must allow the observer to exist.''
\footnote{\cite{Merriam-Webster:2003uq}, Weak Anthropic Principle.}

\vspace*{2cm}
{\it This thesis is dedicated to my parents, Paola and Mario,
and to my brothers, Lorenzo, Emanuele and Marcello.}

\vspace*{5cm}
\vfill
\end{minipage}

%% file: chapters/introduction.tex
\chapter{Introduction and structure}


\lettrine[lines=4]{T}{his thesis}
addresses the problem of interferometer-based gravitational wave (GW) detection in space.
The problem of detecting GW and decoupling them from the static gravitational background
is an intricate one and can be viewed at least as a three-fold issue:
\begin{enumerate}
\item it implies a careful definition of a {\bf reference system}. It is necessary to build a set
of clocks and rulers in space to unequivocally measure radiative space-time variation of
the Riemann tensor embedding the metric;
\item it demands the use of a {\bf detector}. Pairs of particles in free-fall are the only reliable
probe in this case, and then it is the ability of defining {\bf free-fall} and detecting {\bf residual acceleration} which need to be discussed carefully;
\item it calls for detailed knowledge of {\bf noise} versus sensitivity, not to miss the wave
signal or mistake noise for a signal.
\end{enumerate}

The European Space Agency (ESA) and the National Aeronautics and Space Administration (NASA) are planning
the Laser Interferometer Space Antenna (LISA) mission in order to detect GW.
The need of accurate testing of free-fall and
knowledge of noise in a space environment similar to LISA's is considered mandatory
a pre-phase for the project and therefore the LISA Pathfinder on the
Small Mission for Advanced Research in Technology 2 (SMART-2) has been
designed by ESA to fly the LISA Technology Package (LTP).

LTP will be blind to GW. By design, in order to detect any other
disturbance which could jeopardise LISA's sensitivity to GW themselves. Its goal
will be to test free-fall by measuring the residual acceleration between
two test-bodies in the dynamical scheme we address as ``drag-free'', where the satellite is
weakly coupled to one of the proof bodies and follows the motion of the other. The satellite
is supposed to act as a shield to external disturbances and not to introduce too much
noise by its internal devices. The spectral map of the residual acceleration as function of
frequency will convey information on the local noise level, thus producing a
picture of the environmental working conditions of LISA itself.

We're going to show the following:
\begin{enumerate}
\item that construction of a freely-falling global reference frame is possible in theoretical
terms, and laser detection is the utmost sensitive tool both for seeing GW - given a large
baseline detector - and for mapping residual accelerations and noise (with a short baseline);
\item that a dedicated experiment can be designed fully by means of Newtonian mechanics
and control theory. Carefully studied signals will be built as time-estimators of gauge-invariant
observables;
\item that it just won't be enough to send a probe to na{\^i}vely measure correlators of distance variation
in outer space and deduce a spectral figure. It is necessary to design and project noise shapes, make
educated ``guesses'' of
spectra spelling all possible sources, carefully sum them to obtain overall estimates.
\end{enumerate}

The description and contributions to the former tasks will be distributed as follows in the present thesis.

{\bf Chapter \ref{chap:refsys}} starts from simple theoretical arguments and tries to clarify the
idea of rulers and clocks as markers of $4$-locations in $4$-dimensional space-time.
Using only Lorentz group local generators, we'll show an absolute ruler may
be built between two fiduciary mirrors out of a laser beam and that the phase variation
$\Delta \theta_{\text{laser}}(t)$
of the laser light path
is an unbiased estimator of the GW strain as:
\begin{shadefundtheory}
\beq
\frac{\de \Delta \theta_{\text{laser}}(t)}{\de t} \simeq \frac{\pi c}{\lambda_{\text{laser}}}  \left(h(t)-h\left(t -\frac{2 L}{c}\right)\right)\,,
\eeq
\end{shadefundtheory}
\noindent which is valid to $O\left(\omega _{\text{GW}}^2\right)$ where $\omega _{\text{GW}}$ is the
GW pulsation, $h(t)$ the GW strain and $L$ is the detector baseline. $\lambda_{\text{laser}}$ is the laser
wavelength and $c$ is the speed of light in vacuo.

\begin{figure}
\begin{center}
\includegraphics[width=\textwidth]{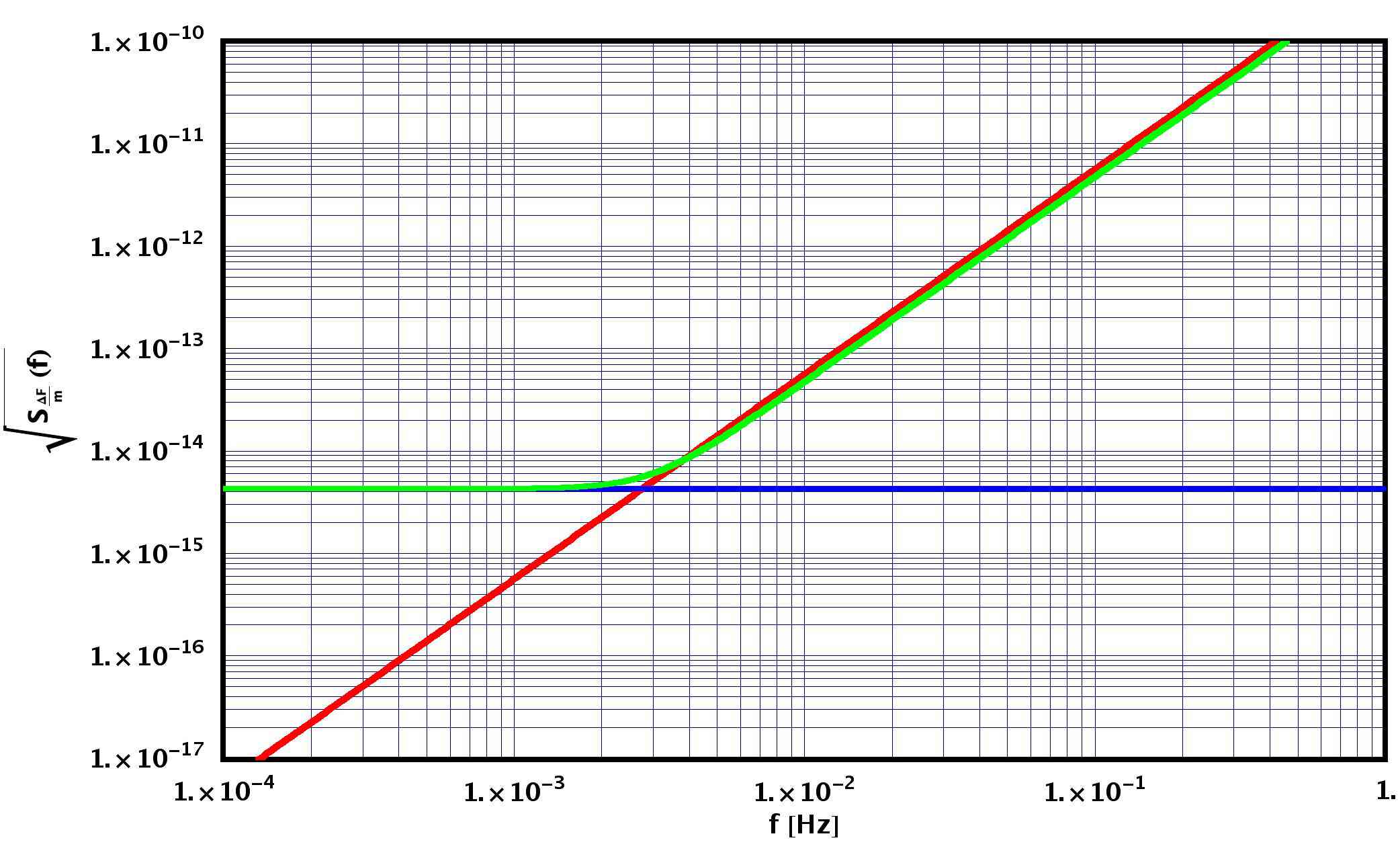}
\caption{Noise PSDs in $\nicefrac{\Delta F}{m}$ for forces difference (blue), interferometer (red) and relaxed noise
requirement of forces difference (green). Green line represents LISA's targeted sensitivity.}
\label{fig:noisepsdsintro}
\end{center}
\end{figure}

We'll shift to power spectral density (PSD) representation and describe the main sources of noise which can
deceive this ``interferometric eye''. Free-fall is replaced by drag-free and motivations
are discussed. The final outcome of the chapter will be an estimate of the
precision needed by the LISA detector in terms of the residual acceleration quality, which we may hereby
summarise as:
\begin{shadefundnumber}
\beq
\begin{split}
S^{\nicefrac{1}{2}}_{\nicefrac{\Delta F}{m},\text{LISA}}(\omega) =& \sqrt{2}\times 3\times 10^{-15} \left(1+\left(
\frac{\omega}{2\pi \times 3\, \unit{mHz}}\right)^{4}
\right)^{\nicefrac{1}{2}}\,\accPSDunit\,\\
\simeq & \sqrt{2}\times 3\times 10^{-15}\,\accPSDunit\,@\,1\,\unit{mHz}\,,
\end{split}
\eeq
\end{shadefundnumber}
\noindent a picture of this is shown in figure \ref{fig:noisepsdsintro} together with the interferometer acceleration noise.
The LISA mission aims at revealing GWs by employing high-precision interferometer detection
in space. Its Pathfinder will be a technology demonstrator to test free-fall
and our knowledge of acceleration noise. The chapter ends with a thorough description of both
and with a simplified uni-dimensional drag-free model to illustrate the features of
the main interferometer measure channel and the physical discussion of measure modes.

LTP sensitivity is worsened by roughly a factor of $7$ with respect to the LISA goal,
the measured acceleration will be differential and this is likely to be a worst case since the residual forces on the
test-masses are considered as correlated over a short baseline:
\begin{shadefundnumber}
\beq
S^{\nicefrac{1}{2}}_{\nicefrac{\Delta F}{m},\text{LTP}}(\omega)=3\times 10^{-14} \left(1+\left(
\frac{\omega}{2\pi \times 3\, \unit{mHz}}\right)^{4}
\right)^{\nicefrac{1}{2}}\,\unitfrac{m}{s^{2}\sqrt{Hz}}\,.
\eeq
\end{shadefundnumber}

{\bf Chapter \ref{chap:ltp}} will complicate the simple mechanical model of chapter \ref{chap:refsys} and
build the LTP dynamics from the ground up. Newtonian
dynamics is employed to write down the equations of motion for the test-masses (TMs) and
spacecraft (SC), with the purpose of deducing the dynamical behaviour of position/attitude
variables and introduce the relative signal estimators. Controls, operating modes
and limiting forms of signals are evaluated, their properties discussed and graphical
behaviour sketched. The chapter is a mandatory deduction to connect the figures of chapter
\ref{chap:refsys} with the world of noise in chapter \ref{chap:noise}. As we said, only the laser phase is regarded as
the observable mapping the gauge-invariant Riemann variation into a distance fluctuation. The
main interferometer signal, whose property we will derive in this chapter looks like:
\begin{shadefundtheory}
\beq
\begin{split}
\text{IFO}\left(x_{2}-x_{1}\right) \simeq
\frac{1}{\omega _{\text{lfs},x}^2-\omega ^2}
\Bigl( & g_{2,x}-g_{1,x}-\text{IFO}_{n}(x_{1})\omega^{2} +(\delta x_{2}-\delta x_{1})\omega_{\text{p},2}^{2}+\\
&+\left(\omega_{\text{p},2}^{2}-\omega_{\text{p},1}^{2}\right)
  \left( \text{IFO}_{n}(x_{1}) + \frac{g_{\text{SC},x}+z_0 g_{\text{SC},\eta}}{\omega^{2}_{\text{df},x}}\right)
\Bigr)\,,
\end{split}
\eeq
\end{shadefundtheory}
\noindent where residual local accelerations are marked with the letter $g$, stiffness with $\omega_{\text{p},i}^{2}$,
deformations as ${\delta}x_{i}$. Noise in readout is embedded in the term $\text{IFO}_{n}(x_{1})$ while
the terms $\omega_{\text{df}}^{2}$ and $\omega_{\text{lfs}}^{2}$ are drag-free (DF) and low-frequency suspension (LFS)
transfer functions. The former signal carries the information we want, as:
\begin{shadefundtheory}
\beq
\begin{split}
S_{\nicefrac{\Delta F}{m}}^{\nicefrac{1}{2}} &= \frac{\omega^{2}}{m} S_{\Delta x}^{\nicefrac{1}{2}}
\simeq \frac{\omega^{2}}{m} S_{\text{IFO}\left(\Delta x\right)}^{\nicefrac{1}{2}} =\\
&\simeq \frac{\omega^{2}}{m} \frac{\lambda_{\text{laser}}}{2\pi} S_{\text{IFO}\left(\Delta \phi_{\text{laser}}\right)}^{\nicefrac{1}{2}}
\simeq \frac{\omega^{2}}{\omega_{\text{lfs},x}^2-\omega^2} S_{{\Delta}g_{x}}^{\nicefrac{1}{2}}
\underset{\omega_{\text{lfs},x}^{2}\ll \omega^{2}}{\simeq} S_{{\Delta}g_{x}}^{\nicefrac{1}{2}}\,,
\end{split}
\eeq
\end{shadefundtheory}
\noindent where we denoted the difference of acceleration on the TMs along $\hat x$ with the symbol ${\Delta}g_{x}$.
It is a very important step to impose the laser mapping in order to guarantee that a gauge-invariant measure is performed.
This very signal is valid for mapping $\nicefrac{{\Delta}F}{m}$ on LTP but also for detecting a wave-strain
$\nicefrac{{\Delta}L}{L}$ on a long-baseline interferometer mission such as LISA.

\begin{figure}
\begin{center}
\includegraphics[width=\textwidth]{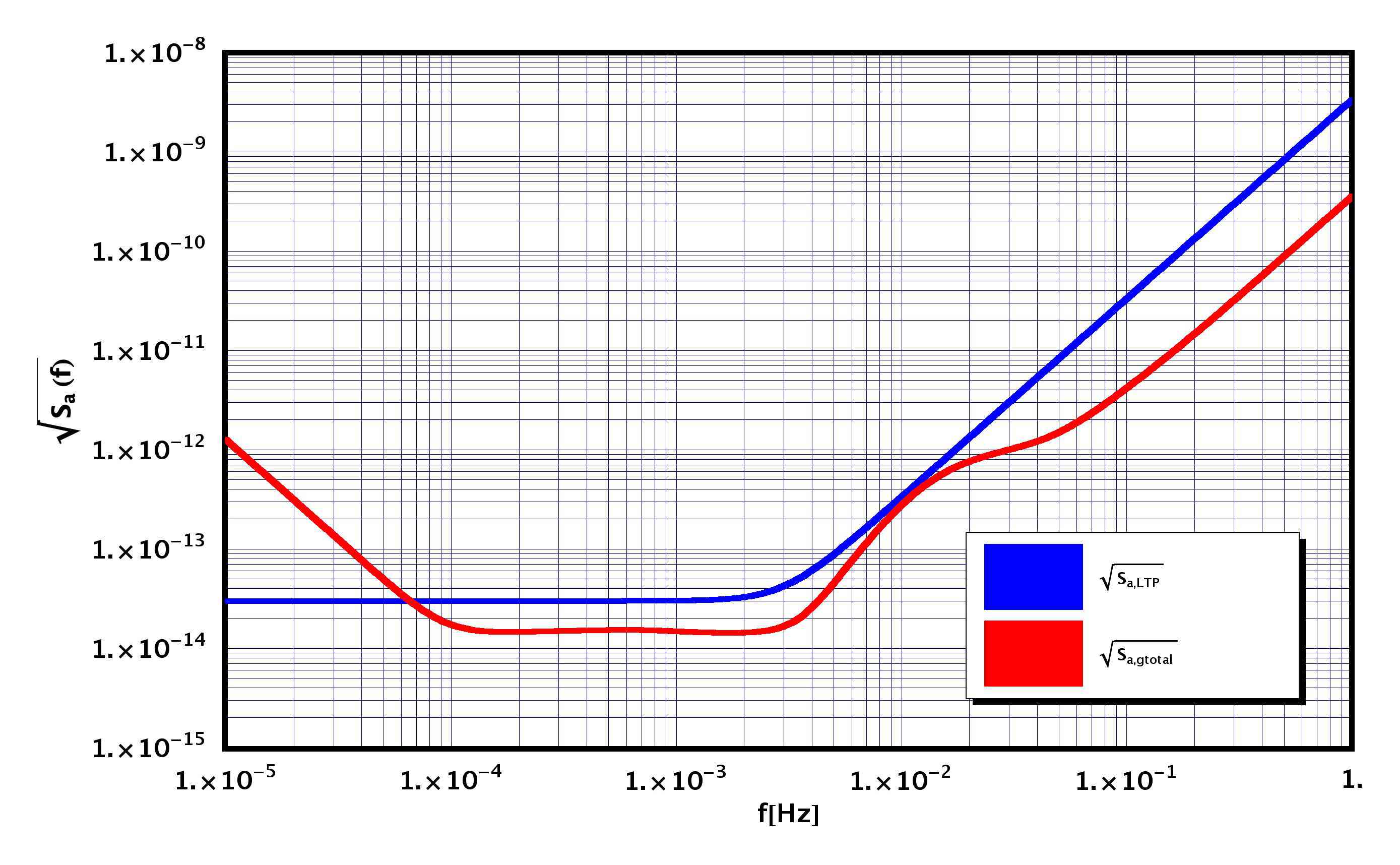}
\caption{Grand total of acceleration noise (red) versus LTP requirement sensitivity curve (blue).}
\label{fig:noisegraphversusltpsensintro}
\end{center}
\end{figure}

Noise will be dealt with in {\bf chapter \ref{chap:noise}}. Every possible recognised form of noise contribution
will be spelled out and analysed and its functional form and dependence upon position,
distribution and sources will be identified. In writing we tried to be the as encyclopedic
possible; hopefully the reader will be able to find derivations for formulae, critical numbers
for constants, tables of spectra and the way to add them. The purpose of
the chapter is in fact to provide an estimate on the acceleration noise for the proof-masses
and to compare it to the figures of chapter \ref{chap:refsys}. The achievable
quality of free-fall at current status is deeply related to such an estimate. A list of noise contributions
is reported in table \ref{tab:sumnoiseintro}, along with a graph of the noise grand-total versus the LTP sensitivity curve
in picture \ref{fig:noisegraphversusltpsensintro} showing that the whole noise is forecasted to be well below
the allowed threshold over the entire measurement band-width (MBW) ranging between $0.1\,\unit{mHz}$ and $1\,\unit{Hz}$.

\begin{table}
\begin{center}
\begin{shadefundnumber}
\begin{tabular}{r|l|l}
Description & Name & Value $\accPSDunit$\\
\hline\rule{0pt}{0.4cm}\noindent
Drag-free &  $S^{\nicefrac{1}{2}}_{a,\text{dragfree}}$  & $1.36\times 10^{-15}$\\
Readout noise & $S^{\nicefrac{1}{2}}_{a,\text{readout}}$ & $1.09 \times 10^{-17}$\\
Thermal effects &  $S^{\nicefrac{1}{2}}_{a,\text{thermal}}$  & $4.97\times 10^{-15}$\\ 
Brownian Noise &  $S^{\nicefrac{1}{2}}_{a,\text{Brownian}}$  & $9.36\times 10^{-16}$\\ 
Magnetics SC &  $S^{\nicefrac{1}{2}}_{a,\text{magnSC}}$  & $8.9\times 10^{-15}$\\ 
Magnetics Interplanetary &  $S^{\nicefrac{1}{2}}_{a,\text{magnIP}}$  & $3.25\times 10^{-16}$\\ 
Random charging and voltage &  $S^{\nicefrac{1}{2}}_{a,\text{charge}}$  & $3.61\times 10^{-15}$\\
Cross-talk &  $S^{\nicefrac{1}{2}}_{a,\text{crosstalk}}$  & $6.12\times 10^{-15}$\\ 
Miscellanea &  $S^{\nicefrac{1}{2}}_{a,\text{misc}}$  & $6.04\times 10^{-15}$\\ 
\hline\rule{0pt}{0.4cm}\noindent
Total &  $S^{\nicefrac{1}{2}}_{a,\text{total}}$  & $1.39\times 10^{-14}$\\ 
Measurement noise &  $S^{\nicefrac{1}{2}}_{a,\text{meas}}$  & $5.06\times 10^{-15}$\\
\hline\rule{0pt}{0.4cm}\noindent
Grand Total &  $S^{\nicefrac{1}{2}}_{a,\text{gtotal}}$  & $1.48\times 10^{-14}$\\
\end{tabular}
\end{shadefundnumber}
\end{center}
\caption{Acceleration noise at $f=1\,\unit{mHz}$, summary.}
\label{tab:sumnoiseintro}
\end{table}

Once the Pathfinder technology has been described, its dynamics and signals at play, the
predictable sources of noise located, {\bf chapter \ref{chap:experim}} will be devoted to reviewing the experiment
from an overall perspective, pointing out the main experimental tasks, the sequence
of tests as a ``run list'', providing a scheme and description of
the envisioned measurements.
The chapter clarifies priorities in the perspective of LTP
as a noise-probe facility, with the main task of gaining knowledge of residual noise
in view of LISA.

Here we deal with the importance of reducing the residual static gravitational imbalances, particularly
along $\hat x$. Such a worry arises in minimising the disturbances produced by the electrostatic actuation
forces - dominated by the additional
electrostatic stiffness and actuation force noise - needed to compensate the gravitational imbalance.
Static compensation of gravity imbalance is mandatory to reduce 
the static parasitic stiffness and to lower the related acceleration noise.
Therefore, maximal budgets have been assigned to each stiffness and noise contribution.

In the chapter we design a set of static compensation
masses, whose effect to counteract the formerly described forces, without introducing
excessive undesired stiffness. A simple Newtonian analysis, with the aid of some
rotational geometry and the wise use of meshing software will be our tools.

In addition to self-gravity compensation, the issue of calibrating the force applied to a TM
is not a minor one, its precision being of primary importance for control and feed-back application and, as such, it is
addressed here as well.

At the end of the chapter we'll present the measurement
of the charge accumulated on the TMs to extend one of the main points in the ``run list''.
Such a feature is of paramount importance, being
a fundamental prerequisite for the gravitational reference sensor (GRS) to operate properly. 

{\bf Chapter \ref{chap:gphysltp}} will briefly review and summarise tests of fundamental physics of
gravity which may be carried on with the LTP as a high-performance
accelerometer, other than a detector of acceleration noise for LISA. We hereby present the measure of $G$,
violations of the inverse square law (ISL) and a discussion on modified dynamics (MOND). As an independent source
of gravity stimulus, the originally planned
NASA parallel experiment Space Technology 7 (ST-7), which was to host the Disturbance Reduction System (DRS) device,
will be thought of as still in place. We confess here
that at the present status of the mission planning, these measurements represent
more an exercise of style than a real part of LTP forecasted schedule. We hope
the gedanken-experiments form we chose shall please and inspire the reader.

In {\bf appendix \ref{chap:gwtheory}} the usual theory of GWs will be refreshed, together with mechanisms
of production and sources, basic figures and examples. In contrast with the highly-geometrized approach
of chapter \ref{chap:refsys}, this appendix provides a perspective tailored more towards an audience with a shallower
training in theoretical physics, to guarantee that the basics will be understood anyway.

{\bf Appendix \ref{chap:geodesics}} re-deduces the main TT-gauge properties starting from the metric
and the connections. A brief discussion of the geodesics deviation
equation is carried on from two different standing points. This background constitutes a sound basis for venturing
in the first part of chapter \ref{chap:refsys}.

{\bf Conclusions} shall tie together the idea presented here and list a number of open issues, but what we can
state here is that - to our understanding -
the present work shows that drag-free is achievable in good experimental TT-gauge
conditions, such as to guarantee a precision measurement of acceleration noise.

This effort is done
in order to clearly pave the way for LISA, map and model the noise landscape, confirm
figures for future detection of GWs, a goal which is clearly
moving away from science-fiction and towards realisation.

The thesis provides a review on several subjects together with original research material of the author.
It seems wise here to shed light on who-is-who and what we also did during the PhD course which doesn't appear in this work.

A considerable time was dedicated to the problem of compensating static gravity. This work appears in chapter \ref{chap:experim} and
it became an article \cite{Armano:2005ut} presented as a talk at the 5th International LISA Symposium, held from July 12th to July 15th 2004 at ESTEC, Noordvijk.
The contribution has become a milestone and resulted in a gravitational control protocol document \cite{ltpgravprot}.

We dedicated a large amount of time in contributing to the development of a theory of cross-talk for the LTP experiment \cite{LTPcrosstalk}.
Cross-talk is a very important piece of noise budged, and can be found in section \ref{sec:crosstalk}.

Furthermore, we were asked to provide a thorough construction of the laser detection procedure starting from GR and differential geometry
arguments; chapter \ref{chap:refsys} extends the work we published in \cite{Anza:2005td}; effort was put in pointing out
the physical motivations for the choices we made. The chapter is somewhat complicated and
we tried to condensate some textbook material into appendices \ref{chap:gwtheory} and \ref{chap:geodesics} with more standard notations.
In this perspective the thesis is meant as a tool for the Group and the Collaboration, and we really hope to have provided some service.
The first part of chapter \ref{chap:refsys} is probably bound to become a new publication.

To our knowledge, a detailed description of LTP dynamics such as that found in chapter \ref{chap:ltp} doesn't exist in literature. The same
can be said for chapter \ref{chap:noise}, but the reader should be aware that we didn't invent anything here, but rather have just extended, reorganized and produced
an introduction to describe noise as a global phenomenon with derivations when needed.

The calibration of force to displacement in section \ref{sec:calibforcedisp} and the measurement of charging and discharging of the test-mass
in section \ref{sec:chargedisctm} are the outcome \cite{tateothesis} of a collaboration work with Nicola Tateo, friend and
then Masters student we assisted across last year's work.

In section \ref{sec:gmeas} we coalesced our contributions to the project of measure of $G$ onboard LTP. The Science Team created across
Trento, ESA and Imperial College London worked hard to understand LTP capabilities in this perspective; as witness and collaborator I
decided to address this subject in a vaster chapter about fundamental physics with LTP, chapter \ref{chap:gphysltp}.

Outside the thesis, we contributed to the writing of the LTP Operation Master Plan \cite{LTPmaster},
and the presently used high-speed real-time
driver for the RS422 serial port for the engineering model of LTP front-end electronics is our creation.

We employed colours in shadings to help the reader focus the main results. Thus, fundamental theoretical
formulae or high-level computational choices will be shaded as follows:
\begin{shadefundtheory}
\beq
R_{\mu\nu}-\frac{1}{2}R g_{\mu\nu} = \frac{8 \pi G}{c^{2}} T_{\mu\nu}\,,
\eeq
\end{shadefundtheory}
\noindent while requirements and very important numerical estimates will get the colour:
\begin{shadefundnumber}
\beq
S^{\nicefrac{1}{2}}_{\nicefrac{\Delta F}{m},\text{LTP}}(\omega)=3\times 10^{-14} \left(1+\left(
\frac{\omega}{2\pi \times 3\, \unit{mHz}}\right)^{4}
\right)^{\nicefrac{1}{2}}\,\unitfrac{m}{s^{2}\sqrt{Hz}}\,.
\eeq
\end{shadefundnumber}
\noindent Especially in the noise section, but in other several places too, numbers and figures less fundamental for the
global picture are seeded. They are underlined as:
\begin{shademinornumber}
\beq
S^{\nicefrac{1}{2}}_{a,\text{dragfree}} = \left|{\Delta}\omega_{\text{p},x}^2\right| S_{x,\text{tot}}^{\nicefrac{1}{2}}\,.
\eeq
\end{shademinornumber}

A table of fundamental constants in Physics follows, together with a list of acronyms. I always found it so annoying to be
left alone in the uncertainty of where to find these that I thought it better to place them in the preface,
where they're easy to retrieve.

The thesis was realized entirely in \LaTeX, the majority of the graphs in {\sf Mathematica} \textregistered. The document is originally produced as
a PDF with navigable links; an electronic version is downloadable from
\href{http://www.science.unitn.it/~armano/michele\_armano\_phd\_thesis.pdf}{http://www.science.unitn.it/{\textasciitilde}armano/michele\_armano\_phd\_thesis.pdf}.

%% file: chapters/fundconstants.tex
\chapter{Table of fundamental constants}

\begin{center}
\begin{tabular}{r|l|l}
Description & Name & Value\\
\hline\rule{0pt}{0.4cm}\noindent
Speed of light in vacuo & $c$ & $2.9979\times 10^{8}\,\unitfrac{m}{s}$ \\
Newton gravitational constant & $G$ & $6.67\times 10^{-11}\,\unitfrac{m^{2} N}{\text{kg}^{2}}$ \\
Planck constant & $h$ & $6.63\times 10^{-34}\,\unit{J s}$ \\
Vacuum electric permittivity & $\epsilon_{0}$ & $8.85\times 10^{-12}\,\unitfrac{A s}{m V}$ \\
Vacuum magnetic permeability & $\mu_{0}$ & $1.26\times 10^{-6}\,\unitfrac{s V}{A m}$ \\
Boltzmann constant & $k_B$ & $1.38\times 10^{-23}\,\unitfrac{J}{K}$ \\
Stefan constant & $\sigma$  & $5.67\times 10^{-8}\,\unitfrac{W}{K^4 m^{2}}$ \\
Electron charge & $q_e$ & $1.6\times 10^{-19}\,\unit{C}$ \\
Earth mass & $M_{\text{Earth}}$ & $5.97\times 10^{24}\,\unit{kg}$ \\
Earth radius & $R_{\text{Earth}}$ & $6.38\times 10^{6}\,\unit{m}$ \\
Gravity acceleration on Earth & $g$ & $9.81\,\unitfrac{m}{s^{2}}$
 \end{tabular}
\end{center}

%% file: chapters/acronyms.tex
\chapter{List of acronyms}

\begin{center}
\begin{tabular}{r|l}
Acronym & Description \\
\hline\rule{0pt}{0.4cm}\noindent
AC & Alternate Current \\
CDR & Critical Design Review \\
CmpMs & Compensation Masses \\
DC & Direct Current \\
DF(df) & Drag-Free\\
DOF & Degree(s) of Freedom \\
DRS & Disturbance Reduction System \\
EH & Electrode Housing \\
EM & Electro-Magnetic \\
ESA & European Space Agency \\
FEEP & Field Emission Electric Propulsion \\
GRS & Gravitational Reference System \\
GSR & Gravitational System Review \\
GW & Gravitational Waves \\
IFO & Interferometer (Output) \\
IS & Inertial Sensor \\
ISL & Inverse Square Law \\
LFS(lfs) & Low Frequency Suspension\\
l.h.s. & Left Hand Side\\
LISA & Laser Interferometer Space Antenna \\
LTP & LISA Technology Package \\
M1 & Nominal Mode \\
M3 & Science Mode \\
MBW & Measurement Bandwidth\\
MOND & Modified Newtonian Dynamics \\
NASA & National Aeronautics and Space Administration \\
OB & Optical Bench \\
PSD & Power Spectral Density \\
r.h.s. & Right Hand Side\\
SC & Space-Craft \\
SGI & Static Gravitational Imbalances\\
SMART-2 & Small Mission for Advanced Research in Technology 2 \\
SP & Saddle Point \\
ST-7 & Space Technology 7 \\
STOC & Science and Technology Operation Centre\\
TM & Test Mass \\
TT & Transverse-Traceless \\
VE & Vacuum Enclosure
\end{tabular}
\end{center}

%% file: chapters/refsys.tex
\chapter{LISA, LTP and the practical construction of TT-gauged set of coordinates}
\label{chap:refsys}

\lettrine[lines=4]{A}{ popular}
gauge choice widely employed to deal with GWs is the so called ``TT'' - for Transverse
and Traceless - gauge. Coupled with the global radiation gauge it permits to get rid of
unphysical degrees of freedom of the theory and focus on measurable observables.

In this chapter we'll try to describe carefully the concept of fiduciary measurement points in
free-fall, relate it to a geometrical description of space-time (a congruence of geodesics),
and build an arbitrary-sized ensemble of tetrads, evolving in time, to mark space with
a rigid ruler and a reliable clock. Photons will be taken as detectors carrying
the effects of radiative metric perturbations, their phase made the observable we seek for.

The Laser Interferometer Space Antenna (LISA) and its Pathfinder (LTP) will be described and their features carefully discussed. A
simple model of a one-dimensional drag-free device mimicking LTP's behaviour follows,
with the purpose of giving a simplified description and introducing signals,
control modes and the physics behind them.


\newpage

\section{A local observer}


The absence or annihilation of local gravitational acceleration is the condition usually referred to as ``free-fall'',
in other words an object is is free-fall when it is in geodesic motion in the gravitational field.
To claim that we can annihilate local gravitational acceleration, Newton's theory is more than enough
\cite{weinberggrav, dinverno}.
We state a body is
accelerated with constant acceleration $g$ if, simplifying to a uni-dimensional case \cite{olesengr} we can write:
\beq
m_{\text{i}}\ddot x = m_{\text{g}}g\,,
\eeq
where $m_{\text{i}}$ is the inertial mass and $m_{\text{g}}$ is the gravitational mass.
We are free nonetheless to co-move with the body, by choosing proper
coordinates:
\beq
\label{eq:naivechangecoord}
y\dot = m_{\text{i}} x - \frac{1}{2} m_{\text{g}} g t^{2}\,,
\eeq
so that
\beq
\ddot y = m_{\text{i}}\ddot x - m_{\text{g}} g = 0\,.
\eeq
We assume therefore the complete physical equivalence of a gravitational field and a corresponding acceleration
of the reference system: free fall is inertial motion.

The weak equivalence principle, also known as the universality of free fall, will be assumed:
the trajectory of a falling test body depends only on its initial position and velocity, and is independent of its composition,
or all bodies at the same spacetime point in a given gravitational field will undergo the same
acceleration ($m_{\text{i}}=m_{\text{g}}$). The concept can be extended by stating that
every system of coordinate is good for
a description of the physical reality, provided it is Lorentz invariant.


Assuming the gravitational
field to be metric and geometric accounts for its instantaneous potential to be smooth and
Taylor-expandable in the position itself \cite{Thorne:1980ru}:
\beq
\label{eq:taylorpotential}
\Phi(\vc x)=\Phi_{0}-\sum_{j} g_{j} x_{j} + \sum_{j,k}\frac{1}{2}R_{j0k0}x_{j}x_{k}+\ldots\,.
\eeq

By changing coordinates in a similar fashion as we mentioned in \eqref{eq:naivechangecoord}, only
contributions of tidal nature shall remain in the local frame ($\Phi_{0}$ is an immaterial term
representing $0$-point potential). To use
the theory of GR at full power, the only true accelerations left
are geodesic deviations: mutual accelerations between world-lines whose dynamics
is imputable to the true metric invariant object at play, the Riemann tensor, some
components of which appear as second order derivatives in \eqref{eq:taylorpotential}.



According to this simple pieces of information, if we'd like to describe and build an apparatus which we could define
to be ``almost intertial'' or sensitive to tidal stress, we'd
need some ingredients:
\begin{enumerate}
\item a suitable choice of coordinates to null the unphysical contributions of 
the Christoffel connection $\Gamma^{\mu}_{\phantom{\mu}\nu\sigma}$ in Einstein's equations of gravitation:
some of these are fictitious combinations of the metric degrees of freedom (DOF), carrying gauge nature.
\item Free-fall at its best, to get rid of
all the local $g_{j}$-like contributions to the potential, in the spirit of \eqref{eq:naivechangecoord}
and \eqref{eq:taylorpotential}. The better the quality of free-fall, the 
smaller the  $g_{j}$ residual accelerations.
\item An electromagnetic noise reduction strategy. This takes the form of a shield from external sources which could
introduce some little EM noise while shielding larger effects. Such noise is easy to disguise as a gravitational one
as it would perturb geodesics just the same (see \eqref{eq:geodesylowspeed}). Moreover the shielding
guarantees the system to remain quasi-inertial.
\item An intrinsic high-fidelity detection tool: if geometry and gravity are so tightly tied by Einstein's equations so that clocks and
rods get deformed, the only way out is choosing a set of clocks and rods with intrinsic spatial relation. By means of
their energy-momentum light-like ties, photons wave $4$-vectors $k_{\mu}$ fulfil the relation:
\begin{shadefundtheory}
\beq
k_{\mu}k^{\mu}=0\,,
\eeq
\end{shadefundtheory}
\noindent which is Lorentz invariant and locally defines a dispersion relation as $c=\lambda\nu$, given the
frequency $\nu$ and wavelength $\lambda$ for a monochromatic beam\footnote{This is true under the conditions of
free-fall of the observer and distance $\ll$ curvature radius. $c$ is the velocity of light in vacuo but we remind the
reader that such a constant is in fact locally defined and does not have a global value.}.

We'll debate on this in the following, but intuitively we can state that a photon beam
has absolute clock given by its constant velocity $c$ and carries absolute metrology by the former relation. It is
thus the perfect carrier for residual acceleration information as well as tidal stress of curvature.
\end{enumerate}

What if, then, we'd decide to place a mirror in space, and claim it's freely falling. First we'd have to answer to the question
of coordinates: freely falling with regard to what? As a matter of fact, we'd need two mirrors in free fall, one
to be employed as a measuring fiduciary ``zero'', and the other to get real difference metrology from.
Whatever the disposition of the mirrors, we can always claim without any loss
of generality to place them face-to-face; there exists then a unique ``straight'' line connecting them.
In absence of external forces a body keeps moving with constant velocity or,
better to say, in absence of external curvature\index{Curvature} of space-time, the body follows an unperturbed geodesic:
unbending world-lines in Minkowski space-time will describe the geodesic curves.

Paradoxically, a point-like body placed idle in a universe with no masses but itself, will
stay idle forever but we'd still need another body to state this. The two then would have reciprocal world-lines
in the relative coordinates $x^{\mu}=x^{\mu}(\tau)$ (where $\tau$ is the proper time) such that:
\beq
\nabla_{\vc V} x^{\mu} = 0\,,
\label{eq:coordfreefall}
\eeq
where $V^{\mu}=\nicefrac{\de x^{\mu}}{\de\tau}$; but, in presence of any
Riemann curvature (background, induced on one another, by gravitational radiation...), the two bodies
(and their world-lines) would accelerate
and bend according to the geodesics acceleration\index{Geodesics acceleration} formula
(see appendix \ref{chap:geodesics} for a demonstration):
\beq
\nabla_{\vc V}\nabla_{\vc V} W^{\mu}=R^{\mu}_{\phantom{\mu}\nu\beta\alpha}V^{\nu}V^{\beta}W^{\alpha}\,,
\label{eq:geoaccriemann}
\eeq
where $W^{\mu}=\nicefrac{\de x^{\mu}}{\de\zeta}$. A congruential hyper-surface of
geodesics $x^{\mu}=x^{\mu}\left(\tau,\zeta\right)$ is built, and the ones we stated
in \eqref{eq:geoaccriemann} are their equations of motion.


Let's now formalise this picture. No matter what the choice of coordinates would be, a freely falling mirror
can be equipped with intrinsic axes: the Fermi-Walker tetrad\index{Fermi-Walker tetrad} associated with the body;
the zero
of the axes will be placed in the centre of mass of the object. In general notation, if the mirror is sitting on
the abstract point $P$, this is a function of $\tau$, the proper time which defines the emanation point by $\tau=0$, a
direction parameter $\vc n$ to tell on which geodesic we are moving, and an elongation parameter $\sigma$,
a proper distance to tell where we are on the geodesic \cite{MTW}:
\beq
P=P\left(\tau, \vc n=\frac{\de P}{\de \sigma}=n^{j}\vc e_{j}, \sigma\right)\,.
\eeq
This is true for both mirrors. How to
relate the two points to one another is the matter of defining an observer with a reference frame,
whose r\^{o}le is in fact casting coordinates while he/she moves or jitter:
\beq
x^{\mu}\left(P(\tau,\vc n,\sigma)\right)=\left\{\tau,\sigma n^{j}\right\}\,,
\eeq
this happens because while moving the observer carries an orthonormal tetrad with himself such that:
\beq
\vc e_{0} \doteq \vc u = \frac{\de P_{0}(\tau)}{\de \tau}\,\qquad \vc e_{\alpha}\cdot\vc e_{\beta}=\eta_{\alpha\beta}\,,
\eeq
where we defined $\vc u$ as the 4-velocity of the observer, tangent to the $P_{0}$ world-line. If the
tetrad is parallel-transported along the world-line its equations of motion are \cite{MTW}:
\beq
\nabla_{\vc u} \vc e^{\alpha} = -\Omega^{\alpha\beta} \vc e_{\beta}\,,
\eeq
with
\begin{shadefundtheory}
\beq
\Omega^{\alpha\beta}=a^{\alpha} u^{\beta}-u^{\alpha}a^{\beta}+u_{\mu}\omega_{\nu}\epsilon^{\mu\nu\alpha\beta}\,,
\eeq
\end{shadefundtheory}
\noindent is the (fully antisymmetric) generator of infinitesimal local Lorentz transformations comprised by $a^{\alpha}=\nicefrac{\de u^{\alpha}}{\de\sigma}$, the 4-acceleration, and $u^{\beta}$, the 4-velocity, already
mentioned, and $\omega^{\alpha}$, the angular velocity of rotation of spatial vector basis $\vc e_{j}$ relative to inertial guidance
gyroscopes\footnote{We say a tetrad is Fermi-Walker transported if $\vc\omega=\vc 0$ and recover the idea of
geodesic transport if $\vc a=\vc\omega=\vc 0$, so that $\nabla_{\vc u}\vc e_{\alpha}=0$. Fermi-Walker transporting a tetrad
means to allow it undergo a general Lorentz transformation but not space rotations. The only infinitesimal transformation which
doesn't allow spatial rotation is such that $\Omega^{\mu\nu}\omega_{\nu}=0$, leaving:
\begin{align*}
\Omega^{\mu\nu}_{\text{SR}}&=u_{\alpha}\omega_{\beta}\epsilon^{\alpha\beta\mu\nu}=0\,,\\
\Omega^{\mu\nu}\to\Omega^{\mu\nu}_{\text{FW}}&=a^{\mu}u^{\nu}-a^{\nu}u^{\mu}\,,
\end{align*}
where the suffix SR stands for ``spatial rotations'' and FW for ``Fermi-Walker''.

Notice also:
\begin{align*}
a_{\mu}u^{\mu}&=\frac{\de u_{\mu}}{\de\sigma}u^{\mu}=\frac{1}{2}\frac{\de}{\de\sigma}\left(u_{\mu}u^{\mu}\right)=0\,,\\
\Omega^{\mu\nu}u_{\nu}&=a^{\mu}u^{\nu}u_{\nu}-u_{\nu}a^{\nu}u^{\mu}
  +u_{\alpha}\omega_{\beta}u_{\nu}\epsilon^{\alpha\beta\mu\nu}=-a^{\mu}\,,
\end{align*}
}.

\begin{figure}
\begin{center}
\includegraphics[width=\textwidth]{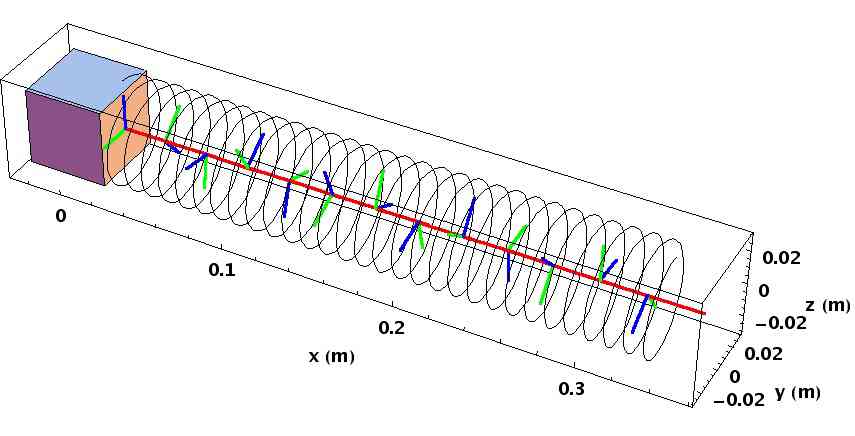}
\caption{Space-like versors of orthonormal tetrads associated with the laser-beam shone from a mirror placed in a
fiduciary point in space-time. Tetrads rotate co-moving with the laser polarisation vector, thus mapping space with the
photons natural pace $\lambda=\nicefrac{c}{\nu}$.}
\label{fig:laser1}
\end{center}
\end{figure}

We may now think our first mirror as an observer sitting in space, with its free tetrad emanating in space-time. Along
the $\hat x$ space direction we may shine a laser beam. To each and every instant of the photons world-line, a tetrad may
be attached, one space axis collinear with $\hat x$, say $\vc e_{1}$, another, say $\vc e_{2}$ rigidly attached to
the polarisation vector, the third space-line thus forced to belong to the polarisation plane. The frequency of rotation
of the tetrad gets connected to the light frequency by $\nicefrac{2\pi c}{\lambda_{\text{laser}}}=\omega$ and its velocity
is tied to be $c$ by the light dispersion relation\index{Dispersion relation}. Picture (\ref{fig:laser1}) may illustrate the point.

\begin{figure}
\begin{center}
\includegraphics[width=\textwidth]{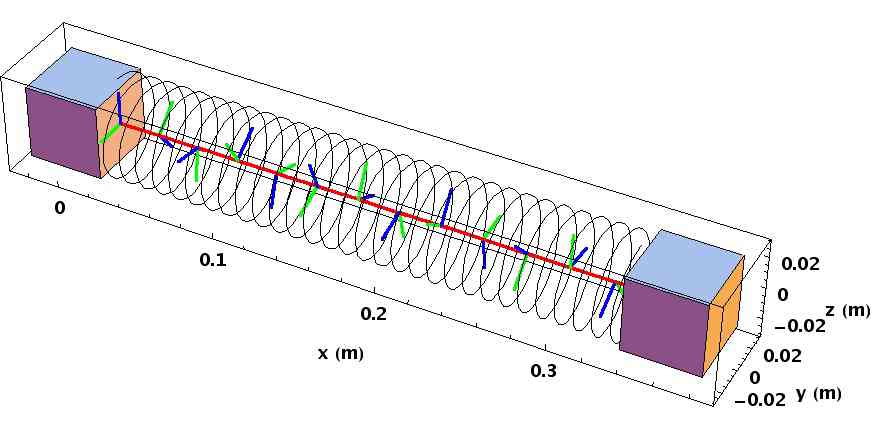}
\caption{See figure \ref{fig:laser1}. Reflecting mirror added.}
\label{fig:laser2}
\end{center}
\end{figure}

\begin{figure}
\begin{center}
\includegraphics[width=\textwidth]{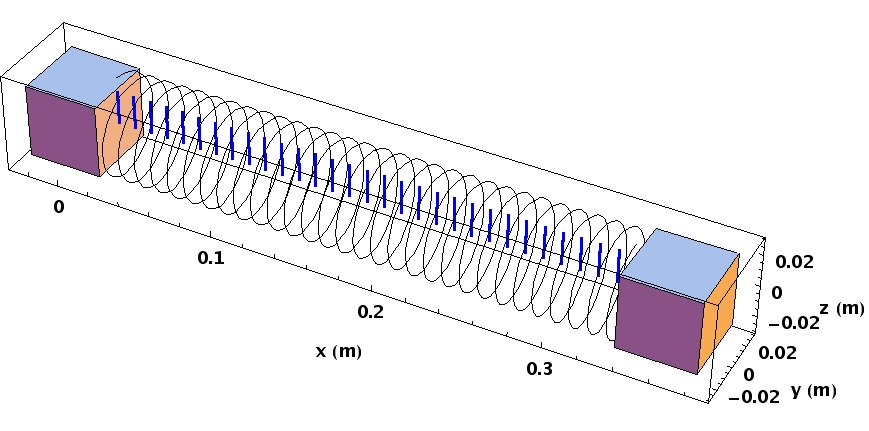}
\caption{See figure \ref{fig:laser1} and \ref{fig:laser2}. Projecting out the tetrad with space step $\lambda$.}
\label{fig:laser3}
\end{center}
\end{figure}

\begin{figure}
\begin{center}
\includegraphics[width=\textwidth]{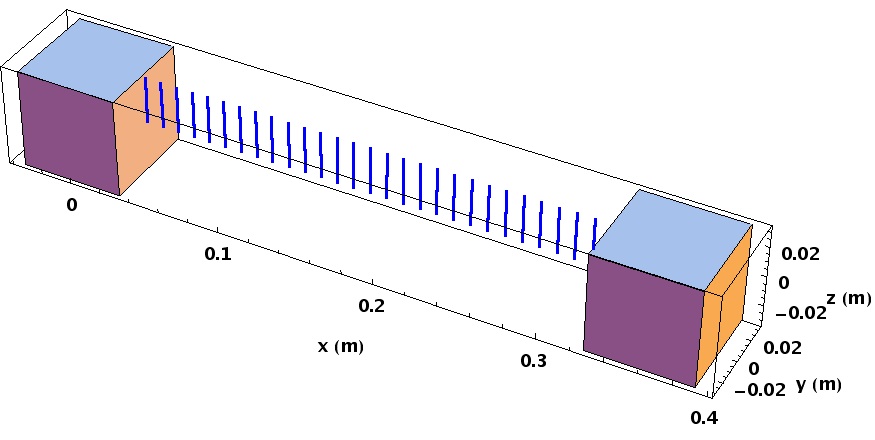}
\caption{See figure \ref{fig:laser1} and \ref{fig:laser2} and \ref{fig:laser3}. Ruler left by the projection.}
\label{fig:laser4}
\end{center}
\end{figure}

Accordingly, the situation is unvaried if we place the second mirror facing the first at some
distance; see figure (\ref{fig:laser2}). If now we decide to choose a clock, that can be the photon's, its ``$0$'' time
being the laser time when leaving the mirror surface, uniquely characterised by placing the polarisation vector on
the surface itself. Subsequent laser time pace is given by the parallel transport of that tetrad for frequency shifts
equal to $\nu=\nicefrac{c}{\lambda}$ or rather by Fermi-Walker transporting the tetrad rotating with
pulsation $\omega=\nicefrac{2\pi c}{\lambda}$ projecting fiduciary points on the path every wavelength. The reference picture is now
(\ref{fig:laser3}). We are then left with a bona-fide ruler in space and a reliable clock in time! See figure (\ref{fig:laser4}).

\section{Gauge fixing, the GW metric}

If space-time is nearly flat, we can assume the metric to be written as
\begin{shadefundtheory}
\beq
g_{\alpha\beta}=\eta_{\alpha\beta}+h_{\alpha\beta}\,,
\eeq
\end{shadefundtheory}
\noindent where $\eta=\diag(-1,1,1,1)$ and $h$ is a perturbation such that $|h_{\alpha\beta}|\ll 1$.
Notice component by component $\eta^{\alpha\beta}=\eta_{\alpha\beta}$ denotes the inverse.

We fix the global gauge by choosing harmonic gauge (see eq. \eqref{eq:harmonicgauge}), so that:
\beq
h^{\alpha}_{\phantom{\alpha}\beta,\alpha}-\frac{1}{2}h^{\eta}_{\phantom{\eta}\eta,\beta} = 0\,,
\eeq
and name TT-gauge that specific local gauge-fixing of metric DOF such that:
\beq
\begin{split}
h_{\mu 0}&=0\,,\\
\eta_{ij}h^{ij}=h_{i}^{i}&=0\,,\\
h_{ij;j}\simeq h_{ij,j}&=0\,,
\end{split}
\eeq
obviously the $h$ tensor retains only spatial components, it's traceless and transverse (TT)\index{TT-gauge}.
We can always employ this choice, without loss of generality, since it won't change the form of the
physical observables made out of $R^{\mu}_{\phantom{\mu}\nu\sigma\eta}$.

We remind now that the connection for a nearly-linear theory is expressed by
formula \eqref{eq:christoffellin}, and
if the gauge choice is TT, most of the mixed components of $\Gamma^{\mu}_{\phantom{\mu}\alpha\beta}$
vanish or get simplified \cite{Teukolsky:1982nz}:
\begin{align}
\Gamma^{i}_{\phantom{i}00}=\Gamma^{0}_{\phantom{0}00}=\Gamma^{0}_{\phantom{0}0j}&=0\,,\\
\Gamma^{0}_{\phantom{0}jk} &=-\frac{1}{2}h_{jk,0}\,.\\
\Gamma^{i}_{\phantom{i}0j} &=\frac{1}{2}h_{j\phantom{i},0}^{\phantom{j}i}\,.
\end{align}
only a few terms will survive due to the mentioned simplifications, to get from \eqref{eq:coordfreefall}
(see appendix \ref{chap:geodesics}):
\begin{shadefundtheory}
\beq
\frac{\de^{2} x^{i}}{\de t^{2}}=
   \left(
   -2 \Gamma^{i}_{\phantom{i}0j}-\Gamma^{i}_{\phantom{i}jk}v^{k}
   +\Gamma^{0}_{\phantom{i}jk}v^{i}v^{k}
   \right)v^{j}\,.
\eeq
\end{shadefundtheory}
\noindent Thus in TT-gauge,
in absence of external forces particles at rest ($v^{i}=0$) remain at rest forever, since
they never accelerate. Hence their coordinates are
good markers of position and time. We'll never stress enough the point that we are now talking about
coordinates; conversely distances are relative objects locally governed by geodesics acceleration equations,
the presence of a tidal field may stretch or shrink them in this scenario according to \eqref{eq:geoaccriemann}.

Let's summarise the ingredients we have collected so far:
\begin{enumerate}
\item we placed two bodies shielded
from external disturbances in space and in free-fall (relative velocities are $\simeq 0$). Moreover
this view of the coordinates is global and it's just a choice, i.e. doesn't modify physical observables, which
are gauge-invariant functions of the Riemann tensor mapped by the laser phase \cite{Garfinkle:2005qf}.
\item We equipped space and time with rods and clocks independent on the presence of gravitational
perturbations. Of course the situation will get more and more complicated the more curved space-time is: geodesics
may cross and eclectic phenomena may appear. Nevertheless in the case of a small perturbation of the metric $h_{\mu\nu}$
we claim this to be suitable to our purposes.
\item If a gravitational radiative phenomenon occurs so to produce GW to travel till being
plane in the premises of such a detection apparatus, tidal contributions to $h_{ij}$ will show up in adherence to Einstein's theory.
Hence variation of curvature will change the laser phase by changing its optical path.
\end{enumerate}

If the perturbation were not there, the metric would be simply flat: $g_{\mu\nu}=\eta_{\mu\nu}$.
With reference to the idle tetrad, the generator of infinitesimal motion along $\vc e_{1}$, mapped
by the proper parameter $\sigma$ and by the laser beam, would get the following form:
\beq
\Omega^{\alpha\beta}=u_{\mu}\omega_{\nu}\epsilon^{\mu\nu\alpha\beta}\,,
\eeq
since no acceleration is induced in that direction and we allow the moving tetrad to whirl with
angular velocity $\omega$. The vector parameters are chosen such that:
\beq
\vc u = \left[\begin{matrix}c \\ 0 \\ 0 \\ 0\end{matrix}\right]\,,
\qquad
\vc \omega = \xi \left[\begin{matrix}0 \\ \nicefrac{2\pi c}{\lambda} \\ 0 \\ 0\end{matrix}\right]\,,
\eeq
indeed: $\vc u \parallel \vc e_{0}$ and $\vc \omega \parallel \vc e_{1}$. We can then normalise
$\vc u$ so that $\vc u = \vc e_{0}$. $\xi$ is an infinitesimal parameter. The rotating tetrad picks then the form:
\beq
\vc e_{0} = \left[\begin{matrix}1 \\ 0 \\ 0 \\ 0\end{matrix}\right]\,,\quad
\vc e_{1} = \left[\begin{matrix}0 \\ 1 \\ 0 \\ 0\end{matrix}\right]\,,\quad
\vc e_{2} = \left[\begin{matrix}0 \\ 0 \\ \cos\omega t \\ \sin\omega t \end{matrix}\right]\,,\quad
\vc e_{3} = \left[\begin{matrix}0 \\ 0 \\ -\sin\omega t \\ \cos\omega t \end{matrix}\right]\,,
\eeq
where vectors have been properly normalised. Application of the $\Omega$ operator to the
vectors give the infinitesimal variation of the vectors themselves:
\begin{shadefundtheory}
\begin{align}
\nabla_{\vc u} \vc e_{0}&=\nabla_{\vc u} \vc u = \vc 0\,,\\
\nabla_{\vc u} \vc e_{1}&=\vc 0\,,\\
\nabla_{\vc u} \vc e_{2}&=\xi\frac{2\pi c}{\lambda}\vc e_{3}\,,\\
\nabla_{\vc u} \vc e_{3}&=-\xi\frac{2\pi c}{\lambda}\vc e_{2}\,,
\end{align}
\end{shadefundtheory}
\noindent as expected $\nabla_{\vc u}$ acts as an infinitesimal transporter\index{Transporter} of the tetrad along $\vc u$; moreover, its exponentiation
will give the result for a finite parameter $\xi$. Let's compute it for the evolution of $\vc e_{2}$:
\beq
\begin{split}
\left(\exp\nabla_{\vc u}\right) \vc e_{2} &= \sum_{j=0}^{\infty} \frac{1}{j!}\left(\nabla_{\vc u}\right)^{j}\vc e_{2} = \\
&= \sum_{j=0}^{\infty} \frac{1}{(2j)!}\left(\nabla_{\vc u}\right)^{2j}\vc e_{2} +
\sum_{j=0}^{\infty} \frac{1}{(2j+1)!}\left(\nabla_{\vc u}\right)^{2j+1}\vc e_{2}\,,
\end{split}
\eeq
we now employ the facts:
\beq
\begin{split}
\left(\nabla_{\vc u}\right)^{2} \vc e_{2} &= - \left(\frac{2\pi c}{\lambda}\xi\right)^{2} \vc e_{2}\,,\\
\left(\nabla_{\vc u}\right)^{2} \vc e_{3} &= \left(\nabla_{\vc u}\right)^{2} \nabla_{\vc u}\vc e_{2} = - \left(\frac{2\pi c}{\lambda}\xi\right)^{2} \vc e_{3}\,,
\end{split}
\eeq
to get
\beq
\left(\exp\nabla_{\vc u}\right) \vc e_{2} = 
\vc e_{2} \sum_{j=0}^{\infty} \frac{(-1)^{j}}{(2j)!}\left(\frac{2\pi c}{\lambda}\xi\right)^{2j} +
\vc e_{3}\sum_{j=0}^{\infty} \frac{(-1)^{j}}{(2j+1)!}\left(\frac{2\pi c}{\lambda}\xi\right)^{2j+1}\,.
\eeq
Finally, according to Taylor's theorem and the series for sine and cosine, we get:
\begin{shadefundtheory}
\beq
\vc e_{2}'=\left(\exp\nabla_{\vc u}\right) \vc e_{2} =
\cos \left(\frac{2\pi c}{\lambda}\xi\right) \vc e_{2} + \sin \left(\frac{2\pi c}{\lambda}\xi\right) \vc e_{3}\,,
\eeq
\end{shadefundtheory}
\noindent therefore, if $\xi$ is an integer multiple of the ratio $\nicefrac{\lambda}{c}$, we get that the rotating tetrad is coincident with the
reference idle one by imposing $\vc e_{2}'=\vc e_{2}$. Hence we can build a set of fiduciary points marking a ruler
with pace $\lambda$, as planned.

Suppose a GW would come along direction $\vc e_{3}$, with reference to the idle tetrad placed
along the first mirror surface. In TT-gauge, its relevant DOF can be
expressed by means of two amplitudes $h_{\times}$ and $h_{+}$, properly added to the unperturbed
flat-metric, to build an overall tensor as:
\begin{shadefundtheory}
\beq
g_{\mu\nu}=\eta_{\mu\nu}
  +\left(\delta_{\mu 1}\delta_{\nu 1} - \delta_{\mu 2}\delta_{\nu 2}\right) h_{+}
  +\left(\delta_{\mu 2}\delta_{\nu 1} + \delta_{\mu 1}\delta_{\nu 2}\right) h_{\times}\,,
\eeq
\end{shadefundtheory}

We remind that the gravity perturbation would act on the tetrad system as follows:
\beq
\begin{split}
\Omega^{\mu\nu}\vc e_{\nu} &= \Omega^{\mu\nu}g_{\nu\chi}\vc e^{\chi}=\\
&= \Omega^{\mu\nu}\eta_{\nu\chi}\vc e^{\chi} + \Omega^{\mu\nu}h_{\nu\chi}\vc e^{\chi}\,,
\end{split}
\eeq
in fact, when crossing the space-time area deformed by the presence of a GW, a correction is added
to the standard transporter, in the form of a Lorentz-rotational operator. In TT-gauge, by contracting
the vectors $\vc e_{2}$ and $\vc e_{3}$ with the new transporter expression we'd get the eigenvalues
equations:
\beq
\begin{split}
\left(\nabla_{\vc u}\right)^{2} \vc e_{2} &= - \left(\frac{2\pi c}{\lambda}\right)^{2} \left(1-h_{\times}\right)\vc e_{2}\,,\\
\left(\nabla_{\vc u}\right)^{2} \vc e_{3} &= - \left(\frac{2\pi c}{\lambda}\right)^{2} \left(1-h_{\times}\right) \vc e_{3}\,,
\end{split}
\eeq
then, the net effect on the rotating tetrad for a finite elongation can be recovered by exponentiation; no computation
needed, provided we make the parameter shift:
\beq
\xi\to\xi \sqrt{1-h_{\times}}\,,
\eeq
to get:
\begin{shadefundtheory}
\beq
\label{eq:exptranspwithgravity}
\vc e_{2}'=\left(\exp\nabla_{\vc u}\right) \vc e_{2} =
\cos \left(\frac{2\pi c}{\lambda}\sqrt{1-h_{\times}}\xi\right) \vc e_{2} + \sin \left(\frac{2\pi c}{\lambda}\sqrt{1-h_{\times}}\xi\right) \vc e_{3}\,.
\eeq
\end{shadefundtheory}
Since $h_{\times}$ is small we get:
\beq
\begin{split}
\vc e_{2}'=&
\left(\cos \left(\frac{2  \pi  c }{\lambda }\xi \right)+h_{\times}\frac{\pi  c}{\lambda} \xi  \sin
   \left(\frac{2 \pi c }{\lambda }\xi\right) \right)\vc e_{2}+\\
   &+\left(\sin \left(\frac{2  \pi  c }{\lambda }\xi \right)-h_{\times}\frac{\pi  c}{\lambda} \xi  \cos
   \left(\frac{2 \pi c }{\lambda }\xi\right)\right)\vc e_{3}\,,
\end{split}
\eeq
assume then for simplicity a multiple integer value $k$ of $\nicefrac{\lambda}{c}$ for the parameter $\xi$;
we'll then get:
\beq
\vc e_{2}'= \vc e_{2} -h_{\times} k \pi \vc e_{3}\,,
\eeq
and the final effect over $k$ wavelengths amounts on summing the space-time strain $h_{\times}$
acting as a phase shift over the tetrad. Notice this extra phase is what we can really measure by laser
interferometry, and since $k=\left[\nicefrac{L}{\lambda}\right]$ - where $L$ now represents the flat-space
distance between the mirrors and the square brackets designate integer part - it is straightforward to see that the longer the detection arm, the highest the precision
in measuring the strain.

\begin{figure}
\includegraphics[width=\textwidth]{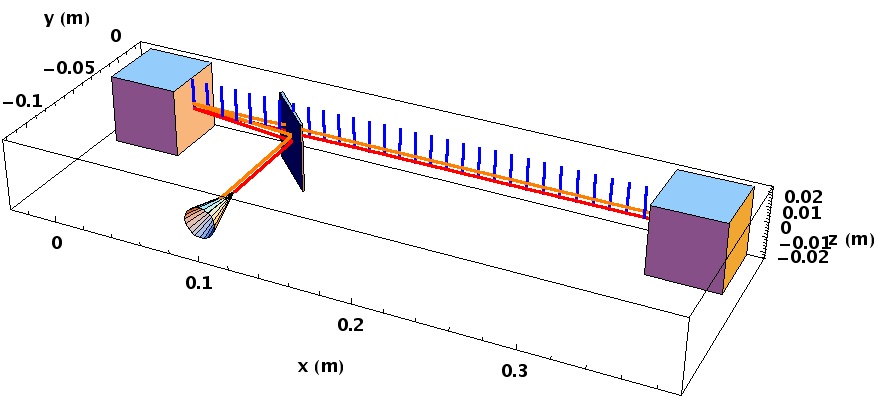}
\caption{Interferometric measure of distance.}
\label{fig:laser5}
\end{figure}

\section{Spurious effects}

A tetrad attached to the photons in the light
beam will be rigidly tied to the reference mirror surface along $\vc e_{1}$. In spite of any shielding we
may put around the mirror, residual stray forces as well as electromagnetic couplings may still be there, though
reduced: the effect would be to add an effective acceleration to the mirror, inducing an extra phase-shift
to the laser-beam. In formulae, a spurious Fermi-Walker\footnote{We may still think the tetrad starting
orientation along the mirror surface to be fixed and unaffected by mirror jitters around $\vc e_{1}$}
transporter gets added to the original one:
\beq
\Omega^{\alpha\beta}=\tilde a^{\alpha}u^{\beta}-\tilde a^{\beta}u^{\alpha}\,,
\eeq
where the $4$-acceleration $\tilde {\vc a}$ may be taken in the form:
\begin{shadefundtheory}
\beq
\frac{e}{m} F^{\mu}_{\phantom{\mu}\chi}u^{\chi}+\frac{1}{m}f^{\mu}\,,
\eeq
\end{shadefundtheory}
\noindent thus embedding EM forces and couplings in the Faraday stress-tensor $F^{\mu\nu}$ \index{Faraday tensor}
and stray, residual
couplings into $f^{\mu}$. These last can of course be of any origin, from mechanical to static gravitational,
to gradients of temperatures.

It is not customary to consider the problem on such a perspective. More often one solves the Einstein
equations for a given energy-momentum distribution, deduces the form of the $g^{\mu\nu}$ metric and the
related connection $\Gamma^{\mu}_{\phantom{\mu}\nu\sigma}$ and then calls ``geodesics'' the
solution to the null geodesic equation in the given metric. If the ``extra'' accelerations are small as perturbations,
the two ways are equivalent. We can still call geodesics those curves described in proper time by
bodies in free fall in the globally unperturbed metric, and study the sources of acceleration noise
causing the oscillations around these ``ground state'' geodesics.

\begin{figure}
\includegraphics[width=0.49\textwidth]{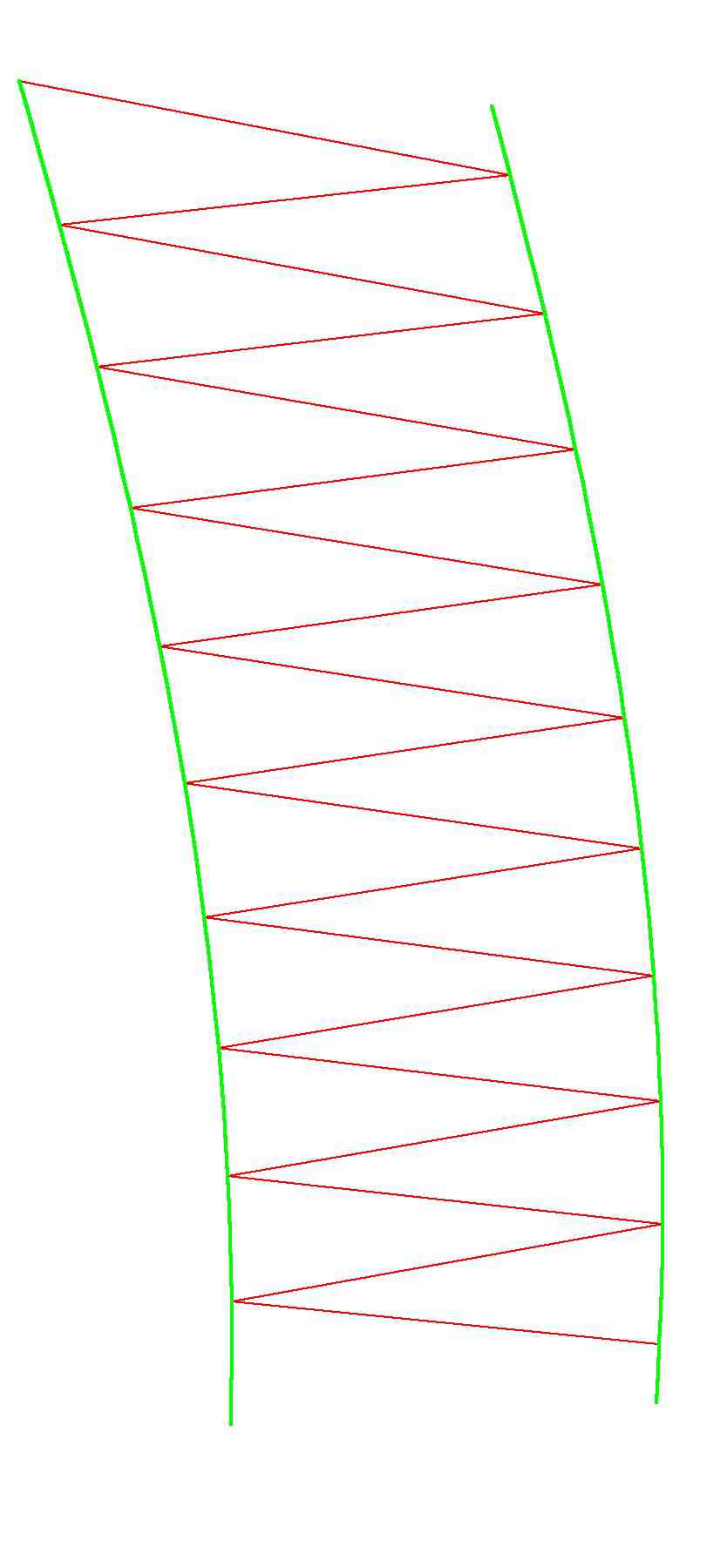}%
\includegraphics[width=0.50\textwidth]{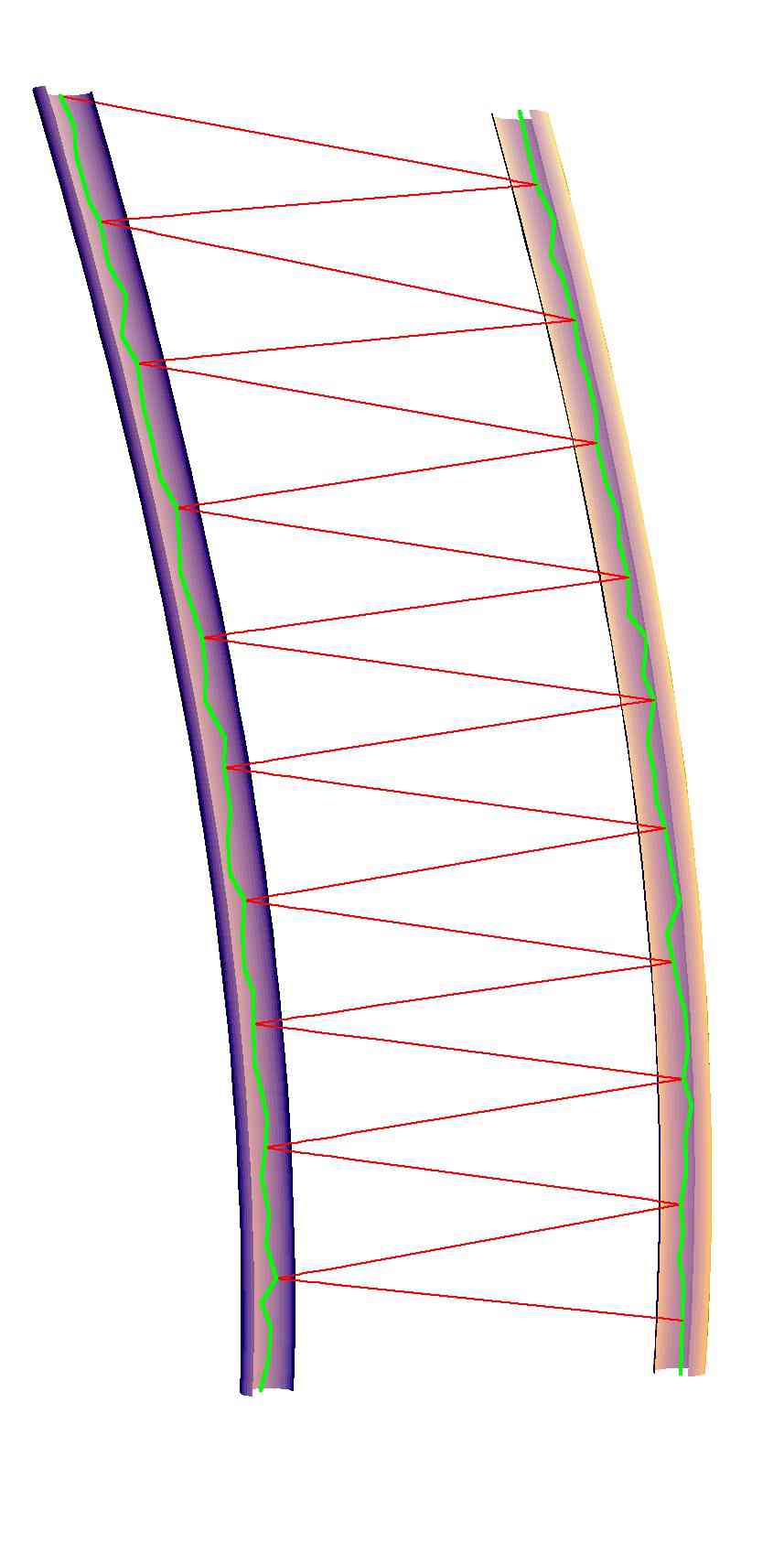}
\caption{Geodesics of freely falling mirrors in space-time. Simplified picture in $3$ dimensions; geodesics are traced in space-time as
curves (green) parametrised along the up-down direction on proper time. A laser beam shone from the first bounces on the second and
then back. Left: unperturbed geodesics in absence of any residual acceleration. Right: noisy geodesics embedded into maximal space-like
circles whose norm is the perturbation scale; as function of the proper time the embedding takes the shape of a tube per each mirror.}
\label{fig:tubes}
\end{figure}

To build up the EM spurious acceleration term, a Faraday stress tensor term must be coupled with a
time-like vector representing a velocity. This last can be thought as composed by a drifting one, having
a specific static orientation and a random one, highly variable in orientation: they both couple to
high-frequency and low-frequency parts of $F^{\mu\nu}$.
The new geodesics oscillate around the unperturbed ones; the effect on the spatial components can be upper limited
by the norm of the random perturbation on a small time-scale (rapid oscillations) so that we can encompass it in a ``circle''
 at fixed proper
time. Along the curve on the proper time parameter the geodesic perturbation is thus embedded in a tube.
For reference, see figure (\ref{fig:tubes}).

The nature of the spurious acceleration needs to be discussed more thoroughly:
\begin{enumerate}
\item it's strongly space-dependent and localised: both the static and dynamic components depend
on EM charge and currents distributions surrounding of the mirrors and sourcing electric and magnetic fields, and
even for external causes (say, for instance the interplanetary magnetic field) the effect is rendered local by parasitic currents
induced in conductors or in the mirrors themselves. Local charges, of static (DC) electric nature and parasitic currents
will dominate the scene, justifying a low-velocity approximation in the estimate and
a predominance of space derivatives and related momenta over the time ones:
\beq
\left|\partial_{x}\right| \gg \left|\partial_{t}\right|\,.
\eeq
\item The static contribution of EM and mechanical nature is highly predominant, thus the ``drift'' problem cannot be ignored.
Therefore the concept of free-fall is not suitable to build an experiment under these circumstances: it is rather preferable
to guarantee local motion to be almost-inertial,
actively compensating for any drifting term by controlling the motion of the (first) reference mirror, ``suspending'' it
to reduce the parasitic spring coupling to the shield too.
Inertiality of this ``spacecraft'' can be naturally provided by going to outer space and
exploiting the Keplerian gravitation around the Sun in the Solar System:
motion is in fact inertial in the $5$ Lagrangian points of the Earth-Sun gravitational potential.
Obviously for the noise and drag issue dedicated tactics must be planned
and the device purposefully studied.
\item Dynamic contributions are faster Fourier components on the background of the static ones: basically magnetic
or electric transients in nature, can be nevertheless thought to be suppressed by the $\nicefrac{e}{m}$ dependence. The larger and more
sophisticated the conductors, the more unpredictable the correlated effect could be; we'll have more space to discuss these effects in
the noise chapter and we'll retain only first-order components in the velocities here.
\end{enumerate}

Our conclusion on the spurious acceleration is the following:
\begin{itemize}
\item it may be thought as an additional acceleration operator, whose effect is adding an undesired spatial offset,
variable with time, to the positions of the mirrors. In other words, curvature picks up a locally generated term
originated by many sources mainly of EM and static gravitational nature.
The detection of tidal effects by perturbed freely-falling mirrors is thus jeopardised by an extra phase,
variable with time, picked up by the laser on its travel and due to the real, gauge-invariant change in position of
the mirrors: we will describe it as an effect
in the mirrors velocities;
\item according to Mach's principle it is impossible to distinguish the local from the non-local source since
tidal effects sum linearly; nevertheless EM fields are gauge invariant objects, therefore observable and
measurable: this is valid for fields generated by local distributions of charges and currents but also for external ones.
The same occurs for local gravitational contributions, which can be very well approximated up to quadrupole
expansions. In summary this part can be detected and ``projected'' out of the noise picture.
\item Limiting the jitter and compensating for the drift are the only feasible methods to mimic a condition
of motion similar to theoretical free-fall. We call the first strategy ``noise reduction'' and the second ``drag-free''.
\end{itemize}

In the low velocity approximation the perturbed geodesic acceleration
equation becomes (see appendix for a proof, eq. \eqref{eq:geodesylowspeed}):
\begin{shadefundtheory}
\beq
\frac{\de^{2} x^{i}}{\de t^{2}}=
   \left(
   -\Gamma^{i}_{\phantom{i}00}-2 \Gamma^{i}_{\phantom{i}0j}v^{j}
   +\Gamma^{0}_{\phantom{0}00}v^{i}
   \right)+
e \left(F^{i}_{\phantom{i}0}+F^{i}_{\phantom{i}k}v^{k}\right)
+\left(f^{i}-f^{0}v^{i}\right)\,.
\eeq
\end{shadefundtheory}
\noindent and if the gauge choice is TT,
\beq
\begin{split}
\label{eq:ttgaugedmotion}
\frac{\de^{2} x^{i}}{\de t^{2}}=\frac{\de v^{i}}{\de t} &=
a^{i}-2 \Gamma^{i}_{\phantom{i}0j}v^{j}
+\frac{e}{m} \left(F^{i}_{\phantom{i}0}+F^{i}_{\phantom{i}k}v^{k}\right)
+\left(f^{i}-f^{0}v^{i}\right) =\\
&= a^{i}-2 \Gamma^{i}_{\phantom{i}0j}v^{j} +\frac{e}{m} \left(E^{i} + \epsilon^{ijk}v_{j}B_{k}\right)
+\left(f^{i}-f^{0}v^{i}\right)\,,
\end{split}
\eeq
where we expanded the expression of the Faraday tensor and introduced the electric $\vc E$ and magnetic $\vc B$ fields.
The term obviously hides all the geometric
peculiarities of the machinery around the mirror, but we can deal with the whole of it later, in the noise section. Notice this
set of terms embodies our former choices: they are strongly distributional-dependent, slowly varying, locally sourced.

We can now introduce the definition of correlator\index{Convolution}:
\beq
C_{\Delta\theta}(\tau)=\int_{-\infty}^{\infty}\Delta\theta(t)\Delta\theta(\tau - t)\de t\,,
\label{eq:convolution}
\eeq
so that the squared Power Spectral Density (PSD)\index{Power spectral density} associated with the phase correlator
is the Fourier transform of the last expression:
\beq
S_{\Delta\theta}(\omega) = \int_{-\infty}^{\infty} e^{-\imag \omega \tau} C_{\Delta\theta}(\tau)\de\tau\,.
\label{eq:generalPSD}
\eeq
Hence, \eqref{eq:ttgaugedmotion} may be rearranged and converted to squared PSD, assuming the terms are all uncorrelated as: 
\beq
S_{a} \simeq \omega^{2} S_{v} + S_{\text{GW, tidal}}+S_{\text{EM, noise}}+S_{\text{extra, noise}}\,,
\eeq
as it can be seen, several competitors concur to the residual acceleration PSD in the former equation.
%

\section{Detection}

According to formula \eqref{eq:exptranspwithgravity} and by simple geometric arguments,
the cosine of the angle $\theta(\xi)$ instantaneously spanned between $\vc e_{2}$ and $\vc e_{2}'$ is:
\beq
\cos \theta \left(\xi\right) = \vc e_{2}'\cdot \vc e_{2} = 
\cos \left(\frac{2\pi c}{\lambda}\sqrt{1-h_{\times}}\xi\right)\,,
\eeq
so that, now replacing $h_{\times}$ with a generic wave-strain\index{Gravitational wave strain} $h$:
\begin{shadefundtheory}
\beq
\theta \left(\xi\right) = \frac{2\pi c}{\lambda}\sqrt{1-h}\xi
= \frac{2 \pi c \xi}{\lambda }-\frac{ \pi c \xi }{\lambda }h + O\left(h^{2}\right)\,
\eeq
\end{shadefundtheory}
\noindent by dimensional power counting, we deduce now that $\xi$ must be a time, in fact, since
$h$ is dimensionless and the argument of transcendental functions have to be
dimensionless too, we have $\left[\xi\right] \sim \left[\nicefrac{\lambda}{c}\right] = \left[\text{time}\right]$
and identify $\xi\equiv t$.
Moreover, we may very well think $h$ to be small, but we need to consider
arbitrary lengths of the laser beam path, or conversely arbitrary frequency span of 
the GW perturbation, at least in principle. We may think of $h=h(t)=h_{0}\cos \omega_{\text{GW}}t$,
where $h_{0}$ can be thought as a slowly varying function of time so to be considered almost constant over
a large multiple of $\lambda_{\text{GW}}$. In this sort of rapidly rotating wave approximation
we get:
\beq
\theta \left(t\right) \simeq
\frac{2 \pi  c}{\lambda }t-\frac{\pi  c}{\lambda } t h_{0}\left(t\right)\cos \left(\omega _{\text{GW}}t\right)\,
\eeq
by time derivative we get to order $h$:
\beq
\begin{split}
\frac{\de \theta(t)}{\de t} =&
-\frac{\pi c }{\lambda }h_0(t)\cos \left(\omega _{\text{GW}}t\right)
+\frac{\pi c}{\lambda} \omega _{\text{GW}} t\,h_0(t) \sin \left(\omega _{\text{GW}}t \right) +\\
   &-\frac{\pi c}{\lambda }t\,h_0'(t) \cos \left(\omega _{\text{GW}}t\right)
   +\frac{2\pi  c}{\lambda }\,.
\end{split}
\eeq
We now wish to evaluate this last across one full reflection period between the mirrors, i.e.
from $t=\tau$, till $t=\tau+\Delta T_{\text{flight}}$ and this latter to $t=\tau+2 \Delta T_{\text{flight}}$,
where $\Delta T_{\text{flight}}=\nicefrac{L}{c}$ is the laser flight time between unperturbed
mirrors in flat curvature. We get:
\beq
\begin{split}
\frac{\de \theta}{\de t}&\Bigr|_{\tau-\frac{L}{c}}-\frac{\de \theta}{\de t}\Bigr|_{\tau}+
\frac{\de \theta}{\de t}\Bigr|_{\tau-2\frac{L}{c}}-\frac{\de \theta}{\de t}\Bigr|_{\tau-\frac{L}{c}}=\\
=&\frac{\pi c}{\lambda }
  \Bigl(  h_0(\tau ) \cos \left( \omega _{\text{GW}}\tau \right) 
-\tau \omega _{\text{GW}} h_0(\tau ) \sin \left(\omega _{\text{GW}}\tau \right)  +\\
&-h_0\left(\tau-\frac{2 L}{c}\right) \cos \left( \omega _{\text{GW}}\left(\tau -\frac{2 L}{c}\right)\right) +\\
&+\left(\tau -\frac{2 L}{c}\right) \omega _{\text{GW}} h_0\left(\tau -\frac{2 L}{c}\right)\sin \left(\left(\tau -\frac{2 L}{c}\right) \omega _{\text{GW}}\right)
\Bigr)\,,
\end{split}
\eeq
where we took $h_{0}'(t)\simeq 0$. Finally, in the low frequency GW approximation, we get, back from $\tau$ to $t$:
\begin{shadefundtheory}
\beq
\label{eq:dDeltatheta}
\frac{\de \Delta \theta(t)}{\de t} =\frac{\pi c}{\lambda }  \left(h(t)-h\left(t -\frac{2 L}{c}\right)\right)+O\left(\omega _{\text{GW}}^2\right)\,.
\eeq
\end{shadefundtheory}
\noindent For low-frequency almost-plane GWs a laser beam shone between two mirrors in drag-free motion with respect to each-other suffers
a phase shift whose instantaneous time derivative depends only on the wave strain evaluated at the laser shining point. If we now
name $\omega_{0}=\nicefrac{2\pi c}{\lambda}$, the former equation shows that the relative variation of
pulsation is only a function of the causal strain difference \cite{Anza:2005td}:
\beq
\frac{\Delta\omega(t)}{\omega_{0}}\simeq \frac{1}{2}\left(h(t)-h\left(t -\frac{2 L}{c}\right)\right)\,.
\label{eq:deltaomegaphizero}
\eeq
In fact, such an estimate is true for any polarisation of the incoming GW. To display the formula in its
full glory we may define an angle $\phi$ span on the common plane by the laser beam and the ``Poynting''
vector of the GW, to write:
\beq
\begin{split}
\frac{\Delta\omega(t)}{\omega_{0}}\simeq & \frac{1}{2} \left(h_{+}(t)-h_{+}\left(t -\frac{2 L}{c}\right)\right)\cos 2\phi +\\
&+ \frac{1}{2} \left(h_{\times}(t)-h_{\times}\left(t -\frac{2 L}{c}\right)\right) \sin 2\phi\,,
\end{split}
\eeq
which reduces to \eqref{eq:deltaomegaphizero} for optimal orientation of the detector, $\phi=0$.

\section{Noise}

As mentioned by means of general arguments, if the two mirrors are in motion, i.e. their velocity is not null with respect to
one another, then there's a real shift in position and hence in mutual distance with respect to the perfect TT frame. Obviously
the accuracy of the TT-gauge definition in itself is not affected by such a motion, but the arising acceleration competes with the
curvature induced by the GW perturbation $h$ and focused in eq. \eqref{eq:dDeltatheta}. In other words, if we'd like to
depict the scene on the pure accounts of coordinates, detected by some sophisticated readout device as our laser
(or an electrostatic capacitance detector), the shift in phase can be easily evaluated from the shift in position:
\beq
\begin{split}
\Delta\theta(t)=\frac{2\pi}{\lambda}\Bigl(&x_{1}(t)-x_{2}\left(t-\frac{L}{c}\right)+\\
  +&x_{1}\left(t-\frac{2 L}{c}\right) -x_{2}\left(t-\frac{L}{c}\right)\Bigr)
  \label{eq:motionshift}\,,
\end{split}
\eeq
where $x_{1}(t)$ is the coordinate of the mirror sending and collecting the laser beam, while $x_{2}(t)$ is that of
the reflecting one. The shift is calculated to first order in $\nicefrac{v}{c}$, and if the frequency of measurement is $\ll
\nicefrac{c}{L}$ we may approximate \eqref{eq:motionshift} to
\beq
\Delta\theta(t)=\frac{4\pi}{\lambda} \Delta L(t)\,.
\eeq
where $\Delta L(t)=x_{1}(t)-x_{2}(t)$.

The former argument is somehow of questionable value when crossed with the TT-gauge
demands. It is not easy to deal with observables and measurability in GR, but
one sure thing is that distance is not an observable quantity. That's why we've been spending so much time
in building a distance estimator not relying on any absolute distance,
but the fixed velocity scale $c$. Conversely the laser phase shift or better its pulsation shift is
a directly measurable object, and a causal carrier of the effect of a gravitational distortion in space-time.
We'd prefer to convert the former argument into one on velocities and phases: according
to the geodesic equation in TT-gauge if bodies are idle to start with, they pick up no further acceleration in time;
estimates on velocities and pulsations are more reliable and in the correct spirit though.
The equivalent of \eqref{eq:motionshift} is thus:
\begin{shadefundtheory}
\beq
\frac{\Delta\omega(t)}{\omega_{0}}=\frac{1}{c}\Bigl(v_{1}(t)- 2 v_{2}\left(t-\frac{L}{c}\right)
  +v_{1}\left(t-\frac{2 L}{c}\right) \Bigr) \,,
\label{eq:deltaomegavel}
\eeq
\end{shadefundtheory}
\noindent and to very low frequency with respect to $\nicefrac{c}{L}$ we get:
\beq
\frac{\Delta\omega(t)}{\omega_{0}}\simeq\frac{2\Delta v(t)}{c}\,.
\eeq
where now $\Delta v(t)\doteq v_{1}(t)-v_{2}(t)$.

In order to open a detailed discussion on noise, correlators and PSDs of the phase shift must be built in time. 
Similar quantities can be built for each velocity signal $v_{i}(t),\,i=1,2$ and for the instantaneous velocity difference
$\Delta v(t)$: $S_{\Delta v}(\omega)$, $S_{v_{i}}(\omega),\,i=1,2$, are the PSD of the sub-indicated quantities at the
frequency $\omega$. We are implicitly assuming velocities to be joint stationary random processes, so that from the
Doppler shift in equation \eqref{eq:deltaomegavel} we deduce a squared PSD
due to non-tidal (non-GW) motion with the following form:
\begin{shadefundtheory}
\beq
\begin{split}
S_{\nicefrac{\Delta \omega}{\omega_{0}}}(\omega) =& \frac{4 S_{\Delta v}(\omega)}{c^{2}} \cos \left( \frac{\omega L}{c}\right)+ \\
  &+ 8 \sin^2 \left( \frac{\omega L}{2 c}\right) \left(
    \frac{S_{v_{2}}(\omega)}{c^{2}} - \cos \left( \frac{\omega L}{c}\right) \frac{S_{v_{1}}(\omega)}{c^{2}}
    \right)
 = \\
\simeq & \frac{4 S_{\Delta v}(\omega)}{c^{2}}\,, \label{eq:psdlowfreqvel}
\end{split}
\eeq
\end{shadefundtheory}
\noindent and the last approximation holds for $\nicefrac{\omega L}{c} \ll 1$.

Summarising the difference of forces acting on the mirrors as $\Delta F = F\left(\vc x_{2}\right) - F\left(\vc x_{1}\right)$,
according to the geodesic deviation equations \eqref{eq:geodevcovar} or
\eqref{eq:geodevextrasimple} to order $h$ and up to order $v$ in the velocities we'd get:
\beq
m \frac{\de v^{i}(t)}{\de t}\simeq -m\dot h_{j}^{\phantom{j}i} v^{j}(t)+ \Delta F(t)\,,
\label{eq:vmotion}
\eeq
assume then the usual form for an incoming GW: $h(t)=h_{0}\cos \omega_{\text{GW}} t$, where $h_{0}$ is
a profile function so slowly varying with time we can consider it almost constant, very small in amplitude. The
Fourier-space implicit propagator obtained from eq. \eqref{eq:vmotion} looks like:
\beq
\frac{\Delta F(\omega)}{m} = \frac{1}{2} \imag \left(-2 \omega  v(\omega )- h_0 \omega _{\text{GW}} v\left(\omega -\omega
   _{\text{GW}}\right)+ h_0 \omega _{\text{GW}} v\left(\omega +\omega _{\text{GW}}\right)\right)
\eeq
and in the limit of small $h_{0}$ amplitude we get for the square modulus of the velocity in Fourier space:
\beq
\left|v(\omega)\right|^{2}=\frac{\Delta F^{2}(\omega )}{m^2 \omega^2}+O\left(h_0\right)\,,
\eeq
thus resulting in a velocity squared PSD:
\beq
S_{v}(\omega)\simeq\frac{S_{\Delta F}(\omega )}{m^2 \omega^2}\,.
\eeq
Employing the relation between velocity and variation of pulsation in Fourier domain, eq. \eqref{eq:psdlowfreqvel},
we can deduce the conversion relation between force PSD and laser pulsation variation PSD:
\begin{shadefundtheory}
\beq
S_{\nicefrac{\Delta\omega}{\omega_{0}}}\simeq \frac{4 S_{\Delta v}(\omega)}{c^{2}} = 
\frac{4}{c^{2}} \frac{S_{\nicefrac{\Delta F}{m}}(\omega )}{\omega^2}\,,
\label{eq:forcesshiftpsd}
\eeq
\end{shadefundtheory}
\noindent thus any difference of force acting on the mirrors would induce a phase variation in the laser according to
the latter expression. Notice the effect is suppressed by a factor $\nicefrac{1}{\omega^{2}}$. We conclude that a TT-gauged
frame can be built in space-time by means of drag-free mirrors, provided external forces in difference are
suppressed in the measurement bandwidth to the point of being considered negligible. The real observable in this
scenario is the laser phase; apart from passing-by considerations, we never introduced or relied in absolute space
or distances with the exception of the laser wavelength, a space elongation marked in fact by a phase.

One final remark here concerns the real definition of a length and time 
standard on-board a space mission. These are available from the interferometer and the time stamp of the data.
 
Again, the key feature for
 absolute calibration of the difference of displacement signal is the conversion relation between laser phase
 and equivalent displacement by
\beq
{\Delta}x = \frac{{\Delta}\phi}{2\pi}\lambda\,,
\eeq
where $\lambda$ is the laser wavelength.
The absolute distance calibration is then limited by the 
combined accuracy of the knowledge of $\lambda$ and that of the conversion of the phase meter signal into 
radians or cycles, the latter being evaluated to $\nicefrac{50}{10^{6}}$ accuracy\footnote{Danzmann, K. and Heintzel, G., private communication.}.

The scale factor of the time tagging of data is the other intervening factor. Noticeably what matters is the 
absolute accuracy on time intervals and not the definition of an absolute universal time. On purpose the LTP experiment
will carry clocks on-board with accuracy $\sigma_{\nicefrac{{\Delta}t}{t}} \simeq 10^{-6}$, which will
provide a definition of experimental time beats and a time length comparable to the intrinsic laser $\nicefrac{1}{\nu} = \nicefrac{\lambda}{c}$.

\section{Laser interferometers and phase shift}

The suppression of forces $\Delta F$ with PSD $S_{\Delta F}$ with non-gravitational origin, local
or non-local be their nature, is mandatory in the measurement bandwidth to ensure the dominance of GW spectrum.
We may now clear the smoke and start calling the mirrors and their envelope as LISA or LTP, since these missions
will be embodying the abstract concepts we put at play so far.
It's impossible to annihilate every disturbance aboard the LISA space-crafts and hence a requirement over forces PSD
has been cast, demanding:
\begin{shadefundnumber}
\beq
S^{\nicefrac{1}{2}}_{\nicefrac{\Delta F}{m},\,\text{LISA}}(\omega)
  \leq \sqrt{2}\times 3\times 10^{-15}\,\accPSDunit
\label{eq:lisamainreq}\,,
\eeq
\end{shadefundnumber}
\noindent for a frequency $f > 0.1\,\unit{mHz}$. This corresponds to a relative pulsation shift PSD in adherence to
\eqref{eq:forcesshiftpsd} like:
\beq
S^{\nicefrac{1}{2}}_{\nicefrac{\Delta\omega}{\omega_{0}},\,\text{LISA}}
  = \frac{c\omega}{2} S^{\nicefrac{1}{2}}_{\nicefrac{\Delta F}{m},\,\text{LISA}}(\omega)
  \simeq \frac{2.83\times 10^{-23}}{\omega\,\nicefrac{1}{Hz}}\,\unitfrac{1}{\sqrt{Hz}}\,.
\label{eq:forcesshiftpsdfinal}
\eeq

The interferometer itself provides some measurement noise, expressed as an equivalent optical path fluctuation
${\delta}x$ for each passage of the light through the interferometer arm. A single arm interferometer hence suffers a
relative pulsation shift per pass:
\beq
\frac{\Delta \omega(t)}{\omega_{0}}\simeq \frac{1}{c}\frac{\de {\delta}x(t)}{\de t} = \frac{\omega}{c} {\delta}x(t)\,,
\eeq
so that, back and forth the added equivalent PSD square of the noise will be
\beq
S_{\nicefrac{\Delta \omega}{\omega_{0}},\,\text{laser}}(\omega)\simeq 2 \frac{\omega^{2}}{c^{2}} S_{{\delta}x}(\omega)
\label{eq:interfshiftpsd}\,,
\eeq
or, in terms of accelerations, by virtue of \eqref{eq:forcesshiftpsd}:
\begin{shadefundtheory}
\beq
S^{\nicefrac{1}{2}}_{\nicefrac{\Delta F}{m},\,\text{laser}}(\omega)
\simeq \frac{\omega^{2}}{\sqrt{2}} S_{{\delta}x}^{\nicefrac{1}{2}}(\omega)\,.
\label{eq:interfdeltafovermpsd}
\eeq
\end{shadefundtheory}
\noindent The corresponding requirement for the interferometer is to achieve a path length noise\index{Interferometer path noise}
of $S_{{\delta}x}^{\nicefrac{1}{2}}\leq 20\,\unitfrac{pm}{\sqrt{Hz}}$, so that finally
from \eqref{eq:interfshiftpsd} or \eqref{eq:interfdeltafovermpsd} we may deduce the following figures:
\begin{align}
S^{\nicefrac{1}{2}}_{\nicefrac{\Delta \omega}{\omega_{0}},\,\text{laser}}(\omega)
  &\simeq 9.43\times 10^{-20}\frac{\omega}{\unit{Hz}}\unitfrac{1}{\sqrt{Hz}}
\label{eq:interfshiftpsdfinal}\,,\\
S^{\nicefrac{1}{2}}_{\nicefrac{\Delta F}{m},\,\text{laser}}(\omega)
  &\simeq 1.41\times 10^{-11} \frac{\omega^{2}}{\unit{Hz}^{2}} \accPSDunit\,.
\end{align}

\begin{figure}
\begin{center}
\includegraphics[width=\textwidth]{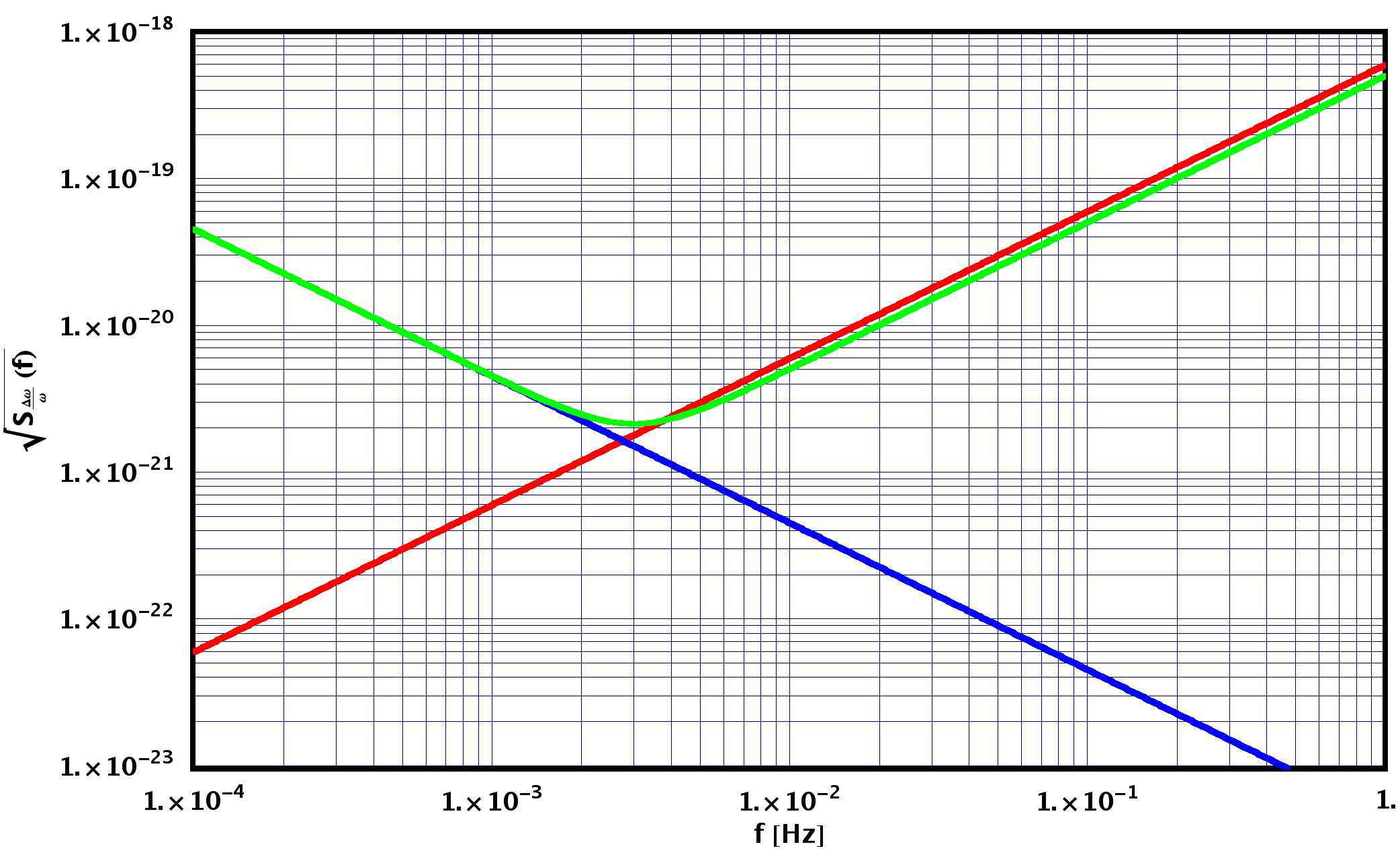}
\includegraphics[width=\textwidth]{figures/noisepsdforce.jpg}
\caption{Upper: noise PSDs in $\nicefrac{\Delta\omega}{\omega_{0}}$ for forces difference (blue), interferometer (red) and relaxed noise
requirement of forces difference (green). Green line represents LISA's targeted sensitivity. Lower: noise PSDs in $\nicefrac{\Delta F}{m}$,
same colour codes.}
\label{fig:noisepsds}
\end{center}
\end{figure}

The force noise in \eqref{eq:lisamainreq} and the interferometer one in
\eqref{eq:interfdeltafovermpsd}
cross at $\omega\simeq 2\pi\times 2.75 \, \unit{mHz} \sim 2\pi\times 3\, \unit{mHz} \doteq \omega_{c}$, thus allowing to relax the
requirement in \eqref{eq:lisamainreq} to:
\begin{shadefundnumber}
\beq
\begin{split}
S^{\nicefrac{1}{2}}_{\nicefrac{\Delta F}{m},\,\text{LISA}}(\omega) &\to
S^{\nicefrac{1}{2}}_{\nicefrac{\Delta F}{m},\,\text{LISA}}(\omega)\left(1+\left(
    \frac{\omega}{\omega_{c}}\right)^{4}
  \right)^{\nicefrac{1}{2}} =\\
&=\sqrt{2}\times 3\times 10^{-15} \left(1+\left(
    \frac{\omega}{2\pi \times 3\, \unit{mHz}}\right)^{4}
  \right)^{\nicefrac{1}{2}}\,\accPSDunit\,.
\label{eq:deltafoverfrelax}
\end{split}
\eeq
\end{shadefundnumber}
\noindent This last formula is in fact a minimal lower interpolation of  \eqref{eq:lisamainreq} and
\eqref{eq:interfdeltafovermpsd}. Graphs of these latter together with \eqref{eq:deltafoverfrelax}
and their $S_{\nicefrac{\Delta\omega}{\omega_{0}}}^{\nicefrac{1}{2}}$ equivalents
as functions of frequency can be inspected in figure \ref{fig:noisepsds}.
In limiting cases we'd get from \eqref{eq:deltafoverfrelax}:
\beq
\begin{split}
S^{\nicefrac{1}{2}}_{\nicefrac{\Delta F}{m},\,\text{LISA}}(\omega) \underset{\omega \ll 2\pi\times 3\,\unit{mHz}}{=}
 & 4.24\times 10^{-15}\,\accPSDunit+O\left(\omega ^2\right)\\
S^{\nicefrac{1}{2}}_{\nicefrac{\Delta F}{m},\,\text{LISA}}(\omega) \underset{\omega \gg 2\pi\times 3\,\unit{mHz}}{=}
 & 1.20\times 10^{-11} \left(\frac{\omega}{\unit{Hz}}\right)^{2}\,\accPSDunit+O\left(\left(\frac{1}{\omega }\right)^2\right)
\end{split}
\eeq
This requirement needs to be qualified and it is not testable on ground \cite{Carbone:2003nb,CarbonePhD}.
By virtue of \eqref{eq:psdlowfreqvel} we
deduce also that limiting the noise in speed difference only by limiting forces in difference may become inaccurate
for frequencies larger than $3-4\,\unit{mHz}$. However, the assumptions to get to \eqref{eq:psdlowfreqvel} are
very reliable and if the fluctuations in velocity of the mirrors are independent, the ``difference of forces'' approach
represents a worst-case occurrence. If the noise would be partly correlated, the dangerous part of it would still be the
one mimicking residual differential acceleration. 

We'd like finally to give one more link between PSD and curvature as follows. The above discussion has been cast in
terms of speed frames and frequency shifts; the requirements can nevertheless be restated in terms of components of
the Riemann tensor only. Back to eq. \eqref{eq:deltaomegaphizero}, we can write
\beq
\frac{\Delta\omega(\omega)}{\omega_{0}} = \frac{1}{2} \left(1-e^{-\imag\frac{2  L \omega }{c}}\right) h(\omega )
\underset{\omega\ll \nicefrac{c}{2 L}}{=}
\imag \omega \frac{ L}{c}h(\omega)+O\left(\omega ^2\right)\,,
\eeq
to find, by means of \eqref{eq:forcesshiftpsd}:
\begin{shadefundtheory}
\beq
S^{\nicefrac{1}{2}}_{h}(\omega) \simeq \frac{2}{L\omega^{2}}S^{\nicefrac{1}{2}}_{\nicefrac{\Delta F}{m}}\,.
\label{eq:hvsforcepsd}
\eeq
\end{shadefundtheory}

\noindent If we consider now the expression for the linearised Riemann tensor, we'd have:
\beq
R_{\mu\nu\rho\sigma} \simeq
   \frac{1}{2}\left(h_{\sigma\mu,\nu,\rho}
    -h_{\sigma\nu,\mu,\rho}
    +h_{\rho\nu,\mu,\sigma}
   -h_{\rho\mu,\nu,\sigma}
    \right)\,,
\eeq
specialising to radiation and TT-gauge, the only survivors would be (see \cite{dinverno} or appendix \ref{chap:gwtheory}):
\beq
R_{0k0j}(t)=-\frac{1}{2}h_{kj,0,0} (t)\,,
\eeq
therefore in Fourier space (we bring back the $c$ constant by dimensional arguments):
\begin{shadefundtheory}
\beq
R_{0k0j}(\omega)=-\frac{\omega^{2}}{2 c^{2}}h_{kj} (\omega)\,.
\label{eq:riemannandh}
\eeq
\end{shadefundtheory}
\noindent By joining \eqref{eq:hvsforcepsd} and \eqref{eq:riemannandh} we can thus deduce that every differential force
$\Delta F$ mimics a curvature noise\index{Curvature noise} with PSD:
\beq
S^{\nicefrac{1}{2}}_{R}(\omega) = \frac{\omega^{2}}{2 c^{2}} S^{\nicefrac{1}{2}}_{h}(\omega)
\simeq \frac{1}{c^{2} L} S^{\nicefrac{1}{2}}_{\nicefrac{\Delta F}{m}}(\omega)\,,
\eeq
the pre-factor can be calculated in our conditions to give:
\begin{shadefundtheory}
\beq
S^{\nicefrac{1}{2}}_{R}(\omega) \simeq 2.2\times 10^{-27}\,\unit{\frac{s^{2}}{m^{3}}}
  \times S^{\nicefrac{1}{2}}_{\nicefrac{\Delta F}{m}}(\omega)\,.
\eeq
\end{shadefundtheory}

The requirement in \eqref{eq:deltafoverfrelax} transforms into a curvature resolution of order
$10^{-41}\,\unitfrac{1}{m^{2}\,\sqrt{Hz}}$, i.e. for a signal at $0.1\,\unit{mHz}$
integrated over a cycle, this gives a resolution of order $10^{-43}\,\unitfrac{1}{m^{2}}$, a figure which
may be compared to the scale of the curvature scalar exerted by the Sun field at LISA location, about $10^{-30}\,\unitfrac{1}{m^{2}}$.

\section{The Laser Interferometer Space Antenna}

The Laser Interferometer Space Antenna (LISA) will be launched in 2017 by the combined
efforts of the ESA and NASA. Nevertheless the concept of building off-ground
interferometric detectors of GWs dates back to the 70's; quite a variety of designs were advanced at the
time \cite{Sumner:2004xf}.

More recently, laser technology allowed for designing very long baseline
detectors, and ESA received plans for the Laser Antenna for Gravitational-radiation Observation in Space
(LAGOS) project, which considered
a constellation of three drag-free satellites orbiting around the Sun at $1\,\unit{AU}$. In fact this
project looks quite similar to LISA, but the arm-lengths ranged $10^{6}\,\unit{km}$.

Seeking for alternative designs in order to validate the mission, ESA considered two
parallel proposals: LISA and SAGITTARIUS, the former orbiting around the Sun,
the latter around Earth, both extending the number of satellites to $6$.
LISA was dropped at start, probably because of the complicated space-crafts setup, each of which
hosting a test mirror, flying coupled
in pairs, with a laser arm to control mutual motion, and the spare, long-baseline one to detect
GW. Thanks to the effort of the ``Team X'' at Jet Propulsion Laboratory, conclusion was
drawn that the constellation could be reduced to $3$ satellites each hosting $2$ mirrors. Eventually
this simplification brought LISA back to the attention of the agency, where it was validated and chosen
as effective mission.

\begin{figure}
\begin{center}
\includegraphics[width=0.9\textwidth]{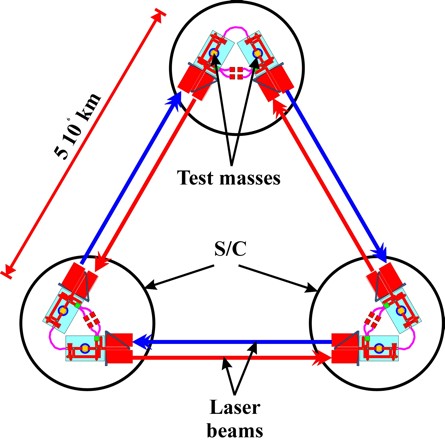}
\end{center}
\caption{Closeup on LISA constellation and laser beams across the space-probes.}
\label{fig:lisalaser}
\end{figure}

LISA is then a constellation of $3$ space-crafts (SC) orbiting at $\nicefrac{D}{2}=1\,\unit{AU}$ from the sun, sharing Earth's
orbit with some $20$ degrees delay. The space-probes form an equilateral triangle and - as mentioned already - each of them
hosts a couple of test-masses (TM) in free fall. An Electrode Housing (EH) and
a set of capacitive Gravitational Reference Sensors (GRS)
surround each TM and constitutes an Inertial Sensor (IS) capable of monitoring TM position and angular attitude.
Each IS is coupled to a telescope and a laser
and shares with the other on-board an interferometer Optical Bench (OB).
A laser beam is shone from each satellite towards the independent far satellites, gets captured by
the proper telescope there and
sent back in ``phase locking'' after hitting the ``alien'' TMs.
Figure \ref{fig:lisalaser} may help focusing the picture.

Each laser's phase is locked either to its companion on 
the same SC, forming the equivalent of a beam-splitter, or to the incoming 
light from the distant SC, forming the equivalent of an amplifying mirror, 
or light transponder.
The overall effect is that the three SC function as a Michelson
interferometer with a redundant third arm. The arm-length size, ranging $5\times 10^{6}\,\unit{km}$,
was chosen to optimise 
the sensitivity at the frequencies of known sources: increasing the arm-length
improves sensitivity to low frequency GW strain (coming from massive black
holes, for example).

\begin{figure}
\begin{center}
\includegraphics[width=\textwidth]{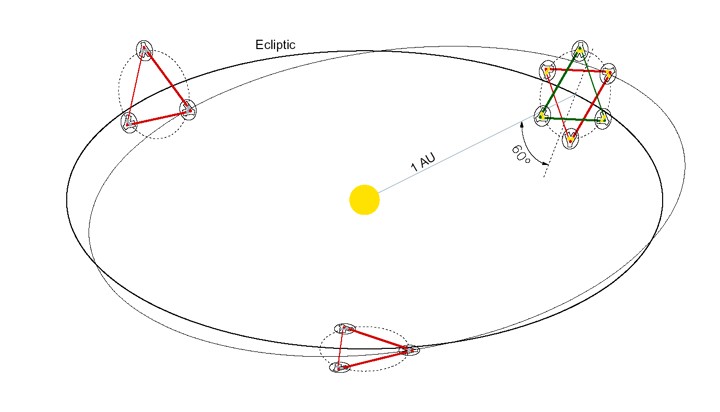}
\end{center}
\caption{LISA's orbit will be the same as Earth's, following the planet by some $20$ degrees delay. The equilateral constellation
will be rotating along its centre of mass while revolving around the Sun.}
\label{fig:lisaorbit}
\end{figure}

\begin{figure}
\begin{center}
\includegraphics[width=\textwidth, angle=-90]{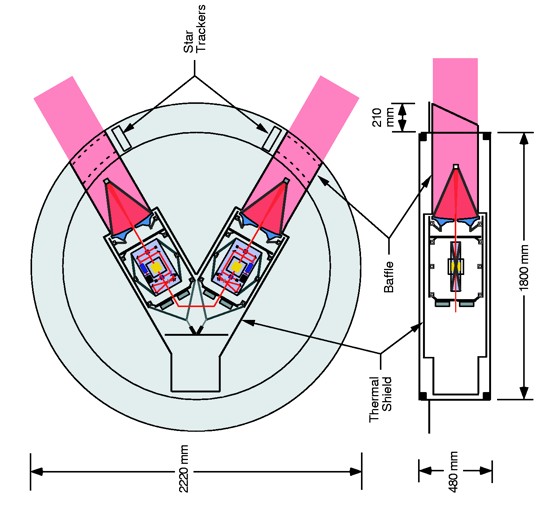}
\end{center}
\caption{LISA's SC internal structure. The so-called ``Y''-shape is critical for accomplishing the mission demands.
Top view shows the SC from top, details of the test-masses can be seen inside the telescopes. Bottom view shows the SC from the side.}
\label{fig:lisayshape}
\end{figure}

Each SC is meant as a protection against external disturbances for the TMs. Inside the SC,
the ISs and relative TMs are obviously oriented with a mutual angle of $\nicefrac{\pi}{3}\,\unit{rad}$. This
non-orthogonality of the reference allows for the so called ``drag-free'' control of SC
(see figure \ref{fig:lisayshape}): each SC is free to chase
both the TMs motion along the bisector of the ``sensing directions'' (the laser beam ones) and can re-adjust
the TMs positions by virtue of capacitance actuation voltages.

\begin{figure}
\begin{center}
\begin{tabular}{rl}
\includegraphics[height=4cm]{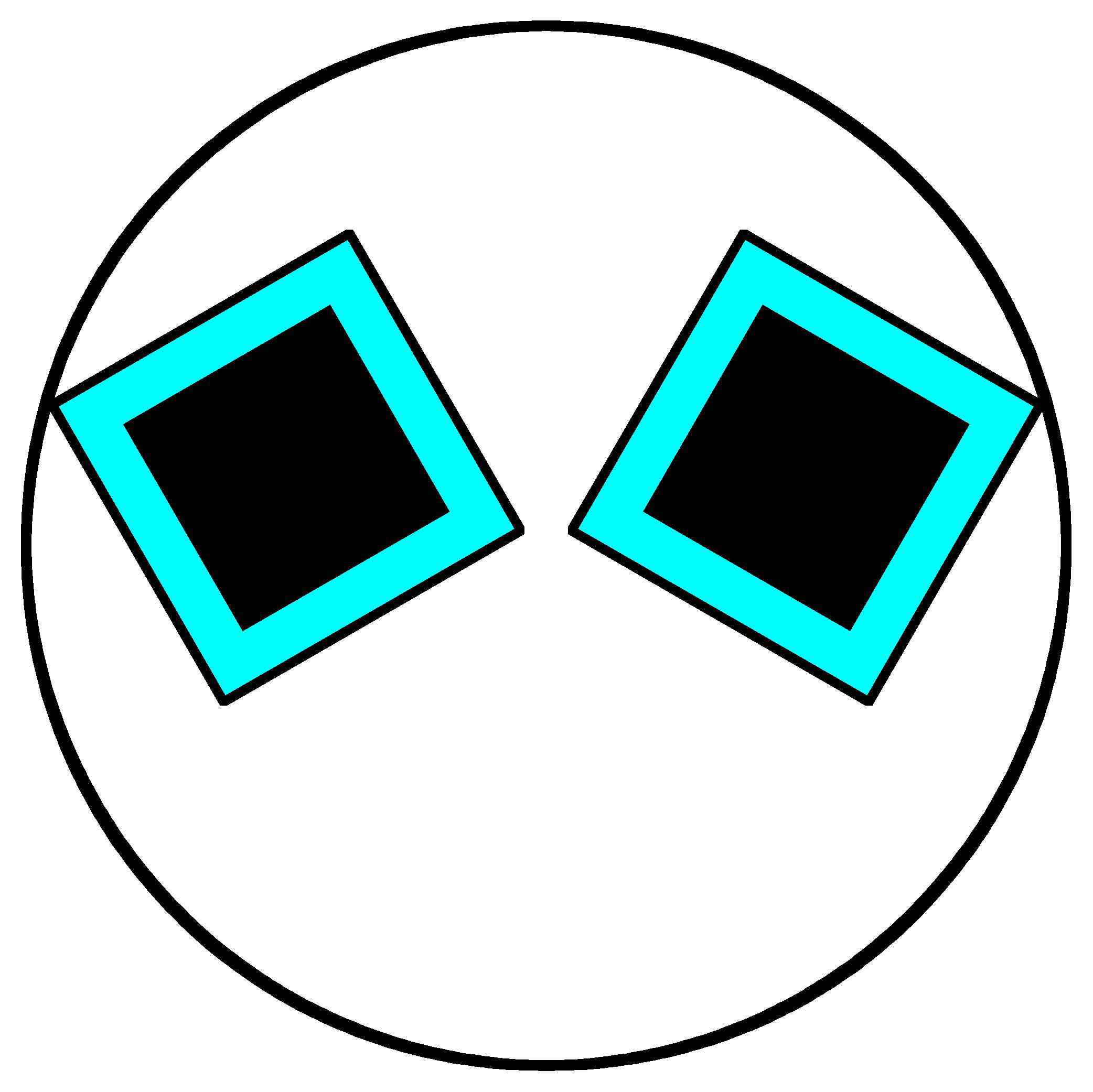} (a) & (b) \includegraphics[height=4cm]{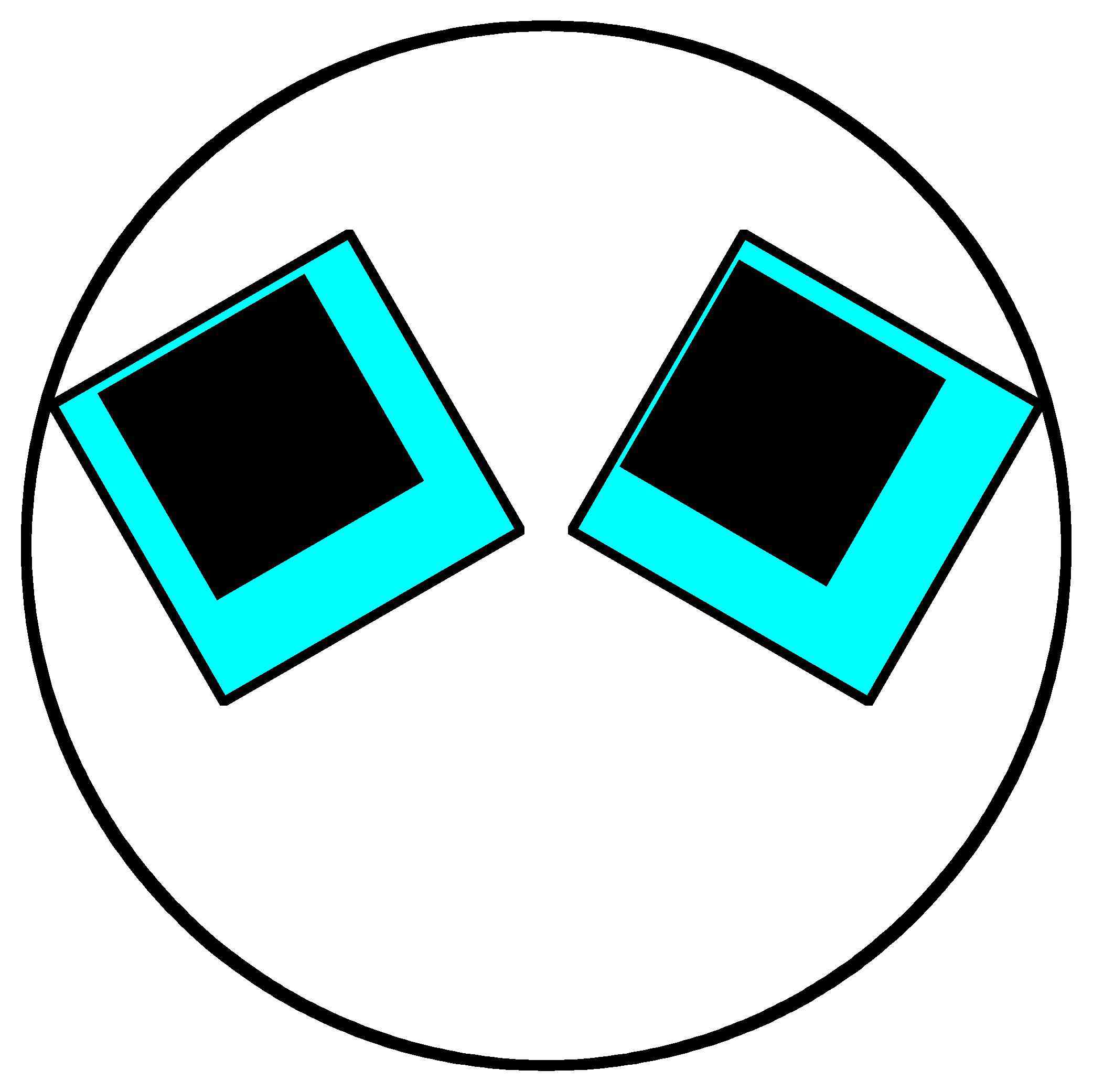} \\
\includegraphics[height=4cm]{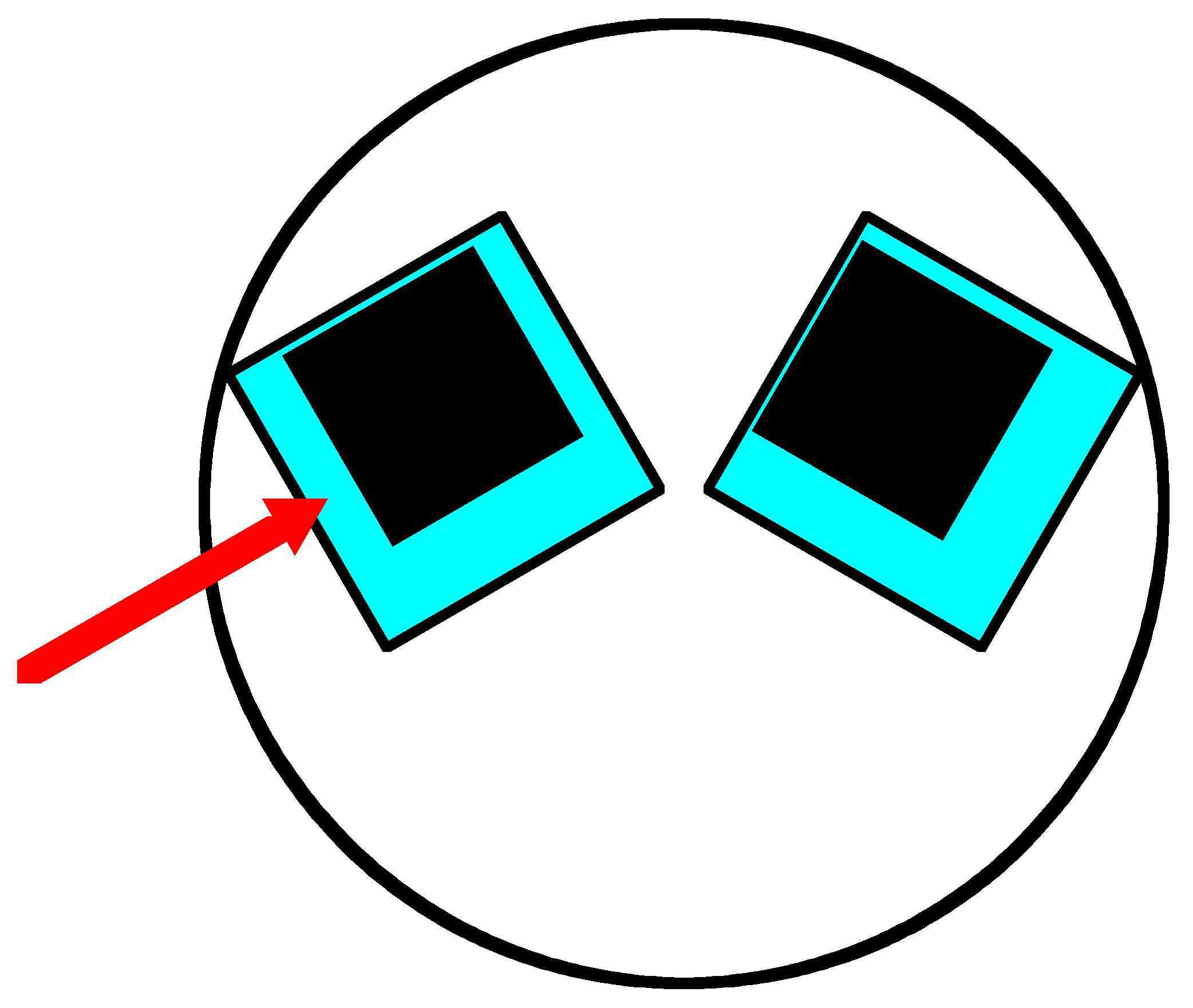} (c) & (d) \includegraphics[height=4cm]{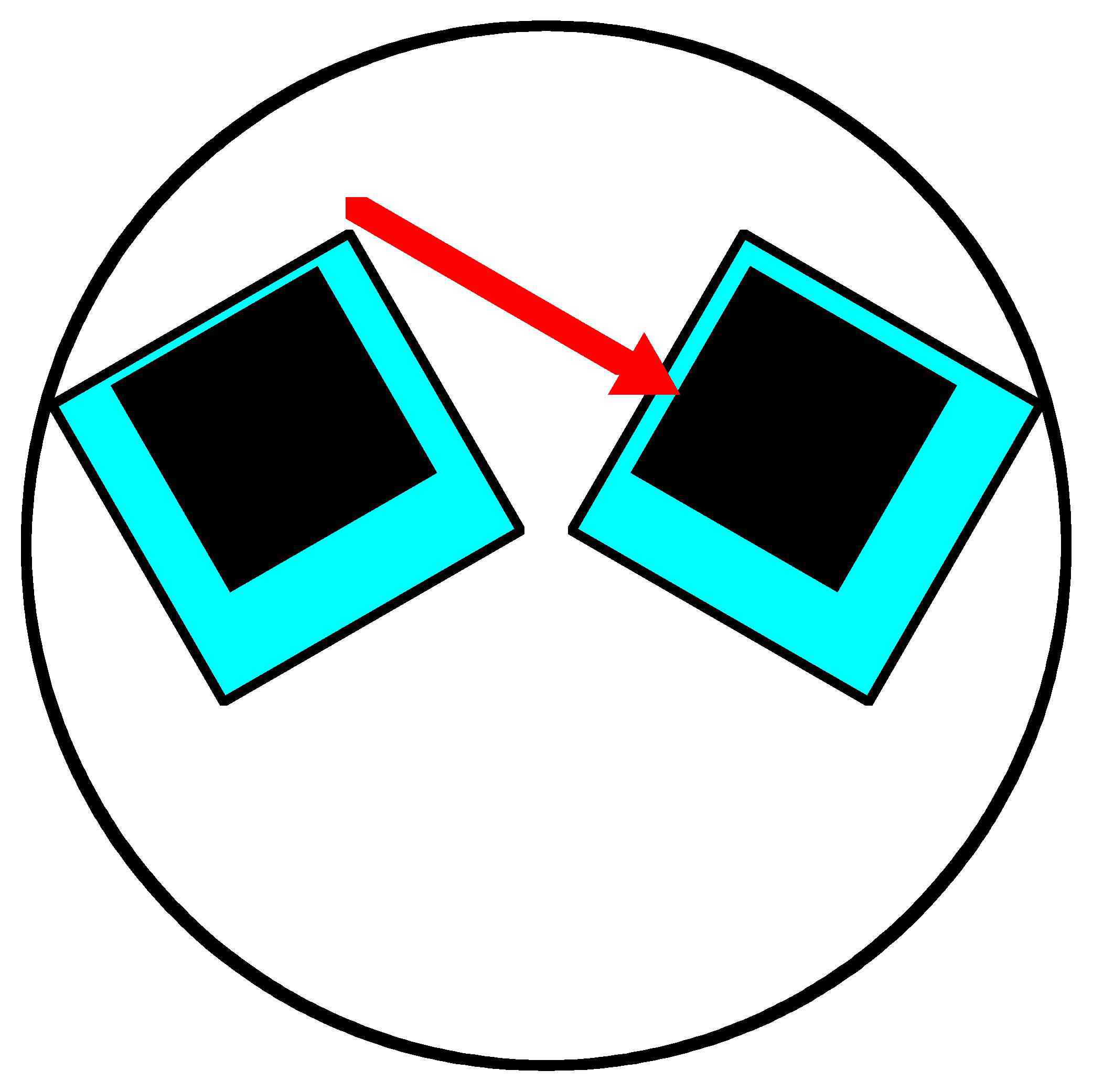} \\
\includegraphics[height=4cm]{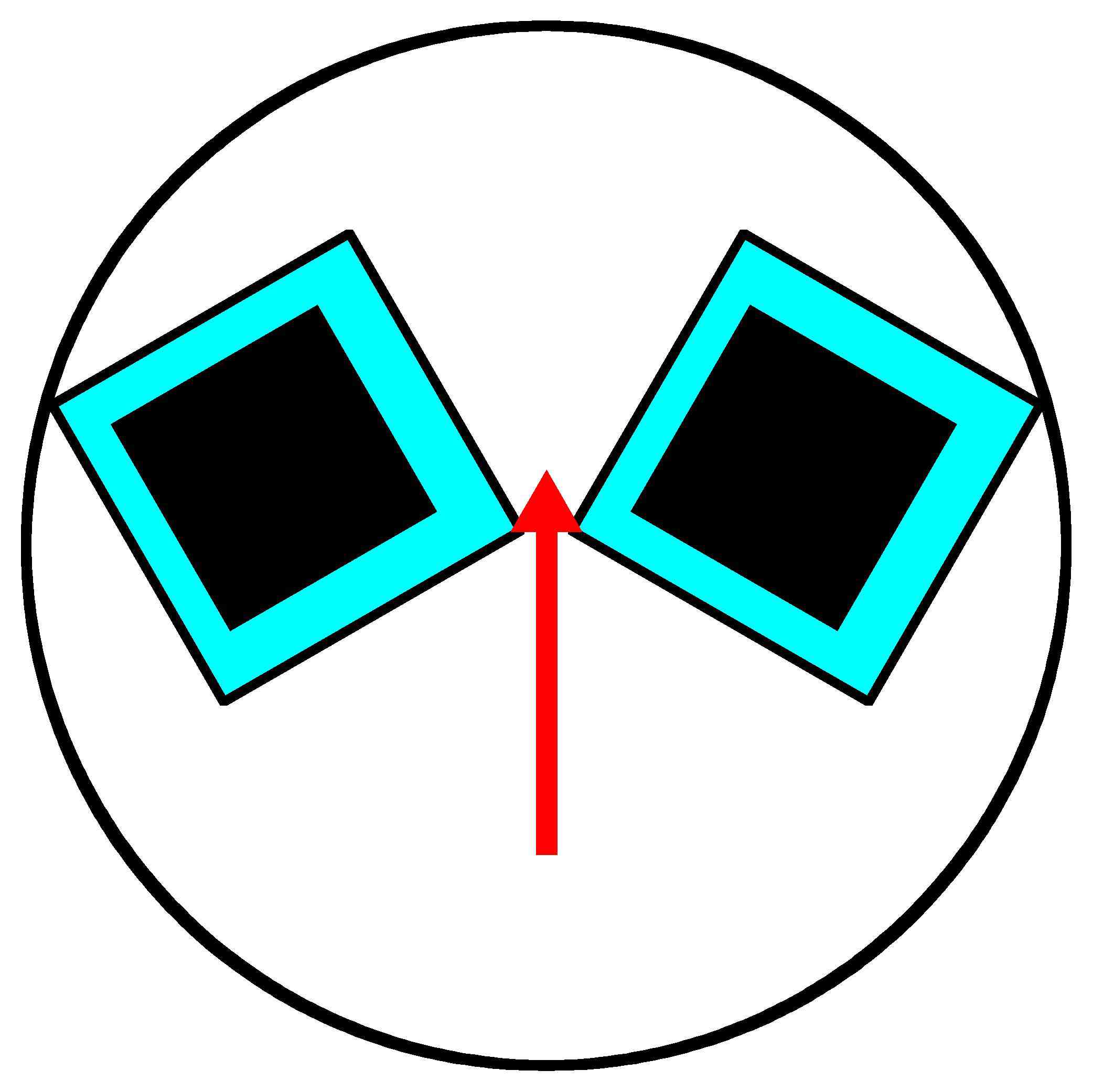} (e) & (f) \includegraphics[height=4cm]{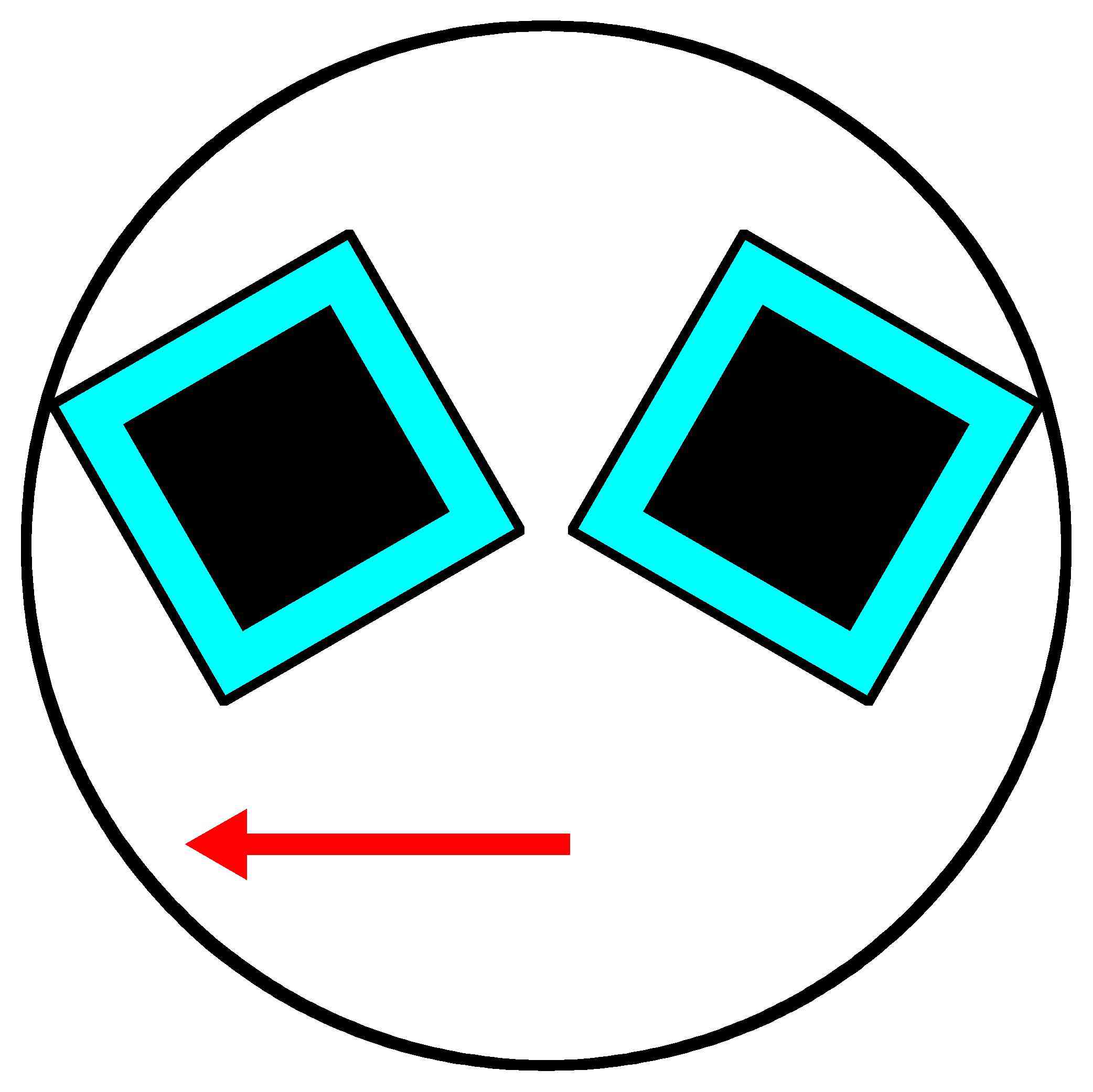}
\end{tabular}
\end{center}
\caption{LISA's control strategy per each SC: (a) shows the nominal position of TMs inside the ISs, in (b) the masses
get arbitrarily displaced (no rotation for simplicity), on (c) and (d) steps the GRSs actuate the TMs in directions orthogonal
to the sensing ones, across (e) and (f) the SC moves to recenter the TMs.}
\label{fig:lisacontrol}
\end{figure}

The SCs constellation rotates around its centre of mass on a plane tilted by $\nicefrac{\pi}{3}\,\unit{rad}$
with respect to the ecliptic (see figure \ref{fig:lisaorbit}). A clever choice of orbit
will allow the formation to complete a full rotation when completing a full revolution
around the Sun. Due to the tilting of the rotating plane, the revolution orbit gets eccentric, with a relative
factor $e\sim L = D \sqrt{3} \sim 10^{-2}$ and inclination to the ecliptic $\phi = \nicefrac{L}{D}\sim 1$ degree.
This special choice of orbits ensures the triangular geometry of the constellation
to remain reliable for a prolonged time over the mission timescale, and the rotation provides
some angular resolution. The orbital motion shall induce Doppler shift on the detected signal
and modulate its amplitude thus allowing angular definition of the source.
LISA's sources (from very distant massive
black holes) should be resolvable to better than an arc-minute; and even the weaker
sources (Galactic binaries) should be positioned to within one degree throughout 
the entire Galaxy. Table \ref{fig:lisacalibbin} provides features and numbers of the
so called ``standard candles''
which will be used to calibrate LISA. We strongly point out anyway that once placed on orbit
in the proper conditions, LISA will gravitate according to the orbits we described, not much
can be done to change or correct it by the controlling thrusters and by itself the constellation will ``breathe'' radially
about some $6\times 10^{4}\,\unit{km}$ length. The motion will be anyway at extremely low
frequency, well outside the MBW, on the scale of months.

\begin{table}
\begin{center}
\begin{tabularx}{\textwidth}{r|l|l|X|l|l|l|X} 
Class & Source & Dist $(\unit{pc})$ & $f$
  & $\nicefrac{M_{1}}{M_{\odot}}$ & $\nicefrac{M_{1}}{M_{\odot}}$ & $h$ & SNR \\
  \hline\rule{0pt}{0.4cm}
\multirow{4}{*}{WD+WD} & WD 0957-666 & $100$ & $0.38$ & $0.37$ & $0.32$ & $4.00\times 10^{-22}$ & $4.1$ \\
 & WD1101+364 & $100$ & $0.16$ & $0.31$ & $0.36$ & $2.00\times 10^{-22}$ & $0.4$ \\
 & WD1704+481 & $100$ & $0.16$ & $0.39$ & $0.56$ & $4.00\times 10^{-22}$ & $0.7$ \\
 & WD2331+290 & $100$ & $0.14$ & $0.39$ & $>0.32$ & $2.00\times 10^{-22}$ & $0.3$ \\
 \hline\rule{0pt}{0.4cm}
\multirow{2}{*}{WD+sdB} & KPD0422+4521 & $100$ & $0.26$ & $0.51$ & $0.53$ & $6.00\times 10^{-22}$ & $2.9$ \\
 & KPD1930+2752 & $100$ & $0.24$ & $0.5$ & $0.97$ & $1.00\times 10^{-21}$ & $4.1$ \\
 \hline\rule{0pt}{0.4cm}
\multirow{9}{*}{AM CVn} & RXJ0806.3+1527 & $300$ & $6.2$ & $0.4$ & $0.12$ & $4.00\times 10^{-22}$ & $173.2$ \\
 & RXJ1914+245 & $100$ & $3.5$ & $0.6$ & $0.07$ & $6.00\times 10^{-22}$ & $195.0$ \\
 & KUV05184-0939 & $1000$ & $3.2$ & $0.7$ & $0.092$ & $9.00\times 10^{-23}$ & $27.3$ \\
 & AM CV n & $100$ & $1.94$ & $0.5$ & $0.033$ & $2.00\times 10^{-22}$ & $35.6$ \\
 & HP Lib & $100$ & $1.79$ & $0.6$ & $0.03$ & $2.00\times 10^{-22}$ & $32.0$ \\
 & CR Boo & $100$ & $1.36$ & $0.6$ & $0.02$ & $1.00\times 10^{-22}$ & $10.6$ \\
 & V803 Cen & $100$ & $1.24$ & $0.6$ & $0.02$ & $1.00\times 10^{-22}$ & $9.2$ \\
 & CP Eri & $200$ & $1.16$ & $0.6$ & $0.02$ & $4.00\times 10^{-23}$ & $3.3$ \\
 & GP Com & $200$ & $0.72$ & $0.5$ & $0.02$ & $3.00\times 10^{-23}$ & $1.1$ \\
 \hline\rule{0pt}{0.4cm}
\multirow{2}{*}{LMXB} & 4U1820-30 & $8100$ & $3$ & $1.4$ & $< 0.1$ & $2.00\times 10^{-23}$ & $5.7$ \\
 & 4U1626-67 & $<8000$ & $0.79$ & $1.4$ & $< 0.03$ & $6.00\times 10^{-24}$ & $0.2$ \\
 \hline\rule{0pt}{0.4cm}
W UM a & CC Com & $90$ & $0.105$ & $0.7$ & $0.7$ & $6.00\times 10^{-22}$ & $0.5$
\end{tabularx}
\end{center}
\caption{LISA calibration binaries. Notice $f=\nicefrac{2}{T}\,(\unit{mHz})$, where $T$ is the period.
Signal to noise ratio SNR is averaged over $1\,\unit{Year}$.}
\label{fig:lisacalibbin}
\end{table}

The TMs relative motion will provide the scientific data, since the masses themselves act
as mirrors for the laser light. Each TM will be a $46\,\unit{mm}$ sided cube, weighting $1.96\,\unit{kg}$,
made of Pt-Au alloy
to guarantee very low magnetic susceptibility. Weight and sizer are details from the LTP design, but likely to
be accepted for LISA as well. By inspection of picture \ref{fig:lisayshape}
the reader might see that each TM is hosted in a separate section of the ``Y-tube'' as for obvious reasons
the internal cavity of LISA is called.

In principle, the SC shall be able to follow both proof-masses with the technique described before. In
practise such a picture needs continuous dynamical adjustment: the capacitive sensors forming the GRS system
continuously monitor the TMs position with the weakest electrostatic coupling possible, while rotational degrees
of freedom are adjusted at low-frequency with the technique of wavefront sensing: each telescope concentrates
the light coming from far SCs on a quadrant photo-diode capable of angular resolution of the source and each SC is thus
slowly ``chasing'' the others to reduce minimise variation of the wavefront angle from the nominal value of zero.

The GRS is mounted on the OB, a rigid structure
made of ultra-low expansion material, about $350\,\unit{mm}$ by $200\,\unit{mm}$ by $40\,\unit{mm}$.
By virtue of optical fibres preserving polarisation the laser light is conducted to the OB after 
bouncing off the proof mass. Here it is brought to interference with a
fraction of the internally generated laser light. As shown, phase noise appears just like a bona-fide GW
signal, therefore lasers must be highly efficient, stable in frequency and amplitude.
Solid-state diode-pumped monolithic miniature Nd:YAG ring lasers have been chosen
for the mission; such a kind of laser generates a continuous $1\,\unit{W}$ infrared beam
with a wavelength of $1.064\,\unit{\mu m}$, relatively immune to refraction by the interplanetary medium.
Each SC has two operational $1\,\unit{W}$ lasers, one per telescope. One laser is switched on first
and acts as matrix: a
fraction of its light ($10\,\unit{mW}$) is reflected from the back surface of the relative proof mass,
and its phase used as a reference for the other local laser.

Hence, the main beams going out along each arm can be considered as a single
laser carrier. It is shone through the telescope, which also collects the incoming light from the spare SC.
The telescope widens the diameter of the beam from a few $\unit{mm}$ to $30\,\unit{cm}$. The transmitting and
receiving telescopes are improved Cassegrain, including an integral matching lens; both are protected by a thermal shield.

The primary mirror has a diameter of $30\,\unit{cm}$ and a focal length of $30\,\unit{cm}$. The secondary mirror
is mounted $27.6\,\unit{cm}$ from the primary and has a diameter of $3.2\,\unit{cm}$ and a
focal length of $2.6\,\unit{cm}$. It is very likely that active focus control will be necessary to
compensate for deformations, in case temperature drifts or other phenomena will create any.
Notice a change of about one micron already deforms the outgoing
wavefront by the specified tolerance $\nicefrac{f}{10}$, hence the temperature fluctuations
at the telescope must be less than $10\,\unitfrac{K}{Hz}$ at $10^{-3}\,\unit{Hz}$.

Each SC will be disk shaped, carrying surface solar cells: LISA will have constant illumination from the
Sun with an angle of $30$ degrees, which in turn provide a very stable environment from the thermal
point of view. A set of Field Emission Electric Propulsion (FEEP) devices are employed as thrusters
in order to move the SC.

A Delta IV carrier will host the three LISA SCs for 
launch. After separation from
the rocket, the three SCs - equipped with own extra-propulsion rocket - will separate and transfer to
solar orbit. Once the constellation is established the propulsion systems are discarded and the FEEPs take over as the
only remaining propulsion system.

\section{The LISA Pathfinder}

\subsection{Noise identification}

Achieving pure geodesic motion at the level requested for LISA, $\sqrt{2}\times 3\times 10^{-15}\,\unitfrac{m\,s}{\sqrt{Hz}}$ at
$0.1\,\unit{mHz}$, is considered a challenging technological task \cite{LTPdefdoc,LTPscrd,Anza:2005td,Vitale:2002qv}.
The goal of the SMART-2 test planned by ESA is
demonstrate geodetic motion within one order of magnitude from the LISA performance to confirm the formerly elucidated
TT-construction and that the shown noise figures are compatible with the LISA demands.

SMART-2 will launch in 2009; on-board the LTP is designed to demonstrate new technologies that
have significant application to LISA and other future Space Science missions. Three primary technologies are included on
LTP/SMART-2: Gravitational Sensors, Interferometers and Micro-thrusters.

Within
the LTP, two LISA-like TMs located inside a single SC are tracked by a laser
interferometer. This minimal instrument is deemed to contain the essence of the construction procedure needed for LISA
and thus to demonstrate its feasibility. This demonstration requires two steps:
\begin{enumerate}
\item first, based on former noise models \cite{Stebbins:2004sy} and the current one in this publication, the mission is
designed so that any differential parasitic acceleration noise of the TMs is kept below the
requirements. For the LTP these requirements are relaxed to $3\times 10^{-14}\,\unitfrac{m}{s^{2} \sqrt{Hz}}$ 
a factor $\simeq 7$
worse than what is required in
LISA.  In addition this performance is only required for frequencies larger than $1\,\unit{mHz}$:
\begin{shadefundtheory}
\beq
S^{\nicefrac{1}{2}}_{\nicefrac{\Delta F}{m},\,\text{LTP}}(\omega)
  =3\times 10^{-14} \left(1+\left(
    \frac{\omega}{2\pi \times 3\, \unit{mHz}}\right)^{4}
  \right)^{\nicefrac{1}{2}}\,\accPSDunit\,.
\label{eq:ltpsensitivity}
\eeq
\end{shadefundtheory}

This relaxation of
performance is accepted since the mission will make use of one single satellite and two probe masses sharing
the sensing axis. With such a configuration, actuation is needed to hold one mass and it's quite unlikely to
reach LISA's precision given actuation and all the disturbances at play. The choice of a single satellite
was made in view of cost and time saving. Notice LTP will measure residual acceleration difference
between the two TMs, therefore though the deemed precision is reduced by one order of magnitude, the test
is highly representative of LISA's TMs behaviour. Moreover, it would be careless to venture into further design
phases of LISA without testing the part of technology which is absolutely mandatory for it to work. The ideal test would
imply the use of $2$ SCs to verify drag-free and depict noise in a situation more closely matching LISA's; nevertheless
a single satellite mission would be order of magnitudes cheaper and much less time-consuming on the design front.

As both for LISA and for the LTP this level of performance cannot be verified on ground due to the presence of the large
Earth gravity, the verification is mostly relying on the measurements of key parameters of the noise model of the
instrument \cite{Carbone:2003nb,Carbone:2004gk,CarbonePhD,Carbone:2004se,Carbone:2003ya}.
In addition an upper limit to
all parasitic forces that act at the proof-mass surface
(electrostatics and electromagnetics, thermal and pressure effects etc.) has been put and keeps being updated by means
of a torsion pendulum test bench \cite{Cavalleri:2001ur,Hueller:2002dq}. In this instrument a hollow version of the proof mass hangs from the
torsion fibre of the pendulum so that it can freely move in a horizontal plane within a housing which is representative
of flight conditions. Current limits on torque noise has been measured that
would amount to $3\times 10^{-13}\,\unitfrac{m\,s}{\sqrt{Hz}}$ \cite{Carbone:2003nb},
when translated into an equivalent differential acceleration.
Such a figure is encouraging and calls for an off-ground testing 

\item Second, once in orbit the residual
differential acceleration noise of the proof
masses is measured. The noise model \cite{Bortoluzzi:2004cz,Bortoluzzi:2003ua} predicts that the total PSD is contributed by sources of three broad
categories:
\begin{enumerate}
\item those sources whose effects can be identified and suppressed by a proper adjustment of selected instrument
parameters. An example of this is the force due to residual coupling of TMs to the SC. By regulating
and eventually matching, throughout the application of electric field, the stiffness of this coupling for both proof
masses, this source of noise can be first highlighted, then measured, and eventually suppressed.

\item Noise sources
connected to measurable fluctuations of some physical parameter. Forces due to magnetic fields or to thermal gradients
are typical examples. The transfer function between these fluctuations and the corresponding differential proof mass
acceleration fluctuations will be measured by purposely enhancing the variation of the physical parameter under
investigation and by measuring the corresponding acceleration response: for instance the LTP carries magnetic coils
to apply comparatively large magnetic field signals and heaters to induce  time varying thermal gradients.

In addition
the LTP also carries sensors to measure the fluctuation of the above physical disturbances while measuring the residual
differential acceleration noise in the absence of any applied perturbation. Magnetometers
and thermometers, to continue with the examples above. By multiplying the measured transfer function by the measured
disturbance fluctuations, an acceleration noise data stream can be computed and subtracted from the main
differential acceleration data stream.

This way the contribution of these noise sources are suppressed and the residual
acceleration PSD decreased. This possible subtraction can relax some difficult requirements, like expensive magnetic ``cleanliness'', or thermal
stabilisation programs.

\item Noise sources that cannot be removed by any of the above methods. The residual differential
acceleration noise must be accounted for by these sources. To be able to do the required comparison, some of the
noise model parameters must and will be measured in flight. One example for all, the charged particle flux due to cosmic
rays will be continuously monitored by a particle detector.
\end{enumerate}
\end{enumerate}

The result of the above procedure is the validation of the
noise model for LISA and the demonstration that no unforeseen source of disturbance is present that exceeds the residual
uncertainty on the measured PSD. The following sections, after describing some details of the experiment, will discuss
the expected amount of this residual uncertainty.

\subsection{The instrument}

\begin{figure}
\begin{center}
\includegraphics[width=.8\textwidth]{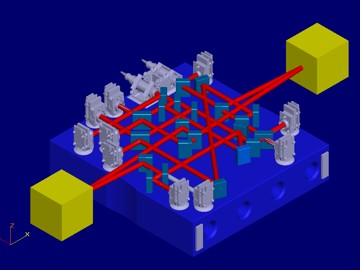}
\end{center}
\caption{The concept of the LTP. The distance between 2 cubic, free floating TMs is measured by a heterodyne
laser interferometer. Each proof mass is surrounded by a set of electrodes
that are used to readout the mass position and orientation relative to the SC. This
measurement is obtained as the motion of the proof mass varies the capacitance's between the electrodes and the proof
mass itself. The same set of electrodes is also used to apply electrostatic forces to the TMs.}
\label{fig:LTPconcept}
\end{figure}

The basic scheme of the LTP \cite{Vitale:2002qv}
is shown in figure \ref{fig:LTPconcept}: two free floating TMs are hosted within a single SC
and the relative motion along a common sensitive axis, the $\hat x$-axis, is measured by means of a laser interferometer.
The
TMs are made of a Gold-Platinum, low magnetic susceptibility alloy, have a mass of $m=1.96\,\unit{kg}$ and are
separated
by a nominal distance of $0.376\,\unit{m}$.

Differential capacitance variations are parametrically read out by a front end electronics composed of high accuracy
differential inductive bridges excited at about $100\,\unit{kHz}$, and synchronously detected via a phase sensitive
detector \cite{2003SPIE.4856...31W, Dolesi:2003ub}. Sensitivity depends on the DOF: for the $\hat x$-axis
it is better than $1.8\,\unitfrac{nm}{\sqrt{Hz}}$ at $1\,\unit{mHz}$.
Angular sensitivities are better than $200\,\unitfrac{nrad}{\sqrt{Hz}}$. Forces and torques on the TMs required
during science
operation are applied through the same front end electronics by modulating the amplitude of an ac carrier applied to the
electrodes. The frequency of the carrier is high enough to prevent the application in the measurement band of unwanted forces by mixing with low
frequency fluctuating random voltages. The front end electronics is also used to apply all voltages required by specific
experiments. Each proof mass, with its own electrode housing, is enclosed in a high vacuum chamber which is pumped down
to $10^{-5}\,\unit{Pa}$ by a set of getter pumps. The laser interferometer light crosses the vacuum chamber wall through an
optical window. 

As the proof mass has no mechanical contact to its surrounding, its electrical charge continues to build up due to
cosmic rays. To discharge the proof mass, an ultra violet light is shone on it and/or on the surrounding electrodes \cite{Sumner:2004ss}.
Depending on the illumination scheme, the generated photo-electrons can be deposited on or extracted from the proof mass to
achieve electrical neutrality. The absence of a mechanical contact also requires that a blocking mechanism keep the
mass fixed during launch and is able to release it once in orbit, overcoming the residual adhesion. This release must leave the proof
mass with low enough linear momentum to allow the control system described in the following to bring it at rest in the
nominal operational working point. The system formed by one proof mass, its electrode housing, the vacuum enclosure and
the other subsystems is called in the following the gravity reference sensor.

The interferometer system includes many measurement channels. It provides:
\begin{enumerate}
\item heterodyne measurement of the relative
position of TMs along the sensitive axis.
\item Heterodyne measurement of the position of one of the proof-masses
(proof mass 1) relative to the optical bench.
\item Differential wave front sensing of the relative orientations of the
proof-masses around the $\hat y$ and $\hat z$ axes.
\item Differential wave front sensing of the orientation of proof-mass 1 around the
$\hat y$ and $\hat z$ axes.
\end{enumerate}

Sensitivities at $\unit{mHz}$ frequency are in the range of $10\,\unitfrac{pm}{\sqrt{Hz}}$
for displacement and of $10\,\unitfrac{nrad}{\sqrt{Hz}}$
on rotation. Interferometry is performed by a front-end electronics largely based on Field Programmable Gate Arrays. Final
combination of phases to produce motion signals is performed by the LTP instrument computer. 
The LTP computer also drives and reads-out the set of subsidiary sensors and actuators needed to apply the already
mentioned selected perturbations to the TMs and to measure the fluctuations of the disturbing fields. Actuators
include coils used to generate magnetic field and magnetic field gradients and heaters to vary temperature and
temperature differences at selected points of the Gravity Reference Sensor and of the optical bench. Sensors include
magnetometers, thermometers, particle detectors and monitors for the voltage stability of the electrical supplies.

\begin{figure}
\begin{center}
\includegraphics[width=.49\textwidth]{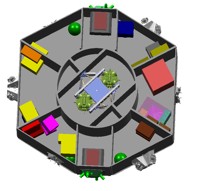}%
\includegraphics[width=.50\textwidth]{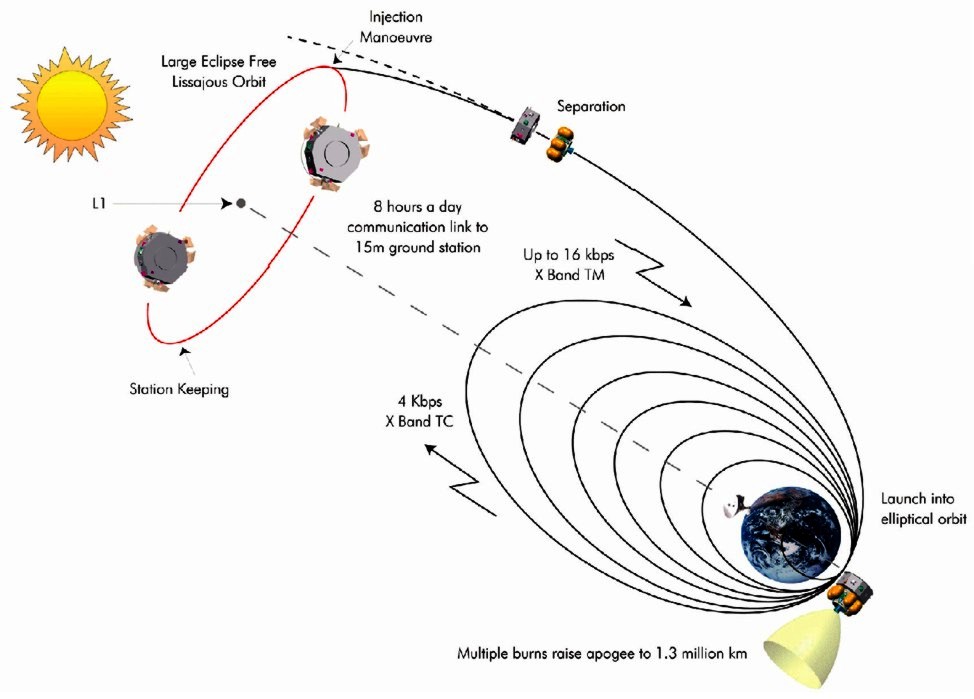}
\end{center}
\caption{Left: the LTP accommodated within the central section of the LISA Pathfinder SC. Right: the injection
of LISA Pathfinder in the final orbit around L1.}
\label{fig:LTPsc}
\end{figure}

\begin{figure}
\begin{center}
\includegraphics[width=\textwidth]{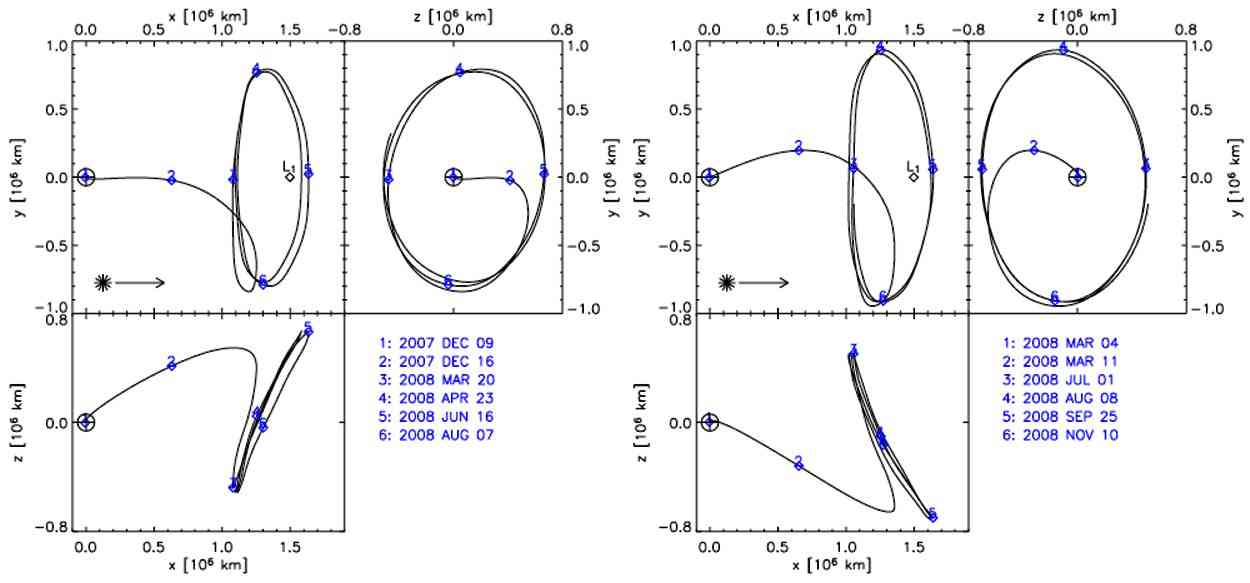}
\end{center}
\caption{Lissajous orbit for LTP around L1.}
\label{fig:LTPorbit}
\end{figure}

LTP will be hosted in the central section of the SC (see figure \ref{fig:LTPsc}, left), where gravitational
disturbances are minimised, and will operate in a Lissajous orbit (see figure \ref{fig:LTPsc}, right and figure \ref{fig:LTPorbit})
\cite{Landgraf:2004gm} around the Lagrange
point 1 of the Sun-Earth system.

\subsection{A simplified model}

The sensitivity performance estimated before is limited at low frequencies by stray forces perturbing the
TMs out of their geodesics. Better would be to say that the presence of perturbations due to non-gravitational
interactions in the energy-momentum tensor generates a deformed geometry in space-time, thus perturbing the
``natural'' geodesics the TMs would follow in vacuo.

In contrast with the usual view of a space-probe dragging along its content, drag-free reverses
the scenario and it's the TMs inside the satellite which dictate the motion of the latter.

\begin{figure}
\begin{center}
\includegraphics[width=0.8\textwidth]{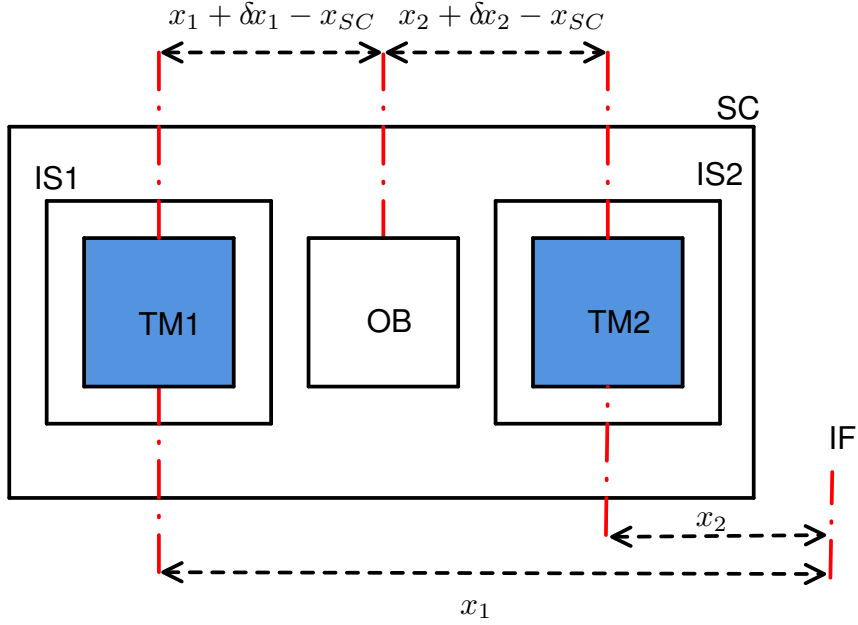}
\end{center}
\caption{Simple scheme of LTP.}
\label{fig:simpleLTP}
\end{figure}

\begin{figure}
\begin{center}
\includegraphics[width=0.8\textwidth]{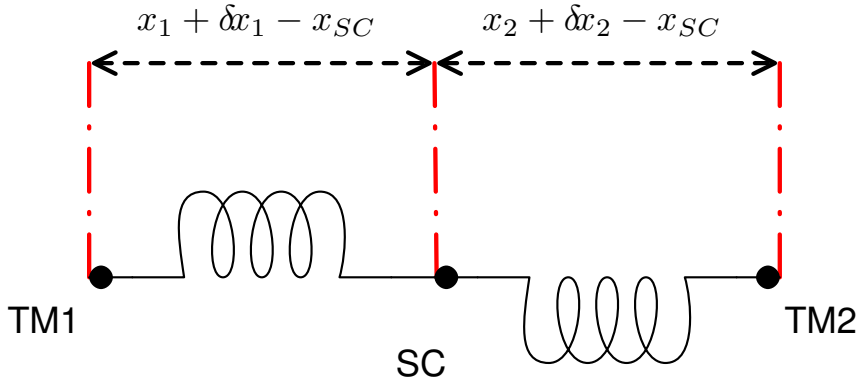}
\end{center}
\caption{Spring and particles model for LTP.}
\label{fig:LTPspringsndots}
\end{figure}

To illustrate the features of drag-free technique and relative detectable signals, we proceed now to illustrate
a simple one-dimensional model of two TMs coupled to a SC. Let $m$ be each TM mass, $m_{\text{SC}}$
the SC mass, $k_{i}$, $i=1,2$ two spring constants summarising the Hooke-like coupling of the various masses.
We let external forces, generally named after $f_{i,x}$, with $i=1,2,\text{SC}$ act on the respective body.
By Newton's law, the dynamics can be written as:
\begin{align}
m \ddot x_1+k_1 \left(x_1-x_{\text{SC}}\right) &= f_{1,x}\,,\\
m \ddot x_2+k_2 \left(x_2-x_{\text{SC}}\right) &= f_{2,x}\,,\\
m_{\text{SC}}\ddot x_{\text{SC}} -k_1 \left(x_1-x_{\text{SC}}\right)-k_2 \left(x_2-x_{\text{SC}}\right)
 &= f_{\text{SC},x}\,.
\end{align}
Each force can be thought as a force per unit mass $m$ and separated into an external contribution and
a feed-back term, accounting for our desire to realise a mechanical control loop:
\beq
\begin{split}
f_{1,x}\to m \left(g_{1,x}+ g_{\text{fb1},x}\right)\,,\\
f_{2,x}\to m \left(g_{2,x}+ g_{\text{fb2},x}\right)\,,\\
f_{\text{SC},x}\to m_{\text{SC}} \left(g_{\text{fbSC},x} + g_{\text{SC},x}\right)\,,
\end{split}
\eeq
here $g_{i,x},\, i=1,2,\text{SC}$ is the external acceleration acting on the $i$-th TM or on the SC, while
$g_{\text{fb}i,x}$ is the feed-back force per unit mass we'd like to apply to realise a certain control strategy.
Moreover, couplings can be translated into elastic stiffness terms, per unit mass, being the
DOF at play linear:
\beq
k_i \to m \omega_{p,i}^2\,,\qquad i=1,2\,.
\eeq
We'll work in the approximation of very large SC mass, and introduce a mass scale parameter $\mu$
as follows:
\beq
\mu = \frac{m}{m_{\text{SC}}}\,.
\eeq
Every uncertainty in the TMs position or every deformation of the bench hosting optical or electrostatic measuring
device may induce
undesired error in position detection and will be summarised into two ${\delta}x_{i}$ ($i=1,2$)
variables, so that in the equations of motion and in the feed-back laws the following substitution
will take place:
\beq
x_i\to x_i+{\delta}x_i\,, \qquad i=1,2\,.
\eeq
Notice we'll assume these deformations to be stationary, to get ${\delta}\dot x_{i} \simeq 0$. Anyway
materials will always be chosen so to ensure these deformation to be small with respect to displacement, in spectral form:
\beq
S_{{\delta}x_{i}}^{\nicefrac{1}{2}} \ll S_{x_{i}}^{\nicefrac{1}{2}}\,.
\eeq
Finally, we'll work in Laplace space from now on, and make the substitution:
\beq
\frac{\de x_{i}}{\de t} \to s x_{i}\,,\qquad i=1,2\,,
\eeq
when needed. In turn, we'll switch to Fourier space by placing $s\doteq -\imag \omega$.
Finally, the equations of motion display like:
\begin{shadefundtheory}
\begin{align}
-m x_1 \omega^2+m \left(x_1-x_{\text{SC}}+{\delta}x_1\right) \omega_{p,1}^2 &= m g_{1,x}+m g_{\text{fb1},x}\,,\\
-m x_2 \omega^2+m \left(x_2-x_{\text{SC}}+{\delta}x_2\right) \omega_{p,2}^2 &= m g_{2,x}+m g_{\text{fb2},x}\,,\\
-\frac{m}{\mu} x_{\text{SC}} \omega^2
 -m \left(x_1-x_{\text{SC}}+{\delta}x_1\right) \omega_{p,1}^2
 -m\left(x_2-x_{\text{SC}}+{\delta}x_2\right) \omega_{p,2}^2
 &= \frac{m}{\mu} g_{\text{fbSC},x}+\frac{m}{\mu}g_{\text{SC},x}\,.
\end{align}
\end{shadefundtheory}

A drag-free strategy is a map whose task is enslaving the satellite motion to the TMs. This can be
achieved in many ways, but since it's impossible to follow the motion of both TMs along
one common axis, two strategies are left unique as solutions:
\begin{enumerate}
\item the SC follows TM1, and TM2 is held in position by continuously servoing its position
with respect to the SC itself, i.e. for our simple system:
\begin{shadefundtheory}
\beq
\begin{split}
g_{\text{fb1},x} &\to 0\,,\\
g_{\text{fb2},x} &\to -\left(x_2-x_{\text{SC}}+x_{n,2}\right) \omega_{\text{lfs},x}^2\,,\\
g_{\text{fbSC},x} &\to \left(x_1-x_{\text{SC}}+x_{n,1}\right) \omega_{\text{df},x}^2\,,
\end{split}
\eeq
\end{shadefundtheory}
\noindent where we introduced ``position readout'' noise for the channels $x_{1}$ and $x_{2}$ and named it
$x_{n,1}$ and $x_{n,2}$ respectively. We may think of this noise as being provided by some electrostatic
readout circuitry. We also decided to complicate our picture by taking
control of the ``gain'' of the feed-back: instead of being $1$, the multiplication constant is
a function of the complex frequency, namely $\omega_{\text{lfs},x}^2(s)$, LFS meaning ``low
frequency suspension'' and $\omega_{\text{df},x}^2(s)$, DF meaning ``drag-free''.

\item On the other hand, we may choose to pursue TM1 with the SC and to hold TM2 fixed
on the distance to TM1 itself, in formulae:
\begin{shadefundtheory}
\beq
\begin{split}
g_{\text{fb1},x} &\to 0\,,\\
g_{\text{fb2},x} &\to -\left(x_2-x_1+{\Delta}x_n\right) \omega_{\text{lfs},x}^2\,,\\
g_{\text{fbSC},x} &\to \left(x_1-x_{\text{SC}}+x_{n,1}\right) \omega_{\text{df},x}^2\,,
\end{split}
\eeq
\end{shadefundtheory}
\noindent where now the new noise ${\Delta}x_{n}$, typical of the difference channel, was introduced\footnote{An interferometer
is quite likely to be the only low-noise detector in town able to perform such a difference measurement. Thus
${\Delta}x_{n}$ will be also called ``interferometer noise''}
\end{enumerate}

If we'd choose the first approach, we could solve the equations of motion in the approximation
of $\mu \to 0$. Moreover, we can decide to take a very severe drag-free control policy,
and take also $\left|\omega_{\text{df},x}^2\right| \gg \left|\omega^{2}\right|$ and larger than every other frequency at play.
The solution of the problem is analytic but quite tedious - it can be computed with the help of any
symbolic algebraic program - and we'll state here only the result for the main difference channel
as function of $\omega$:
\beq
\begin{split}
x_{2}-x_{1}+{\Delta}x_{n} \underset{\substack{\mu\to 0  \\ \omega_{\text{df},x}^2 \to \infty}}{\simeq}
\frac{1}{-\omega^2+\omega_{\text{p},2}^2+\omega_{\text{lfs},x}^2} \Bigl(
&-x_{n,2} \omega_{\text{lfs},x}^2+g_{2,x}-g_{1,x}+\\
&+\left({\delta}x_1 - {\delta}x_2 \right)\left(\omega_{\text{p},2}^2+\omega_{\text{lfs},x}^2\right)+\\
&+\left(x_{n,1}+\frac{g_{\text{SC},x}}{\omega_{\text{df},x}^2}\right)
 \left(-\omega_{\text{p},1}^2+\omega_{\text{p},2}^2+\omega_{\text{lfs},x}^2\right)\Bigr)+\\
 &+{\Delta}x_n\,.
\end{split}
\label{eq:simpleltpdeltaxm1}
\eeq
Furthermore, in the approximation of very low frequency suspension $\omega_{\text{lfs},x}^2$
to compensate the intrinsic static stiffness $\omega_{\text{p},2}^{2}$, we may think the following
approximations to hold. Notice $\omega_{\text{p},2}^{2} \simeq 2\times 10^{-6}$ is a believable value for
the parasitic stiffness \cite{LTPscrd}, versus $\omega_{\text{lfs},x}^{2}$
whose value must be kept small or it would amplify the noise source represented by
$\sim \omega_{\text{lfs},x}^{2} \left(x_{n,1}+\nicefrac{g_{\text{SC},x}}{\omega_{\text{df},x}^{2}}\right)$
which might in turn become dominant over the rest of the expression \eqref{eq:simpleltpdeltaxm1}. Keeping
$\left|\omega_{\text{lfs},x}^{2}\right| \ll \left|\omega_{\text{p},2}^{2}-\omega_{\text{p},1}^{2}\right|$
prevents that every
time the SC suffers jitter or displacement - be it unwilling or induced by thrusters - the same shaking won't affect the TM
due to the tight coupling. Conversely, the value of $\left|\omega_{\text{lfs},x}^2\right|$ must be kept
$\simeq 2\left|\omega_{\text{p},2}^{2}\right|$ to achieve control stability (the LFS acts as a positive spring whose
value must be double the negative one to compensate for). As we can see, a delicate balance is at play. Finally
over a large scale of frequencies, $\omega$ is larger than the parasitic couplings and the feed-back gains
but the drag-free (MBW,
$10^{-3}\,\unit{Hz} \leq \nicefrac{\omega}{2\pi} \leq 1\,\unit{Hz}$):
\beq
\omega^{2} \gg \left|\omega_{\text{p},2}^{2}+\omega_{\text{lfs},x}^2\right|\,.
\eeq
In summary, the front filter becomes:
\beq
\frac{1}{-\omega^2+\omega_{\text{p},2}^2+\omega_{\text{lfs},x}^2} \simeq \frac{1}{-\omega^2}\,,
\eeq
and in the end:
\begin{shadefundtheory}
\beq
\begin{split}
x_{2}-x_{1}+{\Delta}x_{n} \underset{\left|\omega^2\right| \gg \left|\omega_{\text{lfs},x}^2 + \omega_{\text{p},2}^2\right|}{\simeq}
 -\frac{1}{\omega^2}\Bigl(&-x_{n,2} \omega_{\text{lfs},x}^2+g_{2,x}-g_{1,x}+\\
 &+\left({\delta}x_1-{\delta}x_2\right) \left(\omega_{\text{p},2}^2+\omega_{\text{lfs},x}^2\right)+\\
 &+\left(x_{n,1}+\frac{g_{\text{SC},x}}{\omega_{\text{df},x}^2}\right)
 \left(\omega_{\text{p},2}^2-\omega_{\text{p},1}^2+\omega_{\text{lfs},x}^2\right)
 \Bigr) + {\Delta}x_n\,.
\end{split}
\eeq
\end{shadefundtheory}
\noindent In the limit of very low coupling this control mode has thus a natural self-calibration property between force and displacement signal, being
purely inverse proportional to the frequency squared. Undoubtedly, this feature may be of great use
in absence of deep knowledge on a more complicated device with many DOF. In chapter \ref{chap:ltp} we'll
complicate this simple model and the special character of this mode will be discussed and employed.
We'll call this mode ``nominal'' (formerly M1) and will discuss it thoroughly in
section \ref{sec:modes} and \ref{sec:nommode}.

The mentioned signal is anyway a good estimator of the acceleration difference acting on the TMs: $g_{2,x}-g_{1,x}$,
provided a good matching of LFS and parasitic stiffness could be performed
$\left(\omega_{\text{lfs},x}^2\simeq \omega_{\text{p},2}^2\right)$
and drag-free gain could damp SC jitter to a good level $\left(\nicefrac{g_{\text{SC}}}{\omega_{\text{df},x}^2}\ll 1\right)$.

Notice, conversely, that this readout signal carries along the ${\Delta}x_{n}$ noise term fully unabridged,
independent on the frequency applied. It is therefore transparent that this mode will be intrinsically noisier
than other solutions unless we guarantee that ${\Delta}x_{n} \ll x_{n,i}$, another point to choose interferometer detection
for mutual displacement of the TMs.

Equal coupling of the two TMs to the SC can result in a ``common mode'' excitation as response of the
two masses. As an effect, the high-sensitivity interferometric signal will be rendered blind by the coupled
dynamics. The optimal feedback $\omega_{\text{lfs},x}^{2}$ is designed to unbalance the coupling acting
as a control spring and giving a differential coupling like
\beq
\left|\omega_{\text{p},1}^{2}-\left(\omega_{\text{p},2}^{2}+\omega_{\text{lfs}}^{2}\right)\right|
\simeq 2 \left|\omega_{\text{p},2}^{2}\right|\,.
\eeq
This differential coupling may be measured by modulating the drag-free control set-point and tuned
via $\omega_{\text{lfs},x}^{2}$ to distinguish SC coupling noise from random force noise.

This control mode may present very large mechanical transients (long relaxation time for TM2 motion to stabilise),
since $\left|\omega_{\text{lfs},x}^{2}\right|$ cannot be tightened, for all the mentioned motivations. Therefore
this mode might have very poor experimental times, the largest part of it being wasted.

In LISA one single direction will be pursued by the SC, i.e. the mid-line between the directions spanned by the
optical sensing lines. It is impossible to pursue both the TMs in LTP, being they coaxial along the sensing
direction, as stated. Nevertheless this control mode is highly representative of LISA, whose dynamical picture
we shall mimic at maximal level to gain knowledge about forces and noise behaviour \cite{LTPscrd, LTPdfacsM3}.

If conversely we'd use the locking onto the $x_{2}-x_{1}$ distance, in the usual $\mu\to 0$ and
high drag-free gain approximations, we'd find for the distance signal itself:
\begin{shadefundtheory}
\beq
\begin{split}
x_{2}-x_{1}+{\Delta}x_{n} \simeq \frac{1}{\omega_{\text{lfs},x}^2+\omega_{\text{p},2}^2-\omega^{2}} \Bigl(&
\left(\omega_{\text{p},2}^{2}-\omega^{2}\right){\Delta}x_n +g_{2,x}-g_{1,x}+\\
 &+\left(\omega_{\text{p},2}^2-\omega_{\text{p},1}^2\right) \left(x_{n,1}+\frac{g_{\text{SC},x}}{\omega_{\text{df},x}^2}\right)\\
 &+\left({\delta}x_1 - {\delta}x_2\right)\left(\omega_{\text{p},2}^2+\omega_{\text{lfs},x}^2\right)\Bigr)\,.
 \label{eq:scimodesimple}
\end{split}
\eeq
\end{shadefundtheory}
\noindent In this case, apart from fulfilling stability issues the value of $\left|\omega_{\text{lfs},x}^2\right|$
doesn't need to be small since it doesn't amplify any noise or jitter apart $\left|{\delta}x_{2}-{\delta}x_{1}\right|$.
Since the deformations
difference $\left|{\delta}x_{2}-{\delta}x_{1}\right|$ may be thought as small \cite{LTPdefdoc},
the signal is a very good estimator of the distance $x_{2}-x_{1}$ as function of the acceleration difference $g_{2,x}-g_{1,x}$.

This mode doesn't allow for self-calibration, but the instrumental noise can be modulated by the frequency
and the intrinsic stiffness difference is a constant independent of frequency itself, highly damped by DF gain. It is
therefore an extremely clean mode as for readout: no LFS gain appears in the noise sources on the r.h.s. of the
expression, only as a global tuning term in the foremost propagator. Strengthening the grip of the low
frequency suspension is henceforth a technique to damp TM2 motion which may be applied at will
within common-sense boundaries. On this side, this control mode won't waste experimental time in
waiting for transients to elapse.

By electrostatically tuning $\omega_{\text{p},1}^{2}\simeq \omega_{\text{p},2}^{2}$ a source of noise characteristic of LISA will be annihilated, i.e.
\beq
\left(\omega_{\text{p},2}^2-\omega_{\text{p},1}^2\right) \left(x_{n,1}+\frac{g_{\text{SC},x}}{\omega_{\text{df},x}^2}\right)\simeq 0\,,
\label{eq:matchingstiffness}
\eeq
this configuration allows for a clear measurement of $g_{2,x}-g_{1,x}$;
in turn, alien noise sources can then be mapped and henceforth subtracted.

In a word in this control mode the mutual distance between the test-masses is held constant. This laser-locking procedure will be
much less noisy than the nominal one and is hence defined as ``science'' mode. The main interferometer
signal booking the distance variation between the masses is used and its value kept fixed: no need for the satellite to fire the thrusters
as the displacement measurement can be deduced from the applied actuation gains. Besides, the drawback of this mode is the more complicated transfer function from
displacement to force to real pull on the SC, a feature which demands precise calibration.

We point out the equivalence between measuring the real laser phase variation (or voltage variation on the capacitors)
holding one mass and letting the other fly, and measuring the variation in electrostatic force needed to hold both
the masses in position while keeping their distance fixed. In one case, a direct measurement of distance variation is
done, on the other is the force needed to compensate for the motion which is estimated.

This simple model shall motivate and guide the reader in venturing into the LTP dynamics chapter and understanding
the main characters of drag-free on a simplified canvas.

One last remark at this basic level is of course about low frequency noise.
When the controls are forced to exert a direct, non-alternate current in-band
via the capacitors, there the noisiest contribution shows up. We remind to chapter \ref{chap:noise}
for a thorough explanation, but here we can say that with an electrostatic sensor, applying
a force also induces a gradient and thus a spring which couples any relative motion into a noisy force:
whenever additional, spurious or constant accelerations arise, these need to be compensated and the price to pay
for a dynamical, electrostatic compensation is the increase of the coupling between mass and SC, quadratic in the voltage.
It is true on the converse that there's need of a positive spring mechanism to stabilise the dynamics.
This aim can be pursued only via a low frequency suspension applied through the electrostatics.

In addition to the electrostatic actuation stiffness, one of the most dangerous sources of DC forces aboard SMART-2 is the action of
static self-gravity of the space-probe itself.
The presence of uncompensated local static gravity pulls would force the electrostatic system to exert continuous, noisy DC anti-forces \cite{Armano:2005ut}.
Basic requirements and a full gravitational control protocol \cite{ltpgravprot} have been written to pave the way
and solve this delicate issue for both LTP and LISA. A dedicated section of this thesis summarises the large load of work
carried on to study the problem of gravitational compensation, the reader may find about it in section \ref{sec:gravitcomp}.

\subsection{Experiment performance and sensitivity, similarities and differences with LISA.}
\label{sec:lisaltpdiff}
%
By virtue of a control scheme similar to that we described LISA's SCs actively follow the proof mass located
within each of them. If the loop gain is high enough,
the difference in acceleration between two masses sitting in two different SCs can be
measured by the interferometer as:
\beq
{\Delta}a = {\Delta}g -\omega_{p,2}^{2}{\delta}x_{2}+\omega_{p,1}^{2}{\delta}x_{1}
+\omega^{2} {\Delta}x_n\,,\label{eq:LISAdeltaF}
\eeq
where $\Delta g$ is the difference of position independent, fluctuating forces per unit mass directly acting on TMs
and $\omega_{p,i}^{2}\,i=1,2$ is the stiffness
per unit mass of parasitic spring coupling proof mass 1(2) to the SC. ${\delta}x_{i}\,i=1,2$ is the residual
jitter of the same proof-mass relative to the SC. Here as before ${\Delta}x_n \sim \text{IFO}_{n}({\Delta}x)$
is typically the interferometer noise difference along the $\hat x$ channel.

As we saw the LTP experiment uses a similar drag free control scheme. However here both masses sit in one SC and cannot be
simultaneously followed by. We saw formerly two control schemes and
we refer to the dynamics and noise sections for detailed equations including all DOF and
cross-talk.
%
%

In the LTP both TMs are spring coupled to the same SC and both feel then the relative jitter between
this one and the drag free reference proof mass 1. Therefore we can restate the problem by including this modification together
with the
effect of the control loop transfer function, recapping \eqref{eq:scimodesimple} by means of \eqref{eq:matchingstiffness}:
\beq
{\Delta}a \simeq \frac{\omega^{2}}{\omega_{\text{lfs},x}^2+\omega_{\text{p},2}^2-\omega^{2}} \Bigl(
-\omega^{2}{\Delta}x_n +{\Delta}g
+\left({\delta}x_1 - {\delta}x_2\right)\left(\omega_{\text{p},2}^2+\omega_{\text{lfs},x}^2\right)\Bigr)\,.
\eeq
By setting $\omega^{2}_{p,1}\simeq \omega^{2}_{p,2}$ a
substantial
residual jitter ${\delta}x_{1}$ might get unobserved thus bringing to optimistic underestimate of the noise in
\eqref{eq:LISAdeltaF}.
This is
easily avoided by a detailed sequence of  measurements of both $\omega^{2}_{p,1}$ and $\omega^{2}_{p,2}$ that has been
described \cite{Bortoluzzi:2004cz}. Provided this procedure, the measure of ${\Delta}g$ can be carried on successfully.

Correlation of
disturbances on different proof-masses may play a different role in LISA than in LTP. In LISA proof-masses within the
same interferometer arm belong to different SCs and are located $5\times 10^{9}\,\unit{m}$ apart. The only correlated
disturbances
one can think of are connected to the coupling to Sun: magnetic field fluctuations, fluctuation of the flux of charged
particles in solar flares and the fluctuation of solar radiation intensity that may induce correlated thermal
fluctuations in distant SC. These correlations will only slightly affect the error budget and will have no
profound consequences on the experiment itself. 

In the LTP all sources of noise that share the same source for both TMs are correlated. Magnetic noise
generated on board the SC, thermal fluctuations and gravity noise due to thermo-elastic distortion of SC
constitute a few examples. The major concern with correlated noise is that, by affecting the proof-masses the same way,
it might subtract from the differential measurement. Such a subtraction would not occur in LISA and thus would bring to
an optimistic underestimate of the total noise by LTP. This possibility can reasonably be avoided for almost all candidate
effects by a careful study (like the following, see chapter \ref{chap:noise}) and dedicated procedures.

An exception is constituted by the gravitational noise for which the response, due to the equivalence principle, cannot
be changed. However realistic assumptions about thermal distortion make the event of a gravity fluctuation affecting
both proof-masses with the same force along the $\hat x$ axis very unlikely. Moreover, the TMs are localised and
receive pulls by localised mass distributions as well, therefore is quite likely that we'll have different static components
per unit mass.



As an instrument to measure ${\Delta}g$, the LTP
is then limited at high frequency by the laser interferometer noise. For this last the laser path length noise
is still $S_{{\delta}x}^{\nicefrac{1}{2}}\leq 20\,\unitfrac{pm}{\sqrt{Hz}}$ while
the interferometric apparatus is deemed to achieve a sensitivity of:
\begin{shadefundnumber}
\beq
S_{{\delta}x,\,\text{LTP}}^{\nicefrac{1}{2}} = 9
  \left(1+\left(\frac{\omega}{2\pi\times 3\,\unit{mHz}}\right)^{-4}\right)^{\nicefrac{1}{2}}\,
  \unitfrac{pm}{\sqrt{Hz}}\,,
\eeq
\end{shadefundnumber}
\noindent which can be converted into an acceleration noise as before, by virtue of \eqref{eq:interfdeltafovermpsd}.
In the MBW we don't really need to be so strict to ask such a complicated function of $\omega^{2}$,
more than enough is the effective behaviour:
\begin{shadefundnumber}
\beq
\begin{split}
S_{\nicefrac{{\Delta}F}{m},\,\text{laser, LTP}}^{\nicefrac{1}{2}} &= \frac{\omega^{2}}{\sqrt{2}}
S_{{\delta}x,\,\text{LTP}}^{\nicefrac{1}{2}} =\\
&\simeq 3.2\times 10^{-15}
\left(1+\left(\frac{\omega}{2\pi\times 3\,\unit{mHz}}\right)^{2}\right)\,\accPSDunit\,.
\label{eq:LTPinterfsens}
\end{split}
\eeq
\end{shadefundnumber}
At lower frequencies an additional force noise adds up to
mask the parasitic forces due to other sources. This force noise is due to the fluctuations of the gain of the
electrostatic suspension loop. Indeed the electrostatic suspension must also cope with any static force acting on the
TMs. If the force stays constant but the gain fluctuates, the feedback force fluctuates consequently, adding a
noise source that is expected to limit the sensitivity at the lowest frequencies. This effect only appears in the LTP as
in LISA static forces are compensated just by the drag-free loop, and no electrostatic suspension is envisaged. The
largest expected source of gain fluctuations is the fluctuation of the DC voltage which is used to stabilise the
actuation electronics.

Stability should not worsen faster
than $\nicefrac{1}{\omega}$ at lower frequencies down to $0.1\,\unit{mHz}$. If this goal can be reached, suspension gain
fluctuations may be modelled with the following PSD \cite{Anza:2005td}:

\begin{shadefundnumber}
\beq
S^{\nicefrac{1}{2}}_{\nicefrac{{\Delta}F}{m},\,\text{susp, LTP}} \simeq 1.8\times 10^{-15}
\accPSDunit
\left(1+\left(\frac{\omega}{2\pi\times 1\,\unit{mHz}}\right)^{-2}\right)\,.
\label{eq:LTPsenssusp}
\eeq
\end{shadefundnumber}

The effective sensitivity curve of LTP - at the level we want to introduce it here - is
thus obtained combining \eqref{eq:LTPinterfsens} with \eqref{eq:LTPsenssusp}:
\beq
S_{\nicefrac{{\Delta}F}{m},\,\text{sens, LTP}} \simeq S_{\nicefrac{{\Delta}F}{m},\,\text{laser, LTP}} +
S_{\nicefrac{{\Delta}F}{m},\,\text{susp, LTP}}\,.
\eeq

\begin{figure}
\begin{center}
\includegraphics[width=.9\textwidth]{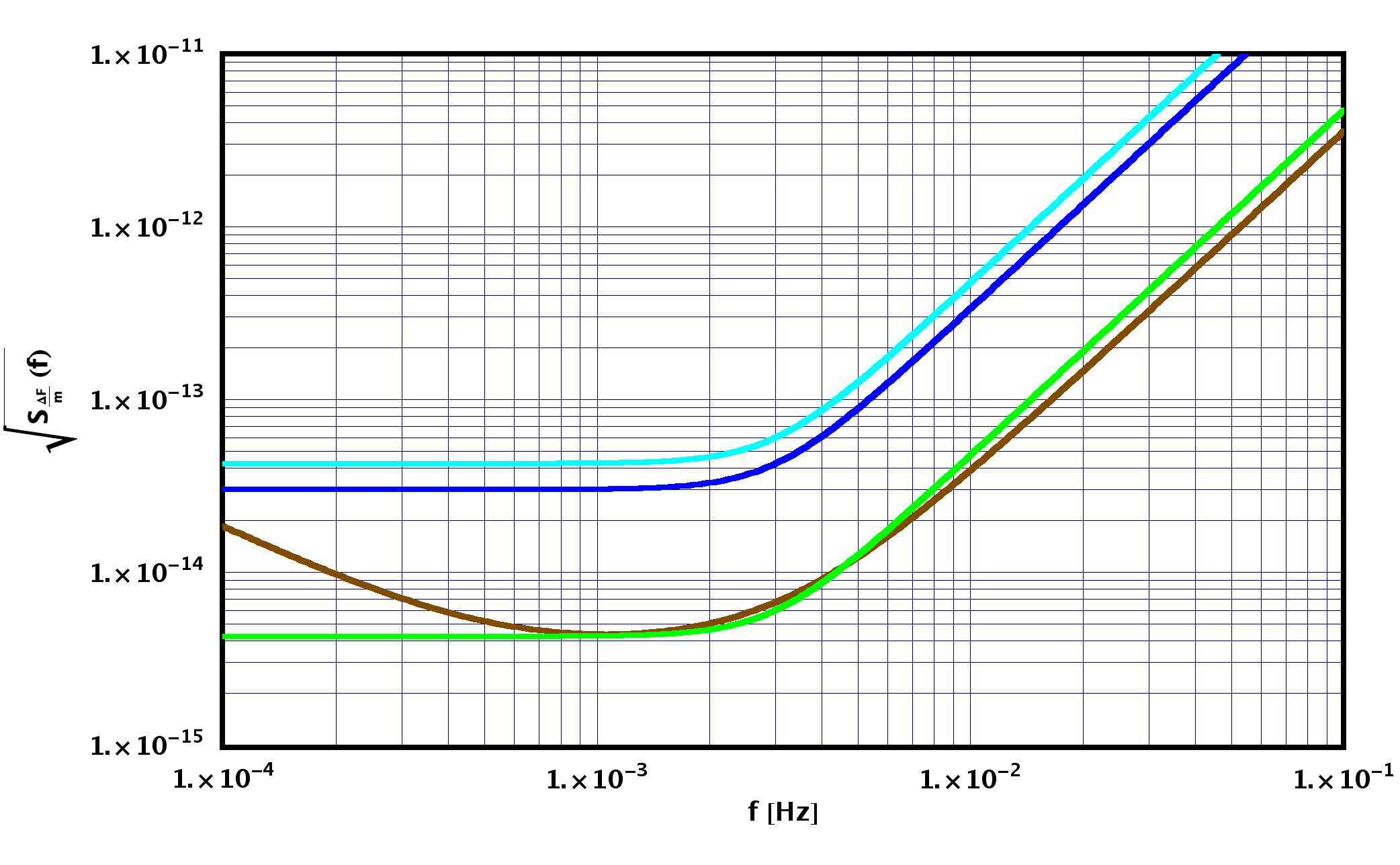}
\end{center}
\caption{Brown curve: projected sensitivity for differential force measurement of the LTP experiment, $S_{\nicefrac{{\Delta}F}{m},\,\text{sens, LTP}}^{\nicefrac{1}{2}}$.
Blue: required
maximum differential acceleration noise for the LTP. Green: LISA requirements. Cyan: LISA
minimum mission requirements.}
\label{fig:LTPspec}
\end{figure}

More will be said on noise in chapter \ref{chap:noise}, here we'd like to give some flavour by
showing a sensitivity prediction figure, obtained by plotting $S_{\nicefrac{{\Delta}F}{m},\,\text{sens, LTP}}^{\nicefrac{1}{2}}$
with the requirement expressions for LISA, eq. \eqref{eq:deltafoverfrelax} and for LTP, eq. \eqref{eq:ltpsensitivity}:
see figure \ref{fig:LTPspec}.
%


Figure \ref{fig:LTPspec} shows that the ultimate uncertainty on the differential acceleration PSD can be potentially
constrained by the
LISA Pathfinder mission below a factor $5$ above LISA requirements at $0.1\,\unit{mHz}$, and near LISA requirements
at $1\,\unit{mHz}$ or above.
In addition, within the entire frequency range, the LISA Pathfinder mission will constrain the acceleration noise
somewhere in the range between $1$ and $10\,\unitfrac{fm}{s^{2}\sqrt{Hz}}$ well below the requirements of LISA minimum
mission, thus strongly
reducing the risk of a LISA failure. Notice that the resulting TT frame, a frame where free particles at rest remain at
rest, is a very close approximation to the classical concept of inertial frame, and would indeed be inertial, within the
MBW, wouldn't it be for the presence of the gravitational wave. Thus LISA Pathfinder will demonstrate
the possibility of building an inertial frame in a standard SC orbiting the Sun on a scale of a meter in space
and of a few hours in time at the above mentioned level of absence of spurious accelerations.
%



%% file: chapters/ltp2.tex
\chapter{LTP: dynamics and signals}
\label{chap:ltp}

\lettrine[lines=4]{H}{aving}
inspected the main peculiarities of the simplified model for the LTP
dynamics in chapter one, we'll now proceed from ground up in writing down
the full LTP dynamics. The purpose is elucidate the procedure of building
the signals and motivate control modes from the signals themselves.

Reference systems will be conjured, pictures and renderings of the devices
will be provided. The equations of motion have been deduced using Newtonian
dynamics. Perhaps a Lagrangian treatment would have been more elegant, but
we thought it better to choose the former because noise and feedback are more natural
to be introduced in this scenario.

Special cases for the signals have been singled out and a ``propagator''-like
approach chosen for transfer functions. Feed-backs and suspensions are discussed
and their graphical behaviour sketched.
The chapter is somehow thought as preparatory for the noise detailed treatment
which follows.

\newpage

\section{Layout, coordinates and frames}
\begin{figure}
\begin{center}
\includegraphics[width=\textwidth]{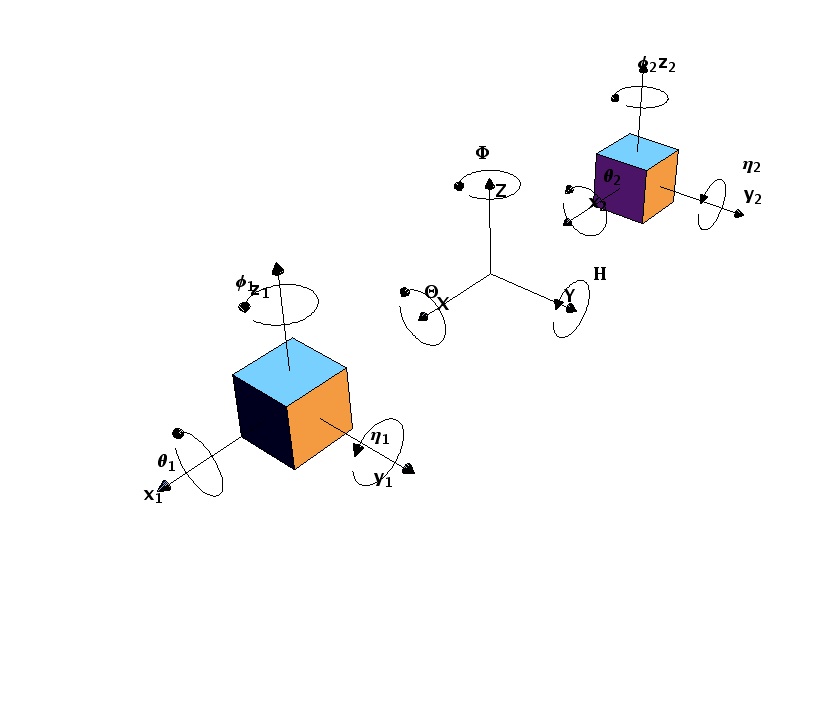}
\caption{Fundamental reference systems on board LTP: TM1, TM2 and SC centre of mass coordinates are
expressed as $6\times 3$ DOF, linear and conjugated angular.}
\label{fig:ltpsysref}
\end{center}
\end{figure}

Our concept of LTP to start with is somehow idealised in comparison to reality. Nevertheless the model can be
complicated up to whatsoever detail level, by introducing non-linearity and additional features \cite{LTPdefdoc}. LTP can simply be
described in terms of the two test masses (TM1, TM2), separated by a distance $r_{0}$ along the $\hat x$-axis (see for
reference, figure \ref{fig:ltpsysref}). Each TM is hosted inside a sensing facility designed in a similar manner as
in ground testing experiments.

\begin{figure}
\begin{center}
\includegraphics[width=\textwidth]{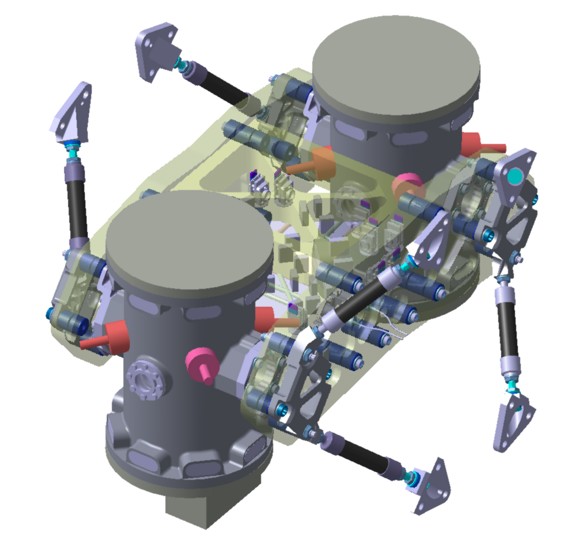}
\caption{Rendering of external apparatus's comprising Vacuum Enclosures (VE) of the Inertial Sensors (cylinders aside),
Optical Bench (midway, semi-transparent), struts and fittings. The VE hold the Electrode Housings which in turn
contain the TMs providing autonomous ultra high
vacuum around the TMs (non visible). The optical bench in between the TMs (in grey) supports the
interferometry that reads out the distance between the masses. The interferometer laser beam hits each TM by
crossing the vacuum enclosures through an optical window. The entire supporting structure is made out of glass-ceramics
for high therm mechanical stability. Optical fibres carrying UV light are used for
contact less discharging of TMs.}
\label{fig:ltpisrender}
\end{center}
\end{figure}

The following reference frames may be considered:
\begin{enumerate}
\item $\text{SF}_\text{CM}$: the system of the SC centre of mass,
\item $\text{SF}_\text{BF}$: the body fixed reference frame of the SC,
\item $\text{ISF}i,\,i=1,2$: the body fixed reference frame of the IS electrode housing of TM1 and TM2 respectively.
\end{enumerate}
In addition, a generic inertial frame IF may appear and in most cases all body fixed reference frames only differ for
very small perturbations in displacement and rotations so that transformations of coordinates may be assumed as linear;
Euler cross-terms will always be neglected in this spirit.

According to the mentioned frames, relevant systems of coordinates can be forged. The general dynamics - and
hence all the signals - will be written as functions of these:
\begin{enumerate}
\item the TM-$i$ coordinates in the IF:
\beq
\vc  x_{i} = \left[\begin{matrix}x\\y\\z\\\theta\\\eta\\\phi\end{matrix}\right]_{i}\,,\, i=1,2\,,
\eeq
whose origins are taken to be the instantaneous positions of the nominal centres of the electrode housings; sometimes
we may refer to the two translational and angular subsets of coordinates:
\beq
\vc  x^{T}_{i} = \left[\begin{matrix}x\\y\\z\end{matrix}\right]_{i}\,,\qquad
\vc  x^{R}_{i} = \left[\begin{matrix}\theta\\\eta\\\phi\end{matrix}\right]_{i}\,.
\eeq
Notice $\vc x_{i} = \vc  x^{T}_{i} ~\otimes~ \left[\begin{matrix}1\\0\end{matrix}\right]
+ \vc  x^{R}_{i} ~\otimes~ \left[\begin{matrix}0\\1\end{matrix}\right]$, for $i=1,2$.
\item The SC coordinates in the IF:
\beq
\vc  X = \left[\begin{matrix}X\\Y\\Z\\\Theta\\H\\\Phi\end{matrix}\right]\,,
\eeq
whose origin is the instantaneous position of the SC centre of mass.
\item The TMs coordinates in the ISF1 and 2:
\beq
\vc x^{\text{SC}}_{i} = \left[\begin{matrix}x\\y\\z\\\theta\\\eta\\\phi\end{matrix}\right]^{\text{SC}}_{i}\,,\, i=1,2\,,
\eeq
origins are assumed in the nominal centres of the respective electrode housings. Their difference with the coordinates
in the $SF_\text{BF}$ is a simple vector only if there's no distortions at play. $\vc x^{\text{SC}}_{i}$ and $\vc x_{i}$ are
related by a roto-translational transformation which will be clarified in the following.
\item The distortions of ISF1,2 relative to $SF_\text{BF}$:
\beq
\vc  {\delta x}^{\text{SC}}_{i} = \left[\begin{matrix}\delta x\\\delta y\\\delta
z\\\delta\theta\\\delta\eta\\\delta\phi\end{matrix}\right]_{i}\,,
\eeq
which won't be used extensively if not for sensitivity and cross-talk analysis.
\end{enumerate}

Moreover, a caveat must be given at this time: TMs and SC are extended bodies, not point-like masses.
Therefore, a complicated set of problems arise which are not present in the standard Newtonian approach
for low velocity particles. Cross-talk summarises part of these, and carries dynamical, geometrical (affine)
and electrostatics features: effective rotational arms between SC and TMs need to be taken into account,
errors in positioning translating into rotational jitter, effective spring or magnetic couplings resulting from
the extensiveness of the bodies. In section \ref{sec:crosstalk} we'll deal with cross-talk in more
detailed way.

Self-gravity acquires also multi-source features: the static imbalance of solid structures of the SC acts on the
TMs in a very nontrivial way, not separable into a couple of point-like sources: relative stiffness and
static gravity gradients demand a dedicated treatment and solution which we'll tackle in section \ref{sec:gravitcomp}.

\section{Signals}

\begin{figure}
\begin{center}
\includegraphics[width=0.49\textwidth]{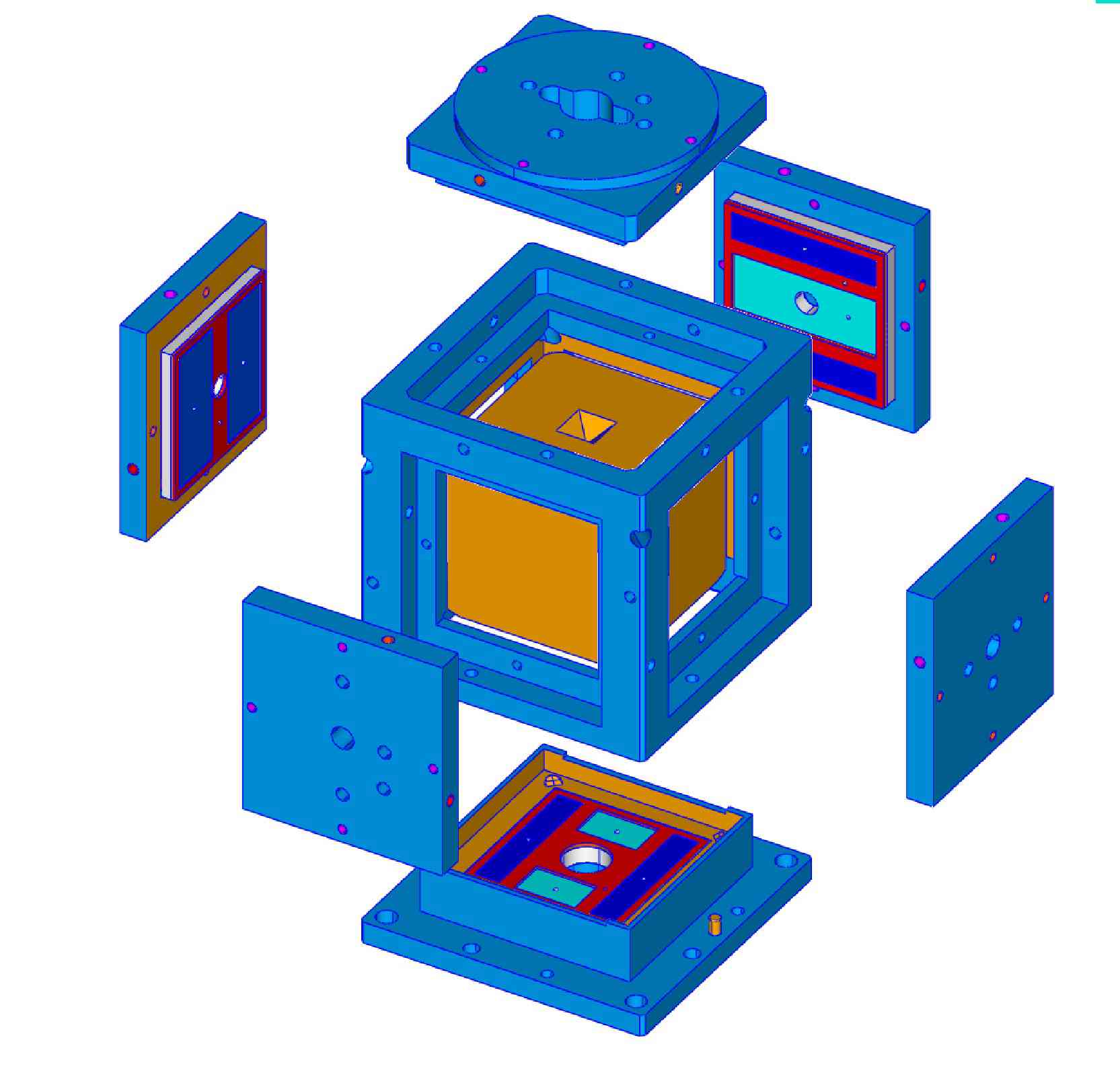}%
\includegraphics[width=0.50\textwidth]{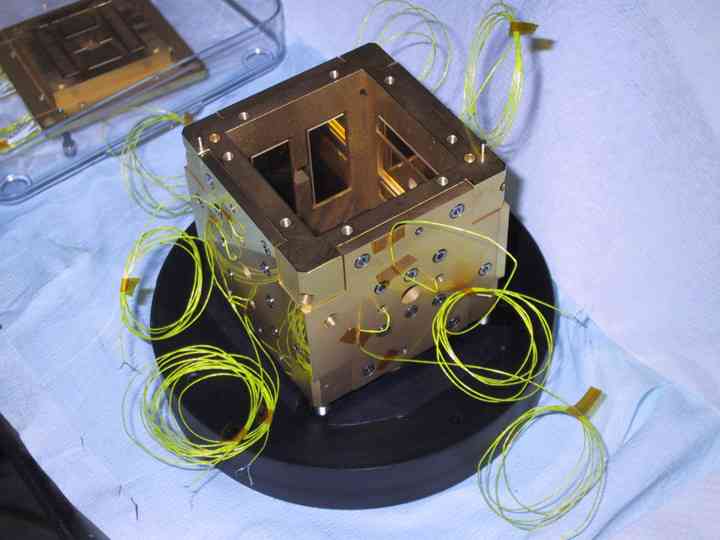}
\includegraphics[width=\textwidth]{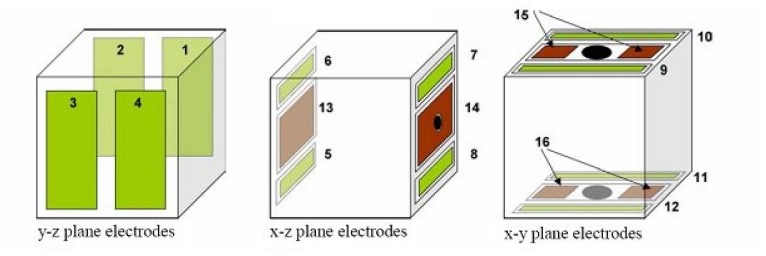}
\caption{Top left: scheme of the LTP readout and actuation electrodes (GRS) per TM: GRS electrodes
are held by the frame of the Electrode Housing and are characterised by many capacitance sleeves. The TM
can be seen inside the frame structure. Top right: engineering model of housing on a table in the Low Temperature
and Experimental Gravity Laboratory in Trento. GRS capacitors glitter inside the cube. Below: detail of
electrodes per direction.}
\label{fig:ltptmeh}
\end{center}
\end{figure}

The generalised vectors $\vc x^{\text{SC}}_{i}$ are internal state vectors of LTP, while $\vc X$ is the state vector of
the SC. Each TM is surrounded by electrodes (see figure \ref{fig:ltptmeh}) which can provide electrostatic readout of the
position as well as induce motion by voltage actuation. From the
state vectors we can generate signals encompassing noise in the definition:
\begin{shadefundtheory}
\beq
\text{GRS}\left(\vc x_{i}\right)=\vc x_{i}^{\text{SC}}+\text{GRS}_{n}\left(\vc x_{i}\right)\quad i=1,2\,,
\eeq
\end{shadefundtheory}
\noindent where we defined the noise vectors $\text{GRS}_{n}\left(\vc x_{i}\right)$, whose characteristic upper
limits can be deduced from the capacitance
sensors properties and, for the specific case of LTP are pointed out in table \ref{tab:capacsensreadoutnoise}.
Anywhere in the following, ``GRS'' designates the readout signal from
the capacitance electronics and stands for Gravitational Reference Sensor.

\begin{table}
\begin{center}
\begin{shadefundnumber}
\begin{tabular}{c|c}
GRS disturbance & maximum at $1\,\unit{mHz}\leq f\leq 1 \unit{Hz}$ \\
\hline\rule{0pt}{0.4cm}
$x^{n}_{i}$, $y^{n}_{i}$, $z^{n}_{i}$ & $1.8\,\nicefrac{\unit{nm}}{\unit{\sqrt{Hz}}}$\\
$\phi^{n}_{i}$, $\eta^{n}_{i}$, $\theta^{n}_{i}$ & $200\,\nicefrac{\unit{nrad}}{\unit{\sqrt{Hz}}}$\\
\end{tabular}
\end{shadefundnumber}
\end{center}
\caption{Expected noise levels in GRS capacitive readout.}
\label{tab:capacsensreadoutnoise}
\end{table}

\begin{figure}
\begin{center}
\includegraphics[width=\textwidth]{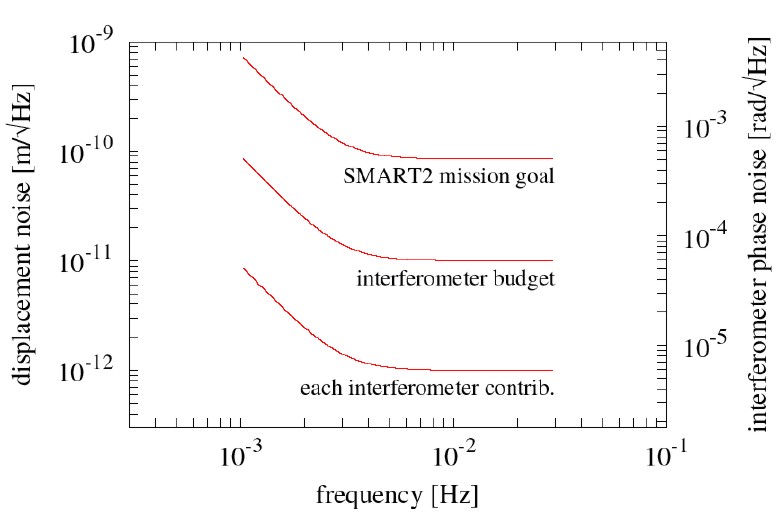}
\caption{Laser interferometer noise (lowest curve) as angular and linear displacement PSDs.}
\label{fig:ltplasernoise}
\end{center}
\end{figure}

More sensitive than the electrostatic readout, a laser metrology equipment is placed aboard
LTP in order to read 3 positions and 4 attitudes by means of quadrant photo-diodes. The ability of measurement with
such a device is the core of the experiment, i.e. to detect the difference between the two $x^{\text{SC}}_{i}$ of the TMs. The signal
vector is defined as:
\begin{shadefundtheory}
\beq
\text{IFO}(
\left[\begin{matrix} x_{1}\\ \phi_{1}\\
\eta_{1}\\x_{2}\\\phi_{2}\\\eta_{2}\\x_{2}-x_{1}\end{matrix}\right]
)
= \left[\begin{matrix} x_{1}\\ \phi_{1}\\
\eta_{1}\\x_{2}\\\phi_{2}\\\eta_{2}\\x_{2}-x_{1}\end{matrix}\right]^{\text{SC}}
+\text{IFO}_{n}(
\left[\begin{matrix} x_{1}\\ \phi_{1}\\
\eta_{1}\\x_{2}\\\phi_{2}\\\eta_{2}\\ x \end{matrix}\right]
)\,,
\eeq
\end{shadefundtheory}
\noindent where we put the subscript $n$ to the laser noise. Laser sensitivity can be given as a power
spectral density (PSD) function of frequency and can be viewed in figure \ref{fig:ltplasernoise}. For attitude readout
the photo-diode sensitivity on the angular signals and in frequency range $1\,\unit{mHz}\leq f\leq 1\,\unit{Hz}$
amounts to a maximum uncertainty of $10\,\nicefrac{\unit{nrad}}{\unit{\sqrt{Hz}}}$.
The name ``IFO'' comes from InterFerometer Output.

Last signals are originated by star-trackers, high-precision pointers directed to
reference stars. These signals shall control rotational modes of the SC, which in principle
wouldn't be otherwise prevented to continuously rotate around the $\hat x$ sensing axis,
wasting fuel and introducing more undesired couplings. Moreover, the solar panels of LTP must
always be facing the Sun to maximise the power outcome \cite{Landgraf:2004gm},
another reason to prevent
LTP from spinning around $\hat x$. Therefore we define:
\begin{shadefundtheory}
\beq
\text{ST}(\left[\begin{matrix} \Theta\\ H\\ \Phi\end{matrix}\right])
=\left[\begin{matrix} \Theta\\ H\\ \Phi\end{matrix}\right]^{\text{SC}}+
\text{ST}_{n}(\left[\begin{matrix} \Theta\\ H\\ \Phi\end{matrix}\right])\,,
\eeq
\end{shadefundtheory}
\noindent whose angular PSD error for $f\leq 5 \times 10^{-4}\,\unit{Hz}$ is
$\leq 10^{-4}\,\nicefrac{\unit{nrad}}{\unit{\sqrt{Hz}}}$.
These give reference attitude of the SC with respect to ``fixed'' stars\footnote{We make a
big fuzz about relativity and inter-changeability of reference frames and then we stick to
the idea that far away huge plasma-gas balls slowly moving can be taken as good references!
But in fact we lock to their light signals, which is reliable, indeed.} and shall provide the
needed control level.

\section{Equations of motion}
Given the coordinates systems and signals of the former section, a set of equations
describing the unperturbed dynamics of LTP can be written. It is quite obvious that the kinetic
matrix of the whole will be manifestly invariant with respect to any coupling we'd choose or any
self-force. Besides, the form of the interaction terms and the couplings between signals and
coordinates will change due to the control strategy we'd apply, other than the reference chosen.

For each TM and for the SC a set of generalised mass matrices can be defined:
\beq
M_{j}=\left[\begin{matrix}
M_{j}^{T} & 0 \\
0 & M_{j}^{R}
\end{matrix}\right]\,,
\eeq
where $j=1,2,{\text{SC}}$ and the superscript $T,\,R$ in the sub-mass matrices
define whether we are talking about a linear or angular DOF, to be multiplied by
accelerations or velocities (linear or angular) in Newton or Lagrange equations. The $M_{j}^{R}$ are
of course, principal moments of inertia matrices:
\beq
M_{j,ik}^{T}=m\delta_{ik}\,,\qquad M_{j,ik}^{R}=m\frac{L^{2}}{6}\delta_{ik}\,,
\eeq
where $i,k=1..3$, $j=1,2$ and $\delta$ is the usual Kronecker delta indicator. We assumed
perfect cubes with side length $L$ for the TMs shapes, having mass $m$.
On the other hand, our picture of the SC is closer
to a cylinder with height $\simeq$ radius $(R_{\text{SC}})$: the moments of inertia will be axis-wise
different and we'll get
\beq
M_{{\text{SC}},ik}^{T}=m_{\text{SC}}\delta_{ik}\,,
  \qquad M_{\text{SC}}^{R}=m_{\text{SC}} R^{2}_{\text{SC}}
     \left[\begin{matrix}
\frac{1}{3} & 0 & 0 \\
0 & \frac{1}{3} & 0 \\
0 & 0 & \frac{1}{2}
\end{matrix}\right]\,,
\eeq
$m_{\text{SC}}$ being the SC mass.
Stiffness matrices can be defined for both TMs, similarly:
\beq
K_{j}=\left[\begin{matrix}
K_{j}^{T} & 0 \\
0 & K_{j}^{R}
\end{matrix}\right]\,,
\eeq
where
\beq
\begin{split}
K_{j}^{T} &= m\left[\begin{matrix}
\omega_{x_{j},x_{j}}^{2} & 0 & 0 \\
0 & \omega_{y_{j},y_{j}}^{2} & 0 \\
0 & 0 & \omega_{z_{j},z_{j}}^{2}
\end{matrix}\right]\,,\\
K_{j}^{R} &= m\frac{L^{2}}{6}\left[\begin{matrix}
\omega_{\theta_{j},\theta_{j}}^{2} & 0 & 0 \\
0 & \omega_{\eta_{j},\eta_{j}}^{2} & 0 \\
0 & 0 & \omega_{\phi_{j},\phi_{j}}^{2}
\end{matrix}\right]\,,
\end{split}
\eeq
conversely, no need or opportunity to define similar matrices for the SC, which has no reference to
be defined ``dynamically stiff'' to. In a word these matrices express the spring-like linear coupling of the TMs to
the whole SC. Dimensions of the coefficients $\omega_{\hat i,\hat i}^{2}$ is $\nicefrac{1}{\unit{s^{2}}}$.

External stimuli acting on each TM can be written as follows, along the linear
and angular conjugated directions:
\beq
\begin{split}
\vc F_{i} &= m \left[\begin{matrix}g_{x} \\ g_{y} \\ g_{z}\end{matrix}\right]_{i}\,,\\
\vc\gamma_{i} &=m \frac{L^{2}}{6}
  \left[\begin{matrix}g_{\theta} \\ g_{\eta} \\ g_{\phi}\end{matrix}\right]_{i}\,,
\end{split}
\eeq
while for the SC similar expressions hold, provided the following substitutions:
\beq
m\to m_{\text{SC}}\,,\qquad \frac{L^{2}}{6}\to R^{2}_{\text{SC}}\left[\begin{matrix}
\frac{1}{3} & 0 & 0 \\
0 & \frac{1}{3} & 0 \\
0 & 0 & \frac{1}{2}
\end{matrix}\right]\,.
\eeq
From time to time we may be grouping forces and torques into generalised forces as
$\vc f_{i}=F_{i}~\otimes~
\left[\begin{array}{c}1\\0\end{array}\right] +
\gamma_{i}~\otimes~
\left[\begin{array}{c}0\\1\end{array}\right]$ and at even higher level we'll define an
indice-less $\vc f=\vc f_{1}~\otimes~
\left[\begin{array}{c}1\\0\end{array}\right] +
\vc f_{2}~\otimes~
\left[\begin{array}{c}0\\1\end{array}\right]$.
The special vectors
\beq
\vc r_{i}=\left[\begin{matrix}(-1)^{i} \frac{r_{0}}{2}\\0\\z_{0}\end{matrix}\right]
\eeq
define the unperturbed direction connecting the SC centre of mass to the TM-$i$ nominal positions.
$r_{0}$ is the mutual distance between the two TMs and $z_{0}$ is the (vertical) distance between the TMs
mutual $\hat x$ axis and the SC centre of mass.

The equations of motion in the TMs IF look then like:
\begin{shadefundtheory}
\begin{align}
M^{T}_{j} \ddot{\vc x}^{T}_{j} + K^{T}_{j} \vc x^{T}_{j} &= \vc F_{j}\qquad j=1,2\,,\\
M^{R}_{j} \ddot{\vc x}^{R}_{j} + K^{R}_{j} \vc x^{R}_{j}&= \vc \gamma_{j}\qquad j=1,2\,,\\
M^{T}_{\text{SC}} \ddot{\vc x}^{T}_{\text{SC}} &= \vc F_{\text{SC}}\,, \label{eq:scdynmotion-t}\\
M^{R}_{\text{SC}} \ddot{\vc x}^{R}_{\text{SC}} &= \vc \gamma_{\text{SC}}\,\label{eq:scdynmotion-r},
\end{align}
\end{shadefundtheory}
\noindent and no sum is implied over repeated indice. As stated already, Euler self-coupling terms are and will be neglected
in the following since they are small and can be dynamically compensated for by the actuation loops.

Since we want to keep track of things in the inertial frame of each TM, the SC motion must
be taken into account. The relative change of coordinate in both translational and rotational frames
can be accounted for with the transformation:
\begin{shadefundtheory}
\beq
\begin{split}
\vc F_{j} &\to \vc F_{j} - M^{T}_{j} \left(\ddot{\vc x}^{T}_{\text{SC}}
  +\ddot{\vc x}^{R}_{\text{SC}}\wedge {\vc r}_{j}\right)\,,\\
\vc \gamma_{j} &\to \vc \gamma_{j} - M^{R}_{j} \ddot{\vc x}^{R}_{\text{SC}}\,.
\end{split}
\label{eq:changerefframes}
\eeq
\end{shadefundtheory}

Due to the wide application of control theory in Fourier and Laplace spaces to analyse the dynamics of
the bodies, for most of the time we'll write the equation in the frequency domain, i.e. we'll make the
association:
\beq
\dot {\vc x} \to s \vc x\,,\qquad s \doteq \imag \omega\,,
\eeq
where $s(\omega)$ is the Laplace(Fourier) transformed time and we assume convergence in
integration unless otherwise discussed. Our equations of motion are natively linear and being
treated in Fourier space will lead us to deducing signals as ``propagators'' in the frequency domain.
This is natural in control theory and permits a more transparent discussion on critical frequencies
as well as measurement bandwidth and resonances. When discussing noise, we'll make wide use
of Power Spectral Densities (PSD), Fourier transforms of correlators, it seemed thus a wise choice to
work in frequency space from here.

Expanding the unperturbed equations of motion from matrix to scalar form, we get the following set of
coupled equations, $18$ in number; namely those for TM1:
\begin{align}
m x_1 s^2+m \left(x_{\text{SC}}+z_0 \eta _{\text{SC}}\right) s^2+m x_1 \omega _{x_1,x_1}^2-m g_{1,x} &=0\,,\\
m y_1 s^2+m \left(y_{\text{SC}}-z_0 \theta _{\text{SC}}-\frac{r_o \phi _{\text{SC}}}{2}\right) s^2+m y_1 \omega _{y_1,y_1}^2-m g_{1,y} &=0\,,\\
m z_1 s^2+m \left(z_{\text{SC}}+\frac{r_o \eta _{\text{SC}}}{2}\right) s^2+m z_1 \omega _{z_1,z_1}^2-m g_{1,z} &=0\,,\\
\frac{1}{6} m \theta _1 \omega _{\theta _1,\theta _1}^2 L^2+\frac{1}{6} m s^2 \theta _1 L^2+\frac{1}{6} m s^2 \theta _{\text{SC}} L^2-\frac{1}{6} m g_{1,\theta} L^2 &=0\,,\\
\frac{1}{6} m \eta _1 \omega _{\eta _1,\eta _1}^2 L^2+\frac{1}{6} m s^2 \eta _1 L^2+\frac{1}{6} m s^2 \eta _{\text{SC}} L^2-\frac{1}{6} m g_{1,\eta} L^2 &=0\,,\\
\frac{1}{6} m \phi _1 \omega _{\phi _1,\phi _1}^2 L^2+\frac{1}{6} m s^2 \phi _1 L^2+\frac{1}{6} m s^2 \phi _{\text{SC}} L^2-\frac{1}{6} m g_{1,\phi} L^2 &=0\,,
\end{align}
those for TM2:
\begin{align}
m x_2 s^2+m \left(x_{\text{SC}}+z_0 \eta _{\text{SC}}\right) s^2+m x_2 \omega _{x_2,x_2}^2-m g_{2,x} &=0\,,\\
m y_2 s^2+m \left(y_{\text{SC}}-z_0 \theta _{\text{SC}}+\frac{r_o \phi _{\text{SC}}}{2}\right) s^2+m y_2 \omega _{y_2,y_2}^2-m g_{2,y} &=0\,,\\
m z_2 s^2+m \left(z_{\text{SC}}-\frac{r_o \eta _{\text{SC}}}{2}\right) s^2+m z_2 \omega _{z_2,z_2}^2-m g_{2,z} &=0\,,\\
\frac{1}{6} m \theta _2 \omega _{\theta _2,\theta _2}^2 L^2+\frac{1}{6} m s^2 \theta _2 L^2+\frac{1}{6} m s^2 \theta _{\text{SC}} L^2-\frac{1}{6} m g_{2,\theta} L^2 &=0\,,\\
\frac{1}{6} m \eta _2 \omega _{\eta _2,\eta _2}^2 L^2+\frac{1}{6} m s^2 \eta _2 L^2+\frac{1}{6} m s^2 \eta _{\text{SC}} L^2-\frac{1}{6} m g_{2,\eta} L^2 &=0\,,\\
\frac{1}{6} m \phi _2 \omega _{\phi _2,\phi _2}^2 L^2+\frac{1}{6} m s^2 \phi _2 L^2+\frac{1}{6} m s^2 \phi _{\text{SC}} L^2-\frac{1}{6} m g_{2,\phi} L^2 &=0\,,
\end{align}
and those for the SC:
\begin{align}
s^2 m_{\text{SC}} x_{\text{SC}}-m_{\text{SC}} g_{\text{SC},x} &=0\,,\\
s^2 m_{\text{SC}} y_{\text{SC}}-m_{\text{SC}} g_{\text{SC},y} &=0\,,\\
s^2 m_{\text{SC}} z_{\text{SC}}-m_{\text{SC}} g_{\text{SC},z} &=0\,,\\
\frac{1}{3} s^2 m_{\text{SC}} R_{\text{SC}}^2 \theta _{\text{SC}}-\frac{1}{3} m_{\text{SC}} R_{\text{SC}}^2 g_{\text{SC},\theta} &=0\,,\\
\frac{1}{3} s^2 m_{\text{SC}} R_{\text{SC}}^2 \eta _{\text{SC}}-\frac{1}{3} m_{\text{SC}} R_{\text{SC}}^2 g_{\text{SC},\eta} &=0\,,\\
\frac{1}{2} s^2 m_{\text{SC}} R_{\text{SC}}^2 \phi _{\text{SC}}-\frac{1}{2} m_{\text{SC}} R_{\text{SC}}^2 g_{\text{SC},\phi} &=0\,.
\end{align}

The true reduced dynamic is represented in terms of the TMs variables, the SC coordinates can
be rid of by linear substitution into the equations, a process we'll eventually get through after having elucidated
something more about the signals.

\section{Operation modes}
\label{sec:modes}

The practical realisation and implementation of the control strategy give rise to what we call an operation mode
\cite{LTPdfacsgen, LTPdfacsfun, LTPcrosstalk}. The name comes from the fact that a certain strategy will be employed for
TMs and SC control during a time-slice
of the experiment, and sets of measures will be taken under those circumstances. Therefore the laws of motion in
the specific situation will become a command mode for the space-probe. Each mode is thus qualified by three
properties:
\begin{enumerate}
\item the mentioned control strategy, consisting in a set of feed-back laws,
\item the characteristic frequencies at play, together with actuation laws which will effectively
provide signal to the controlling apparatus's,
\item the gains, determining whether a control shall be called soft or hard, drag-free or suspension.
\end{enumerate}

For each mode the former points will be carefully discussed, several approximation may come at hand
and will be motivated.

A choice of principal signals must be made due to redundancy ($x_{1}$ can be measured by GRS or IFO, for
example) in order to describe the dynamics and analyse the LTP frequency behaviour: a selection of
signals can be extracted from the TMs DOF
$\vc x=x_{1}~\otimes~
\left[\begin{array}{c}1\\0\end{array}\right] +
x_{2}~\otimes~
\left[\begin{array}{c}0\\1\end{array}\right]$ with $\Omega \left( \vc x \right)$ being the
readout operator matrix in \eqref{eq:ssmat}.

\begin{figure*}
\begin{center}
\begin{sideways}
\begin{minipage}{1.3\linewidth}
\beq
\Omega=\left[\begin{array}{cccccccccccc}
 \text{IFO} & 0 & 0 & 0 & 0 & 0 & 0 & 0 & 0 & 0 & 0 & 0 \\
 0 & \text{GRS} & 0 & 0 & 0 & 0 & 0 & 0 & 0 & 0 & 0 & 0 \\
 0 & 0 & \text{GRS} & 0 & 0 & 0 & 0 & 0 & 0 & 0 & 0 & 0 \\
 0 & 0 & 0 & \text{GRS} & 0 & 0 & 0 & 0 & 0 & 0 & 0 & 0 \\
 0 & 0 & 0 & 0 & \text{IFO} & 0 & 0 & 0 & 0 & 0 & 0 & 0 \\
 0 & 0 & 0 & 0 & 0 & \text{IFO} & 0 & 0 & 0 & 0 & 0 & 0 \\
 -\text{IFO} & 0 & 0 & 0 & 0 & 0 & \text{IFO} & 0 & 0 & 0 & 0 & 0 \\
 0 & 0 & 0 & 0 & 0 & 0 & 0 & \text{GRS} & 0 & 0 & 0 & 0 \\
 0 & 0 & 0 & 0 & 0 & 0 & 0 & 0 & \text{GRS} & 0 & 0 & 0 \\
 0 & 0 & 0 & 0 & 0 & 0 & 0 & 0 & 0 & \text{GRS} & 0 & 0 \\
 0 & 0 & 0 & 0 & -\text{IFO} & 0 & 0 & 0 & 0 & 0 & \text{IFO} & 0 \\
 0 & 0 & 0 & 0 & 0 & -\text{IFO} & 0 & 0 & 0 & 0 & 0 & \text{IFO}
\end{array}\right](\cdot)\,.
\label{eq:ssmat}
\eeq
\end{minipage}
\end{sideways}
\end{center}
\end{figure*}

Notice this is a particularly wise choice of readout which maximises the use of interferometer signals. We point
out that along the main ``science'' DOF redundancy of the signals is a key feature. Moreover,
electrostatic force can be exerted on the TMs via the capacitance (GRS) system, while main readout in
this scenario is IFO-based, thus reducing contamination. We get then a set of fundamental (and minimal) readout
signals, whose expressions will look like:
\begin{shadefundtheory}
\begin{align}
\text{IFO}\left(x_1\right) &=x_1+\text{IFO}_{n}\left(x_1\right)\,,\\
\text{GRS}\left(y_1\right) &=y_1+\text{GRS}_{n}\left(y_1\right)\,,\\
\text{GRS}\left(z_1\right) &=z_1+\text{GRS}_{n}\left(z_1\right)\,,\\
\text{GRS}\left(\theta _1\right) &=\theta _1+\text{GRS}_{n}\left(\theta _1\right)\,,\\
\text{IFO}\left(\eta _1\right) &=\eta _1+\text{IFO}_{n}\left(\eta _1\right)\,,\\
\text{IFO}\left(\phi _1\right) &=\phi _1+\text{IFO}_{n}\left(\phi _1\right)\,,\\
\text{IFO}\left(x_2-x_1\right) &=x_2-x_1+\text{IFO}_{n}\left(x_2-x_1\right)\,,\\
\text{GRS}\left(y_2\right) &=y_2+\text{GRS}_{n}\left(y_2\right)\,,\\
\text{GRS}\left(z_2\right) &=z_2+\text{GRS}_{n}\left(z_2\right)\,,\\
\text{GRS}\left(\theta _2\right) &=\theta _2+\text{GRS}_{n}\left(\theta _2\right)\,,\\
\text{IFO}\left(\eta _2-\eta _1\right) &=\eta _2-\eta _1+\text{IFO}_{n}\left(\eta _2-\eta _1\right)\,,\\
\text{IFO}\left(\phi _2-\phi _1\right) &=\phi _2-\phi _1+\text{IFO}_{n}\left(\phi _2-\phi _1\right)\,.
\end{align}
\end{shadefundtheory}

The concept of feedback is then put at play. Drag-free is achieved by controlling the SC motion enslaved to
TMs displacement, given some rules and a servo-control system. The most general expression we can write
renders also transparent the full strategy: forces and torques acting on the SC are felt by Newton law as accelerations
which jeopardise the inertiality of the TMs inside the LTP shielding. Such forces and torques are
re-injected as additive terms of the general forces and torques acting on the TMs as functions of the SC displacement
and attitude. Those terms can be seen as functions of the forces and torques acting on the SC.
In formulae, we call control strategy the map:
\begin{shadefundtheory}
\beq
\begin{split}
\vc F_{j} &\to \vc F_{j} + \vc {\hat F}\left(\vc x_{\text{SC}},\,\text{GRS}(\vc x_{i}),\,\text{IFO}(\vc x_{i}),\,\dots\right)\,,\\
\vc \gamma_{j} &\to \vc \gamma_{j} + \vc{\hat \gamma} \left(\vc x_{\text{SC}},\,\text{GRS}(\vc x_{i}),\,\text{IFO}(\vc x_{i}),\,\dots\right)\,,
\end{split}
\eeq
\end{shadefundtheory}
where the dependence on the signals has been highlighted, meaning that control strategies will be essentially
based on our knowledge of positions and therefore will naturally bring readout noise inside the equations.

\subsection{Science mode}
\label{sec:scimode}

The main operating mode is christened ``science mode'', formerly known as M3 \cite{LTPdfacsM3}. (In the real mission, unless otherwise
specified, in this mode the control loops are driven
by signals obtained from inertial sensors capacitive readout (GRS) or interferometer. We chose here to maximise the use of IFO readout whenever
possible).
Both in the IFO
MBW and below, the following control scheme is employed \cite{LTPscrd}:
\begin{itemize}
\item The SC is controlled in translation along $\hat x$ on TM1  This means that a force is applied to
the SC by properly firing the thrusters:
\begin{shadefundtheory}
\beq
F^{\text{thrusters}}_{\text{SC},x}=m_{\text{SC}}\omega^{2}_{\text{df,x}} x_{1}\,,
\eeq
\end{shadefundtheory}
\item The gain $\omega^{2}_{\text{df,x}}$ must be representative of that of LISA in the entire
control bandwidth $0\leq f\leq f_{0}$, with $f_{0}\geq 30\,\unit{mHz}$:
\begin{shadefundnumber}
\beq
m_{\text{SC}}\omega^{2}_{\text{df,x}} \geq 2\times 10^{2}\,\unitfrac{N}{m}\,.
\eeq
\end{shadefundnumber}
\item TM2 is subject to a low frequency suspension loop along $\hat x$ in order both to compensate 
for dc-forces and to stabilise its intrinsic negative stiffness. TM2 is controlled using the main
laser channel so that it's forced to follow TM1; the actuation force applied to TM2 by means of the
capacitive system is:
\begin{shadefundtheory}
\beq
F^{\text{act}}_{2,x}=-m\omega^{2}_{\text{lfs},x} (x_{2}-x_{1})\,.
\eeq
\end{shadefundtheory}
The open-loop gain $m\omega^{2}_{\text{lfs},x}$ of the low frequency suspension (LFS, lfs in indice from now on)
must guarantee that
forces between DC and the lower end of the measuring bandwidth do not displace the TM by more than
$1\,\unit{\mu m}$ from the nominal working point. The absolute value of the open-loop gain of the LFS
within the MBW must be adjustable to be $\left|m\omega^{2}_{\text{lfs},x}\right|\leq 3\times 10^{-5}\,\unitfrac{N}{m}$
during calibration of the transfer function of the forces on TM to the IFO$(x_{2}-x_{1})$ signal: this way
the gain is $\leq m\omega^{2}/2$ a the lowest corner of the MBW.
\item The SC is controlled in translation along $\hat y$ on TM2 the same way as it is controlled on
$\hat x$ for TM1.
\item Similarly, the SC is controlled in translation along $\hat z$ on TM1.
\item The SC is controlled in rotation around $\hat z$ and $\hat y$, within the MBW, by using the difference 
of the readouts of TM1 and TM2 along $\hat y$ and $\hat z$ respectively.
\item TM1 and TM2 have no electrostatic suspension along $\hat y$ and $\hat z$ within the MBW.
\item TM1 is subject to an electrostatic suspension along $\hat y$ below the MBW.
\item TM2 is subject to an electrostatic suspension along $\hat z$ below the MBW.
\item Rotation around $\hat x$, $\hat y$ and $\hat z$ below the MBW is controlled by star trackers.
\item Rotation around $\hat x$ within the MBW is controlled on $\theta$ rotation of TM1.
\item TM1 is subject to an electrostatic suspension for attitude control of $\hat \phi$  and $\hat \eta$.
\item TM1 is subject to an electrostatic suspension for attitude control of $\hat\theta$  below the MBW.
\item TM2 is subject to an electrostatic suspension for attitude control of $\hat\phi$,  $\hat\eta$ and $\hat\theta$.
\item The primary measurement goal is the PSD of the laser metrology output along $\hat x$, within the MBW.
\end{itemize}
The last statement is of course considered as the main mission goal. The aforementioned control-suspension strategy
can be put at work by means of the signals we already defined, so that in the MBW we get the scheme as in
table \ref{tab:M3mode}.

\begin{table}
\begin{center}
\begin{shadefundtheory}
\begin{tabular}{c|c|c|c|c}
& \multicolumn{2}{c}{$0\,\unit{mHz} \leq f \leq 0.5\,\unit{mHz}$} & \multicolumn{2}{|c}{MBW, $0.5\,\unit{mHz} \leq f \leq 1\,\unit{Hz}$} \\
\hline\rule{0pt}{0.4cm}\noindent
State variable & Control signal & Gain & Control signal & Gain\\
\hline\rule{0pt}{0.4cm}\noindent
$x_{1}$ & - & $0$ & - & $0$ \\
$y_{1}$ & GRS$(y_{1})$ & $\omega_{\text{df},y}^{2}$ & - & $0$ \\
$z_{1}$ & - & $0$ & - & $0$ \\
$\theta_{1}$ & GRS$(\theta_{1})$ & $\omega_{\text{df},\theta}^{2}$ & - & $0$ \\
$\eta_{1}$ & IFO$(\eta_{1})$ & $\omega_{\text{df},\eta}^{2}$ & IFO$(\eta_{1})$ & $\omega_{\text{lfs},\eta_{1}}^{2}$ \\
$\phi_{1}$ & IFO$(\phi_{1})$ & $\omega_{\text{df},\phi}^{2}$ & IFO$(\phi_{1})$ & $\omega_{\text{lfs},\phi_{1}}^{2}$ \\
\hline\rule{0pt}{0.4cm}\noindent
$x_{2}$ & IFO$(x_{2}-x_{1})$ & $\omega_{\text{df},x}^{2}$ & IFO$(x_{2}-x_{1})$ & $\omega_{\text{lfs},x}^{2}$ \\
$y_{2}$ & - & $0$ & - & $0$ \\
$z_{2}$ & GRS$(z_{2})$ & $\omega_{\text{df},z}^{2}$ & - & $0$ \\
$\theta_{2}$ & GRS$(\theta_{2})$ & $\omega_{\text{df},\theta}^{2}$ & GRS$(\theta_{2})$ & $\omega_{\text{lfs},\theta}^{2}$ \\
$\eta_{2}$ & IFO$(\eta_{2})$ & $\omega_{\text{df},\eta}^{2}$ & IFO$(\eta_{2})$ & $\omega_{\text{lfs},\eta_{2}}^{2}$ \\
$\phi_{2}$ & IFO$(\phi_{2})$ & $\omega_{\text{df},\phi}^{2}$ & IFO$(\phi_{2})$ & $\omega_{\text{lfs},\phi_{2}}^{2}$ \\
\hline\rule{0pt}{0.4cm}\noindent
$X$ & IFO$(x_{1})$ & $\omega_{\text{df},x}^{2}$ & IFO$(x_{1})$ & $\omega_{\text{df},x}^{2}$ \\
$Y$ & GRS$(y_{2})$ & $\omega_{\text{df},y}^{2}$ & GRS$(y_{2})$ & $\omega_{\text{df},y}^{2}$ \\
$Z$ & GRS$(z_{1})$ & $\omega_{\text{df},z}^{2}$ & GRS$(z_{1})$ & $\omega_{\text{df},z}^{2}$ \\
$\Theta$ & ST$(\Theta)$ & $\omega_{\text{df},\theta}^{2}$ & IFO$(\theta_{1})$  & $\omega_{\text{df},\theta}^{2}$ \\
$H$ & ST$(H)$ & $\omega_{\text{df},\eta}^{2}$ & GRS$(z_{2}-z_{1})$ & $\omega_{\text{df},\eta}^{2}$ \\
$\Phi$ & ST$(\Phi)$ & $\omega_{\text{df},\phi}^{2}$ & GRS$(y_{2}-y_{1})$& $\omega_{\text{df},\phi}^{2}$
\end{tabular}
\end{shadefundtheory}
\end{center}
\caption{Science (M3) mode: control logic and gain factors of suspensions and filters.}
\label{tab:M3mode}
\end{table}

By inspecting the table we can see there's two different regions in the frequency
spectrum to be handled, the TMs are left as free as possible in the MBW, while ancillary
DOF are ``suspended'' below it and overall the gain factors are tightened.

\begin{figure*}
\begin{center}
\begin{sideways}
\begin{minipage}{1.4\linewidth}
\begin{align}
\label{eq:ccone}
\Lambda_{1}=m&\left[
\begin{array}{cccccccccccc}
 0 & 0 & 0 & 0 & 0 & 0 & 0 & 0 & 0 & 0 & 0 & 0 \\
 0 & 0 & 0 & 0 & 0 & 0 & 0 & 0 & 0 & 0 & 0 & 0 \\
 0 & 0 & 0 & 0 & 0 & 0 & 0 & 0 & 0 & 0 & 0 & 0 \\
 0 & 0 & 0 & 0 & 0 & 0 & 0 & 0 & 0 & 0 & 0 & 0 \\
 0 & 0 & 0 & 0 & \frac{1}{6} L^2  \omega _{\text{lfs},\eta _1}^2 & 0 & 0 & 0 & 0 & 0 & 0 & 0 \\
 0 & 0 & 0 & 0 & 0 & \frac{1}{6} L^2  \omega _{\text{lfs},\phi _1}^2 & 0 & 0 & 0 & 0 & 0 & 0
\end{array}
\right]\,\\
\label{eq:cctwo}
\Lambda_{2}=m&\left[
\begin{array}{cccccccccccc}
 0 & 0 & 0 & 0 & 0 & 0 &  \omega _{\text{lfs},x}^2 & 0 & 0 & 0 & 0 & 0 \\
 0 & 0 & 0 & 0 & 0 & 0 & 0 & 0 & 0 & 0 & 0 & 0 \\
 0 & 0 & 0 & 0 & 0 & 0 & 0 & 0 & 0 & 0 & 0 & 0 \\
 0 & 0 & 0 & 0 & 0 & 0 & 0 & 0 & 0 & \frac{1}{6} L^2  \omega _{\text{lfs},\theta}^2 & 0 & 0 \\
 0 & 0 & 0 & 0 &0 & 0 & 0 & 0 & 0 & 0 & \frac{1}{6} L^2  \omega_{\text{lfs},\eta _2}^2 & 0 \\
 0 & 0 & 0 & 0 & 0 & 0 & 0 & 0 & 0 & 0 & 0 & \frac{1}{6} L^2  \omega_{\text{lfs},\phi _2}^2
\end{array}
\right]\,,\\
\label{eq:ccsc}
\Lambda_{\text{SC}}=m_{\text{SC}}&\left[
\begin{array}{cccccccccccc}
 - \omega _{\text{df},x}^2 & 0 & 0 & 0 & 0 & 0 & 0 & 0 & 0 & 0 & 0 & 0 \\
 0 & - \omega _{\text{df},y}^2 & 0 & 0 & 0 & 0 & 0 & 0 & 0 & 0 & 0 & 0 \\
 0 & 0 & - \omega _{\text{df},z}^2 & 0 & 0 & 0 & 0 & 0 & 0 & 0 & 0 & 0 \\
 0 & 0 & 0 & -\frac{1}{3}  R_{\text{SC}}^2 \omega _{\text{df},\theta}^2 & 0 & 0 & 0 & 0 & 0 & 0 & 0 & 0 \\
 0 & 0 & -\frac{ R_{\text{SC}}^2 \omega _{\text{df},\eta}^2}{3 r_o} & 0 & 0 & 0 & 0 & 0 & \frac{
   R_{\text{SC}}^2 \omega _{\text{df},\eta}^2}{3 r_o} & 0 & 0 & 0 \\
 0 & \frac{ R_{\text{SC}}^2 \omega _{\text{df},\phi}^2}{2 r_o} & 0 & 0 & 0 & 0 & 0 & -\frac{
   R_{\text{SC}}^2 \omega _{\text{df},\phi}^2}{2 r_o} & 0 & 0 & 0 & 0
\end{array}
\right]\,.
\end{align}
\end{minipage}
\end{sideways}
\end{center}
\end{figure*}

Due to linearity, the feed-back action is obtained as the application of the counteraction
matrices on the TMs coordinates $\vc x$, as follows:
\beq
\begin{split}
\left[\begin{array}{c}\vc F_{j}\\\vc\gamma_{j}\end{array}\right]
&=-\Lambda_{j}\cdot \vc x\,,\qquad j=1,2\,,\\
\left[\begin{array}{c}\vc F_{\text{SC}}\\\vc\gamma_{\text{SC}}\end{array}\right]
&=-\Lambda_{\text{SC}}\cdot \vc x\,,
\end{split}
\eeq
where the form of the matrices is given in equations \eqref{eq:ccone}, \eqref{eq:cctwo} and 
\eqref{eq:ccsc}. The resulting set of feed-back forces looks as follows, 
for TM1:
\begin{align}
F_{1,x} &= 0\,, \\
 F_{1,y} &= 0\,, \\
 F_{1,z} &= 0\,, \\
 \gamma _{1,\theta} &= 0 \\
 \gamma _{1,\eta} &= -\frac{1}{6} L^2 m \eta _1 \omega _{\text{lfs},\eta _1}^2\,, \\
 \gamma _{1,\phi} &= -\frac{1}{6} L^2 m \phi _1 \omega _{\text{lfs},\phi _1}^2\,,
\end{align}
for TM2:
\begin{align}
 F_{2,x} &= m \left(x_1-x_2\right) \omega _{\text{lfs},x}^2\,, \\
 F_{2,y} &= 0\,, \\
 F_{2,z} &= 0\,, \\
 \gamma _{2,\theta} &= -\frac{1}{6} L^2 m \theta _2 \omega _{\text{lfs},\theta}^2\,, \\
 \gamma _{2,\eta} &= -\frac{1}{6} L^2 m \eta _2 \omega _{\text{lfs},\eta _2}^2\,, \\
 \gamma _{2,\phi} &= -\frac{1}{6} L^2 m \phi _2 \omega _{\text{lfs},\phi _2}^2\,,
\end{align}
and finally for the SC:
\begin{align}
 F_{\text{SC},x} &= m_{\text{SC}} x_1 \omega _{\text{df},x}^2\,, \\
 F_{\text{SC},y} &= m_{\text{SC}} y_1 \omega _{\text{df},y}^2\,, \\
 F_{\text{SC},z} &= m_{\text{SC}} z_1 \omega _{\text{df},z}^2\,, \\
 \gamma _{\text{SC},\theta} &= \frac{1}{3} m_{\text{SC}} R_{\text{SC}}^2 \theta _1 \omega _{\text{df},\theta}^2\,, \\
 \gamma _{\text{SC},\eta} &= \frac{m_{\text{SC}} R_{\text{SC}}^2 \left(z_1-z_2\right) \omega _{\text{df},\eta}^2}{3 r_o}\,, \\
 \gamma _{\text{SC},\phi} &= \frac{m_{\text{SC}} R_{\text{SC}}^2 \left(y_2-y_1\right) \omega _{\text{df},\phi}^2}{2 r_o}\,.
\end{align}

A list of explanations is mandatory at this point:
\begin{enumerate}
\item the feed-back strategy is not unique. The main purpose of the set of counteractions at play in the $\Lambda_{j},\,j=1,2,\text{SC}$ is
controlling the SC motion servo-ed to the TMs motions, leaving TM1 motion free along the $\hat x$ direction and interfering the least the better
along the other directions.
\item The readout strategy is thought in this frame too: sensing along the $\hat x$ direction and anti-conjugated\footnote{The proper conjugated
angular axis of $\hat x$ would be $\theta$. By definition every rotation around $\hat x$ of unity vector $\hat \theta$ won't produce any
motion along $\hat x$, being thus the fixed axis of the transformation. Hence the name anti-conjugated to define rotations around $\hat y$ of $\hat z$
which would tilt each TM-$j$ along $\hat x$ and thus produce angular displacement in the $\hat x$ direction.}
 angular directions $\hat\eta,\,\hat\phi$ is interferometer-based and electrostatic, with obvious noise-reduction-driven preferential choice of the
 first. Differential position between the TMs, differential angular and absolute position displacement of TM1 are thus IFO signals and
 considered as main mission signals. No DC force shall be applied in the $\hat x,\,\hat\eta,\,\hat\phi$ direction for TM1, and what is
 applicable in those directions for TM2 is strictly targeted to binding the TM2 motion to the SC.
 \item Hidden in the symbols $\omega^{2}_{\text{lfs},i}$ and $\omega^{2}_{\text{df},i}$ we can retrieve then readout frequencies as
 well as control transfer functions, other than actuation filters. This strategy provides an enormous flexibility in assisting free-fall of TM1 by means
 of dragging the satellite apart, the reverse side of the medal is nevertheless twofold:
 \begin{description}
\item[noise] coming out of measurement and actuation devices mix, complicating our model,
\item[actuation] requires fine-tuning and orthogonalization to reduce cross-talk and optimisation of the effect: part of this process is
dynamical and cannot be simply designed on-ground. Stiffness measurement strategies must be accounted for.
\end{description}
\end{enumerate}

In view of a careful analysis of cross-talk and to group the results till now for the readers trained in control-theory, 
the overall dynamics can then be thought of as follows:
\begin{shadefundtheory}
\beq
\begin{split}
\left(M s^{2} - K\right)\cdot \vc x +s^{2} J \vc x_{\text{SC}}= -\Lambda\cdot \vc x\,,\\
M_{\text{SC}} s^{2} \vc x_{\text{SC}} = -\Lambda_{\text{SC}} \vc x\,,
\end{split}
\label{eq:fullltpdynmatrix}
\eeq
\end{shadefundtheory}
\noindent where we tensor-grouped the whole kinetic, stiffness, control and readout matrices for the TMs as:
\beq
\begin{split}
M&=\left[\begin{array}{cc}M_{1} & 0 \\
0 & M_{2}\end{array}\right]\,,\qquad 
K=\left[\begin{array}{cc}K_{1} & 0 \\
0 & K_{2}\end{array}\right]\,,\\
\Lambda&= \Lambda_{1}~\otimes~
\left[\begin{array}{c}1\\0\end{array}\right] +
\Lambda_{2}~\otimes~
\left[\begin{array}{c}0\\1\end{array}\right]\,.
\end{split}
\label{eq:tensorgroupmat}
\eeq
And $J$ is a matrix containing torsion arms coefficients to embody the change of reference in
expression \eqref{eq:changerefframes}.
The second of \eqref{eq:fullltpdynmatrix} can be solved in $\vc x_{\text{SC}}$ to give:
\beq
\vc x_{\text{SC}} = -\frac{1}{s^{2}} M_{\text{SC}}^{-1}\Lambda_{\text{SC}}\vc x
\eeq
and back-substituted into \eqref{eq:fullltpdynmatrix} to get rid of the SC variables as follows:
\beq
\left(M s^{2} - K\right)\vc x = \left(-\Lambda + J M_{\text{SC}}^{-1}\Lambda_{\text{SC}}\right)\vc x
\doteq \hat\Lambda \vc x\,.
\eeq

We now process all equations with the $\Omega(\cdot)$ operator, which has the effect to switch from
deterministic variables to signals, introducing the readout strategy and the proper noise coupled to
each detector.

At the end of the process, we are left with
the following set of coupled linear equations:

\begin{center}
\begin{sideways}
\begin{minipage}{\textheight}
\fontsize{10pt}{10pt}\selectfont 
\begin{align}
\frac{m \text{GRS}\left(z_1\right) z_0 \omega _{\text{df},\eta}^2}{r_o}-\frac{m \text{GRS}\left(z_2\right) z_0 \omega
   _{\text{df},\eta}^2}{r_o}+m \text{IFO}\left(x_1\right) \left(s^2+\omega _{\text{df},x}^2+\omega _{x_1,x_1}^2\right) =
   m\left(g_{1,x}-g_{\text{SC},x}-z_0 g_{\text{SC},\eta}\right)+m \left(s^2+\omega _{x_1,x_1}^2\right)
   \text{IFO}_n\left(x_1\right)&\,,  \\
 -m \text{GRS}\left(\theta _1\right) z_0 \omega _{\text{df},\theta}^2-\frac{1}{2} m \text{GRS}\left(y_2\right) \omega
   _{\text{df},\phi}^2+\frac{1}{2} m \text{GRS}\left(y_1\right) \left(2 \omega _{\text{df},y}^2+\omega _{\text{df},\phi}^2+2
   \left(s^2+\omega _{y_1,y_1}^2\right)\right)=\qquad&\nonumber\\
   \frac{1}{2} m \left(2 g_{1,y}-2 g_{\text{SC},y}+2 z_0 g_{\text{SC},\theta}+r_o g_{\text{SC},\phi}\right) +m \left(s^2+\omega _{y_1,y_1}^2\right)
   \text{GRS}_n\left(y_1\right)&\,, \\
 \frac{1}{2} m \text{GRS}\left(z_1\right) \left(2 \omega _{\text{df},z}^2+\omega _{\text{df},\eta}^2+2 \left(s^2+\omega
   _{z_1,z_1}^2\right)\right)-\frac{1}{2} m \text{GRS}\left(z_2\right) \omega _{\text{df},\eta}^2=m g_{1,z}-\frac{1}{2}
   m \left(2 g_{\text{SC},z}+r_o g_{\text{SC},\eta}\right)+m \left(s^2+\omega _{z_1,z_1}^2\right)
   \text{GRS}_n\left(z_1\right)&\,, \\
 \frac{1}{6} L^2 m \text{GRS}\left(\theta _1\right) \left(s^2+\omega _{\text{df},\theta}^2+\omega _{\theta _1,\theta
   _1}^2\right)=\frac{1}{6} m \left(g_{1,\theta}-g_{\text{SC},\theta}\right) L^2+\frac{1}{6} m
   \left(s^2+\omega _{\theta _1,\theta _1}^2\right) \text{GRS}_n\left(\theta _1\right) L^2&\,, \\
 \frac{m \text{GRS}\left(z_1\right) \omega _{\text{df},\eta}^2 L^2}{6 r_o}-\frac{m \text{GRS}\left(z_2\right) \omega
   _{\text{df},\eta}^2 L^2}{6 r_o}+\frac{1}{6} m \text{IFO}\left(\eta _1\right) \left(s^2+\omega _{\text{lfs},\eta _1}^2+\omega
   _{\eta _1,\eta _1}^2\right) L^2=\frac{1}{6} m \left(g_{1,\eta}-g_{\text{SC},\eta}\right)
   L^2+\frac{1}{6} m \left(s^2+\omega _{\eta _1,\eta _1}^2\right) \text{IFO}_n\left(\eta _1\right) L^2&\,, \\
 -\frac{m \text{GRS}\left(y_1\right) \omega _{\text{df},\phi}^2 L^2}{6 r_o}+\frac{m \text{GRS}\left(y_2\right) \omega
   _{\text{df},\phi}^2 L^2}{6 r_o}+\frac{1}{6} m \text{IFO}\left(\phi _1\right) \left(s^2+\omega _{\text{lfs},\phi _1}^2+\omega
   _{\phi _1,\phi _1}^2\right) L^2=\frac{1}{6} m \left(g_{1,\phi}-g_{\text{SC},\phi}\right)
   L^2+\frac{1}{6} m \left(s^2+\omega _{\phi _1,\phi _1}^2\right) \text{IFO}_n\left(\phi _1\right) L^2& \,,
\end{align}
\end{minipage}
\end{sideways}
\end{center}

\begin{center}
\begin{sideways}
\begin{minipage}{\textheight}
\fontsize{10pt}{10pt}\selectfont 
\begin{align}
 \frac{m \text{GRS}\left(z_1\right) z_0 \omega _{\text{df},\eta}^2}{r_o}-\frac{m \text{GRS}\left(z_2\right) z_0 \omega
   _{\text{df},\eta}^2}{r_o}+
   m \text{IFO}\left(x_1\right) \left(s^2+\omega _{\text{df},x}^2+\omega _{x_2,x_2}^2\right)+m
   \text{IFO}\left(x_2-x_1\right) \left(s^2+\omega _{\text{lfs},x}^2+\omega _{x_2,x_2}^2\right)=\qquad&\nonumber\\
   m\left(g_{2,x}-g_{\text{SC},x}-z_0 g_{\text{SC},\eta}\right)+m \left(s^2+\omega _{x_2,x_2}^2\right)
   \text{IFO}_n\left(x_1\right)+ m \left(s^2+\omega _{x_2,x_2}^2\right) \text{IFO}_n\left(x_2-x_1\right) &\,, \\
 -m \text{GRS}\left(\theta _1\right) z_0 \omega _{\text{df},\theta}^2+\text{GRS}\left(y_1\right) \left(m \omega
   _{\text{df},y}^2-\frac{1}{2} m \omega _{\text{df},\phi}^2\right)+\frac{1}{2} m \text{GRS}\left(y_2\right) \left(\omega
   _{\text{df},\phi}^2+2 \left(s^2+\omega _{y_2,y_2}^2\right)\right)=\qquad&\nonumber\\
   m g_{2,y}-m g_{\text{SC},y}+m z_0 g_{\text{SC},\theta}-\frac{1}{2} m r_o g_{\text{SC},\phi}+m \left(s^2+\omega _{y_2,y_2}^2\right)
   \text{GRS}_n\left(y_2\right) &\,, \\
 \text{GRS}\left(z_1\right) \left(m \omega _{\text{df},z}^2-\frac{1}{2} m \omega _{\text{df},\eta}^2\right)+\frac{1}{2} m
   \text{GRS}\left(z_2\right) \left(\omega _{\text{df},\eta}^2+2 \left(s^2+\omega _{z_2,z_2}^2\right)\right)=\frac{1}{2} m
   \left(2 g_{2,z}-2 g_{\text{SC},z}+r_o g_{\text{SC},\eta}\right)+m \left(s^2+\omega _{z_2,z_2}^2\right)
   \text{GRS}_n\left(z_2\right) &\,, \\
 \frac{1}{6} m \text{GRS}\left(\theta _1\right) \omega _{\text{df},\theta}^2 L^2+\frac{1}{6} m \text{GRS}\left(\theta _2\right)
   \left(s^2+\omega _{\text{lfs},\theta}^2+\omega _{\theta _2,\theta _2}^2\right) L^2=
   \frac{1}{6} m \left(g_{2,\theta}-g_{\text{SC},\theta}\right) L^2+\frac{1}{6} m \left(s^2+\omega _{\theta _2,\theta
   _2}^2\right) \text{GRS}_n\left(\theta _2\right) L^2 &\,, \\
 \frac{m \text{GRS}\left(z_1\right) \omega _{\text{df},\eta}^2 L^2}{6 r_o}-\frac{m \text{GRS}\left(z_2\right) \omega
   _{\text{df},\eta}^2 L^2}{6 r_o}+\frac{1}{6} m \text{IFO}\left(\eta _1\right) \left(s^2+\omega _{\text{lfs},\eta _2}^2+\omega
   _{\eta _2,\eta _2}^2\right) L^2+
   \frac{1}{6} m \text{IFO}\left(\eta _2-\eta _1\right) \left(s^2+\omega _{\text{lfs},\eta
   _2}^2+\omega _{\eta _2,\eta _2}^2\right) L^2=\qquad&\nonumber\\
   \frac{1}{6} m \left(g_{2,\eta}-g_{\text{SC},\eta}\right) L^2+\frac{1}{6} m \left(s^2+\omega _{\eta _2,\eta _2}^2\right) \text{IFO}_n\left(\eta _1\right)
   L^2
   \frac{1}{6} m \left(s^2+\omega _{\eta _2,\eta _2}^2\right) \text{IFO}_n\left(\eta _2-\eta _1\right) L^2 &\,, \\
 -\frac{m \text{GRS}\left(y_1\right) \omega _{\text{df},\phi}^2 L^2}{6 r_o}+\frac{m \text{GRS}\left(y_2\right) \omega
   _{\text{df},\phi}^2 L^2}{6 r_o}+\frac{1}{6} m \text{IFO}\left(\phi _1\right) \left(s^2+\omega _{\text{lfs},\phi _2}^2+\omega
   _{\phi _2,\phi _2}^2\right) L^2+
   \frac{1}{6} m \text{IFO}\left(\phi _2-\phi _1\right) \left(s^2+\omega _{\text{lfs},\phi
   _2}^2+\omega _{\phi _2,\phi _2}^2\right) L^2=\qquad&\nonumber\\
   \frac{1}{6} m \left(g_{2,\phi}-g_{\text{SC},\phi}\right) L^2+\frac{1}{6} m \left(s^2+\omega _{\phi _2,\phi _2}^2\right) \text{IFO}_n\left(\phi _1\right)
   L^2+
   \frac{1}{6} m \left(s^2+\omega _{\phi _2,\phi _2}^2\right) \text{IFO}_n\left(\phi _2-\phi _1\right) L^2&\,.
\end{align}
\end{minipage}
\end{sideways}
\end{center}

After diagonalising we can express the signals explicitly; not to harass on the reader we decided to write down only the IFO signals and
the GRS signals will be displayed at need only; here, IFO$(\Delta x)$:
\begin{center}
\begin{sideways}
\begin{minipage}{0.8\textheight}
\beq
\begin{split}
\text{IFO}&\left(x_2-x_1\right) = h_{x,x_2,x_2}^{\text{lfs}}(\omega )h_1(\omega ) h_{x,x_1,x_1}(\omega )\\
&\Bigl(-\frac{\omega _{\text{df},x}^2 \left(\omega _{x_1,x_1}^2-\omega _{x_2,x_2}^2\right)}{h_1(\omega )}    \text{IFO}_n\left(x_1\right)
-\frac{2 z_0 \omega _{\text{df},\eta}^2 \left(\omega _{x_1,x_1}^2-\omega _{x_2,x_2}^2\right)}{r_0 h_{0,z_1,z_1}(\omega ) h_{z,z_2,z_2}(\omega )}    \text{GRS}_n\left(z_1\right)\\
&+\frac{1}{h_1(\omega ) h_{0,x_2,x_2}(\omega ) h_{x,x_1,x_1}(\omega )}    \text{IFO}_n\left(x_2-x_1\right)
+\frac{2 z_0 \omega _{\text{df},\eta}^2 \left(\omega _{x_1,x_1}^2-\omega _{x_2,x_2}^2\right)}{r_0 h_{0,z_2,z_2}(\omega ) h_{z,z_1,z_1}(\omega )}    \text{GRS}_n\left(z_2\right)\\
&-\frac{1}{h_1(\omega ) h_{x,x_2,x_2}(\omega )}    g_{1,x}
-\frac{2 z_0 \omega _{\text{df},\eta}^2 \left(\omega _{x_1,x_1}^2-\omega _{x_2,x_2}^2\right)}{r_0 h_{z,z_2,z_2}(\omega )}    g_{1,z}
+\frac{1}{h_1(\omega ) h_{x,x_1,x_1}(\omega )}    g_{2,x}
+\frac{2 z_0 \omega _{\text{df},\eta}^2 \left(\omega _{x_1,x_1}^2-\omega _{x_2,x_2}^2\right)}{r_0 h_{z,z_1,z_1}(\omega )}    g_{2,z}\\
&-\frac{\omega _{x_1,x_1}^2-\omega _{x_2,x_2}^2}{h_1(\omega )}    g_{\text{SC},x}
-\frac{2 z_0 \omega _{\text{df},\eta}^2 \left(\omega _{x_1,x_1}^2-\omega _{x_2,x_2}^2\right) \left(\omega _{z_1,z_1}^2-\omega _{z_2,z_2}^2\right)}{r_0}    g_{\text{SC},z}
-\frac{2 z_0 \left(\omega _{x_1,x_1}^2-\omega _{x_2,x_2}^2\right)}{h_{0,z_2,z_2}(\omega ) h_{z,z_1,z_1}(\omega )}    g_{\text{SC},\eta}
\Bigr)\,,
\label{eq:deltaxifofull}
\end{split}
\eeq
where
\begin{align}
\frac{1}{h_{1}(\omega)}&=\frac{2 \omega _{\text{df},z}^2}{h_{\eta ,z_2,z_2}(\omega )}+\omega _{\text{df},\eta}^2
   \left(\frac{1}{h_{0,z_2,z_2}(\omega )}+\frac{1}{h_{0,z_1,z_1}(\omega
   )}\right)+\frac{2}{h_{0,z_1,z_1}(\omega ) h_{0,z_2,z_2}(\omega )}\,,\\
\frac{1}{h_{2}(\omega)}&=\frac{2 \omega
   _{\text{df},y}^2}{h_{\phi ,y_2,y_2}(\omega )}+\omega _{\text{df},\phi}^2
   \left(\frac{1}{h_{0,y_2,y_2}(\omega )}+\frac{1}{h_{0,y_1,y_1}(\omega
   )}\right)+\frac{2}{h_{0,y_1,y_1}(\omega ) h_{0,y_2,y_2}(\omega )}\,,
\end{align}
\end{minipage}
\end{sideways}
\end{center}
and the general form of the filter functions $h$ can be written down as:
\begin{shadefundtheory}
\beq
h_{x_{\hat i},x_{\hat j},x_{\hat k}}(\omega)=\frac{1}
{\omega_{\text{df},x_{\hat i}}^{2}+\omega^{2}_{x_{\hat j},x_{\hat k}}-\omega^{2}}\,,
\eeq
\end{shadefundtheory}
\noindent meaning the function is characterised by a drag-free high-gain transfer function, namely $\omega_{\text{df},x_{\hat i}}^{2}$, along
the $\hat i$-th DOF, but also coupled to other axes or angular variables by means of a correction
$\omega^{2}_{x_{\hat j},x_{\hat k}}$ which doesn't necessarily need to be small at this level. In this chapter we'll
consider only diagonal couplings, so that $\hat j \equiv \hat k$, but we leave the set of symbols open for a more
careful discussion about cross-talk, to be placed in the noise and disturbances chapter.

LFS functions will take the form:
\begin{shadefundtheory}
\beq
h^{\text{lfs}}_{x_{\hat i},x_{\hat j},x_{\hat k}}(\omega)=\frac{1}
{\omega_{\text{lfs},x_{\hat i}}^{2}+\omega^{2}_{x_{\hat j},x_{\hat k}}-\omega^{2}}\,,
\eeq
\end{shadefundtheory}
\noindent differing from the drag-free ones in the values of the LFS transfer filters.
We used a sloppier notation for the special cases when constants happen to be null, in absence of suspension or control, namely:
\beq
h_{x_{\hat i},x_{\hat j},x_{\hat k}}(\omega) \underset{\omega_{\text{df},x_{\hat i}}^{2}\to 0}{\to} h_{0,x_{\hat j},x_{\hat k}}(\omega)
\equiv  h^{\text{lfs}}_{0,x_{\hat j},x_{\hat k}}(\omega)\,,
\eeq
more limiting cases occur when the additional couplings can be neglected or considered small enough to be perturbations, so
that all filters collapse into drag-free parametrised families:
\beq
h_{x_{\hat i},x_{\hat j},x_{\hat k}}(\omega) \underset{\omega^{2}_{x_{\hat j},x_{\hat k}}\to 0}{\to} h_{x_{\hat i}}(\omega)
\doteq \frac{1}
{\omega_{\text{df},x_{\hat i}}^{2}-\omega^{2}}\,,
\eeq
and obviously perturbations can be analysed to first order at need around these former considered as ``zeroes'':
\beq
h_{x_{\hat i},x_{\hat j},x_{\hat k}}(\omega) \underset{\omega^{2}_{x_{\hat j},x_{\hat k}}\ll 1}{\simeq}
h_{x_{\hat i}}(\omega) \left( 1 - h_{x_{\hat i}}(\omega) \omega^{2}_{x_{\hat j},x_{\hat k}}\right)\,.
\eeq

These approximations can be put at play to inspect the signal formulae and cast a deeper glance to the
arguments leading to the choice of operation modes. To leading order, in absence of any noise and any additional
coupling apart from the LFS, we can write for the main signal as:
\begin{shadefundtheory}
\beq
\text{IFO}\left(x_{2}-x_{1}\right) \simeq
\frac{g_{2,x}-g_{1,x}}{\omega _{\text{lfs},x}^2-\omega ^2}\,,
\label{eq:mainsignalsupersimple}
\eeq
\end{shadefundtheory}
\noindent i.e. as aforementioned, the main mission interferometric signal represents a readout on the relative acceleration of
the TMs, properly scaled to the modulation functions cast by the control and laser devices. We can study the main
science signal more carefully, in the limit when the parasitic stiffness $\omega^{2}_{\text{p},1}\equiv \omega^{2}_{x_{1},x_{1}}$
and $\omega^{2}_{\text{p},2}\equiv \omega^{2}_{x_{2},x_{2}}$ are small compared to $\omega^{2}_{\text{df},x}$
and to leading order in $\nicefrac{\omega^{2}_{\text{df},x}}{\omega^{2}}$ within the MBW, to get:
\begin{shadefundtheory}
\beq
\begin{split}
\text{IFO}\left(x_{2}-x_{1}\right) \simeq
\frac{1}{\omega _{\text{lfs},x}^2-\omega ^2}
\Bigl( & g_{2,x}-g_{1,x}-\text{IFO}_{n}(x_{1})\omega^{2} +(\delta x_{2}-\delta x_{1})\omega_{\text{p},2}^{2}+\\
&+\left(\omega_{\text{p},2}^{2}-\omega_{\text{p},1}^{2}\right)
  \left( \text{IFO}_{n}(x_{1}) + \frac{g_{\text{SC},x}+z_0 g_{\text{SC},\eta}}{\omega^{2}_{\text{df},x}}\right)
\Bigr)\,.
\label{eq:mainsignalsimple}
\end{split}
\eeq
\end{shadefundtheory}
\noindent This fundamental formula has been recast in this shape for its paramount importance\footnote{Notice the interferometer
noise IFO$_{n}(x_{1})$ appears here twice due to the sensing role of the interferometer for the $\Delta x$ channel
and since we decided to use interferometer readout for the $x_{1}$ position too. More properly we shall name IFO$_{n}(x_{2}-x_{1})$
the noise multiplying $\omega^{2}$. In case GRS signal would be used for $x_{1}$ then we'd have IFO$_{n}(x_{1})\to$ GRS$_{n}(x_{1})$
in the second brackets.}. It will be employed widely on discussing
the noise contributions and apportioning and it's the most reliable laser phase estimator we have: in the spirit of chapter \ref{chap:refsys}
this signal describes our real ability to measure differential accelerations and thus to build a TT-gauged frame of reference.
The deformation contribution $(\delta x_{2}-\delta x_{1})\omega_{\text{p},2}^{2}$ has been placed in \eqref{eq:mainsignalsimple}
by hand, the presence of all deformations in the analytical deduction of the dynamical model would have engorged the
formulae lessening the already nightmarish legibility.

In absence of external forces ($g_{\text{SC},x}+z_{0}g_{\text{SC},\eta}=0$) the signal
is in fact representative of the acceleration noise provided the dynamical contribution of deformation is negligible
with respect to the readout noise:
\beq
\left|{\delta}x_{2}-{\delta}x_{1}\right| \ll \text{IFO}_{n}(x_{1})\,,
\eeq
besides, laser noise contribution to \eqref{eq:mainsignalsimple} is a very small fraction of the noise budget.

The interferometer
signal on the $x_{1}$ DOF looks like:
\begin{shadefundtheory}
\beq
\begin{split}
\text{IFO}\left(x_{1}\right) \simeq &
\frac{1}{\left(\omega ^2-\omega _{\text{df},x}^2\right) \left(\omega ^2-\omega _{\text{df},\eta}^2\right)}\\
&\Bigl(
-\frac{z_0 g_{1,z} \omega _{\text{df},\eta}^2}{r_0}
+\frac{z_0 g_{2,z} \omega _{\text{df},\eta}^2}{r_0}+g_{1,x} \left(\omega _{\text{df},\eta}^2-\omega ^2\right)+\\
   &+z_0 g_{\text{SC},\eta} \omega ^2+g_{\text{SC},x} \left(\omega ^2-\omega _{\text{df},\eta}^2\right)\\
   &+\omega ^2\left(\omega^{2}- \omega _{\text{df},\eta}^2\right) \text{IFO}_{n}\left(x_{1}\right)
   +\frac{\omega ^2 z_0 \omega _{\text{df},\eta}^2}{r_0}
     \left(\text{GRS}_{n}\left(z_{1}\right)-\text{GRS}_{n}\left(z_{2}\right)\right)
\Bigr) = \\
\simeq &\frac{1}{\omega^{2}_{\text{df},x}}\Bigl(
-\frac{z_0 g_{1,z}}{r_0} +\frac{z_0 g_{2,z}}{r_0}+g_{1,x}-g_{\text{SC},x}\\
&+\left(\omega_{\text{p},1}^{2}-\omega^{2}\right) \left(\text{IFO}_{n}(x_{1}) - \delta x_{1}\right)
\Bigr)=\\
\simeq & \frac{g_{\text{SC},x}}{\omega^{2}_{\text{df},x}}\,,
\label{eq:ifox1simple}
\end{split}
\eeq
\end{shadefundtheory}
\noindent where the already mentioned approximations have been used in cascade. The last passage emerges taking $\omega^{2}_{\text{df},x}$
to leading order and assuming the forces acting on the SC along $\hat x$ to be dominant. Analogously we get for
$z_{1}$:
\begin{shadefundtheory}
\beq
\begin{split}
\text{GRS}\left(z_{1}\right) \simeq&
\frac{1}{\omega _{\text{df},\eta}^2}\left(\frac{g_{1,z}}{2}-\frac{g_{2,z}}{2}
-\frac{r_0 g_{\text{SC},\eta}}{2}\right)
-\frac{1}{\omega^2}\left(\frac{g_{1,z}}{2}+\frac{g_{2,z}}{2}-g_{\text{SC},z}\right)\\
&+\frac{1}{2}\text{GRS}_n\left(z_1\right)+\frac{1}{2} \text{GRS}_n\left(z_2\right)\,,
\end{split}
\eeq
\end{shadefundtheory}
\noindent and for $\theta_{1}$:
\begin{shadefundtheory}
\beq
\text{GRS}\left(\theta_{1}\right) \simeq
\frac{1}{\omega _{\text{df$\theta $}}^2}
\left(g_{1,\theta}-g_{\text{SC},\theta}
  -\omega^{2}\text{GRS}_n\left(\theta _1\right)\right)\,,
\eeq
\end{shadefundtheory}
\noindent and finally for $\phi_{1}$:
\begin{shadefundtheory}
\beq
\begin{split}
\text{IFO}\left(\phi_{1}\right) \simeq&
  \frac{1}{r_0 \omega _{\text{df},\phi}^2}\left(g_{2,y}-g_{1,y}
  -r_{0}g_{\text{SC},\phi}\right)
  +  \frac{1}{r_0 \omega^2}\left(g_{2,y}-g_{1,y}
  -r_{0}g_{1,\phi}\right)\\
  &+\frac{1}{r_0}\left(\text{GRS}_n\left(y_1\right)-\text{GRS}_n\left(y_2\right)\right)
  +\text{IFO}_n\left(\phi_1\right)\,.
\end{split}
\eeq
\end{shadefundtheory}

These signals are good estimators of noise and main signals contributions and can be very powerful when used
together. For example \eqref{eq:ifox1simple} together with:
\begin{shadefundtheory}
\beq
\text{GRS}(x_{2})\simeq \text{GRS}_{n}(x_{2})-\text{GRS}_{n}(x_{1})
  -\frac{g_{\text{SC},x}}{\omega^{2}_{\text{df},x}}
  +\text{IFO}_{n}(x_{2}-x_{1}) \frac{\omega^{2}_{\text{lfs},x}}{\omega^{2}_{\text{lfs},x}-\omega^{2}}\,,
\label{eq:grsx2}
\eeq
\end{shadefundtheory}
\noindent and
\begin{shadefundtheory}
\beq
\text{GRS}(x_{1}) \simeq -\frac{g_{\text{SC},x}}{\omega^{2}_{\text{df},x}}-\text{GRS}_{n}(x_{1})\,,
\label{eq:grsx1}
\eeq
\end{shadefundtheory}
\noindent can be used to independently estimate $g_{\text{SC},x}$, $\text{GRS}_{n}(x_{1})$ and $\text{GRS}_{n}(x_{2})$.
Since we decided to illustrate the science mode version with maximised optical readout - i.e. whenever possible we switched
from signal acquisition via GRS to IFO signals - we didn't explicitly deduce formulae \eqref{eq:grsx2} and \eqref{eq:grsx1}
but assuming the capacitance electronics to be rigidly co-moving with the SC and the optical bench, they are
form invariant with regard to the redundant IFO ones, suffice it to switch from IFO to GRS when needed.

\begin{center}
\begin{sideways}
\begin{minipage}{\textheight}
\beq
\begin{split}
\text{IFO}\left(x_1\right) = &h_{x,x_1,x_1}(\omega )h_1(\omega )\\
&\Bigl(\frac{1}{h_1(\omega ) h_{0,x_1,x_1}(\omega )}    \text{IFO}_n\left(x_1\right)
-\frac{2 z_0 \omega _{\text{df},\eta}^2}{r_0 h_{0,z_1,z_1}(\omega ) h_{z,z_2,z_2}(\omega )}    \text{GRS}_n\left(z_1\right)
+\frac{2 z_0 \omega _{\text{df},\eta}^2}{r_0 h_{0,z_2,z_2}(\omega ) h_{z,z_1,z_1}(\omega )}    \text{GRS}_n\left(z_2\right)\\
&+\frac{1}{h_1(\omega )}    g_{1,x}
-\frac{2 z_0 \omega _{\text{df},\eta}^2}{r_0 h_{z,z_2,z_2}(\omega )}    g_{1,z}
+\frac{2 z_0 \omega _{\text{df},\eta}^2}{r_0 h_{z,z_1,z_1}(\omega )}    g_{2,z}
-\frac{1}{h_1(\omega )}    g_{\text{SC},x}
-\frac{2 z_0 \omega _{\text{df},\eta}^2 \left(\omega _{z_1,z_1}^2-\omega _{z_2,z_2}^2\right)}{r_0}    g_{\text{SC},z}\\
&-\frac{2 z_0}{h_{0,z_2,z_2}(\omega ) h_{z,z_1,z_1}(\omega )}    g_{\text{SC},\eta}
\Bigr)\,,
\end{split}
\label{eq:x1ifofull}
\eeq

\beq
\begin{split}
\text{IFO}&\left(\eta _2-\eta _1\right) = h_1(\omega ) h_{\eta _1,\eta _1,\eta _1}^{\text{lfs}}(\omega) h_{\eta _2,\eta _2,\eta _2}^{\text{lfs}}(\omega)\\
&\Bigl(-\frac{2 \omega _{\text{df},\eta}^2 \left(\omega _{\text{lfs},\eta _1}^2-\omega _{\text{lfs},\eta _2}^2+\omega _{\eta _1,\eta _1}^2-\omega _{\eta _2,\eta _2}^2\right)}{r_0 h_{0,z_1,z_1}(\omega ) h_{z,z_2,z_2}(\omega )}    \text{GRS}_n\left(z_1\right)
+\frac{\frac{\omega _{\text{lfs},\eta _1}^2}{h_{0,\eta _2,\eta _2}(\omega )}-\frac{\omega _{\text{lfs},\eta _2}^2}{h_{0,\eta _1,\eta _1}(\omega )}}{h_1(\omega )}    \text{IFO}_n\left(\eta _1\right)\\
&+\frac{2 \omega _{\text{df},\eta}^2 \left(\omega _{\text{lfs},\eta _1}^2-\omega _{\text{lfs},\eta _2}^2+\omega _{\eta _1,\eta _1}^2-\omega _{\eta _2,\eta _2}^2\right)}{r_0 h_{0,z_2,z_2}(\omega ) h_{z,z_1,z_1}(\omega )}    \text{GRS}_n\left(z_2\right)
+\frac{1}{h_1(\omega ) h_{0,\eta _2,\eta _2}(\omega ) h_{\eta _1,\eta _1,\eta _1}^{\text{lfs}}( \omega )}    \text{IFO}_n\left(\eta _2-\eta _1\right)\\
&+\frac{2 \omega _{\text{df},\eta}^2 \left(-\omega _{\text{lfs},\eta _1}^2+\omega _{\text{lfs},\eta _2}^2-\omega _{\eta _1,\eta _1}^2+\omega _{\eta _2,\eta _2}^2\right)}{r_0 h_{z,z_2,z_2}(\omega )}    g_{1,z}
+\frac{2 \omega _{\text{df},\eta}^2 \left(\omega _{\text{lfs},\eta _1}^2-\omega _{\text{lfs},\eta _2}^2+\omega _{\eta _1,\eta _1}^2-\omega _{\eta _2,\eta _2}^2\right)}{r_0 h_{z,z_1,z_1}(\omega )}    g_{2,z}\\
&-\frac{1}{h_1(\omega ) h_{\eta _2,\eta _2,\eta _2}^{\text{lfs}}( \omega )}    g_{1,\eta}
+\frac{1}{h_1(\omega ) h_{\eta _1,\eta _1,\eta _1}^{\text{lfs}}( \omega )}    g_{2,\eta}\\
&-\frac{2 \omega _{\text{df},\eta}^2 \left(\omega _{z_1,z_1}^2-\omega _{z_2,z_2}^2\right) \left(\omega _{\text{lfs},\eta _1}^2-\omega _{\text{lfs},\eta _2}^2+\omega _{\eta _1,\eta _1}^2-\omega _{\eta _2,\eta _2}^2\right)}{r_0}    g_{\text{SC},z}
-\frac{2 \left(\omega _{\text{lfs},\eta _1}^2-\omega _{\text{lfs},\eta _2}^2+\omega _{\eta _1,\eta _1}^2-\omega _{\eta _2,\eta _2}^2\right)}{h_{0,z_2,z_2}(\omega ) h_{z,z_1,z_1}(\omega )}    g_{\text{SC},\eta}
\Bigr)\,,
\end{split}
\eeq
\end{minipage}
\end{sideways}
\end{center}

\begin{center}
\begin{sideways}
\begin{minipage}{\textheight}
\beq
\begin{split}
\text{IFO}\left(\eta _1\right) = &h_{\eta _1,\eta _1,\eta _1}^{\text{lfs}}(\omega) h_1(\omega )\\
&\Bigl(-\frac{2 \omega _{\text{df},\eta}^2}{r_0 h_{0,z_1,z_1}(\omega ) h_{z,z_2,z_2}(\omega )}    \text{GRS}_n\left(z_1\right)
+\frac{1}{h_1(\omega ) h_{0,\eta _1,\eta _1}(\omega )}    \text{IFO}_n\left(\eta _1\right)
+\frac{2 \omega _{\text{df},\eta}^2}{r_0 h_{0,z_2,z_2}(\omega ) h_{z,z_1,z_1}(\omega )}    \text{GRS}_n\left(z_2\right)\\
&-\frac{2 \omega _{\text{df},\eta}^2}{r_0 h_{z,z_2,z_2}(\omega )}    g_{1,z}
+\frac{2 \omega _{\text{df},\eta}^2}{r_0 h_{z,z_1,z_1}(\omega )}    g_{2,z}\\
&+\frac{1}{h_1(\omega )}    g_{1,\eta}
-\frac{2 \omega _{\text{df},\eta}^2 \left(\omega _{z_1,z_1}^2-\omega _{z_2,z_2}^2\right)}{r_0}    g_{\text{SC},z}
-\frac{2}{h_{0,z_2,z_2}(\omega ) h_{z,z_1,z_1}(\omega )}    g_{\text{SC},\eta}
\Bigr)\,,
\end{split}
\eeq

\beq
\begin{split}
\text{IFO}\left(\phi _1\right) = &h_{\phi _1,\phi _1,\phi _1}^{\text{lfs}}(\omega )h_2(\omega )\\
&\Bigl(\frac{1}{h_2(\omega ) h_{0,\phi _1,\phi _1}(\omega )}    \text{IFO}_n\left(\phi _1\right)\\
&+\frac{2 \omega _{\text{df},\phi}^2}{r_0 h_{0,y_1,y_1}(\omega ) h_{y,y_2,y_2}(\omega )}    \text{GRS}_n\left(y_1\right)
-\frac{2 \omega _{\text{df},\phi}^2}{r_0 h_{0,y_2,y_2}(\omega ) h_{y,y_1,y_1}(\omega )}    \text{GRS}_n\left(y_2\right)
-\frac{2 z_0 \omega _{\text{df},\theta}^2 \omega _{\text{df},\phi}^2 \left(\omega _{y_1,y_1}^2-\omega _{y_2,y_2}^2\right) h_{\theta ,\theta _1,\theta _1}(\omega )}{r_0 h_{0,\theta _1,\theta _1}(\omega )}    \text{GRS}_n\left(\theta _1\right)\\
&+\frac{2 \omega _{\text{df},\phi}^2}{r_0 h_{y,y_2,y_2}(\omega )}    g_{1,y}
-\frac{2 \omega _{\text{df},\phi}^2}{r_0 h_{y,y_1,y_1}(\omega )}    g_{2,y}
+\frac{2 \omega _{\text{df},\phi}^2 \left(\omega _{y_1,y_1}^2-\omega _{y_2,y_2}^2\right)}{r_0}    g_{\text{SC},y}\\
&-\frac{2 z_0 \omega _{\text{df},\theta}^2 \omega _{\text{df},\phi}^2 \left(\omega _{y_1,y_1}^2-\omega _{y_2,y_2}^2\right) h_{\theta ,\theta _1,\theta _1}(\omega )}{r_0}    g_{1,\theta}
+\frac{1}{h_2(\omega )}    g_{1,\phi}\\
&-\frac{2 z_0 \omega _{\text{df},\phi}^2 \left(\omega _{y_1,y_1}^2-\omega _{y_2,y_2}^2\right) h_{\theta ,\theta _1,\theta _1}(\omega )}{r_0 h_{0,\theta _1,\theta _1}(\omega )}    g_{\text{SC},\theta}
-\frac{2}{h_{0,y_2,y_2}(\omega ) h_{y,y_1,y_1}(\omega )}    g_{\text{SC},\phi}
\Bigr)\,,
\end{split}
\eeq
\end{minipage}
\end{sideways}
\end{center}

\begin{center}
\begin{sideways}
\begin{minipage}{\textheight}
\beq
\begin{split}
\text{IFO}&\left(\phi _2-\phi _1\right) = h_2(\omega ) h_{\phi _1,\phi _1,\phi _1}^{\text{lfs}}( \omega ) h_{\phi _2,\phi _2,\phi _2}^{\text{lfs}}( \omega )\\
&\Bigl(\frac{\frac{\omega _{\text{lfs},\phi _1}^2}{h_{0,\phi _2,\phi _2}(\omega )}-\frac{\omega _{\text{lfs},\phi _2}^2}{h_{0,\phi _1,\phi _1}(\omega )}}{h_2(\omega )}    \text{IFO}_n\left(\phi _1\right)
+\frac{1}{h_2(\omega ) h_{0,\phi _2,\phi _2}(\omega ) h_{\phi _1,\phi _1,\phi _1}^{\text{lfs}}( \omega )}    \text{IFO}_n\left(\phi _2-\phi _1\right)\\
&+\frac{2 \omega _{\text{df},\phi}^2 \left(\omega _{\text{lfs},\phi _1}^2-\omega _{\text{lfs},\phi _2}^2+\omega _{\phi _1,\phi _1}^2-\omega _{\phi _2,\phi _2}^2\right)}{r_0 h_{0,y_1,y_1}(\omega ) h_{y,y_2,y_2}(\omega )}    \text{GRS}_n\left(y_1\right)
-\frac{2 \omega _{\text{df},\phi}^2 \left(\omega _{\text{lfs},\phi _1}^2-\omega _{\text{lfs},\phi _2}^2+\omega _{\phi _1,\phi _1}^2-\omega _{\phi _2,\phi _2}^2\right)}{r_0 h_{0,y_2,y_2}(\omega ) h_{y,y_1,y_1}(\omega )}    \text{GRS}_n\left(y_2\right)\\
&-\frac{2 z_0 \omega _{\text{df},\theta}^2 \omega _{\text{df},\phi}^2 \left(\omega _{y_1,y_1}^2-\omega _{y_2,y_2}^2\right) \left(\omega _{\text{lfs},\phi _1}^2-\omega _{\text{lfs},\phi _2}^2+\omega _{\phi _1,\phi _1}^2-\omega _{\phi _2,\phi _2}^2\right) h_{\theta ,\theta _1,\theta _1}(\omega )}{r_0 h_{0,\theta _1,\theta _1}(\omega )}    \text{GRS}_n\left(\theta _1\right)\\
&+\frac{2 \omega _{\text{df},\phi}^2 \left(\omega _{\text{lfs},\phi _1}^2-\omega _{\text{lfs},\phi _2}^2+\omega _{\phi _1,\phi _1}^2-\omega _{\phi _2,\phi _2}^2\right)}{r_0 h_{y,y_2,y_2}(\omega )}    g_{1,y}
+\frac{2 \omega _{\text{df},\phi}^2 \left(-\omega _{\text{lfs},\phi _1}^2+\omega _{\text{lfs},\phi _2}^2-\omega _{\phi _1,\phi _1}^2+\omega _{\phi _2,\phi _2}^2\right)}{r_0 h_{y,y_1,y_1}(\omega )}    g_{2,y}\\
&+\frac{2 \omega _{\text{df},\phi}^2 \left(\omega _{y_1,y_1}^2-\omega _{y_2,y_2}^2\right) \left(\omega _{\text{lfs},\phi _1}^2-\omega _{\text{lfs},\phi _2}^2+\omega _{\phi _1,\phi _1}^2-\omega _{\phi _2,\phi _2}^2\right)}{r_0}    g_{\text{SC},y}\\
&-\frac{2 z_0 \omega _{\text{df},\theta}^2 \omega _{\text{df},\phi}^2 \left(\omega _{y_1,y_1}^2-\omega _{y_2,y_2}^2\right) \left(\omega _{\text{lfs},\phi _1}^2-\omega _{\text{lfs},\phi _2}^2+\omega _{\phi _1,\phi _1}^2-\omega _{\phi _2,\phi _2}^2\right) h_{\theta ,\theta _1,\theta _1}(\omega )}{r_0}    g_{1,\theta}
-\frac{1}{h_2(\omega ) h_{\phi _2,\phi _2,\phi _2}^{\text{lfs}}( \omega )}    g_{1,\phi}
+\frac{1}{h_2(\omega ) h_{\phi _1,\phi _1,\phi _1}^{\text{lfs}}( \omega )}    g_{2,\phi}\\
&-\frac{2 z_0 \omega _{\text{df},\phi}^2 \left(\omega _{y_1,y_1}^2-\omega _{y_2,y_2}^2\right) \left(\omega _{\text{lfs},\phi _1}^2-\omega _{\text{lfs},\phi _2}^2+\omega _{\phi _1,\phi _1}^2-\omega _{\phi _2,\phi _2}^2\right) h_{\theta ,\theta _1,\theta _1}(\omega )}{r_0 h_{0,\theta _1,\theta _1}(\omega )}    g_{\text{SC},\theta}
+\frac{2 \left(-\omega _{\text{lfs},\phi _1}^2+\omega _{\text{lfs},\phi _2}^2-\omega _{\phi _1,\phi _1}^2+\omega _{\phi _2,\phi _2}^2\right)}{h_{0,y_2,y_2}(\omega ) h_{y,y_1,y_1}(\omega )}    g_{\text{SC},\phi}
\Bigr)\,,
\end{split}
\eeq
\end{minipage}
\end{sideways}
\end{center}


Science mode variants can contemplate SC to be controlled in translation along $\hat y$ on TM1, interchangeability of TM1 and TM2
and mixed use of any redundant metrology sensor.

\subsection{Nominal mode}
\label{sec:nommode}

The ``nominal'' (formerly M1) \cite{LTPdfacsfun,LTPdfacsgen,LTPdfacsM1}
mode was defined at the beginning of the study
and has been used as reference to define
the goals and requirements of LTP. The mode is very similar to the science mode, the only difference being that
in this mode the TM2 is controlled by using GRS$(x_{2})$.

Thus all the requirements valid for
science mode apply, with the exception of those concerning TM2 being locked on the laser signal. Those
are replaced by the following \cite{LTPscrd}:
\begin{enumerate}
\item TM2 is subject to a low frequency suspension loop along $\hat x$ in order to
compensate for DC forces and to stabilise its intrinsic negative stiffness. The force on
TM2 along $\hat x$ is nominally:
\begin{shadefundtheory}
\beq
F^{\text{act}}_{2,x} = -m \omega_{\text{lfs},x}^{2} x_{2}
\label{lfsm1}
\eeq
\end{shadefundtheory}
\item The absolute value of the open loop gain $\left|m \omega_{\text{lfs},x}^{2}\right|$
of the lfs in eq. \eqref{lfsm1} within the MBW must be the minimum possible value that
guarantees stable operation for maximum allowed perturbations during science measurements. Stability for
exceptional events is not foreseen.
\end{enumerate}

The control strategy in nominal mode can be read out of table \ref{tab:m1controls}.

Neglecting any cross-talk, the laser readout gives the following signal for LTP TMs $\Delta x$:
\begin{shadefundtheory}
\beq
\begin{split}
\text{IFO}\left(x_{2}-x_{1}\right) \simeq
\frac{1}{\omega ^2}
\Bigl( & \text{IFO}_{n}(x_{1})\omega^{2} + g_{2,x}-g_{1,x} +(\delta x_{2}-\delta x_{1})
  \left(\omega_{\text{lfs},x}^{2}+\omega_{\text{p},2}^{2}\right)+\\
&+\left(\omega_{\text{lfs},x}^{2}+\omega_{\text{p},2}^{2}-\omega_{\text{p},1}^{2}\right)
  \left( \text{GRS}_{n}(x_{1}) + \frac{g_{\text{SC},x}+z_0 g_{\text{SC},\eta}}{\omega^{2}_{\text{df},x}}\right)
\Bigr)\,,
\label{eq:mainsignalsimplem1}
\end{split}
\eeq
\end{shadefundtheory}
\noindent as before, this signal is a good estimator of residual acceleration in difference between the two TMs provided that
$\left|{\delta}x_{2}-{\delta}x_{1}\right|\ll \text{GRS}_{n}(x_{1})$ (this time we chose the capacitive readout)
and that the laser noise contributes only a small fraction of the noise budget. Moreover, need is to keep the term
$\left|\omega_{\text{lfs},x}^{2}+\omega_{\text{p},2}^{2}-\omega_{\text{p},1}^{2}\right|$ small enough
so that the last term of \eqref{eq:mainsignalsimplem1} wouldn't dominate on the spare ones.

Similarly as in the science mode scenario, more signals are available:
\begin{shadefundtheory}
\beq
\text{GRS}(x_{1}) \simeq \frac{1}{\omega_{\text{df},x}^{2}} \left(g_{x,1}-g_{\text{SC},x}
  -\omega^{2}\left(\text{GRS}_{n}(x_{1})-\delta x_{1}\right)\right)
\simeq -\frac{g_{\text{SC},x}}{\omega^{2}_{\text{df},x}}\,,
\label{eq:grsx1m1}
\eeq
\end{shadefundtheory}

\noindent and

\beq
\begin{split}
\text{GRS}(x_{2})\simeq & -\frac{1}{\omega^{2}}
\Bigl(g_{x,2}-g_{x,1}-\omega^{2}\Bigl(\text{GRS}_{n}(x_{2})-\text{GRS}_{n}(x_{1})\\
  &+\delta x_{2}-\delta x_{1}-\frac{g_{\text{SC},x}}{\omega_{\text{df},x}^{2}}\Bigr) \Bigr)\,,
\end{split}
\label{eq:grsx2m1}
\eeq

\noindent which gives:
\begin{shadefundtheory}
\beq
\text{GRS}(x_{2}) \simeq \text{GRS}_{n}(x_{2})-\text{GRS}_{n}(x_{1}) - \frac{g_{\text{SC},x}}{\omega_{\text{df},x}^{2}}\,,
\eeq
\end{shadefundtheory}
\noindent other than:

\begin{shadefundtheory}
\beq
\text{IFO}\left(x_{1}\right)
\simeq -\frac{g_{\text{SC},x}}{\omega^{2}_{\text{df},x}}+\text{IFO}_{n}(x_{1})-\text{GRS}_{n}(x_{1})\,,
\label{eq:ifox1simplem1}
\eeq
\end{shadefundtheory}

\noindent and these can be used as before to estimate $g_{\text{SC},x}$, $\text{GRS}_{n}(x_{1})$ and $\text{GRS}_{n}(x_{2})$.

By inspection of the main science signal in nominal mode \eqref{eq:mainsignalsimplem1}, we see that it's a worse estimator of
the noise sources in comparison to \eqref{eq:mainsignalsimple} for the science mode: the actuation control loop gain $\omega_{\text{lfs},x}^{2}$
affects the signal to noise ratio among the various terms and adds extra noise by directly coupling TM2 to the SC. This is why science mode has
been promoted to main operation mode.

However, the transfer function from force to displacement \eqref{eq:mainsignalsimplem1} shows multiplicative dependence only
on $\omega^{2}$, being thus self-calibrating in comparison to science mode \eqref{eq:mainsignalsimple} whose
pre-factor is $\nicefrac{1}{(\omega_{\text{lfs},x}^{2}-\omega^{2})}$.
Consequently, some experimental runs are performed in this mode for the sake of calibration of cross-check.

\begin{table}
\begin{center}
\begin{shadefundtheory}
\begin{tabular}{c|c|c|c|c}
& \multicolumn{2}{c}{$0\,\unit{mHz} \leq f \leq 0.5\,\unit{mHz}$} & \multicolumn{2}{|c}{MBW, $0.5\,\unit{mHz} \leq f \leq 1\,\unit{Hz}$} \\
\hline\rule{0pt}{0.4cm}\noindent
State variable & Control signal & Gain & Control signal & Gain\\
\hline\rule{0pt}{0.4cm}\noindent
$x_{1}$ & - & $0$ & - & $0$ \\
$y_{1}$ & GRS$(y_{1})$ & $\omega^{2}_{\text{df},y}$ & - & $0$ \\
$z_{1}$ & - & $0$ & - & $0$ \\
$\theta_{1}$ & GRS$(\theta_{1})$ & $\omega^{2}_{\text{df},\theta}$ & - & $0$ \\
$\eta_{1}$ & GRS$(\eta_{1})$ & $\omega^{2}_{\text{df},\eta}$ & GRS$(\eta_{1})$ & $\omega^{2}_{\text{lfs},\eta_{1}}$ \\
$\phi_{1}$ & GRS$(\phi_{1})$ & $\omega^{2}_{\text{df},\phi}$ & GRS$(\phi_{1})$ & $\omega^{2}_{\text{lfs},\phi_{1}}$ \\
\hline\rule{0pt}{0.4cm}\noindent
$x_{2}$ & GRS$(x_{2})$ & $\omega^{2}_{\text{df},x}$ & GRS$(x_{2})$ & $\omega^{2}_{\text{lfs},x}$ \\
$y_{2}$ & - & $0$ & - & $0$ \\
$z_{2}$ & GRS$(z_{2})$ & $\omega^{2}_{\text{df},z}$ & - & $0$ \\
$\theta_{2}$ & GRS$(\theta_{2})$ & $\omega^{2}_{\text{df},\theta}$ & GRS$(\theta_{2})$ & $\omega^{2}_{\text{lfs},\theta}$ \\
$\eta_{2}$ & GRS$(\eta_{2})$ & $\omega^{2}_{\text{df},\eta}$ & GRS$(\eta_{2})$ & $\omega^{2}_{\text{lfs},\eta_{2}}$ \\
$\phi_{2}$ & GRS$(\phi_{2})$ & $\omega^{2}_{\text{df},\phi}$ & GRS$(\phi_{2})$ & $\omega^{2}_{\text{lfs},\phi_{2}}$ \\
\hline\rule{0pt}{0.4cm}\noindent
$X$ & GRS$(x_{1})$ & $\omega^{2}_{\text{df},x}$ & GRS$(x_{1})$ & $\omega^{2}_{\text{df},x}$ \\
$Y$ & GRS$(y_{2})$ & $\omega^{2}_{\text{df},y}$ & GRS$(y_{2})$ & $\omega^{2}_{\text{df},y}$ \\
$Z$ & GRS$(z_{1})$ & $\omega^{2}_{\text{df},z}$ & GRS$(z_{1})$ & $\omega^{2}_{\text{df},z}$ \\
$\Theta$ & ST$(\Theta)$ & $\omega^{2}_{\text{df},\theta}$ & GRS$(\theta_{1})$ & $\omega^{2}_{\text{df},\theta}$ \\
$H$ & ST$(H)$ & $\omega^{2}_{\text{df},\eta}$ & GRS$(z_{2}-z_{1})$ & $\omega^{2}_{\text{df},\eta}$ \\
$\Phi$ & ST$(\Phi)$ & $\omega^{2}_{\text{df},\phi}$ & GRS$(y_{2}-y_{1})$ & $\omega^{2}_{\text{df},\phi}$
\end{tabular}
\end{shadefundtheory}
\end{center}
\caption{Nominal (M1) mode: control logic and gain factors of suspensions and filters.}
\label{tab:m1controls}
\end{table}

\section{Suspensions and feedback}

In spite of the difference in control of the $\hat x_{2}$ DOF, both the control schemes in nominal and science mode
largely share the same behaviour \cite{LTPdfacsgen}. At high and low frequency the SC linear motion $\hat X,\,\hat Y,\,\hat Z$,
is servo-ed to $\hat x_{1}$, $\hat y_{2}$ and $\hat z_{1}$, while the angular motion along
$\hat\Theta$, $\hat H$, $\hat\Phi$ is controlled at high frequency by star-trackers and by
$\Delta y$, $\Delta z$ signals at low frequency. In our simplified approach, the same transfer function
will be assumed for $\hat y_{1}$ and $\hat y_{2}$, controlling $\Phi$; the same for
 $\hat z_{1}$ and $\hat z_{2}$, controlling $H$.
 
Obviously, the $\hat x_{1}$ (or $\hat x_{2}$, in case of TM switching) DOF is affected only by readout AC
GRS voltages or IFO laser pressure, but never controlled or actuated.

Suspensions or low-gain filters may be applied as additional forces by means of the electrostatic
capacitance system with the purpose of compensating for negative stiffness and instabilities.
``Control'' in this case is a placeholder for a filter function whose functional shape is specified by the order
of the differential equation governing the dynamics, and whose dynamical stiffness is in turn a high-order
rational filter optimised on stability and response. To serve SC motion along, say, $\hat X$ to $\hat x_{1}$
can be achieved in two ways: by injection of the signal directly to the thrusters which will fire and move the SC,
or by adding the signal to the GRS readout and let the feedback loop transfer it to the SC motion. The result is practically
the same, but control cleanliness demands the second choice to be made: the filter will thus remain local and
its global effect propagated by other filters.

The actuation force is exerted linearly in the readout position-attitude vector $\vc x$ as follows:
\begin{shadefundtheory}
\beq
\vc f = -\left(I+{\delta}A\right)\cdot \hat\Lambda\cdot \vc x\,,
\eeq
\end{shadefundtheory}
\noindent where the matrix $A$ may contain actuation cross-talk and is supposed to be small. $I$ is the identity matrix
and $\hat\Lambda$ is the already built control matrix whose elements are transfer functions.
Obviously the form of $\hat\Lambda$ changes on the selected mode of operation. For the science and nominal mode
the form is the same, since the control strategy is one and the matrix $\hat\Lambda$ comes out in the following ``effective form'':

\beq
\hat\Lambda_{\text{eff}}=\left[
\begin{array}{cccccccccccc}
0 & 0 & & & & &\ldots & & & & & 0 \\
0 & G_{y_{1}}^{*} & & & & & & & & & &\\
 & & 0 & & & & & &  & & & \\
 & & & G_{\theta_{1}}^{*} & & & &  & & & &\vdots\\
 & & & & G_{\eta_{1}} & & & & &  & &\\
 & & & & & G_{\phi_{1}} & & & &  & &\\
 \vdots & & & & & & G_{x_{2}} & & & & &\\
 & & & & & & & 0 & & & & \\
 & & & & & & & & G_{z_{2}}^{*} & & & \\
 & & & & & & & & & G_{\theta_{2}} & & \\
 & & & & & & & & & & G_{\eta_{2}} & 0\\
 0 & & & & & \ldots & & & & & 0 & G_{\phi_{2}}
\end{array}
\right]\,,
\eeq
where the zero diagonal elements represent DOF for which no suspension is foreseen
and the elements marked with ``*'', i.e. for $y_{1}$, $\theta_{1}$, $z_{2}$ correspond to
those DOF needing suspension below the MBW.

The generic form of the low-frequency suspensions and drag-free transfer functions is a second-order
integration propagator \cite{Fichter:2018fk, LTPdfacsfun}: 
\beq
\frac{G_{\hat i}}{m} = c_{G,\hat i} \frac{1}{s^{2}-s_{0,\hat i}^{2}}\,,
\eeq
where $c_{G,\hat i}$ is called ``gain''\footnote{Though usually the whole transfer function may be referred to
as gain.} and after optimisation in Laplace space ($s$ is complex) the substitution $s\to \imath \omega$
is performed, together with $s_{0,\hat i}\to \imath \omega_{0,\hat i}$. For angular DOF $m$
must be changed into moment of inertia. $\omega_{0,i}^{2}$ is the filter function which gets optimised to achieve optimal control. The procedure is 
quite involved and it's highly demanding in terms of engineering skills and won't be discussed any further here.
It is nevertheless sensitive to assume a rational polynomial form for $\omega_{0,\hat i}$:
\begin{shadefundtheory}
\beq
\omega_{0,\hat i}^{2}=\frac{\displaystyle\sum_{j=0}^{n_{1}}a_{j}s^{j}}{\displaystyle\sum_{l=0}^{n_{2}}b_{l}s^{l}}\,,
\label{eq:suspfun}
\eeq
\end{shadefundtheory}
\noindent the reason being polynomial fractional functions are well behaved in Laplace space, their main
properties depending only on poles structure.
The order of the polynomials $n_{1}$ and $n_{2}$ depend on the optimisation of control, as well
as the precision of the coefficients $a_{j}$ and $b_{l}$. The coefficient of $b_{0}$ and $b_{n_{2}}$ are usually $0$
and $1$ respectively and the frequency dimensionality of the shape of the filter when $\Ree s \to \infty$
determines the filter degree\footnote{In field theory the same concept of power counting leads to
definition of anomalous dimension functions and renormalisation flows. Filters here are optimised and well
behaved at the frequency scale they are supposed to work.} and if $\omega_{0,\hat i}^{2} \sim \nicefrac{1}{s}$ it is
said to be an integrator, if $\omega_{0,\hat i}^{2} \sim s$ it's a differentiator, finally it's named after
``pure gain'' if $\omega_{0,\hat i}^{2} \sim 1$. Units of the coefficients are always so to
guarantee that $\omega_{0,\hat i}^{2}$ is in $\unit{Hz}^{2}$.

\begin{table}
\begin{center}
\begin{shadefundnumber}
$\begin{array}{r|l|l|l}
& \omega _{\text{lfs},\eta _1}^2 = \omega _{\text{lfs},\eta _2}^2 = \omega _{\text{lfs},\theta}^2
& \omega _{\text{lfs},\phi _1}^2 = \omega _{\text{lfs},\phi _2}^2 & \omega_{\text{lfs},x}^2\\
\hline\rule{0pt}{0.4cm}\noindent
 a_0 & 7.937\times 10^{\text{-9}} & 6.428\times 10^{\text{-11}} & 8.381\times 10^{\text{-10}} \\
 a_1 & 0.00001524 & 9.259\times 10^{\text{-8}} & 1.303\times 10^{\text{-7}} \\
 a_2 & 0.002255 & 0.00002521 & 0.00001665 \\
 a_3 & 0.004739 & 0.00003985 & -2.726\times 10^{\text{-7}} \\
 \hline\rule{0pt}{0.4cm}\noindent
 b_0 & 0 & 0 & 0 \\
 b_1 & 0.1575 & 0.004056 & 0.0007803 \\
 b_2 & 0.6979 & 0.283 & 0.01922 \\
 b_3 & 1.206 & 0.9639 & 0.2189 \\
 b_4 & 1 & 1 & 1
\end{array}$
\end{shadefundnumber}
\end{center}
\caption{Low frequency suspensions coefficients for various DOF. Functional form is
  \eqref{eq:suspfun}. Overall gain is set to $1$ and reabsorbed in the $a_{j}$ coefficients.}
  \label{tab:lfscoeff}
\end{table}

\begin{table}
\begin{center}
\begin{shadefundnumber}
$\begin{array}{r|l|l|l|l}
 & \omega _{\text{df},x}^2 & \omega _{\text{df},y}^2 = \omega _{\text{df},\phi}^2 
 & \omega _{\text{df},z}^2  = \omega _{\text{df},\eta}^2 &\omega _{\text{df},\theta}^2 \\
\hline\rule{0pt}{0.4cm}\noindent
a_0 & 0.00004403 & 0.00001689 & 0.00001837 & 1.612\times 10^{\text{-6}} \\
 a_1 & 0.002978 & 0.001349 & 0.001169 & 0.0001624 \\
 a_2 & 0.07449 & 0.04012 & 0.02967 & 0.006071 \\
 a_3 & 0.8304 & 0.5207 & 0.3791 & 0.1071 \\
 a_4 & 4.37 & 3.234 & 2.598 & 0.9253 \\
 a_5 & 0.1349 & 0.1809 & 0.06991 & 0.01732 \\
 a_6 & 0.0004659 & 0.0009752 & 0.0002122 & 0.0001224 \\
 \hline\rule{0pt}{0.4cm}\noindent
 b_0 & 0 & 0 & 0 & 0 \\
 b_1 & 3.401\times 10^{\text{-6}} & 0.0000308 & 0.0002115 & 1.363\times 10^{\text{-6}} \\
 b_2 & 0.01221 & 0.03377 & 0.08387 & 0.004636 \\
 b_3 & 11.05 & 9.304 & 8.428 & 3.963 \\
 b_4 & 9.609 & 10.34 & 9.685 & 6.775 \\
 b_5 & 5.046 & 5.405 & 5.333 & 4.616 \\
 b_6 & 1 & 1 & 1 & 1
\end{array}$
\end{shadefundnumber}
\end{center}
\caption{Drag-free transfer functions coefficients for various DOF. Functional form is
  \eqref{eq:suspfun}. Overall gain is set to $1$ and reabsorbed in the $a_{j}$ coefficients. Approach
  for the $\hat\phi$ and $\hat\eta$ controls has been simplified and the same control filter assumed
  for both TMs along $\hat y$ and $\hat z$.}
  \label{tab:dfcoeff}
\end{table}

\begin{table}
\begin{center}
\begin{shadefundnumber}
$\begin{array}{r|l|l|l}
 & \omega _{\Theta}^2 & \omega _H^2 & \omega _{\Phi}^2\\
\hline\rule{0pt}{0.4cm}\noindent
 a_0 & -5.169\times 10^{\text{-25}} & -4.192\times 10^{\text{-26}} & -2.133\times 10^{\text{-26}} \\
 a_1 & -7.6\times 10^{\text{-21}} & -8.304\times 10^{\text{-22}} & -5.509\times 10^{\text{-22}} \\
 a_2 & -4.223\times 10^{\text{-17}} & -6.043\times 10^{\text{-18}} & -5.121\times 10^{\text{-18}} \\
 a_3 & -1.97\times 10^{\text{-16}} & -2.22\times 10^{\text{-17}} & -1.999\times 10^{\text{-17}} \\
 a_4 & 2.092\times 10^{\text{-15}} & -1.528\times 10^{\text{-17}} & 1.486\times 10^{\text{-17}} \\
 a_5 & -9.807\times 10^{\text{-14}} & -4.714\times 10^{\text{-16}} & -2.185\times 10^{\text{-15}} \\
 \hline\rule{0pt}{0.4cm}\noindent
 b_0 & 0 & 0 & 0 \\
 b_1 & 1.024\times 10^{\text{-13}} & 1.962\times 10^{\text{-14}} & 2.004\times 10^{\text{-14}} \\
 b_2 & 1.114\times 10^{\text{-10}} & 2.935\times 10^{\text{-11}} & 3.102\times 10^{\text{-11}} \\
 b_3 & 7.209\times 10^{\text{-8}} & 2.645\times 10^{\text{-8}} & 2.794\times 10^{\text{-8}} \\
 b_4 & 0.00003021 & 0.00001545 & 0.00001611 \\
 b_5 & 0.007802 & 0.005582 & 0.005695 \\
 b_6 & 1 & 1 & 1
\end{array}$
\end{shadefundnumber}
\end{center}
\caption{Attitude control functions coefficients for angular DOF of the SC. Functional form is
  \eqref{eq:suspfun}. Overall gain is set to $1$ and reabsorbed in the $a_{j}$ coefficients.}
\end{table}

\begin{figure}
\begin{center}
\includegraphics[width=\textwidth]{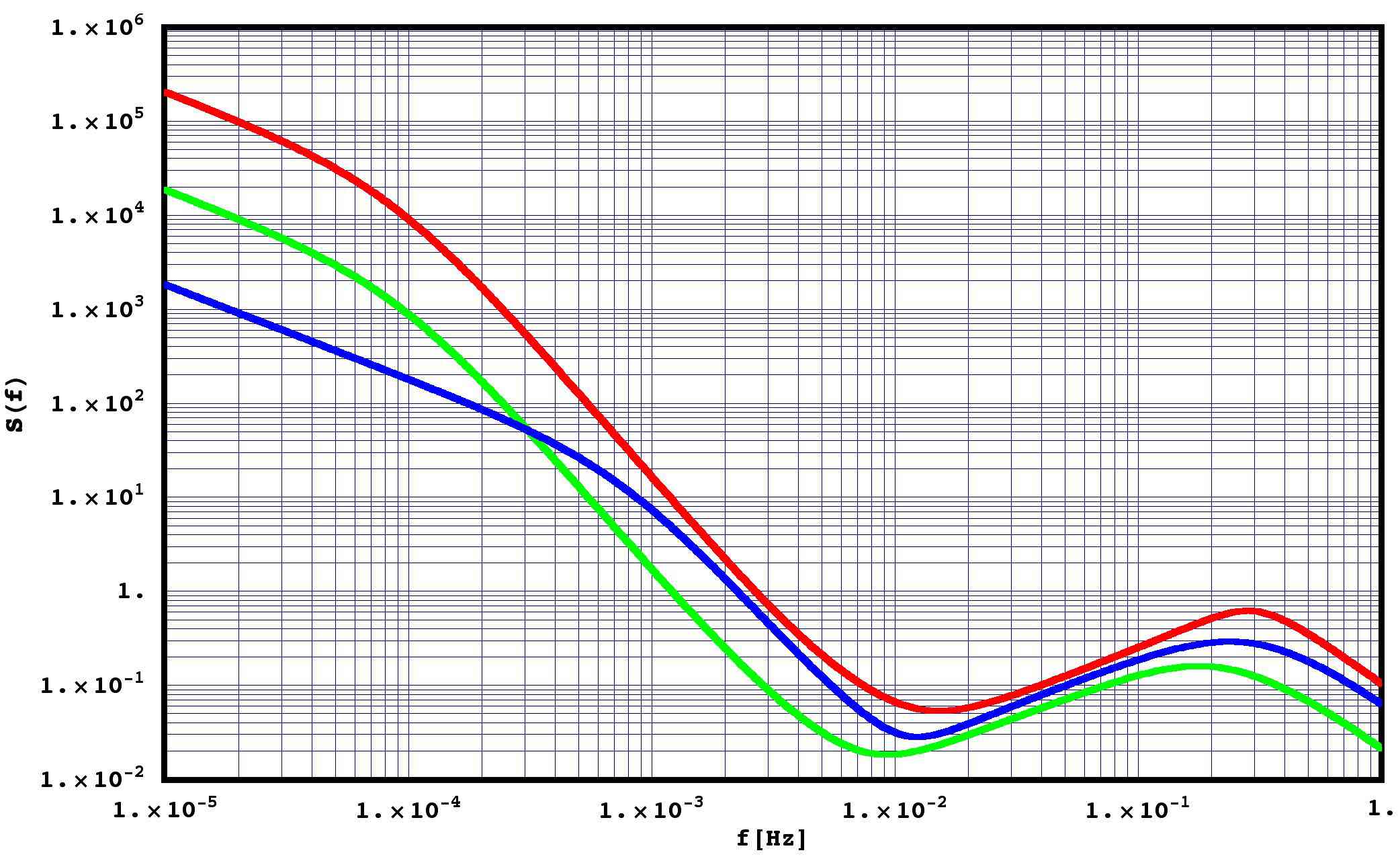}
\caption{$\left|\omega_{\text{df}}^{2}\right|$ as function of frequency $f=\nicefrac{\omega}{2\pi}$
for a frequency range spanning the entire MBW and beyond. Red: $\left|\omega_{\text{df},x_{1}}^{2}\right|$,
green: $\left|\omega_{\text{df},\theta_{1}}^{2}\right|$,
blue: $\left|\omega_{\text{df},z_{2}}^{2}\right|$}.
\end{center}
\end{figure}

\begin{figure}
\begin{center}
\includegraphics[width=\textwidth]{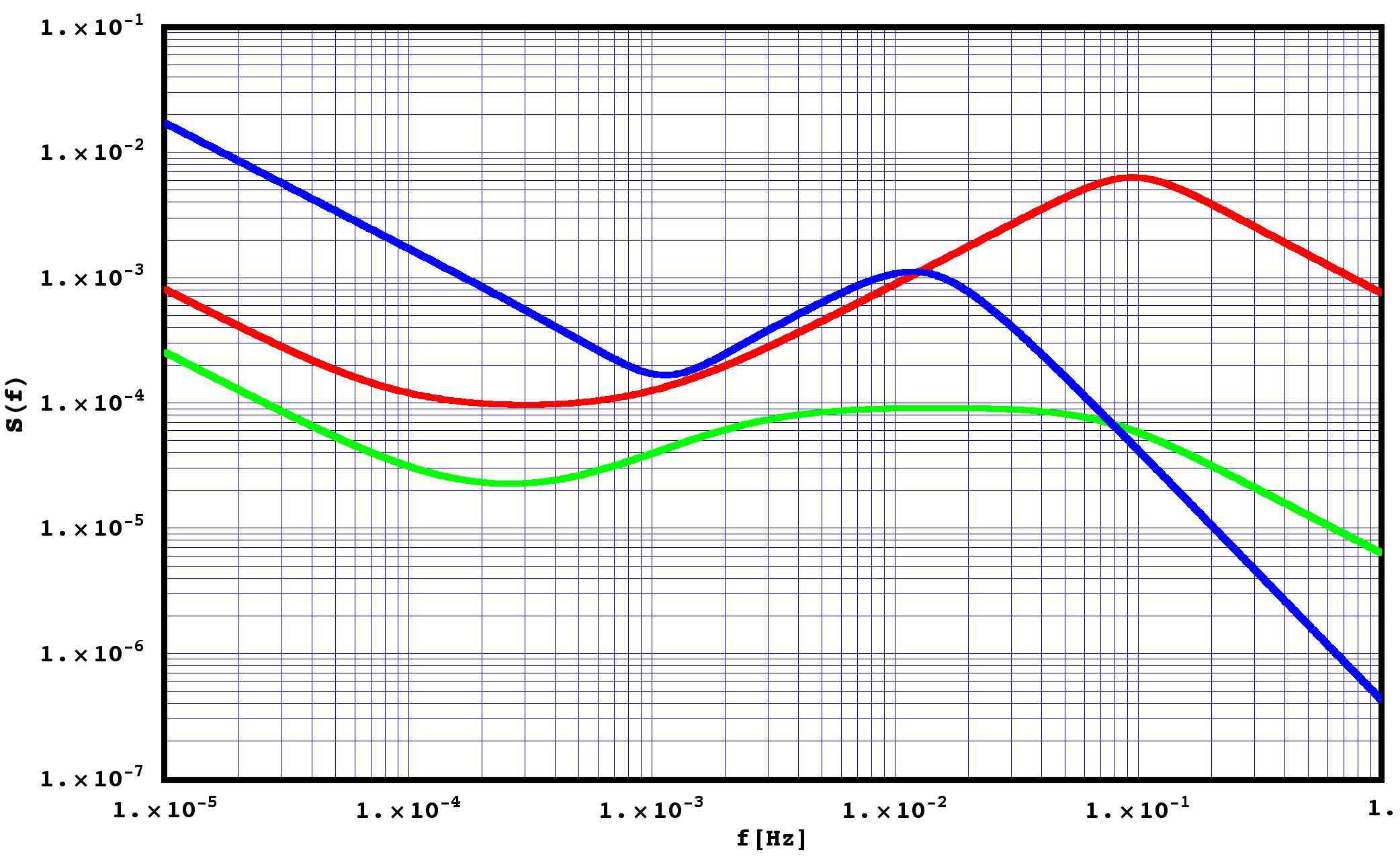}
\caption{$\left|\omega_{\text{lfs}}^{2}\right|$ as function of frequency $f=\nicefrac{\omega}{2\pi}$
for a frequency range spanning the entire MBW and beyond. Red: $\left|\omega_{\text{lfs},\eta_{1}}^{2}\right|$,
green: $\left|\omega_{\text{lfs},\phi_{1}}^{2}\right|$,
blue: $\left|\omega_{\text{lfs},x_{2}}^{2}\right|$}.
\end{center}
\end{figure}

\begin{figure}
\begin{center}
\includegraphics[width=\textwidth]{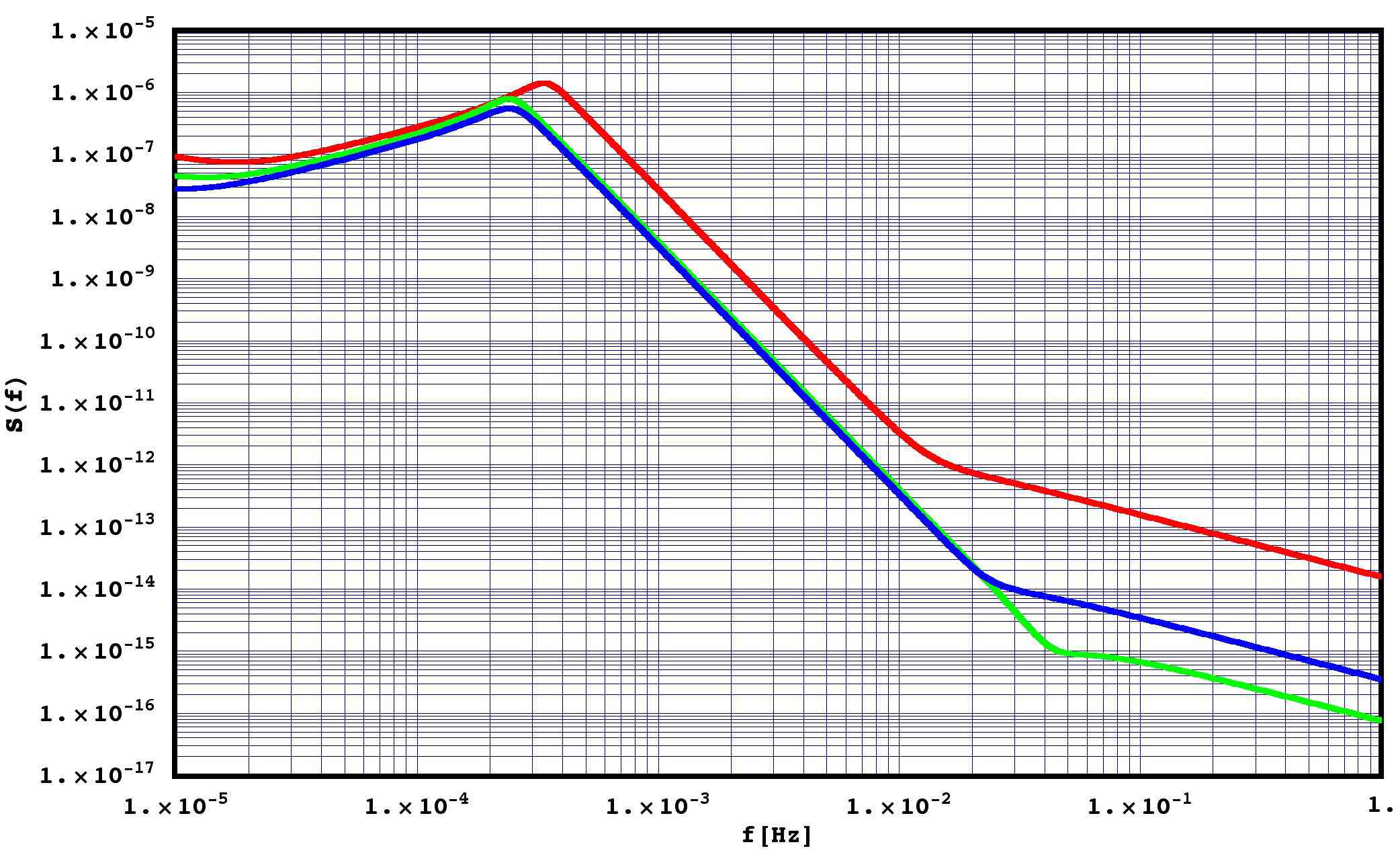}
\caption{Red: $\left|\omega_{\Theta}^{2}\right|$,
green: $\left|\omega_{H}^{2}\right|$,
blue: $\left|\omega_{\Phi}^{2}\right|$. All filters
drawn as function of frequency $f=\nicefrac{\omega}{2\pi}$
for a frequency range spanning the entire MBW and beyond.}
\end{center}
\end{figure}

%

It is moreover necessary to say that in this case the low frequency suspension does not cure the intrinsic
instability coming from the negative stiffness coupling, a task which is left to the drag-free. Besides, forces at play
are already extremely small at the frequencies where the instability needs to be compensated for; it is very unlikely
that the instability would get worse under such circumstances.

Out of the science mode LTP will be configured to run in ``accelerometer mode''. The mode is requested because
the low-frequency suspension can only apply a limited force to the TMs and, trying to minimise the coupling to
the SC, can only exert a very limited damping on TMs motion. Of course part of this process is designed on
purpose to decouple each TM from the SC itself and provide a more reliable freely-falling frame for GW detection,
nevertheless more tough suspensions are needed to face larger forces.

The force is limited by the relation between the parasitic stiffness $\omega^{2}_{\text{p}}$ resulting from the voltages
applied to the electrodes and the force itself: in order to jeep the parasitic stiffness constant for whatever value of the applied
force, this needs to be limited to
\beq
F_{\text{max}} \simeq \omega^{2}_{\text{p}} \times d\,,
\eeq
where $d$ is the effective gap between the TM and the electrodes.

A different actuation loop is needed in order to apply forces larger than the $10^{-9}\,\unit{N}$ foreseen
for the science phase.

Moreover, low frequency suspension is highly under-damped. The typical time-scale for transitory relaxation
spans from a few hours to one day. The accelerometer mode provide the requested fast damping to prepare the TM
for science mode within timescales compatible with the experiment running time.

All DOF can be controlled in accelerometer mode since no measurement is fore-casted. In each signal,
$\omega_{\text{p}}^{2}$ is rendered larger to match the exertion of a larger force. There's an obvious limitation to the amount of
gain displayed by the control loop, coming from the limited dynamical range for the linear behaviour of the device.


Transition amongst modes requires specific algorithms. For instance the transition between the accelerometer mode and
any of the science modes requires that both control laws have the same $0$-frequency transfer function in order not to trigger
very long-lived, high amplitude transients that would make the measurement time intolerably long.


%% file: chapters/noise.tex
\chapter{Noise}
\label{chap:noise}

\lettrine[lines=4]{N}{oise},
especially in such a complicated experiment as LTP, is a delicate subject and can easily slide out of hand
and become an unreadable list of contributions. We'd like to point out a set of common-sense rules and a global scenario so
that the reader won't eventually drown in the flood of formulae and micro-models which will follow.

We use a pedagogical approach whenever possible and interesting. Only a limited number of
contributions are really relevant in amplitude in a high-level noise budget analysis: therefore some of the
minor effects will be grouped in lists and no real derivation will be given.

For all the relevant formulae a complete or sketched deduction will be provided. Characteristic constants, critical
frequencies and rational motivation will be elucidated. The goal is to properly sum the noise PSDs from every source
and show that the achieved noise level is well within tolerances, to demonstrate feasibility of the LTP experiment set
in-folio, in-silico and - whenever possible to show - from ground testing inspired models and results.

\newpage

\section{Introduction, LTP ``master'' equation}

As stated many times, LISA aims at detecting gravitational wave strains in the relative modulation of distances
between freely-falling test masses. However, because of the $\nicefrac{1}{\omega^{2}}$ conversion from force
to displacement, detection of displacement caused by GWs from a background of acceleration noise becomes increasingly
difficult at lower frequencies. LISA's target sensitivity at low frequencies, stretching down to $0.1\times 10^{-3}\,\unit{Hz}$, requires
the test mass acceleration noise be less than $3\times 10^{-15}\,\accPSDunit$ (we may address this result as ``LISA's drag-free goal''
in the following).

%

Though the LTP test will be considered successful if it demonstrates acceleration noise
$10$ times above the LISA goal, to give a truly representative test the IS is designed to satisfy the LISA drag-free goal
down to $0.1\,\unit{mHz}$.
Of course there's a price to pay: the one-axis configuration of LTP requires actuation forces to be applied on TM2
to compensate for differential accelerations ${\Delta}a_{x}\simeq g_{2,x}-g_{1,x}$. This will introduce a parasitic stiffness\index{Parasitic stiffness}
due to actuation $\omega_{\text{p,\,act}}^{2}$ immaterial in LISA. The parasitic stiffness requirement
aboard LISA without actuation amounts to\footnote{From dynamical arguments, given the residual jitter of
$2.5\,\unitfrac{nm}{\sqrt{Hz}}$ and the acceleration goal of $10^{-15}\,\unitfrac{m}{s^{2}\,\sqrt{Hz}}$,
the ratio gives a spring constant value like \eqref{eq:omeganonactfigure}}:
\begin{shadefundnumber}
\beq
\left|\omega_{\text{\text{p}}}^{2}\right| \simeq 4\times 10^{-7}\,\secminsqunit\,.
\label{eq:omeganonactfigure}
\eeq
\end{shadefundnumber}
\noindent together with actuation stiffness it will rise up to $6.5\times 10^{-7}\,\unitfrac{1}{s^{2}}$. Aboard LTP
the $\omega_{\text{\text{p}}}^{2}$ (no actuation) value has been relaxed to $2\times 10^{-6}\,\unitfrac{1}{s^{2}}$

In the chapter devoted to kinematics and analytical description of LTP, we derived the expression
of the signal IFO$(x_{2}-x_{1})$: this displacement differential signal needs to be analysed carefully
in terms of residual
acceleration of the TM in the case of a  single-axis control loop \cite{2003SPIE.4856...31W,Weber:2002dp}.
Without loss of generality equation \eqref{eq:mainsignalsimple} can be restated in a form suggestive for our discussion:
\begin{shadefundtheory}
\beq
\begin{split}
\text{IFO}(x_{2}-x_{1}) =& x_{2}-x_{1}+\text{IFO}_{n}({\Delta}x) = \\
  =& h(\omega) \Bigl(g_{2,x}-g_{1,x}+\left(\omega^{2}_{\text{p},2}-\omega^{2}_{\text{p},1}\right) 
  \left(\text{GRS}_{n}(x_{1})+\frac{g_{\text{SC,x}}}{\omega_{\text{df},x}^{2}}\right)+\\
  &+\left({\delta}x_{1}-{\delta}x_{2}\right) \left(\omega_{\text{p},2}^{2} + \omega_{\text{lfs},x}^{2}\right)
 +\left(\omega_{\text{p},2}^{2}-\omega^{2}\right) \text{IFO}_{n}({\Delta}x)
  \Bigr)
\end{split}
\label{eq:mainifosignalfornoise}
\eeq
\end{shadefundtheory}
\noindent with $g_{i,x}$ being the residual acceleration of the TMs and switching from the notation of chapter \ref{chap:ltp}
we made the associations
$\omega_{x_{1},x_{1}}^{2}\to \omega_{\text{p},1}^{2}$ and $\omega_{x_{2},x_{2}}^{2}\to \omega_{\text{p},2}^{2}$.
We also neglected SC jitters on the $\hat\eta$ direction by placing $\dot\Omega_{\text{SC},\eta}=0$.
The transfer function $h(\omega)$ gets the form:
\beq
h(\omega)=\frac{1}{\omega^{2}-\left(\omega^{2}_{\text{p},2}+\omega_{\text{lfs},x}^{2}\right)}
\eeq
The residual coupling of the TM to the SC is summarised by the $\omega_{\text{p}}^{2}$ factors, regarded as the
natural frequencies of oscillation of the TMs relative to the SC. The gain and transfer functions
of the control loop per unit is represented in $\omega_{\text{lfs},x}^{2}$.

Disturbing forces have been split into three contributions:
\begin{enumerate}
\item those applied to the SC along the $\hat x$ direction, $g_{\text{SC},x}$, which also include the thrusters' noise, and the
difference in gravitational acceleration between the TM and the SC centre of mass,
\item the forces coupling the TM and the SC, of thermal, pressure, DC electric origin which are going to
contribute to the $\omega^{2}_{\text{p},i}$ parasitic coupling factors and their difference;
\item the contribution coming from the sensing, grouped into the GRS$_{n}$ and IFO$_{n}$ terms.
Notice in \eqref{eq:mainifosignalfornoise} the first of these, $\text{GRS}_{n}(x_{1})$:, is going to
get amplified by the factor $\left(\omega^{2}_{\text{p},2}-\omega^{2}_{\text{p},1}\right)$ and
filtered through $h(\omega)$ to become an acceleration noise, but it is a pure displacement noise to begin with,
mostly determined by circuitry readout characteristics. Conversely $\text{IFO}_{n}({\Delta}x)$
comes from interferometer readout and gets filtered through $h(\omega)\left(\omega_{\text{p},2}^{2}-\omega^{2}\right)$,
but this coupling to the SC is peculiar of LTP, therefore the parasitic term in this noise contribution won't be a feature of LISA,
which will retain only the $\sim \omega^{2}$-dependent factor.
\end{enumerate}

By means of the identification ${\Delta}a_{x}=-\omega^{2}{\Delta}x$, any argument or discussion
concerning $a_{x}$ can be further transferred to residual displacement. To demonstrate
the proposed noise goal at $1\,\unit{mHz}$, the total measured displacement noise ${\Delta}x$
must not exceed $1\,\unitfrac{nm}{\sqrt{Hz}}$. Additive force noise associated with
the optical readout, baseline distortions of the optical bench ${\delta}x$, can be
held below $0.1\,\unitfrac{nm}{\sqrt{Hz}}$ level. The measurement noise and stray force for TM2
are additive sources of noise as well but will be indistinguishable from $x_{1, n}$ and $g_{1,x}$
for TM1.

Notice that while coherent noise in $g_{1,x}$ and $g_{2,x}$ can cancel without changing the measured optical
path length, the most important un-modelled electrostatic and Brownian noise sources are likely to be uncorrelated
between IS1 and IS2 and thus produce a measurable and meaningful noise in ${\Delta}x$.

\section{Sources of noise}

In the following the main - and known - sources of noise will be described and analysed \cite{LTPnoisebud}.
We intend to deduce every formula from first principles. The purpose is to clearly identify
the sources and their characteristics and then group them according to the physical process they
are originated by. This approach is helpful on many sides:
\begin{enumerate}
\item it provides a list of formulae and numerical estimates based on well-known physical
processes. Error checking and extension of the list shall be more straightforward.
\item It enables a more effective location of the source on-board and in which
physical phenomenon does it originate (for instance the magnetic field or the temperature
fluctuation or eddy currents flowing on some conductor surface). 
\item Correlated or uncorrelated combination of the various effect comes naturally at the end
of this list. Grouping by phenomenon is the only way to consistently understand which
combination is more meaningful.
\end{enumerate}
In the impossibility of knowing the energy-momentum tensor point-by-point in the neighbourhood
of the fiduciary points we'll use as detectors, power spectral densities (PSD) can be deduced via
fields correlators. In the spirit of qualifying the measurement device upon certain requirements
the clear effort will be to evaluate relevant noise components in the MBW, their subtractibility and time dependence
to give modelling and final confidence bounds to them.
This last procedure leads to apportioning of the different
contributions and noise reduction.

We identify the following noise sources:
\begin{enumerate}
\item inertial sensor readout displacement noise: due to transformer, amplifier, actuation circuitry and
  force noise acting on the SC: this is converted into a force noise via the drag-free control 
loop gain and the difference of parasitic stiffness coupling both TMs to the S/C, with
reference to \eqref{eq:mainifosignalfornoise}:
\beq
S_{a,\text{dragfree}}^{\nicefrac{1}{2}} = h(\omega)\left(\omega^{2}_{\text{p},2}-\omega^{2}_{\text{p},1}\right) 
  \left(\text{GRS}_{n}(x_{1})+\frac{g_{\text{SC,x}}}{\omega_{\text{df},x}^{2}}\right)\,.
\eeq
 In science mode (M3)
only the noise from the readout of TM1 matters, while in
nominal mode (M1) the sensor noise of IS2 is converted into a force noise via the low frequency suspension gain. Notice here that the
displacement noise induced by forces on the SC converts to a force on the TM via the DF loop, as the open 
loop gain is not high enough to suppress them entirely. What's left results in a residual 
displacement between the TM and the SC.

\item Readout back-action. This is split into a part correlated to the former IS displacement noise (same sources) and an
uncorrelated one, i.e. back-action forces from readout that have no displacement noise counterpart.
The correlated  part is the back action force of readout due to source 
of disturbance within the readout that contribute both to displacement noise and a direct force 
disturbance onto the TM. For each source, the product of displacement noise and stiffness and the 
direct back-action force must be added coherently before estimating the spectral density. In science mode these sources are 
only relevant for IS1. In nominal mode also those for IS2 must be added coherently to the contribution of 
the sensor noise coupled through the low frequency suspension. The argument can be made clear with
an example: one of such sources would produce a displacement noise
$x_{n,\text{corr}}$ and an acceleration noise $g_{n,\text{corr}} \doteq \omega^{2}_{\text{corr}}x_{n,\text{corr}}$.
By inspecting \eqref{eq:mainifosignalfornoise} we see that in science mode the contributions will add like:
\beq
g_{n,\text{tot}} = g_{n,\text{corr}}
  + \left(\omega^{2}_{\text{p},2}-\omega^{2}_{\text{p},1}\right) x_{n,\text{corr}}
= \left(\omega^{2}_{\text{corr}}+\omega^{2}_{\text{p},2}-\omega^{2}_{\text{p},1}\right)x_{n,\text{corr}}\,.
\eeq
Thus giving a squared PSD:
\beq
\begin{split}
S_{g,n,\text{tot}} 
&= \left(\left|\omega^{2}_{\text{corr}}\right|^{2}
+\left|\omega^{2}_{\text{p},2}-\omega^{2}_{\text{p},1}\right|^{2}
+ 2 \left|\omega^{2}_{\text{corr}}\right| \left|\omega^{2}_{\text{p},2}-\omega^{2}_{\text{p},1}\right|\cos\phi
\right)S_{x,n,\text{corr}} = \\
&= S_{g,n,\text{corr}} \left(
1 + 2 \frac{ \left|\omega^{2}_{\text{p},2}-\omega^{2}_{\text{p},1}\right|}{\left|\omega^{2}_{\text{corr}}\right|}\cos\phi
\right)
+\left|\omega^{2}_{\text{p},2}-\omega^{2}_{\text{p},1}\right|^{2} S_{x,n,\text{corr}}\,,
\end{split}
\eeq
where $\phi$ is the difference of phase angle between $\omega^{2}_{\text{p},2}-\omega^{2}_{\text{p},1}$
and $\omega^{2}_{\text{corr}}$. In practise usually the effects are tiny, and the correlation term gets neglected to give:
\begin{shadefundtheory}
\beq
S_{g,n,\text{tot}} \simeq S_{g,n,\text{corr}} + \left|\omega^{2}_{\text{p},2}-\omega^{2}_{\text{p},1}\right|^{2} S_{x,n,\text{corr}}\,.
\eeq
\end{shadefundtheory}

\item Thermal effects: forces due to various effects related to temperature and temperature gradients 
fluctuations within the IS, adding coherently. They include: radiometer effect, thermal distortion of housing and optical bench,
fluctuation of thermal radiation pressure difference across the TM,
thermal fluctuation of out-gassing flow difference across the TM,
and the gravitational force induced by thermal distortion of IS.

\item Brownian noise: thermal noise due to several mechanisms: 
dissipation due to dielectric losses in sensing capacitors,
dissipation due to interaction of eddy currents within the test-mass and the  magnetic 
field gradient, dissipation due to magnetic losses within magnetic impurities in the TM.

\item Cross-talk: forces nominally applied to other DOF may leak into the sensitive axis. 
We may list: 
cross-talk of actuation force/torque along other DOF into a force along $\hat x$. The 
sources are the geometric imperfection and the imperfections in balancing actuation 
voltages to the requested electrode pairs;
cross-talk of displacement/rotation of other DOF into the $\hat x$-channel capacitive 
sensor that is used for drag-free and/or electrostatic suspension; 
non diagonal terms of parasitic stiffness matrix;
rotation of DC-forces with the TM applied along $\hat y$ and $\hat z$. This couples the angular jitter of 
TM into force noise along $\hat x$. 

\item Magnetic disturbances  within SC: due to magnetic field and magnetic field 
gradients due to sources within S/C. Field and gradients form the same source are assumed to 
be totally correlated. These forces are due to:
interaction of magnetic field gradient fluctuations within MBW with permanent and 
induced DC magnetisation,
interaction of fluctuating part of induced magnetisation, within MBW,  with DC value of gradient,
magnetic field  fluctuation at frequencies above the MBW, with low frequency 
amplitude modulation and non-zero gradient.

\item Magnetic disturbances due to interplanetary field fluctuation, assumed to be of negligible 
gradient. Susceptibility and moment leftovers of TM1 and TM2 may be different enough to 
prevent cancellation within  the difference of force. They act by: 
inducing a fluctuating moment within the TM that interacts with DC-field gradient 
or by the fluctuating electric field due to Lorentz transformation of magnetic field 
values to the SC reference frame.

\item Random charging: shot noise due to cosmic rays charge interacting with stray DC-voltage 
on electrodes.

\item Fluctuation of stray voltages: due to charged patches.

\item Various: like fluctuation of local gravitational field due to distortion of the system components 
and laser pressure variation.

\item Additional sources of noise - which are not relevant to LISA - that enter into the total noise budget due
to the profound design differences between the apparatus's (see \ref{sec:lisaltpdiff}). In this
case the scene will be dominated by electrostatic actuation noise\footnote{
Any voltage fluctuation is bound to produce undesired stiffness, since:
\beq
{\delta}F  = \frac{\partial F}{\partial V} {\delta}V = \frac{C_{0}}{d} V{\delta}V\,,
\eeq
so that in terms of relative variations:
\begin{shadefundtheory}
\beq
\frac{{\delta}F}{F} = 2 \frac{{\delta}V}{V}\,,
\label{eq:deltaelecforceoverforce}
\eeq
\end{shadefundtheory}
\noindent hence:
\beq
S_{F}^{\nicefrac{1}{2}} = 2 F S_{\nicefrac{{\Delta}V}{V}}^{\nicefrac{1}{2}}\,,
\eeq
and \eqref{eq:dcaccpsdvsvolt} follows.}:
\beq
\omega^{2}_{\text{p,act}} = 2 S_{\nicefrac{{\Delta}V}{V}}^{\nicefrac{1}{2}} a_{\text{DC}}\,.
\label{eq:dcaccpsdvsvolt}
\eeq

\end{enumerate}

Many arguments will be deduced directly in terms of acceleration PSD, nevertheless many
will be derived in terms of voltage, current or field PSD fluctuations. The linking relation between
the expressions is of derivative nature; for example, we can express
a voltage squared PSD in terms of a displacement squared PSD like:
\begin{shadefundtheory}
\beq
S_{V}=\left|\frac{\partial V}{\partial x}\right|^{2} S_{x}
  = \left|\frac{\partial V}{\partial C}\right|^{2}\left|\frac{\partial C}{\partial x}\right|^{2} S_{x}\,.
  \label{eq:voltdisppsd}
\eeq
\end{shadefundtheory}

\section{Electrostatics, magnetics and stiffness}

\subsection{Electrostatics in general}

\begin{table}
\begin{center}
\begin{tabular}{r|l|l|l}
Description & Name & Value & Dimensions\\
\hline\rule{0pt}{0.4cm}\noindent
TM mass & $m$ & $1.96$ & $\unit{kg}$\\ 
TM edge & $L$ & $4.6\times 10^{-2}$ & $\metresunit$\\ 
TM face area & $A$ & $0.046^{2}$ & $\metresunit^2$\\ 
Electrical conductance & $\sigma_0$ & $3.33\times 10^{6}$ & $\unitfrac{N}{s\, V^2}$
\end{tabular}
\end{center}
\caption{Test Masses characteristics}
\label{tab:tmspecs}
\end{table}

Whenever considering electrostatic sources of noise, the following guidelines and basic formulae must be kept in mind :
\begin{enumerate}
\item all formulae follow from a certain number of given constants which may be easily retrieved in tables and will
be pointed out as needed. Conversely, derived constants will be introduced and discussed in order of appearance. Suffice
it to say that all the basic instantaneous electrostatic equations follow from the expression of the potential and force.
The electrostatic energy\index{Electrostatic energy} of the system is given by:
\beq
W = \frac{1}{2}\sum_{j}C_{j}\left(V_{j}-V_{\text{TM}}\right)^{2}\,
\eeq
hence the force along the $\hat x$ direction:
\begin{shadefundtheory}
\beq
F_{x} = -\frac{\partial W}{\partial x} =
   \frac{1}{2}\sum_{j}\frac{\partial C_{j}}{\partial x}
   \left(V_{j}-V_{\text{TM}}\right)^{2}\,.
\label{eq:electroforce}
\eeq
\end{shadefundtheory}
\noindent In reality, the battery restores potential in the circuitry by its energy $W_{\text{batt}}$ and changes the sign of the potential
energy.
For each TM the
following electrostatic potential balance equation holds:
\begin{shadefundtheory}
\beq
C_{\text{tot}} V_{\text{TM}} = \sum_{i} C_{i}V_{i} + Q_{\text{TM}}\,,
\label{eq:electrocapacbalance}
\eeq
\end{shadefundtheory}
\noindent with $C_{\text{tot}} \doteq \sum_{i} C_{i}$, $Q_{\text{TM}}$ is the TM charge
and the index $i$ ranges over all conductors around the TM with non zero capacitance $C_{i} = C_{i}(\vc x)$
or potential $V_{i}$, assumed slowly varying in position. In practise only the GRS electrodes will count in this game.
\item A simple infinite parallel-plate model is used for capacitors, with an infinite wedge model for the angular derivatives. As
such, each electrode capacitance as a function of the displacement $x$ is:
\beq
\label{eq:capx}
C\simeq \epsilon_{0} \frac{A}{d \pm x}=C_{0}\frac{1}{1 \pm \frac{x}{d}}\,,
\eeq
with $C_{0}=\nicefrac{\epsilon_{0}A}{d}$, to give:
\begin{shademinornumber}
\beq
\frac{\partial C}{\partial x} = - C_{0} \frac{1}{\left(1 \pm \frac{x}{d}\right)^{2}}
  \left(\pm \frac{1}{d}\right) \underset{x\to 0}{\to} \mp\frac{C_{0}}{d}\,,
  \label{eq:czero}
\eeq
\end{shademinornumber}
similarly, for the second derivative:
\begin{shademinornumber}
\beq
\frac{\partial^{2} C}{\partial x^{2}} = - C_{0} \frac{2}{\left(1 \pm \frac{x}{d}\right)^{3}}
  \left(\pm \frac{1}{d}\right)^{2} \underset{x\to 0}{\to} \frac{2 C_{0}}{d^{2}}\,,
  \label{eq:d2czero}
\eeq
\end{shademinornumber}

\item In presence of stray voltages on the $j$-th surface and non-zero TM charge $Q_{\text{TM}}$, we
can always write \eqref{eq:electroforce} as:
\beq
F_{x} = 
   \frac{1}{2}\sum_{j}\frac{\partial C_{j}}{\partial x}
   \left(V_{\text{stray},j}-\frac{Q_{\text{TM}}}{C_{\text{tot}}}-V_{\text{TM},0}\right)^{2}\,,
\label{eq:fxstray}
\eeq
where $V_{\text{TM},0}$ represents a $0$-point reference potential and can always be put to $0$.
Conductive surfaces must be seen at this level like complicated patchworks of domains whose conductive
Fermi levels are not necessarily equal: the effect may be regarded as a network of short-circuits
and stray potentials which lead to the creation of stray voltages on electrode surfaces.

Expanding \eqref{eq:fxstray} we get:
\beq
F_{x} = 
   \frac{1}{2}\sum_{j}\frac{\partial C_{j}}{\partial x} V_{\text{stray},j}^{2}
   +\frac{1}{2}\frac{Q^{2}_{\text{TM}}}{C^{2}_{\text{tot}}}\sum_{j}\frac{\partial C_{j}}{\partial x}
   -\frac{Q_{\text{TM}}}{C_{\text{tot}}}\sum_{j}\frac{\partial C_{j}}{\partial x} V_{\text{stray},j}\,.
\eeq
Notice the first term in the latter is very small, since we assume each $V_{\text{stray},j}$ to be so. Anyway the
term may be neglected since it shall roughly cancel out if the potential fluctuations can be considered isotropic on the
conductors. Moreover, assuming the same capacity for each electrode, the second term is null by symmetry,
due to the electrodes configuration and \eqref{eq:czero}. Therefore we are left with:
\beq
F_{x} = - \frac{Q_{\text{TM}}}{C_{\text{tot}}}\sum_{j}\frac{\partial C_{j}}{\partial x} V_{\text{stray},j}\,,
\label{eq:vstraydcbias}
\eeq
the factor $\sum_{j}\frac{\partial C_{j}}{\partial x} V_{\text{stray},j}$ is called ``DC-bias'', a name which is
transferred to the effect as a whole.

To derive an expression for the parasitic stiffness we just need to differentiate \eqref{eq:fxstray} with respect to $x$
and expand:
\beq
\begin{split}
\omega_{\text{p,act},x}^2 = -\frac{1}{m}\frac{\partial F_{x}}{\partial x} =&
   \frac{1}{2 m}\sum_{j}\frac{\partial^{2} C_{j}}{\partial x^{2}} V_{\text{stray},j}^{2}
   +\frac{1}{2 m}\frac{Q^{2}_{\text{TM}}}{C^{2}_{\text{tot}}}\sum_{j}\frac{\partial^{2} C_{j}}{\partial x^{2}}+\\
   &-\frac{Q_{\text{TM}}}{m C_{\text{tot}}}\sum_{j}\frac{\partial^{2} C_{j}}{\partial x^{2}} V_{\text{stray},j}\,.
\end{split}
\eeq
The first term is small in $V_{\text{stray},j}^{2}$, but the value of $\nicefrac{\partial^{2} C_{j}}{\partial x^{2}}$
is always positive, and the contribution can be relevant; the second term provides a net, DC contribution
depending only on the electrode geometrical configuration while the third may be neglected, since upon collection of the
capacitance derivative we may think the mean value of $\sum_{j} V_{\text{stray},j}$ to be $\simeq 0$. By substituting
the values of the capacitance derivatives \eqref{eq:czero} and \eqref{eq:d2czero}
we can write\footnote{In practise, the formula we used
to compute values is:
\beq
\omega^2_{\text{p,DC}}=\frac{1}{m d_{x}}
\left(\frac{1}{3\epsilon_{0} A} q_{e}^{2} q_{0}^{2}
  + \frac{13}{9} \frac{\epsilon_{0} A}{d_{x}^{2}} V_{\text{stray}}^{2}\right)\,,
\eeq
which takes into account geometrical corrections and the real sizes and number of the electrode plates.
}:
\begin{shademinornumber}
\beq
\omega_{\text{p,act},x}^2 \simeq
   6 \frac{\epsilon_{0} A}{m d^{3}} V_{\text{stray}}^{2} + \frac{1}{3}\frac{Q^{2}}{m \epsilon_{0} A d}\,,
\eeq
\end{shademinornumber}
\noindent where we considered $6$ electrodes with average stray voltage $V_{\text{stray}}$ and equal capacitance.

\item Modulation of the sensing bridge introduces stiffness due to the injected voltage $V_{\text{inj}}$
at $f_{\text{inj}}=100\,\unit{kHz}$. In absence of other voltages at $f_{\text{inj}}$, at zero charge and zero reference potential
we get from \eqref{eq:electrocapacbalance} that $V_{\text{TM}} = V_{\text{inj}}$, so that, from
\eqref{eq:electroforce}:
\begin{shademinornumber}
\beq
\omega^2_{\text{p,sens}}=\frac{1}{m}
  \frac{1}{2}\frac{\partial^{2} C_{\text{inj}}}{\partial x^{2}} V_{\text{inj}}^{2}
=\frac{1}{m} \frac{C_{0}}{d^{2}}V_{\text{inj}}^{2}
=\frac{1}{m} \frac{\epsilon_{0} A}{d^{3}}V_{\text{inj}}^{2}\,.
  \label{eq:deduceksens}
\eeq
\end{shademinornumber}

\item If the voltage is exerted to control the TM in DC actuation, given the force as $m a_{x}$,
then the last-but-one passage in \eqref{eq:deduceksens}
together with \eqref{eq:czero} gives the DC stiffness:
\begin{shademinornumber}
\beq
\omega^2_{\text{p,DC}} = \frac{2 a_{x}}{d}\,.
\eeq
\end{shademinornumber}

Angular sensing stiffness can be calculated in analogous way, suffice it to assume:
\beq
{\delta}x \simeq \frac{L}{2} {\delta}\phi\,,
\eeq
i.e. any angular displacement can be thought as a linear one with effective torque arm of $\nicefrac{1}{2}$ the
TM side. hence for any rotational DOF:
\beq
\omega^2_{\text{p,sens}} \to \frac{L}{2} \omega^2_{\text{p,sens}}\,.
\eeq
Rotational stiffness actuation to compensate a residual angular acceleration, say $g_{\text{DC},\phi}$
would get the following expression:
\begin{shademinornumber}
\beq
\omega^2_{\text{p,rot},\phi}= \frac{g_{\text{DC},\phi} L}{d_{c(\phi)}}\,,
\eeq
\end{shademinornumber}
where we designated as $c(\phi)$ the linear direction of the $\hat\phi$-rotation controlling
electrodes, one of the two anti-conjugated DOF in a $3$-dimensional space: the
electrodes can be inspected in figure \ref{fig:ltptmeh} and a DOF conjugation scheme is in table \ref{tab:conjang}.


\end{enumerate}

\subsection{Actuation at constant stiffness}

The constant stiffness actuation model permits to apply AC voltages in order to move or bias
the TM by keeping electrostatic induced extra spring coupling constant. 

\begin{figure}
\begin{center}
\includegraphics[width=0.5\textwidth]{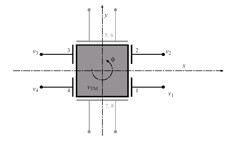}
\caption{$\hat x-\hat y$ electrodes configuration around each TM.}
\label{fig:xyTMelec}
\end{center}
\end{figure}

If we assume the simple configuration in figure \ref{fig:xyTMelec} with equal capacitors ($C_{i}=C,\,\forall i$)
and to apply voltages $\pm V_{x,1}$ to the electrodes on the right
and $\pm V_{x,2}$ to the left, provided we null the charge $Q_{\text{TM}}$ in advance, we get
from \eqref{eq:electrocapacbalance}:
\beq
\left(\sum_{j}C_{j}\right)V_{\text{TM}}= C\sum_{i}V_{i}=0\,,
\label{eq:sumpotentialstozero}
\eeq
hence the change in $V_{\text{TM}}$ caused by actuation voltages is $0$. Notice both the electrodes configuration, left and right, exert a
pulling force on the TM by virtue of electrostatic induction creating odd-signed charges on the
surface facing the electrodes; moreover, if we'd take $V_{x,1}=V_{x,2}$ no motion would be
induced on the TM, but as soon as they are different, the TM moves along $\hat x$ and the
capacitance varies increasing on the ``winning'' side and decreasing on the other by the same amount
according to \eqref{eq:czero}.

\begin{figure}
\begin{center}
\includegraphics[width=0.8\textwidth]{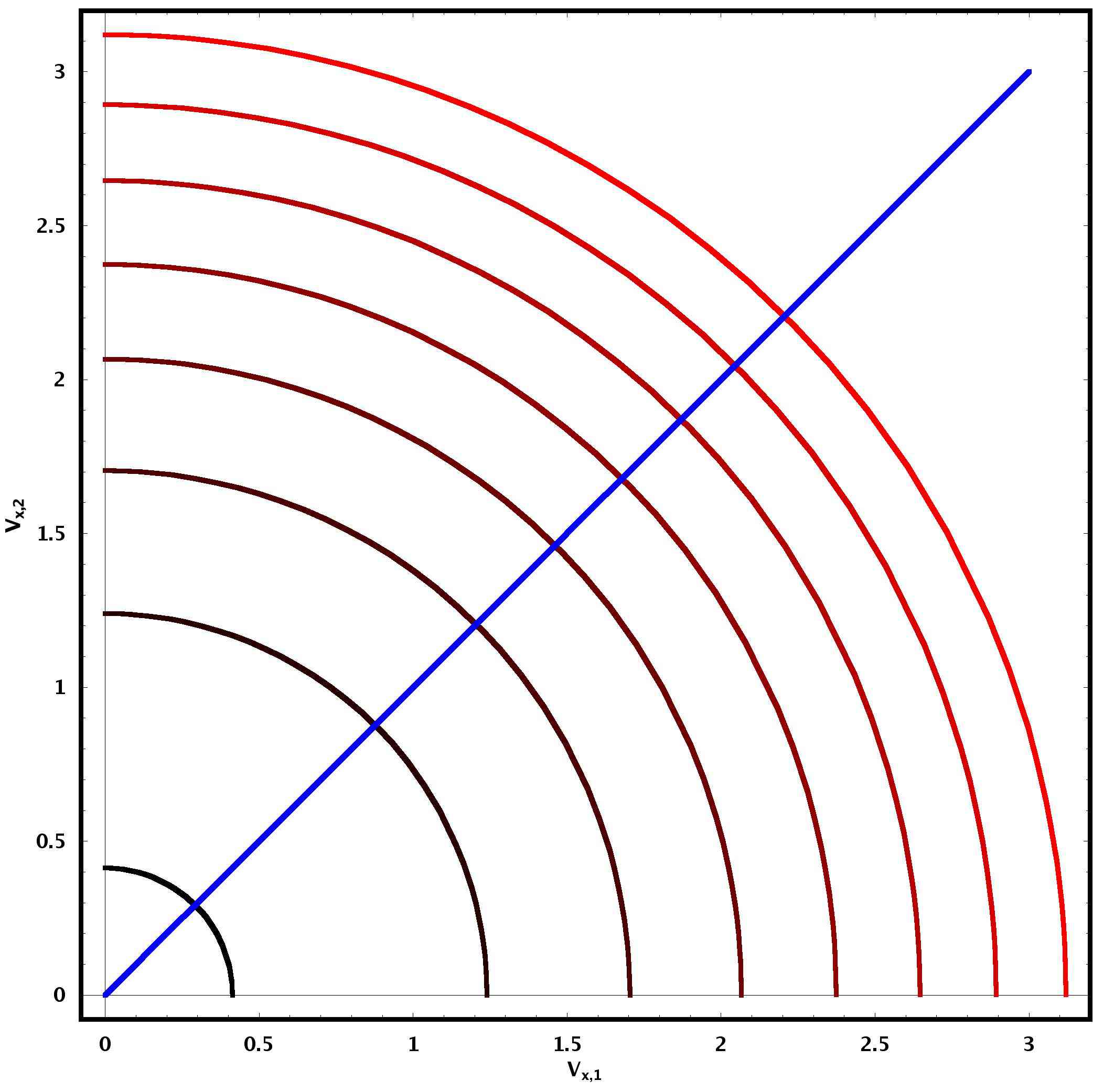}
\caption{Graph of equi-stiffness curves in the electrode potentials $V_{x,i}$. The larger the stiffness,
the more red-tinged the curves. The blue line represents equal voltages - therefore no forces - applied. Employed value
of capacitance is the sensor's.}
\label{fig:actconststiff}
\end{center}
\end{figure}

Therefore from \eqref{eq:electroforce}, assuming $V_{x,2}>V_{x,1}$:
\beq
F_{x} = \left|\frac{\partial C}{\partial x}\right| \left( V_{x,1}^{2}-V_{x,2}^{2}\right)\,,
\label{eq:fcsimple}
\eeq
on the other hand, we can assume the usual spring-like coupling to model the stiffness as:
\beq
\omega_{\text{p,act},x}^2 = - \frac{1}{m}\frac{\partial F}{\partial x} = 
  \frac{1}{2}\sum_{j}\frac{\partial^{2} C_{j}}{\partial x^{2}}
   \left(V_{j}-V_{\text{TM}}\right)^{2}\,,
\label{eq:kfull}
\eeq
and under the same assumptions as before:
\beq
\omega_{\text{p,act},x}^2 = \frac{1}{m} \left|\frac{\partial^{2} C}{\partial x^{2}}\right| \left( V_{x,1}^{2}+V_{x,2}^{2}\right)\,,
\label{eq:ksimple}
\eeq
because the second derivative of the capacitance has the same sign for both sides, no matter
the dominance of $V_{x,1}$ or $V_{x,2}$. The last equation defines a family of circles with radius of constant stiffness
$\sqrt{\left|\omega_{\text{p,act},x}^2\right|}$ in the $V_{x,i}$ space (see picture \ref{fig:actconststiff}) and
 if we'd take e.g. $V_{x,1}=V_{\text{max}}$ and $V_{x,2}=0$, we'd get from
\eqref{eq:fcsimple} and \eqref{eq:ksimple}:
\begin{align}
F_{x,\text{max}} &= \left|\frac{\partial C}{\partial x}\right| V_{\text{max}}^{2}\,,\\
\omega_{\text{p,act},x}^2 &= \frac{1}{m}\left|\frac{\partial^{2} C}{\partial x^{2}}\right| V_{\text{max}}^{2}\,,
\end{align}
hence
\beq
\omega_{\text{p,act},x}^2 = \frac{F_{x,\text{max}}}{m}
  \frac{\left|\frac{\partial^{2} C}{\partial x^{2}}\right|}{\left|\frac{\partial C}{\partial x}\right|}\,.
\label{eq:actistiffratio}
\eeq
By placing the expressions \eqref{eq:czero} and \eqref{eq:d2czero} for the first and second derivatives of $C$ into the latter, we'd get:
\beq
\omega_{\text{p,act},x}^2 = a_{x,\text{max}} \frac{2}{d_{x}}\,.
\label{eq:kappavsfsimple}
\eeq

The former construction can be extended to the other orthogonal electrodes configurations around each TM
and allows for multiple choice of AC potentials induced on the electrodes, providing the sum of voltages
amounts to $0$. Therefore any solution in $V_{j},\,\forall j$ of \eqref{eq:sumpotentialstozero} with the
constraint \eqref{eq:kfull} is valid.

We can point out a number of remarks.
\begin{enumerate}
\item Expression \eqref{eq:kappavsfsimple} clearly shows that balance shall be made between effectiveness
of the actuation and induced stiffness - independently on the DC force value - both
inversely proportional to the capacitance gaps.
\item In general the TMs will carry charge $Q_{\text{TM}}\neq 0$, and in turn will have reference
potential $V_{\text{TM}} \neq 0$. This facts contributes to the whole with a constant DC term in the expressions,
but such an effect is not as troublesome as it might seem.
\item A tempting solution to \eqref{eq:sumpotentialstozero} and \eqref{eq:kfull} would be the AC voltage one:
\beq
V_{x,1} = V_{\text{max}} \sin \omega t\,,\qquad V_{x,2} = V_{\text{max}} \cos \omega t\,,
\eeq
given a pulsation $\omega$. Notice in presence of non-zero charge of the TM the quadratic dependence 
of force and stiffness on the voltage foresees the creating of the mentioned DC term, plus an extra AC
term at frequency $2\omega$. At high frequency neither of them is capable of inducing rotation or
spurious dynamical effects
on the TM due to TM inertia. Care must be taken then to bias
the TM with a voltage whose frequency be outside the MBW.
For LISA and LTP the MBW ranges roughly between $1\,\unit{mHz}$ and $1\,\unit{Hz}$,
biasing with $\omega > 100\,\unit{Hz}$ ensures respecting the constraints.
\item A very important feature of the strategy, which is also transparent from formulae, is that
when $V_{x,1}=V_{x,2}$ no force will be present, but positive stiffness spring will be there anyway. Such a
noteworthy property is extremely useful in compensating negative stiffness. Moreover, the parasitic stiffness
of both TMs-ISs can be matched ($\omega_{\text{p},1}^{2} = \omega_{\text{p},2}^{2}$)
by means of voltage application: if one TM will be servo-ed ($F=0$) and the other suspended ($F\neq 0$)
respecting \eqref{eq:fcsimple} and \eqref{eq:ksimple} they will be subject to non-zero stiffness springs
and tuning of $\omega^{2}_{\text{p},i}$ can be performed. As a direct result, in science mode the
${\Delta}\omega_{\text{p}}^{2}$-modulated term in \eqref{eq:scimodesimple} can be annihilated
as \eqref{eq:matchingstiffness} shows and the residual acceleration ${\Delta}g_{x}$ measurement
directly performed on $\text{IFO}({\Delta}x)$.
\end{enumerate}

\begin{figure}
\begin{center}
\includegraphics[width=0.4\textwidth]{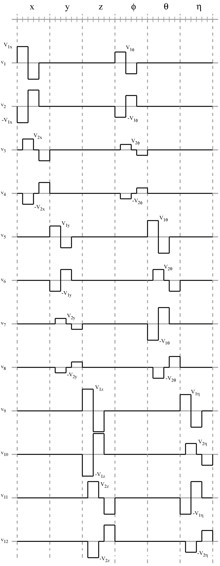}
\caption{Full GRS baseline actuation sequence around slave TM.}
\label{fig:baselineactseq}
\end{center}
\end{figure}

The procedure we described, to order one in stray voltages and upon completion with
a careful rotational DOF treatment (see \cite{LTPact}), is named after ``actuation at constant stiffness strategy''
and it is highly relevant for both missions. In fact the GRS must bias the TMs to sense their position or
measure their charge and
actuate them to move on non-interferometer sensing, but this needs to be done at constant electrostatic
stiffness (and minimal). By describing the stiffness manifold as a quadratic function of the potentials
allows for finding time arrays of solutions, assuming to ``fire'' the capacitors with AC voltages on orthogonal
directions periodically over an actuation period, a fundamental feature to reduce cross-talk holding control
of the TMs. Obviously the simple $\omega_{\text{p,act},x}^2$ constant becomes a full stiffness matrix and a thorough
optimisation of the solution is needed to ensure near-to-null convolution of the actuation signals over an
actuation period.
Figure \ref{fig:baselineactseq} illustrates the carriers shape embedding sine and cosines voltage pulses for a
designed control strategy for LTP \cite{LTPact}.

\subsection{Magnetics and stiffness}

\begin{enumerate}
\item Finally, magnetic stiffness \cite{Hueller:2004zm, HuellerPhD} can be derived by means
of the usual Hooke-like arguments:
\begin{shademinornumber}
\beq
\omega^2_{\text{p,mag}}=\frac{1}{m}
\left(\sqrt{6}\frac{\partial B_{x}}{\partial x}\left(\mu_{x}+\frac{\chi B_{x}L^{3}}{\mu_{0}}\right)
+\sqrt{3}\frac{\partial^{2} B_{x}}{\partial x^{2}}\frac{\chi L^{3}}{\mu_{0}}\right)\,,
\label{eq:magstiff}
\eeq
\end{shademinornumber}
where $B_{x}$ is the $\hat x$ component of the $\vc B$ field and the spare constants are understood. The expression can be
derived from the expression of the magnetic energy density:
\beq
W=\vc\mu\cdot \vc B+\frac{\chi}{2 \mu_{0}}\vc B\cdot \vc B\,,
\eeq
by differentiating it with respect to $x$ we derive the spring constant from the definition:
\beq
\omega^2_{\text{p,mag}}=\frac{1}{m} \frac{\partial^{2} W}{\partial x^{2}}\,.
\eeq
Expression \eqref{eq:magstiff} is thus retrieved assuming isotropy of the permanent magnetic
dipole components, so that:
\begin{shadefundtheory}
\beq
\vc\mu^{2}=\sum_{i=1}^{3}\mu_{i}^{2}\simeq 3 \mu_{x}^{2}\,,
\eeq
\end{shadefundtheory}
hence $\mu_{i}\simeq \sqrt{3}\mu_{x}$. After differentiation we integrate over the TM volume
and we assume we can measure the macroscopic magnetic field and its first and second derivative. to define volume
averages:
\beq
\begin{split}
\left\langle B_{x} \right\rangle L^{3} \simeq & \int_{V_{\text{TM}}} B_{x}(x)\de^{3} x\,,\\
\left\langle B_{x,x}\right\rangle L^{3}\simeq & \int_{V_{\text{TM}}} \frac{\partial}{\partial x}B_{x}(x)\de^{3} x\,,\\
\left\langle B_{x,xx}\right\rangle L^{3}\simeq & \int_{V_{\text{TM}}} \frac{\partial^{2}}{\partial x^{2}}B_{x}(x)\de^{3} x\,,
\end{split}
\eeq
besides, we assume to measure $\mu_{x}$ over the whole TM volume, its dimensionality therefore
being already multiplied by a factor $\metresunit^{3}$. Putting all together \eqref{eq:magstiff} can be
easily found.
\end{enumerate}

\subsection{Summary on stiffness}

\begin{table}
\begin{center}
\begin{tabular}{c|c|c|c|c|c|c}
$\hat i$ & $c(\hat i)$ & $\pi(\hat i)$ & $s(\hat i)$ & $\omega_{\text{p,sens},\hat i}^{2}$
  & $\omega_{\text{p,act},\hat i}^{2}$ & $\omega_{\text{p,DC},\hat i}^{2}$\\
\hline\rule{0pt}{0.4cm}\noindent
$\hat x_{1}$ & $\hat \phi_{1}$ & $\hat \theta_{1}$ & $\hat y_{1}$, $\hat z_{1}$ & $\omega_{\text{p,sens},x}^{2}$ & $\omega_{\text{p,act},\phi_{1}}^{2}$ & $\omega_{\text{p,DC},x}^{2}$\\
$\hat y_{1}$ & $\hat \theta_{1}$ & $\hat \eta_{1}$ & $\hat z_{1}$, $\hat x_{1}$ & $\omega_{\text{p,sens},y}^{2}$ & $\omega_{\text{p,act},\theta_{1}}^{2}$ & $\omega_{\text{p,DC},y}^{2}$\\
$\hat z_{1}$ & $\hat \eta_{1}$ & $\hat \phi_{1}$ & $\hat x_{1}$, $\hat y_{1}$ & $\omega_{\text{p,sens},z}^{2}$ & $\omega_{\text{p,act},\eta_{1}}^{2}$ & $\omega_{\text{p,DC},z}^{2}$\\
$\hat \theta_{1}$ & $\hat y_{1}$ & $\hat x_{1}$ & $\hat z_{1}$, $\hat x_{1}$ & $\nicefrac{3}{2}\sum_{s(\hat i)}\omega_{\text{p,sens},s(\hat i)}^{2}$
  & $\omega_{\text{p,act},\theta_{1}}^{2}$ & $\nicefrac{3}{2}\sum_{s(\hat i)}\omega_{\text{p,DC},s(\hat i)}^{2}$\\
$\hat \eta_{1}$ & $\hat z_{1}$ & $\hat y_{1}$ & $\hat x_{1}$, $\hat y_{1}$ & $\nicefrac{3}{2}\sum_{s(\hat i)}\omega_{\text{p,sens},s(\hat i)}^{2}$
  & $\omega_{\text{p,act},\eta_{1}}^{2}$ & $\nicefrac{3}{2}\sum_{s(\hat i)}\omega_{\text{p,DC},s(\hat i)}^{2}$\\
$\hat \phi_{1}$ & $\hat x_{1}$ & $\hat z_{1}$ & $\hat y_{1}$, $\hat z_{1}$ & $\nicefrac{3}{2}\sum_{s(\hat i)}\omega_{\text{p,sens},s(\hat i)}^{2}$
  & $\omega_{\text{p,act},\phi_{1}}^{2}$ & $\nicefrac{3}{2}\sum_{s(\hat i)}\omega_{\text{p,DC},s(\hat i)}^{2}$\\
$\hat x_{2}$ & $\hat \phi_{2}$ & $\hat \theta_{2}$ & $\hat y_{2}$, $\hat z_{2}$ & $\omega_{\text{p,sens},x}^{2}$ & $\omega_{\text{p,act},\phi_{2}}^{2}+\omega_{\text{p,act},x_{2}}^{2}$ & $\omega_{\text{p,DC},x}^{2}$\\
$\hat y_{2}$ & $\hat \theta_{2}$ & $\hat \eta_{2}$ & $\hat z_{2}$, $\hat x_{2}$ & $\omega_{\text{p,sens},y}^{2}$ & $\omega_{\text{p,act},\theta_{2}}^{2}+\omega_{\text{p,act},y_{2}}^{2}$ & $\omega_{\text{p,DC},y}^{2}$\\
$\hat z_{2}$ & $\hat \eta_{2}$ & $\hat \phi_{2}$ & $\hat x_{2}$, $\hat y_{2}$ & $\omega_{\text{p,sens},z}^{2}$ & $\omega_{\text{p,act},\eta_{2}}^{2}+\omega_{\text{p,act},z_{2}}^{2}$ & $\omega_{\text{p,DC},z}^{2}$\\
$\hat \theta_{2}$ & $\hat y_{2}$ & $\hat x_{2}$ & $\hat z_{2}$, $\hat x_{2}$ & $\nicefrac{3}{2}\sum_{s(\hat i)}\omega_{\text{p,sens},s(\hat i)}^{2}$
  & $\omega_{\text{p,act},\theta_{2}}^{2}+\nicefrac{3}{2}\sum_{s(\hat i)}\omega_{\text{p,act},s(\hat i)}^{2}$ & $\nicefrac{3}{2}\sum_{s(\hat i)}\omega_{\text{p,DC},s(\hat i)}^{2}$\\
$\hat \eta_{2}$ & $\hat z_{2}$ & $\hat y_{2}$ & $\hat x_{2}$, $\hat y_{2}$ & $\nicefrac{3}{2}\sum_{s(\hat i)}\omega_{\text{p,sens},s(\hat i)}^{2}$
  & $\omega_{\text{p,act},\eta_{2}}^{2}+\nicefrac{3}{2}\sum_{s(\hat i)}\omega_{\text{p,act},s(\hat i)}^{2}$ & $\nicefrac{3}{2}\sum_{s(\hat i)}\omega_{\text{p,DC},s(\hat i)}^{2}$\\
$\hat \phi_{2}$ & $\hat x_{2}$ & $\hat z_{2}$ & $\hat y_{2}$, $\hat z_{2}$ & $\nicefrac{3}{2}\sum_{s(\hat i)}\omega_{\text{p,sens},s(\hat i)}^{2}$
  & $\omega_{\text{p,act},\phi_{2}}^{2}+\nicefrac{3}{2}\sum_{s(\hat i)}\omega_{\text{p,act},s(\hat i)}^{2}$ & $\nicefrac{3}{2}\sum_{s(\hat i)}\omega_{\text{p,DC},s(\hat i)}^{2}$
\end{tabular}
\end{center}
\caption{Conjugated and co-sensed DOF versus stiffness. From left to right: the $\hat i$ column represents the variable, $c(\hat i)$ the
one which is sensed by the same GRS electrode surface $s(\hat i)$, $\pi(\hat i)$ is the dynamically conjugated DOF. Sensing, actuation
and DC-force stiffness depends on the choice of electrodes and the conjugation.}
\label{tab:conjang}
\end{table}

We can write, for every DOF, the following general formula with reference to table \ref{tab:conjang}:
\begin{shadefundtheory}
\beq
\omega_{\text{p},\hat i,\hat i}^{2} \doteq
  \omega_{\text{p,mag}}^{2}+\omega^2_{\text{p,grav},\hat i,\hat i}+ \omega_{\text{p,act},\hat i}^{2}
  +\omega_{\text{p,DC},\hat i}^{2}+\omega_{\text{p,sens},\hat i}^{2}\,,
  \label{eq:kappatot1}
\eeq
\end{shadefundtheory}
\noindent so that, e.g. for the $\hat x_{1}$ full stiffness we'd get:
\beq
\omega_{\text{p},x_{1},x_{1}}^{2} =
  \omega_{\text{p,mag}}^{2}+\omega^2_{\text{p,grav},xx}+ \omega_{\text{p,act},\theta_{1}}^{2}
  +\omega_{\text{p,DC},x_{1}}^{2}+\omega_{\text{p,sens},x_{1}}^{2}\,,
\eeq
or, for $\hat \eta_{2}$:
\beq
\begin{split}
\omega_{\text{p},\eta_{2},\eta_{2}}^{2} =&
  \omega_{\text{p,mag}}^{2}+\omega^2_{\text{p,grav},\eta_{2}\eta_{2}}
  +\omega_{\text{p,act},\eta_{2}}^{2}+\frac{3}{2}\left(\omega_{\text{p,act},x_{2}}^{2}+\omega_{\text{p,act},y_{2}}^{2}\right)+\\
  &+\frac{3}{2}\left(\omega_{\text{p,DC},x_{2}}^{2}+\omega_{\text{p,DC},y_{2}}^{2}\right)
  +\frac{3}{2}\left(\omega_{\text{p,sens},x_{2}}^{2}+\omega_{\text{p,sens},y_{2}}^{2}\right)\,.
\end{split}
\eeq
To evaluate the difference in stiffness we cannot assume it to be due to actuation only. In this exposition we are interested to show
this difference only for linear DOF, and we can write in general:
\begin{shadefundtheory}
\beq
{\Delta}\omega_{\text{p},\hat i}^{2} \doteq \lambda_{\hat i} \left(
 \frac{\omega^2_{\text{p,grav},\hat i,\hat i}}{5}+ \sqrt{2}\left(\omega_{\text{p,act},c(\hat i)}^{2}+\omega_{\text{p,mag}}^{2}
  +\omega_{\text{p,DC},\hat i}^{2}\right)+\frac{\omega_{\text{p,sens},\hat i}^{2}}{10}+\omega_{\text{p,act},\hat i}^{2}\right)\,,
\eeq
\end{shadefundtheory}
where $\lambda_{\hat i}$ is a weight which is $\nicefrac{1}{2}$ along $\hat y$ and $\hat z$ but $1$ for $\hat x$ where we
demand a stricter performance. As seen we assume to compensate gravity gradients up to $20\%$, sensing stiffness to $10\%$ but
we have to retain DC, magnetics and angular-actuation stiffness effects, though correlated. Actuation stiffness is taken as it is.
For the $\hat x$ direction we can apply the former and write:
\beq
{\Delta}\omega_{\text{p},x}^{2} =
 \frac{\omega^2_{\text{p,grav},xx}}{5}+ \sqrt{2}\left(\omega_{\text{p,act},\phi}^{2}+\omega_{\text{p,mag}}^{2}
  +\omega_{\text{p,DC},x}^{2}\right)+\frac{\omega_{\text{p,sens},x}^{2}}{10}+\omega_{\text{p,act},x}^{2}\,,
  \label{eq:deltaomparx}
\eeq
With reference to the main signal equation \eqref{eq:mainifosignalfornoise} we can see that
the r{\^o}le of difference of stiffness cannot be neglected or thought as being $0$, due to the term
in $\nicefrac{g_{\text{SC,x}}}{{\omega_{\text{df},x}^{2}}}$ and noise in readout.



\begin{table}
\begin{center}
\begin{tabular}{>{\raggedleft}m{4.5cm}|l|l|l}
Description & Name & Value & Dimensions\\\hline\rule{0pt}{0.4cm}\noindent
Sensing stiffness & $\omega^2_{\text{p,sens}}$ & $0.442\times 10^{-7}$ & $\secminsqunit$\\ 
Actuation stiffness & $\omega^2_{\text{p,act}}$ & $0.501\times 10^{-6}$ & $\secminsqunit$\\ 
Rotation actuation stiffness & $\omega^2_{\text{p,rot}}$ & $0.767\times 10^{-8}$ & $\secminsqunit$\\ 
Magnetic stiffness & $\omega^2_{\text{p,mag}}$ & $0.578\times 10^{-8}$ & $\secminsqunit$\\ 
DC voltage stiffness & $\omega^2_{\text{p,DC}}$ & $0.726\times 10^{-8}$ & $\secminsqunit$\\ 
Total stiffness, TM1, nominal & $\omega^2_{\text{p,1}}$ & $0.565\times 10^{-6}$ & $\secminsqunit$\\ 
Total stiffness, TM2, nominal & $\omega^2_{\text{p,2}}$ & $0.107\times 10^{-5}$ & $\secminsqunit$\\ 
Difference of stiffness & ${\Delta}\omega^2_{\text{p,tot}}$ & $0.574\times 10^{-6}$ & $\secminsqunit$
\end{tabular}
\end{center}
\caption{Stiffness, summary}
\label{tab:allthestiffness}
\end{table}

\section{Inertial sensor displacement noise}

\begin{table}
\begin{center}
\begin{tabular}{r|l|l|l}
Description & Name & Value & Dimensions\\
\hline\rule{0pt}{0.4cm}\noindent
DC differential acceleration $\hat x$ & $g_{\text{DC},x}$ & $1.\times 10^{-9}$ & $\unitfrac{m}{s^2}$\\ 
DC differential acceleration $\hat y$ & $g_{\text{DC},y}$ & $5.\times 10^{-10}$ & $\unitfrac{m}{s^2}$\\ 
DC differential acceleration $\hat z$ & $g_{\text{DC},z}$ & $5.\times 10^{-10}$ & $\unitfrac{m}{s^2}$\\ 
DC  torque/moment of inertia $\hat\theta$ & $g_{\text{DC},\theta}$ & $1.\times 10^{-9}$ & $\secminsqunit$\\ 
DC  torque/moment of inertia $\hat\eta$ & $g_{\text{DC},\eta}$ & $2.\times 10^{-9}$ & $\secminsqunit$\\ 
DC  torque/moment of inertia $\hat\phi$ & $g_{\text{DC},\phi}$ & $1.\times 10^{-9}$ & $\secminsqunit$\\ 
\end{tabular}
\end{center}
\caption{Tolerable maximal differential DC accelerations for linear DOF and maximal DC torques per
unit moment of inertia.}
\label{tab:resaccfull}
\end{table}

\begin{table}
\begin{center}
\begin{tabular}{r|l|l|l}
Description & Name & Value & Dimensions\\\hline\rule{0pt}{0.4cm}\noindent
Gravity gradient $\hat x$ &  $\omega^2_{\text{p,grav},\text{xx}}$  & $5.\times 10^{-7}$ & $\secminsqunit$\\
Gravity gradient $\hat y$ &  $\omega^2_{\text{p,grav},\text{yy}}$  & $5.\times 10^{-8}$ & $\secminsqunit$\\
Gravity gradient $\hat z$ &  $\omega^2_{\text{p,grav},\text{zz}}$  & $-5.5\times 10^{-7}$ & $\secminsqunit$\\
Gravity gradient $\hat \theta$ &  $\omega^2_{\text{p,grav},\theta\theta}$  & $1.\times 10^{-8}$ & $\secminsqunit$\\
Gravity gradient $\hat \eta$ &  $\omega^2_{\text{p,grav},\eta\eta}$  & $1.\times 10^{-8}$ & $\secminsqunit$\\
Gravity gradient $\hat \phi$ &  $\omega^2_{\text{p,grav},\phi\phi}$  & $1.\times 10^{-8}$ & $\secminsqunit$
\end{tabular}
\end{center}
\caption{Gravity gradients}
\label{tab:gravgrad}
\end{table}

The uncertainty coming from the position detectors induces a displacement noise.
Due to the existence of the drag-free control loop, this can be converted into a force noise
via proper transfer functions. In this scenario it is of capital importance to gain knowledge
about the parasitic stiffness that couples both the TMs to the SC. Moreover, the control
strategy is important to understand the origin of the noise: in science mode, since only one TM is
actuated and served by the SC, only the noise coming from the sensing TM matters.

Dealing with an electrostatic detector, we'd like to point out two reading rules for the following formulae:
\begin{enumerate}
\item In general, our electrostatic detector/actuator is nothing but a differential inductive bridge. A capacitive
imbalance caused by a TM displacement creates a differential transformer current according to eq. \eqref{eq:capx},
with the secondary current amplified and read out with lock-in detection. We choose the
bridge injection capacitor $C_{\text{inj}}$ value so to bring the bridge to resonance at $100\,\unit{kHz}$ excitation
frequency, with the main purpose of minimising the amplifier noise. If the model is simplified to
a series RLC circuit, then the complex impedance of the full circuit is
\beq
Z=R+\frac{1}{\imag \omega C_{\text{inj}}}+\imag \omega 2 L_{i}\,,
\eeq
where the factor $2$ comes from the presence of 2 inductors in the bridge and $L_{i}$ is the primary inductance of the single inductor.
Resonance condition is
achieved when $\Imm Z=0$\footnote{One could also calculate the modulus of $Z$, verify it's got a Breit-Wigner resonance shape
and compute when the maximum occurs for $\omega > 0$.}, thus:
\beq
\omega_{\text{inj,res}}^{2}=\frac{1}{2 C_{\text{inj}} L_{i}}\,,
\eeq
from which:
\begin{shadefundtheory}
\beq
\label{eq:cpo}
C_{\text{inj}}=\frac{1}{2 \omega_{\text{inj}}^2 L_i}\,.
\eeq
\end{shadefundtheory}
We named $\omega_{\text{inj}}=2\pi f_{\text{inj}}$ as the
readout bias frequency, such that $f_{\text{inj}}=100\,\unit{kHz}$. Parameters can be checked in table \ref{tab:readout},
a value of $C_{\text{inj}}$ can be checked in table \ref{tab:derivquantities}. In the real circuit (see figure \ref{fig:circuit})
more capacitance's are called in place, their effect being to shift the resonance. We assume nevertheless this correction to be small
and summarise it in the $Q$-factor in the following (see expression \eqref{eq:qfactor}).

\item Concerning thermal noise, dielectric or inductive losses can be caused by many
different phenomena: our general approach will be to discuss them in a sort of intuitive manner
rather then tediously deduce each formula. An example on capacitance may mark the way: according to
Nyquist theory, the thermal power spectrum of a dissipative dynamical system J is a function of 
absolute temperature and impedance $Z_{j}$ as follows:
\beq
S^{\nicefrac{1}{2}}_{\text{J}}=\sqrt{4 k_{\text{B}} T \Ree Z_{\text{J}}}\,,
\eeq
where $k_{\text{B}}$ is Boltzmann's constant. In the case of dielectric loss caused by
electrode surface contamination we can model the presence of impurities and spots by
a macroscopic dimensionless ``loss angle'' $\delta$, such that $C\to C(1+\imag\delta_{C})$, hence, since
a capacitor impedance is
\beq
Z=\frac{1}{\imag \omega C(1+\imag\delta_{C})}\,,
\eeq
we get
\beq
S^{\nicefrac{1}{2}}_{\text{diel}}=\sqrt{4 k_{\text{B}} T \frac{\delta_{C}}{\omega C}}\,.
\eeq
Similar deductions apply for lossy inductance's $L\to L(1+\imag \delta_{L})$. Effects like this perturb the
readout directly affecting position reading (this is the case of patches over the capacitors plates, e.g.).
\end{enumerate}

Readout noise is made of two different kinds of contribution: correlated and uncorrelated. Distinction is made
on the basis of the source of the phenomena and these latter may or may not share correlations
to the same source.
Moreover, a
distinction will be made on the phase shift the current will pick up at $V_{\text{out}}$
with respect to the original phase
at $V_{\text{inj}}$ as a result of the capacitance or inductive electronics it will pass through in the
circuitry. Notice at this level that magnetic flux conservation and the fact that primary and secondary
inductors are equal in the employed transformer doesn't create any phase delay due to induction.

\begin{table}
\begin{center}
\begin{tabular}{>{\raggedleft}m{4.5cm}|l|l|l}
Description & Name & Value & Dimensions\\\hline\rule{0pt}{0.4cm}\noindent
Parallel capacitance to ground & $C_{\text{inj}}$ & $0.486\times 10^{-9}$ & $\unit{F}$\\ 
Readout Q factor & $Q$ & $0.985\times 10^{2}$ & $1$
\end{tabular}
\end{center}
\caption{Summary of derived quantities}
\label{tab:derivquantities}
\end{table}

\subsection{Electric correlated}

Due to the specific form of the electrostatic readout devices three main sources can be spelled in this section.
The readout bridge can be split into three subsystems: the capacitors with the transformer, the amplifier and
the actuation circuitry. Hence there's noise - in phase - produced in the
transformer, deeply caused by thermal excitation of matter states, voltage noise in
the amplifier, and actuation noise.

\begin{description}

\item[Transformer thermal noise, in phase.] In spite of its small contribution the thermal noise induced by the transformer
has a PSD expression which is worth discussing because of the functional form.
Let's write it down:
\beq
\label{eq:trip}
S^{\nicefrac{1}{2}}_{x,\text{trip}}=\frac{d_{x}}{\sqrt{2}}\frac{1}{V_{\text{inj}} Q}
\sqrt{\frac{4 k_BT}{\omega_{\text{inj}} C_{\text{sens}}}\left(\frac{1}{2\omega_{\text{inj}}^{2} L_{i} C_{\text{sens}}Q}\right)}\,,
\eeq
where
\begin{description}
\item[$d_{x}$] is the sensing gap along the $\hat x$ direction,
\item[$C_{\text{sens}}$] is the single electrode sensing capacity,
\item[$V_{\text{inj}}$] is the TM sensing bias voltage amplitude (at $100\,\unit{kHz}$),
\item[$L_{i}$] is the single inductor primary inductance,
\item[$T$] is the absolute temperature,
\item[$k_{B}$] is Boltzmann constant,
\item[$C_{\text{inj}}$] parallel capacitance to ground, expressed at resonance by formula \eqref{eq:cpo}.
\end{description}
Nominal values of the former constants can be found in tables \ref{tab:electrode} and \ref{tab:readout}.
The value of $S^{\nicefrac{1}{2}}_{x,\text{trip}}$ is shown in table \ref{tab:sumdisplnoise}.

We already discussed
the relation between voltage PSD and displacement PSD in \eqref{eq:voltdisppsd}. The reason of the $d_{x}$ pre-factor lies there.
We then recognise the dependence on $\sqrt{4 k_{\text{B}} T}$ and the square root of the real part of the impedance:
intuitively we expect an inverse dependence on $V_{\text{inj}}$, something we'll retrieve in all
thermal voltage noise formulae in the following; the ratio $\nicefrac{C_{\text{sens}}}{C_{\text{inj}}}$ is another common feature,
a pure number hinting to how strong is the capacitance ratio between ground and the bridge.

$Q$ is the readout $Q$-factor, expressed by:
\beq
Q=2 \omega_{\text{inj}}^2 L_i \left(C_{\text{cable}} \delta_{\text{cable}}+C_{\text{par}} \delta_{\text{par}}\right) +\delta_{L_{i}}\,,
\eeq
by employing expression \eqref{eq:cpo} for $C_{\text{inj}}$, we find the algebraic inverse of $Q$ gets an
interesting functional form:
\begin{shadefundtheory}
\beq
\frac{1}{Q}=\frac{C_{\text{cable}} \delta_{\text{cable}}}{C_{\text{inj}}}+\delta_{L_{i}}+\frac{C_{\text{par}}
   \delta_{\text{par}}}{C_{\text{inj}}}\,.
   \label{eq:qfactor}
\eeq
\end{shadefundtheory}
We'd like to spend a couple more words on this expression. It's a combination of the following variables:
\begin{description}
\item[$ \delta_{\text{par}}$] parasitic capacitance loss angle,
\item[$C_{\text{par}}$] electrode parasitic capacitance to ground.
\end{description}
It is clear then that the $Q$-factor expresses an average loss factor, such as $C_{\text{inj}}Q^{-1}=\sum_{j}C_{j}\delta_{j}$,
where $j$ ranges over all the conductors at play. The name ``quality factor'' comes on the line of the introductory discussion we
went through at the beginning of the chapter. Notice the PSD is continuous in $Q$, there would be no noise if there wouldn't be
any dissipation.

\begin{table}
\begin{center}
\begin{tabular}{>{\raggedleft}m{4.5cm}|c|l|l}
Description & Name & Value & Dimensions\\
\hline\rule{0pt}{0.4cm}\noindent
Single electrode sensing capacity & $C_{\text{sens}}$ & $1.15\times 10^{-12}$ & $\unit{F}$\\
Sensing capacity loss angle & $\delta_{\text{sens}}$ & $1.\times 10^{-5}$ & $1$\\ 
Electrode parasitic capacitance to ground & $C_{\text{par}}$ & $2.\times 10^{-11}$ & $\unit{F}$\\ 
Parasitic capacitance loss angle & $\delta_{\text{par}}$ & $2.\times 10^{-2}$ & $1$\\
\hline\rule{0pt}{0.4cm}\noindent
Sensing gap x & $d_{x}$ & $4.\times 10^{-3}$ & $\metresunit$\\ 
Sensing gap y & $d_{y}$ & $2.9\times 10^{-3}$ & $\metresunit$\\ 
Sensing gap z & $d_{z}$ & $3.5\times 10^{-3}$ & $\metresunit$\\ 
\end{tabular}
\end{center}
\caption{Electrode characteristics}
\label{tab:electrode}
\end{table}

\begin{table}
\begin{center}
\begin{tabular}{>{\raggedleft}m{4.5cm}|l|l|l}
Description & Name & Value & Dimensions\\\hline\rule{0pt}{0.4cm}\noindent
TM sensing bias voltage amplitude ($100\,\unit{kHz}$) & $V_{\text{inj}}$ & $0.6$ & $\unit{V}$\\ 
Readout bias frequency & $f_{\text{inj}}=\nicefrac{\omega_{\text{inj}}}{2\pi}$ & $1.\times 10^{5}$ & $\unit{Hz}$\\ 
AC-bias relative amplitude fluctuation (@$\omega$) & $S_{\nicefrac{{\Delta}V_{\text{AC}}}{V_{\text{AC}}}}$ &
  $\left(1.\times 10^{-4}\right)^{2} \left(\frac{2\pi\times10^{-3}\,\unit{Hz}}{\omega}\right)^{2}$ & $\unitfrac{1}{Hz}$\\ 
Single inductor primary inductance & $L_i$ & $2.61\times 10^{-3}$ & $\unit{Henry}$\\ 
Transformer turn ratio & $n_o$ & $1$ & $1$\\ 
Transformer core loss angle & $\delta_{L_{i}}$ & $\nicefrac{1}{110}$ & $1$\\
Transformer imbalance fluctuations & $S_{\nicefrac{{\Delta}L_{i}}{L_{i}}}$ &
  $\left(1.\times 10^{-7}\right)^{2} \left(\frac{2\pi\times10^{-3}\,\unit{Hz}}{\omega}\right)^{2}$ & $\unitfrac{1}{Hz}$\\ 
Cable parasitic capacitance & $C_{\text{cable}}$ & $3.\times 10^{-12}$ & $\unit{F}$\\ 
Cable loss angle & $\delta_{\text{cable}}$ & $4.\times 10^{-2}$ & $1$\\ 
Amplifier voltage noise & $S_{V_{\text{amp}}}$ &
  $\left(2.\times 10^{-9}\right)^{2} \left(\frac{2\pi\times10^{-3}\,\unit{Hz}}{\omega}\right)^{2}$ & $\unitfrac{V^2}{Hz}$\\ 
Amplifier current noise & $S_{I_{\text{amp}}}$ &
  $\left(1.\times 10^{-14}\right)^{2} \left(\frac{2\pi\times10^{-3}\,\unit{Hz}}{\omega}\right)^{2}$ & $\unitfrac{A^2}{Hz}$\\ 
Feedback capacitor & $C_{\text{fb}}$ & $1.\times 10^{-11}$ & $\unit{F}$\\ 
Feedback capacitor loss & $\delta_{C_{\text{fb}}}$ & $1.\times 10^{-2}$ & $1$\\ 
Feedback capacitance relative fluctuation & $S_{\nicefrac{{\Delta}C_{\text{fb}}}{C_{\text{fb}}}}$ & $\left(1.\times 10^{-5}\right)^{2}$ & $\unitfrac{1}{Hz}$\\ 
Effective bridge output offset & ${\Delta}x_{\text{bo}}$ & $2.\times 10^{-6}$ & $\metresunit$\\ 
\end{tabular}
\end{center}
\caption{Readout characteristics}
\label{tab:readout}
\end{table}

\item[Amplifier voltage noise, in phase.] The same arguments applied before to build the transformer thermal
noise apply here as well: hence the turn ratio factor will appear and
the noise PSD will be linearly dependent upon it, i.e. if the voltage get amplified, any noise voltage will be as well
by means of the same physics.
The distance/capacitance pre-factor is unmodified but - as can be seen in the circuit scheme \ref{fig:circuit} - a feedback
capacitor is inserted and its capacitance modulates the noise linearly. In formula:
\beq
S^{\nicefrac{1}{2}}_{x,\text{ampip}}=\frac{d_{x}}{\sqrt{2}} \frac{1}{V_{\text{inj}}}
  \frac{C_{\text{fb}}}{\sqrt{2} C_{\text{sens}}} n_o \sqrt{S_{V_{\text{amp}}}}\,,
\eeq
where - apart from already mentioned variables - we can distinguish between:
\begin{description}
\item[$n_{0}$] representing the transformer turn ratio,
\item[$C_{\text{fb}}$] being the feedback capacitance,
\item[$S_{V_{\text{amp}}}$] the amplifier voltage noise.
\end{description}
Again, values can be retrieved from tables \ref{tab:electrode} and \ref{tab:readout}. The value
of $S^{\nicefrac{1}{2}}_{x,\text{ampip}}$ is shown in table \ref{tab:sumdisplnoise}.

\item[Actuation noise at $100\,\unit{kHz}$.]
Actuation noise at $100\,\unit{kHz}$ is a contribution one order of magnitude larger than the former two mentioned
in this section. No wonder in deducing the formula, which we state as:
\beq
S^{\nicefrac{1}{2}}_{x,\text{act100}}=\frac{d_{x}}{\sqrt{2}} \frac{1}{V_{\text{inj}}}
  \frac{C_{\text{inj}}}{C_{\text{sens}}} 
  \gamma_{o,100\unit{kHz}}\sqrt{S_{V_{\text{act}}}}\,,
\eeq
where 
\begin{description}
\item[$S_{V_{\text{act}}}$] is the actuation voltage noise at output ($100\,\unit{kHz}$ and MBW);
\item[$\gamma_{o,100\unit{kHz}}$] is the actuation filter open loop transfer. Transfer functions always
appear unabridged homogeneously multiplying PSDs, dimensionless ones.
\end{description}
Values can be retrieved from tables \ref{tab:electrode} and \ref{tab:readout} together with \ref{tab:actuation}.
The value of $S^{\nicefrac{1}{2}}_{x,\text{act100}}$ is to be found in table \ref{tab:sumdisplnoise}.

\begin{table}
\begin{center}
\begin{tabular}{>{\raggedleft}m{4.5cm}|l|l|l}
Description & Name & Value & Dimensions\\
\hline\rule{0pt}{0.4cm}\noindent
Actuation amplitude relative fluctuation & $S_{\nicefrac{{\Delta}V_{\text{act}}}{V_{\text{act}}}}$ &
  $\left(2.\times 10^{-6}\right)^{2} \left(\frac{2\pi\times10^{-3}\,\unit{Hz}}{\omega}\right)^{2}$ & $\unitfrac{1}{Hz}$\\ 
Actuation voltage noise at output($100\,\unit{kHz}$ and MBW) & $S_{V_{\text{act}}}$ &
  $\left(1.\times 10^{-6}\right)^{2} \left(\frac{2\pi\times10^{-3}\,\unit{Hz}}{\omega}\right)^{2}$ & $\unitfrac{V^2}{Hz}$\\ 
Actuation filter open loop transfer & $\gamma_{o,100 \unit{kHz}}$ & $1.\times 10^{-4}$ & $1$\\ 
Actuation filter closed loop transfer & $\gamma_{c,100 \unit{kHz}}$ & $1.\times 10^{-4}$ & $1$\\ 
Actuation filter impedance & $Z_{\text{act}}$ & $10.$ & $\unit{\Omega}$\\ 
\end{tabular}
\end{center}
\caption{Actuation characteristics}
\label{tab:actuation}
\end{table}

\end{description}

\begin{figure}
\begin{center}
\includegraphics[width=\textwidth]{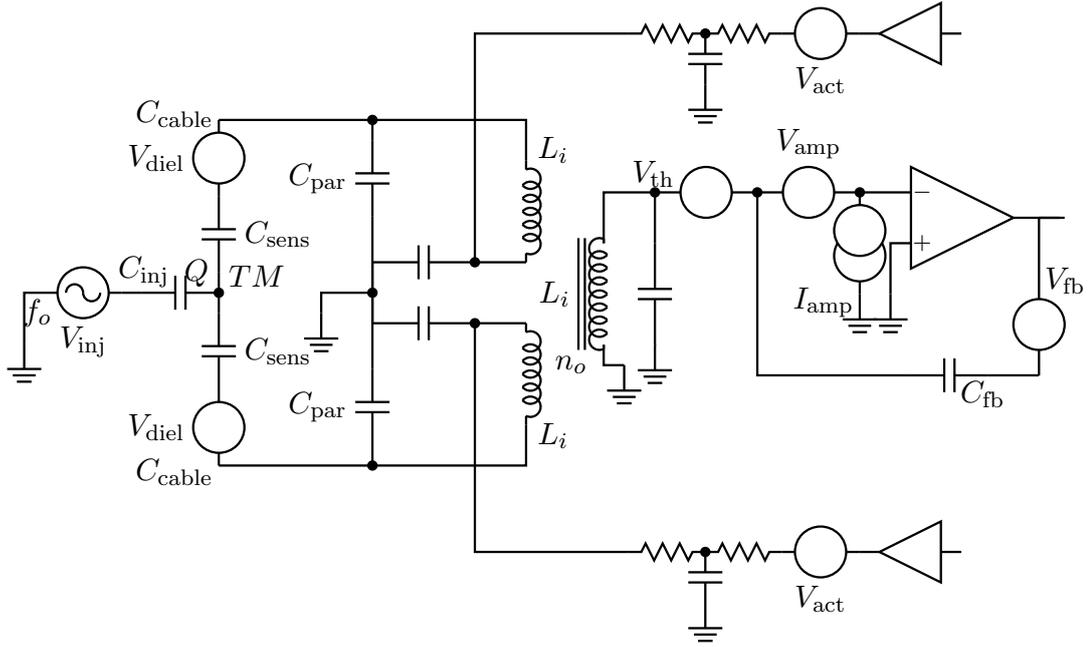}
\caption{Sensing bridge of the type LTP will be equipped with and similar to sensing devices of torsion
pendulum facility at the University of Trento. The set of electrodes on the right mimics the TM, with charge $q$,
the transformer in the middle with turn ratio $n_{0}$, the amplifier on the right. The actuation circuit is shown
at top and bottom.}
\label{fig:circuit}
\end{center}
\end{figure}



\subsection{Electric uncorrelated}

All values can be retrieved from table \ref{tab:sumdisplnoise}. Here we have:

\begin{description}

\item[Transformer thermal noise, out of phase.]
The out of phase transformer thermal noise is a highly relevant contribution to the noise PSD, amounting
to magnitude $10^{-9}$. No difference in deducing or building the formula rather than eq. \eqref{eq:trip},
but obviously we get rid of the $\nicefrac{1}{Q}$ pre-factor since we are now looking for out-of-phase contributions:
\beq
\label{eq:trop}
S^{\nicefrac{1}{2}}_{x,\text{trop}}=\frac{d_{x}}{\sqrt{2} V_{\text{inj}}}
\sqrt{\frac{4 k_BT}{\omega_{\text{inj}} C_{\text{sens}}}\left(\frac{1}{2\omega_{\text{inj}}^{2} L_{i} C_{\text{sens}}Q}\right)}\,,
\eeq


\item[Amplifier voltage noise, out of phase]
\beq
S^{\nicefrac{1}{2}}_{x,\text{ampop}} = \frac{d_{x}}{\sqrt{2}}
\frac{1}{V_{\text{inj}}} \frac{C_{\text{fb}}}{\sqrt{2} C_{\text{sens}}} 
\frac{1}{\left(n_o \omega_{\text{inj}}^2 L_i C_{\text{fb}} Q\right)}\sqrt{S_{V_{\text{amp}}}}\,,
\eeq
\item[Amplifier current noise, out of phase]
\beq
S^{\nicefrac{1}{2}}_{x,\text{curop}} =
\frac{d_{x}}{\sqrt{2}}\frac{\sqrt{S_{I_{\text{amp}}}}}{V_{\text{inj}}}\frac{n_o}
{\sqrt{2}\omega_{\text{inj}}C_{\text{sens}}}\,,
\eeq
\item[Feedback capacitor current noise, in phase]
\beq
S^{\nicefrac{1}{2}}_{x,\text{cufb}} =
\frac{d_{x}}{\sqrt{2}}\frac{n_o}{V_{\text{inj}}}\sqrt{\frac{2 k_BT \delta_{C_{\text{fb}}}}
{\omega_{\text{inj}} C_{\text{fb}}}}\,,
\eeq
\item[Differential transformer imbalance, in phase]
\beq
S^{\nicefrac{1}{2}}_{x,\text{dt}} = \frac{d_{x}}{\sqrt{2}}\frac{\sqrt{S_{\nicefrac{{\Delta}L_{i}}{L_{i}}}}}{4}\,,
\eeq
\item[Oscillator amplifier amplitude noise]
\beq
S^{\nicefrac{1}{2}}_{x,\text{osc}} = {\Delta}x_{\text{bo}}\sqrt{S_{\nicefrac{{\Delta}V_{\text{AC}}}{V_{\text{AC}}}}}\,,
\eeq
\item[Feedback capacitor instability noise]
\beq
S^{\nicefrac{1}{2}}_{x,\text{Cfb}} = \frac{{\Delta}x_{\text{bo}}}{\sqrt{2}}\sqrt{S_{\nicefrac{{\Delta}C_{\text{fb}}}{C_{\text{fb}}}}}\,,
\eeq
\end{description}

\subsection{Thermal correlated distortion}

The thermal displacement noise formula is easily deductible from thermal dilatation principles. Given a linear
dilatation model:
\beq
{\Delta}x \simeq \alpha L_{0} {\Delta}T\,,
\eeq
where ${\Delta}x$ is the dilatation, $L_{0}$ the starting length, $\alpha$ some dilatation coefficient and ${\Delta}T$
the temperature excursion, we simply have to switch to PSD space as follows: 
\beq
S^{\nicefrac{1}{2}}_{x,\text{th}}=\alpha_{\text{th}}\left(\frac{L}{2}\right)\sqrt{S_{{\Delta}T}}\,.
\eeq

Constants and descriptions can be found in table \ref{tab:presstherm}. Notice since in this case we are
dealing with a temperature difference fluctuation $S_{{\Delta}T}$, we consider
a geometric factor of $\nicefrac{1}{2}$ in the baseline length. The final value of $S^{\nicefrac{1}{2}}_{x,\text{th}}$
is to be found in table \ref{tab:sumdisplnoise}.

\subsection{Contribution from forces on the SC}
Again a careful inspection of \eqref{eq:mainifosignalfornoise} reveals the contributions of forces coming
from the SC to sensor displacement noise can be summarised in the term:
\beq
\left(\omega^{2}_{\text{p},2}-\omega^{2}_{\text{p},1}\right)\frac{g_{\text{SC,x}}}{\omega_{\text{df},x}^{2}}\,.
\eeq
Therefore, in terms of the PSD of forces acting on the SC:
\beq
S^{\nicefrac{1}{2}}_{x,\text{SC}}=
  \frac{\left|\omega^{2}_{\text{p},2}-\omega^{2}_{\text{p},1}\right|}{\left|\omega_{\text{df},x}^{2}\right|}
  \frac{\sqrt{S_{\text{SC}}}}{m_{\text{SC}}}\,.
\eeq
The value is reported in table \ref{tab:sumdisplnoise},
values of constants and their meaning can be found in table \ref{tab:spacecraftchars},
the form of $\omega^{2}_{\text{df},x}$ can be retrieved from \eqref{eq:suspfun} and table \ref{tab:dfcoeff}.

\begin{table}
\begin{center}
\begin{tabular}{r|l|l|l}
Description & Name & Value & Dimensions\\
\hline\rule{0pt}{0.4cm}\noindent
Forces on SC  & $S_{\text{SC}}$ & $(5.0\times 10^{-6})^{2}\left(2\pi\times \frac{10^{-3}\,\unit{Hz}}{\omega}\right)^{2}$
& $\unitfrac{N^{2}}{Hz}$\\
SC mass & $m_{\text{SC}}$ & $476.$ & $\unit{kg}$\\
SC effective radius & $R_{\text{SC}}$ & $1$ & $\text{m}$
\end{tabular}
\end{center}
\caption{SC characteristics and estimate of external forces}
\label{tab:spacecraftchars}
\end{table}


\subsection{Summary of displacement noise, drag-free noise}

At the end of the analysis the contributions will be summed quadratically, assuming they are
basically uncorrelated or weakly correlated anyway:
\begin{shadefundtheory}
\beq
S^{\nicefrac{1}{2}}_{x,j}S^{\nicefrac{1}{2}}_{x,k} \ll S_{x,j} + S_{x,k}\,.
\eeq
\end{shadefundtheory}
\noindent It turns out, in estimating the total sensor noise, that this procedure gives the worst
estimate of the effects, therefore being preferable in planning phase; moreover, due to the very local nature of
the disturbances, this is also physically meaningful. We have then, for the total noise on the sensor:
\begin{shadefundtheory}
\beq
S_{x,\text{sens}} = S_{x,\text{corr}} + S_{x,\text{uncorr}}+S_{x,\text{th}}\,,
\eeq
\end{shadefundtheory}
\noindent where
\begin{shadefundtheory}
\beq
S_{x,\text{corr}}=S_{x,\text{trip}}+S_{x,\text{ampip}}+S_{x,\text{act100}}\,,
\eeq
\end{shadefundtheory}
\noindent and
\begin{shadefundtheory}
\beq
\begin{split}
S_{x,\text{uncorr}}=&S_{x,\text{trop}}+S_{x,\text{ampop}}+S_{x,\text{curop}}+\\
&+S_{x,\text{cufb}}+S_{x,\text{dt}}+S_{x,\text{osc}}+S_{x,\text{Cfb}}\,,
\end{split}
\eeq
\end{shadefundtheory}
\noindent while, if we take into account the forces acting on the SC, we can compute the total
drag-free displacement noise:
\begin{shadefundtheory}
\beq
S_{x,\text{tot}}=S_{x,\text{SC}}+S_{x,\text{sens}}\,.
\eeq
\end{shadefundtheory}
\noindent Figures are to be retrieved from table \ref{tab:sumdisplnoise}.

This long discussion about displacement noise formerly carried on is motivated by the drag-free control loop.
Acceleration noise is enlarged by displacement noise converted into acceleration via the difference of
stiffness constant ${\Delta}\omega_{\text{p}}^2$ - see eq. \eqref{eq:deltaomparx} - so that the displacement spectrum gets converted into an acceleration
one:
\begin{shademinornumber}
\beq
S^{\nicefrac{1}{2}}_{a,\text{dragfree}} = \left|{\Delta}\omega_{\text{p},x}^2\right| S_{x,\text{tot}}^{\nicefrac{1}{2}}\,,
\eeq
\end{shademinornumber}
\noindent whose value can be found in table \ref{tab:sumnoise}.

\begin{table}
\begin{center}
\begin{tabular}{>{\raggedleft}m{4.5cm}|l|l|l}
Description & Name & Value & Dimensions\\
\hline\rule{0pt}{0.4cm}\noindent
Stray DC electrode potential & $V_{\text{stray}}$ & $3.\times 10^{-2}$ & $\unit{V}$\\ 
Charge events rate & $\lambda $ & $5.\times 10^{2}$ & $\unit{Hz}$\\ 
Test-mass charge/electron charge & $q_{0}$ & $1.\times 10^{7}$ & $1$\\ 
Shielding factor & $\alpha_{\text{sh}}$ & $1.\times 10^{-3}$ & $1$\\ 
In-band voltage fluctuations & $S_{V_{\text{ib}}}$ & $\left(1.\times 10^{-4}\right)^{2} \left(2\pi\times \frac{10^{-3}\,\unit{Hz}}{\omega}\right)^{2}$
& $\unitfrac{V^2}{Hz}$\\ 
Maximum AC voltage within electrodes & $V_{\text{AC}}$ & $1.$ & $\unit{V}$\\ 
AC voltage noise & $S_{V_{\text{AC}}}$ & $\left(1.\times 10^{-7}\right)^{2} \left(2\pi\times \frac{10^{-3}\,\unit{Hz}}{\omega}\right)^{2}$ &
$\unitfrac{V^2}{Hz}$\\ 
\end{tabular}
\end{center}
\caption{Voltage and charge characteristics}
\label{tab:voltage}
\end{table}

\begin{table}
\begin{center}
\begin{shademinornumber}
\begin{tabular}{r|l|l}
Description & Name & Value $(\unitfrac{m}{\sqrt{Hz}})$\\
\hline\rule{0pt}{0.4cm}\noindent
Transformer thermal noise. In-phase &  $S^{\nicefrac{1}{2}}_{x,\text{trip}}$  & $1.49\times 10^{-11}$\\ 
Amplifier voltage noise. In-phase &  $S^{\nicefrac{1}{2}}_{x,\text{ampip}}$  & $5.79\times 10^{-11}$\\ 
Actuation noise at 100 kHz &  $S^{\nicefrac{1}{2}}_{x,\text{act100}}$  & $1.99\times 10^{-10}$\\ 
Transformer thermal noise. Out-of-phase &  $S^{\nicefrac{1}{2}}_{x,\text{trop}}$  & $1.46\times 10^{-9}$\\ 
Amplifier voltage noise. Out-of-phase &  $S^{\nicefrac{1}{2}}_{x,\text{ampop}}$  & $5.72\times 10^{-11}$\\ 
Amplifier current noise. Out-of-phase &  $S^{\nicefrac{1}{2}}_{x,\text{curop}}$  & $4.61\times 10^{-11}$\\ 
Feedback capacitor current noise. In-phase &  $S^{\nicefrac{1}{2}}_{x,\text{cufb}}$  & $1.69\times 10^{-11}$\\ 
Differential transformer imbalance. In phase &  $S^{\nicefrac{1}{2}}_{x,\text{dt}}$  & $7.07\times 10^{-11}$\\ 
Oscillator amplifier amplitude noise &  $S^{\nicefrac{1}{2}}_{x,\text{osc}}$  & $2.\times 10^{-10}$\\ 
Feedback capacitor instability noise &  $S^{\nicefrac{1}{2}}_{x,\text{Cfb}}$  & $1.41\times 10^{-11}$\\ 
Thermal distortion &  $S^{\nicefrac{1}{2}}_{x,\text{th}}$  & $1.15\times 10^{-11}$\\
\hline\rule{0pt}{0.4cm}\noindent
Total sensor noise &  $S^{\nicefrac{1}{2}}_{x,\text{sens}}$  & $1.49\times 10^{-9}$\\ 
Effect of forces on SC &  $S^{\nicefrac{1}{2}}_{x,\text{SC}}$  & $6.43\times 10^{-10}$\\
\hline\rule{0pt}{0.4cm}\noindent
Total drag-free &  $S^{\nicefrac{1}{2}}_{x,\text{tot}}$  & $1.63\times 10^{-9}$\\ 
\end{tabular}
\end{shademinornumber}
\end{center}
\caption{Summary of displacement noise}
\label{tab:sumdisplnoise}
\end{table}

\section{Inertial sensor acceleration noise}
The following contributions act directly on the TM as force noises. What we'll designate with the
symbol $S^{\nicefrac{1}{2}}_{a}$ with some additional lower index specifying which
the origin will be. Everywhere in the following we'll always mean forces per unit mass of the TM.

\subsection{Readout circuitry back-action}

In every GW detector the effect of readout back-action is very important to estimate. The fundamental
source of such an effect is hidden in the very form of the detector, which is basically an electrostatic
bridge coupled to an amplifier. The bridge needs to be powered to work, and the presence of such a modulation
voltage creates a thermal back-acting current whose squared PSD is function of the real part of the impedance,
the thermal Nyquist factor $4k_{\text{B}}T$, the voltage itself and its modulation frequency. According
to Callen and Welton \cite{PhysRev.83.34, PhysRev.83.1231} we have for the transformer thermal noise:
\begin{shadefundtheory}
\beq
S_{V}(\omega) = 2 \Ree Z \hbar \omega \left(\frac{1}{2}+ \frac{1}{e^{\frac{\hbar \omega}{k_{\text{B}}T}}-1} \right)\,,
\eeq
\end{shadefundtheory}
\noindent whose limits are:
\beq
S_{V}(\omega) \to \begin{cases}
2 k_{\text{B}}T \Ree Z\,, &\text{for}\, k_{\text{B}}T \gg \hbar \omega\,,\\
\hbar \omega \Ree Z\,, &\text{for}\, k_{\text{B}}T \ll \hbar \omega\,.
\end{cases}
\eeq

The lower limit of such a noise is of quantum nature, as it could be expected. Therefore, the product
of the current noise PSD times the voltage noise PSD, being the energy fluctuation PSD divided by frequency
(inverse of time), shall be larger or equal than $\nicefrac{\hbar}{2}$ \cite{PhysRevD.26.1817}:
\begin{shadefundtheory}
\beq
\frac{S_{I}^{\nicefrac{1}{2}} S_{V}^{\nicefrac{1}{2}}}{\omega} \geq \frac{\hbar}{2}\,.
\label{eq:noisequantumlimit}
\eeq
\end{shadefundtheory}

In resonant bar detectors like AURIGA \cite{Vinante:2006uk} the circuitry back-action is a very important issue, since the
readout is purely electrostatic. Nevertheless, even in interferometer ground-based detectors as VIRGO \cite{Acernese:2006np} the need
of a very high-frequency modulation voltage for the bridge - motivated by winning over Newtonian ground
noise in sensitivity amplification - brings the quantum limit closer, as \eqref{eq:noisequantumlimit} scaling with frequency would suggest.

Conversely, LISA doesn't need to power the bridge at such a high frequency, keeping $V_{\text{inj}}$ at $f_{\text{inj}}=100\,\unit{kHz}$,
and can employ full laser detection. This argument, together with the following numbers,
shall convince the reader that LISA's readout back-action is well under control and highly negligible within a full noise budget
analysis.

Notice, to close the introduction, that due to the amplifier configuration we chose, our transformer thermal noise
doesn't produce back-action (all back-effects from the transformer are somehow shielded due to its very high impedance modulus),
and that the amplifier back action is well beyond the quantum limit. In addition, we state that
- in case - we could even tolerate a worse figure given the laser metrology sensitivity
spectrum $S_{x}^{\nicefrac{1}{2}} \sim 10^{-9}\,\unitfrac{m}{\sqrt{Hz}}$. We included all potential contributions
nevertheless for the sake of completeness.
Another form of \eqref{eq:noisequantumlimit} in terms of displacement and acceleration spectra would state:
\begin{shadefundtheory}
\beq
S_{x}^{\nicefrac{1}{2}} S_{F}^{\nicefrac{1}{2}} \geq \frac{\hbar}{2}\,,
\eeq
\end{shadefundtheory}
\noindent therefore, using a popular value for $\hbar \sim 10^{-34}\,\unitfrac{kg\,m^{2}}{s}$ we'd get
\beq
S_{F} ^{\nicefrac{1}{2}} \geq 10^{-25}\,\unitfrac{N}{\sqrt{Hz}}\,,
\eeq
which is the ``force quantum limit'' for an interferometer detection apparatus. The amplifier back-action is
much worse, but not as big as $10^{-15}\,\unitfrac{N}{\sqrt{Hz}}$, our binding threshold for LISA and LTP.

\subsubsection{Correlated}
We mention here three contributions:
\begin{description}
\item[the transformer thermal noise, in phase:]
\beq
S^{\nicefrac{1}{2}}_{a,\text{trip}} =
\frac{\sqrt{2}}{m d_{x}} V_{\text{inj}}C_{\text{sens}}\sqrt{2 k_BT \omega_{\text{inj}}L_i Q}\,,
\eeq
\item[the amplifier voltage noise, in phase:]
\beq
S^{\nicefrac{1}{2}}_{a,\text{ampip}} =
\frac{\sqrt{2}}{m d_{x}}\frac{V_{\text{inj}}C_{\text{sens}}\sqrt{S_{V_{\text{amp}}}}}{\sqrt{2}n_o}\,,
\eeq
\item[the actuation noise at $100\,\unit{kHz}$:]
\beq
S^{\nicefrac{1}{2}}_{a,\text{act100}} =
\frac{\sqrt{2}}{m d_{x}}\frac{V_{\text{inj}} C_{\text{sens}}\gamma_{c, 100\unit{kHz}}}{4}
 \sqrt{S_{V_{\text{act}}}}\,.
\eeq
\end{description}
These will be gathered quadratically for the two TMs in a correlated readout noise term:
\begin{shadefundtheory}
\beq
S_{a,\text{corr}}=2\left(S_{a,\text{trip}}+S_{a,\text{ampip}}+S_{a,\text{act100}}\right)\,.
\eeq
\end{shadefundtheory}
\noindent Values are retrievable from table \ref{tab:corrisnoise}.
The sum of correlated readout noise $S_{a,\text{corr}}^{\nicefrac{1}{2}}$ is
summarised in table \ref{tab:sumnoise}.

\begin{table}
\begin{center}
\begin{shademinornumber}
\begin{tabular}{>{\raggedleft}m{6.0cm}|l|l}
Description & Name & Value $\accPSDunit$\\
\hline\rule{0pt}{0.4cm}\noindent
Transformer thermal noise. In-phase & $S^{\nicefrac{1}{2}}_{a,\text{trip}}$ & $4.50\times 10^{-18}$ \\
Amplifier voltage noise. In-phase & $S^{\nicefrac{1}{2}}_{a,\text{ampip}}$ & $1.76\times 10^{-19}$ \\
Actuation noise at $100\,\unit{kHz}$ & $S^{\nicefrac{1}{2}}_{a,\text{act100}}$ & $3.11\times 10^{-21}$\\
\hline\rule{0pt}{0.4cm}\noindent
Total correlated readout back-action & $S^{\nicefrac{1}{2}}_{a,\text{corr}}$ & $6.36\times 10^{-18}$
\end{tabular}
\end{shademinornumber}
\end{center}
\caption{Summary of correlated readout force noise}
\label{tab:corrisnoise}
\end{table}

\subsubsection{Uncorrelated}
Uncorrelated readout noise $S^{\nicefrac{1}{2}}_{a,\text{unc}}$ has two contributors:
\begin{description}
\item[actuation noise in MBW]
\beq
S^{\nicefrac{1}{2}}_{a,\text{act0}} =
\frac{\sqrt{2}}{m d_{x}}C_{\text{sens}}V_{\text{stray}}\sqrt{S_{V_{\text{act}}}}\,,
\eeq
\item[thermal noise at actuation frequency]
\beq
S^{\nicefrac{1}{2}}_{a,\text{actth}} =
\sqrt{2}\sqrt{\frac{g_{\text{DC},x}}{m}\frac{C_{\text{sens}}}{d_{x}}4 k_BT Z_{\text{act}}}\,.
\eeq
\end{description}
In the latter we assumed the need of compensating for maximal allowed DC acceleration along $\hat x$ (see table \ref{tab:resaccfull}).
Values are expressed in table \ref{tab:uncorrisnoise}. Notice the
contribution are quadratically summed per TM:
\begin{shadefundtheory}
\beq
S_{a,\text{unc}}=2\left(S_{a,\text{act0}}+S_{a,\text{actth}}\right)\,,
\eeq
\end{shadefundtheory}
whose value can be found in table \ref{tab:sumnoise}.

\begin{table}
\begin{center}
\begin{shademinornumber}
\begin{tabular}{>{\raggedleft}m{6.0cm}|l|l}
Description & Name & Value $\accPSDunit$\\
\hline\rule{0pt}{0.4cm}\noindent
Actuation noise in MBW & $S^{\nicefrac{1}{2}}_{a,\text{act0}}$ & $6.22\times 10^{-18}$ \\
Thermal noise at actuation frequency & $S^{\nicefrac{1}{2}}_{a,\text{actth}}$ & $2.18\times 10^{-19}$\\
\hline\rule{0pt}{0.4cm}\noindent
Total uncorrelated readout back-action & $S^{\nicefrac{1}{2}}_{a,\text{unc}}$ & $8.81\times 10^{-18}$
\end{tabular}
\end{shademinornumber}
\end{center}
\caption{Summary of uncorrelated readout force noise}
\label{tab:uncorrisnoise}
\end{table}

\subsubsection{Total readout back-action noise}
The two contributions $S^{\nicefrac{1}{2}}_{a,\text{corr}}$ and $S^{\nicefrac{1}{2}}_{a,\text{unc}}$ may be
summed quadratically to get the PSD for the total readout circuitry acceleration noise as:
\begin{shadefundtheory}
\beq
S_{a,\text{readout}} = S_{a,\text{corr}} + S_{a,\text{unc}}\,.
\eeq
\end{shadefundtheory}

\subsection{Thermal effects}

\begin{table}
\begin{center}
\begin{tabular}{>{\raggedleft}m{4.5cm}|l|l|l}
Description & Name & Value & Dimensions\\
\hline\rule{0pt}{0.4cm}\noindent
Pressure in EH & $P$ & $1.\times 10^{-5}$ & $\unit{Pa}$\\ 
Temperature & $T$ & $293.$ & $\unit{K}$\\ 
Temperature fluctuation & $S_T$ & $\left(1.\times 10^{-4}\right)^{2}\left(\frac{2\pi\times 10^{-3}\,\unit{Hz}}{\omega}\right)^{2}$ &
$\unitfrac{K^2}{Hz}$\\ 
Temperature difference fluctuation & $S_{{\Delta}T}$ & $\left(1.\times 10^{-4}\right)^{2} \left(\frac{2\pi\times 10^{-3}\,\unit{Hz}}{\omega}\right)^{2}$ & $\unitfrac{K^2}{Hz}$\\ 
Conductance of venting holes & $C_{\text{hole}}$ & $4.3\times 10^{-3}$ & $\unitfrac{m^3}{s}$\\ 
Ratio of conductance: y & $\sigma_y$ & $1.33$ & $1$\\ 
Ratio of conductance: z & $\sigma_z$ & $1.40$ & $1$\\ 
Activation temperature & $\Theta_o$ & $3.\times 10^{4}$ & $\unit{K}$\\ 
Out-gassing area(1 face) & $A_{\text{og}}$ & $\left(0.053\right)^{2}$ & $\metresunit^2$\\ 
Electrode Housing Linear thermal expansion & $\alpha_{\text{th}}$ & $5.\times 10^{-6}$ & $\unitfrac{1}{K}$ 
\end{tabular}
\end{center}
\caption{Pressure and thermal characteristics}
\label{tab:presstherm}
\end{table}

\subsubsection{Radiometric effects}

Radiometric effects occur in connection to behaviour proper of radiometer gauges \cite{Roth:1982uq}. A
connection of such a type is characterised by two plates $A_{i}\,i=1,2$ at temperatures $T_{i}$
respectively. The average speed of particles leaving the surfaces may be written as $v_{\text{av},i}$,
while their root-mean-square velocities as $v_{\text{r},i}$. $n_{i}$ is the number density of molecules
at any instant. Notice, for a Maxwell-distributed velocity that:
\beq
\frac{v_{\text{r}}}{v_{\text{av}}}= \sqrt{\frac{3\pi}{8}}\,,
\label{eq:maxw1}
\eeq
and that
\beq
n_{1} v_{\text{av},1} = n_{2} v_{\text{av},2}\,.
\label{eq:maxw2}
\eeq
We write the pressure between $A_{1}$ and $A_{2}$ as:
\beq
P_{12}=\frac{1}{3}m n_{1} v_{\text{r},1}^{2}+\frac{1}{3}m n_{2} v_{\text{r},2}^{2}\,,
\label{eq:P12}
\eeq
and state:
\beq
\frac{1}{4}n v_{\text{av}} \doteq \frac{1}{4}n_{1} v_{\text{av},1}+\frac{1}{4}n_{2} v_{\text{av},2}\,.
\eeq
By means of \eqref{eq:maxw2} we get from the former that:
\beq
n_{1} v_{\text{av},1} = n_{2} v_{\text{av},2} = \frac{1}{2} n v_{\text{av}}\,,
\eeq
so that, back to the pressure expression \eqref{eq:P12} we get:
\beq
\begin{split}
P_{12}&=\frac{1}{3}m \frac{1}{2}n \left(
   v_{\text{r},1}^{2} \frac{v_{\text{av}}}{v_{\text{av,1}}}
  +v_{\text{r},2}^{2} \frac{v_{\text{av}}}{v_{\text{av,2}}}
  \right) =\\
  &=\frac{1}{6}m n v_{\text{r}}^{2} \left(\frac{v_{\text{r},1}+v_{\text{r},2}}{v_{\text{r}}}\right) = \\
  &=\frac{1}{2}P\left(\sqrt{\frac{T_{1}}{T}}+\sqrt{\frac{T_{2}}{T}}\right)\,.
\end{split}
\eeq
Where $T$ was introduced as average temperature.
By placing a third plate $A_{3}$ between the former two, the pressure between $A_{2}$ and $A_{3}$ is analogously:
\beq
P_{23}=\frac{1}{2}P\left(\sqrt{\frac{T_{2}}{T}}+\sqrt{\frac{T_{3}}{T}}\right)\,,
\eeq
thus, the resulting pressure difference on $A_{2}$ is
\beq
{\Delta}P = P_{12}-P_{23} = \frac{1}{2}P\left(\sqrt{\frac{T_{1}}{T}}-\sqrt{\frac{T_{3}}{T}}\right)\,,
\eeq
and it's independent of $T_{2}$. Let $T_{3} = T$ and $T_{1} \to T+{\delta}T$ in the last expression; we'll have
then:
\beq
{\delta}P \simeq \frac{1}{2} P \left(\sqrt{\frac{T+{\delta}T}{T}}-1\right) = \frac{P {\delta}T}{4 T}+O\left( {\delta}T^2\right)\,.
\eeq
%
Notice in this whole treatment the mid-plate mimics the TM in the EH; moreover the plate (or TM) must be kept isothermal for the last relation to
hold. Hence, per unit mass and naming the temperature difference fluctuation spectrum
of ${\delta}T$ as $S_{{\Delta}T}$, we get:
\beq
S^{\nicefrac{1}{2}}_{a,\text{rad}}=\frac{A P}{4 m T}\sqrt{S_{{\Delta}T}}\,.
\eeq

\subsubsection{Radiation pressure asymmetry}

The radiation pressure may be given as the ratio between force and area as:
\beq
P=\frac{F}{A}=\frac{1}{A}\frac{\de p}{\de t}\,,
\eeq
where $p$ is the linear momentum. Since for photons we have the simple relation for energy $W=p c$, where $c$
is the light speed in vacuo, we get:
\beq
P=\frac{1}{c}\rho(W)\,,
\eeq
where $\rho(W)$ is the spectral radiance per unit time and surface. For relativistic particles the radiation energy and the radiation density
are related by a factor of $\nicefrac{1}{3}$. Hence, integrating the Planck radiation density we obtain the
Stefan law, with a pre-factor of $8$ accounting for the octant integration on frequency \cite{Eisberg:1985fk}:
\beq
\rho(W)=\frac{8}{3}\sigma T^{4}\,,
\eeq
so that
\beq
P=\frac{8}{3}\frac{1}{c}\sigma T^{4}\,.
\eeq
By variation we get:
\beq
{\delta}P=\frac{8}{3}\frac{1}{c}\sigma T^{3} {\delta}T\,,
\eeq
and then a PSD for acceleration as:
\beq
S^{\nicefrac{1}{2}}_{a,\text{radpr}}=\frac{8 A \sigma}{3 m c}T^3\sqrt{S_{{\Delta}T}}\,.
\eeq

\subsubsection{Asymmetric out-gassing}

Out-gassing is potentially a major source for residual internal pressure within the EH. Molecules of
gas can be thought as trapped on the surface of the housing like in a potential well. Thermal energy
occasionally excites the molecule above the energy barrier, hence the molecule is released (out-gassed)
in free space. The flow of gas is the then modelled as a decaying process, with a given $\infty$-temperature
flow $I_{0}$ and activation temperature $\Theta_{0}$ (see table \ref{tab:presstherm} for a value):
\beq
I_{\text{og}}=I_{0} \exp\left( -\frac{\Theta_{0}}{T}\right)\,.
\eeq
By simple variations:
\beq
{\delta}I_{\text{og}} = I_{0} \frac{\Theta_{0}}{T^{2}} \exp \left(-\frac{\Theta_{0}}{T}\right) {\delta}T\,,
\eeq
so that the relative variation is:
\beq
\frac{{\delta}I_{\text{og}}}{I_{\text{og}}} = \frac{\Theta}{T} \frac{{\delta}T}{T}\,.
\eeq
A temperature gradient can cause then a molecular outflow, and fluctuation of the former will induce fluctuation of the
latter. In presence of asymmetry of the venting holes, pumps and outflow channels an asymmetry pre-factor
$\alpha_{\text{og}}$ appears which can be computed from the expressions of pressure gradients
as a function of conductance's and their ratios as \cite{Vitale:2003fk}:
\beq
\alpha_{\text{og}}=\frac{1}
{2\left(\sigma_y+\sigma_z\right)+1}\,,
\eeq
where $\sigma_{y}$, $\sigma_{z}$ are relative surface conductance's.
Notice we can form a pressure out of $\alpha_{\text{og}}$ and $I_{\text{og}}$, via
\beq
P=\alpha_{\text{og}} \frac{A_{\text{og}} I_{\text{og}}}{C_{\text{hole}}}\,,
\eeq
where $A_{\text{og}}$ is the effective out-gassing surface, $C_{\text{hole}}$
is the hole conductance. Values can be found in table
\ref{tab:presstherm} and \ref{tab:gasderivquantities}.

From the expression of the pressure, the acceleration is easy to compute as $a=\nicefrac{A P}{m}$,
so that the fluctuation is given by:
\beq
{\delta}a = \frac{A}{m} {\delta}P = \alpha_{\text{og}} \frac{A}{m} \frac{ (A_{\text{og}} {\delta}I_{\text{og}})}{C_{\text{hole}}}
= \alpha_{\text{og}} \frac{A}{m} \frac{A_{\text{og}}}{C_{\text{hole}}} I_{\text{og}} \frac{\Theta_{0}}{T^{2}} {\delta}T\,,
\eeq
from which the PSD:
\beq
S^{\nicefrac{1}{2}}_{a,\text{og}}=\alpha_{\text{og}}
  \frac{A}{m} \frac{A_{\text{og}}}{C_{\text{hole}}} I_{\text{og}} \frac{\Theta_{0}}{T^{2}} \sqrt{S_{{\Delta}T}}\,.
\eeq

\begin{table}
\begin{center}
\begin{tabular}{>{\raggedleft}m{4.5cm}|l|l|l}
Description & Name & Value & Dimensions\\
\hline\rule{0pt}{0.4cm}\noindent
Residual molecular gas mass &  $m_{\text{gas}}$  & $6.69\times 10^{-26}$ & $\unit{kg}$\\ 
Out-gas factor & $\alpha_{\text{og}}$ & $3.07\times 10^{6}$ & $1$\\
Out-gassing rate & $I_{\text{og}}$ & $5.\times 10^{-7}$ & $\unitfrac{kg}{s^3}$\\ 
Gas damping time & $\tau$  & $0.228\times 10^{11}$ & $\unitfrac{1}{Hz}$
\end{tabular}
\end{center}
\caption{Gas phenomena derived quantities, summary}
\label{tab:gasderivquantities}
\end{table}

\subsubsection{Thermal distortion}
Thermal distortion may be accounted for by means of standard dilatation formulae applied to the
$\hat x$ direction of the force per unit mass. Given the maximal tolerable DC differential acceleration along $\hat x$, $g_{\text{DC},x}$
the fluctuation in acceleration is thus given by the same times a thermal dilatation coefficient $\alpha_{\text{th}}$
times the temperature fluctuation. Namely, in spectral form:
\beq
S^{\nicefrac{1}{2}}_{a,\text{th}}=g_{\text{DC},x}\alpha_{\text{th}}\sqrt{S_{{\Delta}T}}\,.
\label{eq:thermaldistor}
\eeq

\subsubsection{Gravitational distortion of IS}
No matter how complicated the geometry of the SC and housing surrounding the sensor are, the
net effect of the gravity contribution coming from these mentioned shells may be embedded into an
effective acceleration coefficient ${\delta}g_{\text{th}}$. We refer to the appendix for a careful explanation on
how compensating DC effects of the self-gravity. We assume here a behaviour similar to equation \eqref{eq:thermaldistor}:
\beq
S^{\nicefrac{1}{2}}_{a,\text{gravIS}}={\delta}g_{\text{th}} \alpha_{\text{th}}\sqrt{S_{{\Delta}T}}\,.
\eeq

\subsubsection{Total thermal effects noise}
All the mentioned contributions will be considered coherent between one-another and therefore summed
linearly in modulus. This accounts for the highly correlated nature of them, if a radiometric phenomenon or thermal distortion
phenomenon occurs, the source of it may be the same for both TMs and thus be incident in terms of acceleration more like
twice the absolute value rather than the square average. Therefore
\begin{shadefundtheory}
\beq
S_{a,\text{thermal}}^{\nicefrac{1}{2}} = 2 \left(S^{\nicefrac{1}{2}}_{a,\text{rad}}+S^{\nicefrac{1}{2}}_{a,\text{radpr}}
  +S^{\nicefrac{1}{2}}_{a,\text{og}}+S^{\nicefrac{1}{2}}_{a,\text{th}}+S^{\nicefrac{1}{2}}_{a,\text{gravIS}}\right)\,.
\eeq
\end{shadefundtheory}
\noindent A summary of all the thermal noises is presented in table \ref{tab:sumthermal}, together with the total,
which can be inspected in the acceleration summary table, \ref{tab:sumnoise}.

\begin{table}
\begin{center}
\begin{shademinornumber}
\begin{tabular}{r|l|l}
Description & Name & Value $\accPSDunit$\\
\hline\rule{0pt}{0.4cm}\noindent
Radiometric effect &  $S^{\nicefrac{1}{2}}_{a,\text{rad}}$  & $9.21\times 10^{-16}$\\ 
Radiation pressure asymmetry &  $S^{\nicefrac{1}{2}}_{a,\text{rdapr}}$  & $1.37\times 10^{-15}$\\ 
Asymmetric out-gassing &  $S^{\nicefrac{1}{2}}_{a,\text{og}}$  & $1.91\times 10^{-16}$\\ 
Thermal distortion &  $S^{\nicefrac{1}{2}}_{a,\text{th}}$  & $5.\times 10^{-19}$\\ 
Gravitational distortion of IS &  $S^{\nicefrac{1}{2}}_{a,\text{gravIS}}$  & $5.\times 10^{-18}$\\ 
\hline\rule{0pt}{0.4cm}\noindent
Thermal effects, total &  $S^{\nicefrac{1}{2}}_{a,\text{thermal}}$  & $4.97\times 10^{-15}$\\ 
\end{tabular}
\end{shademinornumber}
\end{center}
\caption{Thermal effects, summary}
\label{tab:sumthermal}
\end{table}

\subsection{Brownian noise}

\begin{table}
\begin{center}
\begin{tabular}{>{\raggedleft}m{4.5cm}|l|l|l}
Description & Name & Value & Dimensions\\
\hline\rule{0pt}{0.4cm}\noindent
DC-magnetic field component & $\left\langle B_x\right\rangle$ & $2.\times 10^{-6}$ & $\unit{T}$\\ 
DC-magnetic gradient & $\left\langle B_{\text{x,x}}\right\rangle$ & $5.\times 10^{-6}$ & $\unitfrac{T}{m}$\\ 
DC-magnetic second derivative & $\left\langle B_{\text{x,xx}}\right\rangle$ & $0.02$ & $\unitfrac{T}{m^2}$\\
AC-magnetic field maximum value & $B_{\text{AC},\max}$ & $5.\times 10^{-7}$ & $\unit{T}$\\ 
Magnetic field fluctuation & $S_{B_x}$ &
  $\left(1.\times 10^{-7}\right)^{2}\left(\frac{2\pi\times 10^{-3}\,\unit{Hz}}{\omega}\right)^{2}$ &
  $\unitfrac{T^{2}}{Hz}$\\ 
Magnetic field fluctuation interplanetary & $S_{B_{\xi}}$ & $ \left(0.3\times 10^{-7}\right)^{2}
\left(\frac{10^{-3}\,\secminoneunit}{f_e}\right)^{2}$ & $\unitfrac{T^{2}}{Hz}$\\ 
Magnetic gradient fluctuation  & $S_{B_{\text{x,x}}}$ & $ \left(2.5\times 10^{-7}\right)^{2}
\left(\frac{10^{-3}\,\secminoneunit}{f_e}\right)^{2}$ & $\unitfrac{T^{2}}{m^{2}\,Hz}$\\ 
Magnetic field fluctuation above MBW & $S_{B_{x,\text{AC}}}$ & $\left(10.\times 10^{-8}\right)^{2}$ & $\unitfrac{T^{2}}{Hz}$\\ 
Magnetic susceptibility & $\chi $ & $2.\times 10^{-5}$ & $1$\\ 
Imaginary susceptibility & $\delta_{\chi} $ & $3.\times 10^{-7}$ & $1$\\ 
Permanent magnetic moment & $\mu_x$ & $2.\times 10^{-8}$ & $\unitfrac{J}{T}$\\ 
\end{tabular}
\end{center}
\caption{Magnetics characteristics}
\label{tab:magnetics}
\end{table}

\subsubsection{Dielectric losses}

The expression of the acceleration noise PSD for dielectric losses can be computed
by adding the contribution coming from stray voltages to the static charge
and multiplying by the Nyquist power spectrum at temperature $T$: $4 k_{\text{B}} T \Ree Z$. Here $Z$ is the
impedance between the electrodes and the ground. The
real part of the circuitry impedance is given in this case by the lossy part of the sensing
capacitance $\nicefrac{\delta_{\text{sens}}}{C_{\text{sens}}}$
and must be divided by the frequency $\omega$:
\beq
S^{\nicefrac{1}{2}}_{a,\text{diel}}=\left(\sqrt{2}\frac{V_{\text{stray}}
   C_{\text{sens}}}{d_{x}m}+\frac{1}{6}\frac{ q_{e}q_{0}}{d_{x}m}\right)
   \sqrt{\frac{8 k_BT \delta_{\text{sens}}}{\omega C_{\text{sens}}}}
\eeq

\subsubsection{Residual gas}
The residual gas around the TM behaves accordingly to Maxwell distribution. Hence
calling $m_{\text{gas}}$ the molecular mass of the gas in the VE, assumed to be Argon,
we can derive from the kinetic theory of gases that:
\beq
v  = \sqrt{\frac{k_{\text{B}}T}{m_{\text{gas}}}}\,.
\label{eq:vgasres}
\eeq
Assuming a Stokes-like behaviour for the corresponding force $F=-\beta v$,
we can solve the first cardinal equation of dynamics $-\beta v = m a = m \nicefrac{\de v}{\de t}$ and name the
damping time constant as $\tau$ to get:
\beq
\tau= \frac{m}{\beta}\,,
\eeq
then,
by the definition of pressure:
\beq
P A = \left|F\right| = \frac{m}{\tau} v\,,
\label{eq:presstau}
\eeq
we get
\beq
\tau = \frac{m v}{P A} = \frac{m}{P A}\sqrt{\frac{k_{\text{B}}T}{m_{\text{gas}}}}\,,
\eeq
and finally we can employ the
fluctuation-dissipation theorem and state that the squared PSD of the force itself
is given by
\beq
S_{F} = m^{2} S_{a} = 4 k_{\text{B}} T \beta\,.
\label{eq:flucdisspsd}
\eeq
We get in the end:
\beq
S^{\nicefrac{1}{2}}_{a,\text{gas}}=\sqrt{\frac{4 k_BT}{m \tau}}\,.
\eeq

\subsubsection{Magnetic damping}

Magnetic damping is a phenomenon close to Foucault (currents) in the effect. The Lorentz force given by
a current is:
\beq
\vc F = \vc j \wedge \vc B\,,
\eeq
by virtue of Maxwell-Amp\'ere equation:
\beq
\vc\nabla \wedge \vc E = \frac{\partial \vc B}{\partial t}\,,
\eeq
we can write:
\beq
\frac{1}{\sigma} \vc\nabla \wedge \vc j = \sum_{i} \frac{\partial \vc B}{\partial x^{i}}\frac{\partial x^{i}}{\partial t}
= \left(\vc v \cdot \vc \nabla \right) \vc B\,,
\eeq
which can be inverted if the gradient of the field is homogeneous, therefore:
\beq
\vc F = \sigma \left(\vc v \cdot \vc \nabla \right) \vc B \wedge \vc B\,,
\eeq
and it can be seen there's a Stokes-like dependence of the force from the velocity. Hence the fluctuation-dissipation
theorem can be employed again, this time with a $\beta$ factor depending on the magnetic field gradient. We won't
bother the reader with the details, suffice it to say that after volume integration only one term survives along $\hat x$
and finally we get:
\beq
S^{\nicefrac{1}{2}}_{a,\text{magdmp}}=\frac{\left\langle B_{\text{x,x}}\right\rangle}{m}
\sqrt{\frac{3^{2/3} L^5 \sigma_0 k_B T}{5\times 2^{1/3} \pi^{2/3}}}\,.
\eeq

\subsubsection{Magnetic impurities}

The presence of an imaginary susceptibility component such that $\chi \to \chi \left(1+\imag\delta_{\chi}\right)$
gives, via the fluctuation-dissipation theorem a fluctuating magnetic dipole whose
squared PSD is
\beq
S_{\mu} = \frac{4\pi k_{\text{B}} T}{\mu_{0}}\frac{\delta_{\chi}}{\omega}\,.
\eeq
Given the relation between force and field as:
\beq
F_{x} = -\vc\mu\cdot \partial_{x} \vc B\,,
\eeq
we'd get, for a sphere of radius $R$ or volume-equivalent cube of side $L$ that:
\beq
S_{F} = \frac{4}{3}\pi R^{3} \frac{4\pi k_{\text{B}} T}{\mu_{0}}\frac{\delta_{\chi}}{\omega}
  \left(\partial_{x} \vc B\right)^{2}\,,
\eeq
to get, finally:
\beq
S^{\nicefrac{1}{2}}_{a,\text{magimp}}=\frac{\left\langle B_{\text{x,x}}\right\rangle}{m}
  \sqrt{\frac{8 k_B T L^{3}\delta_{\chi}}{\omega \pi  \mu_o}}
\eeq

\subsubsection{Total Brownian noise}
In contrast with the approach taken to deal with thermal effects, here highly incoherent effects between one-another showed up.
The uncorrelated nature of them, as stochastic processes allows us to sum them quadratically and multiply then by two as
the multiplicity of the TMs. The deep nature of a large number of Bernoulli processes makes them almost-Gaussian, moreover
the quadratic sum of PSDs for Gaussian processes is distributed like $\chi^{2}$ in the DOF; absence of
correlation factors is just the ``simplified'' reading of this scenario.
We get:
\begin{shadefundtheory}
\beq
S_{a,\text{Brownian}} = 2 \left(S_{a,\text{diel}}+S_{a,\text{gas}}
  +S_{a,\text{magdmp}}+S_{a,\text{magimp}}\right)\,,
\eeq
\end{shadefundtheory}
\noindent whose value is presented in table \ref{tab:sumbrownian} with the various contributions. Find the 
value in table \ref{tab:sumnoise} for a summary.

\begin{table}
\begin{center}
\begin{shademinornumber}
\begin{tabular}{r|l|l}
Description & Name & Value $\accPSDunit$\\
\hline\rule{0pt}{0.4cm}\noindent
Dielectric losses &  $S^{\nicefrac{1}{2}}_{a,\text{diel}}$  & $2.7\times 10^{-16}$\\ 
Residual gas & $S^{\nicefrac{1}{2}}_{a,\text{gas}}$  & $6.02\times 10^{-16}$\\ 
Magnetic damping &  $S^{\nicefrac{1}{2}}_{a,\text{magdmp}}$  & $5.27\times 10^{-17}$\\
Magnetic impurities &  $S^{\nicefrac{1}{2}}_{a,\text{magimp}}$  & $1.57\times 10^{-17}$\\ 
\hline\rule{0pt}{0.4cm}\noindent
Brownian noise &  $S^{\nicefrac{1}{2}}_{a,\text{Brownian}}$  & $9.36\times 10^{-16}$
\end{tabular}
\end{shademinornumber}
\end{center}
\caption{Brownian effects, summary}
\label{tab:sumbrownian}
\end{table}

\subsection{Magnetics from the SC}

Generally speaking, the energy of a magnetic dipole $\vc \mu$ crossing a magnetic field $\vc B$
can be locally expressed by a permanent
contribution plus a self-energy of the field itself, $\sim B^{2}$:
\beq
W=\int_{\text{TM}}\de^{3}x\left(\vc\mu\cdot\vc B+\frac{\chi}{2 \mu_{0}} \vc B\cdot \vc B\right)\,,
\label{eq:magenergy1}
\eeq
where $\chi$ is the magnetic susceptibility and $\mu_{0}$ the permeability in vacuo.
We may think of the $\vc\mu$ term as being formed by a permanent magnetic moment dipole
term, plus an induction term, as follows:
\beq
\vc\mu_{\text{eff}}=\vc\mu_{\text{perm}}+\frac{\chi}{2 \mu_{0}} \vc B\,,
\eeq
The force is hence given as the counter-gradient
of the energy: $\vc F = -\vc\nabla W$, to get - specialising to the situation of a cubic TM, with side $L$:
\beq
F_{j} = \left(\mu_{\text{perm}\,i}+\frac{\chi}{\mu_{0}} B_{i}\right)\partial_{j}B_{i}\,,
\eeq
summation over repeated indice is understood. Suppose now the contribution in the sum comes
from a roughly isotropic field distribution. Taking the the $\hat x$-term as representative, we get a homogeneity
factor of $\sqrt{3}$. We'd like to inspect the force per unit mass over the TM, thus getting:
\beq
a_{x}=\frac{\sqrt{3}}{m}\left(\mu_{\text{perm}\,x}+\frac{\chi}{\mu_{0}} B_{x}\right)\partial_{x}B_{x}\,.
\label{eq:magindacc}
\eeq
Mainly two quantities can oscillate in the former expression, namely the field $B_{x}$ and its gradient $\partial_{x}B_{x}$;
in turn the field may pick-up another noise contribution from down-converted alternate currents contributions on
the surfaces of the TMs. Field and gradient will be thought as having nominal values in the zero centring position of the TM,
and will be regarded as constants plus oscillatory term at need.

A recent success of ground testing by means of a torsion pendulum\footnote{Weber, W.J. and Vitale, S., {\it Private communication}.} consists in the retrieval of eddy currents
on the TM surface. Eddy (Foucault) currents effect can be deduced as follows: let's consider the Maxwell equation correspondent to
Amp\'{e}re theorem:
\beq
\vc\nabla\wedge \vc E = - \frac{\partial \vc B}{\partial t}\,,
\eeq
by placing $\vc j=\sigma_{0} \vc E$ and inverting the curl in presence of an isotropic field, we get:
\beq
\vc j = \sigma_{0} \dot {\vc B} \wedge \vc r \simeq \sigma_{0} \omega \vc B \wedge \vc r \,,
\eeq
where the dot stands for time derivative and $\vc r$ is the linear position vector and $\omega=2\pi f$.
From the definition of magnetic dipole:
\beq
\vc\mu = \int_{\partial \text{TM}} j  \hat n \de^{2} x\,,
\eeq
where the integral is carried on the TM surface and $\hat n$ is the surface orientation vector. By retaining only
the $\hat x$ component:
\beq
j \simeq \sigma_{0} \omega \left\langle B_{x}\right\rangle  L^{2}\,,
\eeq
after carrying the volume integration out, the extra term picks the form:
\beq
\mu_{x,\text{Foucault}} \simeq \frac{1}{24} L^{3} \sigma_{0}\omega \left\langle B_{x}\right\rangle  L^{2}
\eeq

\subsubsection{Magnetic field fluctuations}

Suppose in the expression of the magnetic-fluctuation induced acceleration \eqref{eq:magindacc} we'd let
oscillate the field only; then, by varying $a_{x}$ with respect to $B_{x}$, we'd get:
\beq
{\delta}a_{x}=\frac{\sqrt{3}}{m}\left(
  \frac{\chi}{\mu_{0}}L^{3}
  +\frac{1}{24} L^{5} \sigma_{0} \omega
  \right) \left(\partial_{x}B_{x}\right){\delta}B_{x}\,,
\eeq
and switching to spectral representation:
\beq
S_{a,B}^{\nicefrac{1}{2}} = \frac{\sqrt{3}}{m}\left(
  \frac{\chi}{\mu_{0}}L^{3}
  +\frac{1}{24} L^{5} \sigma_{0} \omega
  \right) \left\langle B_{\text{x,x}}\right\rangle \sqrt{S_{B_x}}\,,
\eeq
which is the noise contribution.

\subsubsection{Magnetic gradient fluctuations}

Conversely, if it is the gradient variation that we seek, from \eqref{eq:magindacc} we'd get:
\beq
{\delta}a_{x}=\frac{\sqrt{3}}{m}\left(\mu_{\text{perm}\,x}+\frac{\chi}{\mu_{0}}L^{3} \left\langle B_{x}\right\rangle \right)
  \delta\left(\partial_{x}B_{x}\right)\,,
\eeq
to give in spectral form:
\beq
S_{a,{\Delta}B}^{\nicefrac{1}{2}} = \frac{\sqrt{3}}{m}
 \left(\mu_x+\frac{\chi \left\langle B_{x}\right\rangle L^3}{\mu_o}\right) \sqrt{S_{B_{\text{x,x}}}}
\eeq

\subsubsection{Down-converted AC magnetic field}
The expression of the effective magnetic spectral noise as down-converted from the currents looping on the TM surfaces is in principle
not different from the expression given by pure magnetic field fluctuation.
No, better use derivative of flux...
\beq
S_{a,\text{BAC}}^{\nicefrac{1}{2}} = \frac{5 L^2 \chi  B_{\text{AC},\max} \sqrt{S_{B_{x,\text{AC}}}}}{m \mu_o}
\eeq

\subsubsection{Total magnetic SC noise}
The effects shown could have common sources, hence they can be taken as coherent and correlated effects between one-another.
We sum them linearly in the square-roots, with a factor 2 in front of the whole:
\begin{shadefundtheory}
\beq
S^{\nicefrac{1}{2}}_{a,\text{magnSC}} = \sqrt{2} \left(S^{\nicefrac{1}{2}}_{a,B}+S^{\nicefrac{1}{2}}_{a,{\Delta}B}
+S^{\nicefrac{1}{2}}_{a,\text{BAC}}\right)\,,
\eeq
\end{shadefundtheory}
\noindent whose value is presented in table \ref{tab:sumintmagnetic} with the various contributions.

\begin{table}
\begin{center}
\begin{shademinornumber}
\begin{tabular}{r|l|l}
Description & Name & Value $\accPSDunit$\\
\hline\rule{0pt}{0.4cm}\noindent
Magnetic field fluctuations &  $S^{\nicefrac{1}{2}}_{a,B}$  & $7.64\times 10^{-16}$\\ 
Magnetic gradient fluctuations &  $S^{\nicefrac{1}{2}}_{a,{\Delta}B}$  & $5.1\times 10^{-15}$\\ 
Down-converted AC magnetic field &  $S^{\nicefrac{1}{2}}_{a,\text{BAC}}$  & $4.3\times 10^{-16}$\\
\hline\rule{0pt}{0.4cm}\noindent
Magnetics SC &  $S^{\nicefrac{1}{2}}_{a,\text{magnSC}}$  & $8.9\times 10^{-15}$\\ 
\end{tabular}
\end{shademinornumber}
\end{center}
\caption{Internal magnetic field effects, summary.}
\label{tab:sumintmagnetic}
\end{table}

\subsection{Magnetics interplanetary}

\subsubsection{Magnetic field fluctuations}
The magnetic field self-energy formula can be extended in presence of an external field by taking
\beq
\vc B \to \vc B+\vc B_{\text{ext}}\,,
\eeq
with obvious meaning of the symbols. Hence $\vc B$ now contains the effect of the permanent field - if any - and
the induced one. In the expression of the energy \eqref{eq:magenergy1} we get then two more terms:
\beq
W=\int\de^{3}x\left(\vc\mu\cdot\vc B+\vc\mu\cdot\vc B_{\text{ext}}+\frac{\chi}{2 \mu_{0}} \left(\vc B\cdot \vc B
+2 \vc B\cdot \vc B_{\text{ext}}+\vc B_{\text{ext}}\cdot\vc B_{\text{ext}}
\right)
\right)\,,
\eeq
but the external field self-energy can be thought as a point-$0$ energy and subtracted away - in fact its contribution is immaterial to our
purposes. By employing the same procedure as above, we vary the external field, to get, along $\hat x$, in the approximation of isotropic
contributions:
\beq
{\delta}a_{x}=\frac{\sqrt{3}}{m}
\left(\frac{\chi}{\mu_{0}}L^{3}+\frac{1}{24} L^{5} \sigma_{0} \omega\right)
\left(\partial_{x}B_{x}\right){\delta}B_{\text{ext}\,x}\,,
\eeq
where we considered the eddy currents contribution. In spectral form, we get:
\beq
S_{a,\text{Bi}}^{\nicefrac{1}{2}} = \frac{\sqrt{3}}{m}
\left(\frac{\chi}{\mu_{0}}L^{3}+\frac{1}{24} L^{5} \sigma_{0} \omega\right)
\left\langle B_{\text{x,x}}\right\rangle \sqrt{S_{B_{\xi}}}\,.
\eeq

\subsubsection{Lorenz force}

We can derive the expression of the Lorentz force acting on the TM from the usual definition of
the force itself:
\beq
F=\left| q\vc v \wedge \vc B\right|\,,
\eeq
where $q$ represents the charge, $\vec v$ the velocity of the particle or the body crossing the magnetic field $\vec B$. This force
depends on the reference given for the velocity; the SC shields outer EM disturbances and only locally generated ones count.
Moreover we don't know precisely which will be the average velocity
inside the SC. Given this situation,
we get $q\to q_{e}q_{0}$ and $v\to \alpha_{\text{sh}} v_{\text{orbit}}$, representing a very small fraction of the mean orbital velocity of the SC.
The magnetic field is the
average interplanetary field $B_{\xi}$.
The result per unit mass may be given in spectral form as follows:
\beq
S_{a,\text{Lz}}^{\nicefrac{1}{2}} = \frac{ q_{e}q_{0} \sqrt{S_{B_{\xi}}} v_{\text{orbit}} \alpha_{\text{sh}}}{m}
\eeq

\subsubsection{Total magnetics interplanetary noise}
As before:
\begin{shadefundtheory}
\beq
S^{\nicefrac{1}{2}}_{a,\text{magnIP}} = \sqrt{2} \left(S^{\nicefrac{1}{2}}_{a,\text{Bi}}+S^{\nicefrac{1}{2}}_{a,\text{Lz}}\right)\,,
\eeq
\end{shadefundtheory}
table \ref{tab:sumextmagnetic} displays the sub-contributions and the total.

\begin{table}
\begin{center}
\begin{shademinornumber}
\begin{tabular}{r|l|l}
Description & Name & Value $\accPSDunit$\\
\hline\rule{0pt}{0.4cm}\noindent
Magnetic field fluctuations &  $S^{\nicefrac{1}{2}}_{a,\text{Bi}}$  & $2.29\times 10^{-16}$\\ 
Lorenz force &  $S^{\nicefrac{1}{2}}_{a,\text{Lz}}$  & $7.36\times 10^{-19}$\\
\hline\rule{0pt}{0.4cm}\noindent
Magnetics Interplanetary &  $S^{\nicefrac{1}{2}}_{a,\text{magnIP}}$  & $3.25\times 10^{-16}$\\ 
\end{tabular}
\end{shademinornumber}
\end{center}
\caption{External magnetic field effects, summary.}
\label{tab:sumextmagnetic}
\end{table}

\subsection{Charging  and voltage effects}

\subsubsection{Random charge}
We can assume a Poisson model for the random charge events hitting or depositing on the surface of the TM.
Hence the total charge collected at time $t$ looks like:
\beq
q(t)=\sum_{j}q_{e}\Theta(t-t_{j})\,,
\eeq
where the set $\left\{t_{j}\right\},\,j=0..\infty$ represents the ensemble of charge arrival times. From standard stochastic
processes techniques we acknowledge the distribution has only one parameter $\lambda$, such that:
\beq
\begin{split}
\langle q(t) \rangle &= \lambda t\,, \\
\langle q(t')q(t) \rangle &= \lambda t_{<} (\lambda t_{>}+1)\,,\\
\langle q(t')q(t) \rangle - \langle q(t') \rangle \langle q(t)\rangle &= \lambda t_{<}\,,
\end{split}
\eeq
where in the second expression and in the third, representing the connected auto-correlation $C(t',t)$,
we took $t_{<}=\min(t',t)$ and $t_{>}=\max(t',t)$. We get then, by definition of standard deviation:
\beq
C(t,t)=\lambda t = \sigma^{2}_{q}\,,
\eeq
and the random charging process shows then fluctuations of order $\sqrt{\lambda}$, where $\lambda$
expresses the charge rate. The PSD of the process is thus:
\beq
S_{q}^{\nicefrac{1}{2}} = \frac{\sqrt{2 \lambda}q_{e}}{\omega}\,,
\eeq


while in terms of the electrostatics developed at the beginning of the chapter, we can
take \eqref{eq:vstraydcbias} for a single electrode in presence of stray voltages, to get in spectral form:
\beq
S_{F}^{\nicefrac{1}{2}} \simeq \frac{\partial C}{\partial x} V_{\text{stray}} \frac{S_{q}^{\nicefrac{1}{2}}}{C}
= \frac{V_{\text{stray}}}{d_{x}} S_{q}^{\nicefrac{1}{2}}\,.
\eeq
by easy substitution we'd write:
\beq
S_{a,q}^{\nicefrac{1}{2}} = \frac{V_{\text{stray}} \sqrt{2 \lambda}}{m \omega d_{x}}\,,
\label{eq:chargeeffect}
\eeq
whose value can be retrieved in table \ref{tab:sumcharging}, for $\omega=2\pi \times 1\,\unit{mHz}$.

\subsubsection{Other voltage fluctuation in the measurement bandwidth}

We collect here contributions coming from potentially unknown or un-modelled contributions of voltage nature.
$S_{V_{\text{ib}}}$ describes a generic vibrational PSD oscillating in the MBW:
\beq
S_{a,\text{vs}}^{\nicefrac{1}{2}} = \frac{\sqrt{2} A V_{\text{stray}}\epsilon_o \sqrt{S_{V_{\text{ib}}}}}{m d_{x}^2}
\eeq

\subsubsection{Summary of charge and voltage noise}

Random charging effects and voltage fluctuations in the MBW can be considered stochastic uncorrelated processes, hence:
\begin{shadefundtheory}
\beq
S_{a,\text{charge}}^{\nicefrac{1}{2}}=\sqrt{2}\left(S_{a,q}^{\nicefrac{1}{2}}+S_{a,\text{vs}}^{\nicefrac{1}{2}}\right)\,.
\eeq
\end{shadefundtheory}
\noindent Read this total and the two mentioned contributions in table \ref{tab:sumcharging}.

\begin{table}
\begin{center}
\begin{shademinornumber}
\begin{tabular}{r|l|l}
Description & Name & Value $\accPSDunit$\\
\hline\rule{0pt}{0.4cm}\noindent
Random charge &  $S_{a,q}^{\nicefrac{1}{2}}$  & $2.81\times 10^{-16}$\\ 
Other voltage fluctuation in MBW &  $S_{a,\text{vs}}^{\nicefrac{1}{2}}$  & $2.53\times 10^{-15}$\\ 
\hline\rule{0pt}{0.4cm}\noindent
Charging and voltage &  $S_{a,\text{charge}}^{\nicefrac{1}{2}}$  & $3.61\times 10^{-15}$
\end{tabular}
\end{shademinornumber}
\end{center}
\caption{Charging and voltage effects, summary.}
\label{tab:sumcharging}
\end{table}

\subsection{Cross-talk}
\label{sec:crosstalk}

As we sketched during the construction of the LTP dynamics equations, cross-talk is a phenomenon
arising from extensivity of the TMs and sensing, controlling, hosting (ISs, SC) structures. The effect of
uncertainties always at play in any fundamental physics experiment - whose r{\^o}le is diminished
in gedanken-experiment assuming point-like shapes for sources and proof-masses -
gets magnified and self-coupled by the action of the fundamental mechanisms we listed already:
\begin{enumerate}
\item noise in sensing gets naturally multiplied by the control strategy matrix, thus a first effect is the
creation of unwanted forces and cross-terms potentially along every direction (i.e. the pure control
pushes on e.g. direction $\hat x$ but the effect is felt also along e.g. $\hat y$);
\item the solar wind push on the SC and the consequent compensating actuation on one of the TMs (in both control modes)
may be skewed-directed by virtue of different orientations of the IS - rigid with the SC - and the TM.
Effective rotational arms get created, thus creating unwanted DC forces at play over the TM
itself;
\item uncertainty in the knowledge of the value of stiffness create underestimation of parasitic spring-like
coupling leading to under(over)sized gain factors in control filters.
\end{enumerate}

To simplify the situation in a funny way, if the reader won't be bored or upset, let's think the SC as an elephant,
the two TMs as its eardrums (tympanic membranes), and the cross-talk effects as a group of monkeys rattling
on the elephant's back.
The presence of the monkeys as individuals is of no bother to the elephant, but what is they'll start humming
into its ears from both sides, grouping together on a single side of its back, or bouncing rhythmically? Well, the
coherent effects we listed are examples of coupling of the readout, undesired DC side torque, unwanted stiffness.
Maybe the example is a queer one, but we think it has some value.

Back to science, building a cross-talk model for TM1 requires a delicate ensemble of arguments at play \cite{LTPcrosstalk}.

We may split the contributions into those coming from the coupling to the SC and those coming from the
coupling to TM2. Of course TM2 will be coupled to the SC as well, but this effect can be reduced by the control loop
and what's left can be put in budget with the TM1-TM2 and TM1-SC couplings. Moreover, a third part of cross-talk
will come from the dynamical coupling of TM1 with itself.

The generic form of the LTP dynamics equations can be inspected in \eqref{eq:fullltpdynmatrix}. That
form can be complicated now to include cross-talk effects as follows:
\beq
(M s^{2} + K) \cdot \vc x = \vc f\,,
\label{eq:fulldyn}
\eeq
with
\beq
\begin{split}
\vc f &= I_{\text{aff}}\cdot\left(\vc f_{0}-\hat\Lambda\cdot \vc o\right) + \vc f_{n}\,,\\
\vc o &= \Omega\cdot \vc x + \vc o_{0} + \vc o_{n}\,,
\label{eq:ancillfo}
\end{split}
\eeq
where the matrix $I_{\text{aff}}$ plays the r\^{o}le of the perfect orthogonality matrix of reference systems,
being the identity in absence of any deformation or skewness; $\vc f_{0}$ is the vector
of intrinsic non-zero forces acting on the TMs (non-actuation, out-of-loop forces). The $K$ matrix embodies
stiffness and the control loop is embedded in the matrix $\hat\Lambda$ (cfr. \eqref{eq:ccone}, \eqref{eq:cctwo}, \eqref{eq:ccsc}
and \eqref{eq:tensorgroupmat}).
The readout signals are hereby summarised in the $\vc o$ vector together with the readout choice matrix $\Omega$.
The compound effect of these produces the feed-back force $\vc f$ on the r.h.s. of \eqref{eq:fulldyn}.

Obviously - we spent quite some time to deal with it - readout carries some offsets,
represented in $\vc o_{0}$, and some noise specific of the sensors, $\vc o_{n}$.

The former set of linear equations could be solved in the readout vector $\vc o$ and the unperturbed
- but for noise - result would correspond to the set of signal equations we found already. Notice the SC variables
can always be eliminated and the reduced set of equations brought to $12$ in number. We
now proceed to complicate the scenario as follows:
\begin{description}
\item[the stiffness] matrix $K$ expresses intrinsic rigidity and spring coupling of the TMs with the SC. As we saw
it's block-wise and got $0$'s on the main diagonal. A number of non-ideality coefficient can thus be introduced per DOF
to mimic our ignorance and uncertainty on the stiffness:
\begin{shademinornumber}
\beq
K \to K +{\delta}K\,,
\eeq
\end{shademinornumber}
\noindent and one can easily count that the total number of non-ideal stiffness coefficients amounts to
$2\times(6\times 6 - 6)=60$. An example may be given for the TM1 sector of the matrix \eqref{eq:stiffnessmatrix}.

\begin{figure*}
\begin{center}
\begin{sideways}
\begin{minipage}{1.5\linewidth}
\beq
{\delta}K_{i=1..6,j=1..6}=
\left[
\begin{array}{cccccc}
 0 & m {\delta}\omega_{x_1,y_1}^2 \omega_{y_1,y_1}^2 & m {\delta}\omega_{x_1,z_1}^2 \omega_{z_1,z_1}^2 &
 L m {\delta}\omega_{x_1,\theta_1}^2 \omega_{x_1,x_1}^2 & L m {\delta}\omega_{x_1,\eta_1}^2 \omega_{x_1,x_1}^2 &
 L m {\delta}\omega_{x_1,\phi_1}^2 \omega_{x_1,x_1}^2 \\
 m {\delta}\omega_{y_1,x_1}^2 \omega_{x_1,x_1}^2 & 0 & m {\delta}\omega_{y_1,z_1}^2 \omega_{z_1,z_1}^2 &
 L m {\delta}\omega_{y_1,\theta_1}^2 \omega_{y_1,y_1}^2 & L m {\delta}\omega_{y_1,\eta_1}^2 \omega_{y_1,y_1}^2 &
 L m {\delta}\omega_{y_1,\phi_1}^2 \omega_{y_1,y_1}^2 \\
 m {\delta}\omega_{z_1,x_1}^2 \omega_{x_1,x_1}^2 & m {\delta}\omega_{z_1,y_1}^2 \omega_{y_1,y_1}^2 & 0 &
 L m {\delta}\omega_{z_1,\theta_1}^2 \omega_{z_1,z_1}^2 & L m {\delta}\omega_{z_1,\eta_1}^2 \omega_{z_1,z_1}^2 &
 L m {\delta}\omega_{z_1,\phi_1}^2 \omega_{z_1,z_1}^2 \\
 \frac{I {\delta}\omega_{\theta_1,x_1}^2 \omega_{\theta_1,\theta_1}^2}{L} &
 \frac{I {\delta}\omega_{\theta_1,y_1}^2 \omega_{\theta_1,\theta_1}^2}{L} &
 \frac{I {\delta}\omega_{\theta_1,z_1}^2 \omega_{\theta_1,\theta_1}^2}{L} &
 0 & I {\delta}\omega_{\theta_1,\eta_1}^2 \omega_{\eta_1,\eta_1}^2 & I {\delta}\omega_{\theta_1,\phi_1}^2 \omega_{\phi_1,\phi_1}^2\\
 \frac{I {\delta}\omega_{\eta_1,x_1}^2 \omega_{\eta_1,\eta_1}^2}{L} &
 \frac{I {\delta}\omega_{\eta_1,y_1}^2 \omega_{\eta_1,\eta_1}^2}{L} &
 \frac{I {\delta}\omega_{\eta_1,z_1}^2 \omega_{\eta_1,\eta_1}^2}{L} &
 I {\delta}\omega_{\eta_1,\theta_1}^2 \omega_{\theta_1,\theta_1}^2 & 0 & I {\delta}\omega_{\eta_1,\phi_1}^2 \omega_{\phi_1,\phi_1}^2 \\
 \frac{I {\delta}\omega_{\phi_1,x_1}^2 \omega_{\phi_1,\phi_1}^2}{L} &
 \frac{I {\delta}\omega_{\phi_1,y_1}^2 \omega_{\phi_1,\phi_1}^2}{L} &
 \frac{I {\delta}\omega_{\phi_1,z_1}^2 \omega_{\phi_1,\phi_1}^2}{L} &
 I {\delta}\omega_{\phi_1,\theta_1}^2 \omega_{\theta_1,\theta_1}^2 & I {\delta}\omega_{\phi_1,\eta_1}^2 \omega_{\eta_1,\eta_1}^2 & 0
\end{array}
\right]\,.
\label{eq:stiffnessmatrix}
\eeq
\end{minipage}
\end{sideways}
\end{center}
\end{figure*}

Each correction is the result of the mass or moment of inertia component (here respectively $m$ or $I$, assuming all principal moment of inertia
to be the same for a cubic TM), times the stiffness factor proper of the DOF scaled by a factor $L$ or $\nicefrac{1}{L}$
in case of linear-angular or angular-linear coupling. Therefore, each ${\delta}\omega^{2}$ correction is rendered dimensionless.

Notice only coefficients coupling each $\hat x_{i}\,i=1,2$ direction to any other non-$\hat x_{i}$ for the same TM will be of interest
in our discussion of perturbations affecting each $x_{i}$ variable only. Thus
the total number of relevant corrections is $2\times 6 - 2 = 10$. We estimated the relative uncertainty to be of order $5\times 10^{-3}$,
values are in table \ref{tab:deltaK}.

\item [Signals] are embedded in the $\Omega$ matrix which - apart from the subtraction factor for IFO($x_{2}-x_{1}$)
is essentially diagonal. Non ideality in the readout can be introduced as ${\delta}\Omega_{i,j}$ coefficients so that the
$\Omega$ operator extends from dimensions $12\times 12$ to $18\times 12$, to take care of the whole signals. Namely:
\begin{shademinornumber}
\beq
\Omega\to \Omega+{\delta}\Omega\,.
\eeq
\end{shademinornumber}
\noindent The TM1 sector
of the perturbation matrix is given in \eqref{eq:fulldeltas}, where we renamed IFO to I for space reasons. The $\delta
S$ matrix carries a total of $108$ coefficients and follows the same scaling rules as for multiplication by $L$ or $\nicefrac{1}{L}$
as ${\delta}K$.

\beq
\begin{split}
{\delta}\Omega&_{i=1..6,j=1..6}=\\
&\left[
\begin{array}{cccccc}
 {\delta}\Omega_{I,{\Delta}x,\bar{x}} & {\delta}\Omega_{I,{\Delta}x,y_1} & {\delta}\Omega_{I,{\Delta}x,z_1} &
 L {\delta}\Omega_{I,{\Delta}x,\theta_1} & L {\delta}\Omega_{I,{\Delta}x,\eta_1} & L {\delta}\Omega_{I,{\Delta}x,\phi_1} \\
 {\delta}\Omega_{I,\eta_2,x_1} & {\delta}\Omega_{I,\eta_2,y_1} & {\delta}\Omega_{I,\eta_2,z_1} &
 {\delta}\Omega_{I,\eta_2,\theta_1} & {\delta}\Omega_{I,\eta_2,\eta_1} & {\delta}\Omega_{I,\eta_2,\phi_1} \\
 {\delta}\Omega_{I,\phi_2,x_1} & {\delta}\Omega_{I,\phi_2,y_1} & {\delta}\Omega_{I,\phi_2,z_1} &
 {\delta}\Omega_{I,\phi_2,\theta_1} & {\delta}\Omega_{I,\phi_2,\eta_1} & {\delta}\Omega_{I,\phi_2,\phi_1} \\
 0 & {\delta}\Omega_{I,x_1,y_1} & {\delta}\Omega_{I,x_1,z_1} & L {\delta}\Omega_{I,x_1,\theta_1} &
 L {\delta}\Omega_{I,x_1,\eta_1} & L {\delta}\Omega_{I,x_1,\phi_1} \\
 \frac{{\delta}\Omega_{I,\eta_1,x_1}}{L} & \frac{{\delta}\Omega_{I,\eta_1,y_1}}{L} & \frac{{\delta}\Omega_{I,\eta_1,z_1}}{L} &
   {\delta}\Omega_{I,\eta_1,\theta_1} & 0 & {\delta}\Omega_{I,\eta_1,\phi_1} \\
 \frac{{\delta}\Omega_{I,\phi_1,x_1}}{L} & \frac{{\delta}\Omega_{I,\phi_1,y_1}}{L} & \frac{{\delta}\Omega_{I,\phi_1,z_1}}{L} &
   {\delta}\Omega_{I,\phi_1,\theta_1} & {\delta}\Omega_{I,\phi_1,\eta_1} & 0
\end{array}
\right]\,.
\label{eq:fulldeltas}
\end{split}
\eeq

\item[Affinity] of the actuation system must be taken into account. An affinity matrix is then defined for each TM and
for the SC, and the three combined tensor-wise into a resulting perturbation affinity matrix which can then be reduced
as usual to an effective $12\times 12$ whose form is nevertheless much less transparent and won't be printed here. With reference
to \eqref{eq:ancillfo} the following substitution is made:
\begin{shademinornumber}
\beq
I_{\text{aff}}\to I + {\delta}A
\eeq
\end{shademinornumber}
\noindent With $I$ the identity matrix. The ${\delta}A$ matrix introduces $3\times (6 \times 6) = 108$ ${\delta}A_{i,j}$ coefficients, which may look like
in \eqref{eq:deltaafftm1} for the TM1 sector.

\beq
\begin{split}
{\delta}A&_{i=1..6,j=1..6}=\\
&\left[
\begin{array}{cccccc}
 0 & {\delta}A_{x_1,y_1} & {\delta}A_{x_1,z_1} & \frac{{\delta}A_{x_1,\theta_1}}{L} &
   \frac{{\delta}A_{x_1,\eta_1}}{L} & \frac{{\delta}A_{x_1,\phi_1}}{L} \\
 {\delta}A_{y_1,x_1} & 0 & {\delta}A_{y_1,z_1} & \frac{{\delta}A_{y_1,\theta_1}}{L} &
   \frac{{\delta}A_{y_1,\eta_1}}{L} & \frac{{\delta}A_{y_1,\phi_1}}{L} \\
 {\delta}A_{z_1,x_1} & {\delta}A_{z_1,y_1} & 0 & \frac{{\delta}A_{z_1,\theta_1}}{L} &
   \frac{{\delta}A_{z_1,\eta_1}}{L} & \frac{{\delta}A_{z_1,\phi_1}}{L} \\
 L {\delta}A_{\theta_1,x_1} & L {\delta}A_{\theta_1,y_1} & L {\delta}A_{\theta_1,z_1} & 0 &
   {\delta}A_{\theta_1,\eta_1} & {\delta}A_{\theta_1,\phi_1} \\
 L {\delta}A_{\eta_1,x_1} & L {\delta}A_{\eta_1,y_1} & L {\delta}A_{\eta_1,z_1} &
 {\delta}A_{\eta_1,\theta_1} & 0 & {\delta}A_{\eta_1,\phi_1} \\
 L {\delta}A_{\phi_1,x_1} & L {\delta}A_{\phi_1,y_1} & L {\delta}A_{\phi_1,z_1} &
 {\delta}A_{\phi_1,\theta_1} & {\delta}A_{\phi_1,\eta_1} & 0
\end{array}
\right]\,.
\label{eq:deltaafftm1}
\end{split}
\eeq
Notice here rules of effective arms multiplication are inverted because the matrix will
multiply forces and torques already scaled by $L$ or $\nicefrac{1}{L}$. Actuation uncertainty factors
are not requirements, the values which can be found in \ref{tab:affcoeff} are thus confidence bounds.

\item[DC] additional cross-talk stiffness is called in place by the control strategy when
responding to rotational motion on the SC induced by solar wind pull, orbiting revolution or other sources. Each TM
brings the electric field along when the SC rotates because the field is always orthogonal to conductors surfaces, hence
a set of $g_{\text{DC},i}$ ($\dot\Omega_{\text{DC},i}$ for torques)
parameters will couple to $\vc x$ in skew-symmetric form (the control is
always exerted
by electrodes counter-acting rotor-like to actuate rotation and pull-like to displace) and block-wise in matrix form. The
effects can be grouped in a ${\delta}\Lambda_{\text{DC}}$ matrix so that:
\begin{shademinornumber}
\beq
\vc f_{n} \to \vc f_{n} + {\delta}\Lambda_{\text{DC}}\cdot \vc x\,,
\eeq
\end{shademinornumber}
\noindent for TM1 the matrix would look like:
\beq
{\delta}\Lambda_{\text{DC},i=1..6,j=1..6}=\left[
\begin{array}{cccccc}
0 & 0 & 0 & 0 & m g_{\text{DC},z_1} & -m g_{\text{DC},y_1} \\
 0 & 0 & 0 & -m g_{\text{DC},z_1} & 0 & m g_{\text{DC},x_1} \\
 0 & 0 & 0 & m g_{\text{DC},y_1} & -m g_{\text{DC},x_1} & 0 \\
 0 & 0 & 0 & 0 & I g_{\text{DC},\phi_1} & -I g_{\text{DC},\eta_1} \\
 0 & 0 & 0 & -I g_{\text{DC},\phi_1} & 0 & I g_{\text{DC},\theta_1} \\
 0 & 0 & 0 & I g_{\text{DC},\eta_1} & -I g_{\text{DC},\theta_1} & 0
\end{array}
\right]\,.
\eeq
To count them, notice we'll have $3$ block matrices like the former, each having $2$ relevant blocks,
skew-symmetric (therefore ranking $\nicefrac{N\times (N-1)}{2}$ generators per sub-block with $N=3$),
thus amounting to a number of $18$. Though representing a stiffness matrix, no need for scaling by length factors
is deemed, since only the sector of the matrix multiplying angular variables is non-zero; the DC force or torque
elements in table \ref{tab:CDC} have been written per unit radian for consistency.

\item[Coordinate variations] can be introduced for by assuming the first order expansion:
\begin{shademinornumber}
\beq
\vc x \to \vc x_{0} + \vc{{\delta}x}\,.
\eeq
\end{shademinornumber}
\end{description}

With the former substitutions in place, the set of equations \eqref{eq:fulldyn} and \eqref{eq:ancillfo}
become:
\begin{align}
(M s^{2} + K+{\delta}K) \cdot (\vc x_{0} + \vc{{\delta}x}) &= (I+{\delta}A) \cdot \left(\vc f_{0}-\hat\Lambda \cdot \vc o\right)
  + \vc f_{n} + {\delta}\Lambda_{\text{DC}}\cdot (\vc x_{0} + \vc{{\delta}x})\,,\\
\vc o &= \left(\Omega+{\delta}\Omega\right)\cdot (\vc x_{0} + \vc{{\delta}x}) + \vc o_{0} + \vc o_{n}\,, \label{eq:osignals}
\end{align}
and further expansion to first order in the computations gives:
\beq
\begin{split}
\left(M s^{2}+K+{\delta}K\right)& \cdot \vc x_{0} + \left(M s^{2}+K\right) \cdot \vc{{\delta}x} =
\vc f_{0}-\hat\Lambda \cdot \Omega \cdot \left(\vc x_{0}+\vc{{\delta}x}\right)+\hat\Lambda\cdot {\delta}\Omega\cdot \vc x_{0}+\\
&-\hat\Lambda\cdot\left(\vc o_{0}+\vc o_{n}\right)+\vc f_{n}+{\delta}\Lambda_{\text{DC}}\cdot \vc x_{0}\\
&+{\delta}A\cdot \left(\vc f_{0}-\hat\Lambda\cdot \Omega\cdot \vc x_{0}-\hat\Lambda\cdot \left(\vc o_{0}+\vc o_{n}\right)\right)
\end{split}
\eeq
to order $0$ in the deformations the unperturbed dynamics can be read out as:
\beq
D_{0} \cdot \vc x_{0} = \vc f_0+\vc f_n-\hat\Lambda\cdot \left(\vc o_0+\vc o_n\right)\,,
\label{eq:dynx0}
\eeq
where we defined the unperturbed dynamical matrix $D_{0}\equiv M s^2+K+\hat\Lambda\cdot \Omega$. To first
order in the deformations we conversely read the evolution equation in the coordinates variation:
\beq
\begin{split}
D_{0}\cdot \vc{{\delta}x} =& \left({\delta}K+\hat\Lambda\cdot {\delta}\Omega -{\delta}\Lambda_{\text{DC}}+{\delta}A\cdot \hat\Lambda\cdot \Omega\right)\cdot \vc
x_{0}\\
&+{\delta}A \cdot \left(\vc f_0 - \hat\Lambda\cdot \left(\vc o_0+\vc o_n\right)\right) = \\
&\left({\delta}K+\hat\Lambda\cdot {\delta}\Omega -{\delta}\Lambda_{\text{DC}}+{\delta}A\cdot \hat\Lambda\cdot \Omega\right)\cdot \vc
x_{0}+{\delta}A \cdot D_{0}\cdot \vc x_{0} - {\delta}A\cdot \vc f_{n} = \\
&\left({\delta}K+\hat\Lambda\cdot {\delta}\Omega -{\delta}\Lambda_{\text{DC}} - {\delta}A\cdot \left(M s^{2} - K\right)\right)\cdot \vc
x_{0} - {\delta}A\cdot \vc f_{n}\,,
\label{eq:deltaxevol}
\end{split}
\eeq
where in the last-but-one passage we added and subtracted a term ${\delta}A\cdot \vc f_{n}$ and used \eqref{eq:dynx0}.
The first order expression for the
readout signals $\vc o$ in \eqref{eq:osignals} reads:
\beq
\vc o = \Omega\cdot \left(\vc x_{0}+\vc{{\delta}x}\right) + {\delta}\Omega\cdot \vc x_{0} + \vc o_{0} + \vc o_{n}\,,
\eeq
this can be
grouped into three contributions as:
\beq
\vc o = \vc o_{00} + \vc{{\delta}o} + \vc{{\delta}o}_{n}\,.
\eeq
By means of \eqref{eq:deltaxevol} and \eqref{eq:dynx0} we proceed to elucidate term by term: the $0$-th order is of
course independent on perturbations:
\beq
\begin{split}
\vc o_{00}&=\vc o_0+\vc o_n+\Omega \cdot \vc x_{0} = \\
 &=\vc o_0+\vc o_n+\Omega \cdot D_{0}^{-1}\left(\vc f_0+\vc f_n-\hat\Lambda \left(\vc o_0+\vc o_n\right)\right)\,,
\end{split}
\eeq
conversely, for the first order corrections:
\beq
\begin{split}
\vc{{\delta}o} + \vc{{\delta}o}_{n} &= \Omega\cdot \vc{{\delta}x} + {\delta}\Omega\cdot \vc x_{0} = \\
&=\Delta \cdot \vc x_{0} - \Omega\cdot D_{0}^{-1}\cdot {\delta}A\cdot \vc f_{n}\,,
\end{split}
\eeq
where
\beq
\Delta = {\delta}\Omega - \Omega\cdot D_{0}^{-1}\cdot \left({\delta}K+\hat\Lambda\cdot {\delta}\Omega -{\delta}\Lambda_{\text{DC}} - {\delta}A\cdot
\left(M s^{2} - K\right)\right)\,.
\eeq
By substituting the expression for $\vc x_{0}$ we can split the cross-contributions into the $0$-point force one:
\beq
\vc{{\delta}o}=\Delta  \cdot D_{0}^{-1} \left(\vc f_0-\hat\Lambda \cdot \vc o_0\right)\,,
\label{eq:deltao}
\eeq
and the omnipresent noise:
\beq
\vc{{\delta}o}_{n}=\Delta  \cdot D_{0}^{-1}\left(\vc f_n-\hat\Lambda \cdot \vc o_n\right)- \Omega\cdot D_{0}^{-1}{\delta}A\cdot \vc f_n\,.
\label{eq:deltaonoise}
\eeq

The comparison of \eqref{eq:deltaonoise} and \eqref{eq:deltao} shows that by applying the appropriate
stimuli $\vc f_0-\hat\Lambda \cdot \vc o_0$ one 
could measure the matrix that converts force and signal noise into cross-talk noise except for the 
extra term $\Omega\cdot D_{0}^{-1}{\delta}A\cdot \vc f_n$ in \eqref{eq:deltaonoise}.
One of the desired measurement approaches is to be able to measure some disturbance and the 
relative transfer function from the disturbance to the acceleration noise. One can then predict the 
contribution of this disturbance to the overall acceleration noise by multiplying the disturbance by 
the transfer function. If this is made in the time domain, via the appropriate convolution, the 
predicted noise can be subtracted from the measured acceleration data with the aim of suppressing 
the noise source.

This approach can hardly be followed with cross-talk. The reason for this is twofold:
\begin{enumerate}
\item the matrix that converts coordinates to signals is not invertible. The measurable coordinates are 
just the $12$ relative ones while the disturbances are $18$ (forces and torques on three bodies),
Notice that also the available signals are $18$, but this is just by chance: the interferometer signals 
are redundant, from a dynamical point of view, relative to the GRS ones;
\item cross-talk is due to forces acting on the TMs and the SC: the signal 
measure displacement plus noise and cannot be inverted back to force.
\end{enumerate}
It is therefore transparent that for a large subsets of cross-talk phenomena the way out is calibrating the
signal on a dynamical perspective and perform a careful budget analysis to shoot down the largest contributions during
designing and mounting phase.

For our purposes, we need to evaluate the worst-case scenario for the cross-talk computation, i.e. estimate
$o_{j} - o_{00,j} = {\delta}o_{j} + {\delta}o_{n,j}$ for $j=\text{IFO}(x_{2}-x_{1})$. Notice
we may discard the contribution from ${\delta}o$, since we may think to reduce to $0$ every intrinsic force
and consider it as pure noise; in fact, the first part of \eqref{eq:deltaonoise} and \eqref{eq:deltao} are
form invariant. The task is then specialise to \eqref{eq:deltaonoise} for the main science channel and
put explicit values of disturbances, deviations and transfer functions to get a figure of cross-talk PSD:
\begin{shadefundtheory}
\beq
\begin{split}
S_\text{IFO, cross-talk, n}({\Delta}x) =& \sum_{j}
  S\left[\Delta \cdot D_{0}^{-1}- \Omega\cdot D_{0}^{-1}{\delta}A\right]_{{\Delta}x, j}S_{g_{\text{n}},j}+\\
  & -\sum_{j}S\left[\Delta \cdot D_{0}^{-1} \hat\Lambda \right]_{{\Delta}x, j} S_{\text{n},o_{j}}\,,
\end{split}
\eeq
\end{shadefundtheory}
\noindent where we switched to (squared) PSDs for the linear operators in square brackets and sum noise
contributions quadratically. We took the liberty of generalising $S_{g_{\text{n}},j}$ which contains also
angular contributions (cfr. table \ref{tab:residualfcs})

This tedious work has been performed with the help of {\sf Mathematica}\textregistered; the outcome is practically unpublishable
for the lengths of the propagators, but the relevant ${\delta}\Omega$, ${\delta}A$, ${\delta}K$ and ${\delta}\Lambda_{\text{DC}}$
components have been summarised across tables \ref{tab:deltaS}, \ref{tab:affcoeff}, \ref{tab:deltaK}, \ref{tab:CDC}.
The noise PSDs for each signal $S_{\text{n},O_{j}}$ may be read in table \ref{tab:noisechannels} and the residual forces(torques)
PSD contributions per unit mass(moment of inertia) $S_{g_{\text{n}},j}$ in table \ref{tab:residualfcs}.

Finally, the cross-talk contribution to the acceleration noise PSD can be computed (at $\omega=2\pi\times f$ with $f=1\,\unit{mHz}$)
as:
\begin{shademinornumber}
\beq
S_{a,\text{cross-talk}}^{\nicefrac{1}{2}}= 6.12\times 10^{-15}\,\accPSDunit\,.
\eeq
\end{shademinornumber}

\begin{table}
\begin{center}
\begin{tabular}{r|l|l|l}
Name & Symbol & Value & Dimensions \\
\hline\rule{0pt}{0.4cm}\noindent
Signal IFO $x\to {\Delta}x$ &  ${\delta}\Omega_{\text{IFO},{\Delta}x,\bar{x}}$  & $10^{-3}$ & $1$\\ 
Signal IFO $y_1\to {\Delta}x$ &  ${\delta}\Omega_{\text{IFO},{\Delta}x,y_1}$  & $10^{-3}$ & $1$\\ 
Signal IFO $\phi_1\to {\Delta}x$ &  ${\delta}\Omega_{\text{IFO},{\Delta}x,\phi_1}$  & $10^{-3}$ & $1$\\ 
Signal IFO $y_2\to {\Delta}x$ &  ${\delta}\Omega_{\text{IFO},{\Delta}x,y_2}$  & $10^{-3}$ & $1$\\ 
Signal IFO $\phi_2\to {\Delta}x$ &  ${\delta}\Omega_{\text{IFO},{\Delta}x,\phi_2}$  & $10^{-3}$ & $1$\\ 
Signal IFO $z_1\to {\Delta}x$ &  ${\delta}\Omega_{\text{IFO},{\Delta}x,z_1}$  & $10^{-3}$ & $1$\\ 
Signal IFO $\eta_1\to {\Delta}x$ &  ${\delta}\Omega_{\text{IFO},{\Delta}x,\eta_1}$  & $10^{-3}$ & $1$\\ 
Signal IFO $z_2\to {\Delta}x$ &  ${\delta}\Omega_{\text{IFO},{\Delta}x,z_2}$  & $10^{-3}$ & $1$\\ 
Signal IFO $\eta_2\to {\Delta}x$ &  ${\delta}\Omega_{\text{IFO},{\Delta}x,\eta_2}$  & $10^{-3}$ & $1$\\ 
Signal IFO $\theta_1\to {\Delta}x$ &  ${\delta}\Omega_{\text{IFO},{\Delta}x,\theta_1}$  & $10^{-4}$ & $1$\\ 
Signal IFO $\theta_2\to {\Delta}x$ &  ${\delta}\Omega_{\text{IFO},{\Delta}x,\theta_2}$  & $10^{-4}$ & $1$\\ 
Signal GRS $y_1\to x_1$ &  ${\delta}\Omega_{\text{GRS},x_1,y_1}$  & $5.\times 10^{-3}$ & $1$\\ 
Signal GRS $\phi_1\to x_1$ &  ${\delta}\Omega_{\text{GRS},x_1,\phi_1}$  & $5.\times 10^{-3}$ & $1$\\ 
Signal GRS $z_1\to x_1$ &  ${\delta}\Omega_{\text{GRS},x_1,z_1}$  & $5.\times 10^{-3}$ & $1$\\ 
Signal GRS $\eta_1\to x_1$ &  ${\delta}\Omega_{\text{GRS},x_1,\eta_1}$  & $5.\times 10^{-3}$ & $1$\\ 
Signal GRS $\theta_1\to x_1$ &  ${\delta}\Omega_{\text{GRS},x_1,\theta_1}$  & $5.\times 10^{-3}$ & $1$\\ 
\end{tabular}
\end{center}
\caption{${\delta}\Omega$ relevant readout perturbations for cross-talk on IFO$(x_{2}-x_{1})$. Names have
been specialised to the proper readout identifier in main science mode.}
\label{tab:deltaS}
\end{table}

\begin{table}
\begin{center}
\begin{tabular}{r|l|l|l}
Name & Symbol & Value & Dimensions \\
\hline\rule{0pt}{0.4cm}\noindent
Actuation $\phi_1\to x_1$ &  ${{\delta}A}_{x_1,\phi_1}$  & $5.\times 10^{-3}$ & $1$\\ 
Actuation $\phi_2\to x_2$ &  ${{\delta}A}_{x_2,\phi_2}$  & $5.\times 10^{-3}$ & $1$\\ 
Actuation $y_\text{SC}\to x_\text{SC}$ &  ${{\delta}A}_{x_{\text{SC}},y_{\text{SC}}}$  & $5.\times 10^{-3}$ & $1$\\ 
Actuation $\phi_\text{SC}\to x_\text{SC}$ &  ${{\delta}A}_{x_{\text{SC}},\phi_{\text{SC}}}$  & $5.\times 10^{-3}$ & $1$\\ 
Actuation $\eta_1\to x_1$ &  ${{\delta}A}_{x_1,\eta_1}$  & $5.\times 10^{-3}$ & $1$\\ 
Actuation $\eta_2\to x_2$ &  ${{\delta}A}_{x_2,\eta_2}$  & $5.\times 10^{-3}$ & $1$\\ 
Actuation $z_\text{SC}\to x_\text{SC}$ &  ${{\delta}A}_{x_{\text{SC}},z_{\text{SC}}}$  & $5.\times 10^{-3}$ & $1$\\ 
Actuation $\eta_\text{SC}\to x_\text{SC}$ &  ${{\delta}A}_{x_{\text{SC}},\eta_{\text{SC}}}$  & $5.\times 10^{-3}$ & $1$\\ 
Actuation $\theta_2\to x_2$ &  ${{\delta}A}_{x_2,\theta_2}$  & $5.\times 10^{-3}$ & $1$\\ 
Actuation $\theta_\text{SC}\to x_\text{SC}$ &  ${{\delta}A}_{x_{\text{SC}},\theta_{\text{SC}}}$  & $5.\times 10^{-3}$ & $1$\\ 
\end{tabular}
\end{center}
\caption{${\delta}A$ relevant actuation affinity perturbations for cross-talk on IFO$(x_{2}-x_{1})$}
\label{tab:affcoeff}
\end{table}

\begin{table}
\begin{center}
\begin{tabular}{r|l|l|l}
Name & Symbol & Value & Dimensions \\
\hline\rule{0pt}{0.4cm}\noindent
Stiffness $y_1\to x_1$ &  ${\delta}\omega^2_{x_1,y_1}$  & $5.\times 10^{-3}$ & $1$\\ 
Stiffness $z_1\to x_1$ &  ${\delta}\omega^2_{x_1,z_1}$  & $5.\times 10^{-3}$ & $1$\\ 
Stiffness $\theta_1\to x_1$ &  ${\delta}\omega^2_{x_1,\theta_1}$  & $5.\times 10^{-3}$ & $1$\\ 
Stiffness $\eta_1\to x_1$ &  ${\delta}\omega^2_{x_1,\eta_1}$  & $5.\times 10^{-3}$ & $1$\\ 
Stiffness $\phi_1\to x_1$ &  ${\delta}\omega^2_{x_1,\phi_1}$  & $5.\times 10^{-3}$ & $1$\\  
Stiffness $y_2\to x_2$ &  ${\delta}\omega^2_{x_2,y_2}$  & $5.\times 10^{-3}$ & $1$\\ 
Stiffness $z_2\to x_2$ &  ${\delta}\omega^2_{x_2,z_2}$  & $5.\times 10^{-3}$ & $1$\\ 
Stiffness $\theta_2\to x_2$ &  ${\delta}\omega^2_{x_2,\theta_2}$  & $5.\times 10^{-3}$ & $1$\\
Stiffness $\eta_2\to x_2$ &  ${\delta}\omega^2_{x_2,\eta_2}$  & $5.\times 10^{-3}$ & $1$\\ 
Stiffness $\phi_2\to x_2$ &  ${\delta}\omega^2_{x_2,\phi_2}$  & $5.\times 10^{-3}$ & $1$\\
\end{tabular}
\end{center}
\caption{${\delta}K$ relevant stiffness perturbations for cross-talk on IFO$(x_{2}-x_{1})$. Values have
been rescaled properly according to linear or angular coupling and renamed to $\delta\omega_{i,j}^{2}$
in adherence to standard policy.}
\label{tab:deltaK}
\end{table}

\begin{table}
\begin{center}
\begin{tabular}{r|l|l|l}
Name & Symbol & Value & Dimensions \\
\hline\rule{0pt}{0.4cm}\noindent
DC force rotation TM1 $\hat y$ &  $g_{\text{DC},y_{1}}$  & $1.\times 10^{-8}$ & $\unitfrac{m}{rad\,s^2}$\\ 
DC force rotation TM2 $\hat y$ &  $g_{\text{DC},y_{2}}$  & $1.\times 10^{-8}$ & $\unitfrac{m}{rad\,s^2}$\\ 
DC force rotation \text{SC} $\hat Y$ &  $g_{\text{DC},Y}$  & $\nicefrac{1.5\times 10^{-6}}{m_{\text{SC}}}$ & $\unitfrac{m}{rad\,s^2}$\\ 
DC force rotation TM1 $\hat z$ &  $g_{\text{DC},z_{1}}$  & $1.\times 10^{-8}$ & $\unitfrac{m}{rad\,s^2}$\\ 
DC force rotation TM2 $\hat z$ &  $g_{\text{DC},z_{2}}$  & $1.\times 10^{-8}$ & $\unitfrac{m}{rad\,s^2}$\\ 
DC force rotation \text{SC} $\hat Z$ &  $g_{\text{DC},Z}$  & $\nicefrac{1.5\times 10^{-6}}{m_{\text{SC}}}$ & $\unitfrac{m}{rad\,s^2}$
\end{tabular}
\end{center}
\caption{${\delta}\Lambda_{\text{DC}}$ relevant control DC perturbations for cross-talk on IFO$(x_{2}-x_{1})$.
Values per unit mass, TMs or SC at occurrence.}
\label{tab:CDC}
\end{table}

\begin{table}
\begin{center}
\begin{tabular}{r|l|l|l}
Name & Symbol & Value & Dimensions \\
\hline\rule{0pt}{0.4cm}\noindent
Linear acceleration noise PSD TM1 $\hat x$ &  $S^{\nicefrac{1}{2}}_{g_{\text{n}},x_1}$  & $3.\times 10^{-14}$ & $\unitfrac{m}{s^{2}\,\sqrt{Hz}}$\\ 
Linear acceleration noise PSD TM1 $\hat y$ &  $S^{\nicefrac{1}{2}}_{g_{\text{n}},y_1}$  & $3.\times 10^{-14}$ & $\unitfrac{m}{s^{2}\,\sqrt{Hz}}$\\ 
Linear acceleration noise PSD TM1 $\hat z$ &  $S^{\nicefrac{1}{2}}_{g_{\text{n}},z_1}$  & $3.\times 10^{-13}$ & $\unitfrac{m}{s^{2}\,\sqrt{Hz}}$\\ 
Torsional acceleration noise PSD TM1 $\hat\theta$ &  $S^{\nicefrac{1}{2}}_{\dot\Omega_{\text{n}},\theta_1}$  & $2.\times 10^{-11}$ & $\unitfrac{1}{s^{2}\,\sqrt{Hz}}$\\ 
Torsional acceleration noise PSD TM1 $\hat\eta$ &  $S^{\nicefrac{1}{2}}_{\dot\Omega_{\text{n}},\eta_1}$  & $2.\times 10^{-11}$ & $\unitfrac{1}{s^{2}\,\sqrt{Hz}}$\\ 
Torsional acceleration noise PSD TM1 $\hat\phi$ &  $S^{\nicefrac{1}{2}}_{\dot\Omega_{\text{n}},\phi_1}$  & $4.\times 10^{-12}$ & $\unitfrac{1}{s^{2}\,\sqrt{Hz}}$\\ 
Linear acceleration noise PSD TM2 $\hat x$ &  $S^{\nicefrac{1}{2}}_{g_{\text{n}},x_2}$  & $3.\times 10^{-14}$ & $\unitfrac{m}{s^{2}\,\sqrt{Hz}}$\\ 
Linear acceleration noise PSD TM2 $\hat y$ &  $S^{\nicefrac{1}{2}}_{g_{\text{n}},y_2}$  & $3.\times 10^{-14}$ & $\unitfrac{m}{s^{2}\,\sqrt{Hz}}$\\ 
Linear acceleration noise PSD TM2 $\hat z$ &  $S^{\nicefrac{1}{2}}_{g_{\text{n}},z_2}$  & $3.\times 10^{-13}$ & $\unitfrac{m}{s^{2}\,\sqrt{Hz}}$\\ 
Torsional acceleration noise PSD TM2 $\hat\theta$ &  $S^{\nicefrac{1}{2}}_{\dot\Omega_{\text{n}},\theta_2}$  & $2.\times 10^{-11}$ & $\unitfrac{1}{s^{2}\,\sqrt{Hz}}$\\ 
Torsional acceleration noise PSD TM2 $\hat\eta$ &  $S^{\nicefrac{1}{2}}_{\dot\Omega_{\text{n}},\eta_2}$  & $2.\times 10^{-11}$ & $\unitfrac{1}{s^{2}\,\sqrt{Hz}}$\\ 
Torsional acceleration noise PSD TM2 $\hat\phi$ &  $S^{\nicefrac{1}{2}}_{\dot\Omega_{\text{n}},\phi_2}$  & $4.\times 10^{-12}$ & $\unitfrac{1}{s^{2}\,\sqrt{Hz}}$\\ 
Linear acceleration noise PSD \text{SC} $\hat X$ &  $S^{\nicefrac{1}{2}}_{g_{\text{n}},X}$  & $\nicefrac{\sqrt{S_{\text{SC}}}}{m_{\text{SC}}}$ & $\unitfrac{m}{s^{2}\,\sqrt{Hz}}$\\ 
Linear acceleration noise PSD \text{SC} $\hat Y$ &  $S^{\nicefrac{1}{2}}_{g_{\text{n}},Y}$  & $\nicefrac{\sqrt{S_{\text{SC}}}}{m_{\text{SC}}}$ & $\unitfrac{m}{s^{2}\,\sqrt{Hz}}$\\ 
Linear acceleration noise PSD \text{SC} $\hat Z$ &  $S^{\nicefrac{1}{2}}_{g_{\text{n}},Z}$  & $\nicefrac{\sqrt{S_{\text{SC}}}}{m_{\text{SC}}}$ & $\unitfrac{m}{s^{2}\,\sqrt{Hz}}$\\ 
Torsional acceleration noise PSD \text{SC} $\hat \Theta$ &  $S^{\nicefrac{1}{2}}_{\dot\Omega_{\text{n}},\Theta}$  & $\nicefrac{\sqrt{S_{\text{SC}}}}{m_{\text{SC}}R_{SC}^{2}}$ & $\unitfrac{1}{s^{2}\,\sqrt{Hz}}$\\ 
Torsional acceleration noise PSD \text{SC} $\hat H$ &  $S^{\nicefrac{1}{2}}_{\dot\Omega_{\text{n}},H}$  & $\nicefrac{\sqrt{S_{\text{SC}}}}{m_{\text{SC}}R_{SC}^{2}}$ & $\unitfrac{1}{s^{2}\,\sqrt{Hz}}$\\ 
Torsional acceleration noise PSD \text{SC} $\hat \Phi$ &  $S^{\nicefrac{1}{2}}_{\dot\Omega_{\text{n}},\Phi}$  & $\nicefrac{\sqrt{S_{\text{SC}}}}{m_{\text{SC}}R_{SC}^{2}}$ & $\unitfrac{1}{s^{2}\,\sqrt{Hz}}$\\ 
\end{tabular}
\end{center}
\caption{Residual forces and torques acting on TM1, TM2 and SC for cross-talk on IFO$(x_{2}-x_{1})$.
Values per unit mass or moment of inertia for TMs. We remind that in general $f_{i} = m_{i} g_{i}$ and $\gamma_{j}=I_{j} \dot\Omega_{j}$,
where no summation is implied and $I_{j}$ represent principal moments of inertia.}
\label{tab:residualfcs}
\end{table}

\begin{table}
\begin{center}
\begin{tabular}{r|l|l|l}
Name & Symbol & Value & Dimensions \\
\hline\rule{0pt}{0.4cm}\noindent
Readout noise PSD IFO ${\Delta}x$ &  $S^{\nicefrac{1}{2}}_{\text{n,IFO}\left({\Delta}x\right)}$  & $8.2\times 10^{-11}$ & $\unitfrac{m}{\sqrt{Hz}}$\\ 
Readout noise PSD IFO $\eta_1$ &  $S^{\nicefrac{1}{2}}_{\text{n,IFO}\left(\eta_1\right)}$  & $5.\times 10^{-8}$ & $\unitfrac{1}{\sqrt{Hz}}$\\ 
Readout noise PSD IFO $\phi_1$ &  $S^{\nicefrac{1}{2}}_{\text{n,IFO}\left(\phi_1\right)}$  & $5.\times 10^{-8}$ & $\unitfrac{1}{\sqrt{Hz}}$\\
Readout noise PSD IFO $\eta_2$ &  $S^{\nicefrac{1}{2}}_{\text{n,IFO}\left(\eta_2\right)}$  & $5.\times 10^{-8}$ & $\unitfrac{1}{\sqrt{Hz}}$\\ 
Readout noise PSD IFO $\phi_2$ &  $S^{\nicefrac{1}{2}}_{\text{n,IFO}\left(\phi_2\right)}$  & $5.\times 10^{-8}$ & $\unitfrac{1}{\sqrt{Hz}}$\\  
Readout noise PSD GRS $x_1$ &  $S^{\nicefrac{1}{2}}_{\text{n,GRS}\left(x_1\right)}$  & $2.\times 10^{-9}$ & $\unitfrac{m}{\sqrt{Hz}}$\\ 
Readout noise PSD GRS $y_1$ &  $S^{\nicefrac{1}{2}}_{\text{n,GRS}\left(y_1\right)}$  & $2.\times 10^{-9}$ & $\unitfrac{m}{\sqrt{Hz}}$\\ 
Readout noise PSD GRS $z_1$ &  $S^{\nicefrac{1}{2}}_{\text{n,GRS}\left(z_1\right)}$  & $3.\times 10^{-9}$ & $\unitfrac{m}{\sqrt{Hz}}$\\ 
Readout noise PSD GRS $\theta_1$ &  $S^{\nicefrac{1}{2}}_{\text{n,GRS}\left(\theta_1\right)}$  & $1.\times 10^{-7}$ & $\unitfrac{1}{\sqrt{Hz}}$\\ 
Readout noise PSD GRS $y_2$ &  $S^{\nicefrac{1}{2}}_{\text{n,GRS}\left(y_2\right)}$  & $2.\times 10^{-9}$ & $\unitfrac{m}{\sqrt{Hz}}$\\ 
Readout noise PSD GRS $z_2$ &  $S^{\nicefrac{1}{2}}_{\text{n,GRS}\left(z_2\right)}$  & $3.\times 10^{-9}$ & $\unitfrac{m}{\sqrt{Hz}}$\\ 
Readout noise PSD GRS $\theta_2$ &  $S^{\nicefrac{1}{2}}_{\text{n,GRS}\left(\theta_2\right)}$  & $1.\times 10^{-7}$ & $\unitfrac{1}{\sqrt{Hz}}$\\ 
\end{tabular}
\end{center}
\caption{Readout noise for relevant channels for cross-talk on IFO$(x_{2}-x_{1})$.}
\label{tab:noisechannels} 
\end{table}


\subsection{Other noise contributions}

\begin{table}
\begin{center}
\begin{tabular}{r|l|l|l}
Description & Name & Value & Dimensions\\\hline\rule{0pt}{0.4cm}\noindent
Laser power fluctuation &  $S_{W_{\text{laser}}}$  &
$\left(10^{-3}\times 10^{-4}\right)^{2}\left(2\pi\times \frac{10^{-3}\,\unit{Hz}}{\omega}\right)^{2}$ & $\unitfrac{J}{Hz}$\\ 
Self-gravity acceleration fluctuation &  $S_{\text{grav}}$  &
$\left(3.\times 10^{-15}\right)^{2}\left(2\pi\times \frac{10^{-3}\,\unit{Hz}}{\omega}\right)^{2}$ & $\unitfrac{m^2}{s^4\,Hz}$\\ 
Orbital velocity &  $v_{\text{orbit}}$  & $3.\times 10^{4}$ & $\unitfrac{m}{s}$\\
Gravitational noise coefficient & ${\delta}g_{\text{th}}$ & $1.\times 10^{-8}$ & $\unitfrac{m}{s^2}$
\end{tabular}
\end{center}
\caption{Miscellaneous constants}
\label{tab:miscconstants}
\end{table}

Miscellaneous effects would demand a more thorough treatment we lack time to deal with. Luckily, they are
easy to deduce from simple dynamical models and we can reduce to sketching their expressions together
with the computed values in table \ref{tab:summisc}. Effects are regarded as uncorrelated, and summed
quadratically:
\begin{shadefundtheory}
\beq
S_{a,\text{misc}} = 2\left( S_{a,\text{VAC}}+S_{a,\text{laser}}+S_{\text{grav}}\right)\,.
\eeq
\end{shadefundtheory}

\begin{description}
\item[AC voltage down-conversion]
\beq
S^{\nicefrac{1}{2}}_{a,\text{VAC}}=
\frac{5 \sqrt{2} C_{\text{sens}} \sqrt{S_{V_{\text{AC}}}} V_{\text{AC}}}{m d_{x}}\,,
\eeq
\item[Laser force noise]
\beq
S^{\nicefrac{1}{2}}_{a,\text{laser}}=\frac{2 \sqrt{S_{W_{\text{laser}}}}}{m c}\,,
\eeq
\item[Self-gravity noise]
\beq
S^{\nicefrac{1}{2}}_{a,\text{grav}} = \sqrt{S_{\text{grav}}}\,.
\eeq
\end{description}

\begin{table}
\begin{center}
\begin{shademinornumber}
\begin{tabular}{>{\raggedleft}m{6.0cm}|l|l}
Description & Name & Value $\accPSDunit$\\
\hline\rule{0pt}{0.4cm}\noindent
AC voltage down-conversion &  $S^{\nicefrac{1}{2}}_{a,\text{VAC}}$ & $1.04\times 10^{-16}$\\ 
Laser force noise &  $S^{\nicefrac{1}{2}}_{a,\text{laser}}$ & $3.4\times 10^{-16}$\\ 
Self-gravity noise &  $S^{\nicefrac{1}{2}}_{a,\text{grav}}$ & $3.0\times 10^{-15}$\\ 
\hline\rule{0pt}{0.4cm}\noindent
Miscellanea &  $S^{\nicefrac{1}{2}}_{a,\text{misc}}$  & $6.04\times 10^{-15}$
\end{tabular}
\end{shademinornumber}
\end{center}
\caption{Miscellaneous effects, summary.}
\label{tab:summisc}
\end{table}

\subsection{Measurement noise}

In spite of its being relegated at the end of the chapter, this section is of capital importance for the LTP mission. Inspection
of \eqref{eq:mainifosignalfornoise} - just for a change - and the comparison we built up in section \ref{sec:lisaltpdiff}
reveal immediately that there's noise specific of the way we measure on LTP which is immaterial in LISA.
Having two TMs on-board the same SC, to measure the laser interference pattern between the two is radically different
from having each ``sensing'' TM on-board different SC. We group these effect into the category ``measurement noise''.

\subsubsection{Actuation amplitude instability}
We can employ the results from the introductory section to get a figure of the instability in actuation. Suppose a residual acceleration along $\hat x$
would be given by $g_{\text{DC},x}$, hence from \eqref{eq:deltaelecforceoverforce} we'd get, per unit mass and in spectral
form the following acceleration noise spectrum:
\beq
S^{\nicefrac{1}{2}}_{a,\text{act}} = 2 g_{\text{DC},x} \sqrt{S_{\nicefrac{{\Delta}V_{\text{act}}}{V_{\text{act}}}}}\,.
\eeq
Such an effect is peculiar of LTP since both the masses feel the same DC acceleration $g_{\text{DC},x}$, a thing which will be
very unlikely to happen in LISA due to the large distance between SCs. Causality forbids gravity perturbations to travel faster
than the speed of light and we may assume gravity DC phenomena to be very local.

\subsubsection{Baseline fluctuation}

Fluctuation of the baseline is a phenomenon physically governed by parasitic coupling and temperature fluctuations. Nevertheless, due to the existence of
the readout, the actuation, magnetic phenomena and what else we discussed in advance, the scenario gets complicated.
One way out is the ``Hooke'' approach: we may consider an incoherent summation of the mentioned effects as spring terms
coupling the baseline length to the other phenomena. Again LTP only feels such a problem, because of the term in
\eqref{eq:mainifosignalfornoise} coupling ${\delta}x_{1}-{\delta}x_{2}$ to the optical bench:
\beq
-\left({\delta}x_{1}-{\delta}x_{2}\right) \omega_{\text{p},2}^{2}
\label{eq:ltpbaselinevar}
\eeq
where we took $\omega_{\text{lfs},x}^{2}=-2\omega_{\text{p},2}^{2}$. This feature
is not proper of LISA, where the OBs sit on different SC and the detector baseline is independent on potential OBs
length variations \cite{LTPdefdoc}.

On LTP, variation of the temperature of the OB will induce variation on the
baseline. We can evaluate this in \eqref{eq:ltpbaselinevar} with a simple thermal expansion law:
the expansion coefficient $\alpha_{\text{OB}}$ will be multiplied by the absolute value of $\omega^2_{\text{p,2}}$
which we already mentioned in the introduction (see \eqref{eq:kappatot1} and following).
Notice anyway thermal stability is very high within the MBW. The requirement
asks for $10^{-4}\,\unitfrac{K}{\sqrt{Hz}}$ \cite{LTPdefdoc}.


\begin{table}
\begin{center}
\begin{tabular}{>{\raggedleft}m{3.5cm}|l|l|l}
Description & Name & Value & Dimensions\\
\hline\rule{0pt}{0.4cm}\noindent
Baseline length & $r_o$ & $0.376$ & $\metresunit$\\ 
Optical bench temperature fluctuation & $S_{T,\text{OB}}$ & $\left(1.\times 10^{-4}\right)^{2}$ & $\unitfrac{K^2}{Hz}$\\ 
Optical bench expansion coefficient & $\alpha_{\text{OB}}$ & $4.\times 10^{-8}$ & $\unitfrac{1}{K}$\\ 
Optical metrology noise & $S_{\text{laser}}$ & $\left(9.\times 10^{-12}\right)^{2}\left(1+\left(\frac{2\pi\times 3\times 10^{-3}\,\unit{Hz}}{\omega}\right)^{4}\right)$ &
$\unitfrac{m^2}{Hz}$\\ 
\end{tabular}
\end{center}
\caption{Optical bench and baseline characteristics}
\label{tab:optbench}
\end{table}

Switching to spectral form, we'd get:
\beq
S_{a,{\Delta}r_o}^{\nicefrac{1}{2}}=\left|\omega^2_{\text{p,2}}\right|r_o \alpha_{\text{OB}} \sqrt{ S_{T,\text{OB}}}
\eeq
where $r_o$ is the $0$-temperature baseline length and $S_{T,\text{OB}}$ is the square PSD temperature
fluctuation of the OB. Both constants can be found in table \ref{tab:optbench}.

\subsubsection{Optical metrology}
The optical metrology term arises from the intrinsic noise of the laser device. It is a quite straightforward laser shot-noise
phenomenon, which in turn gets complicated for LTP in a similar manner as the baseline fluctuation. The last noise term in
\eqref{eq:mainifosignalfornoise} looks like:
\beq
\left(\omega_{\text{p},2}^{2}-\omega^{2}\right) \text{IFO}_{n}({\Delta}x)\,,
\eeq
then, in spectral form:
\beq
S_{a,\text{OM}}^{\nicefrac{1}{2}}=\left|\omega^2_{\text{p},2}+\omega^2\right|\sqrt{S_{\text{laser}}}
\eeq

\subsubsection{Summary}

Actuation, baseline and optical metrology fluctuations may be considered highly uncorrelated and summed
quadratically. No factor of $2$ will be used here, since actuation is exerted on one TM only along $\hat x$ and
the other two features pertain the optical bench only:
\begin{shadefundtheory}
\beq
S_{a,\text{meas}} = S_{a,\text{act}}+S_{a,{\Delta}r_{0}}+S_{a,\text{OM}}\,.
\eeq
\end{shadefundtheory}
\noindent Inspect figures in table \ref{tab:summeas}.

\begin{table}
\begin{center}
\begin{shademinornumber}
\begin{tabular}{r|l|l}
Description & Name & Value $\accPSDunit$\\
\hline\rule{0pt}{0.4cm}\noindent
Actuation amplitude instability &  $S^{\nicefrac{1}{2}}_{a,\text{act}}$  & $4.\times 10^{-15}$\\ 
Baseline fluctuation &  $S^{\nicefrac{1}{2}}_{a,{\Delta}r_o}$  & $2.33\times 10^{-18}$\\ 
Optical Metrology &  $S^{\nicefrac{1}{2}}_{a,\text{OM}}$  & $3.09\times 10^{-15}$\\
\hline\rule{0pt}{0.4cm}\noindent
Measurement noise &  $S^{\nicefrac{1}{2}}_{a,\text{meas}}$  & $5.06\times 10^{-15}$\\ 
\end{tabular}
\end{shademinornumber}
\end{center}
\caption{Measurement noise effects, summary.}
\label{tab:summeas}
\end{table}

\subsection{Summary of acceleration noise}

The structure of the spelling sequence for force noise is not different from the position one, but richer and hence
requires more attention in adding the various contributions. We state then that the total will look like:
\begin{shadefundtheory}
\beq
\begin{split}
S_{a,\text{total}}=& S_{a,\text{readout}}+S_{a,\text{thermal}}+S_{a,\text{Brownian}}+\\
&+S_{a,\text{crosstalk}}+S_{a,\text{dragfree}}+S_{a,\text{magnSC}}\\
&+S_{a,\text{magnIP}}+S_{a,\text{charge}}+S_{a,\text{misc}}\,,
\end{split}
\eeq
\end{shadefundtheory}
\noindent in addition, when measurement noise gets added, the grand-total displays as:
\begin{shadefundtheory}
\beq
S_{a,\text{gtotal}} = S_{a,\text{total}} + S_{a,\text{meas}}\,,
\eeq
\end{shadefundtheory}
\noindent and a summary of values can be read from table \ref{tab:sumnoise}. Note however that
the grand total amounts to $1.48\times 10^{-14}\,\accPSDunit$, well within the sensitivity limit
of LTP itself at $f=1\,\unit{mHz}$.

\begin{table}
\begin{center}
\begin{shadefundnumber}
\begin{tabular}{r|l|l}
Description & Name & Value $\accPSDunit$\\
\hline\rule{0pt}{0.4cm}\noindent
Drag-free &  $S^{\nicefrac{1}{2}}_{a,\text{dragfree}}$  & $1.36\times 10^{-15}$\\
Readout noise & $S^{\nicefrac{1}{2}}_{a,\text{readout}}$ & $1.09 \times 10^{-17}$\\
Thermal effects &  $S^{\nicefrac{1}{2}}_{a,\text{thermal}}$  & $4.97\times 10^{-15}$\\ 
Brownian Noise &  $S^{\nicefrac{1}{2}}_{a,\text{Brownian}}$  & $9.36\times 10^{-16}$\\ 
Magnetics SC &  $S^{\nicefrac{1}{2}}_{a,\text{magnSC}}$  & $8.9\times 10^{-15}$\\ 
Magnetics Interplanetary &  $S^{\nicefrac{1}{2}}_{a,\text{magnIP}}$  & $3.25\times 10^{-16}$\\ 
Random charging and voltage &  $S^{\nicefrac{1}{2}}_{a,\text{charge}}$  & $3.61\times 10^{-15}$\\
Cross-talk &  $S^{\nicefrac{1}{2}}_{a,\text{crosstalk}}$  & $6.12\times 10^{-15}$\\ 
Miscellanea &  $S^{\nicefrac{1}{2}}_{a,\text{misc}}$  & $6.04\times 10^{-15}$\\ 
\hline\rule{0pt}{0.4cm}\noindent
Total &  $S^{\nicefrac{1}{2}}_{a,\text{total}}$  & $1.39\times 10^{-14}$\\ 
Measurement noise &  $S^{\nicefrac{1}{2}}_{a,\text{meas}}$  & $5.06\times 10^{-15}$\\
\hline\rule{0pt}{0.4cm}\noindent
Grand Total &  $S^{\nicefrac{1}{2}}_{a,\text{gtotal}}$  & $1.48\times 10^{-14}$\\
\end{tabular}
\end{shadefundnumber}
\end{center}
\caption{Acceleration noise at $f=1\,\unit{mHz}$, summary.}
\label{tab:sumnoise}
\end{table}

\begin{figure}
\begin{center}
\includegraphics[angle=90,width=\textwidth]{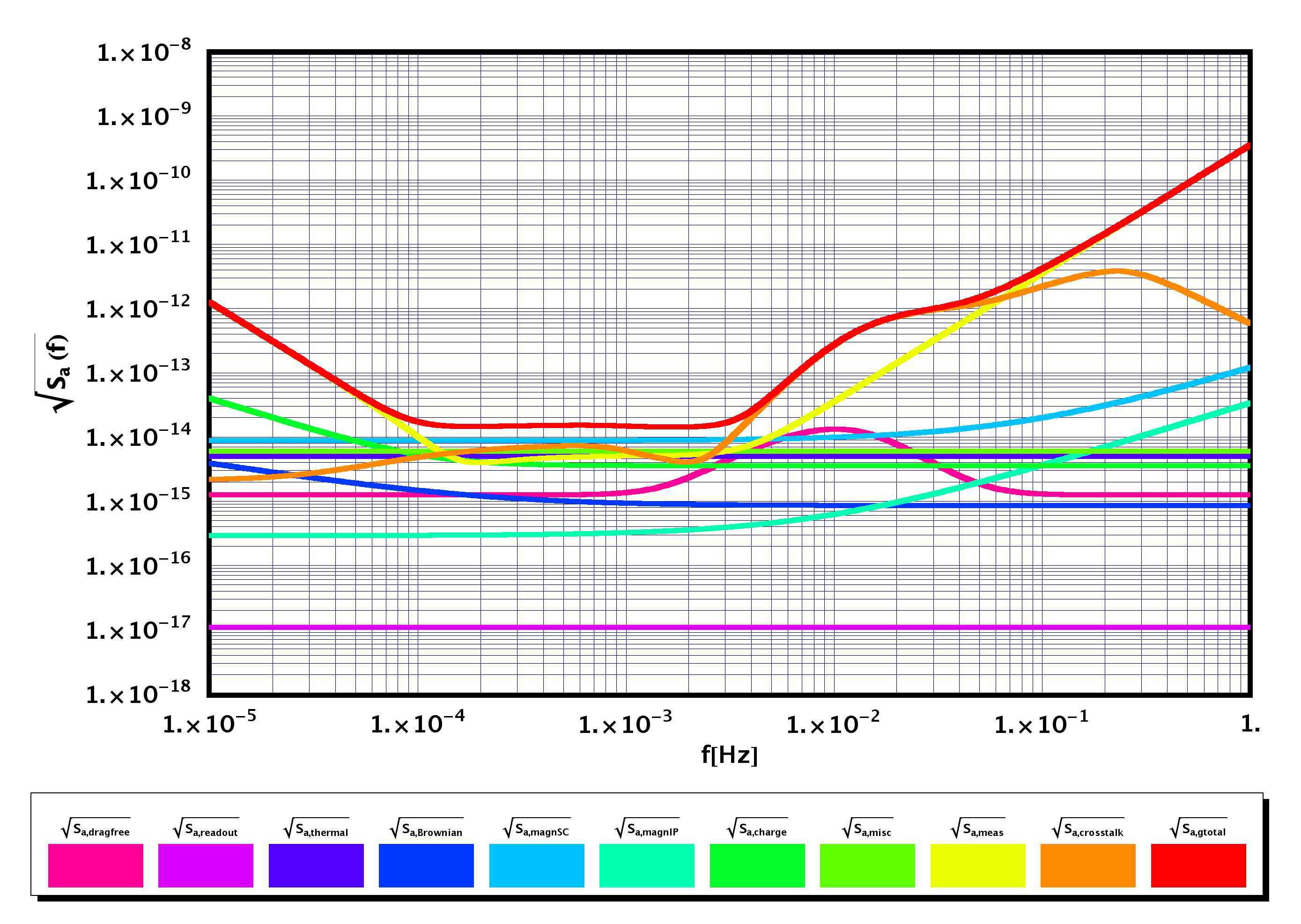}
\caption{Graph of acceleration noise contributions. Grand total is in red.}
\label{fig:noisegraphltp}
\end{center}
\end{figure}

\begin{figure}
\begin{center}
\includegraphics[angle=90,width=\textwidth]{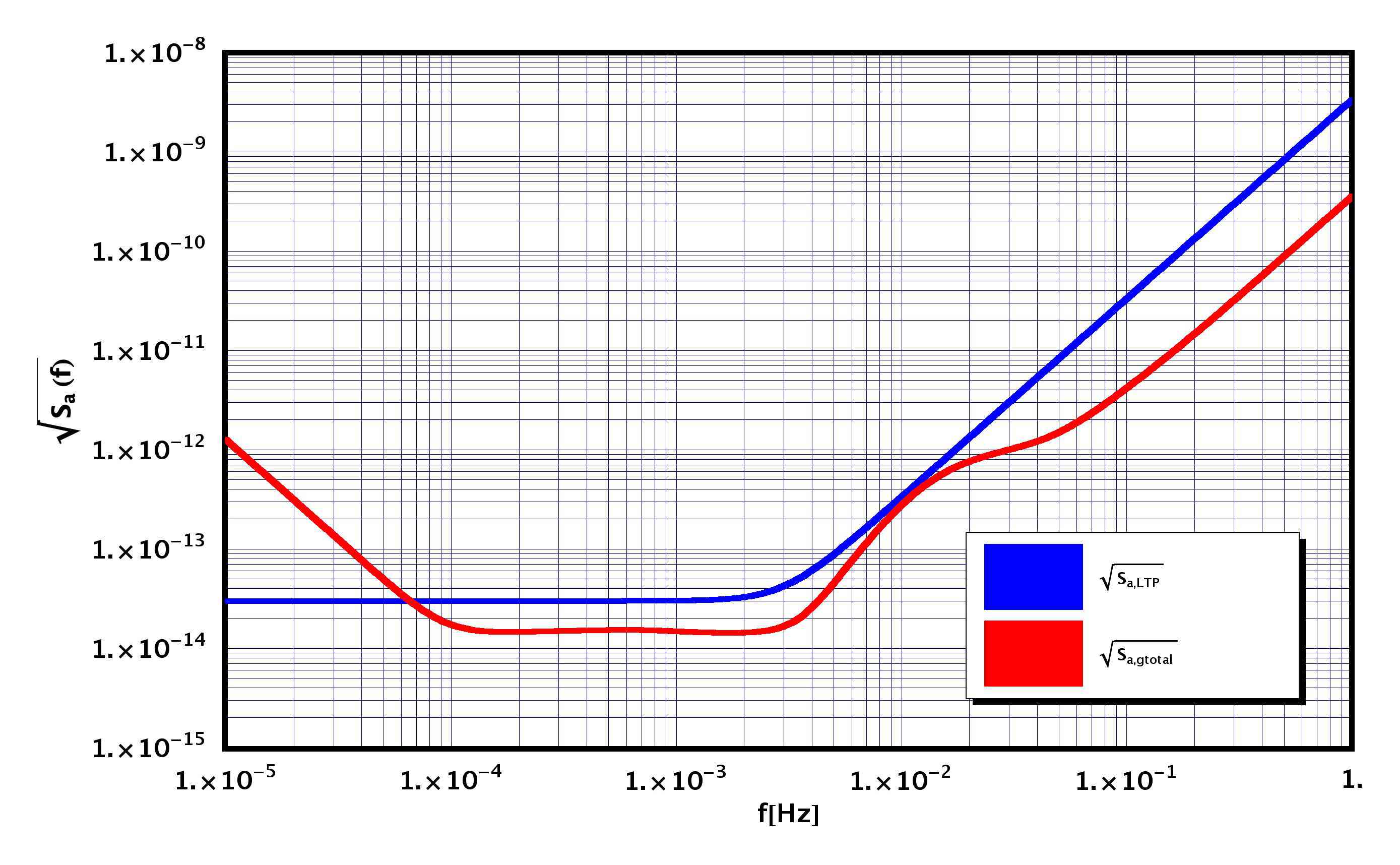}
\caption{Grand total of acceleration noise (red) versus LTP requirement sensitivity curve (blue).}
\label{fig:noisegraphversusltpsens}
\end{center}
\end{figure}

%% file: chapters/experiment.tex
\chapter{Experiment and measures}
\label{chap:experim}

\lettrine[lines=4]{W}{e'd like}
to review here the experiment
from a general side, pointing out the main experimental tasks, the sequence
and priorities and a ``run list'', providing a scheme of what
will be the number of measures and describing them up to some extent.

LTP as a noise-probe facility embodies the main task of gaining knowledge of residual noise
to model it in view of LISA. The chapter clarifies priorities in this perspective.

As a pre-requisite we'll present two contributions: gravitational compensation and calibration
of actuation forces.

The results will show that gravitational compensation can be achieved and it is robust
against displacement and rotations. A discussion on meshing size, reliability and
precision of estimates versus distance is sketched. Special attention was payed to engineering
aspects, in view of the definition of a gravitational control protocol to discipline mass addition and
removal from the SMART-2 satellite.


In the force calibration sections basic filter functions in somewhat simplified conditions will be sketched and
the proper signals employed. Simplifications will not reduce the generality of the
procedure or skip critical issues.

The optimal Kolmogorov filter theory will be used
to retrieve a solution. The method provides a numeric pattern which can be
convoluted with the detected data-set; moreover, the same procedure can be
extended to many other analyses, as well as to similar problems on LISA.

Next, the run list gives insight to the different operations LTP will be asked to perform. This
perspective shall show the reader how complicated and demanding the different tasks are
and how important is therefore to get a clear noise picture reading as a first step, and then
proceed with remaining tests.

We'll present the measurement
of the charge accumulated on the TMs to extend one of the main points in the ``run list''.
Such a feature is of paramount importance being
a fundamental prerequisite for the GRS to operate properly. The measure
will be carried on in science mode,
continuously and compatibly with the science mode requirements; the same be valid for the discharge procedure.
In view of optimising the detection different methods will be employed on purpose and estimates provided.

\newpage

\section{Main experimental task and phases}

It became increasingly clear by inspecting LISA's demands how important is to demonstrate that the
acceleration requirement \eqref{eq:lisamainreq} is met when the TMs are used as fiducial mirrors
for an interferometer based ``local'' metrology system \cite{LTPguideplan,willtheoexp}.
The importance of such a test is very high
in view of localised measurements of TMs position relative to the OB placed in every LISA SC: the final
displacement along one LISA's arm is the sum of the optical signal (with baseline $5\times 10^{6}\,\unit{km}$)
and two OB-to-TM signals, not to be underestimated in their precision.
Thus, the displacement performance measurement at better than $10\,\unitfrac{pm}{\sqrt{Hz}}$,
already included in LTP goals, has become a key science requirement.

LTP level of free-fall performance is set by worsening \eqref{eq:deltafoverfrelax} \cite{LTPscrd}
by a factor of $10$, and a term $\sqrt{2}$ appears
since the measured acceleration will be differential and the residual forces on the
test-masses are considered correlated over a short baseline:

\begin{shadefundtheory}
\beq
S^{\nicefrac{1}{2}}_{\nicefrac{\Delta F}{m}}(\omega)= 3\times 10^{-14} \left(1+\left(
\frac{\omega}{2\pi \times 3\, \unit{mHz}}\right)^{4}
\right)^{\nicefrac{1}{2}}\,\unitfrac{m}{s^{2}\sqrt{Hz}}\,,
\eeq
\end{shadefundtheory}
\noindent however, the major data outcome of the SMART-2 mission will be the physical noise
portrait for the quality of free-fall achievable on orbit around L1 within the limits of the same technology
that will fly on LISA. Its analysis and crossing with the known noise models will
produce a semi-empirical performance and disturbance model to build a realistic LISA sensitivity and environmental
picture upon. The experience on ground-based interferometers is employed at this level for debugging and optimisation.

The general procedure calls for a detailed mathematical model in the frequency domain, and for the measurement
of perturbation effects over the linear transfer functions upon application of distinct stimuli (magnetic fields or gradients,
thermal gradients, electrostatic forces, variation in suspension gains, induced displacements) as functions of frequency.
This allows to build up a deterministic map uniquely coupling sources and effects, other than ``continuous'' manifolds
of parameters cross-correlations in seek for minimisation.

Optimal filtering theory is widely employed to produce discretised filters - by virtue of well-known models - to be
convolved with the real stochastic data. This ``noise projection'' procedure provides thus a set of PSDs for all
known sources, to be combined quadratically (assuming them to be uncorrelated)
to hopefully explain the full acceleration and readout PSD curve, or to isolate unpredicted effects.
Deviations between the semi-empirical model and the physical combined measurement upon projection are the
hint of unfocused correlation and/or unpredicted noise and will demand further investigation.

In view of this scenario (which is physically very likely to occur), a line of priorities was drawn, a base schedule
of measurements planned, and protocols written to support contingency on the formerly described events or in case of systems
breakdown. The main experiment time-line is defined so to minimise the risk of missing the main science goals: a first
noise portrait is taken as soon as minimal experimental conditions are met. Accuracy is improved as a second issue and calibrations
come in the third place. Obviously measuring is a loop-wise procedure, the longer the experimental time and the more
reliable the device proves to be, the highest the accuracy and more the pieces of physical information gathered.

Right after cruise phase (we assume all systems to be on-line from here) the first mission goal is obtaining a noise
spectrum from the IFO and GRS signals even before TMs release. Environmental spectra will be taken together
with the former ones. In the unfortunate possibility of Pathfinder systems failure, these valuable data will be
a solid basis to check models and figures. We spare the reader of the technical details (see \cite{LTPguideplan,willtheoexp})
and just state that ``Basic Working Status'' is reached when the TMs are floating and a reliable IFO signal
providing the main relative TMs displacement ${\Delta}x$ is on-line.

Scan of the $1\,\unit{mHz}$ to $30\,\unit{mHz}$ (and beyond, if possible) frequency domain follows,
book-keeping the $\text{IFO}({\Delta}x)$. We strongly point out that LTP will be the first space-borne mission
to provide relative acceleration data at such a level of accuracy and precision, such a result would be a major advance
in science in itself. A ``Nominal Mode'' (M1) measurement will be performed first, implying an easier control strategy and
the self-calibration issues we discussed already. A ``Science Mode'' (M3) acceleration PSD measurement will follow,
deemed to generate the first off-ground measurement of acceleration between two TMs in quasi-geodesic free-fall. To our
best knowledge, this will be the most precise realisation of a TT-gauged laser-frequency-locked frame of reference.


\subsection{Noise shooting and PSD minimisation phase}

Once the goals of the former phases have been achieved, the next task is to minimise the PSD
possibly to achieve the figure in \eqref{eq:deltafoverfrelax} or beyond.
Therefore the main sources of excess noise are identified and, whenever possible, minimised. The
optimised noise PSD is then compared to the prediction obtained from the noise projections models.

All main diagnostics are at work and all data gathered simultaneously with the $\text{IFO}({\Delta}x)$ channel.
This phase requires absolute calibration of acceleration \cite{LTPscrd} via the
identification and suppression of anomalous physical disturbances and
the identification and optimisation of anomalous direct and cross-coupling transfer functions from physical disturbances
to the $\text{IFO}({\Delta}x)$ channel.

Dedicated experiments are planned on purpose:
\begin{description}
\item[Measurement and ``diagonalising'' of actuation] to displacement or rotation transfer functions. Large amplitude force and torque
drives are applied to TM and SC measuring the outputs of all
channels. By uploading several sets of parameters into the actuation matrix \cite{LTPdfacsfun} and tracking the signals will
result into a large sample of ``reaction'' matrices, to allow diagonalising the references and minimisation of cross-talk
(see section \ref{sec:crosstalk} and \cite{LTPcrosstalk, LTPmaster})
\item[Measurement of the charge on test-mass] and subsequent discharging (non-continuous).
\item[Identification of anomalous stray dc-voltages] on electrodes and their compensation by means of appropriate
voltage biases.
\item[Identification of anomalous magnetic noise] searching for anomalously large magnetic
field and magnetic field gradient fluctuation from malfunctioning devices. The measurement of
permanent magnetic moment of TM may take place to cross-check for variation after launch.
\item[Identification of anomalous thermal effects] involves both temperature measurements and the measurement of the
acceleration response to applied temperature gradients. Anomalous pressure and out-gassing phenomena may be identified.
\item[Coarse measurement of stiffness] and stiffness matching.
\end{description}

Nominal and optimal values will be used to start the experimental phase and further optimised by variation
of the most relevant (or suspected to be) ones. For example TMs position and orientation will be centred
employing values from a-priori modelling, rotation jitter may be applied on a frequency span
monitoring the $\text{IFO}({\Delta}x)$ output
to estimate the ``angular jitter to apparent displacement'' effect as a function of frequency. The converse -
displacing TM and reading angular jitter - gives the transposed cross-talk and both the transfer functions can be used
to minimise the overall effect by redefining offsets or relative gains. To first order, cross-couplings may be assumed
to be independent and evaluated with such procedures; fine-tuning multi-dimensional optimisation may be
needed for higher-order corrections. Some effects may of course be thought as very weakly-coupled to
some sources, e.g. the angular-linear jitter cross-talk is almost independent of temperature.

The system goes then into an iterative procedure where the previously estimated new optimal conditions become
nominal values for the next loop, possibly improving the working point till maximum in matter of two-three loops.
At every step comparison is made of the measured IFO PSD with the projected noise model, differences
analysed and decisions taken on what to tune in the next step.

Ideally the achievement of the goal performance may put an even more stringent upper limit on any excess noise than the
figure in \eqref{eq:deltafoverfrelax}; the computational accuracy of the noise model constitutes anyway a sort of default
level, though the accuracy of the prediction of the noise floor depends
critically on some system parameters to be measured like total magnetic field fluctuations, magnetic
susceptibility, etc.

\subsection{Noise model detailed investigation phase}

A detailed investigation phase will begin right after cross-coupling optimisation has reached a reasonable convergence. In
this new phase the LTP will be perform a dedicated set of experiments to ``demonstrate'' the noise model and allow
further projection for LISA. The list of all planned measurements is discussed later \cite{LTPscrd}.

The sequence of the ``runs'' is cast in logical priority list and tries to make best use of the given (and fore-casted) experimental time-span.
Cross-coupling optimisation is repeated after more information on the noise contributions and parameters tweaking has been
gathered thanks to the experiments sequence.

\subsection{Extended investigation phase}

Other key experiments that are part of the LISA Pathfinder mission are planned to take place in this phase. 
Specifically continuous charge measurements and continuous discharging will be performed, together
with very long data runs to assess the performance of free fall and interferometry down to Fourier frequencies of
$30\,\unit{{\mu}Hz}$,
if possible, covering the full frequency range envisioned for LISA. On purpose, such data runs will have to extend over several days,
to guarantee reliable data averaging as well. 

A planned time slot should be reserved for this period for additional investigations that arise during the
previous phases of the mission and are suggested by the science team and approved by the STOC.

This phase is called ``extended'' because it will add quite a deal of news to the former ones, but its priority
is second with respect to taking a noise portrait and applying the noise projection scheme.

\subsection{Fundamental science phase}

During the Fundamental Science Phase, scientific investigations not directly addressing the verification of LISA
technologies are performed. These investigations will have been suggested by the Science team during mission
preparation. Examples could be investigations of Modified Newtonian Dynamics (MOND) theories, measurements of big-$G$, or
a general validation of the inverse square law.

\subsection{Extended mission phase}

The need for more detailed investigations might have arisen
during the nominal mission phase; these will be tackled in the Extended Mission Phase.

\begin{figure}
\begin{center}
\includegraphics[width=0.8\textwidth]{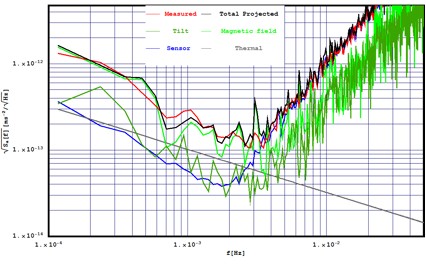}
\caption{Data noise projection. The effect of measured magnetic field noise and
apparatus tilt has
been estimated from independent measurements and cross-spectra. Intrinsic thermal noise contribution has been calculated
from measured pendulum properties. The black line is the sum of all expected contribution to total noise. Projection has
been realised through the multi coherence method \cite{LTPdataan}.}
\label{fig:subtorquepsd}
\end{center}
\end{figure}

\section{Fighting gravitational noise: compensation}
\label{sec:gravitcomp}

\subsection{Introduction}

In the following
we'll name the three residual accelerations $\nicefrac{\vc F}{m}$, the three torques per unit moment of inertia $\nicefrac{\gamma_{i}}{I_{i}},\,i=1..3$
and the $6\times 6$ stiffness $\omega^{2}_{\text{p,grav},ij}$
``static gravitational imbalances'' (SGI) since in our case they have gravitational nature and origin.

The SGI contribution to maximum acceleration and stiffness is significant:
the total requirement for LTP is
$\omega^{2}_{\text{p},\text{tot}, xx} \leq 20 \times 10^{-7}\,\unit{s}^{-2} @\, 1 \unit{mHz}$, while
$\omega^{2}_{\text{p,grav}, xx} \leq 5 \times 10^{-7}\,\unit{s}^{-2}$
(see table \ref{tab:reqgammas} and \cite{Bortoluzzi:2003ua}).

Why don't we choose the easy way of electrostatic compensation? Dynamic compensation
could be achieved by injecting DC voltage to produce a force by the GRS. But as it was shown
in the noise chapter, any applied voltage produce a parasitic coupling in the form
\cite{2003SPIE.4856...31W}:
\begin{shadefundtheory}
\beq
\omega^{2}_{\text{p},\text{act}, xx}=-\Delta a_{\text{DC}}
  \frac{\partial^{2}C_{x}}{\partial x^{2}}
  \left| \frac{\partial C_{x}}{\partial x} \right|^{-1}\,,
\eeq
\end{shadefundtheory}
\noindent where $\Delta a_{\text{DC}}$ is the acceleration to counteract and we specialised the
formula to the $\hat x$ direction of actuation (see \eqref{eq:actistiffratio}).
$C_{x}$ is the actuators capacitance along $\hat x$. Moreover, "in band" fluctuation of the actuation drive voltage amplitude translate into
acceleration noise, according to:
\begin{shadefundtheory}
\beq
a_{\text{act}, n}=2 \Delta a_{\text{DC}}\cdot S^{\nicefrac{1}{2}}_{\nicefrac{{\delta}V}{V_{\text{act}}}}\,.
\eeq
\end{shadefundtheory}
\noindent where $S^{\nicefrac{1}{2}}_{\nicefrac{{\delta}V}{V_{\text{act}}}}$ is the noise spectral density of the actuation voltage.
Both the stiffness and actuation force noise are thus proportional to the residual imbalanced $\Delta a_{\text{DC}}$.

Static compensation of gravity imbalance is thus mandatory to reduce budget on
the amount of static parasitic stiffness and to lower the related acceleration noise.
In this scenario, maximal budgets have been assigned to each stiffness and noise contribution. The maximal
SGI and can be found in tables \ref{tab:reqforces} and \ref{tab:reqgammas}.

In principle, perfect compensation to annihilate all the SGI might be possible, but given the
set of requirements, we need only to find the minimum mass distribution, with best geometry to bring the SGI
values within requirement. Due to engineering needs, these compensation masses (CmpMs) shall be easy to
manufacture and mount and, additionally, should be geometrically as simple as possible.

\begin{table}
\begin{center}
\begin{shadefundnumber}
\begin{tabular}{r@{}l|r@{}l|r@{}l|r@{}l|r@{}l|r@{}l}
\multicolumn{2}{c|}{$\nicefrac{\Delta F_x}{m}$} & \multicolumn{2}{c|}{$\nicefrac{\Delta F_y}{m}$} & \multicolumn{2}{c|}{$\nicefrac{\Delta F_z}{m}$} & \multicolumn{2}{c|}{$\nicefrac{\gamma_{\theta }}{I_{\theta }}$} & \multicolumn{2}{c|}{$\nicefrac{\gamma_{\eta }}{I_{\eta }}$} & \multicolumn{2}{c}{$\nicefrac{\gamma_{\phi }}{I_{\phi }}$}\\
\hline
$1$ & $.1$ & $1$ & $.7$ & $3$ & $.2$ & $14$ & $.0$ & $18$ & $.0$ & $23$ & $.0$\\
\end{tabular}
\end{shadefundnumber}
\end{center}
\caption{Requirements on forces and torques, TM1-TM2 (forces), TM1 only (torques). Absolute values.
  $F_{i}/m$ in $[\unit{nm\,s}^{-2}]$, $T_{i}/I_{i}$ in $[\unit{nrad\,s}^{-2}]$ for each $i$.}
\label{tab:reqforces}
\end{table}

\begin{table}
\begin{center}
\begin{shadefundnumber}
\begin{tabular}{r@{}l|r@{}l|r@{}l|r@{}l|r@{}l|r@{}l|r@{}l}
\multicolumn{2}{c|}{$$} & \multicolumn{2}{c|}{$\nicefrac{F_x}{m}$} & \multicolumn{2}{c|}{$\nicefrac{F_y}{m}$} & \multicolumn{2}{c|}{$\nicefrac{F_z}{m}$} & \multicolumn{2}{c|}{$\nicefrac{\gamma_{\theta }}{I_{\theta }}$} & \multicolumn{2}{c|}{$\nicefrac{\gamma_{\eta }}{I_{\eta }}$} & \multicolumn{2}{c}{$\nicefrac{\gamma_{\phi }}{I_{\phi }}$}\\
\hline
\multicolumn{2}{c|}{${{\partial }_x}$} & $500$ & $.0$ & $7$ & $.0$ & $7$ & $.0$ & \multicolumn{2}{c|}{$-$} & \multicolumn{2}{c|}{$-$} & \multicolumn{2}{c}{$-$}\\
\multicolumn{2}{c|}{${{\partial }_y}$} & $7$ & $.0$ & $500$ & $.0$ & \multicolumn{2}{c|}{$-$} & \multicolumn{2}{c|}{$-$} & \multicolumn{2}{c|}{$-$} & \multicolumn{2}{c}{$-$}\\
\multicolumn{2}{c|}{${{\partial }_z}$} & $7$ & $.0$ & \multicolumn{2}{c|}{$-$} & $1000$ & $.0$ & \multicolumn{2}{c|}{$-$} & \multicolumn{2}{c|}{$-$} & \multicolumn{2}{c}{$-$}\\
\multicolumn{2}{c|}{${{\partial }_{\theta }}$} & $14$ & $.0$ & \multicolumn{2}{c|}{$-$} & \multicolumn{2}{c|}{$-$} & $1960$ & $.0$ & \multicolumn{2}{c|}{$-$} & \multicolumn{2}{c}{$-$}\\
\multicolumn{2}{c|}{${{\partial }_{\eta }}$} & $14$ & $.0$ & \multicolumn{2}{c|}{$-$} & \multicolumn{2}{c|}{$-$} & \multicolumn{2}{c|}{$-$} & $1960$ & $.0$ & \multicolumn{2}{c}{$-$}\\
\multicolumn{2}{c|}{${{\partial }_{\phi }}$} & $14$ & $.0$ & \multicolumn{2}{c|}{$-$} & \multicolumn{2}{c|}{$-$} & \multicolumn{2}{c|}{$-$} & \multicolumn{2}{c|}{$-$} & $1480$ & $.0$\\
\end{tabular}
\end{shadefundnumber}
\end{center}
\caption{Requirement on stiffness, linear-angular and angular-angular gradients over TM1. Absolute values.
  Dimensions for each element are $ 10^{-9} [\unit{s}^{-2}]$. The symbol ``$-$'' means
  no precise requirement is demanded.}
\label{tab:reqgammas}
\end{table}

\begin{figure}
\begin{center}
\includegraphics [width=0.5\textwidth] {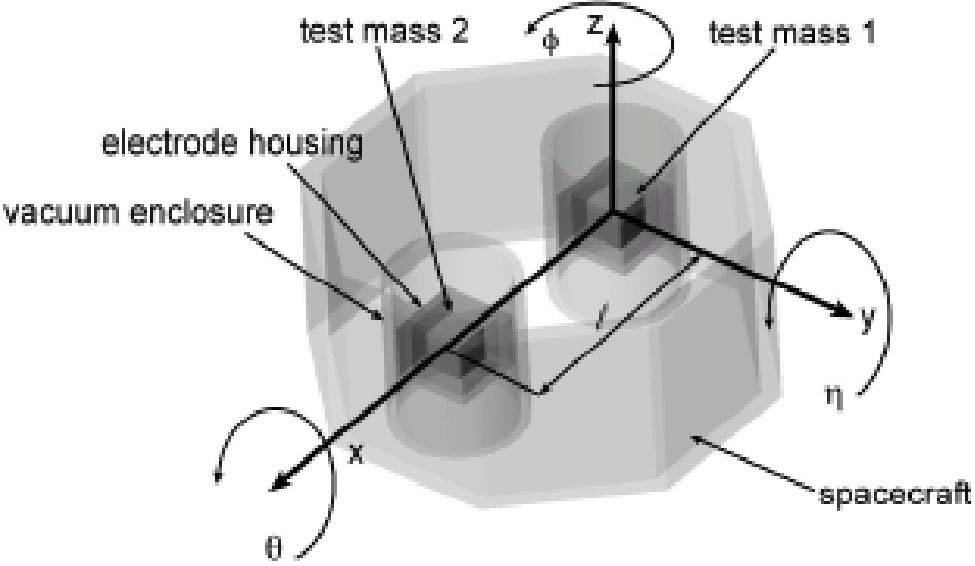}%
\hspace{4mm}\includegraphics [width=0.39\textwidth] {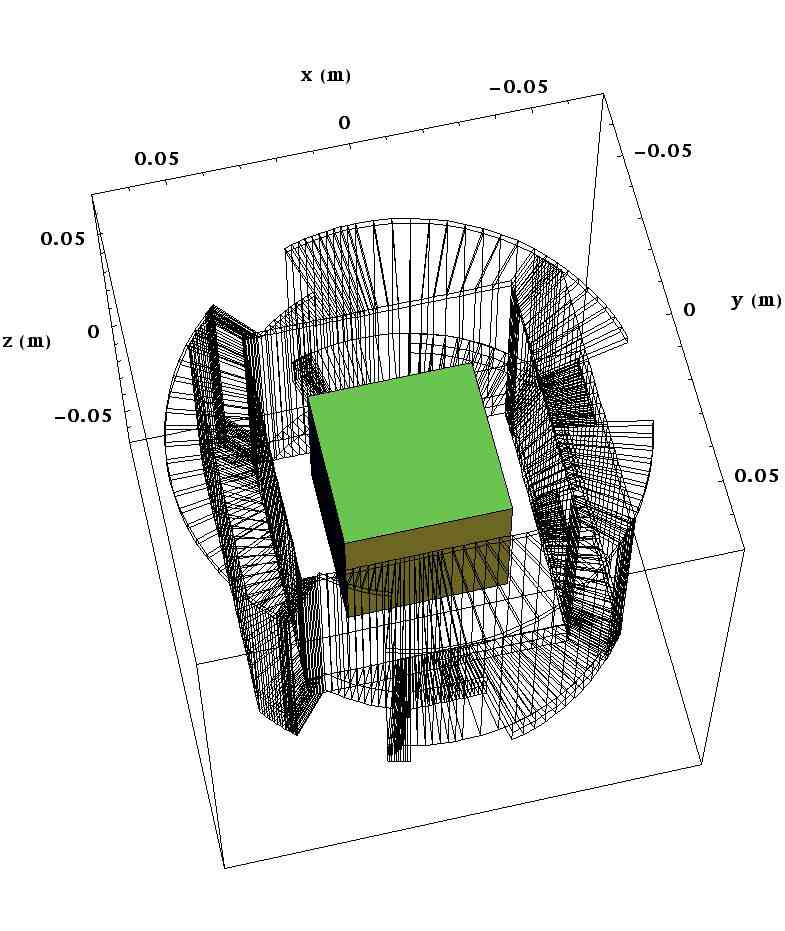}
\caption{On the left: reference system of the LTP. Schematic view of the experimental apparatus's of the LTP
with the main DOF of the freely floating TM. On the right: top-side view of the available room inside the
vacuum enclosure hosting each TM (see \cite{Dolesi:2003ub, ltpgravprot}).} \label{fig:refsys}
\end{center}
\end{figure}

We refer now to the system of coordinates in figure \ref{fig:refsys}, left.
The gravitational force along the $i$-th axis, exerted by a material point with mass
$m$ located on $\vc r = \left\{x,y,z\right\}$ on the homogeneous TM with mass $m$ and volume $V=L_{x}L_{y}L_{z}$
whose centre of mass is chosen as the origin of coordinates, can be analytically calculated as the integral of the usual Newtonian
potential \cite{vit:2004int, ltpgravprot}:
$\Phi  (x, y, z) = \nicefrac{G m}{\left|\vc r\right|}$. Torques follow from the definition:
$\vc \gamma =\vc r \wedge \vc F = \left\{\gamma_{\theta}, \gamma_{\eta}, \gamma_{\phi}\right\}$,
while the linear stiffness matrix can be computed by means of $\omega^{2}_{\text{p,grav},ij}(\vc r) = \partial_{i} F_j(\vc r)$.
Ancillary formulae must be employed to compute
angular-linear, angular-angular gradients due to the spatial extent of the TMs. For example, to calculate
the derivative of the force with respect to the angular DOF conjugated to $\hat z$, i.e. $\hat\phi$, we would use:
\beq
\partial_{\hat\phi} \vc F =
  (\hat \phi \wedge \vc F) - \left((\hat\phi\wedge \vc r) \cdot \vc\nabla_{r}\right) \vc F \label{eq:deriv}\,.
\eeq
In this way, we are left with only translational derivatives to compute. For example:
\begin{align}
\omega^{2}_{\text{p,grav},\phi x} &= \frac{\partial F_{x}}{\partial \phi} = \frac{\partial \gamma_{\phi}}{\partial x}
  = -F_{y}-x\omega^{2}_{\text{p,grav},xy} + y\omega^{2}_{\text{p,grav},xx}\,,  \label{eq:gamma1}\\
\omega^{2}_{\text{p,grav},\phi \theta}  &= \frac{\partial \gamma_{\theta}}{\partial \phi}
  = x F_{x} + y F_{y} - y^{2} \omega^{2}_{\text{p,grav},xx} - x^{2} \omega^{2}_{\text{p,grav},yy} - 2 x y \omega^{2}_{\text{p,grav},xy}\, \label{eq:gamma2},
\end{align}
where we implied $F_i=F_i(\vc r)$, $\gamma_i=\gamma_i(\vc r)$, $\omega^{2}_{\text{p,grav},ij}=\omega^{2}_{\text{p,grav},ij}(\vc r)$ for any $i,j$.

\begin{figure}
\begin{center}
\includegraphics [width=0.39\textwidth] {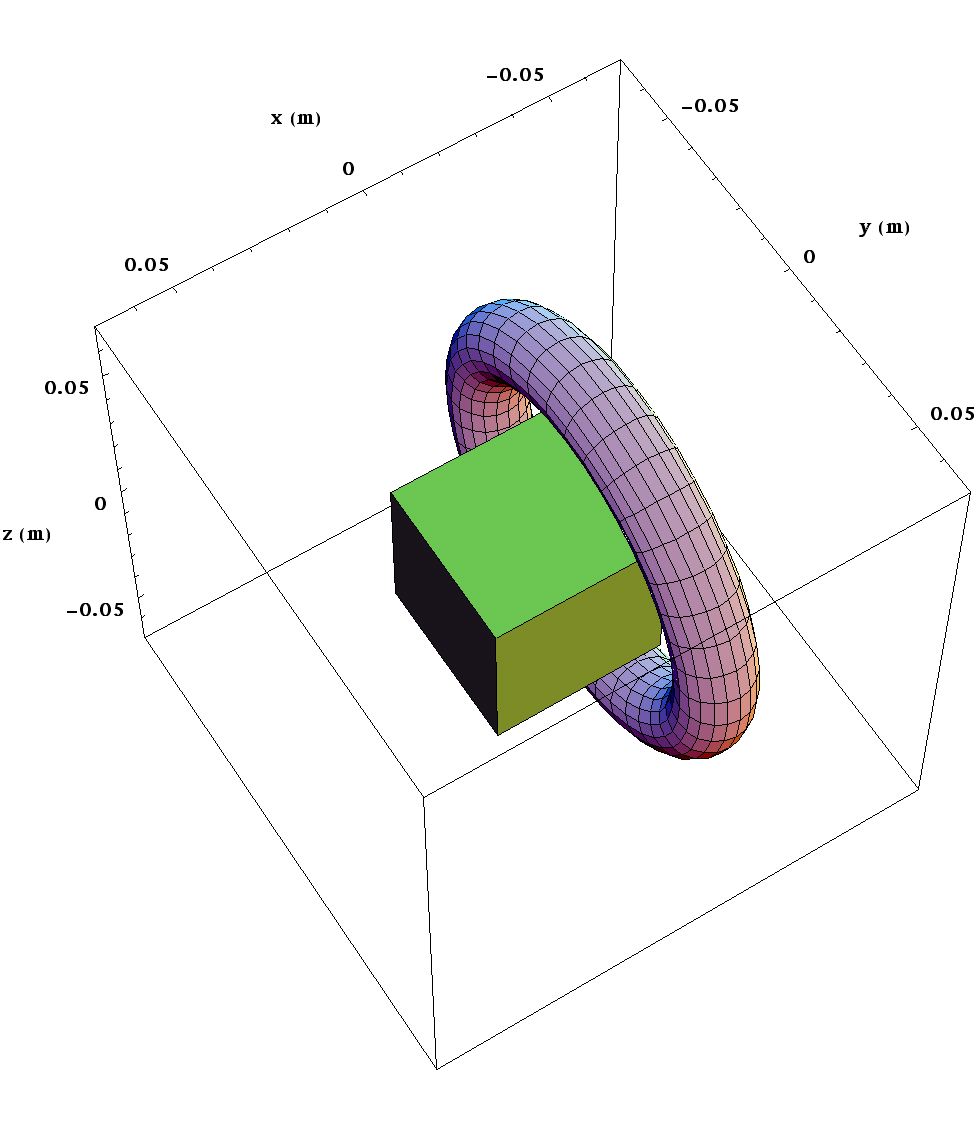}%
\hspace{4mm}\includegraphics [width=0.4\textwidth] {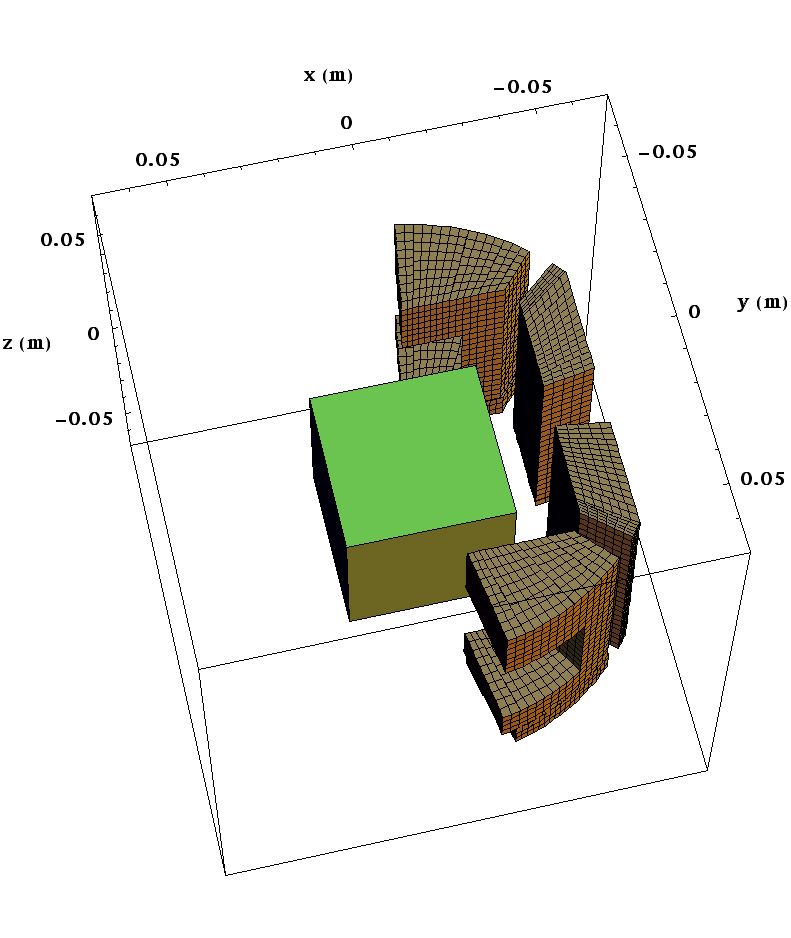}
\caption{Left: tentative shape of compensating ring contributing positive gravitational stiffness. On the right: final shape of compensation
masses (4 lobes) around TM1.} \label{fig:shapes}
\end{center}
\end{figure}

\subsection{Explaining the strategy}

The SGI due to the different components on board (DRS, LTP, SC) have been evaluated by means of
\eqref{eq:gamma1}, \eqref{eq:gamma2} and similar formulae, after meshing
of each structure. A dedicated {\sf Mathematica}\textregistered\, code was written to
mesh simple polyhedron's forms as well as reading nodes-elements {\sf STEP} files, allowing
computation of forces, torques and derivatives from point-like sources on a generic cubic mass.

Meshing was chosen to be adaptive, tetrahedral, with size $\sim 2\,\unit{mm}$ maximum,
while the density and mass distribution of the sub-components is known for each subsystems.
The mesh size is chosen for the different masses as a function of
distance and object size in order to obtain the required precision and
is a compromise with computational time and power constraints. We note
that $2~\unit{mm}$ size is a conservative choice for nearby IS and OB
hardware and could be relaxed for more distant spacecraft components (see beyond for a more
complete discussion and formulae).

Contributions coming from different blocks vary in their relevance; linearity of the SGIs permits us to sum the contributions to the total SGIs,
taking advantage of symmetries and partial cancellations. Resulting values can be found in
tables \ref{tab:extraforces} and \ref{tab:extragammas}. Notice the accelerations $\nicefrac{\Delta F_{x}}{m}$,
$\nicefrac{\Delta F_{y}}{m}$, $\nicefrac{\Delta F_{z}}{m}$
are differential, because only the relative acceleration counts for these DOF on LTP.
On the other hand torques and stiffness are computed with absolute reference to TM1.

After receiving input on boundary conditions and limits, the code performs a weighted choice of the allowed elements in free space
thus giving the sum of the related contributions. The weighing is optimised to completely eliminate $\Delta F_{x}$
while minimising stiffness. Once a solution is found, it may be refined at will, by re-meshing and re-weighing.

Notice from tables \ref{tab:extraforces} and \ref{tab:extragammas} that the
uncompensated value of $\nicefrac{\Delta F_{x}}{m}$ is $60$ times larger than the allowed value; the other values of the force-torques vector
appear to be within requirements. The same for the stiffness matrix, showing a small, negative spring for $\omega^{2}_{\text{p,grav},xx}$
and a value of $\omega^{2}_{\text{p,grav},\eta x}$ just the $3\%$ out of specification.

The positions of the two CmpMs are chosen to minimise the distance to the TMs, to achieve
maximum effect for a given mass (forces scale like $\nicefrac{1}{r^{2}}$, $r$ mutual distance between the bodies).
The largest contiguous portion of available free room around each TMs between the electrode housing (EH)
and vacuum enclosure (VE) has
been considered a suitable location for the CmpMs (see figure \ref{fig:refsys}, right)
and reduced to geometrically simple elements for meshing.

The CmpMs are then assumed to be in Tungsten, due to its high density ($19300\,\unitfrac{kg}{m^{3}}$) and
modelled assuming the TMs in centred, nominal position.

The tentative ideal shape of a torus,
coaxial with the LTP $\hat x$ axis, belt-like around each TMs had been chosen.
The far ring used for compensating will look like a point as seen from each far TM,
while it is a true torus for the closer TM. The rings thus attract the TMs outwards, compensating
$\Delta F_x$ without introducing undesirable $\hat x$-gradient stiffness due to the rotational symmetry
of the tori (of course this is not the case for $\hat y$, $\hat z$).
This analysis on springs constants accounts for the gravitational
budget; electrostatic negative stiffness will still dominate the scenario, but pure gravitational springs
induced by this configuration are positive nevertheless.

A picture of the starting shape can be seen in figure \ref{fig:shapes}, left. Due to lack of space on the upper and lower
parts, each perfect ring gets cut into two lobes, enlarged toward the $\hat x$ axis to embed each TM.
The CmpMs assume the final form of eight lobes, four on each TM, standing on the external
side of the cubes (see figure \ref{fig:shapes}, right),
between the VE and EH. Each CmpMs weights $2.51\pm 0.04\,\unit{kg}$.
Resulting residual value in $\unitfrac{\Delta F_x}{m}$ is highly sensitive to the CmpMs mass:
the imbalance in $\unitfrac{\Delta F_x}{m}$ cannot be compensated with a smaller mass.

The CmpMs gravitational contribution
brings the overall SGI values within requirements as shown in tables \ref{tab:resforces} and \ref{tab:resgammas}
in comparison with tables \ref{tab:reqforces} and \ref{tab:reqgammas}.
We note in comparing tables \ref{tab:resforces} and \ref{tab:extraforces} that we have successfully compensated
$\nicefrac{\Delta F_{x}}{m}$ without significantly disturbing the other forces and torques,
which were already compliant with the requirements even before
compensation.  
Unfortunately this is obtained with the price of a considerable
increase in $\omega^{2}_{\text{p,grav},xx}, \omega^{2}_{\text{p,grav},yy}, \omega^{2}_{\text{p,grav},zz}$ and other DOF in
table \ref{tab:resgammas}. The $\omega^{2}_{\text{p,grav},xx}$ factor is a negative stiffness close in value to the requirement limit,
although this proves to be still reasonable
for LTP \cite{Bortoluzzi:2003ua}. Conversely, the $\omega^{2}_{\text{p,grav},zz}$ factor looks like a positive spring, increasing robustness in $\hat z$.

\begin{table}
\begin{center}
\begin{shadefundnumber}
\begin{tabular}{r@{}l|r@{}l|r@{}l|r@{}l|r@{}l|r@{}l}
\multicolumn{2}{c|}{$\nicefrac{\Delta F_x}{m}$} & \multicolumn{2}{c|}{$\nicefrac{\Delta F_y}{m}$} & \multicolumn{2}{c|}{$\nicefrac{\Delta F_z}{m}$} & \multicolumn{2}{c|}{$\nicefrac{\gamma_{\theta }}{I_{\theta }}$} & \multicolumn{2}{c|}{$\nicefrac{\gamma_{\eta }}{I_{\eta }}$} & \multicolumn{2}{c}{$\nicefrac{\gamma_{\phi }}{I_{\phi }}$}\\
\hline
$65$ & $.7$ & $-0$ & $.13$ & $0$ & $.58$ & $0$ & $.04$ & $0$ & $.36$ & $0$ & $.05$\\
\end{tabular}
\end{shadefundnumber}
\end{center}
\caption{Forces and Torques exerted by the SC, DRS and LTP Path-finder systems over TM1-TM2 (forces) and TM1 only (torques).
  $F_{i}/m$ in $[\unit{nm\,s}^{-2}]$, $T_{i}/I_{i}$ in $[\unit{nrad\,s}^{-2}]$ for each $i$.}
\label{tab:extraforces}
\end{table}

\begin{table}
\begin{center}
\begin{shadefundnumber}
\begin{tabular}{r@{}l|r@{}l|r@{}l|r@{}l|r@{}l|r@{}l|r@{}l}
\multicolumn{2}{c|}{$$} & \multicolumn{2}{c|}{$\nicefrac{F_x}{m}$} & \multicolumn{2}{c|}{$\nicefrac{F_y}{m}$} & \multicolumn{2}{c|}{$\nicefrac{F_z}{m}$} & \multicolumn{2}{c|}{$\nicefrac{\gamma_{\theta }}{I_{\theta }}$} & \multicolumn{2}{c|}{$\nicefrac{\gamma_{\eta }}{I_{\eta }}$} & \multicolumn{2}{c}{$\nicefrac{\gamma_{\phi }}{I_{\phi }}$}\\
\hline
\multicolumn{2}{c|}{${{\partial }_x}$} & $38$ & $.57$ & $4$ & $.83$ & $-0$ & $.87$ & $-7$ & $.02$ & $-39$ & $.29$ & $5$ & $.13$\\
\multicolumn{2}{c|}{${{\partial }_y}$} & $4$ & $.84$ & $-23$ & $.88$ & $-5$ & $.62$ & $137$ & $.17$ & $2$ & $.45$ & $17$ & $.67$\\
\multicolumn{2}{c|}{${{\partial }_z}$} & $-0$ & $.88$ & $-5$ & $.64$ & $-14$ & $.7$ & $11$ & $.53$ & $18$ & $.87$ & $4$ & $.56$\\
\multicolumn{2}{c|}{${{\partial }_{\theta }}$} & $-0$ & $.01$ & $14$ & $.45$ & $-0$ & $.1$ & $31$ & $.42$ & $-1$ & $.16$ & $2$ & $.93$\\
\multicolumn{2}{c|}{${{\partial }_{\eta }}$} & $-14$ & $.43$ & $0$ & $.0$ & $-61$ & $.82$ & $-1$ & $.1$ & $33$ & $.31$ & $0$ & $.28$\\
\multicolumn{2}{c|}{${{\partial }_{\phi }}$} & $0$ & $.1$ & $61$ & $.82$ & $0$ & $.0$ & $2$ & $.57$ & $0$ & $.33$ & $26$ & $.41$\\
\end{tabular}
\end{shadefundnumber}
\end{center}
\caption{Stiffness, linear-angular and angular-angular gradients exerted by the SC, DRS and LTP Path-finder systems over TM1.
  Dimensions for each element are $ 10^{-9} [\unit{s}^{-2}]$.}
\label{tab:extragammas}
\end{table}

\begin{table}
\begin{center}
\begin{shadefundnumber}
\begin{tabular}{r@{}l|r@{}l|r@{}l|r@{}l|r@{}l|r@{}l}
\multicolumn{2}{c|}{$\nicefrac{\Delta F_x}{m}$} & \multicolumn{2}{c|}{$\nicefrac{\Delta F_y}{m}$} & \multicolumn{2}{c|}{$\nicefrac{\Delta F_z}{m}$} & \multicolumn{2}{c|}{$\nicefrac{\gamma_{\theta }}{I_{\theta }}$} & \multicolumn{2}{c|}{$\nicefrac{\gamma_{\eta }}{I_{\eta }}$} & \multicolumn{2}{c}{$\nicefrac{\gamma_{\phi }}{I_{\phi }}$}\\
\hline
$-0$ & $.05$ & $-0$ & $.13$ & $0$ & $.58$ & $0$ & $.05$ & $0$ & $.49$ & $0$ & $.08$\\
\end{tabular}
\end{shadefundnumber}
\end{center}
\caption{Residual forces and torques on TM1-TM2 (forces) and TM1 only (torques).
  $F_{i}/m$ in $[\unit{nm\,s}^{-2}]$, $T_{i}/I_{i}$ in $[\unit{nrad\,s}^{-2}]$ for each $i$.}
\label{tab:resforces}
\end{table}

\begin{table}
\begin{center}
\begin{shadefundnumber}
\begin{tabular}{r@{}l|r@{}l|r@{}l|r@{}l|r@{}l|r@{}l|r@{}l}
\multicolumn{2}{c|}{$$} & \multicolumn{2}{c|}{$\nicefrac{F_x}{m}$} & \multicolumn{2}{c|}{$\nicefrac{F_y}{m}$} & \multicolumn{2}{c|}{$\nicefrac{F_z}{m}$} & \multicolumn{2}{c|}{$\nicefrac{\gamma_{\theta }}{I_{\theta }}$} & \multicolumn{2}{c|}{$\nicefrac{\gamma_{\eta }}{I_{\eta }}$} & \multicolumn{2}{c}{$\nicefrac{\gamma_{\phi }}{I_{\phi }}$}\\
\hline
\multicolumn{2}{c|}{${{\partial }_x}$} & $471$ & $.61$ & $5$ & $.09$ & $-3$ & $.63$ & $-6$ & $.91$ & $-31$ & $.67$ & $3$ & $.96$\\
\multicolumn{2}{c|}{${{\partial }_y}$} & $5$ & $.1$ & $161$ & $.37$ & $-5$ & $.69$ & $140$ & $.49$ & $2$ & $.0$ & $-697$ & $.89$\\
\multicolumn{2}{c|}{${{\partial }_z}$} & $-3$ & $.64$ & $-5$ & $.71$ & $-632$ & $.99$ & $10$ & $.88$ & $61$ & $.83$ & $4$ & $.9$\\
\multicolumn{2}{c|}{${{\partial }_{\theta }}$} & $-0$ & $.01$ & $13$ & $.95$ & $-0$ & $.13$ & $5$ & $.34$ & $-1$ & $.2$ & $2$ & $.65$\\
\multicolumn{2}{c|}{${{\partial }_{\eta }}$} & $-13$ & $.93$ & $0$ & $.0$ & $3$ & $.67$ & $-1$ & $.11$ & $88$ & $.92$ & $0$ & $.29$\\
\multicolumn{2}{c|}{${{\partial }_{\phi }}$} & $0$ & $.13$ & $-3$ & $.91$ & $0$ & $.0$ & $2$ & $.16$ & $0$ & $.34$ & $-2$ & $.09$\\
\end{tabular}
\end{shadefundnumber}
\end{center}
\caption{Residual stiffness, linear-angular and angular-angular gradients over TM1.
  Dimensions for each element are $ 10^{-9} [\unit{s}^{-2}]$.}
\label{tab:resgammas}
\end{table}

\subsection{Robustness and tests}

Several aspects of this strategy need to be clarified:
\begin{itemize}
\item the dependence of the SGI on the source knowledge (shape, position and density) needs further investigation
to render our results robust against small mass variation (due to assembly imprecision or design changes)
at a given distance. The definition of a ``gravitational protocol''
to discipline mass addition/removal from the Path-finder systems is in advanced progress \cite{ltpgravprot}.
In our analysis we crudely assumed
precise knowledge of the positioning of the subsystems blocks as well as their density;
\item inhomogeneity of the TMs has not been investigated, although such a scenario may be mapped
into a suitable point-like mass distribution in the proximity
of each TM generating an effective field to mimic bubbles, cracks
and surface defects. Nonetheless, in the present work we assumed perfect cubic TMs;
\item the compensation reliability depends on the knowledge of the CmpMs mass, shape and positioning. Therefore
we performed tests to shed light on uncertainties in the meshing procedure as well as in the placement
upon translations and rotations.
\end{itemize}

\subsubsection{Rotations and translations}
We displaced CmpM1 by roto-translating it, while CmpM2 was assumed to be perfectly placed.
Residual SGIs remain within specifications under small rotations but become
less robust in the process until exceeding the allowed values
for rotations with Euler angles $(\eta, \theta, \phi)~>~3 \times~10^{-3}\,\unit{rad}$. Stiffness begins to exceed
first along the $xy$, $yz$ couples.
The uncertainty in rotation upon placement of the CmpMs can be checked by measuring the position of the CmpMs sides and
shall not exceed the nominal position more than $300\,\unit{\mu m} = 3\cdot 10^{-3}\,\unit{rad} \cdot R_{VE}$,
with $R_{VE}=10\,\unit{cm}$ being the radius of the VE chamber.

The analysis on translations is similar to the one carried out for rotations.
By moving CmpM1 with a vector whose maximum length is $\sim 300\,\unit{\mu{m}}$,
predominantly in the sensitive $\hat x$ direction, parameters are seen to be within specifications.
Larger translation induce breakdown
either along cross directions (translations with dominant off-$\hat x$ terms) or
along the $\hat x$ direction of the force.

\subsubsection{Density and meshing, placing}
Since Tungsten cannot be easily purified beyond $95\%$ of the nominal density, we assumed
a $5\%$ error on density during simulation, mimicking impurities and defect in production using
an isotropic bubble distribution in the mesh.
Provided the mass is $2.51\pm 0.04\,\unit{kg}$ per CmpMs, compensation can be achieved in spite of the defects.
Of course, a larger volume is needed to reach the mass demand.

To circumvent potential problems of density inhomogeneity, the CmpMs can be
over-sized at first and then trimmed to the appropriate weight.
Maximum needed over-sizing is estimated to be around $2-3\,\unit{mm}$ which compensates for the $5\%$ mass defect;
recommended growing points are the ``back'' and side ``wings'' of the CmpMs (see figure \ref{fig:shapes}, right).

Additional checks on the ideal mesh element size have been performed. Each mesh brick field
is the result of an average of $~1000$ sub-bricks. This way the
size of $\sim 2\,\unit{mm}$ proves to be sufficiently fine for the purpose of our analysis.
Mass loss due to meshing of the volumes is under control and doesn't contribute more than $0.1\%$
of the calculated field values.

Generally speaking, both the problems of meshing and misplacement upon mounting can
be addressed analytically. Ordinary meshing software doesn't encounter any trouble when dealing
with straight corners and sharp surfaces, problems arise when curvature is at work: adherence to
a curved bounding manifold when mapping its volume with meshing bricks is necessarily
approximated. We can model it as follows: choose tetrahedrons as bricks and a sphere as the
target object. The inner volume of the sphere will not bring any trouble, even cubes could map it
correctly up to the surface. The number of external tetrahedrons adhering to the surface
is roughly given (no combinatorial) by the ratio of the sphere surface measure and tetrahedrons base area:
\beq
N \simeq \frac{4 \pi R^{2}}{\frac{\sqrt{3}}{4}L^{2}}\,,
\eeq
where we named $R$ the sphere radius and $L$ the tetrahedron side. In fact, more care
should be placed into combining the shapes on the surface and the integer value of the former must
be taken, but to zero-order the former is not wrong.

We may then assume each tetrahedron to extend till the sphere centre therefore becoming
a pyramid whose long side be $R$; the height can then be computed as:
\beq
h = \sqrt{R^{2}-\left(\frac{2}{3}\frac{\sqrt{3}}{2} L\right)^{2}}\,,
\eeq
and the volume as:
\beq
V=\frac{1}{3} h \frac{\sqrt{3}}{4}L^{2}\,.
\eeq
The relative meshing error is the difference in volume from real to estimated
divided by the real one itself:
\begin{shadefundtheory}
\beq
\epsilon_{\text{mesh}} = \frac{\frac{4}{3}\pi R^{3}-N V}{\frac{4}{3}\pi R^{3}}=
 1-\sqrt{1-\frac{1}{3}\left(\frac{L}{R}\right)^{2}}\,,
 \label{eq:epsmesh}
\eeq
\end{shadefundtheory}
\noindent as expected, the former goes to $0$ as $L\ll R$, and to first order in $\nicefrac{L}{R}$
we get:
\beq
\epsilon_{\text{mesh}} = \frac{1}{6}\left(\frac{L}{R}\right)^{2}
  +O\left(\left(\frac{L}{R}\right)^{2}\right)\,.
\eeq

On the other hand the misplacement error can be computed from the expression of the force. In modulus:
\beq
F = G \frac{m_{1}m_{2}}{r^{2}}\,,
\eeq
we have
\beq
\delta{F} = - 2 G \frac{m_{1}m_{2}}{r^{3}} \delta{r}\,,
\eeq
and
\begin{shadefundtheory}
\beq
\epsilon_{F}=\left|\frac{\delta{F}}{F}\right| = 2 \frac{\delta{r}}{r}\,.
\label{eq:epsf}
\eeq
\end{shadefundtheory}
\noindent Henceforth, the former representing the relative error on force upon displacement - undesired or not -
it also describes the dependence of the meshing length scale with respect of the distance scale
from the observer to confine force estimate error to a given value $\epsilon_{F}$.
Formula  \eqref{eq:epsmesh} together with \eqref{eq:epsf} are powerful tool to estimate
the minimal side of meshing brick and the relevance of mass addiction according to distance
from the TMs.

\subsection{Open issues. Gravitational control protocol}
The results show that the SGI can be compensated within the required levels.
At the end of this investigation, while masses, shapes and sizes appear to be reasonable,
a number of engineering challenges remain:
machinability and trimming, compatibility with both the mass and position of cabling,
and mounting procedures.

The calculation presented here is the first step toward assessing if the following gravitational 
compensation protocol may be followed:
\begin{enumerate}
\item based on nominal design of SC, LTP and DRS, calculation of the gravitational disturbances 
on the test-mass is performed and a first design of compensation masses is provided. This 
design allows the preliminary definition of the mechanical interfaces of the compensation 
masses and to tackle the interference problems;
\item at Critical Design Review (CDR), based on the available design knowledge of
SC and LTP, the design of 
interference-avoidance features and of mechanical interfaces is frozen;
\item a Gravitational System Review (GSR) is introduced within the planning to freeze the final 
trimming of the compensation mass design. This final trimming can only relate to minor 
adjustments  of the outer surfaces of compensation masses and cannot affect their 
mechanical interfaces. A change of these would indeed imply a redesign of the entire inertial sensor;
\item after GSR final manufacturing of compensation masses is performed. From there on 
compensation must be performed by masses outside the VE and a specific mechanical 
interface for that must be designed.
\end{enumerate}


%

\section{Calibrating force to displacement}
\label{sec:calibforcedisp}

\subsection{Calibration of force applied to TM1}
\label{app:calibtm1}

In order to calibrate the force applied to the first TM, expressed by the term $g_{1,x}$ in \eqref{eq:deltaxifofull},
\eqref{eq:x1ifofull} and derived ones, the most convenient signal to employ is in fact \eqref{eq:x1ifofull}. By
setting to $0$ all the spare contribution but for $g_{1,x}$, we see the signal equation reduces to:
\beq
\text{IFO}\left(x_{1}\right) \simeq h_{x,x_{1},x_{1}}(s) g_{1,x}=\frac{1}{s^{2}+\omega_{\text{df},x_{1}}^{2}+\omega_{x_{1},x_{1}}^{2}}\,.
\label{eq:ifox1g1xonly}
\eeq
Where all the terms have been discussed in chapter \ref{chap:ltp}. The former equation neglects the need for any calibration, a thing we know for sure to be false. Had we to complicate
the model so to introduce a bona-fide mimicking of this unbalance, a proportionality factor $\kappa_{\text{TM1}}$
would appear homogeneous to $h_{x,x_{1},x_{1}}$ but would also be placed to multiply $\omega_{\text{df},x_{1}}^{2}$
since the drag-free control loop makes use of the same readout apparatus to exert force, hence suffering the same lack of calibration.
Moreover, we can't assume anymore the full stiffness to be given by $\omega_{x_{1},x_{1}}^{2}$ only, we'll
hence replace the factor with a generic $\tilde\omega_{x_{1},x_{1}}^{2}$.

Expression \eqref{eq:ifox1g1xonly} gets thus modified and displays like:
\begin{shadefundtheory}
\beq
\text{IFO}\left(x_{1}\right) \simeq \frac{\kappa_{\text{TM1}}}{s^{2}+\kappa_{\text{TM1}}\omega_{\text{df},x_{1}}^{2}+\tilde\omega_{x_{1},x_{1}}^{2}}g_{1,x}\,.
\label{eq:ifox1g1xmod}
\eeq
\end{shadefundtheory}
We proceed to linearised the problem by introducing the following approximations:
\begin{shademinornumber}
\beq
\begin{split}
\tilde\omega_{x_{1},x_{1}}^{2}&\simeq \omega_{x_{1},x_{1}}^{2} + {\delta}k_{1}\,,\\
\kappa_{\text{TM1}}&\simeq 1+{\delta}k_{2}\,,
\end{split}
\eeq
\end{shademinornumber}
\noindent where obviously ${\delta}k_{2} \ll 1$ and ${\delta}k_{1}\ll \omega_{\text{df},x}^{2}$
together with ${\delta}k_{1}\ll \omega^{2}=-s^{2}$. By expanding \eqref{eq:ifox1g1xmod} to first order in
${\delta}k_{2}$ and ${\delta}k_{1}$ we find:
\beq
\text{IFO}\left(x_{1}\right) \simeq f_{0}(s)+f_{1}(s)+f_{2}(s)\,,
\eeq
where
\beq
\begin{split}
f_{0}(s) &= H_{0}(s) g_{1,x}\,,\\
f_{1}(s) &= H_{0}(s) H_{1}(s) {\delta}k_{1} g_{1,x} \,,\\
f_{2}(s) &= H_{0}(s) H_{2}(s) {\delta}k_{2} g_{1,x}\,,
\end{split}
\label{eq:ffortm1}
\eeq
and where
\begin{align}
H_{0}(s) &\doteq h_{x,x_{1},x_{1}}(s)\,,\\
H_{1}(s) &\doteq -h_{x,x_{1},x_{1}}(s)\,,\\
H_{2}(s) &\doteq 1-h_{x,x_{1},x_{1}}(s)\omega^{2}_{\text{df},x_{1}}\,,
\end{align}
The first component, $f_{0}$, represents the ideal signal, that would be measured under perfect calibration and accuracy conditions;
the latter two, $f_{1}$ and $f_{2}$ are due to parasitic stiffness and non-ideal conversion from actuation command and real
effect on the TM1. This way the transfer function is written in a form suitable for the application of the optimal filter
theory:
\begin{shademinornumber}
\beq
x = g_{1,x} H_{0}(s)\left(1+{\delta}k_{1} H_{1}(s)+{\delta}k_{2} H_{2}(s)\right)\,.
\eeq
\end{shademinornumber}

According to the Wiener-Kolmogorov optimal filter theory, a set of coefficients ${\delta}k_{i}\,,\,i=1,2$ in a
signal of the form
\beq
 x(t) = n(t)+A_{0}\left(f_{0}(t)+{\delta}k_{1} f_{1}(t)+{\delta}k_{2} f_{2}(t)\right)\,,
\eeq
may be estimated linearly up to optimal precision as follows. By subtracting the predictable $0$ component, we
introduce the deviation:
\beq
 \tilde x(t) \doteq  x(t)-A_{0} f_{0}(t)\,,
\eeq
and the linear combinations
\beq
\hat A_{i} \doteq \int_{\mathbb R} h_{i}(\tau) \tilde x(\tau) \de\tau\,
\eeq
defined so that the average values are homogeneous in the deviation coefficients:
\beq
\left\langle \hat A_{i} \right\rangle = A_{0} {\delta}k_{i}\,.
\label{eq:bonds}
\eeq
The $h_{i}(\tau)$ filter transfer functions are designed to minimise the root mean square errors of the estimates,
as follows:
\beq
\sigma^{2}_{\hat A_{i}} \doteq \int_{{\mathbb R}^{2}} h_{i}(\tau')h_{i}(\tau'')C(\tau'-\tau'')\de\tau'\de\tau''\,,
\label{eq:sigma2}
\eeq
where $C(\tau)$ is the auto-correlation function already defined in chapter \ref{chap:refsys}, linked to
the PSD $S(\omega)$ by the Fourier anti-transform (see \eqref{eq:convolution} and \eqref{eq:generalPSD}).
No summation is implied in the former equation.
By Lagrange multipliers maximisation of \eqref{eq:sigma2}
under the bonds \eqref{eq:bonds} the solution is found for the $h_{i}$ functions as
\begin{shadefundtheory}
\beq
h_{i}(\omega) = \frac{1}{S(\omega)}\sum_{j} \Lambda_{ij} f_{j}(\omega)\,,
\eeq
\end{shadefundtheory}
\noindent where $\Lambda$ is the covariance matrix associated to the optimal solution:
\beq
\Lambda_{ij} = \left\langle \hat A_{i} \hat A_{j} \right\rangle\,,
\eeq
defined as
\begin{shadefundtheory}
\beq
\Lambda_{ij}^{-1} \doteq \frac{1}{2\pi} \int_{\mathbb R}
  \frac{f_{i}(\omega) f_{j}^{*}(\omega)}{S(\omega)} \de\omega\,.
  \label{eq:lambdacovar}
\eeq
\end{shadefundtheory}

The covariance matrix for ${\delta}k_{1}$ and ${\delta}k_{2}$ is thus:
\beq
\Lambda = A_{0}^{2}\left[\begin{array}{cc}
\sigma^{2}_{{\delta}k_{2}} & r \sigma_{{\delta}k_{2}}\sigma_{{\delta}k_{1}} \\
r \sigma_{{\delta}k_{2}}\sigma_{{\delta}k_{1}} & \sigma^{2}_{{\delta}k_{1}}
\end{array}\right]\,,
\eeq
where we introduced the variances for ${\delta}k_{i}\,,\,i=1,2$ and their correlation coefficient $r$.

\begin{figure}
\begin{center}
\includegraphics[width=\textwidth]{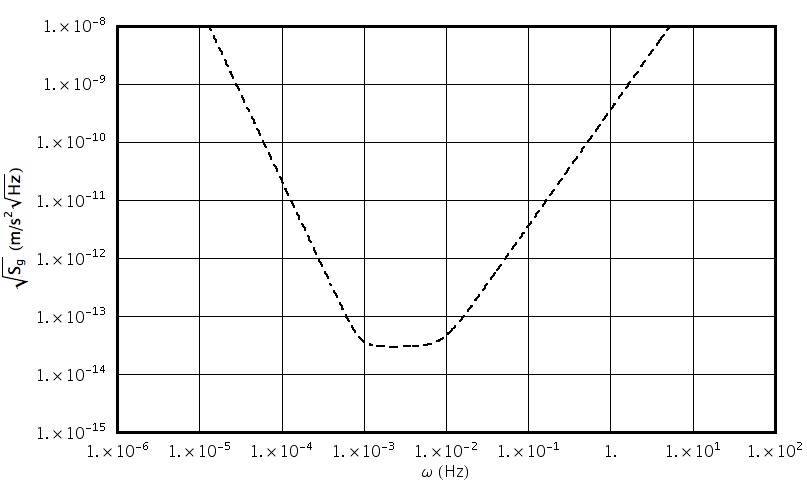}
\caption{Root squared PSD for $S_{g}(\omega)$.}
\label{fig:psdgstim}
\end{center}
\end{figure}

We now apply the optimal filter theory to the formerly defined expressions for $f_{i}$, expressions \eqref{eq:ffortm1}.
We get for the covariance matrix that:
\beq
\Lambda = \left[\frac{1}{2\pi}\int_{\mathbb R} H_{i}(\omega)H_{j}^{*}(\omega)\frac{\left|g_{1,x}(\omega)\right|^{2}}{S_{g}(\omega)}\de\omega\right]^{-1}\,,
\eeq
where
\beq
\begin{split}
S_{g}(\omega)\doteq \frac{S(\omega)}{\left|H_{0}(\omega)\right|^{2}}
 =& \left(3\times 10^{-14}\right)^{2} \unitfrac{m^{2}}{s^{4} Hz}\left(1+\left(\frac{2\pi\times 0.9}{\omega\nicefrac{1}{\unit{mHz}}}\right)^{6}\right)\\
   &+ \left(9\times 10^{-12}\right)^{2} \unitfrac{m^{2}}{Hz} \left(\omega^{2}+\tilde\omega_{x_{1},x_{1}}^{2}\right)^{2}\,.
\end{split}
\eeq
Such a choice depends on the performance of the interferometer ($9\,\nicefrac{\unit{pm}}{\sqrt{\unit{Hz}}}$)
and on the requirement on acceleration noise ($3\times 10^{-14}\,\nicefrac{\unit{m}}{\unit{s^{2}}\sqrt{\unit{Hz}}}$);
the noise in acceleration worsens rapidly below $1\,\unit{mHz}$ as a function of $\left(\nicefrac{\omega}{2\pi}\right)^{-6}$.
A graph may be viewed in picture \ref{fig:psdgstim}.

The calibration signal used for calibration ``in silico'' would be a bi-damped cosine with frequency $\nu_{0}$ and
characteristic damping time $\tau$:
\begin{shademinornumber}
\beq
g(t)=\exp \left(-\frac{\left|t\right|}{\tau}\right)\cos 2\pi\nu_{0}t\,.
\label{eq:push}
\eeq
\end{shademinornumber}
\noindent The functional behaviour of $\omega_{\text{df},x_{1}}^{2}$ can be read out of \eqref{eq:suspfun} together with
table \ref{tab:dfcoeff}.

Finally, $A_{0}$ was chosen so to avoid saturation and retain linearity in the signal response. Practically, this corresponds to the introduction of
two joined conditions:
\begin{shademinornumber}
\beq
\left|A_{0} f_{0}(t)\right| \leq 0.1\,\unit{\mu m}\quad \vee\quad A_{0} \leq 2\times 10^{-10}\,\nicefrac{\unit{m}}{\unit{s}^{2}}\,.
\eeq
\end{shademinornumber}
\noindent A simulation was then performed for a spread of values of the constants $\tau$ and $\nu_{0}$, keeping in mind as primary task
the estimate of ${\delta}k_{2} \simeq \kappa_{\text{TM1}}-1$, i.e. the deviation from unity of the conversion factor between the
wished actuation force and the real one. By finding the minimum of $\sigma_{{\delta}k_{2}}$, we estimate the injected signal parameters
to be:
\begin{shadefundnumber}
\beq
\tau =100\,\unit{s}\,,\qquad\nu_{0} = 2\times 10^{-2}\,\unit{s^{-1}}\,,
\eeq
\end{shadefundnumber}
\noindent with $A_{0}$ being always of the order $2\times 10^{-10}\,\nicefrac{\unit{m}}{\unit{s}^{2}}$ and
the standard deviations scoring:
\begin{shadefundnumber}
\beq
\begin{split}
\sigma_{{\delta}k_{1}} &= 1.02\times 10^{-5}\,\nicefrac{1}{\unit{s}^{-2}}\,,\\
\sigma_{{\delta}k_{2}} &= 2.17\times 10^{-3}\,.
\end{split}
\eeq
\end{shadefundnumber}

\begin{figure}
\begin{center}
\includegraphics[width=\textwidth]{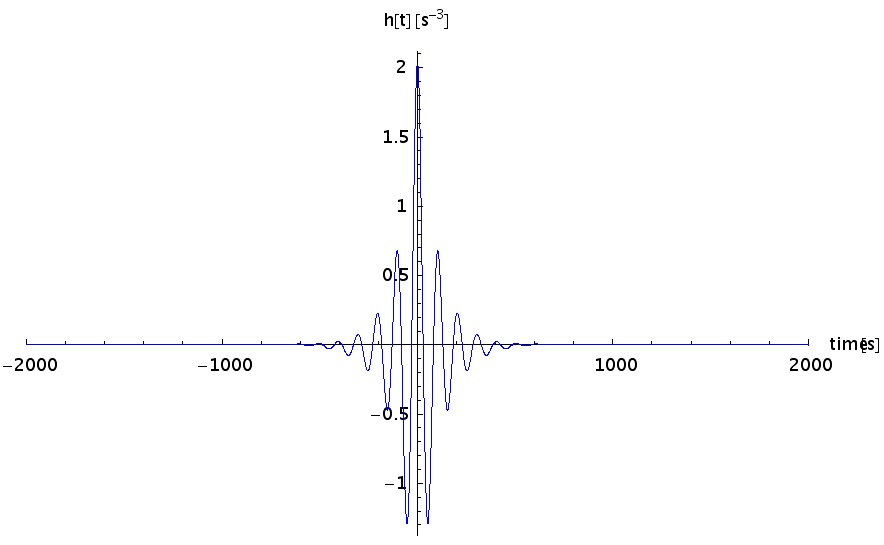}
\includegraphics[width=\textwidth]{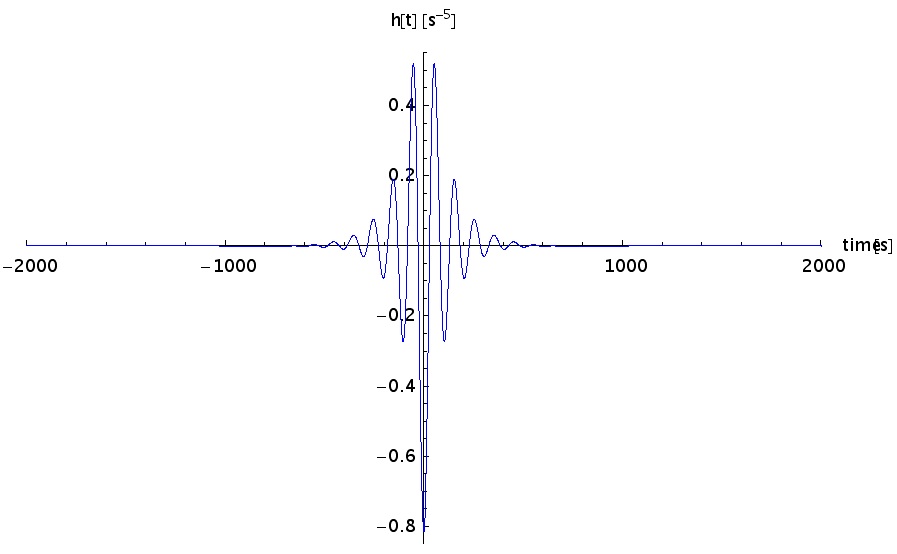}
\caption{Data filters for IFO$(x_{1})$ to obtain $H_{1}$ (filter on the top) and $H_{2}$ (bottom).}
\label{fig:hfilters}
\end{center}
\end{figure}

In practise, when given a real set of data from the channel IFO$(x_{1})$, convolution with the $h_{i}$ filters shown
in figure \ref{fig:hfilters} would obtain the filters $H_{i}$.

The reached precision, $\sigma_{{\delta}k_{2}} = 2.17\times 10^{-3}$, is far from satisfactory since the wished
goal would be below $10^{-4}$. In order to improve the result the signal IFO$(x_{2}-x_{1})$ can be employed
under the same pull $g_{1,x}$; by virtue of \eqref{eq:deltaxifofull} we get:
\begin{shadefundtheory}
\beq
\text{IFO}(x_{2}-x_{1}) \simeq \frac{-\kappa_{\text{TM1}}
\left(s^{2}+\kappa_{\text{TM1}}\omega^{2}_{\text{df},x}+\tilde\omega_{x_{2},x_{2}}\right)}
  {\left(s^{2}+\kappa_{\text{TM1}}\omega^{2}_{\text{df},x}+\tilde\omega^{2}_{x_{1},x_{1}}\right)
   \left(s^{2}+\kappa_{\text{TM2}}\omega^{2}_{\text{lfs},x}+\tilde\omega^{2}_{x_{2},x_{2}}\right)}g_{1,x}\,.
   \label{eq:ifox2-x1g1mod}
\eeq
\end{shadefundtheory}
\noindent More terms appeared and an additional uncertainty term has been produced for TM2 stiffness, together with a
$\kappa_{\text{TM2}}$ conversion factor which couples to the low frequency suspension as $\kappa_{\text{TM1}}$
does to the drag-free control.

We introduce linear deviation for the new terms as:
\begin{shademinornumber}
\beq
\begin{split}
\tilde\omega_{x_{1},x_{1}}^{2}&\simeq \omega_{x_{1},x_{1}}^{2} + {\delta}k_{1}\,,\\
\tilde\omega_{x_{2},x_{2}}^{2}&\simeq \omega_{x_{2},x_{2}}^{2} + {\delta}k_{2}\,,\\
\kappa_{\text{TM1}}&\simeq 1+{\delta}k_{3}\,,\\
\kappa_{\text{TM2}}&\simeq 1+{\delta}k_{4}\,,
\end{split}
\eeq
\end{shademinornumber}
\noindent where again ${\delta}k_{1}\ll \omega_{\text{df},x}^{2},\,\omega^{2}$, 
${\delta}k_{2}\ll \omega_{\text{lfs},x}^{2},\,\omega^{2}$,and ${\delta}k_{i} \ll 1$ for $i=3,4$.
Expression \eqref{eq:ifox2-x1g1mod} thus gives to first order in
${\delta}k_{i}$:
\beq
\Delta x = g_{1,x} H_{0}(s)\left(1+\sum_{i=1}^{4}{\delta}k_{i} H_{i}(s)\right)\,,
\eeq
where
\begin{align}
H_{0}(s)&=\frac{h_{x,x_{1},x_{1}}(s)h^{\text{lfs}}_{x,x_{2},x_{2}}(s)}{h_{x,x_{2},x_{2}}(s)}\,,\\
H_{1}(s)&=h_{x,x_{1},x_{1}}\,,\\
H_{2}(s)&=h^{\text{lfs}}_{x,x_{2},x_{2}}-h_{x,x_{2},x_{2}}\,,\\
H_{3}(s)&=\left(h_{x,x_{1},x_{1}}-h_{x,x_{2},x_{2}}\right)\omega_{\text{df},x}^{2}\,,\\
H_{4}(s)&=h^{\text{lfs}}_{x,x_{2},x_{2}}(s)\omega_{\text{lfs},x}^{2}\,.
\end{align}
The construction of the filter doesn't change, the $\Lambda$ covariance matrix being defined exactly as in
\eqref{eq:lambdacovar}. The expression for $\omega_{\text{lfs},x}^{2}$ can be retrieved from
\eqref{eq:suspfun} together with table \ref{tab:lfscoeff}.

\begin{figure}
\begin{center}
\includegraphics[width=\textwidth]{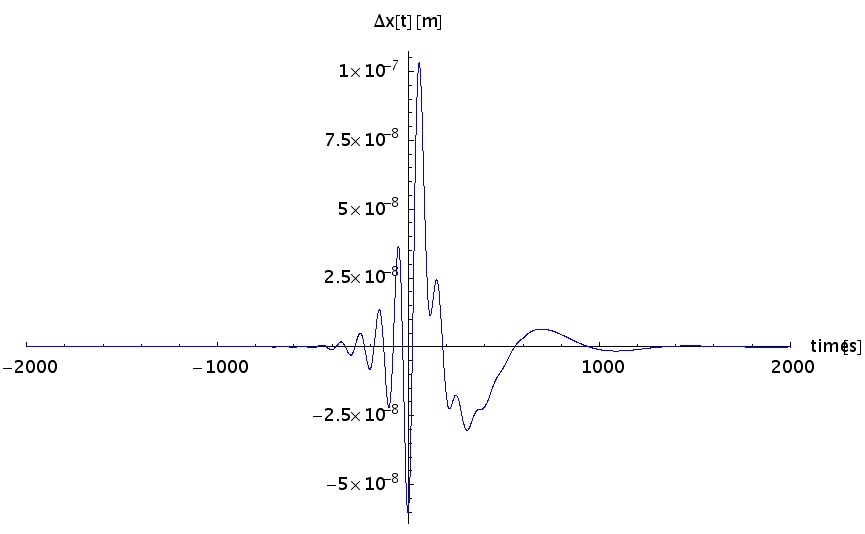}
\caption{IFO$(x_{2}-x_{1})$ signal corresponding to the pulse \eqref{eq:push} with parameters from \eqref{eq:param2}.}
\label{fig:ifox2-x1g1x}
\end{center}
\end{figure}

The minimal value of standard deviation for ${\delta}k_{3}$, correction to unity for the actuation calibration factor,
occurs for the following parameters of the pulse function:
\begin{shadefundnumber}
\beq
\tau=100\,\unit{s}\qquad \nu_{0}=10^{-1}\,\nicefrac{1}{\unit{s}}\,,
\label{eq:param2}
\eeq
\end{shadefundnumber}
\noindent for a correct value of $A_{0}$ within requirements and providing the following deviations:
\begin{shadefundnumber}
\beq
\begin{split}
\sigma_{{\delta}k_{1}} &= 2.72\times 10^{-6}\,\nicefrac{1}{\unit{s}^{2}}\,,\\
\sigma_{{\delta}k_{2}} &= 7.11\times 10^{-8}\,\nicefrac{1}{\unit{s}^{2}}\,,\\
\sigma_{{\delta}k_{3}} &= 5.53\times 10^{-5}\,,\\
\sigma_{{\delta}k_{4}} &= 1.69\times 10^{-4}\,.
\end{split}
\eeq
\end{shadefundnumber}
\noindent The out-signal is shown in figure \ref{fig:ifox2-x1g1x} and shows a larger time is needed to obtain the estimate,
roughly $2000\,\unit{s}$ integration time. Similarly to the IFO$(x_{1})$ case, special filter functions $h_{i}$, $4$
in numbers can be built and used as convolution patterns for real data coming from LTP in order to estimate the $H_{i}$.
Shapes of filters are similar to figure \ref{fig:hfilters} and won't be shown here.

Notice the absolute deviation is largely below threshold, but it might be even improved by using multiple frequencies or
averaging over different measures. 
Other sources of inaccuracy may be enumerated here, including vibrations in the laser wavelength or nonlinear effects, none of
them would nevertheless put at stake this procedure of progressive improvement of the estimate.

\subsection{Calibration of force applied to TM2}

The very same kind of analysis we carried on for TM1 could be applied to TM2 to evaluate the
calibration factor between ideal and real actuation exerted to the TM2.
It is convenient in this case to employ the signal IFO$(x_{2}-x_{1})$
so that the expression \eqref{eq:deltaxifofull} in presence of the sole $g_{2,x}$ stimulus becomes:
\begin{shadefundtheory}
\beq
\text{IFO}\left(x_{2}-x_{1}\right) \simeq \frac{\kappa_{\text{TM2}}}{s^{2}+\kappa_{\text{TM2}}\omega_{\text{lfs},x_{1}}^{2}+\tilde\omega_{x_{2},x_{2}}^{2}}\,.
\label{eq:ifox2-x1g2xmod}
\eeq
\end{shadefundtheory}
\noindent No wonder we can linearise the problem as before by:
\begin{shademinornumber}
\beq
\begin{split}
\tilde\omega_{x_{2},x_{2}}^{2}&\simeq \omega_{x_{2},x_{2}}^{2} + {\delta}k_{1}\,,\\
\kappa_{\text{TM2}}&\simeq 1+{\delta}k_{2}\,,
\end{split}
\eeq
\end{shademinornumber}
\noindent where again ${\delta}k_{2} \ll 1$ and ${\delta}k_{1}\ll \omega_{\text{lfs},x}^{2},\,\omega^{2}$.
To first order in
${\delta}k_{2}$ and ${\delta}k_{1}$ we find from \eqref{eq:ifox2-x1g2xmod}:
\beq
\text{IFO}\left(x_{2}-x_{1}\right) \simeq g_{x,2} H_{0}(s)\left(1+{\delta}k_{1}H_{1}(s)+{\delta}k_{2}H_{2}(s)\right)\,,
\eeq
where
\begin{align}
H_{0}(s) &\doteq h^{\text{lfs}}_{x,x_{2},x_{2}}(s)\,,\\
H_{1}(s) &\doteq -h^{\text{lfs}}_{x,x_{2},x_{2}}(s)\,,\\
H_{2}(s) &\doteq \left(s^{2}+\omega^{2}_{x_{2},x_{2}}\right)h^{\text{lfs}}_{x,x_{2},x_{2}}(s)\,.
\end{align}
What follows in the analysis is completely adherent to section \ref{app:calibtm1}, thus we write the result in damping and
frequency parameters which minimises the standard deviation as:
\begin{shadefundnumber}
\beq
\tau=100\,\unit{s}\qquad \nu_{0}=10^{-2}\,\nicefrac{1}{\unit{s}}\,,
\eeq
\end{shadefundnumber}
\noindent so that the bound on $A_{0}$ is respected and the deviations are:
\begin{shadefundnumber}
\beq
\begin{split}
\sigma_{{\delta}k_{1}} &= 5.37\times 10^{-8}\,\nicefrac{1}{\unit{s}^{-2}}\,,\\
\sigma_{{\delta}k_{2}} &= 2.93\times 10^{-5}\,.
\end{split}
\eeq
\end{shadefundnumber}
\noindent To get the estimate with the mentioned precision a pulse time of $\sim 2000\,\unit{s}$ is necessary. Filters for data processing may be
built as in previous cases, for more details, we refer to \cite{tateothesis}.

\subsection{Calibration of force applied to the SC}

To end this section, the same optimal filtering technique may be applied to calibrate the SC actuators, i.e. the identify
the real conversion factor between the control cycle and the FEEPs.
It is convenient in this case to employ the signal IFO$(x_{1})$ already used for calibrating the TM1
actuation-to-motion factor.

Expression \eqref{eq:ifox1g1xonly} gets contribution only from $g_{\text{SC,x}}$ and displays like:
\begin{shadefundtheory}
\beq
\text{IFO}\left(x_{1}\right) \simeq \frac{-\kappa_{\text{SC}}}{s^{2}+\kappa_{\text{TM1}}\omega_{\text{df},x_{1}}^{2}+\tilde\omega_{x_{1},x_{1}}^{2}}\,.
\label{eq:ifox1gscxmod}
\eeq
\end{shadefundtheory}
\noindent The following approximations will be put at play to linearise the problem:
\begin{shademinornumber}
\beq
\begin{split}
\tilde\omega_{x_{1},x_{1}}^{2}&\simeq \omega_{x_{1},x_{1}}^{2} + {\delta}k_{1}\,,\\
\kappa_{\text{TM1}}&\simeq 1+{\delta}k_{2}\,,\\
\kappa_{\text{SC}}&\simeq 1+{\delta}k_{3}\,,
\end{split}
\eeq
\end{shademinornumber}
\noindent where ${\delta}k_{2}, {\delta}k_{3} \ll 1$ and ${\delta}k_{1}\ll \omega_{\text{df},x}^{2}$
Expression \eqref{eq:ifox1g1xmod} to first order in
${\delta}k_{i}$ may be rewritten as:
\beq
\text{IFO}\left(x_{1}\right) \simeq g_{\text{SC},x} H_{0}(s)\left(1+\sum_{j=1}^{3}{\delta}k_{j} H_{j}(s)\right)\,,
\eeq
with
\begin{align}
H_{0}(s) &\doteq h_{x,x_{1},x_{1}}(s)\,,\\
H_{1}(s) &\doteq H_{0}\,,\\
H_{2}(s) &\doteq -\omega_{\text{df},x}^{2}h_{x,x_{1},x_{1}}(s)\,\\
H_{3}(s) &\doteq 1\,.
\end{align}
The analysis is completely adherent to section \ref{app:calibtm1}, thus we write the result in damping and
frequency parameters which minimises the standard deviation as:
\begin{shadefundnumber}
\beq
\tau=100\,\unit{s}\qquad \nu_{0}=2\times 10^{-2}\,\nicefrac{1}{\unit{s}}\,,
\eeq
\end{shadefundnumber}
\noindent the bound on $A_{0}$ is respected and the deviations are:
\begin{shadefundnumber}
\beq
\begin{split}
\sigma_{{\delta}k_{1}} &= 3.36\times 10^{-5}\,\nicefrac{1}{\unit{s}^{-2}}\,,\\
\sigma_{{\delta}k_{2}} &= 2.33\times 10^{-3}\,\\
\sigma_{{\delta}k_{3}} &= 2.47\times 10^{-3}\,.
\end{split}
\eeq
\end{shadefundnumber}
\noindent To get the estimate with the mentioned precision a pulse time of $\sim 1000\,\unit{s}$ is necessary. Filters for data processing may be
built as in previous cases \cite{tateothesis}.

\section{Experimental runs}

As the test wants to asses the ability to achieve free-fall with the technology envisaged for LISA, drag-free and actuation
control schemes must guarantee that all disturbances affecting the TMs be detectable at required levels \cite{LTPscrd}. The key
design features
of LTP are pointing to improve sensitivity and signal to noise ratio for the metrology but deep care is taken not to unwillingly
suppress sensitivity to disturbances that may affect LISA. For instance at least in some of the tests the coupling
between SC and TM1 will be mismatched with respect to the coupling between SC and TM2, $\omega_{p,1}^{2}\neq \omega_{p,2}^{2}$
in order to keep the sensitivity to the relative jitter of apparatus's.

We follow now with the list of experimental runs.

\subsection{Measurement of total acceleration in science mode}
\label{subs:run1m3}

In science mode scheme, the first measurement ever is the main mission goal, i.e. a measurement of the PSD of IFO$(x_{2}-x_{1})$ across
the entire MBW, taking $\omega_{p,1}^{2}$ minimum within requirements needed for operation without any actuation of TM1
along the $\hat x$ axis. Data must be taken with a rate and for a time span such that frequency resolution be $\leq 1\,\unit{mHz}$
and relative error amplitude of $50\%$ on each frequency sample. Obviously all metrology signals of any type (GRS and IFO)
will be recorded simultaneously - and under same rate and span - for cross-correlation analysis. In particular GRS$(x_{1})$ and
IFO$(x_{1})$ are mandatory in order to measure $S_{\nicefrac{g_{\text{SC},x}}{\omega_{\text{df},x}^{2}}}$ and
$S_{\text{GRS}_{n}(x_{1})}$.

The goal of the test is demonstrate that:
\beq
\left|\omega^{2}-\omega_{\text{lfs}}^{2}\right| S^{\nicefrac{1}{2}}_{\text{IFO}(x_{2}-x_{1})}
\leq 3\times 10^{-14} \left(1+\left(\frac{f}{3\,\unit{mHz}}\right)^{2}\right)\,\unitfrac{m}{s^{2}\sqrt{Hz}}
\,,
\eeq
for $1\,\unit{mHz}\leq f \leq 30\,\unit{mHz}$.
Such a performance must be achieved with $\left|\omega_{p,1}^{2}\right|$ and $\left|\omega_{p,2}^{2}\right|$
having the minimum values compatible with the operation of M3, as a joint condition we may state:
\beq
\left|\omega_{p,2}^{2}-\omega_{p,1}^{2}\right|
\leq 2\times 10^{-6} \left(1+\left(\frac{f}{3\,\unit{mHz}}\right)^{2}\right)\,\unitfrac{1}{s^{2}}\,.
\eeq

As a prerequisite to the measurement the total charge on the TM must be measured before the main measurement. The
TM charge must be then reduced to the value required to meet the noise performance via UV beam injection.

The measurement must be performed again after interchanging the role of TM1 and TM2.

Transfer function calibration from force to IFO$(x_{2}-x_{1})$ in M3 operating conditions must be known
with $5\%$ accuracy within the entire MBW.

\subsection{Measurement of acceleration noise in nominal mode}

In perfect analogy with the former run, a measurement of the acceleration noise must be carried out in nominal mode.
The goal is to demonstrate that
\beq
\omega^{2}S^{\nicefrac{1}{2}}_{\text{IFO}(x_{2}-x_{1})}
\leq 3.6\times 10^{-14} \left(1+\left(\frac{f}{3\,\unit{mHz}}\right)^{2}\right)\,\unitfrac{m}{s^{2}\sqrt{Hz}}
\,,
\eeq
for $1\,\unit{mHz}\leq f \leq 30\,\unit{mHz}$. The advantage of M1 is its self-calibration feature (no additional stiffness
constant in the frequency filter propagator); The requirement has been relaxed to fit the potentially needed relaxation on the
stiffness requirement.

Again, $\left|\omega_{p,1}^{2}\right|$ and $\left|\omega_{p,2}^{2}\right|$
must be kept within minimum values ranges compatible with the operation of M1, moreover, due to the specific form of
\eqref{eq:mainsignalsimplem1}, an additional requirement comes out:
\beq
\left|\omega_{p,2}^{2}-\omega_{p,1}^{2}+\omega_{\text{lfs}}^{2}\right|
\leq 4\times 10^{-6} \left(1+\left(\frac{f}{3\,\unit{mHz}}\right)^{2}\right)\,\unitfrac{1}{s^{2}}\,.
\eeq
All requirements that apply to run 1 in \ref{subs:run1m3} do apply here too.

\subsection{Measurement of internal forces}
\label{subs:intforces}

A measurement of the PSD of IFO$(x_{2}-x_{1})$ across the entire MBW with operating condition such that
$\omega_{p,1}^{2} \simeq \omega_{p,2}^{2}$ in M3 mode is representative of the PSD of internal forces
at play on the TMs. In fact, assuming $\left|{\delta}x_{2}-{\delta}x_{1}\right| \ll \text{IFO}_{n}(x_{2}-x_{1})$,
eq. \eqref{eq:mainsignalsimple} reduces to \eqref{eq:mainsignalsupersimple}.

Parasitic stiffness must be adjusted by applying an AC-voltage bias to TM2 GRS to minimise the difference to
the level:
\beq
\left|\omega_{p,2}^{2}-\omega_{p,1}^{2}\right|
\leq 2\times 10^{-7} \left(1+\left(\frac{f}{3\,\unit{mHz}}\right)^{2}\right)\,\unitfrac{1}{s^{2}}\,.
\eeq

The goal of the measurement is to demonstrate that:
\beq
S^{\nicefrac{1}{2}}_{\text{IFO}(x_{2}-x_{1})}
\leq 2.8\times 10^{-14} \left(1+\left(\frac{f}{3\,\unit{mHz}}\right)^{2}\right)\,\unitfrac{m}{s^{2}\sqrt{Hz}}\,,
\eeq
for $1\,\unit{mHz}\leq f \leq 30\,\unit{mHz}$. Due to the form of \eqref{eq:mainsignalsupersimple} it
is transparent that:
\beq
S^{\nicefrac{1}{2}}_{\text{IFO}(x_{2}-x_{1})} \simeq \frac{1}{\left|\omega _{\text{lfs},x}^2-\omega ^2\right|}
S^{\nicefrac{1}{2}}_{\left(g_{2,x}-g_{1,x}\right)}\,,
\eeq
thus allowing to cast an upper limit on $S^{\nicefrac{1}{2}}_{\left(g_{2,x}-g_{1,x}\right)}$ over all the MBW.

All requirements that apply to run 1 in \ref{subs:run1m3} do apply here too.

\subsection{Stiffness calibration and thrust noise determination}
\label{sec:stiffcalib}

This will be essentially a measurement of $\omega_{p,1}^{2} - \omega_{p,2}^{2}$ in M3 mode.
The measurement can be performed
by adding a sinusoidal drive signal $g_{\text{drv}}$ to GRS$(x_{1})$ and by measuring the coherent
response of IFO$(x_{2}-x_{1})$. By the form of \eqref{eq:mainsignalsimple} with small $\omega$, a drive
signal added to the GRS$(x_{1})$ channel will behave functionally as an additional noise like IFO$_{n}(x_{1})$,
hence:
\beq
\text{IFO}(x_{2}-x_{1}) \simeq \frac{\omega_{p,2}^{2} - \omega_{p,1}^{2}}{\omega_{\text{lfs}}^{2}-\omega^{2}}g_{\text{drv}}\,.
\eeq

The estimator must be evaluated as a function of the AC-voltage bias on TM2 to help separate voltage-dependent from
voltage-independent contributions: at least 4 voltage points must be taken. The test must be repeated at least at $f=3\,\unit{mHz}$
and at $f=30\,\unit{mHz}$, with interchange of TM1 and TM2 and must be better in accuracy than $5\times 10^{-8}\,\unit{s}^{-2}$.

Stiffness on TM1 must be compensated by an electrostatic suspension with a frequency independent loop gain $\omega_{\text{cp}}^{2}$
such that:
\beq
\left|\omega_{p,1}^{2}-\omega_{\text{cp}}^{2}\right| \leq 5\times 10^{-8}\,\unitfrac{1}{s^{2}}\,.
\eeq

\subsection{Measurement of cross-talk}

A measurement of the force induced along the sensitive axis by the motion of TM relative to SC along all other degrees of
freedom is mandatory to gain knowledge of cross-talk effects. The test is performed in M3 mode
 by applying a set of sinusoidal drives, at different frequencies
for different DOF, to electrode or drag-free loops. The coherent component of IFO$(x_{2}-x_{1})$
is then detected at each frequency.

Accuracy must be achieved up to $10\%$ of maximum allowed value for each cross-talk coefficient.

\subsection{Test of continuous charge measurement}
\label{subs:contcharge}

This measure won't differ in strategy from the main acceleration noise measurement in \ref{subs:run1m3}. In addition,
a voltage dither along another DOF is kept on bathing TM1 permanently at the proper frequency. The
coherent response along the same DOF is then measured in order to continuously detect the force
on TM1 due to the interaction of the TM charge with the dither voltage.

A charge resolution of $10^{4}$ electron charges over a measuring time $T=1000\,\unit{s}$ is requested due to
frequency and averaging issues. The coherent response to the voltage dither is also measured within IFO$(x_{2}-x_{1})$
in search for cross-correlation.

\subsection{Test of continuous discharge}

As in sect. \ref{subs:contcharge}, a voltage dither for charge measurement is applied to TM1. UV light is shone
on TM1 and electrode housing of IS1 within a control loop to null the TM charge. The goal is measuring total acceleration
noise in M3 mode as in \ref{subs:run1m3}. Requirements on the discharge control loop are optimised on the basis of
the following strategy:
\begin{itemize}
\item loop must be operating on continuous feed-back action,
\item measurement will be performed with dither light intermittently on to establish rate of charge deposition and
feed-forward,
\item residual charge on TM will be kept within $10^{5}$ electron charges at all times.
\end{itemize}

\subsection{Drift mode}

By means of switching off any low frequency suspension loop controlling one TM per time, a measurement of
residual drift on the LTP TMs can be performed. Control loop gain will be set to $0$ in M3 mode
and estimators of the following quantities will be measured:
\begin{itemize}
\item displacement of the unsuspended TM relative to the one driving the drag-free. The main goal of
this measurement is estimating the DC-force acting on the TM; measurement time is estimated to
be $T\leq 10000\,\unit{s}$ in order to average over a large frequencies span.
\item After estimating uniform acceleration, its contribution will be subtracted from the PSD, thus
providing acceleration fluctuations in MBW.
\end{itemize}

\subsection{Acceleration at different working points}

A static DC-offset can be added to the reference signal, thus inducing a shift in TM position. Measuring total acceleration
noise in M3 mode as in \ref{subs:run1m3} elucidates on breakdown of linearity in the capacitance to force model
and provides information on the correlation between the geometry of the GRS and its actuation capabilities.
Signal addition can be performed both on the drag-free controlling TM or on the low frequency suspension.

TM positions are displaced by up to $100\,\unit{\mu m}$ relative to the unbiased nominal working points of both loops.

\subsection{Acceleration noise measurement at $f < 1\,\unit{mHz}$}

A measurement of the main science signal as is \ref{subs:run1m3}, M3 mode, can be performed upon
MBW ranging from $0.1\,\unit{mHz}$ and $30\,\unit{mHz}$. The goal of the run is to put an upper limit to
disturbances in the reduced frequency range across $0.1\,\unit{mHz}$ to $1\,\unit{mHz}$.

Data must be taken with a rate and for a time span to achieve frequency resolution $\leq 0.1\,\unit{mHz}$ and
relative error amplitude of $50\%$ on each frequency sample.

Naturally all available metrology signals will be recorded simultaneously to the main science IFO for cross-correlation
analysis. Diagnostic of signals compatible with low-noise operation will be recorded too.
In analogy to what discussed in sect. \ref{subs:run1m3}, residual acceleration and readout noise PSDs
will be estimated.

\subsection{Sensitivity to magnetic fields and thermal gradients. Estimate of parasitic DC potential}

By application of the same conditions as in sect. \ref{subs:intforces} a set of measurements can be performed.
Purposeful conjuring of disturbances within MBW sheds light to apparatus sensitivity:
\begin{description}
\item[magnetic field gradients] can be applied at TM position, sufficient to detect TM response at $2\%$ resolution
with integration time $T\simeq 3600\,\unit{s}$;
\item[temperature gradients] may be imposed too, enough to detect TM response to radiometer effect with same
resolution and integration time $T$ as for the magnetic field gradients.
\end{description}

Moreover, a set of measurements can be performed biasing one of the TM motion with a low frequency
dither voltage within MBW applied to injection electrodes. Voltage must be set to give the TM the same
potential to ground as $2\times 10^{7}$ electron charges. Coherent displacement along every $\hat x$
IFO channel will me measured.

Similarly, the amount of DC-bias applied to $\hat x$ electrodes may be made vary to detect phase
changes in the response. The goal of the measurement is to estimate the effective parasitic dc-potential interacting with
the TM charge. Requested resolution in voltage is $1\,\unit{mV}$.

\section{A detailed measurement: charging and discharging the proof-mass}
\label{sec:chargedisctm}

\subsection{Introduction}

TM2 is always subject to
the electrostatic suspension around $\hat \theta_{2}$; we'll deal then
with the verification of charge and discharge procedures by employing the
rotational conjugated $\hat x$ signal for TM2, i.e. GRS$(\theta_{2})$, for
two main reasons:
\begin{enumerate}
\item TM1 is subject to a much higher level of noise, being subdued to drag-free control
along $\hat\theta_{1}$;
\item rotation around $\hat x$ will always be electrostatic controlled in LISA, thus
this case is mostly significant in LTP as prototype of the LISA case.
\end{enumerate}
Obviously nothing prevents to perform the same test on TM1.

In absence of strong correlation between $\hat x$ and $\hat\theta$ induced
by highly non-trivial cross-talk, a continuous roll around $\hat\theta$ should be
almost decoupled from $\hat x$, thus minimising scientific data contamination.

This collection of statements deeply motivates the choice of $\hat\theta_{2}$
as ``charge management'' DOF; nevertheless charge measurement can be carried
on along every IFO direction, taking advantage of the higher sensitivity of
the interferometer.

The measurement can be carried on the two TMs separately or contemporaneously.

\subsection{Tension characteristics}

The charge is measured by biasing the TM via $4$ electrode skew-placed along $\hat y$
to induce the roll around $\hat \theta$, see figure \ref{fig:ltptmeh} for reference, the
electrodes are numbered $5$ to $8$. The applied tension is:
\begin{shademinornumber}
\beq
V_{0}\cos\omega_{0}t\,,
\eeq
\end{shademinornumber}
\noindent so that if a charge $Q$ is located on the surface of the TM, a torque around $\hat x$
shows up. If no residual DC current contribution is left and electrodes be perfectly calibrated
and alike, the torque will suffer no phase shift and might be easily deduced to behave as:
\begin{shademinornumber}
\beq
\gamma_{Q}(t)=Q\frac{4 V_{0}}{C_{\text{tot}}} \partial_{\theta}C\Bigr|_{\theta=0}\cos\omega t\,.
\label{eq:chargetorque}
\eeq
\end{shademinornumber}
\noindent Most of the symbols in the former equation have been clarified in the noise chapter. We remind the measured
value of those as follows:
\beq
\begin{split}
C_{0}&=0.83\,\unit{pF}\,,\\
\partial_{\theta}C_{0} &= 3.1\,\unitfrac{pF}{rad}\,,\\
C_{\text{tot}}&=25.6\,\unit{pF}\,.
\end{split}
\eeq
We may rewrite \eqref{eq:chargetorque} as:
\beq
\gamma_{Q}(t)=N_{e}q_{0}V_{0}\cos\omega t\,,
\eeq
where $N_{e}$ is the number of elementary charges and $q_{0}=7.8\times10^{-20}\,\unit{C}$. In presence
of residual tensions on the electrodes an additional contribution independent of the accumulated charge shows
up, proportional to the effective charge:
\beq
\sum_{k=1}^{N} C_{k} V_{k}\,,
\eeq
where $N$ is the number of electrodes at non-zero potential. Systematic errors may occur as a consequence of
the additional term, conversely this shall be of little influence on measuring charge variations being a constant term.
Moreover, it is always possible to compensate for extra electrodes potential before entering measure phase. The
signal at $2\omega$ may be employed to test for unforecasted effects, such as skewness in the electrodes or capacitance
patches.

\subsection{Angular displacement signal}

\begin{figure}
\begin{center}
\includegraphics[width=\textwidth]{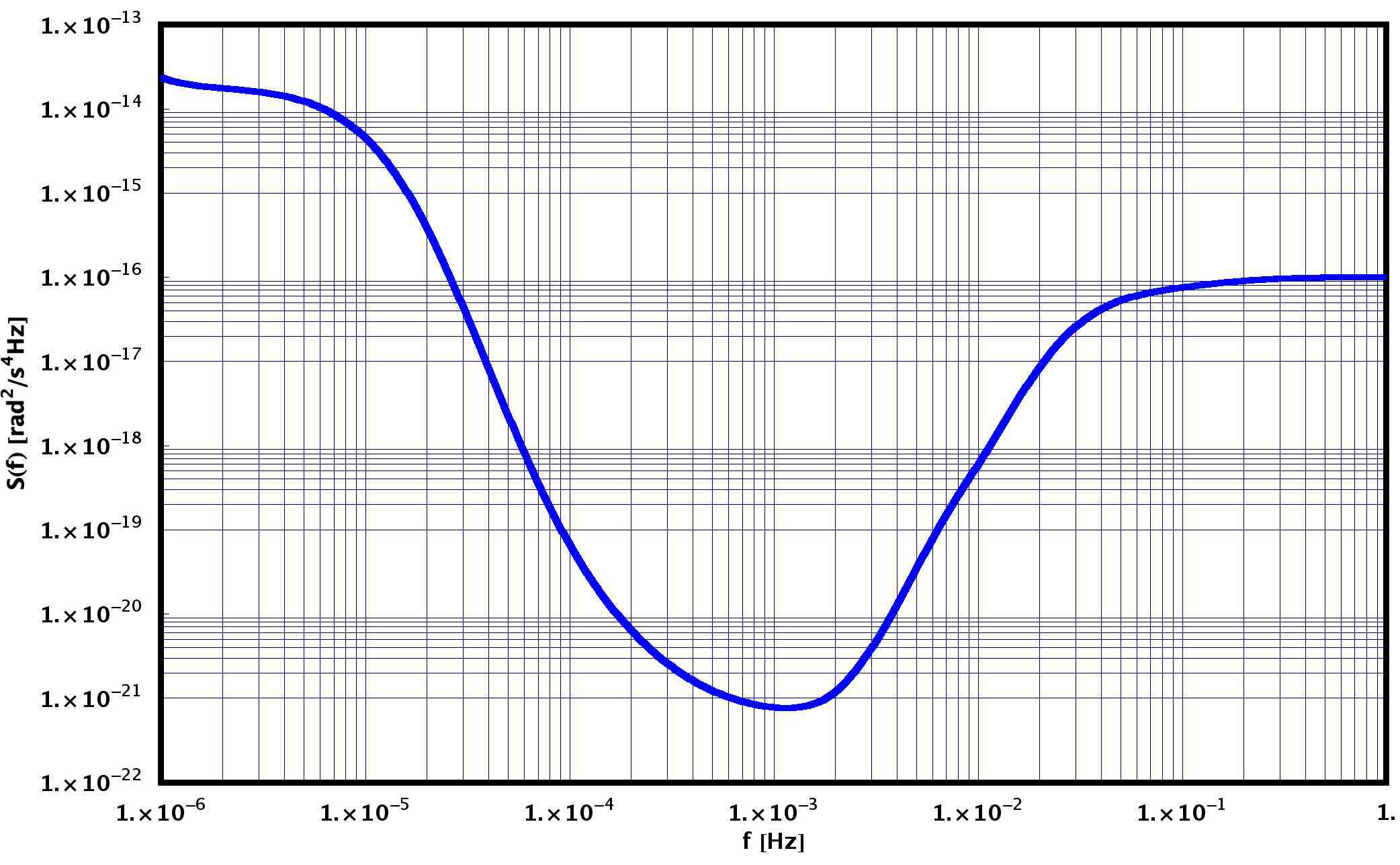}
\caption{PSD of torque noise per unit moment of inertia equivalent to the PSD in \eqref{eq:equivtorspec}.}
\label{fig:psdeqtorque}
\end{center}
\end{figure}

As motivated, the most suitable signal for the charge analysis is the GRS $\hat\theta_{2}$ one, in the form:
\begin{shadefundtheory}
\beq
\begin{split}
\text{GRS}(\theta_{2})(\omega)=&h_{\theta ,\theta _2,\theta_2}^{\text{lfs}}(\omega )
   \Bigl(-\dot{\Omega }_{1,\theta } h_{\theta ,\theta _2,\theta _2}(\omega ) \omega _{\text{df},\theta }^2+\dot{\Omega
   }_{2,\theta }+\left(\omega _{\theta _2,\theta _2}^2-\omega ^2\right) \text{GRS}_n\left(\theta _2\right)\\
   &+\left(\omega
   ^2-\omega _{\theta _1,\theta _1}^2\right) \left(\text{GRS}_n\left(\theta _1\right) \omega _{\text{df},\theta }^2+\dot{\Omega
   }_{\text{SC},\theta }\right) h_{\theta ,\theta _2,\theta _2}(\omega )\Bigr) \,,
\end{split}
\label{eq:grstheta2sig}
\eeq
\end{shadefundtheory}
\noindent the terms and propagators have been deduced and described in chapter \ref{chap:ltp}. The presence of
the torque \eqref{eq:chargetorque} converts into an additional local angular acceleration term whose Fourier transform
per unit moment of inertia $I_{\theta}$ scales like:
\beq
h^{\text{lfs}}_{\theta,\theta_{2},\theta_{2}}\frac{\gamma_{Q}(\omega)}{I_{\theta}},
\eeq
hence, the equivalent PSD may be deduced from \eqref{eq:grstheta2sig}:
\begin{shadefundtheory}
\beq
\begin{split}
S_{\gamma_{Q}}(\omega)&=I^{2}_{\theta}\Bigl(
S_{\dot{\Omega }_{1,\theta }} \left|h_{\theta ,\theta _2,\theta _2}(\omega ) \omega _{\text{df},\theta }^2\right|^{2}
+S_{\dot{\Omega}_{2,\theta }}+\left|\omega _{\theta _2,\theta _2}^2-\omega ^2\right|^{2} S_{\text{GRS}_n\left(\theta _2\right)}\\
   &+\left|\omega
   ^2-\omega _{\theta _1,\theta _1}^2\right|^{2} \left(S_{\text{GRS}_n\left(\theta _1\right)} \left|\omega _{\text{df},\theta }^2\right|^{2}
   +S_{\dot{\Omega}_{\text{SC},\theta }}\right) \left|h_{\theta ,\theta _2,\theta _2}(\omega )\right|^{2}\Bigr) \,,
\end{split}
\label{eq:equivtorspec}
\eeq
\end{shadefundtheory}
\noindent where the drag-free and LFS control functions $\omega^{2}_{\text{df},\theta}$, $\omega^{2}_{\text{lfs},\theta}$ can
be retrieved from \eqref{eq:suspfun} together with table \ref{tab:dfcoeff} and table \ref{tab:lfscoeff}.
A picture of \eqref{eq:equivtorspec} can be found in figure \ref{fig:psdeqtorque}.

More terms appear in \eqref{eq:equivtorspec} and need a comment:
\begin{description}
\item[$S_{\text{GRS}_n\left(\theta _1\right)}$, $S_{\text{GRS}_n\left(\theta _2\right)}$] are the PSDs for the GRS sensors
noise along the angular $\hat\theta_{1}$ and $\hat\theta_{2}$ directions. These may be derived from the sensitivity of the $\hat\theta$ degree
of freedom, multiplied by an educated guess filter to prevent poles mixing and ease numerical estimate below $1\,\unit{mHz}$:
\begin{shadefundnumber}
\beq
S_{\text{GRS}_n\left(\theta _1\right)} = S_{\text{GRS}_n\left(\theta _2\right)}
= 10^{-14}\left(1+
  \frac{\left(1+\frac{\omega_{e,0}^{2}}{\omega_{e,1}^{2}}\right) \left(1+\frac{\omega_{e,0}^{2}}{\omega_{e,2}^{2}}\right)}
  {\left(1+\frac{\omega^{2}}{\omega_{e,1}^{2}}\right) \left(1+\frac{\omega^{2}}{\omega_{e,2}^{2}}\right)}\right)
  \,\unitfrac{rad^{2}}{Hz}\,,
\eeq
\end{shadefundnumber}
\noindent where $\omega_{e,0}=2\pi 0.5\,\unit{mHz}$, $\omega_{e,1}=2\pi 0.10\,\unit{mHz}$, $\omega_{e,2}=2\pi 0.11\,\unit{mHz}$.
A picture of both $S_{\text{GRS}_n\left(\theta _1\right)}$, $S_{\text{GRS}_n\left(\theta _2\right)}$ PSDs together
with the relative noise models as functions of frequency $\omega$ may be inspected in figure \ref{fig:thetanoiseandpsd}.

\begin{figure}
\begin{center}
\includegraphics[width=\textwidth]{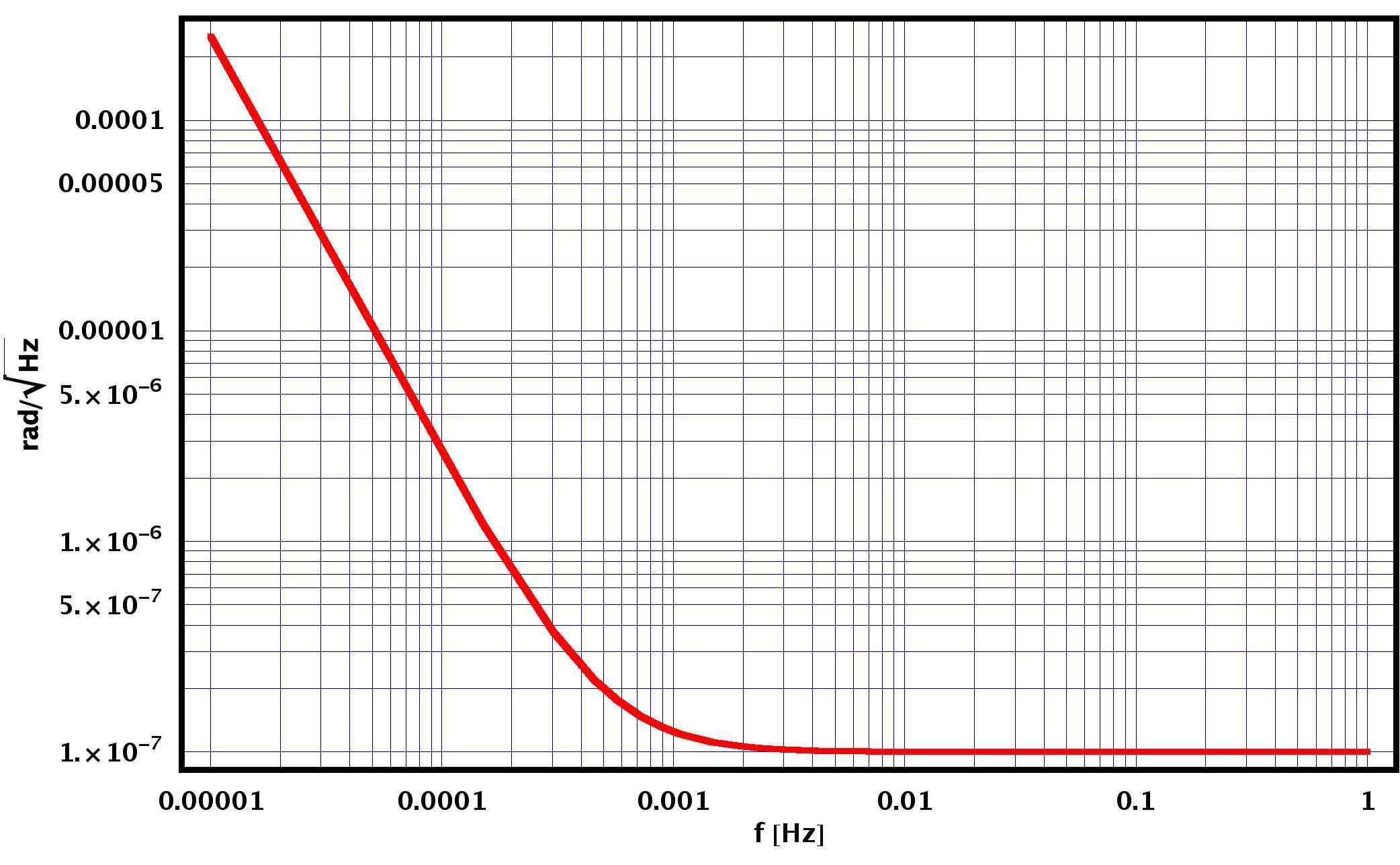}
\includegraphics[width=\textwidth]{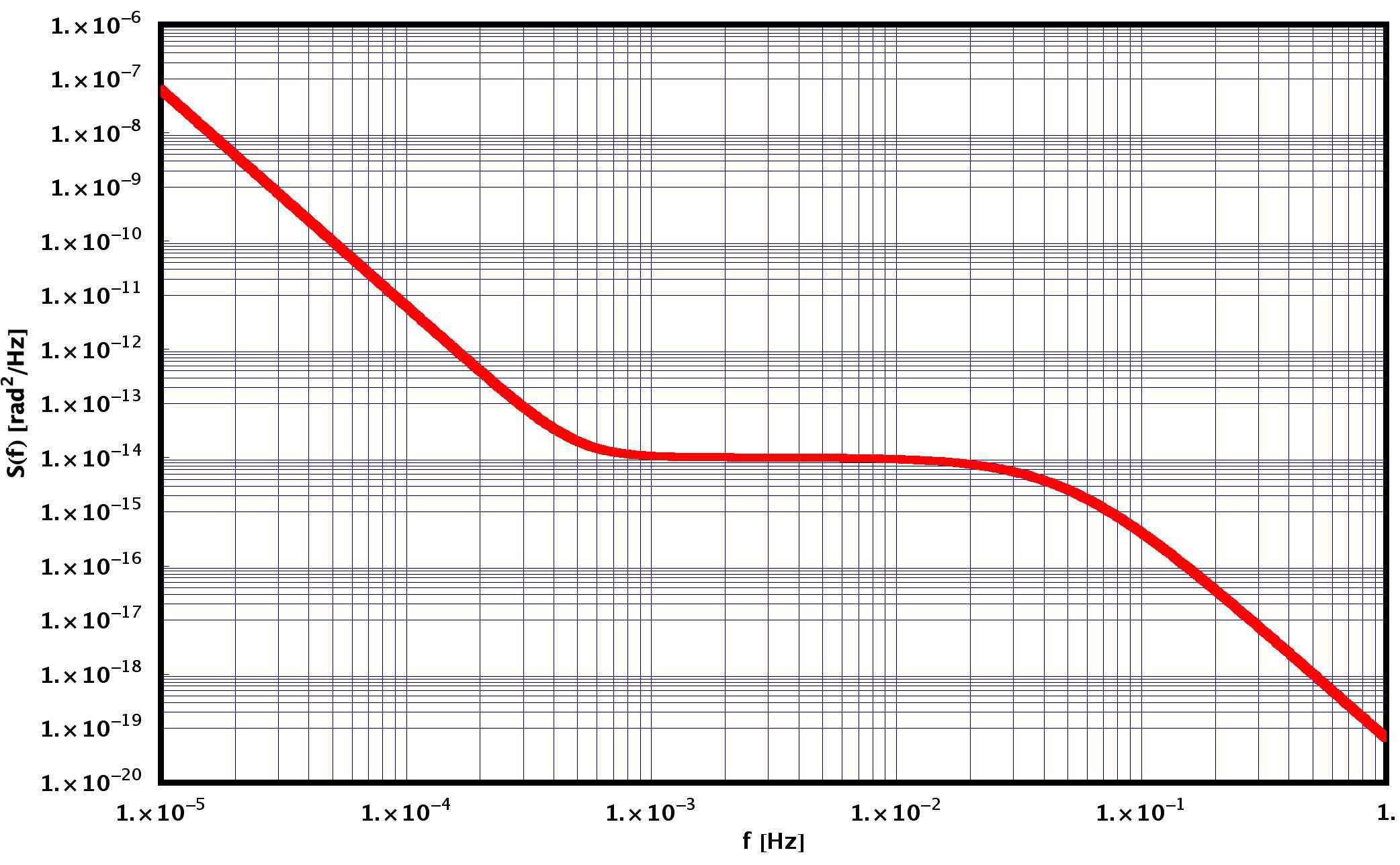}
\caption{Noise from angular displacement sensor (top) and its squared PSD (bottom).}
\label{fig:thetanoiseandpsd}
\end{center}
\end{figure}

\item[$S_{\dot{\Omega }_{1,\theta }}$, $S_{\dot{\Omega }_{2,\theta }}$] may be deduced from the actuation and measurement noises
around $\hat\theta$. The two contributions will be square summed, to get:
\begin{shadefundnumber}
\beq
S_{\dot{\Omega }_{1,\theta }}=S_{\dot{\Omega }_{2,\theta }} =
\frac{25^{2}+10^{2}}{20.35^{2}}
\left|\frac{1.3\times 10^{-11}(s+0.0006)(s+0.0005)^{2}(s+0.0003)}
  {\left(s+9\times 10^{-5}\right)\left(s+9.5\times 10^{-5}\right)^{2}\left(s+0.0001\right)}\right|^{2}\,
  \unitfrac{rad^{2}}{s^{4}\,Hz}\,.
\eeq
\end{shadefundnumber}
\noindent Figure \ref{fig:angactnoiseandpsd} depicts both the PSDs $S_{\dot{\Omega }_{1,\theta }}$, $S_{\dot{\Omega }_{2,\theta }}$
and the actuation noise leading to those.

\begin{figure}
\begin{center}
\includegraphics[width=\textwidth]{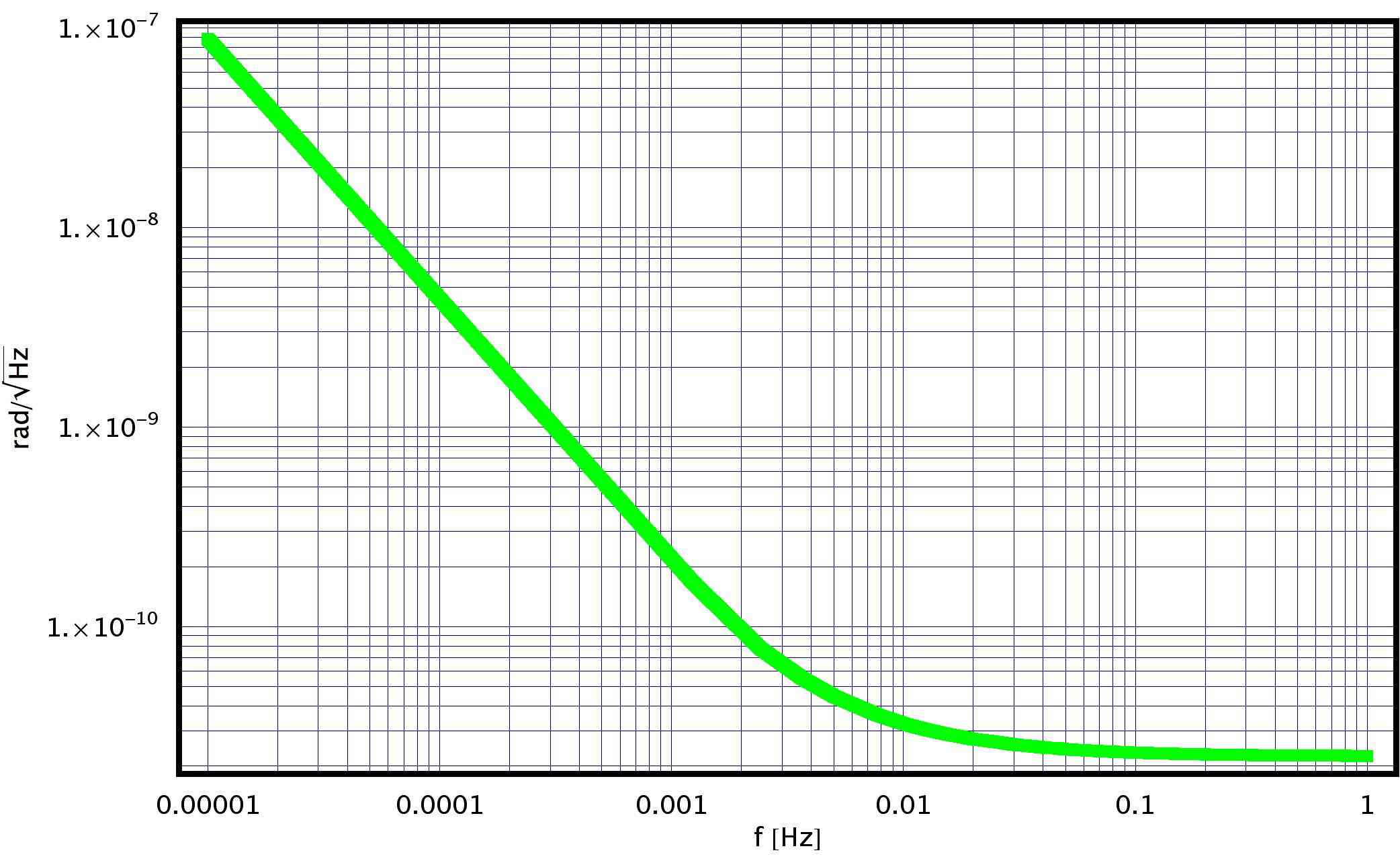}
\includegraphics[width=\textwidth]{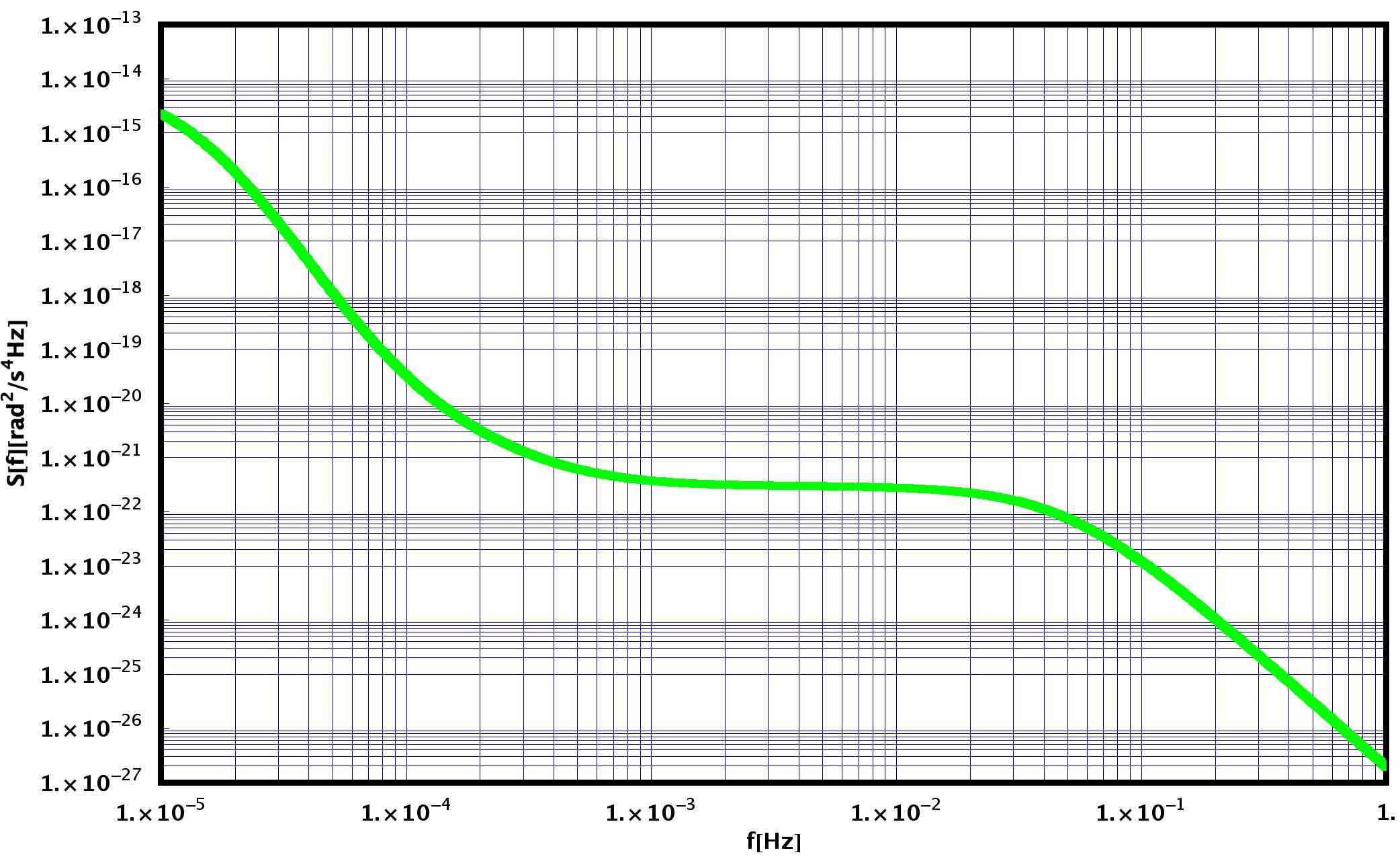}
\caption{Noise from angular actuation (top) and its squared PSD (bottom).}
\label{fig:angactnoiseandpsd}
\end{center}
\end{figure}

\item[$S_{\dot{\Omega}_{\text{SC},\theta }}$] may be calculated by multiplying each thruster noise times the
number of thrusters $(6)$ and considering an effective torsion arm of $1\,\unit{m}$:
\begin{shadefundnumber}
\beq
S_{\dot{\Omega}_{\text{SC},\theta }} = \frac{6\times 1\,\unit{m}}{(197.4\,\unit{kg\,m^{2}})^{2}}
\left(10^{-14}+4\times 10^{-8}\frac{1}
  {\left(1+\frac{\omega^{2}}{\omega_{e,3}^{2}}\right)\left(1+\frac{\omega^{2}}{\omega_{e,4}^{2}}\right)}\right)
  \,\unitfrac{rad^{2}}{s^{4}\,Hz}\,.
\eeq
\end{shadefundnumber}
\noindent The value $197.4\,\unit{kg\,m^{2}}$ is the measured SC moment of inertia and
$\omega_{e,3}=2\pi 0.20\,\unit{mHz}$, $\omega_{e,4}=2\pi 0.21\,\unit{mHz}$.
\item[$\omega^{2}_{\theta_{1},\theta_{1}}$, $\omega^{2}_{\theta_{2},\theta_{2}}$] may be
taken respectively to score $-1.35\times 10^{-6}\,\unitfrac{1}{s^{2}}$ and $-2\times 10^{-6}\,\unitfrac{1}{s^{2}}$.
The reader may find the frequency behaviour of $S_{\dot{\Omega}_{\text{SC},\theta }}$
and the thruster noise figure per single FEEP in picture \ref{fig:thrustnoiseandpsd}.

\begin{figure}
\begin{center}
\includegraphics[width=\textwidth]{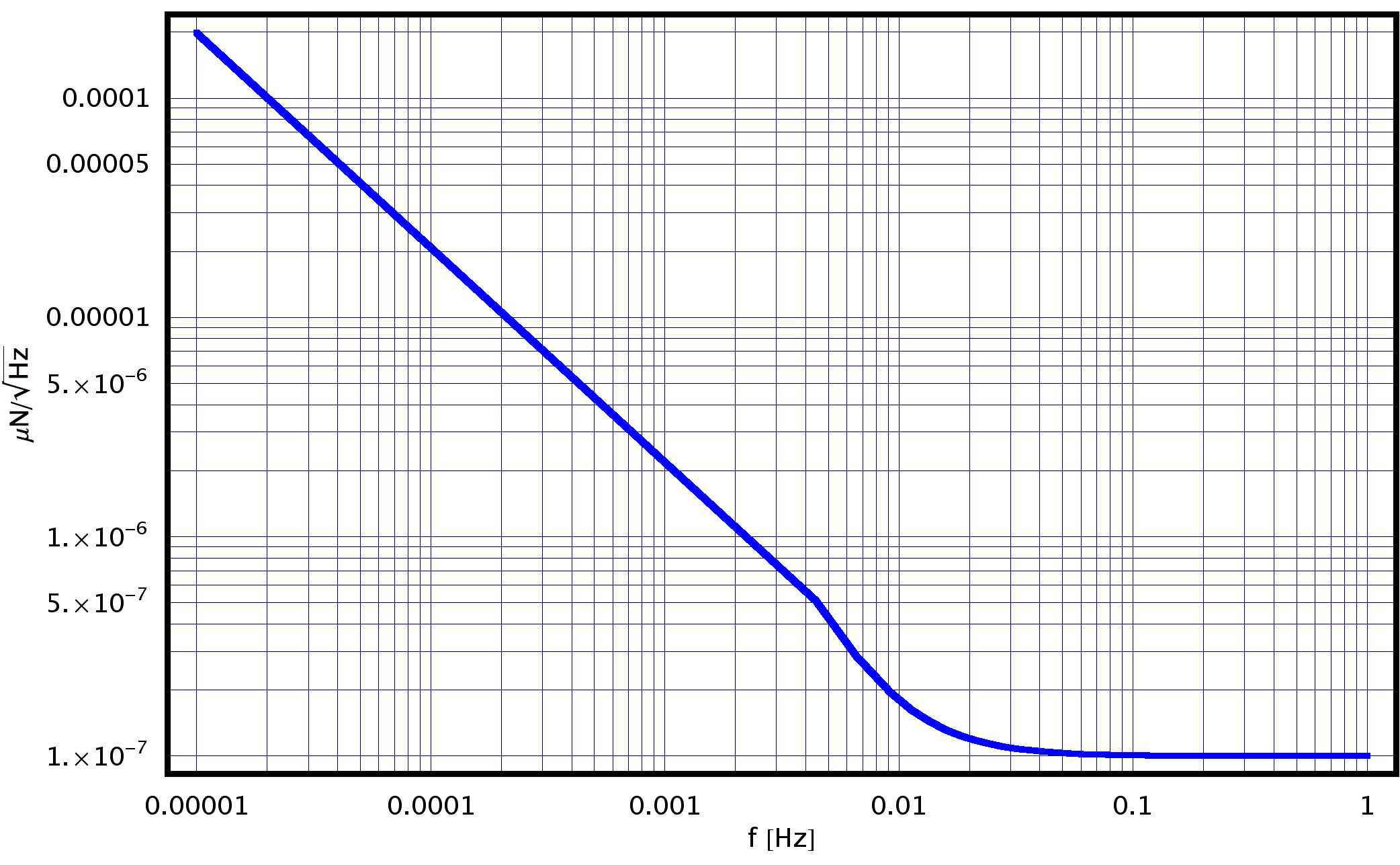}
\includegraphics[width=\textwidth]{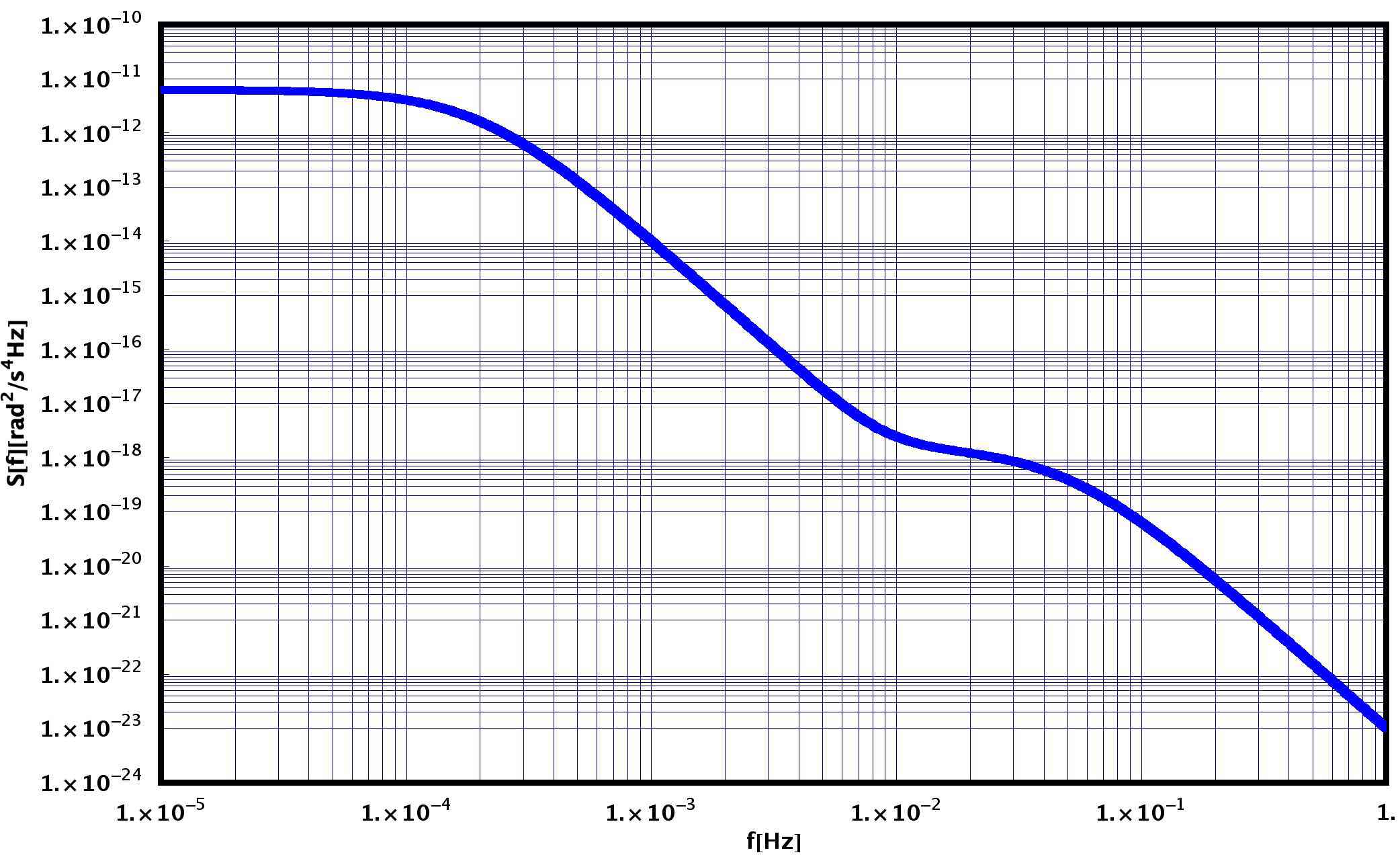}
\caption{Noise from per single thruster (top) and its squared PSD (bottom).}
\label{fig:thrustnoiseandpsd}
\end{center}
\end{figure}

\end{description}

\begin{figure}
\begin{center}
\includegraphics[width=\textwidth]{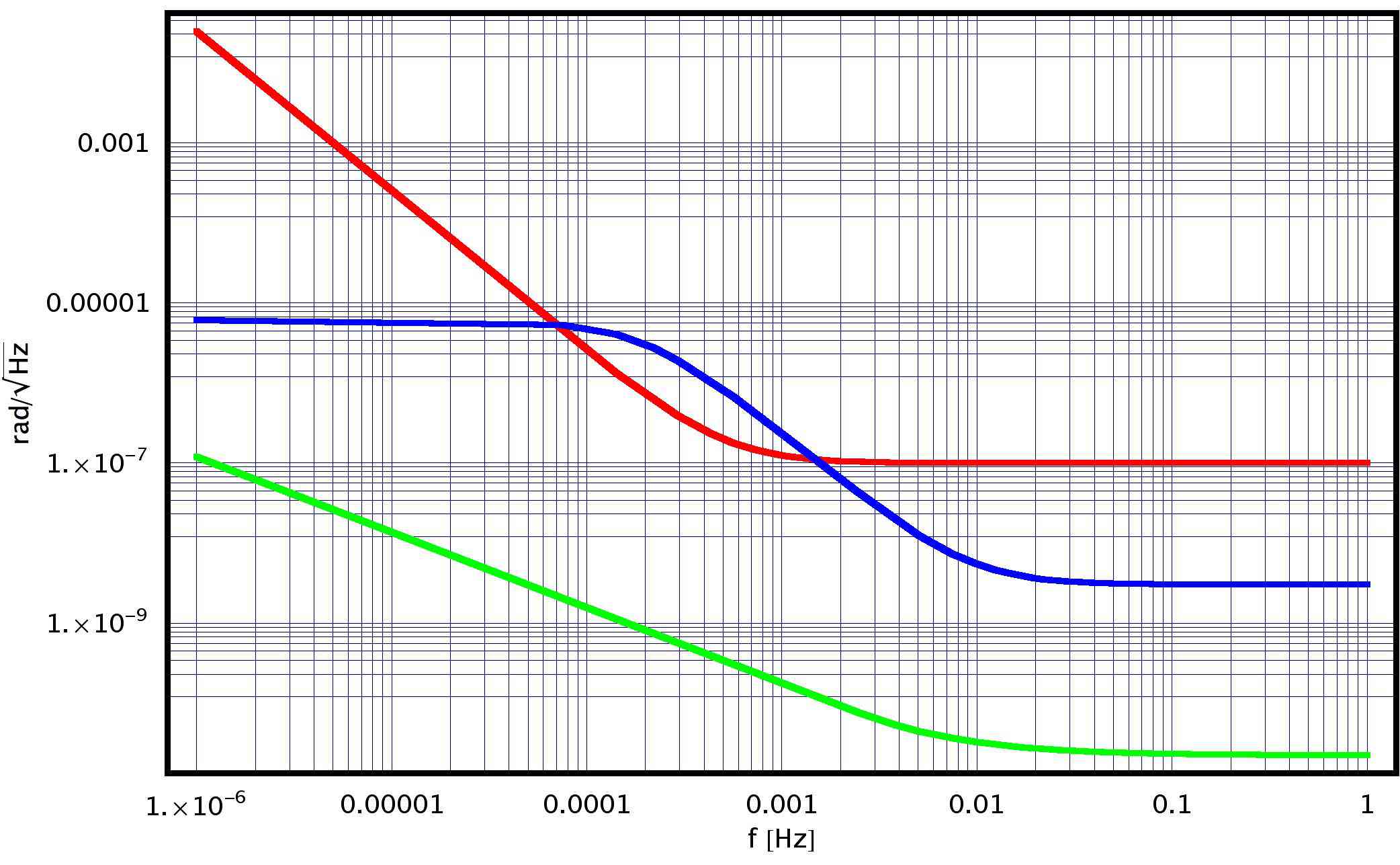}
\includegraphics[width=\textwidth]{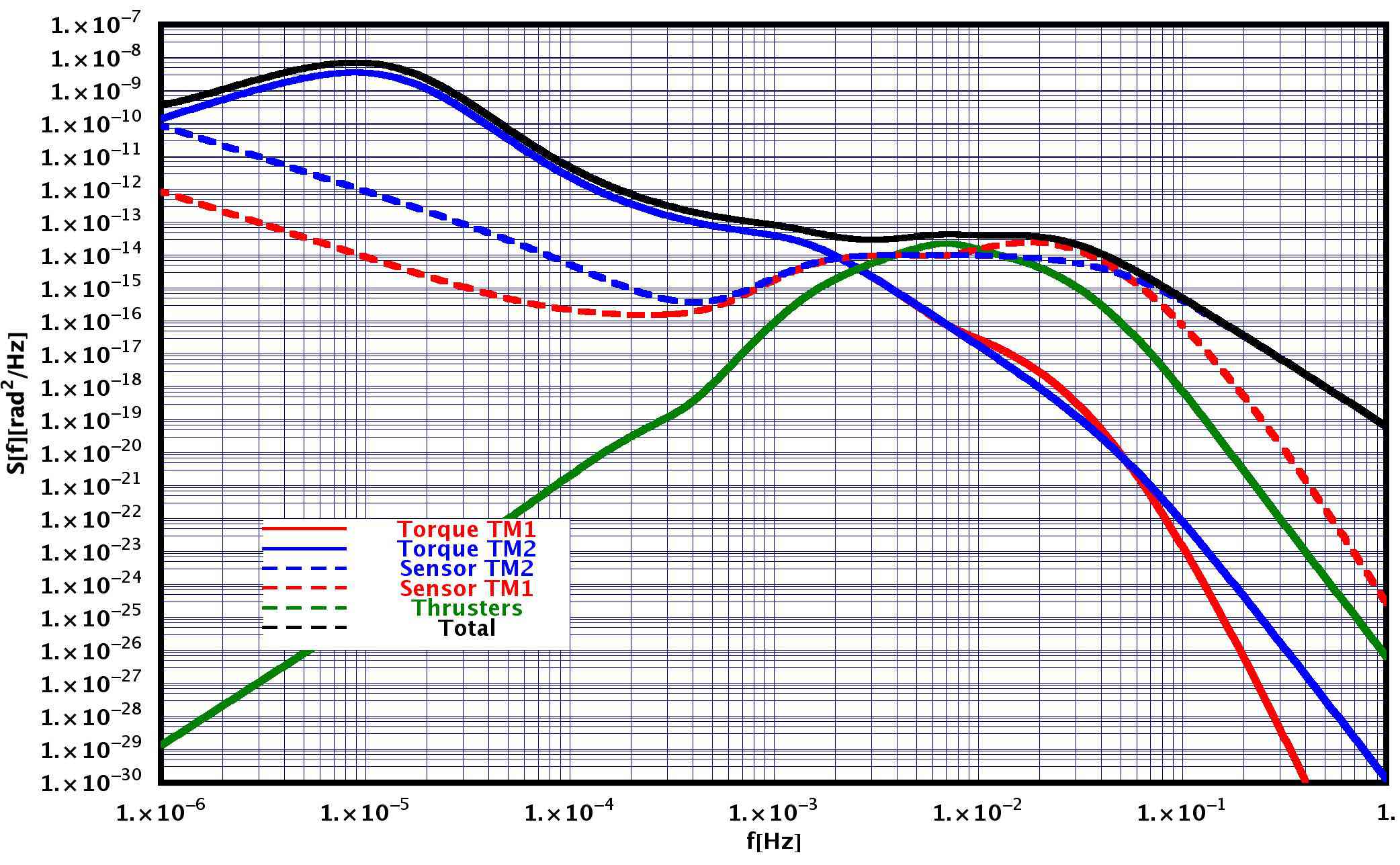}
\caption{Noise summary (top) and its squared PSD (bottom).}
\label{fig:noisetotal}
\end{center}
\end{figure}

A summary of all the noises together with the total can be found in figure \ref{fig:noisetotal}.

\subsection{Algorithms}

In order to estimate the value of accumulated charge an optimal filtering procedure might probably suitable. Nevertheless,
the optimal filter requires a detailed knowledge of the noise model and - being this latter a function of time - may cast
too demanding a task on the on-board facilities. Many alternative filters have thus been used to proceed in the estimate:
\begin{enumerate}
\item continuous time Wiener-Kolmogorov, to evaluate the maximal sensitivity in charge;
\item linear regression to square fitting on a sine signal at discrete time rate, probably being the fastest solution;
\item the former at continuous time, to estimate the data-loss on digitisation;
\item linear regression to square fitting on a sine signal at discrete time rate with a superimposed linear fit;
\item the former at continuous time, same purpose to estimate digitisation loss.
\end{enumerate}

\subsubsection{Wiener-Kolmogorov filtering}

The filter is the same as in section \ref{app:calibtm1}, though we'll require here a somewhat simpler form for the data:
\begin{shadefundtheory}
\beq
\theta(t)=A s(t)+n(t)\,,
\eeq
\end{shadefundtheory}
\noindent $A$ being the amplitude we'd like to estimate for the signal $s(t)$, $n(t)$ the relative noise, assumed to be null-average Gaussian
distributed. The best estimate of $A$ is:
\beq
\hat A = \frac{1}{\sigma^{2}_{\hat A}} \int_{0}^{T}\theta(t) s(t)\de t\,,
\eeq
with
\beq
\sigma_{\hat A}^{2}=\frac{S_{0}}{\int_{0}^{T} s^{2}(t)\de t}\,.
\eeq
Here $S_{0}$ is the noise spectral density, assumed as constant due to the white noise shape (notice it's always possible to filter
the noise so to obtain a white PSD).

\subsubsection{Linear regression}

We can also approximate the data-set in the form of a sine-cosine signal:
\begin{shadefundtheory}
\beq
\theta(t)=c_{s}\sin\omega_{0}t+c_{c}\cos\omega_{0}t\,,
\eeq
\end{shadefundtheory}
\noindent where $\omega_{0}$ is the input signal pulsation. The charge can be deduced in terms of $N_{e}$:
\beq
N_{e}=\frac{\sqrt{c_{s}^{2}+c_{c}^{2}}}{\frac{q_{0}V_{0}}{I_{\theta}}
  \left|h^{\text{lfs}}_{\theta,\theta_{2},\theta_{2}}\right|}\,.
\label{eq:numbercharges}
\eeq
By linear regression fitting, ignoring the noise correlation we can compute the value of the $c_{s}$ and $c_{c}$ coefficients as:
\beq
c_{i}=\sum_{k=1}^{n} w_{i}(k)\theta(k)\,,\,i=c,s\,,
\eeq
where $n$ is the sample size. The functions $w_{i}(k)$ look like:
\beq
\begin{split}
w_{s}(k)=P(n) \Bigl(&\cos\left((2n-(k+1))\omega_{0}\Delta T\right) - \cos\left((k+1)\omega_{0}\Delta T\right)\\
  &+2 n \sin\left(\omega_{0}\Delta T\right)\sin\left(k\omega_{0}\Delta T\right)\Bigr)\,,
\end{split}
\eeq
and:
\beq
\begin{split}
w_{c}(k)=P(n) \Bigl(&(n-1)\sin\left((k+1)\omega_{0}\Delta T\right) - \sin\left((2n-(k+1))\omega_{0}\Delta T\right)\\
  &- n \sin\left((k-1)\omega_{0}\Delta T\right)\Bigr)\,,
\end{split}
\eeq
with
\beq
P(n)=-\frac{2 n \sin\left(\omega_{0}\Delta T\right)}
  {1-n^{2}+n^{2}\cos\left(2\omega_{0}\Delta T\right)-\cos\left(2 n\omega_{0}\Delta T\right)}\,.
\eeq
Obviously what we carried on til here is a discrete-time analysis, by switching summations to integrals in the former expressions
the continuous-time picture can be obtained. This last is used solely for sensitivity purposes.

The linear regression may be complicated with a linear fit superimposed to the sine signal. In this case the
data may be approximated by a function in the form:
\begin{shadefundtheory}
\beq
\theta(t)=c_{s}\sin\omega_{0}t+c_{c}\cos\omega_{0}t+a\frac{t}{T}+b\,,
\eeq
\end{shadefundtheory}
\noindent with obvious meaning of the symbols. In perfect analogy to what stated already in absence of the straight line term, the formula
to determine $N_{e}$ is unvaried from \eqref{eq:numbercharges}, as well as the procedure to estimate the $c_{i}$ coefficients.
We refer to \cite{tateothesis} for details.

\subsection{Results}

\begin{figure}
\begin{center}
\includegraphics[width=\textwidth]{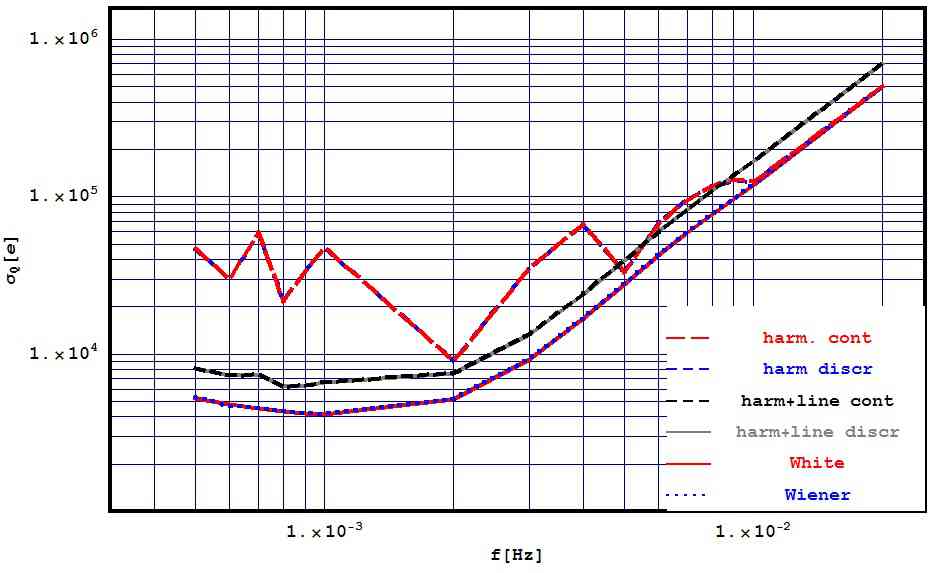}
\caption{Standard deviation per method as function of frequency.}
\label{fig:stddevcharge}
\end{center}
\end{figure}

Results with different filter strategies have been compared. We report the graph corresponding to the final values
of standard deviation per frequency and per strategy. We clearly see the frequency corresponding to the
minimum is around $1\,\unit{mHz}$, see figure \ref{fig:stddevcharge}.

We can also deduce that:
\begin{enumerate}
\item a sampling at $1-5\,\unit{Hz}$ frequency is enough for a driving force at $1\,\unit{mHz}$,
\item the pure sine approximation is inaccurate and the result are strongly dependent on measure duration
and data truncation, this failure is highly reduced by the introduction of the linear fitting line to de-trend
the data.
\end{enumerate}
In spite of the complications, with a driving frequency of $1\,\unit{mHz}$ the sub-optimal filter
resolution is of order $6600$ electron charges for $1$ hour time integration and $1\,\unit{V}$ amplitude. By
accepting a resolution of $10^{5}$ charges the amplitude could be reduced to $50-70\,\unit{mV}$, over a 
measuring time of roughly $2$ hours.

%% file: chapters/g-phys-with-ltp.tex
\chapter{Fundamental physics with LTP}
\label{chap:gphysltp}

\lettrine[lines=4]{I}{t was shown}
in the former chapters that a wave-like gravity perturbation can be accounted
for as a local cumulative effect in a global set of coordinates in the form of tidal stretching and shrinking
of distances. Moreover, it was shown that it's possible to build a correspondence between a TT-gauged
frame of coordinates and a set of local Fermi-Walker tetrads attached to a laser beam metricised by its pulsation
frequency. LTP
is then seen as an experimental demonstrator of our ability to build a fiduciary point in free-fall
with respect to a very low noise reference. The experiment is born as a technology demonstrator
to achieve free-fall up to the level of $3\times 10^{-14}\,\unitfrac{m}{s^{2}\sqrt{Hz}}$, but while this
remains the primary task of the mission, we can explore a more daring perspective.

By means of reverting the scenario - once we've shown that the sensitivity of LTP is the mentioned one and
that the operational band is not shaded or reduced in width by unknown or unforecasted phenomena - LTP
is - as a matter of fact - a high-precision gradiometer, a spacecraft whose free orbit explores the shape of
the gravitational potential, and a gravitational sensor with ability of self-orientation in space. In a word,
the perfect projector from TT-gauge to a freely falling tetrad set.

We present in the following some ideas and gedanken tests for a measure of $G$, short and long range violations of the
inverse square law, MOND corrections to Newton's dynamics. Throughout all the chapter we'll assume a device
similar to LTP will be in place, skewed in direction, with the purpose of calibration or gravity signal-generation.

\newpage

\section{Introduction}

Let's clear some smoke from the view and sketch some wide-range perspectives as well as list
more carefully the tests LTP may be capable of.
\begin{enumerate}
\item Though cancelled from the present state of mission, the idea of an internal bona-fide gravity perturbation
generator aside LTP could still be pursued. The Disturbance Reduction System DRS (lately renamed as ST7)
\cite{hanson:9, folkner:221}
was meant to be a parallel experiment, to confirm
data from LTP aboard the very same SC, and to provide a non-parallel, skewed direction of detection, to test
controls mimicking a scenario closer to the one of LISA, where the directions of detection of the TMs
would be neither collinear nor orthogonal. In brief, the presence of an oscillatory perturbation generator
could provide a carefully tuned signal from a source independent of LTP, useful for calibration purposes but for
fine-level measurements too. A measure of $G$ could be planned with such a tool; alternatively, one of
the two LTP TM can be moved as to induce a gravitational signal on the other TM. In principle the result
would be the same, practically having an independent source would rule out some of the instrumental noise
contamination.
\item Dynamics of moving bodies is carefully described to low velocities and mesoscopic scales by means of
Newton law, stating that the mutual acceleration between two bodies is a linear function of the square inverse
of the mutual distance. Deviations from this law may pick up several forms, but we'll discuss here three
different ones, with links to recent developments from the theory and in adherence to what we think LTP could
measure for real.
\begin{description}
\item[Short range violations] of the inverse square law (ISL) may come out of effective field-theory models
which describe possible corrections having exponential form. Hence the
additional composition-independent Yukawa potential (i.e. just coupling to mass), can be parametrised as
\begin{shadefundtheory}
\beq
\label{eq:yukawaviolation}
V_{\text{N}+\text{Y}}(r)=G\frac{m_{1} m_{2}}{r}
  \left(1+\alpha \exp\left(-\frac{r}{\lambda}\right)\right)\,,
\eeq
\end{shadefundtheory}
\noindent where $\alpha$ characterises the strength of the interaction (relative to gravitation) and $\lambda$ its range. The
resulting force is then given by
\beq
\begin{split}
F_{\text{N}+\text{Y}}(r)&=G\frac{m_{1} m_{2}}{r^{2}}
  \left(1+\alpha \left(1+\frac{r}{\lambda}\right)\exp\left(-\frac{r}{\lambda}\right)\right) = \\
&= G\frac{m_{1} m_{2}}{r^{2}} \left(1+\xi_\text{Y}(r)\right)\,.
  \label{eq:yukawanewtonforce}
\end{split}
\eeq
Short range tests can be carried on by means of using the proof-mass/calibration-mass scheme.

\item[Long range violations] of the ISL may occur as well as s.r. ones, and can be a result of a similar potential
correction.

\item[MOdified Newtonian Dynamics] in the recent and more convincing field-theoretical reformulation of
a bi-metric theory of gravity (TeVeS) is a serious competitor to Dark Matter in contemporary cosmology. Its low-velocities
approximation builds up an alternative scale-dependent theory of dynamics where the physical gravitational potential
$\Phi$ is determined by the modified Poisson equation:
\begin{shadefundtheory}
\beq
\vc\nabla\cdot \left( \tilde\mu\left(\frac{\left|\vc\nabla\Phi\right|}{a_{0}}\right) \vc\nabla\Phi\right)
  = 4\pi G\tilde\rho\,,
\eeq
\end{shadefundtheory}
\noindent where $\tilde\rho$ is the baryon's (only!) mass density, $a_{0}\simeq 10^{-10}\,\unitfrac{m}{s^{2}}$ is
Milgrom's characteristic acceleration, and $\tilde\mu(x)$ is a free function constrained on the extrema as:
\beq
\tilde\mu(x)\underset{x\ll 1}{\to} x\,,\qquad \tilde\mu(x)\underset{x\gg 1}{\to} 1\,.
\eeq
The physical acceleration is retrieved as $\vc a=-\vc\nabla\Phi$, and matches the Newtonian prediction for
$a\gg a_{0}$. The na{\^{i}}ve version of the theory, known as MOND, has been phenomenologically successful
in fitting data from galactic rotational curves and at explaining other anomalies, such as the Pioneer one, without
invoking additional matter distribution in space other than the baryon's one.
Following Bekenstein and Magueijo \cite{PhysRevD.73.103513}
we'll discuss a possible strategy of direct measurement for LTP crossing in a weak MOND region.

Similarly, the orbit of LTP could be used as a dynamical estimator of the underlying gravity potential. When (if?)
crossing close to a Newtonian force saddle point (SP) ($\vc F(x_\text{SP})=\vc 0$), acceleration could be close enough
to the scale $a_{0}$ to reveal MOND phenomena.

LTP is going to be placed in a free Lissajous orbit \cite{Landgraf:2004gm} around L1 (see picture \ref{fig:LTPorbit}), the Lagrangian point between Earth and the Sun. The
space-probe will get there following a pretty much complicated orbit, which we are not going to describe in detail:
picture \ref{fig:LTPsc}, right and \ref{fig:LTPorbit}
can be of some help focusing the scenario. While moving, LTP will experience
forces and gradients characteristic of the solar system both on the Sun-Earth line, on the ecliptic plane as well as
transverse forces and gradients on the Lissajous orbit. A review of these pulls is given in table \ref{tab:gpullslissa}; notice the
total of forces is obviously $0$ being on orbit.
\end{description}
\end{enumerate}

\begin{table}
\begin{center}
\begin{tabular}{>{\raggedleft}m{3.0cm}|r@{.}l|r@{.}l|r@{.}l|r@{.}l}
Heliocentric Orbit - L1 & \multicolumn{2}{c|}{Sun}  & \multicolumn{2}{c|}{Earth} & \multicolumn{2}{c|}{Inertial} & \multicolumn{2}{c}{Total} \\
\hline\rule{0pt}{0.4cm}\noindent
Force $(\unit{N})$ & $-1$&$18\times 10^{-2}$ & $3$&$\times 10^{-4}$ & $1$&$15\times 10^{-2}$ & $0$&$$ \\
Radial gradients $(\unitfrac{N}{m})$ & $1$&$59\times 10^{-13}$ &$ 1$&$65\times 10^{-13}$ & $-7$&$74\times 10^{-14}$ & $2$&$47\times 10^{-13}$\\
\hline\rule{0pt}{0.4cm}\noindent
Lissajous Orbit  & \multicolumn{2}{c}{}  & \multicolumn{2}{c}{} & \multicolumn{2}{c}{} & \multicolumn{2}{c}{} \\
\hline\rule{0pt}{0.4cm}\noindent
Force $(\unit{N})$ & $-5$&$98\times 10^{-5}$ & $-1$&$.24\times 10^{-4}$ & $1$&$83\times 10^{-4}$ & $0$&$$ \\
Radial gradients $(\unitfrac{N}{m})$ & $-7$&$93\times 10^{-14}$ & $-1$&$65\times 10^{-13}$ & $-2$&$45\times 10^{-13}$ & $-4$&$89\times 10^{-13}$
\end{tabular}
\end{center}
\caption{Detail of modulus of forces and their gradients in Heliocentric, L1 premises and Lissajous orbits.
Contributions are split into those given by Sun and Earth and the inertial ones, self-induced by LTP. Totals aside.
Obviously, radial gradients for Heliocentric orbit are computed on the Sun-Earth line, while computation
for the Lissajous orbit is transverse to the former.}
\label{tab:gpullslissa}
\end{table}

\section{Measure of $G$}
\label{sec:gmeas}

The measure of $G$ is based on our classical knowledge of gravitation. Newton's law of
gravitation for extensive sources can be stated as follows:
\begin{shadefundtheory}
\beq
\label{eq:newtonfinite}
\vc F = G \frac{m_{1} m_{2}}{r^{2}} \Pi\left(L,r\right)\frac{\vc r}{r}\,,
\eeq
\end{shadefundtheory}
\noindent where $m_{1}$ and $m_{2}$ are masses, $\vc r$ being the mutual distance vector in
three-dimensional space, and $\Pi$ is a form function accounting for the finiteness of the
proof-masses, whose characteristic side-scale is summarised by the length $L$. Obviously, in general:
\beq
\Pi(L,r)\underset{r/L\to\infty}{\to} 1\,,
\eeq
because of consistency arguments with the point-like version of Newton's law.

$G$, Newton's gravitational constant has current accepted value coming from ground-based
experiments of:
\begin{shadefundnumber}
\beq
G = 6.6742 \times 10^{-11}\,\unitfrac{m^{3}}{kg\, s^{2}}\,,
\label{eq:valueofG}
\eeq
\end{shadefundnumber}
\noindent with relative uncertainty of $1.5\times 10^{-4}$, i.e. absolute measuring error of
$10^{-4}\,\unitfrac{m^{3}}{kg\, s^{2}}$. Such a resolution is quite poor in comparison
of other fundamental constants, such as for example Planck's $h$, known with
relative uncertainty of $1.7\times 10^{-7}$. Hence, a new measurement by
LTP would at least provide the novelty of being carried on in space, off-ground and
could possibly improve precision or being competitive with the existing estimate.

A simple measurement of $G$ can be carried on as follows. Let's consider the two TMs
of LTP (but it may be also one of LTP TMs and an external mass oscillating to produce an
educated disturbance signal), the gravitational force between them can be expressed by eq. \eqref{eq:newtonfinite},
by means of moving TM2 by an amount ${\delta}r$ the shift in gravitational force
in modulus is
\beq
{\delta}F = -2 G \frac{m_1 m_2}{r^3} \Pi(L,r) {\delta}r\,,
\eeq
i.e. as it is well known:
\beq
\frac{{\delta}F}{F} = -2\frac{{\delta}r}{r}\,.
\eeq
Such an imbalance in force induces a motion in TM1 which we can model
in frequency space by means of our usual main spring-coupling model,
introducing an effective spring constant $k_{1}$:
\beq
{\delta}F = \left(m_{1}\omega^{2} - k_{1}^{2}\right){\delta}x\,,
\eeq
where now we named the main detection axis motion after ${\delta}x$.
Grouping and solving, we get that the measurement of $G$ will come from:
\begin{shadefundtheory}
\beq
G=\frac{r^3\left(k_1-\omega ^2 m_1\right)  {\delta}x }{2  \Pi(L,r)  m_1 m_2 {\delta}r}\,.
\label{eq:gvalltp}
\eeq
\end{shadefundtheory}
\noindent Notice to this level that systematically $G=G({\delta}x, {\delta}r, \omega)$, meaning
the values we'll get from the former formula will be strongly dependent on the strategy of
inducing motion in TM1 due to TM2, the detection frequency and more than everything
the sensitivity of the detection apparatus. Identifying ${\delta}x \simeq \text{IFO}(x_{2}-x_{1})$
it is obvious that the residual jitter of TM1 can be a serious source of noise spoiling
the precision of the measurement. In first place then the true value of $G$ will be
a wise averaging $\left\langle G \right\rangle$ on a set of frequencies and
shaking schemes of the source.

It is transparent that in order to be competitive with the existing estimates the precision
on the constants at play must be maximised. By means of straightforward error
propagation analysis, we get the relative variation of $G$ to be:
\beq
\begin{split}
\left(\frac{{\delta}G}{G}\right)^{2} =&
\frac{9 \epsilon_r^2}{r^2}+
\frac{\epsilon_{{\delta}r}^2}{{\delta}r^2}+
\frac{\epsilon_{{\delta}x}^2}{{\delta}x^2}+
\frac{\epsilon_{\Pi }^2}{\Pi ^2}+
\frac{4 \omega ^2 m_1^2 \epsilon_{\omega }^2}{\left(k_1-\omega ^2 m_1\right)^2}+\\
&+\frac{\epsilon_{k_1}^2}{\left(k_1-\omega ^2 m_1\right)^2}+
\frac{k_1^2 \epsilon_{m_1}^2}{m_1^2 \left(\omega ^2 m_1-k_1\right)^2}+\frac{\epsilon_{m_2}^2}{m_2^2}\,,
\end{split}
\label{eq:grelvariat}
\eeq
where we indicated the uncertainties relative to each variable at play with the
symbol $\epsilon$ and the variable name as subscript. The given quantities in this game
are shown in table \ref{tab:givendata}, together with the uncertainty. We can summarise the
main cause inducing relative error as follows:
\begin{description}
\item[masses] cannot be easily estimated beyond
a certain level due to imprecision in measuring distances while machining. Weight -
the first issue in estimating gravitational forces - is thus affected; pre-flight metrology
and an accurate mass weighting is the only way to reduce these contributions;
\item[mutual distances] between LTP components get lower bounds on precision
because of engineering tolerances; absolute TMs placements to  $200\,\unit{nm}$ and a strict pre-flight
policy to measure displacements within $100\,\unit{nm}$ in both sensors is foreseen as
a cleanliness method;
\item[distances] are dynamically measured by the GRS and IFO signals. Design
sensitivity limits are placed in form of spectral sensitivity curves and cast errors on
length estimates, both of the source and of the ``sensor''; nothing on this side
can be done more than working in the most sensitive bands of the spectrum, such
as around frequencies of order $10^{-2}$ till $10^{-3}\,\unit{Hz}$. Time
integration ranges correspondingly from $3\times 10^{11}$ to $3\times 10^{7}\,\unit{s}$
in order to achieve the demanded precision of $10^{-5}$;
\item[stiffness] is dynamically measured with an error induced by the control
loop precision and gets more demanding at lower frequency.
\end{description}

\begin{table}
\begin{center}
\begin{tabular}{r|r@{.}l|r@{.}l|r@{.}l}
Parameter & \multicolumn{2}{c|}{Nominal value}  & \multicolumn{2}{c|}{Relative error} & \multicolumn{2}{c}{Requirement} \\
\hline\rule{0pt}{0.4cm}\noindent
$m_{1}$ & $1$&$95$ \unit{kg} & $6$&$\times 10^{-4}$ & $8$&$\times 10^{-3}$ \\
$m_{2}$ & $1$&$95$ \unit{kg} & $6$&$\times 10^{-4}$ & $2$&$\times 10^{-6}$ \\
$r$ & $0$&$30$ \unit{m} & $3$&$\times 10^{-4}$ & $6$&$\times 10^{-7}$ \\
${\delta}r$ & $2$&$\unitfrac{nm}{\sqrt{Hz}}$ & $2$&$\times 10^{-6}\,\unitfrac{1}{\sqrt{Hz}}$ & $3$&$\times 10^{-6}$ \\
${\delta}x$ & $2$&$\unitfrac{nm}{\sqrt{Hz}}$ & $1$&$7\times 10^{-6}\,\unitfrac{1}{\sqrt{Hz}}$ & $3$&$\times 10^{-6}$ \\
$k_{1}$ & $1$&$\times 10^{-6}$ & $0$&$1$ & $2$&$\times 10^{-2}$
\end{tabular}
\end{center}
\caption{Tolerances and estimated uncertainties in key parameters for measurement of $G$ up to precision $10^{-5}$. The column
of number named as requirements can be multiplied by $10$ if deemed precision is reduced to $10^{-4}$; conversely integration time
can be reduced by two orders of magnitude.}
\label{tab:givendata}
\end{table}

\begin{table}
\begin{center}
\begin{tabular}{r|r@{.}l}
Parameter & \multicolumn{2}{c}{Value}  \\
\hline\rule{0pt}{0.4cm}\noindent
${\delta}r$ & $1$&$0\times 10^{-3}\,\unit{m}$ \\
${\delta}x$ & $1$&$2\times 10^{-9}\,\unit{m}$\\
$F$ & $2$&$8\times 10^{-9}\,\unit{N}$ \\
${\delta}F$ & $9$&$4\times 10^{-12}\,\unit{N}$
\end{tabular}
\end{center}
\caption{Figures of force and its variation for amplitude of modulation or order $10^{-3}\,\unit{m}$. Frequency of
evaluation is $10^{-2}\,\unit{Hz}^{-\nicefrac{1}{2}}$}
\end{table}

The presence of an independent source of educated noise could be of extreme value in the perspective of measurement of $G$ and in 
the other tests we'll mention. The ST-7 (DRS) experiment \cite{folkner:221} was planned aside LTP on the very same spacecraft and has been mentioned
already. In absence, a controlled ``gravity generator'' (like a couple of rotating spheres joined by a rod) could be regarded as a good source.
In place of it, we can consider DRS to be still operational in what follows.

\begin{figure}
\begin{center}
\includegraphics[width=.7\textwidth]{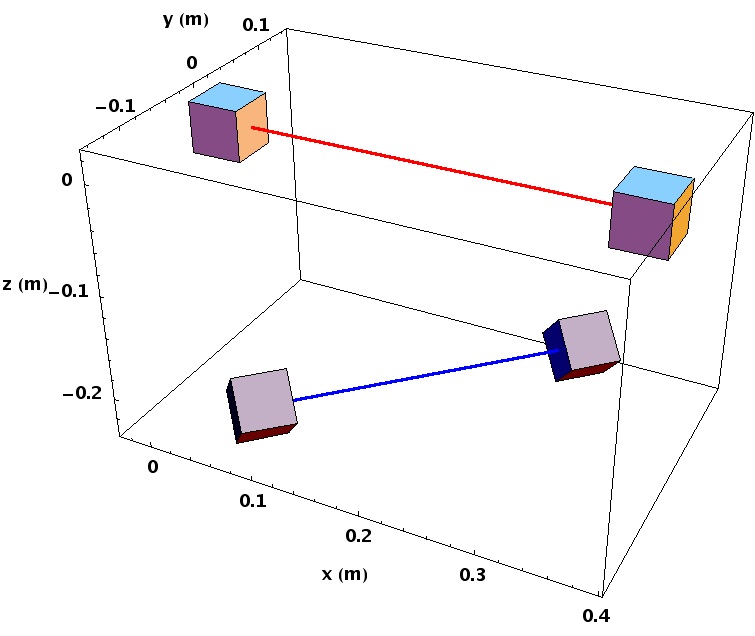}
\caption{LTP and DRS mutual positioning and main laser beams.}
\label{fig:ltpdrslaser}
\end{center}
\end{figure}

\begin{figure}
\begin{center}
\includegraphics[width=.5\textwidth]{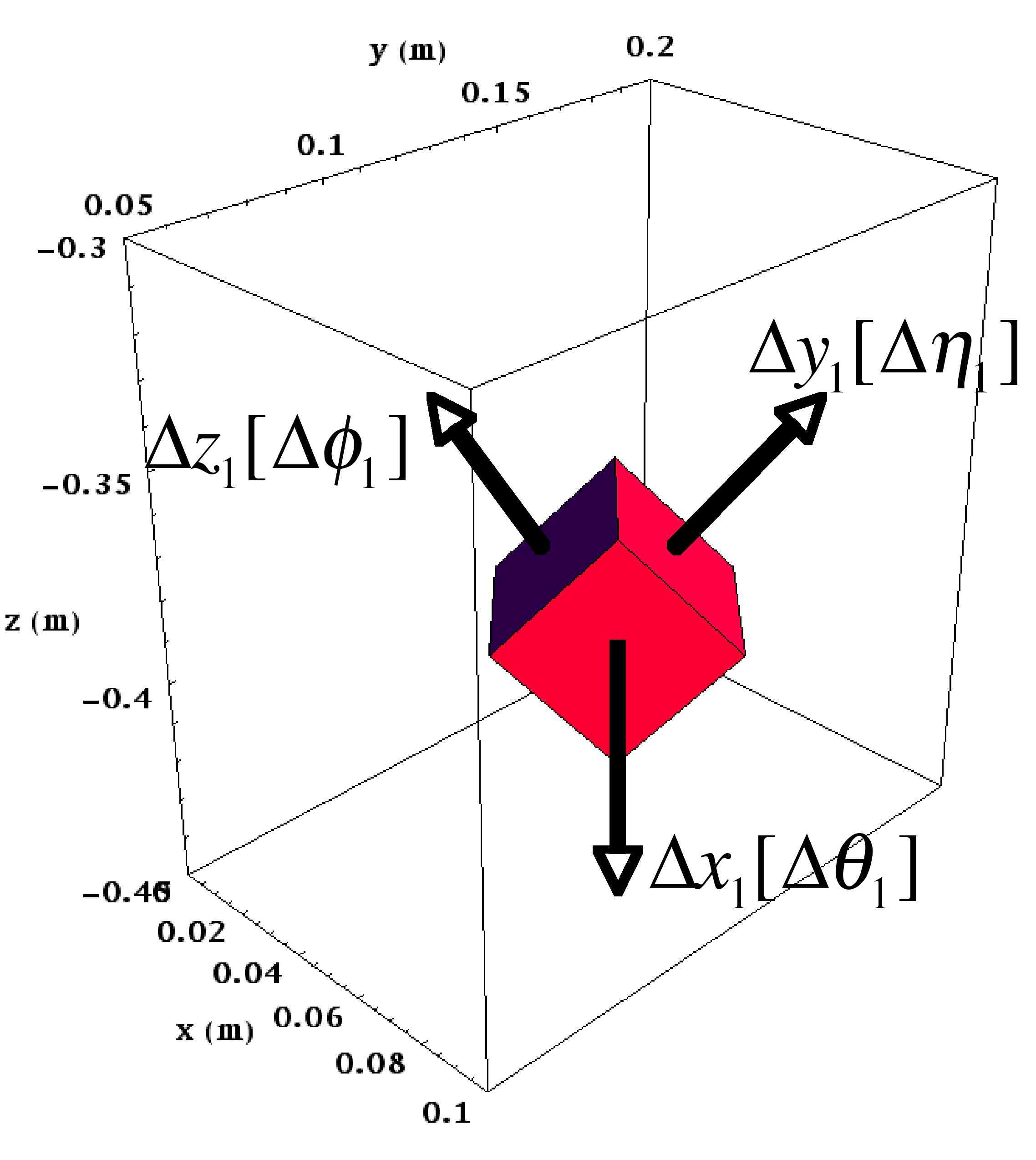}
\caption{DRS TM degrees of freedom.}
\label{fig:drsdof}
\end{center}
\end{figure}

A picture of the mutual DRS-LTP positioning can be inspected in figure \ref{fig:ltpdrslaser}, where laser lines have been emphasised
for dramatisation. A detail of the DOF for each DRS TM are shown in picture \ref{fig:drsdof}.

Measuring $G$ with a DRS+LTP joined operation is not hard to describe. The whole consists into shaking DRS TMs in a proper
manner and reading the effect on LTP's ones. More formally we could say that by a careful control procedure we'd read the phase of
the laser locked on the DRS TMs mutual distance after modulating this latter via its GRS system; together, we'd read the effect on the LTP
TMs, one servo-ed drag-free, the other suspended. The analysis can be carried on to first harmonic approximation if the DRS TMs
motion can be considered like a stable low-frequency sine in the displacement or attitude.

Several simulations were performed and different DOF used for both the devices, even beyond the IFO readout.
We hereby focus on results for a $\Delta x_{\text{DRS}}$ to $\Delta x_{\text{LTP}}$ perturbation: for a motion amplitude
of the first DRS TM amounting to $200\,\unit{\mu m}$ peak, but tables will show numbers for other interesting coupling modes.

As we stated, the motion along
the generic DOF $o_{\text{DRS},i}$ is driven to be sine-like:
\beq
o_{\text{DRS},i} \simeq A \sin t\omega_{D}\,,
\label{eq:estim}
\eeq
with $A>0$ (in metres) and $\omega_{D}$ (in Hertz) the drive frequency.
By virtue of a version of \eqref{eq:newtonfinite} integrated over cubic source masses to point-like observers (see \cite{Armano:2005ut}
and \cite{vit:2004int}, improved recently in \cite{Adelberger:2005bt}) the effect was tracked down the LTP masses along
each DOF of the latter. In all cases, a certain granularity in the motion was assumed to be non-influent by continuity of
the  fields, therefore a linear increment step of $10\,\unit{\mu m}$ was chosen in discretising the sine motion.

The following fitting expansion was chosen in one dimension for the effect on LTP, generic DOF
$o_{\text{LTP},k}$:
\beq
p\left(o_{\text{LTP},k}\right)=\sum_{j=1}^{n} c_{j} \left(o_{\text{DRS},i}\right)^{j}\,.
\label{eq:poli}
\eeq
We assumed the so called ``first harmonic'' result to be reliable, nevertheless checked the result till the fifth (the system is highly causal
due to the distances at play, and we can always assume a sine motion on LTP TMs given a sine motion on DRS TMs apart from a
negligible phase). By taking
\eqref{eq:poli} together with the form \eqref{eq:estim} for the signal and stopping at fifth harmonic ($n=5$), the
amplitude of the induced sine motion looks is non-linear in the stimulus and looks like:
\beq
c_1 A + \frac{3 c_3 A^3}{4}+ \frac{5 c_5 A^5}{8}\,.
\eeq
Notice, because of the bounded nature of the signal that
\beq
\left|p (x)\right| \leq \sum_{j} \left|c_{j}\right| A^{j},
\eeq
so that, in case we'd like to estimate the signal/noise ratio by knowing the error $\sigma$ we can tolerate, we can demand
\beq
\sum_{j} |c_{j}| A^{j} \geq \frac{\sigma}{\sqrt{T}},
\eeq
where $T$ is the integration time we can sustain. To the first harmonic the previous relation gets simplified to
\beq
|c_{1}| A \sqrt{T} \geq \sigma,
\label{eq:visibfirstharm}
\eeq
which can be used as a ``visibility condition'' for the signal, provided the drive displacement (rotation) is small and
the coefficients $c_{i} A^{i}, i>2$ are small compared to $|c_{1}| A$ .
In the specific case of our analysis a fitting error of less than $9$ order of magnitudes w/r to the mean
was reached with $n=5$ in the polynomial degree.

The $c_{1}$ fit coefficients can be found in table \ref{tab:c1coeff} for a driving force with
$\omega_{D}= 2\pi \times 5 \times 10^{-2}\,\unit{Hz}$. For a displacement (or equivalent rotation)
of DRS TMs amounting to $A=200\,\unit{\mu m}$ the product $A c_{1}$ can be found in table \ref{tab:c1acoeff}.
We estimated the error with respect to $5$-th harmonic analysis in table \ref{tab:erroncoeff}. Notice the first
order $c_{1}$ coefficient behaves like a gradient of force (torque), therefore shows the correct symmetries
with respect to the DRS and LTP TMs positioning.

Inspection of the tables shows a promising value for the $\nicefrac{\Delta F_{x}}{m} \sim \omega^{2} {\Delta}x_{\text{LTP}}$
upon the given displacement of $\simeq 2.18\times 10^{-13}\,\unitfrac{m}{s^{2}}$.
By employing \eqref{eq:visibfirstharm} over an integration time $T=3600\,\unit{s}$ and with
$\sigma=\sigma_{\text{LTP},\nicefrac{\Delta F_{x}}{m}}\simeq 3\times 10^{-14}$ we get a signal to noise
ratio of $\sim 435$, very satisfactory given the simplicity of the procedure.

\begin{figure}
\begin{center}
\includegraphics[width=.7\textwidth]{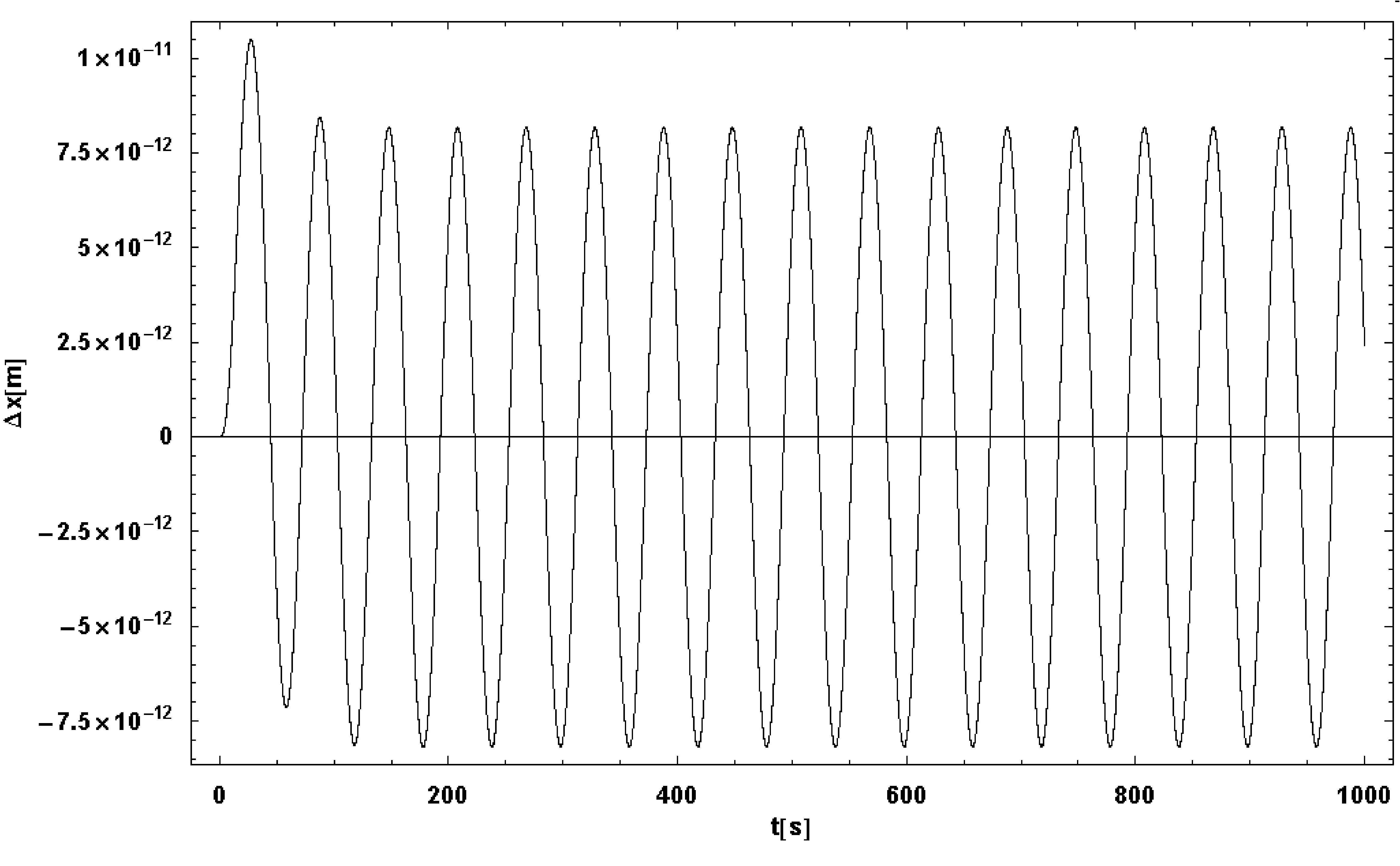}
\caption{Oscillation of LTP TMs mutual distance in response to DRS sine motion. Result from simulation.}
\label{fig:biggoscill}
\end{center}
\end{figure}

This scheme can be complicated by calling in place the control mode transfer functions. The quality of noise projection
boosts up in this case and transients can be plotted and analysed. Picture \ref{fig:biggoscill} shows the case.

\begin{table}
\begin{center}
\begin{tabular}{c|r@{.}l|r@{.}l|r@{.}l}
 &
\multicolumn{2}{c|}{$\nicefrac{{\Delta}F_x}{m}$} &
\multicolumn{2}{c|}{$\nicefrac{{\Delta}\gamma_\eta}{I_\eta}$} &
\multicolumn{2}{c}{$\nicefrac{{\Delta}\gamma_\phi}{I_\phi}$} \\
\hline\rule{0pt}{0.4cm}\noindent
 ${{\Delta}x}_1$ & $1$&$09\times 10^{-9}$ & $2$&$33\times 10^{-11}$ & $1$&$06\times 10^{-13}$ \\
 ${{\Delta}x}_2$ & $1$&$09\times 10^{-9}$ & $2$&$33\times 10^{-11}$ & $-1$&$06\times 10^{-13}$ \\
 ${{\Delta}y}_1$ & $-1$&$16\times 10^{-9}$ & $-1$&$57\times 10^{-11}$ & $-3$&$34\times 10^{-13}$ \\
 ${{\Delta}y}_2$ & $-1$&$16\times 10^{-9}$ & $-1$&$57\times 10^{-11}$ & $3$&$34\times 10^{-13}$ \\
 ${{\Delta}z}_1$ & $4$&$07\times 10^{-10}$ & $-6$&$03\times 10^{-13}$ & $1$&$69\times 10^{-12}$ \\
 ${{\Delta}z}_2$ & $4$&$07\times 10^{-10}$ & $-6$&$03\times 10^{-13}$ & $-1$&$69\times 10^{-12}$ \\
 ${\Delta}\theta_1$ & $-1$&$59\times 10^{-15}$ & $1$&$32\times 10^{-16}$ & $3$&$50\times 10^{-17}$ \\
 ${\Delta}\theta_2$ & $-1$&$59\times 10^{-15}$ & $1$&$32\times 10^{-16}$ & $-3$&$50\times 10^{-17}$ \\
 ${\Delta}\eta_1$ & $-2$&$29\times 10^{-15}$ & $-2$&$87\times 10^{-16}$ & $9$&$62\times 10^{-18}$ \\
 ${\Delta}\eta_2$ & $-2$&$29\times 10^{-15}$ & $-2$&$87\times 10^{-16}$ & $-9$&$62\times 10^{-18}$ \\
 ${\Delta}\phi_1$ & $1$&$04\times 10^{-15}$ & $-1$&$51\times 10^{-17}$ & $3$&$34\times 10^{-17}$ \\
 ${\Delta}\phi_2$ & $1$&$04\times 10^{-15}$ & $-1$&$51\times 10^{-17}$ & $-3$&$34\times 10^{-17}$
\end{tabular}
\end{center}
\caption{$c_{1}$ fitting coefficients in the first harmonic signal \eqref{eq:visibfirstharm}.}
\label{tab:c1coeff}
\end{table}

\begin{table}
\begin{center}
\begin{tabular}{c|r@{.}l|r@{.}l|r@{.}l}
 &
\multicolumn{2}{c|}{$\nicefrac{{\Delta}F_x}{m}$} &
\multicolumn{2}{c|}{$\nicefrac{{\Delta}\gamma_\eta}{I_\eta}$} &
\multicolumn{2}{c}{$\nicefrac{{\Delta}\gamma_\phi}{I_\phi}$} \\
\hline\rule{0pt}{0.4cm}\noindent
 ${{\Delta}x}_1$ & $2$&$18\times 10^{-13}$ & $4$&$65\times 10^{-15}$ & $2$&$13\times 10^{-17}$ \\
 ${{\Delta}x}_2$ & $2$&$18\times 10^{-13}$ & $4$&$65\times 10^{-15}$ & $-2$&$13\times 10^{-17}$ \\
 ${{\Delta}y}_1$ & $-2$&$32\times 10^{-13}$ & $-3$&$14\times 10^{-15}$ & $-6$&$67\times 10^{-17}$ \\
 ${{\Delta}y}_2$ & $-2$&$32\times 10^{-13}$ & $-3$&$14\times 10^{-15}$ & $6$&$67\times 10^{-17}$ \\
 ${{\Delta}z}_1$ & $8$&$13\times 10^{-14}$ & $-1$&$21\times 10^{-16}$ & $3$&$38\times 10^{-16}$ \\
 ${{\Delta}z}_2$ & $8$&$13\times 10^{-14}$ & $-1$&$21\times 10^{-16}$ & $-3$&$38\times 10^{-16}$ \\
 ${\Delta}\theta_1$ & $-3$&$17\times 10^{-19}$ & $2$&$64\times 10^{-20}$ & $6$&$99\times 10^{-21}$ \\
 ${\Delta}\theta_2$ & $-3$&$17\times 10^{-19}$ & $2$&$64\times 10^{-20}$ & $-6$&$99\times 10^{-21}$ \\
 ${\Delta}\eta_1$ & $-4$&$58\times 10^{-19}$ & $-5$&$75\times 10^{-20}$ & $1$&$92\times 10^{-21}$ \\
 ${\Delta}\eta_2$ & $-4$&$58\times 10^{-19}$ & $-5$&$75\times 10^{-20}$ & $-1$&$92\times 10^{-21}$ \\
 ${\Delta}\phi_1$ & $2$&$08\times 10^{-19}$ & $-3$&$02\times 10^{-21}$ & $6$&$67\times 10^{-21}$ \\
 ${\Delta}\phi_2$ & $2$&$08\times 10^{-19}$ & $-3$&$02\times 10^{-21}$ & $-6$&$67\times 10^{-21}$
\end{tabular}
\end{center}
\caption{$c_{1}$ fitting coefficients times $A=200\unit{\mu m}$ in the first harmonic signal \eqref{eq:visibfirstharm}.}
\label{tab:c1acoeff}
\end{table}

\begin{table}
\begin{center}
\begin{tabular}{c|r@{.}l|r@{.}l|r@{.}l}
 &
\multicolumn{2}{c|}{$\nicefrac{{\Delta}F_x}{m}$} &
\multicolumn{2}{c|}{$\nicefrac{{\Delta}\gamma_\eta}{I_\eta}$} &
\multicolumn{2}{c}{$\nicefrac{{\Delta}\gamma_\phi}{I_\phi}$} \\
\hline\rule{0pt}{0.4cm}\noindent
 ${{\Delta}x}_1$ & $-6$&$68\times 10^{-18}$ & $-2$&$87\times 10^{-19}$ & $-2$&$14\times 10^{-20}$ \\
 ${{\Delta}x}_2$ & $-6$&$68\times 10^{-18}$ & $-2$&$87\times 10^{-19}$ & $2$&$14\times 10^{-20}$ \\
 ${{\Delta}y}_1$ & $1$&$87\times 10^{-18}$ & $2$&$33\times 10^{-19}$ & $7$&$26\times 10^{-20}$ \\
 ${{\Delta}y}_2$ & $1$&$87\times 10^{-18}$ & $2$&$33\times 10^{-19}$ & $-7$&$26\times 10^{-20}$ \\
 ${{\Delta}z}_1$ & $-9$&$36\times 10^{-18}$ & $-4$&$32\times 10^{-19}$ & $-7$&$93\times 10^{-21}$ \\
 ${{\Delta}z}_2$ & $-9$&$36\times 10^{-18}$ & $-4$&$32\times 10^{-19}$ & $7$&$93\times 10^{-21}$ \\
 ${\Delta}\theta_1$ & $3$&$17\times 10^{-24}$ & $-2$&$7\times 10^{-25}$ & $-4$&$66\times 10^{-26}$ \\
 ${\Delta}\theta_2$ & $3$&$17\times 10^{-24}$ & $-2$&$7\times 10^{-25}$ & $4$&$66\times 10^{-26}$ \\
 ${\Delta}\eta_1$ & $4$&$57\times 10^{-24}$ & $6$&$03\times 10^{-25}$ & $1$&$54\times 10^{-26}$ \\
 ${\Delta}\eta_2$ & $4$&$57\times 10^{-24}$ & $6$&$03\times 10^{-25}$ & $-1$&$54\times 10^{-26}$ \\
 ${\Delta}\phi_1$ & $-2$&$08\times 10^{-24}$ & $3$&$95\times 10^{-26}$ & $-8$&$32\times 10^{-26}$ \\
 ${\Delta}\phi_2$ & $-2$&$08\times 10^{-24}$ & $3$&$95\times 10^{-26}$ & $8$&$32\times 10^{-26}$
\end{tabular}
\end{center}
\caption{Correction to first harmonic from third and fifth harmonic contributions. Values in table in comparison to
table \ref{tab:c1acoeff} show that the
percentage error is confined to order $10^{-6}$.}
\label{tab:erroncoeff}
\end{table}

\section{Violation of the ISL}

We can consider now the Yukawa-like correction to Newtonian potential expressed by eq. \eqref{eq:yukawaviolation}. In terms of
a field-theoretical background such a potential may be motivated as an effective low-energy theory coming from averaged boson
interactions (reference here), or as a low-energy by-product from bosonic string theory landscapes. Moreover, given the former
potential expression, it's straightforward to see that
\beq
V_{N+Y}\underset{\alpha\to 0}{\to} V_{N}\,,
\eeq
and that the correction to acceleration from Newtonian to Yukawa-improved theory scales like
\begin{shadefundtheory}
\beq
\frac{a_{Y+N}}{a_{N}}= 1 + \alpha\left(1+\frac{r}{\lambda}\right)\exp\left( -\frac{r}{\lambda}\right)\,,
\eeq
\end{shadefundtheory}
\noindent which for a large correlation scale $\lambda$ scales to first order like
\beq
(\alpha +1)+O\left(\left(\frac{1}{\lambda }\right)^2\right)\,,
\eeq
thus amounting to an effective renormalisation of the $G$ constant. We can compute then the ratio of difference of acceleration
according to the ``Y+N'' and pure Newton theories at two different working points $r_{1}$ and $r_{2}$ as follows:
\beq
\begin{split}
&\frac{a_{Y+N}\left(r_{1}\right)-a_{Y+N}\left(r_{2}\right)}{a_{N}\left(r_{1}\right)-a_{N}\left(r_{2}\right)}=\\
&= \frac{
\frac{1}{r_{1}^{2}} \left(1 + \alpha\left(1+\frac{r_{1}}{\lambda}\right)\exp\left( -\frac{r_{1}}{\lambda}\right)\right)
-\frac{1}{r_{2}^{2}} \left(1 + \alpha\left(1+\frac{r_{2}}{\lambda}\right)\exp\left( -\frac{r_{2}}{\lambda}\right)\right)
}{\frac{1}{r_{1}^{2}}-\frac{1}{r_{2}^{2}}} \\
&= 1 + \frac{\alpha}{r_{2}^{2}-r_{1}^{2}}
\left(
r_{2}^{2}\left(1+\frac{r_{1}}{\lambda}\right)\exp\left( -\frac{r_{1}}{\lambda}\right)
-r_{1}^{2}\left(1+\frac{r_{2}}{\lambda}\right)\exp\left( -\frac{r_{2}}{\lambda}\right)
\right)\,,
\end{split}
\eeq
it is easy to see that the ratio goes to $1$ as $\alpha\to 0$. It is hence a very good estimator of potential dearths of the
theory from Newtonian behaviour at least for a band selection of the $\lambda$ parameter. Notice the ratio is independent
on the value of $G$ which we assume independent of the distance. 

A short range test can then be performed aboard LTP through two $G$ measurements using two source masses
at different distances, namely $G_{1}$ and $G_{2}$. Forgetting about the stiffness correction, we can
equate the gravity pull per unit mass to the acceleration in frequency domain (from eq. \eqref{eq:yukawanewtonforce}) to get:
\beq
\frac{G}{r^2}\left(
1+ \alpha  \left(1+\frac{r}{\lambda }\right)\exp\left({-\frac{r}{\lambda }}\right) \right)
= r \omega^{2}\,,
\eeq
solving and substituting $G\to G_{i}$ for $r\to r_{i}$, we can then rescale each measure so
that $G_{i}\to G_{i} r_{i}^{3}$, hence, the $\alpha$ factor can be retrieved as:
\beq
\alpha=\frac{G_1-G_2}{
  G_1 \left(1+\frac{r_1}{\lambda}\right) \exp\left(-\frac{r_1}{\lambda }\right)
  - G_2 \left(1+\frac{r_2}{\lambda}\right) \exp\left(-\frac{r_2}{\lambda }\right)
  }\,.
\eeq

In a similar fashion, a test over long range violations or the ISL can be planned. On purpose
the two LTP masses could be used as a gradiometer aligned with the orbit path. Suppose for the time being
to consider only one mass $m_{0}$ placed along the Sun-Earth axis, at distance $r_{1}$ from the
Sun and $r_{2}$ from Earth. Naming $r_{E}$ the distance Sun-Earth if the Newton $1/r^{2}$
scaling law would apply, we'd get:
\beq
\begin{split}
\Delta a_{S} &= \frac{a_{S}(r_{E})}{r_{1}^{2}} -\frac{a_{S}(r_{1})}{r_{E}^{2}}\,,\\
\Delta a_{E} &= \frac{a_{E}(r_{E})}{r_{2}^{2}}-\frac{a_{E}(r_{2})}{r_{E}^{2}}\,,
\label{eq:deltaas}
\end{split}
\eeq
where $a_{E}(r)$ is the Earth's field acceleration at the radius $r$, $a_{S}(r)$ is the Sun field one. 
Consequently, since the effect is linear in the accelerations, we get:
\beq
\Delta a = \Delta a_{S} - \Delta a_{E}\,.
\label{eq:deltaglob}
\eeq
If the scaling of the accelerations be Newtonian, we'd expect both \eqref{eq:deltaas} to be null, thus enforcing
\eqref{eq:deltaglob} to be null too. Besides, if each acceleration carries scaling corrections in the
form of \eqref{eq:yukawanewtonforce}, i.e.
\beq
a_{i}(r)\to a_{\text{N}+\text{Y},\,i}(r)=
G\frac{m_{i}}{r^{2}}
  \left(1+\alpha \left(1+\frac{r}{\lambda}\right)\exp\left(-\frac{r}{\lambda}\right)\right)\,.
\eeq
then a gradiometer effect may arise from \eqref{eq:deltaglob} as the effect of the modified power scaling in
the distance:
\begin{shadefundtheory}
\beq
\frac{\Delta a}{a_{S}(r_{E})} =
  \frac{1}{r_{1}^{2}}\left(1-\frac{1+\xi_\text{Y}(r_{1})}{1+\xi_\text{Y}(r_{E})}\right)
  -\frac{1}{r_{2}^{2}}\frac{m_{E}}{m_{S}}\left(1-\frac{1+\xi_\text{Y}(r_{2})}{1+\xi_\text{Y}(r_{E})}\right)\,,
\eeq
\end{shadefundtheory}
\noindent where $a_{S}(r_{E})=5.9\times 10^{-3}\,\unitfrac{m}{s^{2}}$ is the Earth acceleration towards the Sun.

Limits in the ISL on large scales can be obtained using SC tracking during transfer mode and during Lissajous orbit. Naturally this
implies some requirements:
\begin{enumerate}
\item drag-free needs to be operational for the TMs be reliable low-noise detector to build the gradiometer on;
\item SC bias (mainly gravitational forces and gradients) shall be known to high level of accuracy in order to use LTP in alignment with the
orbit path and trace gravity gradients;
\item star-trackers in charge of placement of the SC with respect to distant stars carry the most demanding features: in order
to achieve an appreciable signal to noise ratio accuracy must be around $30\,\unit{\mu m}$. This highly demanding figure
seems to place a final shade on the possibility to measure long-range ISL violations.
\end{enumerate}

\section{MOND}

Are MOND \cite{McGaugh:2003qw,PhysRevD.73.103513,Bekenstein:2004ne} effects of importance in the Solar System? Milgrom
was the first to consider the effects of MOND
on orbits of long period comets from the Oort cloud. Later it was observed that anomalously large perihelia precession
of planets fitted predictions from relativistic MOND schemes. The so called ``Pioneer anomaly''\footnote{We
don't argue here on the reality of the effect in itself. Assuming such an anomaly to exist and to have dynamical
or gravitational origin, further investigation is mandatory and deemed to wipe out either the effect or to reabsorb it
into a suitable theoretical scheme. This is the spirit of the following discussion.} drew quite some
attention and a possible MOND origin of the effect was debated \cite{tangen-2006-,jaekel-2005-,anderson-1998-81}.

We'd like to point out some basic facts about MOND:
\begin{enumerate}
\item TeVeS \cite{Bekenstein:2004ne} encapsulates MOND in a bi-metric scalar-tensor-vector theory of
gravity, thus providing a way for the acceleration scale $a_{0}$ to emerge dynamically within the space of parameters as a reference
acceleration caused by fields configuration. TeVeS suggests the standard $g_{\alpha\beta}$ metric to be replaced by
\begin{shadefundtheory}
\beq
\tilde g_{\alpha\beta}=e^{-2\phi}g_{\alpha\beta}-2 V_{\alpha}V_{\beta}\sinh 2\phi\,,
\label{eq:tevesmetric}
\eeq
\end{shadefundtheory}
\noindent $\phi$ being a dilaton-type scalar field and $V_{\alpha}$ a four-vector time-like field: $U_{\alpha}U^{\alpha}=-1$.
The metric \eqref{eq:tevesmetric} collapses to $g_{\alpha\beta}$ for $\phi\to 0$ and the presence of the $V_{\alpha}$
field keeps $\phi$ propagation causal.
\item All the good properties of an action-principle derived theory are respected by TeVeS: locality, invariance
and flat-gravity limit of the theory. Causality and positivity of energy are
respected.
\item Flat space-time Lagrangians are transposed into TeVeS space-time by switching $g_{\alpha\beta}\to \tilde g_{\alpha\beta}$
in indice contractions and by modifying derivatives into co-variant form with respect to $\tilde g_{\alpha\beta}$. Integrals
are rendered invariant by the Jacobi's measure $(-\tilde g)^{\frac{1}{2}}\de^{4} x=e^{-2\phi}(- g)^{\frac{1}{2}}\de^{4} x$,
where $\tilde g=\det \tilde g_{\alpha\beta}$ and $g=\det g_{\alpha\beta}$.
\item Metric ($\tilde g_{\alpha\beta}$), Vector ($V_{\alpha}$) and Scalar ($\phi$) equations of motion are
obtained by variation of the action $S$. The metric action is the Hilbert-Einstein's with Ricci tensor written in terms of $\tilde g_{\alpha\beta}$.
It is of some interest for us only to state that the scalar action carries dependence on the dimensionless parameter $k$ and the length
parameter $l$, via a free dimensionless function $\mu$ chosen so to reproduce the MOND phenomenology.
\end{enumerate}

By assuming the
matter distribution to be an ideal fluid with density $\tilde\rho$ and pressure $\tilde P$ we'd get for the $\phi$ field:
\begin{shadefundtheory}
\beq
\left(\mu\left(k l^{2} \phi_{,\nu}\phi^{,\nu} \right)\phi^{,\beta}\right)_{;\beta}
= k G (\tilde\rho+3\tilde P)e^{-2\phi}\,,
\eeq
\end{shadefundtheory}
\noindent by replacing $g_{\alpha\beta}\to \eta_{\alpha\beta}$ and $e^{-2\phi}\to 1$ and assuming $\tilde P$ to be
small with respect to $\tilde\rho$ we are brought to the set of equations \cite{PhysRevD.73.103513}:
\beq
\begin{split}
\nabla\cdot \vc u = -4\pi G \tilde\rho\,,\\
\nabla\wedge \frac{\vc u}{\mu} = 0\,,
\label{eq:tevesueq}
\end{split}
\eeq
where
\beq
\vc u = -\frac{4\pi}{k}\mu \nabla\phi\,.
\label{eq:defu}
\eeq
Notice $\mu=\mu\left(k l^{2}\left(\nabla\phi\right)^{2}\right)$ is the formerly introduced free-function
whose implicit definition could be taken as
\beq
\frac{\mu}{\sqrt{1-\mu^{4}}}=\frac{k}{4\pi}\frac{\left|\nabla\phi\right|}{a_{0}}\,,
\label{eq:mondfreefun}
\eeq
and Milgrom's acceleration
$a_{0}$ is recovered as $a_{0}=\nicefrac{\sqrt{3 k}}{4\pi l}\simeq 10^{-10}\,\unit{m}{s^{2}}$.

The set of two equations - nonlinear in $\phi$! - \eqref{eq:tevesueq} tell us that $\vc u$ equals the
Newtonian acceleration $\vc F^{(N)}=-\nabla\Phi_{N}$ up to a curl which gets fixed by the
second equation up to a gradient. Thus:
\beq
\vc u=\vc F^{(N)}+\nabla\wedge\vc h\,,
\label{eq:utot}
\eeq
the whole potential thus given by $\Phi=\Phi_{N}+\phi$ and is recovered by inverting \eqref{eq:utot}.

By squaring \eqref{eq:defu}, using it in \eqref{eq:mondfreefun} and
carrying out the curl in the second of \eqref{eq:tevesueq} we are lead to:
\begin{align}
u^{2}=\left(\frac{4\pi}{k}\right)^{2}\mu^{2}\left|\nabla\phi\right|^{2}\,,\\
\frac{u^{2}}{a_{0}^{2}}=\left(\frac{4\pi}{k}\right)^{4}\frac{\mu^{4}}{1-\mu^{4}}\,,\\
\frac{\partial \ln u^{2}}{\partial \ln \mu} u^{2}\nabla\wedge\vc u
  +\vc u\wedge\nabla u^{2}=0\,.
\end{align}
In terms of the dimensionless vector field
\beq
\vc U \equiv \left(\frac{k}{4\pi}\right)^{2} \frac{\vc u}{a_{0}}
\label{eq:nodimvecfld}
\eeq
\eqref{eq:tevesueq} takes the form:
\beq
\begin{split}
\nabla\cdot \vc U &= 0\,,\\
4(1+U^{2})U^{2}\nabla\wedge\vc U + \vc U\wedge \nabla U^{2}&=0\,,
\end{split}
\eeq
where we dropped the source term since we are interested the empty region near the SP.
Once $\vc U$ has been retrieved, $\nabla\phi$ can be recovered by:
\beq
-\nabla\phi = \frac{4\pi a_{0}}{k}(1+U^{2})^{\frac{1}{4}}\frac{\vc U}{U^{\frac{1}{2}}}\,.
\label{eq:nablaphiwithU}
\eeq
The Newtonian limit condition is equivalent to $U\gg 1$, in this case:
\beq
-\nabla \phi \underset{U\gg 1}{\sim} \frac{4\pi a_{0}}{k}\vc U = \frac{k}{4\pi}\vc u\,.
\eeq

The SP of the Sun-Earth system is a potential candidate for MOND corrections to Newtonian gravity.
We spare the reader from the gory details of the deduction, but we'd like now to focus on the premises of the SP itself
and thereby analyse the behaviour of the theory. Till now and in the following, we're just reviewing \cite{PhysRevD.73.103513}
and compacting the notation.

A quasi-Newtonian region is defined when $U^{2}\simeq 1$. An estimate of the size and shape might be given dropping the
curl term in \eqref{eq:utot} and finding the solution to $U^{2} = 1$ by virtue of \eqref{eq:nodimvecfld}. The final
shape is that of an ellipsoid:
\begin{shadefundtheory}
\beq
r^{2}\left(\cos^{2}\psi+\frac{1}{4}\sin^{2}\psi\right)=r_{0}^{2}\,,
\label{eq:finshape}
\eeq
\end{shadefundtheory}
\noindent where $r_{0}=\nicefrac{16\pi^{2}a_{0}}{k^{2} A}$ and $A$ is a parameter hiding the mass and distance peculiarities of the SP \cite{PhysRevD.73.103513}:
\beq
A=2\frac{G M}{r_{s}^{2}}\left(1+\sqrt{\frac{M}{m}}\right)\,.
\label{eq:exprA}
\eeq

In spherical polar coordinates $(r,\,\psi,\,\phi)$ with origin at the SP, we introduce the Ansatz of splitting the $\vc U$ field into a Newtonian and a solenoidal
component:
\beq
\vc U = \vc U_{0} + \vc U_{2}\,,
\eeq
where $\vc U_{2}$ is sourced purely by $\vc U_{0}$ to first order and it lacks the $\phi$ component
being solenoidal. Therefore the following equations hold:
\beq
\begin{split}
\frac{r_{0}}{r}\vc U_{0} &= \frac{1}{4}\left(1+3\cos 2\psi\right)\vc e_{r}-\frac{3}{4}\sin 2\psi \vc e_{\psi}\,,\\
\nabla\cdot \vc U_{2}&=0\,,\\
\nabla\wedge \vc U_{2}&=-\frac{\vc U_{0}\wedge\nabla\left|\vc U_{0}\right|^{2}}{4 \left|\vc U_{0}\right|^{4}}\,.
\end{split}
\label{eq:mondsys}
\eeq
Boundary conditions are of paramount importance: $\vc U_{2}$ vanishes for $r\to\inf$ so that $\vc U\to \vc U_{0}$
and the Newtonian field is restored. The inward part of the boundary shifts from almost-Newtonian behaviour till more severe MOND
regimes. Moreover, the normal component of $\vc U$ must vanish on all boundaries. The solution of the former system
\eqref{eq:mondsys} for the field $\vc U_{2}$ with these boundary conditions is:
\beq
\vc U_{2} = \frac{r_{0}}{r}\left(H_{1}(\psi) \vc e_{r} + H_{2}(\psi) \vc e_{\psi}\right)\,,
\label{eq:u2sol}
\eeq
with
\beq
\begin{split}
H_{1}(\psi)&=\frac{2}{5+3\cos 2\psi}-\frac{\pi}{3\sqrt{3}}\,,\\
H_{2}(\psi)&=\frac{\arctan(\sqrt{3}-2\tan{\frac{\psi}{2}})+\arctan(\sqrt{3}+2\tan{\frac{\psi}{2}})-\frac{\pi}{3}\left(\cos\psi+1\right)}{\sqrt{3}\sin\psi}\,.
\end{split}
\eeq

The relevant conclusion for our purposes is that a SP far away from the strong MOND bubble is characterised by a
Newtonian component proportional to $r$ together with a magnetic-like perturbation falling off like $\nicefrac{1}{r}$.
By combining \eqref{eq:nablaphiwithU} with \eqref{eq:u2sol} and the expression for $\vc U_{0}$ we find
that the extra acceleration felt by test particles is expressed by:
\beq
\vc{{\delta}F}=-\nabla\phi\simeq\frac{4\pi a_{0}}{k}\left(\vc U_{0}+\frac{\vc U_{0}}{4U_{0}^{2}}+\vc
U_{2}+\ldots\right)\,.
\eeq
The contribution proportional to $\vc U_{0}$ are Newtonian in nature, but a new magnetic-like contribution shows up,
of the same order of magnitude as the second Newtonian one.

\begin{table}
\begin{center}
\begin{tabular}{r|l}
System & $r_{0}\,(\unit{km})$\\
\hline
Earth-Sun & $383$ \\
Jupiter-Sun & $9.65\times 10^{5}$\\
Earth-Moon & $140$
\end{tabular}
\end{center}
\caption{MOND bubbles major semi-axis lengths for three relevant binary isolated systems in the Solar System.}
\label{tab:mondbubb}
\end{table}

The ``magnetic'' term we've included in the analysis is hence of great importance.
In defining the ellipsoid \eqref{eq:finshape}, solution of $U^{2}=1$ we also
define a turn-point between linear corrections to Newton's theory and full MOND
regime and we can estimate the relative corrections as:
\begin{shadefundtheory}
\beq
\frac{{\delta}F}{F^{(N)}}\sim \left(\frac{4\pi}{k}\right)^{3}\left(\frac{a_{0}}{A}\right)^{2}
  \frac{1}{r^{2}} = \frac{k}{4\pi}\left(\frac{r_{0}}{r}\right)^{2}\,.
  \label{eq:fraccorr}
\eeq
\end{shadefundtheory}
\noindent We see the correction falls off as $\nicefrac{1}{r^{2}}$ moving away from the SP
and by taking the phenomenological value $k\simeq 0.03$ \cite{Bekenstein:2004ne}
in the nearby of the bubble is amounts to order $0.25\%$.

Given the expression $r_{0}=\nicefrac{16\pi^{2}a_{0}}{k^{2} A}$ and the choice
of $k$ we made, together with formula \eqref{eq:exprA} for $A$, the value of $r_{0}$
can be computed for the SPs between the Earth-Sun, Jupiter-Sun and Earth-Moon systems.
Results can be retrieved in table \ref{tab:mondbubb}.

Of course at least two caveat we need to cast:
\begin{enumerate}
\item this analysis gets more and more na{\^i}ve for severe nonlinear MOND corrections,
where the effect will become much larger than any prediction of \eqref{eq:fraccorr}
and the very expression cannot be believed anymore,
\item the real Solar System has a large number of complications at play, starting from
elliptical orbits to being a many-body problem. It is hence sensitive to say that
the shape and the size of the MOND bubbles may be affected by such perturbations, but
not to leading order, while the location of SPs may change abruptly. Notice locating SPs
is a Newtonian physics problem, independent of MOND.
\end{enumerate}

LTP may target and detect MOND effects in spite of their smallness.
Background tidal stresses of the Sun-Earth system are of
order $A\simeq 10^{-11}\,\unit{s}^{-2}$,
roughly four order of magnitudes above the sensitivity of LTP. \eqref{eq:fraccorr}
displays therefore:
\beq
{\delta}F \simeq F^{(N)} \frac{k}{4\pi} \left(\frac{r_{0}}{r}\right)^{2}
  \simeq 10^{-13} \left(\frac{r_{0}}{r}\right)^{2}\,\unitfrac{1}{s^{2}}\,,
\eeq
solving for $r$ assuming LTP sensitivity for ${\delta}F$ gives the radius the space-probe
shall hang around to detect some effect, amounting to be $\sim 10 r_{0} = 3830\,\unit{km}$
for the Sun-Earth system.

Unfortunately, on a Lissajous orbit around L1 - sharing L1 dynamical environment - the MOND
stresses will be far too small to be detectable: given $r_{L}$ to be L1 distance from the Sun, then
its distance from the Earth will be $R-r_{L} \simeq 1.5\times 10^{6}\,\unit{km}$. Conversely, the SP
of the Sun-Earth system is at $R-r_{s} \simeq 2.6\times 10^{5}\,\unit{km}$ from Earth.
Hence L1 is far from SP some ${\Delta}r=r_{s}-r_{L}\simeq 1.24\times 10^{6}\,\unit{km}$,
bringing the correction in \eqref{eq:fraccorr} to:
\begin{shademinornumber}
\beq
\frac{{\delta}F}{F^{(N)}} \simeq \frac{k}{4\pi}\left(\frac{r_{0}}{r+{\Delta}r}\right)^{2}
\simeq \frac{k}{4\pi}\left(\frac{r_{0}}{{\Delta}r}\right)^{2} \simeq 2.4\times 10^{-10}\,.
\eeq
\end{shademinornumber}
\noindent The background tidal stresses due to Newtonian dynamics at L1 are still high in
comparison to LTP sensitivity; e.g. for the radial component:
\begin{shademinornumber}
\beq
\frac{\partial F^{(N)}}{\partial r}\Bigr|_{\text{L1}} \simeq 3.17\times 10^{-13}\,\unitfrac{1}{s^{2}}\,,
\eeq
\end{shademinornumber}
\noindent which is two order of magnitudes above the figure of $\sim 10^{-15}\,\unitfrac{1}{s^{2}}$.
Therefore the game will be to detect corrections $8$ order of magnitudes beneath
experimental sensitivity, a thing we cannot believe.

\section{Conclusions, measurable effects and limitations}

At the present status of the mission design, the {\bf measure of $G$} seems to be the most
promising test of fundamental physics onboard LTP.

Critical parameters may be inspected in table \ref{tab:givendata} with
reference to formula \eqref{eq:grelvariat}. Let motion be induced by
the spare mass of LTP or by virtue of a DRS-like source, the knowledge of the
distance parameter is the most troublesome issue. Moreover, in lack of an independent source
such as DRS, stiffness may dominate the measure anyway.

The baseline distance $r$ will be known with precision $\sim 10^{-4}$, while
the deemed precision shall be $\sim 10^{-7}$; furthermore, what will happen after launch and
orbit placement to mutual
distance of apparata is predictable only up to some extent. This is not going to affect
a differential acceleration measurement, but unless a dedicated calibration tool will be
introduced - a fact which is very unlikely to happen - the figure of $\sim 10^{-4}$
precision on $r$ is a real one and a heavy one for the measure of $G$.

In the fortunate case of a thorough calibration of the $k_{1}$ constant in \eqref{eq:gvalltp},
the value of $\nicefrac{{\delta}G}{G}$ will be dominated anyway by $\epsilon_{r}$, thus
providing a competitive value for $G$ with regard to the accepted value of \eqref{eq:valueofG}. We fear this
occurrence to be too optimistic, it will be rather more preferable to take profit of a dedicated
source in place of DRS. As stated, this may take the shape of a two-mass slid, or a rotator
with cubic masses, but nothing prevents to think about many little devices to be placed
with careful symmetry around the LTP centre of mass, and activated independently at will.

On the converse, with no calibration and no external source, $\epsilon_{k_{1}}$ will
dominate the relative uncertainty picture, and $\nicefrac{{\delta}G}{G}$ will be
- optimistically - of order $10^{-1}$... the measure will still be the first one in space, but
with no surprises or real scientific value.

As for the {\bf test of ISL}, we discussed them here for the sake of completeness, but we are
not really confident that LTP shall provide news about this argument. Twofold the
counter-proof: the static-gravitation analysis which was carried on for gravitational compensation
is based on the $\nicefrac{1}{r^{2}}$ ISL scaling, in case new effects would arise, it might
be very hard to discriminate them from this SC-induced background, and anyway a more sophisticated
analysis will be needed\footnote{Lobo, A., private communication}.
In addition, unless the $\epsilon_{k_{1}}$ relative error on stiffness dominance or in turn the
$\epsilon_{r}$ distance one will be depressed on the measure of $G$, it is very unlikely the
deemed sensitivity will be reached onboard LTP in order to perform a competitive test of the ISL.
Again a minor task might be to confirm the $\nicefrac{1}{r^{2}}$ scaling and
standard Newtonian behaviour at L1 and across the orbit.

It may be true that indirect {\bf MOND} effects may be felt by the orbit of SMART-2;
an extra acceleration will be present at L1 pointing toward the SP (away from the Sun, in
direction of Earth) may be felt by the SC and seen in the orbit \cite{PhysRevD.73.103513}.

To give our little contribution to this debate, we frankly think that MOND effects will be
very hard to detect by LTP unless a careful numerical estimate of the Sun-Earth SP be made
given the many-body influence of the other planets. Once located, the space-probe may
be sent in the premises and a measure might be taken.

We have nevertheless three arguments we'd like to share at the end of the chapter:
\begin{enumerate}
\item theoretically, MOND doesn't seem very interesting a solution at
mesoscopic level. To clarify the thought, MOND doesn't come out of any natural
quantum scenario and doesn't look like a consistent or cooperative effect of
some fundamental field at play. It is not enough in physics to collect
a bunch of fields with some given constants and claim this is a new paradigm.
Though curious and astonishing in the remarkable phenomenological success,
it is very unlikely that even an experimental test at LTP level would produce
unequivocal data to be interpreted as ``MOND effects''.
\item At experimental level, perhaps a longer baseline interferometric
device could (will?) have better chance to see MOND effects. If used
as gradiometers, the SCs of LISA could for example laser-map bubbles
around the SPs. These do not behave like GW sources, they are an example
of a static, non radiative gravitational deformed region, but whenever a laser beam would
move and cross their section it would experience the $\nicefrac{{\delta}F}{F^{(N)}}$
correction, and the amount of it would be larger the closer to the SP the beam
would shine. At the moment we have no numbers or figures to substantiate the picture,
but the idea seems promising.
\item The introduction of a time-like field $V^{\mu}$ corresponds to
the identification of a preferred time-frame (cosmological, since the action is local and
no difference is cast on the form of $V^{\mu}$ from point to point). The metric
stretching may be therefore anisotropic. Another chance could be then to
perform acceleration difference measurement with the sensing axis of LTP
randomly oriented in space and see whether any difference arises (i.e. by moving the whole SC). The large Lissajous orbit
and the yearly revolution of SMART-2 might provide a natural envelope of
space orientation to explore this scenario.
\end{enumerate}

%% file: chapters/gwtheory.tex
\chapter{Gravitational waves in Einstein General Relativity}
\label{chap:gwtheory}

\lettrine[lines=4]{W}{e hereby}
introduce the concept of gravity waves according to the framework of GR. Equations of motion for
wave-like propagation of gravity in space-time are derived from Einstein's equations in the far-source approximation,
polarisation is discussed in relation to the
choice of a gauge (the TT-gauge), energy density carried by the waves and the lowest-level
quadrupole nature of the sources are introduced and discussed.

Next we'll gauge to the first part in answering the question
on how GWs may be detected and their effect measured by interferometers on small and large scales.

Strains figures and examples will be provided, together with a link to sources frequencies.

The chapter is thought in support to chapter \ref{chap:refsys} and supposed to provide a more human-readable and common introduction, in fact,
what we present here may be regarded as standard text-book material.
%

\newpage

\section{Gravitational waves in Einstein's theory}

The existence of GWs can be deduced starting from two assumptions: Lorentz invariance
and local causality: fields are functions of space and time an the requirement of all signals
(including the gravitational) to be non-superluminal with respect to
the limiting speed scale $c$ forces the propagation in vacuo to be wave-like. If not a full certainty,
these facts must at least induce some suspect on the existence of GWs.

GWs can be deduced \cite{wald} from Einstein's relativistic field equations under the approximation of weak field,
which accounts on having far sources. No assumptions is a-priori made on time dependence of motion and no restriction
is cast on particle motion too.

The weakness of the gravitational field is expressed as our ability to decompose the metric $g_{\mu\nu}$
into flat Minkowski plus a small perturbation $h_{\mu\nu}$, such that $|h_{\mu\nu}| \ll 1 $ (a condition
which must hold throughout all the following!):
\beq
g_{\mu\nu} = \eta_{\mu\nu}+h_{\mu\nu}\,. \label{eq:met1}
\eeq
From now on, we will adopt Einstein's conventions as for Greek indice, i.e. designating four-vectors, Latin indice
to address three-vectors, and implied summation over repeated indice. Indice are normally raised or lowered at need by means
of the full metric $g_{\mu\nu}$ or its inverse unless otherwise specified. In the case of weak field, $\eta_{\mu\nu}$
and/or its inverse will be used. A comma will identify partial derivative whilst a semicolon a $g_{\mu\nu}$-co-variant one.
It can be easily read out of \eqref{eq:met1} that the inverse is
\beq
g^{\mu\nu} = \eta^{\mu\nu}-h^{\mu\nu}\,. \label{eq:met2}\,.
\eeq
Whenever the symbol $\vc x^{2}$ will appear, it will have the meaning of $\sum_{i}x_{i}x_{i}=|\vc x|^{2}$.

We'd like now to proceed finding the Christoffel connection symbols\footnote{We employ symmetryzers and anti-symmetryzers
with the following meaning:
\begin{equation*}
\begin{split}
A_{(\mu,\nu)}&=\frac{1}{2}\left(A_{\mu,\nu}+A_{\nu,\mu}\right)\,,\\
A_{[\mu,\nu]}&=\frac{1}{2}\left(A_{\mu,\nu}-A_{\nu,\mu}\right)\,.
\end{split}
\end{equation*}}:
\begin{align}
\Gamma^{\rho}_{\phantom{\rho}\mu\nu} &= \frac{1}{2}g^{\rho\sigma}
 \left( g_{\nu\sigma,\mu}+g_{\mu\sigma,\nu}-g_{\mu\nu,\sigma} \right) = \\
 &= \frac{1}{2}\eta^{\rho\sigma}
 \left( h_{\nu\sigma,\mu}+h_{\mu\sigma,\nu}-h_{\mu\nu,\sigma} \right)\,,
\label{eq:christoffellin}
\end{align}
and the Riemann tensor expression:
\begin{align}
R_{\mu\nu\rho\sigma} &= g_{\mu\lambda}R^{\lambda}_{\phantom{\lambda}\nu\rho\sigma}
 = g_{\mu\lambda} \left(
 \Gamma^{\lambda}_{\phantom{\lambda}\nu[\rho,\sigma]}
 + \Gamma^{\tau}_{\phantom{\tau}\rho[\nu}\Gamma^{\lambda}_{\phantom{\lambda}\sigma]\tau}
 \right) = \\
 &= \eta_{\mu\lambda}
 \Gamma^{\lambda}_{\phantom{\lambda}\nu[\sigma,\rho]}
 = \\
 &= \frac{1}{2}\left(h_{\sigma[\mu,\nu],\rho}
 +h_{\rho[\nu,\mu],\sigma}
 \right)\,.
\end{align}
The Ricci tensor is obtained by contracting over $\rho$ and $\sigma$, thus giving:
\beq
R_{\mu\nu}=g^{\rho\sigma}R_{\mu\nu\rho\sigma}=\frac{1}{2} \left(
 h^{\sigma}_{\phantom{\sigma}(\mu,\nu),\sigma}
 -h_{,\nu,\mu}-\square h_{\mu\nu}
 \right)\,,
\eeq
where we defined the trace $h=\eta^{\mu\nu}h_{\mu\nu}$ and the D'Alembert operator in Minkowski space $\square =
\eta^{\mu\nu}\partial_{\mu}\partial_{\nu}$.
The Ricci scalar is:
\beq
R=g^{\mu\nu} R_{\mu\nu}=h^{\mu\nu}_{\phantom{\mu\nu},\nu,\mu}-\square h.
\eeq

With these assumptions, the l.h.s. of Einstein' equations:
\begin{shadefundtheory}
\beq
E_{\mu\nu} = R_{\mu\nu}-\frac{1}{2}R g_{\mu\nu} = \frac{8 \pi G T_{\mu\nu}}{c^{2}}\,,
\label{eq:einstein}
\eeq
\end{shadefundtheory}
\noindent become:
\begin{align}
E_{\mu\nu} &\to R_{\mu\nu}-\frac{1}{2}R \eta_{\mu\nu} = \label{eq:lin1}\\
 &= \frac{1}{2} \left(
 h^{\sigma}_{\phantom{\sigma}(\mu,\nu),\sigma}
 -h_{,\nu,\mu}-\square h_{\mu\nu} -\eta_{\mu\nu}h^{\rho\sigma}_{\phantom{\rho\sigma},\sigma,\rho}+\eta_{\mu\nu}\square h
 \right)=\\
 &= \frac{8 \pi G T_{\mu\nu}}{c^{2}}\,.
\end{align}
In \eqref{eq:einstein} $T_{\mu\nu}$ represents the energy-momentum tensor whose expansion will be considered
to zeroth order in $h$. $T_{\mu\nu}$ must be small for the weak-field condition to apply; to lowest order
the conservation of energy and momentum simplifies to $T_{\mu\nu;\mu}\to T_{\mu\nu,\mu} = 0$.

The field equations \eqref{eq:lin1} do not have unique solutions due to general covariance:
the same physical
situation is represented by different choices of coordinates as:
\beq
x^{\mu}\to x^{\mu}+\xi^{\mu}\,,
\eeq
so that the perturbation $h_{\mu\nu}$ is related to a transformed
other, leaving curvature and hence the physical space-time unchanged. Formally:
\beq
h_{\mu\nu} \to h_{\mu\nu}+
\xi_{(\nu,\mu)}
\,.
\eeq
Notice, since $\left|h_{\mu\nu}\right|\ll 1$ that $\xi_{\mu}$ and $\xi_{\mu,\nu}$ must be
(infinitesimal) of order $h_{\mu\nu}$.
The choice of a gauge is important to simplify the form of the final equations. We choose the harmonic
gauge condition:
\beq
0=g^{\mu\nu} \Gamma^{\lambda}_{\phantom{\lambda}\mu\nu}=h^{\lambda}_{\phantom{\lambda}\mu,\lambda}-\frac{1}{2}h_{,\mu}\,,
\label{eq:harmonicgauge}
\eeq
by introducing $\bar h_{\mu\nu}=h_{\mu\nu}-\frac{1}{2}\eta_{\mu\nu}h$ the harmonic gauge
reduces to the Lorentz divergence condition $\bar h^{\mu}_{\phantom{\mu}\lambda,\mu}=0$
and the linearised Einstein' equations simplify to:
\begin{shadefundtheory}
\beq
\square \bar h_{\mu\nu}=-\frac{16 \pi G}{c^{2}} T^{\mu\nu}\,. \label{eq:wave1}
\eeq
\end{shadefundtheory}

Outside the source, in empty space $T^{\mu\nu}=0$ and we may drop the r.h.s. of \eqref{eq:wave1} thus getting:
\beq
\square \bar h_{\mu\nu}=0\,. \label{eq:wave2}
\eeq
Such an equation admits a plane-wave solution of the kind $\bar h_{\mu\nu}=A_{\mu\nu} \exp (i k_{\alpha}x^{\alpha})$ where
$A_{\mu\nu}$ is a constant, symmetric, rank-2 tensor and $k_{\alpha} = (\omega, \vc k)$ is the wave vector. Gravitational waves travel at the speed of light
and this can be proved by plugging the readily-found solution into \eqref{eq:wave2} to find:
\beq
k_{\alpha} k^{\alpha} = 0\,,
\eeq
that is, the world line of a GW is light-like. Moreover, using the linearised harmonic gauge condition we get
\beq
k_{\alpha}A^{\alpha\beta}=0
\eeq
implying trasversality (orthogonality) of the wave front to $k_{\alpha}$. The remaining gauge d.o.f. in $A_{\mu\nu}$ can be
fixed by choosing $\bar h_{00}=0$ and $\eta^{\mu\nu}\bar h_{\mu\nu}=0$, i.e. the tracelessness. In full, this gauge choice is named after TT-gauge,
meaning transverse and traceless. As an effect of the lack of trace we have $\bar h^{TT}_{\mu\nu}=h^{TT}_{\mu\nu}$. Choosing now to meet
the GW along our $\hat z$ direction, we have
\beq
k_{\mu=0..3}\to (\omega, 0, 0, \omega)\,,
\eeq
which implies $A_{\alpha 3=0}$ and we are left with \cite{Chakrabarty:1999aa}
\beq
A_{\mu\nu}^{TT}=\left[\begin{matrix}
0 & 0 & 0 & 0 \\
0 & A_{11} & A_{12} & 0 \\
0 & A_{21} & -A_{11} & 0 \\
0 & 0 & 0 & 0
\end{matrix}\right]=\left[\begin{matrix}
0 & 0 & 0 & 0 \\
0 & h_{+} & h_{\times} & 0 \\
0 & h_{\times} & -h_{+} & 0 \\
0 & 0 & 0 & 0
\end{matrix}\right]\,,\label{eq:tensoramp}
\eeq
and the two a-dimensional lengths $h_{+}$ an $h_{\times}$ completely characterise the wave ``amplitude''.

\section{Sources and energy-momentum}

Back to equation \eqref{eq:wave1}, a general solution can be found by means of Green's function formalism \cite{wald, Flanagan:2005yc},
to give:
\beq
\bar h_{\mu\nu}(t,\vc x)=\frac{4 G}{c^{2}} \int \de^{3} y
 \frac{T_{\mu\nu}(t-\frac{|\vc x-\vc y|}{c},\vc y)}{|\vc x-\vc y|}\,,
 \label{eq:greensol}
\eeq
where the retarded time $t_{R}=t-\nicefrac{|\vc x-\vc y|}{c}$ accounts for the disturbance of the gravitational field at
present coordinates $(t, \vc x)$ to be the superposition of influences from energy and momentum sources at $(t_{R},\vc y)$ on the past
light cone.

Under the assumption of being far from the source, considered point-like in its spatial extent compared to the distance ($|\vc y| \ll |\vc x|$),
we get
\beq
\bar h_{\mu\nu}(t,\vc x)=\frac{4 G}{c^{4} |\vc x|}
 \int T_{\mu\nu}\left(t-\frac{|\vc x|}{c},\vc y\right)\de^{3}y\,,
 \label{eq:greensol2}
\eeq
and by means of the sources identity and definition of quadrupole moment\footnote{This is easily proved
by repetitive use of $T^{\mu\nu}_{\phantom{\mu\nu},\mu}=0$ and by writing
$0=\int x^{k} T^{\mu\nu}_{\phantom{\mu\nu},\mu} \de^{3} x = \frac{\partial}{\partial t} \int x^{k} T^{0\nu}
+\int x^{k} \partial_{m}T^{m\nu}$ and using partial integration and Gauss' theorem.
}
\beq
\int T^{jk} \de^{3} x = \frac{1}{2} \frac{\partial^{2}}{\partial t^{2}} \int T^{00} x^{j} x^{k} \de^{3} x\,,
\eeq
\beq
q_{jk}\doteq \int T^{00} x^{j} x^{k} \de^{3} x\,.
\eeq
Notice anyway that neither side of \eqref{eq:greensol2} is transverse or traceless with respect to the observation
direction. We need therefore to introduce projectors onto the plane with normal $\vc n \doteq \nicefrac{\vc x}{\left|\vc x\right|}$:
\beq
P_{jk} \doteq \delta_{jk}-n_{j} n_{k}\,,
\eeq
and combine them in a TT projector:
\beq
P^{\text{TT}}_{jkmn} \doteq P_{jm}P_{kn} -\frac{1}{2} P_{jk}P_{mn}\,,
\eeq
so that, for the newly defined $q_{jk}$ we'd get:
\beq
q^{\text{TT}}_{jk} = P^{\text{TT}}_{jkmn} q_{mn}\,,
\eeq
and finally from \eqref{eq:greensol2}:
\begin{shadefundtheory}
\beq
\bar h^{\text{TT}}_{jk}(t,\vc x)=\frac{2 G}{c^{4}|\vc x|}\frac{\de^{2}q^{\text{TT}}_{jk}}{\de t^{2}}\Big|_{t-\nicefrac{|\vc x|}{c}}\,,
\eeq
\end{shadefundtheory}
\noindent therefore showing the gravitational field produced by an isolated non-relativistic object is proportional to the
second time derivative of the TT-projected quadrupole moment of the energy density at the point where the past light cone
of the observer intersects the source. The universal nature of gravitation is therefore quadrupolar to lowest order
due to mass and momentum conservation, elucidating also the weakness of the interaction, generally smaller than dipolar
EM one.

The canonical energy-momentum pseudo-tensor (Landau) carried by gravitational waves can be obtained from the usual field-theory
formalism:
\beq
t_{\mu\nu}=\frac{1}{32 \pi G} \left\langle
h^{TT}_{\rho\sigma,\mu} h^{TT\rho\sigma}_{\phantom{TT\rho\sigma},\nu} \right\rangle\,,
\eeq
where the r.h.s. has been averaged over several wavelengths, to circumvent a definition of local gravitational
field - a nonsense for GW due to the existence of Riemann normal coordinates \cite{MTW}. Moreover the averaging has
the practical meaning to capture more information on the physical curvature of a small region to describe a
gauge-invariant measure.

Finally we can compute the luminosity of a source:
\begin{shademinornumber}
\beq
L_{\text{GW}}=-\frac{\de W_{\text{GW}}}{\de t} = \frac{G}{5 c^{5}}\left\langle
\left(\frac{\de^{3}Q_{jk}}{\de t^{3}}\right)^{2}
\right\rangle\Big|_{t-\frac{|\vc x|}{c}}\,,
\eeq
\end{shademinornumber}
\noindent in terms of its traceless (reduced) quadrupole tensor\footnote{Perhaps it is wise to state precise definitions and relations: we already
 defined the quadrupole moment $q_{jk}$, we now introduced the traceless quadrupole tensor $Q_{jk}$; another familiar quantity
 is the inertia tensor:
\beq
 I_{jk}=\int \rho(\vc x) (\vc x^{2}\delta_{jk}-x_{j}x_{k})\de^{3} x.
\eeq
 If $T^{00} \simeq \rho(\vc x)$ the three are related as follows:
 \beq
 \begin{split}
 q_{jk}&=-I_{jk}+\delta_{jk}q\,,\\
 Q_{jk}&=q_{jk}-\frac{1}{3}\delta_{jk}q=-\left(I_{jk}-\frac{1}{3}\delta_{jk}I\right)\,,
\end{split}
 \eeq
 where $q=\tr q_{jk}$ and $I=\tr I_{jk}$.
 }
 \beq
 Q_{jk}=\int \left( x^{j} x^{k} - \frac{1}{3}\vc x^{2} \delta_{jk} \right) T^{00} \de^{3} x\,.
 \eeq

\section{Effects on test particles}

Let's consider the effect of gravitational waves on test particles and moving bodies \cite{Flanagan:2005yc, MTW}
in their proper frame of reference. If we take two
particles in free-fall, described by a single velocity field $U^{\alpha}=\frac{\de x^{\alpha}}{\de \tau}$,
where $\tau$ is the proper time, it is well known that the separation vector $X^{\alpha}$ obeys
the equation of geodesic deviation (see \eqref{eq:geodevcovar}):
\beq
\frac{\de^{2} X^{\mu}}{\de \tau^{2}} + \Gamma^{\mu}_{\phantom{\mu}\sigma\rho} \frac{\de X^{\rho}}{\de \tau} \frac{\de X^{\sigma}}{\de \tau} = R^{\mu}_{\phantom{\mu}\nu\rho\sigma} U^{\nu}U^{\rho}X^{\sigma}\,,
\label{eq:geodev1}
\eeq
considering only the lowest-order components of $U^{\nu}$, we can take $U^{\nu}=\delta^{\nu}_{\phantom{\nu}0}$ pointing in the
time direction $\hat t$ and $X^{\nu}=\delta^{\nu}_{\phantom{\nu}1}$ along $\hat x$. To first order in $h_{\mu\nu}$ we get from
\eqref{eq:geodev1}:
\beq
\frac{\de^{2} X^{\mu}}{\de t^{2}} = R^{\mu}_{\phantom{\mu}00\sigma}X^{\sigma}\,.
\label{eq:geodev2}
\eeq
This expression of the Riemann tensor can be calculated in linearised theory:
\begin{align}
R^{\mu}_{\phantom{\mu}00\sigma} &= \eta^{\mu\alpha}\frac{1}{2}\left(
h^{TT}_{\alpha\sigma,0,0}+h^{TT}_{00,\alpha,\sigma}-h^{TT}_{\mu 0,0,\sigma}-h^{TT}_{\sigma 0,0,\alpha}
\right) = \\
&= \frac{1}{2} \eta^{\mu\alpha}h^{TT}_{\alpha\sigma,0,0}\,,
\end{align}
where we used again $h^{TT}_{\mu 0}=0$. Thus equation \eqref{eq:geodev2} becomes:
\begin{shadefundtheory}
\beq
X^{\mu}_{\phantom{\mu},0,0}=\frac{1}{2} \eta^{\mu\alpha}h^{TT}_{\alpha\sigma,0,0} X^{\sigma}\,.
\eeq
\end{shadefundtheory}
\noindent If the incoming gravitational perturbation is characterised by an amplitude tensor of the form \eqref{eq:tensoramp} and
it has got the form of a plane wave - which is locally likely to be, due to the distance from the sources - the test particles will only
be disturbed in directions $\hat x$ and $\hat y$, orthogonal to the wave vector. Suppose for illustration to choose $h_{\times}=0$
so that:
\beq
\begin{split}
X^{1}_{\phantom{1},0,0} &= \frac{1}{2} X^{1} \left( h_{+}\exp (i k_{\lambda} x^{\lambda}) \right)_{,0,0} \\
X^{2}_{\phantom{2},0,0} &= -\frac{1}{2} X^{2} \left( h_{+}\exp (i k_{\lambda} x^{\lambda}) \right)_{,0,0}\,;
\end{split}
\eeq
To first order in the wave phase, we get:
\beq
\begin{split}
\ddot X^{1} &= \frac{1}{2} X^{1} k_{0}^{2}\left( h_{+} (1+i k_{\lambda} x^{\lambda}) \right)\,, \\
\ddot X^{2} &= -\frac{1}{2} X^{2} k_{0}^{2}\left( h_{+} (1+i k_{\lambda} x^{\lambda}) \right)\,,
\end{split}
\eeq
where we used an over-dot to designate the time derivative.

\begin{figure}
\begin{center}
\includegraphics[width=.8\textwidth]{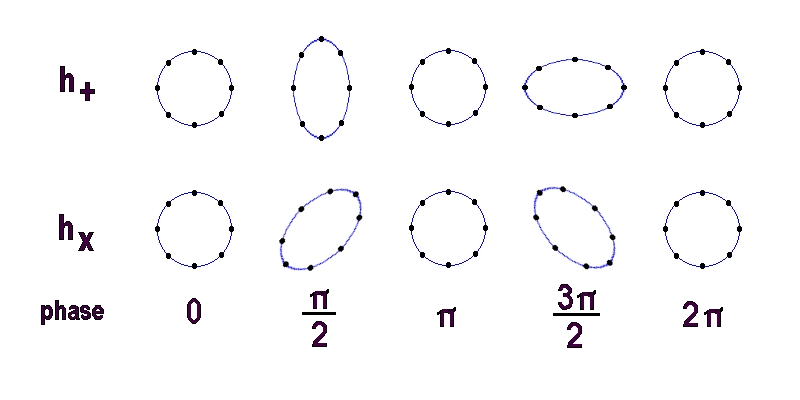}
\end{center}
\caption{Ring of particles reacting to the gravitational wave income. Top: $+$ mode, bottom: $\times$ mode.}
\label{fig:rings}
\end{figure}

These equations show that particles initially apart in the $\hat x$ direction will oscillate along the same and likewise
along $\hat y$. A ring of particles on a $xy$-plane will hence stretch and squeeze bouncing like a volleyball hitting the ground
(see picture \ref{fig:rings}).
In a similar fashion the same ring would bounce with a tilted cross polarisation, had we chosen $h_{+}=0$. The two
amplitudes $h_{+}$ and $h_{\times}$ are independent, therefore any linear combination of the two fundamental polarisation
states is allowed.

General relativity is unique at predicting only two states of polarisation for gravity waves. From fully general symmetry
arguments it may be deduced for a metric theory of gravity that there can be at most $6$ states in $4$
space-time dimensions \cite{Eardley:1974nw}.

\section{Detection of gravitational waves}

To specialise the situation of detection \cite{Flanagan:2005yc}, suppose we have free, unconstrained masses at two points along
the $\hat x$ axis separated by a distance $L$. Measuring a local acceleration between these masses relies in
the ability of shooting light between them measuring the elapsed time. As if it were a problem in
geometric optics, for a light ray connecting the two masses, due to the light-like nature of photons, we have:
\begin{shadefundtheory}
\beq
\de \tau^{2}=g_{\mu\nu}\de x^{\mu} \de x^{\nu} = 0\,.
\eeq
\end{shadefundtheory}
As stated in the previous sections, the nature of a gravitational perturbation of the kind similar to a gravity wave
far from the source is an extremely weak phenomenon, we can therefore assume again $g_{\mu\nu}=\eta_{\mu\nu}+h_{\mu\nu}$.
Under these assumptions, a GW coming from the $\hat z$ direction will perturb the metric as the world-line of
photons gets distorted as follows:
\beq
0=\de \tau^{2}=-c^{2} \de t^{2} + (1+h_{11}(\omega t - k z))\de x^{2}\,,\label{eq:light2}
\eeq
where now $x$ designates the former $x_{1}$ space coordinate, $c$ is the speed of light in vacuo and the wave
vector has been properly chosen, with frequency $\omega$. The effect of the GW is to modulate the distance between the two
fixed coordinate points marked by the two masses, by the fractional amount $h_{11}$. The travel time is
given by integrating \eqref{eq:light2} over the distance $L$:
\beq
\label{eq:forth}
\int_{0}^{T_{out}}\de t=\frac{1}{c}\int_{0}^{L}\sqrt{1+h_{11}(\omega t - k z)} \de x\,.
\eeq
A similar expression can be written fro the return trip:
\beq
\label{eq:back}
\int_{T_{out}}^{T}\de t=-\frac{1}{c}\int_{L}^{0}\sqrt{1+h_{11}(\omega t - k z)} \de x\,,
\eeq
and the total round trip is thus
\beq
T=\frac{2 L}{c}+\frac{1}{c}\int_{0}^{L} h_{11}(\omega t - k z) \de x\,,
\eeq
where the expansion $\sqrt{1+x}\simeq 1+\frac{1}{2}x$ was used, having $|h_{11}| \ll 1$.

If timescales are such that $T$ is short compared with the period of the wave we find that the fluctuation
in $T$ due to the wave is given by
\beq
\Delta T= \frac{L}{c} h_{11}\,,
\eeq
or the effective percent change in separation of the masses is
\begin{shademinornumber}
\beq
\label{eq:lscaling}
\frac{\Delta L}{L}= \frac{h_{11}}{2}\,,
\eeq
\end{shademinornumber}
\noindent therefore $h$ can be interpreted as a physical strain in space.

The former deduction has been carried on bearing in mind $T$ to be larger than the GW
period and assuming perfect positioning of the detector with regard to the angular position
of the source. For cross polarisation a detector with two masses on a joining line tilted by
and angle $\theta$ toward the $\hat z$ direction of the incoming wave, with projection
on the $\hat x-\hat y$ plane rotated by $\phi$ with respect to $\hat x$, the
effective strain will be $h_{11} \sin^{2}\theta \cos 2 \phi$.

The scaling law \eqref{eq:lscaling} won't hold for arbitrary long interferometer arms \cite{saulson}. If
the optical path is so long that $\omega_{\text{GW}} T \ll 1$ is no longer valid, we have
to carry out carefully the integrations in \eqref{eq:forth} and \eqref{eq:back} considering
the ``zeroes'' of the wave, such that $\omega_{\text{GW}} T = 1$: in these cases
the light spends exactly one gravitational wave period in the apparatus, for every part of
its path for which the light crosses through a region of positive $h$, there's an equal
part for which it sees an equal but opposite value of $h$. Hence no net modulation
of the total optical path could be sensed. We need then do drop our assumption of
constancy of $h$ along $T$. The easiest way is to think $h(t)$ a modulated plane wave,
where $h(t)$ is now slowly varying with respect to $\omega_{\text{GW}}$:
\begin{shadefundtheory}
\beq
h(t)\to h(t) \exp\left(\iota \omega_{\text{GW}}t\right)\,,
\eeq
\end{shadefundtheory}
\noindent then the ongoing-path integral can be computed as follows:
\beq
\int_{0}^{T_{out}}\de t \simeq \frac{L}{c}+\frac{h(t)}{2\iota \omega_{\text{GW}}}
\left(\exp\left(\iota\frac{\omega_{\text{GW}}L}{c}\right)-1\right)\,,
\eeq
while the return trip yields
\beq
\int_{T_{out}}^{T}\de t\simeq\frac{L}{c}+\frac{h(t)}{2\iota \omega_{\text{GW}}}
 \exp\left(\iota\frac{2 \omega_{\text{GW}}L}{c}\right)
 \left(1-\exp\left(-\iota\frac{\omega_{\text{GW}}L}{c}\right)\right)\,,
\eeq
we thus find, for the time difference, expressed in terms of lengths now, that:
\beq
\Delta T = \frac{\Delta L}{c} = h(t) \frac{L}{c} \exp\left(-\iota\frac{\omega_{\text{GW}}L}{c}\right)
\frac{\sin\left(\frac{\omega_{\text{GW}}L}{c}\right)}{\frac{L}{c} \omega_{\text{GW}}}\,,
\eeq
so that
\begin{shadefundtheory}
\beq
\frac{\Delta L}{L} = h(t) \exp\left(-\iota\frac{\omega_{\text{GW}}L}{c}\right)
\sinc\left(\frac{\omega_{\text{GW}}L}{c}\right)\,,
\eeq
\end{shadefundtheory}
\noindent where we defined $\sinc x = \nicefrac{\sin x}{x}$, and it's easy to see that the result is continuous
into the na\^{i}ve expression \eqref{eq:lscaling} for $\omega_{\text{GW}}T\ll 1$.

An example of the kind of frequencies at play may be given considering the emission of GW by a binary system
in circular orbit. If the two stars are considered to have both mass $m$ and
keep orbiting at distance $l$, we can write the force between them as:
\beq
G\frac{m^{2}}{l^{2}}=m \omega^{2} \frac{l}{2}\,,\quad \text{with}
\quad \omega^{2}=\frac{2 G m}{l^{3}}.
\eeq
The energy density is then
\beq
T^{00}=\sum_{j=1}^{2} m c^{2} \delta(x-x_{j})\delta(y-y_{j})\delta{z}\,,
\eeq
where we considered the stars to be point-like and the orbit lying on the plane
$z=0$. The coordinates are dictated by
the equations of motion:
\beq
\begin{cases}
x_{1}=\frac{l}{2} \cos \omega t \\
y_{1}=\frac{l}{2} \sin \omega t
\end{cases}\quad\begin{cases}
x_{2}=-x_{1} \\
y_{2}=-y_{1}
\end{cases}\,.
\eeq

After some pain - not so much indeed - all the components of $q_{ij}$ and $Q_{ij}$
can be computed, transposed into TT-gauge and we finally find:
\begin{align}
h^{TT}_{xx}=-h^{TT}_{yy}&=-\frac{G}{2 c^{4} r} m l^{2} (2\omega)^{2}
 \cos 2 \omega \left( \frac{r}{c}-t\right)\,, \\
h^{TT}_{xy}&=\frac{G}{2 c^{4} r} m l^{2} (2\omega)^{2}
 \sin 2 \omega \left( \frac{r}{c}-t\right)\,,
\end{align}
showing the radiation is emitted at twice the orbital frequency, carries both polarisation
and in the specific case it's circularly polarised.

Observational evidence of the theoretical prediction can be provided by many
systems; we choose here the binary pulsar PSR $1913+16$, observed by Taylor and
Weisberg in 1982 \cite{taylor:908, weisberg:1348}. From the data it appears $M_{1}\sim M_{2} \sim 1.4 M_{\odot}$
and $l\simeq 0.19 \times 10^{9}\, m \simeq 2 R_{\odot}$, where $M$ designate
masses, $R$ whenever used radius's and the symbol $\odot$ has been employed to
name the Sun.

Again from observed data we see the orbit shows some eccentricity (factor $\epsilon=0.62$)
but we can assume it's circular since we are only interested in estimates. The
orbital frequency is $\omega\sim 3.7\times 10^{-5}\, \unit{Hz}$, so that the
wavelength of the emitted GW will be:
\beq
\lambda_{\text{GW}}=\frac{c}{2 \omega} \sim 10^{12}\, \unit{m}\,.
\eeq
Indeed, $\lambda_{\text{GW}} \ll l$.

The distance\footnote{We remind $1\, \unit{pc} = 3.08 \times 10^{16}\, \unit{m}$.}
of the system from the Earth if $r=5\, \unit{kpc}=1.5 \times 10^{20}\, \unit{m}$,
resulting in a wave amplitude like:
\beq
\frac{G}{2 c^{4} r} m l^{2} (2\omega)^{2}\sim 5.8 \times 10^{-23}\,.
\eeq

Another example may be the PSR J0737-3039 pulsar, more eccentric in the orbit and
with diversely weighting stars. The estimated amplitude is $h=1.1 \times 10^{-21}$,
for GW emitted at frequency $\omega_{\text{GW}}=2.3 \times 10^{-4}\, \unit{Hz}$:
this may be a very good candidate for LISA.

\begin{figure}
\begin{center}
\includegraphics[width=\textwidth]{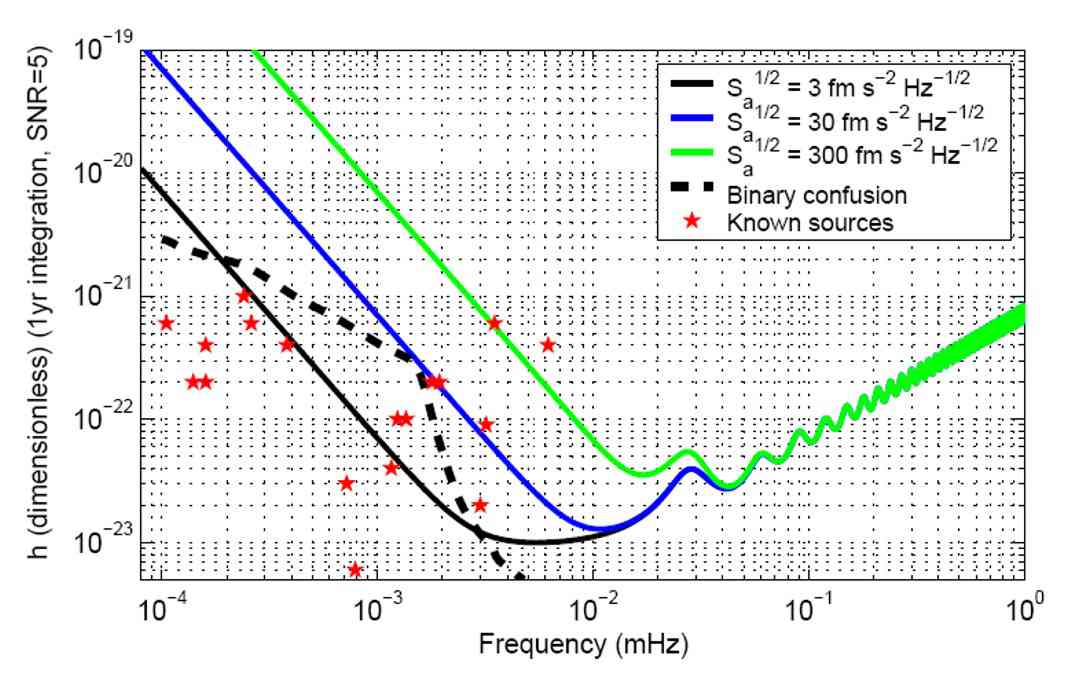}
\end{center}
\caption{Strain sensitivity versus known sources for three values of instrumental resolution.
Values for sources can be found in table \ref{fig:lisacalibbin}.}
\label{fig:sourcesvsfreq}
\end{figure}

As shown, the first experimental problem connected to the detection of GW
is the extreme smallness of of predicted amplitudes and strains in the vicinity of the Earth.
The weakness of the signal makes it mandatory to reduce the noise - from any source - to minimum.
For signals above $\sim 10\,\unit{Hz}$, ground based experiments are possible, but for lower
frequencies local fluctuating gravitational gradients become important and seismic noise
is more problematic, hence detectors for operation in space may be more suitable if not
the unique chance to detect GW. An overview of the characteristic frequencies
with dependence of the sources is shown in picture \ref{fig:sourcesvsfreq}
and table \ref{fig:lisacalibbin}.

A detailed analysis of noise sources, reduction and subtraction will be placed in the
appropriate section.

Let's give an idea about sensitivity and magnitudes. If we were to build a
gravitational wave observatory we could consider test bodies
separated by some distance of the order of kilometres. Suppose the incoming
wave would have a magnitude $h\simeq 10^{-21}$, then the sensitivity
would be of the order
\beq
\Delta L\simeq 10^{-16} \frac{h}{10^{-21}}\frac{L}{\unit{km}}\,\unit{cm}\,,
\eeq
comparable with the size of atoms $a_{0}\simeq 5 \times 10^{-11}\,\unit{m}$.
A gravitational wave observatory will have to be sensitive to changes in distances much smaller than
the size of the constituent atoms out of which the masses have to be made.

Laser interferometers provide a way to perform such an accurate measurement.

%% file: chapters/geodesics.tex
\chapter{Basic facts in differential geometry}
\label{chap:geodesics}

\lettrine[lines=4]{I}{n support}
to chapter \ref{chap:refsys}, this appendix is meant
to review some basic concepts of differential geometry with the purpose
of defining parallel transporters and geodesics, both in absence and
in presence of external forces.

The geodesics mutual acceleration equation and its link to the
Riemann tensor is reviewed in the general case and in weak field approximation.

We'll provide some text-book based demonstrations at need, which may
please the mathematically skilled reader. We invite anyway to get to a more advanced
exposition (books and cited articles) if need were for finer details.

\newpage

\section{Geodesics}

According to differential geometry a \index{geodesic}geodesic is a curve defined by
a parallel transporter along itself.
The curve may be thought as a single parameter string $C=C(\tau)=\left\{x^{\mu}(\tau),\,\mu=0..3\right\}$, so that its tangent vector
component $V^{\mu}$ con be easily obtained as \cite{wald,dinverno}:
\beq
V^{\mu}=\frac{\de x^{\mu}}{\de \tau}\,.
\eeq

Parallel transport along the tangent vector $\vc V$ is put in place by
means of the parallel transport operator $\nabla_{\vc V}$. Naturally, when parallel
transporting a vector along itself, we have no variation of trajectory:
\begin{shadefundtheory}
\beq
\nabla_{\vc V} \vc V=0\,,
\eeq
\end{shadefundtheory}
\noindent where
\beq
\left(\nabla_{\vc V} \vc X\right)^{\mu}=V^{\alpha}X^{\mu}_{\phantom{\mu};\alpha}
=V^{\alpha}\left(X^{\mu}_{\phantom{\mu},\alpha}+\Gamma^{\mu}_{\phantom{\mu}\alpha\beta}X^{\beta}\right)\,.
\eeq
We used the comma to designate standard partial derivative: $X^{\mu}_{\phantom{\mu},\alpha}=\partial_{\alpha}X^{\mu}$
while the semicolon abbreviates the standard co-variant one. Christoffel symbols bearing affine connection are related to the metric as follows:
\begin{shadefundtheory}
\beq
\Gamma^{\rho}_{\phantom{\rho}\mu\nu} = \frac{1}{2}g^{\rho\sigma}
  \left( g_{\nu\sigma,\mu}+g_{\mu\sigma,\nu}-g_{\mu\nu,\sigma} \right)\,.
\eeq
\end{shadefundtheory}
\noindent As a side remark, notice the connection symbols may be not related to a proper metric
tensor, the transport equation being general in nature.

By substitution of the expression of $V^{\mu}$, the geodesic equation reads:
\begin{align}
0=\nabla_{\vc V} \vc V &= V^{\alpha}\left(V^{\mu}_{\phantom{\mu},\alpha}+\Gamma^{\mu}_{\phantom{\mu}\alpha\beta}V^{\beta}\right) =\\
&=\frac{\de x^{\alpha}}{\de \tau}\partial_{\alpha}\frac{\de x^{\mu}}{\de \tau}+\Gamma^{\mu}_{\phantom{\mu}\alpha\beta}\frac{\de x^{\alpha}}{\de \tau}\frac{\de x^{\beta}}{\de \tau} =\\
&=\frac{\de^{2} x^{\alpha}}{\de \tau^{2}}+\Gamma^{\mu}_{\phantom{\mu}\alpha\beta}\frac{\de x^{\alpha}}{\de \tau}\frac{\de x^{\beta}}{\de \tau}\,,
\end{align}
where we used
\beq
\frac{\de x^{\alpha}}{\de \tau}\partial_{\alpha}\frac{\de x^{\mu}}{\de \tau}=
\frac{\de x^{\alpha}}{\de \tau}\frac{\partial \tau}{\partial x^\alpha}\frac{\de }{\de \tau}\frac{\de x^{\mu}}{\de \tau}=
\frac{\de^{2} x^{\mu}}{\de \tau^{2}}\,.
\eeq
Then, there's no acceleration along a geodesic on the line flow dictated by the natural parameter,
i.e. the pure parallel motion is indeed ``freely falling''.

\section{External forces. $3+1$ representation}

In presence of an
external disturbance both EM and generic the geodesics equation gets modified as follows:
\begin{shadefundtheory}
\beq
\frac{\de^{2} x^{\mu}}{\de \tau^{2}}+\Gamma^{\mu}_{\phantom{\mu}\alpha\beta}\frac{\de x^{\alpha}}{\de \tau}\frac{\de x^{\beta}}{\de \tau}
= \frac{e}{m} F^{\mu}_{\phantom{\mu}\sigma}\frac{\de x^{\sigma}}{\de \tau}+ \frac{1}{m} f^{\mu}\,,
\label{eq:geodeqextfcs}
\eeq
\end{shadefundtheory}
\noindent where $F_{\mu\nu}=A_{\mu,\nu}-A_{\nu,\mu}=A_{[\mu,\nu]}$ and $A_{\mu}$ is the usual EM $4$-potential
such that the fields can be calculated as $E_{i}=A_{0,i}$ and $B_{i}=\epsilon_{ijk} A_{j,k}$. $e$ is the electron
charge and $m$ is the mass of the test-body at play.

In general:
\beq
a^{i}=\frac{\de^{2} x^{i}}{\de t^{2}}=\left(\frac{\de \tau}{\de t}\right)^{2} \left(\frac{\de^{2} x^{i}}{\de \tau^{2}}-v^{i}\frac{\de^{2} t}{\de \tau^{2}}\right)\,,
\eeq
where $v^{i}=\nicefrac{\de x^{i}}{\de t}$. substituting from the geodesic equation we get:
\beq
\frac{\de^{2} x^{i}}{\de t^{2}}=\left(\frac{\de \tau}{\de t}\right)^{2}
   \left(
   -\Gamma^{i}_{\phantom{\mu}\alpha\beta}\frac{\de x^{\alpha}}{\de \tau}\frac{\de x^{\beta}}{\de \tau}
   -v^{i}\frac{\de^{2} t}{\de \tau^{2}}
   \right)\,,
\eeq
together with
\beq
\frac{\de^{2} t}{\de \tau^{2}}=-\Gamma^{0}_{\phantom{\mu}\alpha\beta}\frac{\de x^{\alpha}}{\de \tau}\frac{\de x^{\beta}}{\de \tau}\,.
\eeq

Employing the chain rule we can use
$\frac{\de x^{\alpha}}{\de \tau}=\frac{\de x^{\alpha}}{\de t} \frac{\de t}{\de \tau}$
so that the l.h.s. of \eqref{eq:geodeqextfcs} can be expressed recast in $3+1$ representation with the time $t$ as independent variable:
\beq
\frac{\de^{2} x^{i}}{\de t^{2}}=
   \left(
   -\Gamma^{i}_{\phantom{i}00}-2 \Gamma^{i}_{\phantom{i}0j}v^{j}-\Gamma^{i}_{\phantom{i}jk}v^{j}v^{k}
   +v^{i}\left(
     \Gamma^{0}_{\phantom{0}00}+2\Gamma^{0}_{\phantom{i}0j}v^{j}+\Gamma^{0}_{\phantom{i}jk}v^{j}v^{k}
   \right)
   \right)\,,
\eeq

Of course, this is not the case if local forces act on the couple of particles. In this case,
the tidal terms on r.h.s. of the acceleration equation pick up corrections of the
following form
\beq
\frac{e}{m} \left(F^{i}_{\phantom{i}\sigma}v^{\sigma}-F^{0}_{\phantom{0}j}v^{j}v^{i}\right)\frac{\de \tau}{\de t}
+\frac{1}{m}\left(f^{i}-f^{0}v^{i}\right)\left(\frac{\de \tau}{\de t}\right)^{2}\,.
\eeq
If the velocities are small ($|v^{i}| \ll 1$), we get then
\beq
\frac{\de^{2} t}{\de \tau^{2}}\simeq 0 \quad \Rightarrow \quad \frac{\de t}{\de \tau}\simeq \text{const.}\,
\eeq
but if the velocities are small it is also geometrically true that
\beq
\de \tau^{2}=g_{\mu\nu}\de x^{\mu} \de x^{\nu} \simeq g_{00}\de t^{2}\,,
\eeq
hence
\beq
\de t \simeq \frac{\de \tau}{\sqrt{g_{00}}} = \frac{\de \tau}{\sqrt{\eta_{00}+h_{00}}}\simeq\de\tau
\eeq
and finally the true deviations at small velocities are given by
\beq
\frac{e}{m} \left(F^{i}_{\phantom{i}0}+F^{i}_{\phantom{i}k}v^{k}\right)
+\frac{1}{m}\left(f^{i}-f^{0}v^{i}\right)\,.
\eeq

At low velocities then, with no assumptions whatsoever on the form of the metric tensor, we
are brought to the expression of local accelerations in real time (notice we drop the terms quadratic in the velocities):
\begin{shadefundtheory}
\beq
\label{eq:geodesylowspeed}
\frac{\de^{2} x^{i}}{\de t^{2}}=
   \left(
   -\Gamma^{i}_{\phantom{i}00}-2 \Gamma^{i}_{\phantom{i}0j}v^{j}
   +\Gamma^{0}_{\phantom{0}00}v^{i}
   \right)+
e \left(F^{i}_{\phantom{i}0}+F^{i}_{\phantom{i}k}v^{k}\right)
+\left(f^{i}-f^{0}v^{i}\right)+O(v^{2})\,.
\eeq
\end{shadefundtheory}

\section{Congruence of geodesics. Geodesics deviation}

If the curves form a set, they can be parametrised or indexed by a continuous parameter $\sigma$,
hence a second vector acquires meaning, describing the motion orthogonal to the geodesics:
\beq
W^{\mu}=\frac{\de x^{\mu}}{\de \sigma}\,.
\eeq
We proceed now to calculate the Lie derivative of $\vc W$ along $\vc V$:
\begin{align}
{\mathcal L}_{\vc V}W^{\mu}&=V^{\alpha}W^{\mu}_{\phantom{\mu},\alpha}-W^{\alpha}V^{\mu}_{\phantom{\mu},\alpha}=\\
&=\frac{\de x^{\alpha}}{\de \tau}\partial_{\alpha}\frac{\de x^{\mu}}{\de \sigma}
-\frac{\de x^{\alpha}}{\de \sigma}\partial_{\alpha}\frac{\de x^{\mu}}{\de \tau}=\\
&=\frac{\de^{2} x^{\mu}}{\de \tau \de \sigma}-\frac{\de^{2} x^{\mu}}{\de\sigma \de\tau}=0
\end{align}
on the other hand
\begin{align}
{\mathcal L}_{\vc V}W^{\mu}&=V^{\alpha}W^{\mu}_{\phantom{\mu},\alpha}-W^{\alpha}V^{\mu}_{\phantom{\mu},\alpha}
+\Gamma^{\mu}_{\phantom{\mu}\alpha\beta}\left(V^{\alpha}W^{\beta}-V^{\beta}W^{\alpha}\right)=\\
&=V^{\alpha}W^{\mu}_{\phantom{\mu};\alpha}-W^{\alpha}V^{\mu}_{\phantom{\mu};\alpha}=\\
&=\nabla_{\vc V}W^{\mu}-\nabla_{\vc W} V^{\mu}\,,
\end{align}
since $\Gamma^{\mu}_{\phantom{\mu}\alpha\beta}=\Gamma^{\mu}_{\phantom{\mu}\beta\alpha}$.

We showed that
\beq
\nabla_{\vc V}W^{\mu}=\nabla_{\vc W} V^{\mu}
\eeq
and then
\beq
\nabla_{\vc V}\nabla_{\vc W} V^{\mu}=\nabla_{\vc V}\nabla_{\vc V}W^{\mu}\,,
\label{eq:doublegeotau}
\eeq
therefore, the latter expression gives the relative, co-variant acceleration of the separation vector between geodesics.

We now wish to calculate:
\beq
\nabla_{\vc [V}\nabla_{\vc W]} V^{\mu}=\nabla_{\vc V}\nabla_{\vc W} V^{\mu}-\nabla_{\vc W}\nabla_{\vc V} V^{\mu}\,
\eeq
by using $\nabla_{\vc V}V^{\mu}=0$ and $\nabla_{\vc V}\nabla_{\vc W} V^{\mu}=\nabla_{\vc V}\nabla_{\vc V} W^{\mu}$
and relate it to the former acceleration.

We have\footnote{%
We can employ:
\beq
\begin{split}
V^{\mu}_{\phantom{\mu};\alpha;\beta}
  =&V^{\mu}_{\phantom{\mu};\alpha,\beta}
  +\Gamma^{\mu}_{\phantom{\mu}\gamma\beta}V^{\gamma}_{\phantom{\gamma};\alpha}
  +\Gamma^{\sigma}_{\phantom{\sigma}\alpha\beta}V^{\mu}_{\phantom{\mu};\sigma} =\\
  =&\left(V^{\mu}_{\phantom{\mu},\alpha}+\Gamma^{\mu}_{\phantom{\mu}\gamma\alpha}V^{\gamma}\right)_{,\beta}
  +\Gamma^{\mu}_{\phantom{\mu}\gamma\beta}\left(V^{\gamma}_{\phantom{\gamma},\alpha}+\Gamma^{\gamma}_{\phantom{\gamma}\sigma\alpha}V^{\sigma}\right)
  +\Gamma^{\sigma}_{\phantom{\sigma}\alpha\beta}\left(V^{\mu}_{\phantom{\mu},\sigma}+\Gamma^{\mu}_{\phantom{\mu}\eta\sigma}V^{\eta}\right) =\\
  =&V^{\mu}_{\phantom{\mu},\alpha,\beta}+\Gamma^{\mu}_{\phantom{\mu}\gamma\alpha,\beta}V^{\gamma}+\Gamma^{\mu}_{\phantom{\mu}\gamma\alpha}V^{\gamma}_{\phantom{\gamma},\beta}
  +\Gamma^{\mu}_{\phantom{\mu}\gamma\beta}V^{\gamma}_{\phantom{\gamma},\alpha}\\
  &+\Gamma^{\mu}_{\phantom{\mu}\gamma\beta}\Gamma^{\gamma}_{\phantom{\gamma}\sigma\alpha}V^{\sigma}
  +\Gamma^{\sigma}_{\phantom{\sigma}\alpha\beta}V^{\mu}_{\phantom{\mu},\sigma} +\Gamma^{\sigma}_{\phantom{\sigma}\alpha\beta}\Gamma^{\mu}_{\phantom{\mu}\eta\sigma}V^{\eta}\,,\\
\end{split}
\eeq
from which
\beq
V^{\mu}_{\phantom{\mu};\alpha;\beta}=V^{\mu}_{\phantom{\mu};\beta;\alpha}+R^{\mu}_{\phantom{\mu}\nu\alpha\beta}V^{\nu}\,,
\eeq
where as usual the Riemann tensor is defined as
\beq
R^{\mu}_{\phantom{\mu}\nu\beta\alpha}=\Gamma^{\mu}_{\phantom{\mu}\nu[\alpha,\beta]}+\Gamma^{\mu}_{\phantom{\mu}\sigma[\alpha}\Gamma^{\sigma}_{\phantom{\sigma}\beta]\nu}.
\eeq
}:
\begin{align}
&\nabla_{[\vc V}\nabla_{\vc W]} V^{\mu} =\\
=& \nabla_{\vc V}\nabla_{\vc W} V^{\mu} =\\
=& V^{\alpha}\left(W^{\beta}V^{\mu}_{\phantom{\mu};\beta}\right)_{;\alpha} = V^{\alpha}W^{\beta}_{\phantom{\beta};\alpha}V^{\mu}_{\phantom{\mu};\beta}+V^{\alpha}W^{\beta}V^{\mu}_{\phantom{\mu};\beta;\alpha} =\\
=& W^{\alpha}V^{\beta}_{\phantom{\beta};\alpha}V^{\mu}_{\phantom{\mu};\beta}+V^{\alpha}W^{\beta}\left(V^{\mu}_{\phantom{\mu};\alpha;\beta}+R^{\mu}_{\phantom{\mu}\nu\alpha\beta}V^{\nu}\right) =\\
=& W^{\alpha}V^{\beta}_{\phantom{\beta};\alpha}V^{\mu}_{\phantom{\mu};\beta}+W^{\beta}\left(V^{\alpha}V^{\mu}_{\phantom{\mu};\alpha}\right)_{;\beta} - W^{\beta}V^{\alpha}_{\phantom{\mu};\beta} V^{\mu}_{\phantom{\mu};\alpha} +W^{\beta}V^{\alpha}R^{\mu}_{\phantom{\mu}\nu\alpha\beta}V^{\nu} =\\
=& R^{\mu}_{\phantom{\mu}\nu\beta\alpha}V^{\nu}V^{\beta}W^{\alpha}\,.
\end{align}

Finally:
\begin{shadefundtheory}
\beq
\nabla_{\vc V}\nabla_{\vc V} W^{\mu}=R^{\mu}_{\phantom{\mu}\nu\beta\alpha}V^{\nu}V^{\beta}W^{\alpha}\,.
\label{eq:geodevcovar}
\eeq
\end{shadefundtheory}

\section{Further developments. Geodesics deviation equation at low speed}

By means of eq. \eqref{eq:doublegeotau} and the definition of geodesic the latter result can be written also
in the following form \cite{weinberggrav, Leclerc:2006vm}:
\beq
\ddot W^{\mu}=-\Gamma^{\mu}_{\phantom{\mu}\beta\alpha}\dot W^{\alpha}\dot W^{\beta}+R^{\mu}_{\phantom{\mu}\nu\beta\alpha}V^{\nu}V^{\beta}W^{\alpha}\,,
\eeq
where now $\dot W^{\mu}\equiv \de W^{\mu}/\de\tau$. After some simplifications, we get:
\beq
\ddot W^{\mu}=-\Gamma^{\mu}_{\phantom{\mu}\beta\alpha,\gamma}V^{\beta}V^{\alpha} W^{\gamma}
-2 \Gamma^{\mu}_{\phantom{\mu}\beta\alpha}V^{\beta}\dot W^{\alpha}\,.
\label{eq:geodevextrasimple}
\eeq
Notice there's no contradiction between the different forms, since we can always choose the origin of
connection so to make it vanish in the neighbourhood of the origin itself. Hence the second term of
\eqref{eq:geodevextrasimple} would vanish, and the Riemann tensor would simplify til getting the
simple form:
\beq
R^{\mu}_{\phantom{\mu}\nu\beta\alpha}=\Gamma^{\mu}_{\phantom{\mu}\nu\alpha,\beta}-\Gamma^{\mu}_{\phantom{\mu}\nu\beta,\alpha}\,,
\eeq
and the first term of the latter may always be made $0$ along the whole geodesic in the first place but
even more simply just proper-time to proper-time.

Taking now \eqref{eq:geodevextrasimple} and evaluating it in TT-gauge, i.e. with the choices:
\begin{align}
\Gamma^{i}_{\phantom{i}00}=\Gamma^{0}_{\phantom{0}00}=\Gamma^{0}_{\phantom{0}0j}&=0\,,\\
\Gamma^{0}_{\phantom{0}jk}=\frac{1}{2}\eta^{0\nu}\left(h_{\nu j,k}+h_{k\nu,j}-h_{j k,\nu}\right)&=-\frac{1}{2}h_{jk,0}\,.\\
\Gamma^{i}_{\phantom{i}0j}=\frac{1}{2}\eta^{i\nu}\left(h_{\nu 0,j}+h_{j\nu,0}-h_{0 j,\nu}\right)&=\frac{1}{2}h_{j\phantom{i},0}^{\phantom{j}i}\,,
\end{align}
we are left with:
\beq
\begin{split}
\ddot W^{0}&=-\Gamma^{0}_{\phantom{0}ij,\gamma}V^{i}V^{j} W^{\gamma}
-2 \Gamma^{0}_{\phantom{0}ij}V^{i}\dot W^{j}\,,\\
\ddot W^{i}&=-\Gamma^{i}_{\phantom{i}0j,\gamma}V^{0}V^{j} W^{\gamma}
-2 \Gamma^{i}_{\phantom{i}0j}V^{0}\dot W^{j}\,.
\end{split}
\eeq
Assume that now initially $V^{\alpha}=\delta^{\alpha}_{\phantom{\alpha}0}+O(h)$, so that $V^{i}=0\, i=1..3$,
we'd get then:
\beq
\begin{split}
\ddot W^{0}&\simeq 0\,,\\
\ddot W^{i}&\simeq -2 \Gamma^{i}_{\phantom{i}0j}\dot W^{j}\,,
\end{split}
\eeq
and, since we can trade $\tau$ for $t$, we finally get:
\begin{shadefundtheory}
\beq
\frac{\de^{2} W^{i}}{\de t^{2}}\simeq -\dot h_{j}^{\phantom{j}i}\frac{\de W^{j}}{\de t}\,,
\label{eq:geohere}
\eeq
\end{shadefundtheory}
\noindent to order $h$. This result is in contrast to what all standard textbooks claim. In fact, the
usual result:
\begin{shadefundtheory}
\beq
\frac{\de^{2} W^{i}}{\de t^{2}}\simeq \frac{1}{2}\ddot h_{j}^{\phantom{j}i} W^{j}\,,
\label{eq:geobooks}
\eeq
\end{shadefundtheory}
\noindent is derived from \eqref{eq:geodevcovar} under the assumption that
$\Gamma^{\mu}_{\phantom{\mu}\beta\alpha,\gamma}V^{\gamma}=
  \dot \Gamma^{\mu}_{\phantom{\mu}\beta\alpha}=0$, so that $\nabla_{\vc V}\nabla_{\vc V} W^{\mu}=\ddot W^{\mu}$,
but with this very same argument then we'd get $\ddot W^{i}=0$ rather then
\eqref{eq:geobooks}. In practise, \eqref{eq:geobooks} is apparently inconsistent with the Transverse-Traceless
gauge condition \cite{Leclerc:2006vm}. Notice smoke clears and the two equations convey no contradiction if one
points out that \eqref{eq:geobooks} is derived from the gauge-invariant expression
\eqref{eq:geodevcovar} in the proper reference frame of the detector, while \eqref{eq:geohere}
holds in the Lorentz frame of the waves themselves: therefore we have to rely in \eqref{eq:geobooks}
due to its gauge-invariant nature in our reference frame to detect effects of GWs, but and observer sitting on the GWs
would conclude by virtue of \eqref{eq:geohere} that no effect is produced by the waves.

%% file: chapters/conclusions.tex
\chapter{Conclusions}

\lettrine[lines=4]{S}{tarting} this work we listed some problems and issues which needed some clarification on the LTP and LISA
projects. Up to a certain level we reduced the extension of some of these and others were - we think - completely
solved:
\begin{enumerate}
\item the problem of a construction of a TT-gauged frame was shown to be solvable by an ensemble of
tetrads co-moving with the laser beam. Thus, a one-to-one map is built between the perturbed metric
change in space-time and the laser phase;
\item the problem of building signals out of a complicated dynamics and selecting a MBW to guarantee correct estimation
of a differential acceleration observable was addressed;
\item gravitational static unbalance was analysed and a global numerical strategy based on local multipole expansion elucidated;
\item calibration of force-to-displacement actuation was successfully addressed and the method of Wiener-Kolmogorov
proven to be a correct strategy for the task;
\item the problem of charge collection and estimation on the TM surface was tackled
in many ways;
\item cross-talk was reduced to its main causes, the effect carried onto the relevant signals and estimated throughout the whole
MBW on the basis of up-to-date specifications.
\end{enumerate}

A number of open issues emanate from the former successes:
\begin{enumerate}
\item further analyses of static gravity with modified ISL dynamics shall be carried on to test
LTP ability to perform an ISL test;
\item all the remaining operation procedures described in chapter \ref{chap:experim} which were not
deeply inspected (as the charge measurement or the calibration of force to displacement) shall be
analysed carefully.
\item LTP may very well fail in some of the more advanced tasks we listed in chapter \ref{chap:experim}, but the
simple residual acceleration noise picture across some integration time can be guaranteed a result provided
no problem in releasing the TMs will occur and laser metrology won't be damaged. Adherence problems are
now being addressed\footnote{Bortoluzzi, D. et al, private communication} with new experimental
apparata and more tests will be performed on the caging mechanism.
\end{enumerate}

We hope we convinced the reader of the reliability of LTP, both from the point of view of the
theory and as an experimental device. If points were left pending and may seem obscure, we may take the
blame of it as little scientific communicators, perhaps. The time needed by two PhD's courses in sequence
won't be enough to enter all the intricacies of the experiment - at least from our standing position
as junior scientists. If we were allowed to abuse a statement, that will be Kant's ``I can, therefore I must'';
we contributed to showing that LTP will work, this measure can be done, hence we must do it.

We wish
long life to LTP, to serve its purpose for LISA, and we really hope it will bring unexpected news
from space.

%% file: chapters/colophon.tex
\chapter{Colophon}

[Omissis]

%% file: chapters/homage.tex
\chapter{Homage}

\begin{center}
\includegraphics[width=0.7\textwidth]{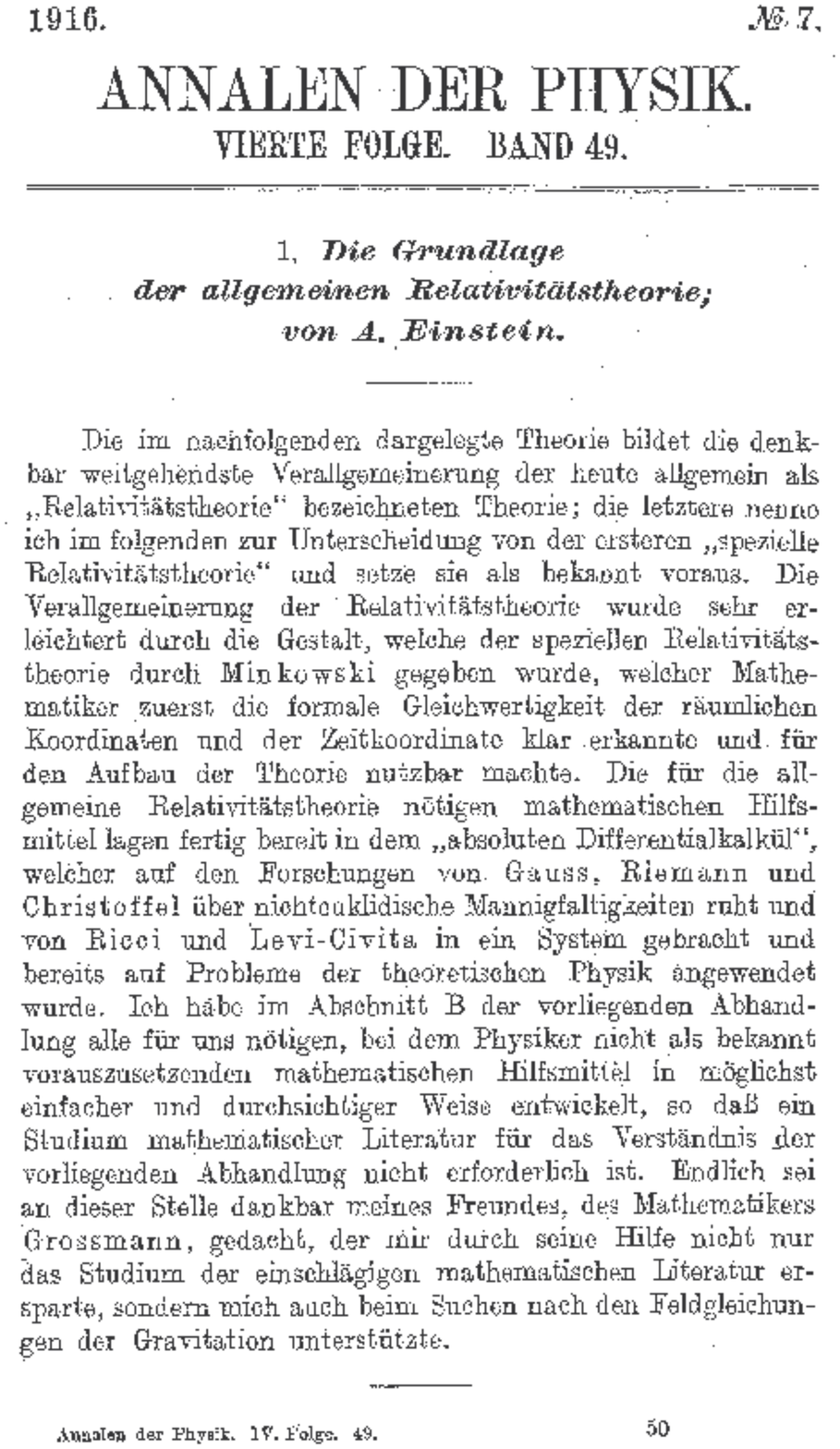}
\end{center}